\documentclass[b5paper,10pt]{book}
\usepackage[top=3 cm, bottom=3 cm, left=2 cm, right=2 cm]{geometry}

\usepackage{pdfpages}
\usepackage{comment}
\usepackage{graphics}
\usepackage{graphicx}
\usepackage{float}
\usepackage{slashed}
\usepackage{amsmath,amssymb}
\numberwithin{equation}{section}
\usepackage{subfigure}
\usepackage{color}
\usepackage{bbm}
\usepackage{bm}
\usepackage[Lenny]{fncychap}
\usepackage{epstopdf}
\usepackage{afterpage}
\usepackage{color}
\usepackage{array}
\usepackage{braket}
\usepackage{enumerate}
\usepackage[final]{feynmp}
\usepackage{hyperref}

\makeatletter
\@addtoreset{footnote}{chapter}
\makeatother

\begin{document}

\title{Collective properties of quantum matter: from Hawking radiation analogues to quantum Hall effect in graphene}

\author{Juan Ram\'on Mu\~noz de Nova}

\date{\today}

\begin{center}
\begin{figure}[t]
\centering
\includegraphics[width=0.5\columnwidth]{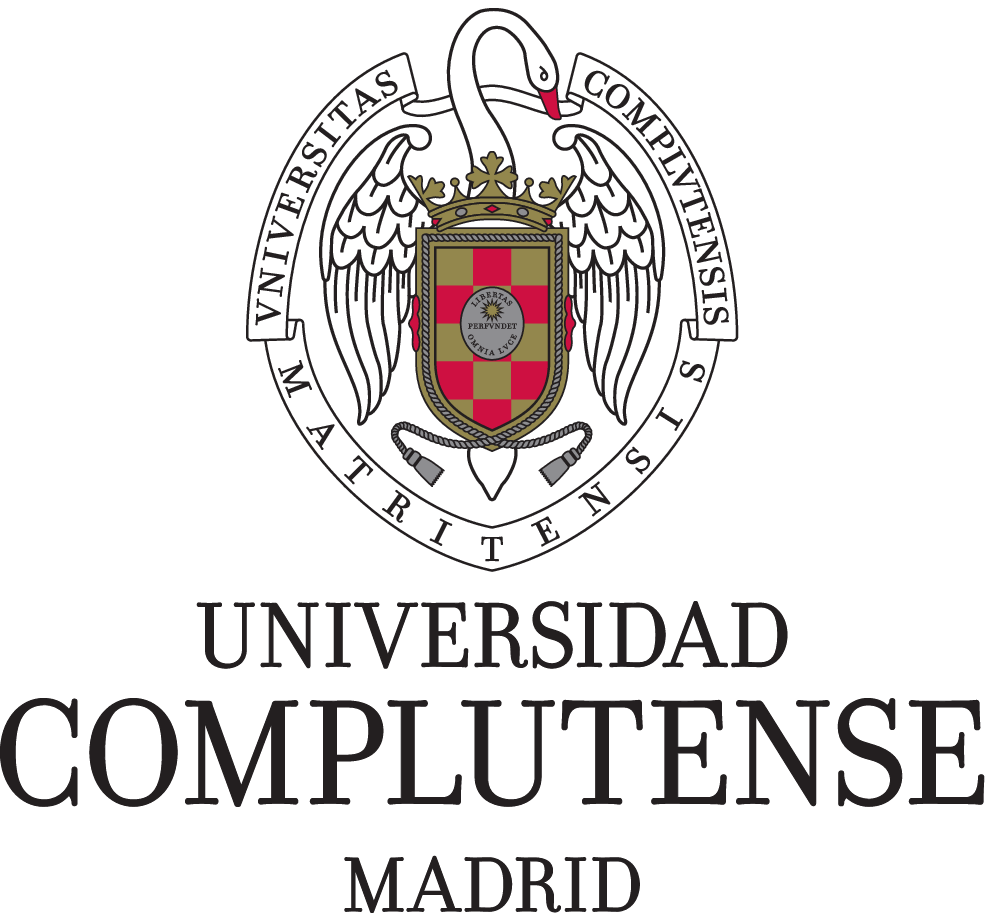}
\end{figure}
{\Large Departamento de F\'isica de Materiales}
~\\[0.5cm]
{\Huge \textbf{Collective properties of quantum matter: from Hawking radiation analogues to quantum Hall effect in graphene}}\\[0.7cm]
{\Huge \textbf{Propiedades colectivas de la materia cu\'antica: desde an\'alogos de radiaci\'on de Hawking hasta efecto Hall cu\'antico en grafeno}}\\[0.5cm]
{\Large {\Large\textbf{Juan Ram\'on Mu\~noz de Nova}} \\ \vspace{0.2cm} Tesis supervisada por Ivar Zapata Olson-Lunde y Fernando Sols Lucia} \\

\end{center}

\pagenumbering{Roman}

\tableofcontents

\frontmatter

\chapter{Resumen en espa\~nol}

\pagenumbering{arabic}
\setcounter{page}{7}

El contenido de la tesis aqu\'i expuesta posee un car\'acter verdaderamente interdisciplinar, ya que se estudian sistemas tan variados como an\'alogos gravitacionales en condensados o efecto Hall cu\'antico en grafeno. De hecho, en esta tesis tratamos sistemas en 1, 2 y 3 dimensiones. En todos los casos, los objetos de estudio son fen\'omenos colectivos de la materia cu\'antica, lo que motiva el t\'itulo de esta tesis.

La tesis est\'a dividida en tres partes. La parte principal de la misma corresponde al estudio de an\'alogos de radiaci\'on de Hawking en condensados de Bose-Einstein, con vistas a una eventual propuesta de realizaci\'on experimental. Los an\'alogos gravitacionales est\'an recibiendo gradualmente una mayor atenci\'on en los \'ultimos a\~nos desde el punto de vista experimental y te\'orico al ser la \'unica manera de estudiar en el laboratorio determinadas predicciones te\'oricas. Nosotros nos centramos en este trabajo en el escenario concreto de an\'alogos en condensados at\'omicos, pero existen tambi\'en an\'alogos en gases de Fermi, trampas de iones o polaritones, por ejemplo. Es por ello un campo tremendamente interdisciplinar ya que aparecen conceptos que involucran desde Relatividad General hasta \'optica cu\'antica pasando por interacciones entre \'atomos fr\'ios, transporte cu\'antico...

De este modo, el primer cap\'itulo est\'a dedicado a dar una introducci\'on general sobre radiaci\'on de Hawking an\'aloga en condensados de Bose-Einstein,  proporcionando los conceptos te\'oricos b\'asicos sobre los que se sustenta el resto del trabajo.

El segundo cap\'itulo est\'a dedicado a un arduo trabajo computacional que representa una de las principales contribuciones de esta tesis. En \'el, se ha estudiado mediante simulaci\'on num\'erica la formaci\'on del an\'alogo de un agujero negro en un condensado de Bose-Einstein dentro de un escenario experimental realista.  Concretamente, partiendo de un condensado inicialmente confinado por una red \'optica, es posible obtener una configuraci\'on de agujero negro cuasi-estacionaria al descender lo suficiente la amplitud de la red \'optica como para permitir el escape de una peque\~na corriente de \'atomos a trav\'es de la misma. Este trabajo est\'a motivado, por un lado, por la ausencia de modelos te\'oricos realistas para an\'alogos de agujeros negros, ya que casi todos consideran escenarios semi-infinitos y estacionarios y no dan cuenta del transitorio hacia dichos reg\'imenes.  Por otro lado, tambi\'en propone un nuevo escenario experimental que complementa las actuales propuestas. Uniendo los resultados num\'ericos para la funci\'on de onda macrosc\'opica a los resultados preliminares para el espectro de radiaci\'on de Hawking, llegamos a la conclusi\'on de que el escenario propuesto es un buen candidato para observar la radiaci\'on de Hawking. Aparte de la realizaci\'on de an\'alogos gravitacionales, el escenario obtenido puede ser de inter\'es en el campo de transporte cu\'antico pues la configuraci\'on cuasi-estacionaria de agujero negro alcanzada es capaz de proporcionar  una corriente supers\'onica con una velocidad muy bien definida. El trabajo de este cap\'itulo ha sido fruto de una colaboraci\'on con D. Gu\'ery-Odelin, al que agradezco la colaboraci\'on realizada. Tambi\'en doy las gracias a R. Parentani y a I. Carusotto por valiosas observaciones as\'i como a A. Malyshev por darnos la referencia de la descomposici\'on QR utilizada en problemas de localizaci\'on de Anderson. La mayor parte de este cap\'itulo ha sido publicada en Ref. \cite{deNova2014a}. Tambi\'en ha sido el contenido de una charla en el V Encuentro de \'Atomos Fr\'ios de Madrid.

El trabajo del tercer cap\'itulo est\'a m\'as enfocado hacia la teor\'ia, donde analizamos posibles criterios para distinguir el genuino efecto Hawking de emisi\'on espont\'anea de fonones de la se\~nal t\'ermica o las propias fluctuaciones coherentes de la funci\'on de onda macrosc\'opica. Partiendo de ideas previamente desarrolladas en el contexto de la \'optica cu\'antica, proponemos la violaci\'on de desigualdades de Cauchy-Schwarz como una se\~nal inequ\'ivoca de la presencia de radiaci\'on de Hawking espont\'anea. Esta propuesta supone otra de las principales contribuciones de la tesis. Asimismo, elaboramos una comparaci\'on con otros trabajos contempor\'aneos, en los que se utiliza la presencia de entrelazamiento cu\'antico para caracterizar el efecto Hawking; en particular, demostramos que, bajo t\'ipicas suposiciones, los dos criterios son equivalentes, unificando as\'i los trabajos presentes en la literatura. Tambi\'en analizamos la posible implementaci\'on experimental de los criterios previamente analizados, llegando a la conclusi\'on de que solamente cierto tipo de violaciones de Cauchy-Schwarz  pueden ser detectadas. Finalmente, presentamos datos num\'ericos para demostrar que los criterios propuestos podr\'ian ser v\'alidos a la hora de detectar la radiaci\'on de Hawking espont\'anea. Los resultados de este cap\'itulo son de una notable importancia de cara al dise\~no de futuros esquemas de detecci\'on en el laboratorio. Tambi\'en pueden resultar interesantes en otros campos tales como la \'optica o la informaci\'on cu\'antica, as\'i como en el campo general de gases condensados.  Para la escritura de este cap\'itulo han resultado de gran utilidad conversaciones con A. Amo, F. Michel, R. Parentani, D. Gu\'ery-Odelin y J. Steinhauer.  El trabajo de este cap\'itulo ha sido publicado en las Refs. \cite{deNova2014,deNova2015}. Tambi\'en he presentado un p\'oster en la FQMT'13 de Praga y he publicado un art\'iculo en sus Proceedings \cite{deNova2015Proc}. Durante este evento, interesantes interacciones han sido llevadas a cabo con D. Crivelli, K. Pawlowski, B. Gardas and K. Hovhannisyan. Tambi\'en agradezco a V. Spicka la magn\'ifica organizaci\'on de la conferencia.

En el cuarto cap\'itulo pasamos a otro tipo de an\'alogo gravitacional relacionado con el anterior: el conocido como efecto l\'aser de agujeros negros, consistente en la aparici\'on de inestabilidades din\'amicas en la regi\'on supers\'onica comprendida entre un par de agujeros negro y blanco. De manera similar al segundo cap\'itulo, hemos realizado simulaciones num\'ericas para observar la evoluci\'on temporal de la inestabilidad l\'aser. Aparte de confirmar las predicciones te\'oricas realizadas para el estado estacionario a tiempos largos, hemos identificado con claridad un r\'egimen en el cual el sistema emite de manera peri\'odica trenes de solitones. Comparando con los l\'aseres \'opticos normales, llegamos a la conclusi\'on de que es este \'ultimo r\'egimen de emisi\'on peri\'odica de solitones el que representa un an\'alogo m\'as fuerte con los l\'aseres \'opticos. La identificaci\'on de dicho r\'egimen de emisi\'on continua de solitones supone otro de los principales resultados de esta tesis. Los resultados de este cap\'itulo pueden ser por un lado de utilidad a la hora de entender la evaporaci\'on de agujeros negros. Por otro lado, el r\'egimen de emisi\'on continua de solitones proporciona una suerte de ``l\'aser de solitones'', lo cual no deja de ser un efecto bastante sorpredente que adem\'as podr\'ia tener potenciales aplicaciones en el campo del trasporte cu\'antico o la atomotr\'onica. El trabajo de este cap\'itulo es el resultado de una estancia de cinco meses en el BEC Center de Trento, al cual agradezco profundamente la hospitalidad y el buen trato recibido. En particular, doy las gracias a I. Carusotto y S. Finazzi por el trabajo realizado conjuntamente y a D. Papoular, T. Congy, L. A. Pe\~na, F. Ramiro y M. Abad por interesantes charlas. Una versi\'on provisional de un art\'iculo venidero ha sido colgada en el ArXiv \cite{deNova2015arxiv}. Asimismo, este trabajo ha sido expuesto en una charla en el LPT de Orsay, Par\'is. En este punto, quisiera dar las gracias a R. Parentani y F. Michel por las valiosas charlas que mantuvimos durante mi visita.

La segunda parte de la tesis encierra un solo cap\'itulo, el quinto, en el que, dentro del mismo contexto de gases de bosones, pasamos al l\'imite opuesto de un nube t\'ermica por encima de la temperatura cr\'itica, de modo que no hay condensado. Concretamente, analizamos el efecto producido por la introducci\'on de un pulso corto de Bragg. Demostramos, usando los formalismos cl\'asico y cu\'antico, que el patr\'on peri\'odico de densidad inducido decae al valor inicial de equilibrio. Sin embargo, en vez de la usual relajaci\'on colisional de los modos colectivos, el mecanismo responsable de dicho decaimiento es el desorden t\'ermico de las part\'iculas, con un tiempo t\'ipico de decaimiento que solo depende de la temperatura, el vector de onda del pulso y la masa de los \'atomos. De hecho, demostramos que dicho efecto es bastante universal. Comparando con datos experimentales, encontramos una gran concordancia con las predicciones te\'oricas. Los resultados aqu\'i presentados pueden ser aplicados de manera directa a otros sistemas tales como la nube t\'ermica de un condensado por debajo de la temperatura cr\'itica o gases de fermiones a temperaturas suficientemente altas. Cabe destacar la gran belleza matem\'atica del trabajo expuesto en este cap\'itulo, en el que aparecen involucradas, de m\'ultiples maneras, funciones tales como polinomios de Hermite, funciones de Bessel, la funci\'on error, la funci\'on sinc, la funci\'on zeta de Riemann... Este trabajo procede de una colaboraci\'on con el grupo Technion de Haifa, concretamente, con S. Rinott y J. Steinhauer, a los que me gustar\'ia agradecer la interacci\'on efectuada.

En la tercera parte de la tesis, cambiamos a un sistema totalmente distinto: electrones en grafeno bajo la presencia de un campo magn\'etico. En particular, analizamos el estado Hall cu\'antico $\nu=0$ de una bicapa de grafeno utilizando la aproximaci\'on de Hartree-Fock dependiente del tiempo. As\'i, tras rederivar el diagrama de fases de campo medio, estudiamos los modos colectivos del sistema y nos fijamos especialmente en los de menor energ\'ia, que describen las posibles inestabilidades asociadas a las transiciones de fase. Entre los resultados m\'as interesantes, cabe destacar la presencia de una simetr\'ia continua residual justo en la transici\'on entre las fases ferromagn\'etica y capa-polarizada, que podr\'ia ser un remanente de una simetr\'ia total $SO(5)$. Por otro lado, las fases ferromagn\'etica inclinada y parcialmente capa-polarizada pueden presentar inestabilidades din\'amicas. Es digno de mencionar el papel clave desempe\~nado por las interacciones de corto alcance en la aparici\'on de dichas inestabilidades. Debido a la fuerte analog\'ia con el Hamiltoniano del grafeno monocapa, podemos extender los resultados previos a dicho material de manera casi directa. Tambi\'en analizamos los efectos derivados de permitir la mezcla de niveles de Landau. Por \'ultimo, hacemos unas peque\~nas observaciones relevantes de cara a futuros escenarios experimentales. Este trabajo ha sido desarrollado durante una estancia de tres meses  en la universidad de Harvard, a la que me gustar\'ia agradecer el buen trato dispensado. Tambi\'en agradezco la colaboraci\'on con E. Demler. Asimismo, doy las gracias a M. Zvonarev y a V. Stojanovic por agradables conversaciones.

Finalmente, me gustar\'ia agradecer a I. Carusotto, J. Steinhauer y a C. Westbrook sus valiosos comentarios sobre la tesis.

\chapter{Abstract}

The content of the thesis here presented has a truly interdisciplinary character since we address system such as different as gravitational analogues in condensed matter or quantum Hall effect on graphene. Indeed, we face systems in 1,2 and 3 dimensions along the whole work. In all cases, we study collective properties of quantum matter, which motivates the title of this thesis.

The thesis is divided in three parts. The main part corresponds to the study of analog Hawking radiation in Bose-Einstein condensates, bearing in mind an eventual experimental implementation. Gravitational analogues are receiving gradually more and more attention from both theoretical and experimental points of view since they represent the only way to study some gravitational predictions. We focus in this work on the specific scenario of analogues in atomic condensates but there are also analog systems in Fermi gases, ion rings or polaritons, for instance. Because of this, it represents a very interdisciplinary field since it involves concepts from General Relativity, quantum transport, cold atoms, Quantum Optics...

In this way, the first chapter is devoted to provide a general introduction about analog Hawking radiation in Bose-Einstein condensates, giving the basic theoretical notions in which the rest of the work is based.

The second chapter shows an extensive computational work that represents one of the main contributions of this thesis. We study numerically the birth of a sonic black hole in a Bose-Einstein condensate within a realistic experimental setup. Specifically, starting from a condensate initially confined by an optical lattice, we find that it is possible to achieve a quasi-stationary black-hole configuration after lowering enough the amplitude of the optical lattice to allow the leaking of a small atom current. This work is motivated, on the one hand, by the absence of realistic theoretical models for analog black holes, since they usually consider semi-infinite media and perfectly stationary configurations and do not describe the transient regime to such scenarios. On the other hand, it proposes a new possible experimental scenario that complements the current proposals. Joining the results from the numerical simulations for the time evolution of the macroscopic wave function with the preliminary values of the Hawking spectrum, we conclude that this scenario is a promising candidate for the detection of the spontaneous Hawking effect. Apart from the obtention of gravitational analogues, this quasi-stationary black hole could be of interest for quantum transport scenarios since it is able to provide a supersonic current with well-defined velocity. This work has arisen as a collaboration with D. Gu\'ery-Odelin, to whom I am thankful for the nice interaction. I also thank I. Carusotto and R. Parentani for valuable discussions as well as A. Malyshev for referring us the QR decomposition used in Anderson localization problems. Most of the part of this chapter has been published in Ref. \cite{deNova2014a}. It has been also the subject of a talk in the V Madrid Cold Atoms Meeting.

The work of the third chapter is more focused on the theoretical side, analyzing possible criteria of detection of the Hawking effect in order to distinguish the genuine spontaneous quantum signal from the stimulated one or from the coherent fluctuations of the macroscopic wave function. Using concepts previously developed within a quantum optics context, we propose the violation of Cauchy-Schwarz type inequalities as an unambiguous signal of the presence of spontaneous Hawking radiation. This proposal is another of the main contributions of this thesis. Also, we compare this criterion with other one that uses entanglement to characterize the Hawking effect; in particular, we show that, under quite general assumptions, both criteria are equivalent. We also analyze the possible experimental detection of the previous criteria, finding that only the violation of certain type of Cauchy-Schwarz inequalities can be detected. Finally, we support the theoretical work with numerical data. The results of this chapter are of remarkable conceptual importance with a view to a future implementation of an experimental detection scheme. They can be also interesting for other fields such as quantum optics, quantum information physics or in the broader topic of bosonic condensates. Valuable discussions with A. Amo, F. Michel, R. Parentani, D. Gu\'ery-Odelin and J. Steinhauer are acknowledged. The characterization of the spontaneous Hawking radiation through the violation of Cauchy-Schwarz inequalities was published in Ref. \cite{deNova2014}. The work of Ref. \cite{deNova2015} analyzes and unifies the current criteria for the detection of the Hawking effect and presents a experimental discussion on their possible implementation. I have also presented a poster in the FQMT'13 conference of Prague and published an article for the Proceedings of the conference \cite{deNova2015Proc}. During this event, nice interactions with D. Crivelli, K. Pawlowski, B. Gardas and K. Hovhannisyan are acknowledged. I also thank V. Spicka for the great organization.

In the fourth chapter we change to another type of gravitational analogue: the so-called black-hole laser, consisting in a black hole-white hole pair that could give rise to a dynamical instability in the internal supersonic region. In the same fashion of the second chapter, we perform numerical simulations in order to study the time-evolution of the laser instability. Apart from confirming previous theoretical predictions for the long-time steady state, we have clearly identified a regime in which the system continuously emits periodic trains of solitons. By comparing with standard laser devices, we conclude that this last self-oscillating regime represents the most strong analogue with optical lasers. The identification of this regime of continuous emission of solitons is another of the main contributions of the thesis. On one hand, the results of this chapter can help to understand the evaporation of gravitational black holes. On the other hand, the regime of continuous emission of solitons provides some kind of ``soliton laser'', which represents a very interesting effect by itself and could have potential applications in the fields of quantum transport or atomtronics. The work of this chapter is the result of a five-month stay in the BEC Center of Trento, to which I would like to acknowledge the kind hospitality. I thank specially I. Carusotto and S. Finazzi for the work developed together. I also acknowledge interesting talks with D. Papoular, T. Congy, L. A. Pe\~na, F. Ramiro and M. Abad. An earlier version of a forthcoming article has been uploaded to ArXiv \cite{deNova2015arxiv}. It has also been the content of a seminary at the LPT of Orsay, Paris. At this point, I would like to acknowledge fruitful discussions with R. Parentani and F. Michel during my visit to Paris.

The second part of the thesis contains only one chapter, where, within the same context of boson gases, we switch to the opposite limit of a thermal cloud above the critical temperature, so there is no condensate. Specifically, we analyze the effect of the introduction of a short Bragg pulse. We show, using classical and quantum formalisms, that the induced periodic density pattern decays to the equilibrium profile. However, instead of the usual collisional relaxation, the mechanism responsible for the decay is the thermal disorder of the particles, with a characteristic time that only depends on the temperature, the mass of the atoms and the wave vector of the lattice potential. In fact, we show that the predicted decay appears under quite general grounds. We find a very good agreement with actual experimental data. The theoretical calculations developed can be extended to other systems such as Fermi gases at sufficiently high temperature or the thermal component of a Bose gas below the critical temperature. It is remarkable the mathematical beauty of this chapter, since the calculations involve, in different ways, special functions such as Hermite polynomials, Bessel functions, error function, sinc function, Riemann zeta function... This work comes from a collaboration with S. Rinott and J. Steinhauer from the Technion group of Haifa. I would like to thank the nice interaction with them.

In the last part of the thesis, we switch to a very different system: electrons in graphene under the presence of a magnetic field. In particular, we have studied the $\nu=0$ quantum Hall state of bilayer graphene using the time-dependent Hartree-Fock approach. So, after re-deriving the corresponding mean-field phase diagram, we compute the collective modes within the zero Landau level, paying special attention to the lowest energy ones, since they describe the energetic instabilities associated to phase transitions. Among the most remarkable results, we have found that at the boundary between the full layer-polarized and the ferromagnetic phases a gapless mode appears resulting from an accidental symmetry that can be regarded as a remanent of a broken $SO(5)$ symmetry. On the other hand, the canted anti-ferromagnetic and partially layer polarized phases can present dynamical instabilities. It is worth noting the crucial role played by the valley/sublattice short-range interactions in the appearance of such instabilities. Due to the strong analogy with the Hamiltonian of monolayer graphene, we can straightforwardly extend the previous results to that system. We also discuss the effects of allowing Landau level mixing. By last, we make some relevant observations for possible experimental scenarios. This work has been developed during a three-month stay in Harvard University, as a collaboration with E. Demler, to whom I am grateful. I would also like to thank the hospitality of the institution. Interesting conversations with V. Stojanovic and M. Zvonarev are also acknowledged.

Finally, I thank I. Carusotto, J. Steinhauer and C. Westbrook for their valuable comments about the thesis.

\mainmatter
\setcounter{page}{17}

\part{Gravitational analogs in Bose-Einstein condensates}

\chapter{Hawking radiation in Bose-Einstein condensates}\label{chapter:Introduction}

\section{Introduction}

The emission of Hawking radiation by a black hole (BH) \cite{Hawking1974} is one of the most remarkable predictions of modern Physics and its observation stills remains a major challenge. The problem is that the effective temperature of emission is so low that its cosmological detection represents an extremely difficult task. However, Unruh \cite{Unruh1976,Unruh1981} noted that an analogue of Hawking radiation (HR) could be observed in the laboratory using quantum fluids at temperatures which, while
still too low, lie within conceivable reach. In particular, an attractive feature of Bose-Einstein condensates (BEC) is that they are good candidates to provide a convenient way of investigating analog black-hole physics in the laboratory \cite{Garay2000}. There are also analogues in Fermi gases \cite{Giovanazzi2005}, ion rings \cite{Horstmann2010} or polaritons \cite{Nguyen2015}.

For a condensate traversing a subsonic-supersonic interface, which is the acoustic analog of an event horizon, the spontaneous correlated emission of phonons into the subsonic and supersonic regions has been predicted \cite{Leonhardt2003,Leonhardt2003a,Balbinot2008,Carusotto2008,Macher2009,Coutant2010}. Sonic event horizons have been experimentally produced by accelerating a Bose-Einstein condensate \cite{Lahav2010}. In a similar setup, the self-amplifying stimulated HR resulting from the black-hole laser effect \cite{Finazzi2010} has been observed in a Bose-Einstein condensate \cite{Steinhauer2014}. However, the emission of spontaneous Hawking radiation still remains undetected. An alternative route to produce analog black-hole scenarios is the quasi-stationary leaking of a large condensate so that the outgoing flux is sufficiently dilute to reach the supersonic regime \cite{Zapata2011,Larre2012}.

In this chapter, we present a review of analog Hawking radiation in Bose-Einstein condensates. We first introduce the physical model that we will use through this work in Sec. \ref{sec:physmodel}. In particular, we revise the physics of condensates near $T=0$ in Sec. \ref{subsec:GPBdG} and explain the achievement of an effective one-dimensional (1D) regime in Sec. \ref{subsec:1Dmeanfield}. After that, we study sonic event horizons in 1D stationary flows in Sec. \ref{subsec:BHBEC}, giving rise to the analog Hawking effect, described in Sec. \ref{subsec:hawkingeffect}. At the end of the chapter, we resume in Sec. \ref{sec:typicalbh} the main theoretical models considered for studying analog black holes in BEC. We present in Appendices \ref{app:1DGP}-\ref{app:SMatrixBehav} the technical details about the solutions of the GP and BdG equations and the computation of the scattering matrix.

\section{Physical model} \label{sec:physmodel}

\subsection{Gross-Pitaevskii and Bogoliubov-de Gennes equations}\label{subsec:GPBdG}

\subsubsection{Time-independent situation}\label{subsec:timeindependent}

The starting point is the usual second quantization Hamiltonian for bosons \cite{Fetter2003,Dickhoff2005}:
\begin{equation}\label{eq:2ndHamiltonian}
\hat{H}=\int\mathrm{d}^3\mathbf{x}~\hat{\Psi}^{\dagger}(\mathbf{x})\left(-\frac{\hbar^2}{2m}\nabla^2+V_{\rm{ext}}(\mathbf{x})\right)\hat{\Psi}(\mathbf{x})\\
+\frac{g_{\rm{3D}}}{2}\hat{\Psi}^{\dagger}(\mathbf{x})\hat{\Psi}^{\dagger}(\mathbf{x})\hat{\Psi}(\mathbf{x})\hat{\Psi}(\mathbf{x})
\end{equation}
wh $V_{\rm{ext}}$ is an external time-independent potential and we have replaced the actual interacting potential by a contact pseudo-potential of the form $V_{\rm{int}}(\mathbf{x})=g_{\rm{3D}}\delta(\mathbf{x})$, with $g_{\rm{3D}}=4\pi\hbar^2a_s/m$ and $a_s$ the $s$-wave scattering length \cite{Pitaevskii2003,Pethick2008}. The time evolution of the field operator in the Heisenberg picture, $\hat{\Psi}(\mathbf{x},t)$, is given by:
\begin{equation}\label{eq:fieldoperatorheisenberg}
i\hbar\partial_t\hat{\Psi}(\mathbf{x},t)=[\hat{\Psi}(\mathbf{x},t),\hat{H}(t)]=\left(-\frac{\hbar^2}{2m}\nabla^2+V_{\rm{ext}}(\mathbf{x})\right)\hat{\Psi}(\mathbf{x},t)+g_{\rm{3D}}\hat{\Psi}^{\dagger}(\mathbf{x},t)\hat{\Psi}(\mathbf{x},t)\hat{\Psi}(\mathbf{x},t)
\end{equation}
where we have used the canonical commutation rules $[\hat{\Psi}(\mathbf{x}),\hat{\Psi}^{\dagger}(\mathbf{x}')]=\delta(\mathbf{x}-\mathbf{x}')$. We now restrict ourselves to the specific case of a Bose-Einstein condensate near $T=0$. Making a mean-field approximation:
\begin{equation}\label{eq:meanfieldGP}
\hat{\Psi}(\mathbf{x},t)=[\Psi_0(\mathbf{x})+\hat{\varphi}(\mathbf{x},t)]e^{-i\frac{\mu}{\hbar}t}~,
\end{equation}
with $\Psi_0$ the so-called Gross-Pitaevskii (GP) wave-function \cite{Pitaevskii2003}, we arrive at:
\begin{eqnarray}\label{eq:TIGP}
\nonumber H_{GP}\Psi_0(\mathbf{x})&=&\mu\Psi_0(\mathbf{x})\\
H_{GP}&=&-\frac{\hbar^2}{2m}\nabla^2+V_{\rm{ext}}(\mathbf{x})+g_{\rm{3D}}|\Psi_0(\mathbf{x})|^2
\end{eqnarray}
and, to first order in the quantum fluctuations, $\hat{\varphi}$,

\begin{eqnarray}\label{eq:Fieldequation}
i\hbar\partial_t\hat{\varphi}&=&G\hat{\varphi}+L\hat{\varphi}^{\dagger},\\
\nonumber G&=&H_{GP}+g_{\rm{3D}}|\Psi_0(\mathbf{x})|^2-\mu,~L=g_{\rm{3D}}\Psi^2_0(\mathbf{x})
\end{eqnarray}

Equation (\ref{eq:TIGP}) is the time-independent GP equation. It describes the stationary condensate wave function, normalized to the total number of particles $N$
\begin{equation}\label{eq:normalization}
N=\int\mathrm{d}^3\mathbf{x}~|\Psi_0(\mathbf{x})|^{2}
\end{equation}
since we neglect the population of the depletion cloud as we are close to $T=0$. Equation (\ref{eq:Fieldequation}) governs the time evolution of the quantum fluctuations above the stationary condensate and it can be rewritten as:
\begin{eqnarray}\label{eq:BdGfieldequation}
\nonumber i\hbar\partial_t\hat{\Phi}&=&M\hat{\Phi},\\
\hat{\Phi}=\left[\begin{array}{c}\hat{\varphi}\\ \hat{\varphi}^{\dagger}\end{array}\right]&,&M=\left[\begin{array}{cc}G & L\\
-L^{*}&-G\end{array}\right]
\end{eqnarray}
This equation is known as the Bogoliubov-de Gennes equation (BdG). First we consider Eq. (\ref{eq:BdGfieldequation}) in a classical context, which amounts to removing all the ``hats". The resulting equation is also useful to describe the linearized collective motion of small perturbations around the stationary GP wave function $\delta\Psi_0(\mathbf{x},t)$, $\Psi(\mathbf{x},t)=[\Psi_0(\mathbf{x})+\delta\Psi_0(\mathbf{x},t)]e^{-i\frac{\mu}{\hbar}t}$. As a linear equation, it admits an expansion in eigenmodes:
\begin{eqnarray}\label{eq:BdGmodesexpansion}
\nonumber\Phi(\mathbf{x},t)&=&\sum_n \gamma_nz_n(\mathbf{x})e^{-i\omega_nt}+\gamma^{*}_n\bar{z}_n(\mathbf{x})e^{i\omega_nt}\\
z_n(\mathbf{x})&=&\left[\begin{array}{c}u_n(\mathbf{x})\\ v_n(\mathbf{x})\end{array}\right],\bar{z}_n(\mathbf{x})=\left[\begin{array}{c}v^*_n(\mathbf{x})\\ u^*_n(\mathbf{x})\end{array}\right]
\end{eqnarray}
where the spinor $z_n$ satisfies the time-independent BdG equation:

\begin{equation}\label{eq:BdGequation}
Mz_n=\epsilon_nz_n, \, \epsilon_n=\hbar\omega_n
\end{equation}

This equation has the property that, if $z_n$ is a mode with eigenvalue $\epsilon_n$, the conjugate $\bar{z}_n$ is also a mode with eigenvalue $-\epsilon^*_n$.
Equation (\ref{eq:BdGequation}) is clearly non-Hermitian and thus can yields complex eigenvalues (see for instance Chapter \ref{chapter:BHL}). Modes with positive imaginary part of the frequency are dynamical instabilities since they represent small perturbations that grow exponentially in time. Another interesting property is that the BdG equations always have a mode with zero energy:

\begin{equation}\label{eq:zeromode}
z_0=\left[\begin{array}{c}\Psi_0(\mathbf{x})\\ -\Psi^*_0(\mathbf{x})\end{array}\right]
\end{equation}
which corresponds to a global phase change in the GP wave function $\Psi_0(\mathbf{x})\rightarrow e^{i\delta}\Psi_0(\mathbf{x})$.

The BdG equation has an associated Klein-Gordon type scalar product given by:
\begin{equation}\label{eq:KGProduct}
(z_1|z_2)\equiv\braket{z_1|\sigma_z|z_2}=\int\mathrm{d}^3\mathbf{x}~z_1^{\dagger}(\mathbf{x})\sigma_zz_2(\mathbf{x})=\int\mathrm{d}^3\mathbf{x}~u^*_1(\mathbf{x})u_2(\mathbf{x})-v^*_1(\mathbf{x})v_2(\mathbf{x}) \, ,
\end{equation}
with $\sigma_z=\rm{diag}(1,-1)$ the corresponding Pauli matrix. It is straightforward to prove the property:
\begin{equation}\label{eq:orthogonal}
(\epsilon_n-\epsilon^*_m)(z_m|z_n)=0
\end{equation}
from which it follows that two modes with different eigenvalues are orthogonal according to this scalar product. Another important property is that the previous scalar product is not positive definite so the norm of a given solution $z$, defined as $(z|z)$, can be positive, negative or zero. In particular, the norm of the conjugate $\bar{z}$ has the opposite sign to that of $z$, $(\bar{z}|\bar{z})=-(z|z)$. We immediately see from Eq. (\ref{eq:orthogonal}) that any eigenmode with complex eigenvalue has zero norm. Also note that the zero mode of Eq. (\ref{eq:zeromode}) has zero norm.

We now go back to the quantum version of the BdG equations. We promote the amplitude of each mode to a quantum operator:
\begin{equation}\label{eq:destructoroperators}
\hat{\gamma}_n=(z_n|\hat{\Phi}) \, ,
\end{equation}
which implies, using  the canonical commutation rules, that $[\hat{\gamma}_n,\hat{\gamma}_{n'}]=-(z_n|\bar{z}_{n'})=0$ and $[\hat{\gamma}_n,\hat{\gamma}^{\dagger}_{n'}]=(z_n,z_{n'})=\pm\delta_{nn'}$ where the sign depends on whether the mode has positive or negative norm. The quantization of the amplitudes of modes with zero norm is done in a different way that we do not discuss here; see Ref. \cite{Pethick2008} for the quantization of the zero mode or Ref. \cite{Finazzi2010} for the quantization of modes with complex eigenvalues in the context of the BH laser. Then, the amplitude of a mode with positive (negative) norm corresponds to an annihilation (creation) type operator. As the conjugate $\bar{z}_n$ has opposite norm, its associated amplitude has the opposite character to that of $z_n$. From now on, when we write $\hat{\gamma}_n$ operators, we are considering them as annihilation operators.

Another way to understand the previous results arises when examining the grand-canonical Hamiltonian $\hat{K}=\hat{H}-\mu\hat{N}$, where $\hat{H}$ is given the Hamiltonian (\ref{eq:2ndHamiltonian}) and $\hat{N}$ is the particle number operator:
\begin{equation}\label{eq:numberparticles}
\hat{N}=\int\mathrm{d}^3\mathbf{x}~\hat{\Psi}^{\dagger}(\mathbf{x})\hat{\Psi}(\mathbf{x})
\end{equation}
Inserting the mean-field approximation (\ref{eq:meanfieldGP}) and keeping up to quadratic terms in the quantum fluctuations $\hat{\varphi}$ in the expression of $\hat{K}$ [which is consistent with keeping up to first order in the quantum fluctuations of the field operator in Eq. (\ref{eq:fieldoperatorheisenberg}), leading to the BdG equations], we find that the grand-canonical Hamiltonian is decomposed as $\hat{K}=K[\Psi_0]+K_V+\hat{K}_B$. Here, $K[\Psi_0]$ is the mean field contribution (i.e., the result of the substitution of $\hat{\Psi}$ by $\Psi_0$ in the expression for $\hat{K}$) which, after taking into account that $\Psi_0(\mathbf{x})$ is solution of Eq. (\ref{eq:TIGP}), gives
\begin{equation}\label{eq:meanfieldhamiltonian}
K[\Psi_0]=-\int\mathrm{d}^3\mathbf{x}~\frac{g_{\rm{3D}}}{2}|\Psi_0(\mathbf{x})|^4
\end{equation}
The contribution from the depletion cloud, $K_V$, is given by:
\begin{equation}\label{eq:depletionhamiltonian}
K_V=-\sum_n\int\mathrm{d}^3\mathbf{x}~\epsilon_n|v_n(\mathbf{x})|^2
\end{equation}
and arises from the non-commutative character of the annihilation and creation operators of the quasiparticles (the eigenmodes of the BdG equations). The Bogoliubov contribution is:
\begin{equation}\label{eq:Bogoliubovhamiltonian}
\hat{K}_B=\sum_n\epsilon_n\hat{\gamma}^{\dagger}_{n}\hat{\gamma}_n
\end{equation}
Hence, in the Bogoliubov approximation, the grand-canonical Hamiltonian is diagonalized by the solutions of the BdG equations.

There is no linear contribution in the $\hat{\gamma}_n$ operators because the GP equation (\ref{eq:TIGP}) is precisely the condition for $\Psi_0$ to be an extreme of the functional $K[\Psi]$:

\begin{equation}\label{eq:Kfunctional}
K[\Psi]=\int\mathrm{d}^3\mathbf{x}~\Psi^{*}(\mathbf{x})\left(-\frac{\hbar^2}{2m}\nabla^2+V_{\rm{ext}}(\mathbf{x})\right)\Psi(\mathbf{x})+\frac{g_{\rm{3D}}}{2}|\Psi(\mathbf{x})|^4-\mu|\Psi(\mathbf{x})|^2
\end{equation}

If $\Psi_0$ is a minimum of the previous functional, it is an energetically stable solution. It is also said that the solution is statically or Landau stable. If it is an extreme but not a local minimum, the solution $\Psi_0$ is Landau unstable since any perturbation would induce the system to decay to a lower energy state. For examining the specific character of $\Psi_0$, we consider linear perturbations around the stationary solution in $K[\Psi]$, $\Psi=\Psi_0+\delta \Psi$, obtaining:
\begin{eqnarray}\label{eq:Landauperturbation}
\nonumber \delta K&=&K[\Psi]-K[\Psi_0]=\int\mathrm{d}^3\mathbf{x}~\frac{1}{2}\delta\Phi^{\dagger}\Lambda\delta\Phi\\
\delta\Phi&=&\left[\begin{array}{c}\delta \Psi\\ \delta \Psi^{*}\end{array}\right],~\Lambda=\sigma_z M=\left[\begin{array}{cc}G & L\\
L^{*}&G\end{array}\right]
\end{eqnarray}

We note that a similar expression arises when expanding the grand-canonical Hamiltonian in the quantum fluctuations, eventually obtaining the contributions $K_V$ and $\hat{K}_B$. Specifically, if we expand $\delta \Phi$ in an analogous way to Eq. (\ref{eq:BdGmodesexpansion}), we would obtain the the classical version of the quadratic contribution $\hat{K}_B$ (note that $K_V$ only appears in a quantum context due to the non-commutativity of annihilation and destruction operators). We compute now the eigenvalues of $\Lambda$, arriving at a similar equation to the time-independent BdG equation (\ref{eq:BdGequation}):
\begin{equation}
\Lambda\left[\begin{array}{c} u \\
v\end{array}\right]=\lambda\left[\begin{array}{c} u \\
v\end{array}\right]
\end{equation}
If all the eigenvalues of the matrix operator $\Lambda$ are positive (except the zero eigenvalue arising from phase transformations), the state $\Psi_0$ is Landau stable and in the opposite case, the state is Landau unstable. In particular, as argued by Landau, supersonic flows are unstable (see Sec. \ref{subsec:BHBEC} for more details).

The operator $\Lambda$ and the energy of the excitations $\epsilon_n$ of the BdG equations are further related for a given mode of the BdG equation through:
\begin{equation}
\braket{z_n|\Lambda|z_n}=\epsilon_n (z_n|z_n)
\end{equation}
Therefore, if the state is Landau stable, there are no complex eigenvalues since then $(z_n|z_n)=0$. Also, for a Landau stable state, if a mode has positive (negative) energy, it also has positive (negative) normalization. Thus, the presence of an {\it anomalous} mode, i.e., a mode with positive (negative) normalization and negative (positive) energy, implies that the system is energetically unstable.

Physically, the minimization of $K$ can be understood as the minimization of the Hamiltonian $H$ with the constraint of fixed total number of particles $N$, with the chemical potential $\mu$ playing the role of Lagrange multiplier.

Importantly, we note that, although they have similar form, $\Lambda$ is an Hermitian matrix operator in contrast to the matrix $M$ appearing in the BdG equations and then $\lambda$ can only take real values in the previous equation.

\subsubsection{Time-dependent situation}\label{subsec:timedependent}

We can extend the previous stationary concepts to time-dependent situations by writing the field operator in Eq. (\ref{eq:Fieldequation}) in a more general mean-field approximation, $\hat{\Psi}(\mathbf{x},t)=\Psi(\mathbf{x},t)+\hat{\varphi}(\mathbf{x},t)$. We obtain for the mean-field wave function the corresponding time-dependent GP equation:

\begin{eqnarray}\label{eq:TDGP}
H_{GP}(t)\Psi(\mathbf{x},t)&=&i\hbar\partial_t\Psi(\mathbf{x},t)\\
\nonumber  H_{GP}(t)&=&-\frac{\hbar^2}{2m}\nabla^2+V_{\rm{ext}}(\mathbf{x},t)+g_{\rm{3D}}|\Psi(\mathbf{x},t)|^2
\end{eqnarray}
where we have allowed for a possible time-dependence of the external potential. The normalization of the GP wave function, Eq. (\ref{eq:normalization}), is guaranteed for every time $t$ by the conserved current:

\begin{equation}\label{eq:GPcurrent}
\mathbf{J}(\mathbf{x},t)=-\frac{i\hbar}{2m}\left(\Psi^*(\mathbf{x},t)\nabla \Psi(\mathbf{x},t)-\Psi(\mathbf{x},t)\nabla \Psi^*(\mathbf{x},t)\right)
\end{equation}
which gives the conservation law:
\begin{equation}\label{eq:GPnormconservation}
\partial_t|\Psi(\mathbf{x},t)|^2+\nabla\mathbf{J}(\mathbf{x},t)=0
\end{equation}

In particular, if we write the wave function in terms of its amplitude and phase, $\Psi(\mathbf{x},t)=A(\mathbf{x},t)e^{i\theta(\mathbf{x},t)}$, we arrive at:

\begin{eqnarray}\label{eq:PhaseAmplitude}
\partial_tn+\nabla(n\mathbf{v})&=&0\\
\nonumber -\hbar \frac{\partial \theta}{\partial t}A&=&-\frac{\hbar^2}{2m}\nabla^2A+\frac{1}{2}m\mathbf{v}^2A+V_{\rm{ext}}(\mathbf{x},t)A+g_{\rm{3D}}A^3\, ,
\end{eqnarray}
where
\begin{eqnarray}\label{eq:density}
n(\mathbf{x},t)&=&A^2(\mathbf{x},t)\\
\label{eq:flowvelocity}\mathbf{v}(\mathbf{x},t)&=&\frac{\hbar\nabla\theta(\mathbf{x},t)}{m}
\end{eqnarray}
are the local mean-field density and local flow velocity, respectively.
The first line of Eq. (\ref{eq:PhaseAmplitude}) is just the conservation law (\ref{eq:GPnormconservation}) rewritten in a more appealing way, since $\mathbf{J}=n\mathbf{v}$. It is often denoted as the continuity equation since it is analog to the continuity equation for a hydrodynamical fluid.

In respect to the quantum fluctuations, we have that their time evolution is given by:
\begin{eqnarray}\label{eq:TDFieldequation}
\nonumber i\hbar\partial_t\hat{\Phi}&=&M(t)\hat{\Phi},\\
M(t)&=&\left[\begin{array}{cc}G(t) & L(t)\\
-L^{*}(t)&-G(t)\end{array}\right]\\
\nonumber G(t)&=&H_{GP}(t)+g_{\rm{3D}}|\Psi(\mathbf{x},t)|^2\\
\nonumber L(t)&=&g_{\rm{3D}}\Psi^2(\mathbf{x},t)
\end{eqnarray}
As in the time-independent case, we can decompose the field operator in terms of a complete set of solutions, arriving at a time-dependent version of the BdG equations for the spinors $z_n(t)$:

\begin{equation}\label{eq:TDBdG}
M(t)z_n(t)=i\hbar\partial_t z_n(t),
\end{equation}

The scalar product (\ref{eq:KGProduct}) of two different solutions $z_1(t),z_2(t)$ of the time-dependent BdG equations is preserved during the time evolution. In particular, the norm of a particular solution $z(\mathbf{x},t)$, $(z|z)$, is conserved. Related to the conservation of this norm norm, there is also an associated quasiparticle current given by:
\begin{equation}\label{eq:quasiparticlecurrent}
\mathbf{J}_z=-\frac{i\hbar}{2m}\left(u^*\nabla u-u\nabla u^*+v^*\nabla v-v\nabla v^*\right)
\end{equation}
where $u(\mathbf{x},t),v(\mathbf{x},t)$ are the components of the spinor $z(\mathbf{x},t)$. This current satisfies a similar conservation law to that of the GP equation:

\begin{equation}\label{eq:BdGnormconservation}
\partial_t(z^{\dagger}\sigma_zz)+\nabla\mathbf{J}_z=0
\end{equation}

As expected, when considering stationary GP solutions $\Psi(\mathbf{x},t)=\Psi_0(\mathbf{x})e^{-i\frac{\mu}{\hbar}t}$, the time-dependent GP equation (\ref{eq:TDGP}) reduces to the time-independent GP equation (\ref{eq:TIGP}). Also, by removing the phase factor $e^{-i\frac{\mu}{\hbar}t}$ from the field fluctuations $\hat{\varphi}$, the time-dependent BdG equation (\ref{eq:TDFieldequation}) is reduced to the stationary BdG equation (\ref{eq:BdGfieldequation}). In particular, the solutions of the time-independent GP and the BdG equations previously considered satisfy, respectively, the conservation laws:

\begin{equation}\label{eq:stationarycurrentssolutions}
\nabla\mathbf{J}=0,~\nabla\mathbf{J}_z=0
\end{equation}

\subsection{1D mean-field regime}\label{subsec:1Dmeanfield}

In order to describe a one-dimensional configuration along the $x$-axis, we consider an external potential of the form $V_{\rm{ext}}(\mathbf{x})=V(x)+\frac{1}{2}m\omega_{\rm{tr}}^2\rho^2$, where $\rho=\sqrt{y^2+z^2}$ is the radial distance to the $x$-axis and $V(x)$ an external potential that only depends on the $x$ coordinate. Using the following adiabatic ansatz for the wave function \cite{Leboeuf2001,Menotti2002},
\begin{equation}
\Psi_0(\mathbf{x})=\psi_0(x)\phi(\rho)
\end{equation}
we obtain an equation for the transverse and longitudinal degrees of freedom:
\begin{eqnarray}\label{eq:TransversalGP}
\nonumber
\epsilon(n)\phi(\rho)&=&\left[-\frac{\hbar^2}{2m}\nabla^2_{\perp}+\frac{1}{2}m\omega_{\rm{tr}}^2\rho^2+g_{\rm{3D}}n(x)|\phi(\rho)|^2\right]\phi(\rho)\\
\mu\psi_0(x)&=&\left[-\frac{\hbar^2}{2m}\partial^2_x+V(x)+\epsilon(n)\right]\psi_0(x)
\end{eqnarray}
where $n(x)=|\psi_0(x)|^2$ is the one-dimensional density and $\epsilon(n)$ is the eigenvalue of the transverse equation, which depends on $n(x)$.

We work in the low-density limit
\begin{equation}\label{eq:1Dmeanfieldregime}
na_s\ll1
\end{equation}
known as the 1D mean-field regime. In this case, the non-linear term in the equation for the transverse coordinated can be treated as a perturbation to the harmonic oscillator ground state yielding:
\begin{eqnarray}
\nonumber\phi(\rho)&=&\phi_0(\rho)=\frac{e^{-\frac{\rho^2}{2a^2_{\rm{tr}}}}}{\sqrt{\pi}a_{\rm{tr}}}\\
\epsilon\left(n\right)&=&\hbar\omega_{\rm{tr}}+g_{\rm{1D}}n(x)\\
\nonumber g_{\rm{1D}}&=&g_{\rm{3D}}\int\mathrm{d}y\mathrm{d}z\, |\phi_0|^4=2\hbar\omega_{\rm{tr}}a_s
\end{eqnarray}
with $a_{\rm{tr}}=\sqrt{\hbar/m\omega_{\rm{tr}}}$ the transversal harmonic oscillator length (see Sec. \ref{subsec:staHO} for a detailed discussion on the eigenstates of the harmonic oscillator). Absorbing the zero-point energy of the harmonic oscillator $\hbar\omega_{\rm{tr}}$ into the chemical potential, we finally arrive at an effective one-dimensional GP equation for $\psi_0(x)$:
\begin{eqnarray}\label{eq:TIGP1D}
\nonumber H_{GP\rm{1D}}\psi_0(x)&=&\mu\psi_0(x)\\
H_{GP\rm{1D}}&=&-\frac{\hbar^2}{2m}\partial_x^2+V
(x)+g_{\rm{1D}}|\psi_0(x)|^2
\end{eqnarray}
Precisely, the condition (\ref{eq:1Dmeanfieldregime}) ensures that $g_{1D}n(x)$ is indeed a small correction to the transverse confinement energy scale. In the same fashion, one can also derive an effective one-dimensional BdG equation for the quantum fluctuations:
\begin{eqnarray}\label{eq:BdGfieldequation1D}
\nonumber i\hbar\partial_t\hat{\Phi}_{1D}&=&M_{\rm{1D}}\hat{\Phi}_{1D},\\
M_{\rm{1D}}&=&\left[\begin{array}{cc}G_{\rm{1D}} & L_{\rm{1D}}\\
-L_{\rm{1D}}^{*}&-G_{\rm{1D}}\end{array}\right]\\
\nonumber G_{\rm{1D}}&=&H_{GP\rm{1D}}+g_{\rm{1D}}|\psi_0|^2-\mu,~L_{\rm{1D}}=g_{\rm{1D}}\psi^2_0
\end{eqnarray}
A mode expansion can be performed for $\hat{\Phi}_{1D}$ similar to that of the 3D case, Eq. (\ref{eq:BdGmodesexpansion}), obtaining the analog 1D BdG equations. This one-dimensional Bogoliubov approximation is valid whenever
\begin{equation}\label{eq:Tonkscondition}
na^2_{\rm{tr}}/a_s\gg1
\end{equation}
Otherwise, we are in a regime known as the Tonks-Girardeau gas \cite{Menotti2002,Dunjko2001}. Hereafter, we drop the index $\rm{1D}$ as we will only focus on 1D configurations. We also identify $\Psi_0(x)$ with $\psi_0(x)$, since the the motion of the transverse degrees of freedom is frozen.

As in Sec. \ref{subsec:timedependent}, the previous effective 1D GP equation can be extended to time-dependent situations by replacing $\mu$ by $i\hbar\partial_t$ in Eq. (\ref{eq:TIGP1D}). {\it Mutatis mutandi}, we can also obtain the effective 1D time-dependent BdG equations.

\subsection{Black holes in Bose-Einstein condensates}\label{subsec:BHBEC}

Once shown how to reach an effective 1D configuration, we proceed to study the stationary flow of a 1D BEC within the 1D mean-field regime previously explained. We consider homogeneous plane wave solutions to the GP equation, $\Psi_0(x)=\sqrt{n}e^{iqx+\theta_0}$. The associated BdG solutions are given in terms of plane waves, characterized by wave vector $k$ and energy $\epsilon=\hbar\omega$ given by the dispersion relation:
\begin{equation}\label{eq:dispersionrelation}
\left[\omega-vk\right]^{2}=\Omega^2=c^{2}k^{2}+\frac{\hbar^2k^{4}}{4m^2}
\end{equation}
where $c=\sqrt{gn/m}$ is the sound velocity, $v=\hbar q/m$ is the constant flow velocity, as given by Eq. (\ref{eq:density}), and $\Omega$ is the comoving frequency, which is defined positive. The previous relation is just the usual phonon Bogoliubov dispersion relation for a condensate at rest, $\Omega(k)$, with the Doppler shift due to the fluid velocity $v$. For convention, we always consider $v>0$. In this work, we define the coherence length as $\xi\equiv\hbar/mc$, which differs from the usual definition by a factor $1/\sqrt{2}$. The system is said to be supersonic when $v>c$ and subsonic when $v<c$. The previous dispersion relation has four solutions for $k$ for a given laboratory frequency $\omega$. Interestingly, the sum and the product of these four wave vectors, labeled with the index $a$, satisfy
\begin{equation}\label{eq:wavevectorsum}
\sum_{a}k_a(\omega)=0,~\Pi_{a} k_a(\omega)=-\frac{4m^2\omega^2}{\hbar^2}
\end{equation}
The corresponding spinor solution of the BdG equations for each wave vector $k_a$ can be written as:
\begin{eqnarray} \label{eq:PlaneWaveSpinors}
s_{a,\omega}\left(x\right) & \equiv & \frac{e^{ik_{a}\left(\omega\right)x}}{\sqrt{2\pi|w_{a}\left(\omega\right)|}}\left[\begin{array}{c}
e^{i(qx+\theta_0)}u_{a}(\omega)\\
e^{-i(qx+\theta_0)}v_{a}(\omega)
\end{array}\right] \nonumber\\
\left[\begin{array}{c}
\nonumber u_{a}(\omega)\\
v_{a}(\omega)
\end{array}\right]&=&N_a\left[\begin{array}{c}
\frac{\hbar k_{a}^{2}\left(\omega\right)}{2m}+[\omega-vk_{a}\left(\omega\right)]\\
\frac{\hbar k_{a}^{2}\left(\omega\right)}{2m}-[\omega-vk_{a}\left(\omega\right)]
\end{array}\right]\\
N_a&=&\left(\frac{m}
{2\hbar k_{a}^{2}\left(\omega\right)\left|\omega-vk_{a}\left(\omega\right)\right|}\right)^{\frac{1}{2}}.
\end{eqnarray}
with $w_{a}\left(\omega\right)\equiv\left[dk_{a}\left(\omega\right)/d\omega\right]^{-1}$ the corresponding group velocity. The group velocity is included in the definition of the solutions in order to normalize the propagating modes (those with purely real wave vector) in frequency domain $(s_{a,\omega}|s_{a,\omega'})=\pm\delta\left(\omega-\omega'\right)$. For solutions with complex wave vectors, all the normalization factors (in particular that including the group velocity) can be removed as they do not play any significant role.

\begin{figure*}[tb!]
\includegraphics[width=1\textwidth]{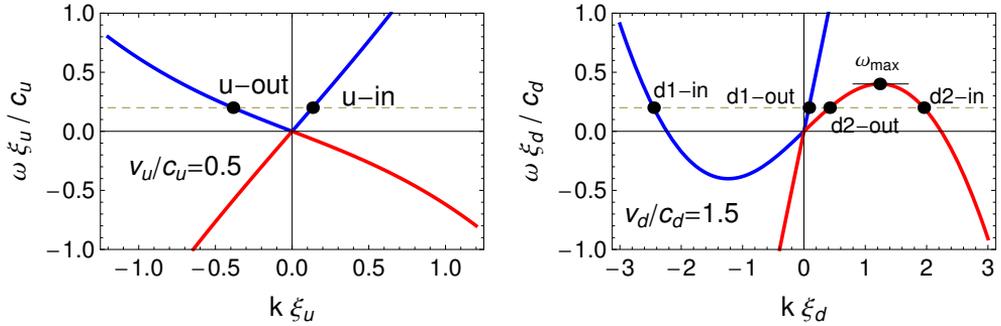}
\caption{Plot of the dispersion relation for quasiparticles given by Eq. (\ref{eq:dispersionrelation}). Left: dispersion relation for subsonic flow, denoted by label $u$. Right: dispersion relation for supersonic flow, denoted by label $d$. The blue (red) curves represent positive (negative)
normalization of the modes. Here, $\xi_{u(d)}$ is the
corresponding healing length, $c_{u(d)}$ and $v_{u(d)}$ the corresponding sound and
flow velocities and $\omega_{{\rm max}}$ is the Hawking frequency.}
\label{figDispRelation}
\end{figure*}

The dispersion relation for a subsonic and supersonic system is plotted in Fig. \ref{figDispRelation}, where ``u" stands for upstream (subsonic) flow and ``d" stands for downstream (supersonic); we explain later the motivation for choosing this convention. Rewriting Eq. (\ref{eq:dispersionrelation}) as $\omega=qk\pm \Omega$, the blue (red) curves represents the sign $+(-)$ in the previous equation. They also correspond to positive (negative) norm of the associated BdG solutions. In the subsonic case, we only have two propagating modes, denoted as $u-{\rm in}$ and $u-{\rm out}$, where ``in" stands for positive group velocity and ``out" for negative group velocity. Both modes have positive normalization and the other two solutions have complex wave vector, being one conjugate of the other so one is exponentially increasing and the other one exponentially decreasing. In the supersonic case, for $0<\omega<\omega_{\rm max}$, all the four modes are propagating, structured into two pairs of modes with positive (negative) normalization which are labeled as $d1(d2)$. The frequency $\omega_{\rm max}$ is the Hawking frequency and represents the upper limit of the HR spectrum. Above $\omega_{\rm max}$, the anomalous $d2$ modes are no longer propagating and thus, there is no Hawking effect (see Sec. \ref{subsec:hawkingeffect}). The presence of anomalous modes results from the Landau instability of supersonic flows, see discussion at the end of Sec. \ref{subsec:timeindependent}. For supersonic flows, the modes are denoted as ``in" when they have negative group velocity and ``out" when they have positive group velocity.

We can extend the previous definition of sound speed to non-homogeneous GP solutions by taking $c(x)\equiv\sqrt{gn(x)/m}$. In an analog way, we say the condensate is subsonic where $v(x)<c(x)$ and supersonic if $v(x)>c(x)$. A black-hole configuration is that with two asymptotic homogeneous regions, one subsonic and one supersonic with flow velocity going from subsonic to supersonic. If the flow velocity is going from supersonic to subsonic, we have a white hole, the time reversal of a black hole (obtained by simply conjugating the wave function). Throughout this work, we only consider black-hole configurations.

By continuity, a black-hole configuration implies that, at least, one sonic horizon exists, i.e., a point where $v(x)=c(x)$. We refer to the region between the two asymptotic regions, where the sonic horizon is placed, as the scattering region. For convention, we place the scattering region near $x=0$. We also take the flow velocity always positive, so the subsonic region is located in the upstream region (labeled as ``u") at $x\rightarrow-\infty$ while the supersonic region is located at $x\rightarrow\infty$ in the downstream region (labeled as ``d"), which matches the notation ``u" and ``d" previously introduced. Moreover, with the convention chosen for ``in" and ``out" for the modes, the ``in" modes are incoming (they travel towards the scattering region located near $x=0$) and the ``out" modes are outgoing (they travel outwards, away from $x=0$). The asymptotic homogeneous values of the momentum, flow velocity, sound velocity and healing length are labeled as $q_{u,d},v_{u,d},c_{u,d},\xi_{u,d}$. The notation followed here is the standard one considered in the literature, see for instance Refs. \cite{Recati2009,Larre2012,Zapata2011}.

The analogy with black holes comes from the fact that acoustic phonons, those with vanishing wave vector for $\omega\rightarrow 0$, are trapped in the supersonic region inside the sonic horizon (since they are dragged away by the flow) as light is trapped in a black hole inside the event horizon. However, we note that, due to the superluminal comoving dispersion relation $\Omega(k)$, the incoming modes of the supersonic region can travel upstream and escape from the acoustic black hole unlike gravitational black holes where nothing escapes.

The solutions to the BdG equations in the asymptotic regions of a BH configuration are combinations of the different plane waves, also called scattering channels in this context. In particular, the retarded (``in") scattering states $z_{a,\omega}(x)\equiv\left[u_{a,\omega}(x),v_{a,\omega}(x)\right]^{T}$ are those solutions with unit amplitude in the incoming channels $a=d1,d2,u-{\rm in}$ and zero amplitude in the other incoming channels. The amplitude of the ``out" scattering channels for these scattering solutions are given by the $S$-matrix, as usual in scattering theory:
\begin{eqnarray}\label{eq:scatteringchannelstate}
z_{d2-\rm{in},\omega}\left(x\rightarrow-\infty\right)&=&S_{ud2}\left(\omega\right)s_{u-\rm{out},\omega}\left(x\right)\\
\nonumber z_{d2-\rm{in},\omega}\left(x\rightarrow\infty\right)&=&s_{d2-\rm{in},\omega}\left(x\right)+S_{d1d2}\left(\omega\right)s_{d1-\rm{out},\omega}\left(x\right)
+S_{d2d2}\left(\omega\right)s_{d2-\rm{out},\omega}\left(x\right)
\end{eqnarray}
The expression for the remaining ``in" scattering states can be written in a similar fashion. The advanced (``out") scattering states are the analog of the ``in" states but changing the index ``in" by ``out", i.e., they have unit amplitude in one outgoing channel and zero in the other outgoing channels.

The quantum fluctuations of the field operator reads, for a BH configuration, in terms of these scattering states as \cite{Zapata2011}
\begin{eqnarray}
\hat{\Phi}(x) & = & \int_{0}^{\infty}d\omega\sum_{a={\rm u-\rm{in},d1-\rm{in}}}[z_{a,\omega}(x)\hat{\gamma}_{a}(\omega)+\bar{z}_{a,\omega}(x)\hat{\gamma}_{a}^{\dag}(\omega)]\\
\nonumber &+&\int_{0}^{\omega_{{\rm max}}}d\omega[z_{{\rm {d2-\rm{in},\omega}}}(x)\hat{\gamma}_{{\rm {d2-\rm{in}}}}^{\dag}(\omega)\nonumber+\bar{z}_{{\rm {d2-\rm{in}},\omega}}(x)\hat{\gamma}_{{\rm {d2-\rm{in}}}}(\omega)]\label{eq:BHFieldOperator} \, .
\end{eqnarray}
We can also express $\hat{\Phi}(x)$ in terms of the ``out'' scattering states, which simply reduces to change the label ``in'' by ``out'' in the previous equation. Here, $\hat{\gamma}_{i-\alpha}$ is the annihilation operator of a quasiparticle in the scattering state $i-\alpha$, with $i=u,d1,d2$ and $\alpha={\rm in,out}$. The frequency dependence will be often implicit through this work. We can see that the order of the creation/annihilation operators is changed for the $d2$ mode because of its anomalous character. In this case, the conjugate $\bar{z}_{{\rm {d2-\rm{in}},\omega}}$ is that with positive norm, $(\bar{z}_{{\rm {d2-\rm{in}},\omega}}|\bar{z}_{{\rm {d2-\rm{in}},\omega}})=1$.
This is due to our choice of taking only the modes with $\omega>0$ to characterize the solutions of BdG equation (the modes with $\omega<0$ can be seen as the conjugates of those with $\omega>0$ by the symmetry $z\rightarrow\bar{z}$, as previously explained).
The relation between the ``out" and the ``in" states is given by the $S$-matrix:
\begin{equation}\label{eq:inoutmodesrelation}
\left[\begin{array}{c}
\hat{\gamma}_{u-\rm{out}}\\
\hat{\gamma}_{d1-\rm{out}}\\
\hat{\gamma}_{d2-\rm{out}}^{\dagger}
\end{array}\right] = \left[\begin{array}{ccc}S_{uu}&S_{ud1}&S_{ud2}\\
S_{d1u}&S_{d1d1}&S_{d1d2}\\
S_{d2u}&S_{d2d1}&S_{d2d2}\end{array}\right]\left[\begin{array}{c}
\hat{\gamma}_{u-\rm{in}}\\
\hat{\gamma}_{d1-\rm{in}}\\
\hat{\gamma}_{d2-\rm{in}}^{\dagger}
\end{array}\right] \, .
\end{equation}
Using the conservation of the quasiparticle current [Eqs. (\ref{eq:quasiparticlecurrent}),(\ref{eq:stationarycurrentssolutions})] for an arbitrary linear combination of ``in" scattering states, it can be shown that the $S$-matrix is pseudo-unitary, i.e.,
\begin{equation}\label{eq:pseudounitarity}
S^{\dagger}\eta S=\eta\equiv{\rm diag}(1,1,-1).
\end{equation}
In terms of group theory, this means that $S\in U(2,1)$. We refer the reader to Appendix \ref{app:parametrization} for a detailed discussion about the structure of the scattering matrix.

\subsection{Hawking effect}\label{subsec:hawkingeffect}

Equation (\ref{eq:inoutmodesrelation}) is a Bogoliubov type relation that mixes creation and annihilation operators, based on the anomalous character of the $d2$ modes. Here lies the origin of the Hawking effect. If we evaluate the expectation value of the number of outgoing $u$ phonons, $\braket{\hat{\gamma}_{u-\rm{out}}^{\dagger}\hat{\gamma}_{u-\rm{out}}}$, in the vacuum of the incoming modes, we obtain
\begin{eqnarray}\label{eq:hawkingef}
\braket{\hat{\gamma}_{u-\rm{out}}^{\dagger}\hat{\gamma}_{u-\rm{out}}}=|S_{ud2}(\omega)|^2\neq 0
\end{eqnarray}

This emission of outgoing quasiparticles into the subsonic region in the presence of the vacuum of incoming quasiparticles is known as spontaneous Hawking radiation and is a genuine quantum effect. It is analog to the spontaneous emission of particles in a gravitational black hole, where the role of the outside of the black hole is played by the subsonic region. We see, from Eq. (\ref{eq:hawkingef}), that the intensity of the spontaneous Hawking signal is given by $|S_{ud2}|^2$.

A qualitative explanation of the Hawking effect can be obtained by examining the grand-canonical Hamiltonian $\hat{K}$. If we focus on the $\omega<\omega_{\rm max}$ sector of the Bogoliubov contribution [given by Eq. (\ref{eq:Bogoliubovhamiltonian})], we find
\begin{equation}\label{eq:BHK}
\int_{0}^{\omega_{\rm max}}d\omega~\hbar\omega(\hat{\gamma}_{a-\rm{in}}^{\dagger}\eta_{ab}\hat{\gamma}_{b-\rm{in}})=\int_{0}^{\omega_{\rm max}}d\omega~\hbar\omega(\hat{\gamma}_{u-\rm{in}}^{\dagger}\hat{\gamma}_{u-\rm{in}}+\hat{\gamma}_{d1-\rm{in}}^{\dagger}\hat{\gamma}_{d1-\rm{in}}-\hat{\gamma}_{d2-\rm{in}}^{\dagger}\hat{\gamma}_{d2-\rm{in}}) \,.
\end{equation}

The minus sign corresponding to the $d2$ modes in the previous equation is due to their negative norm. It can also be seen as the negative energy of the corresponding conjugate modes, that are conventionally normalized. Using Eqs. (\ref{eq:inoutmodesrelation})  and (\ref{eq:pseudounitarity}), we can rewrite Eq. (\ref{eq:BHK}) in terms of the ``out" states, which is the same expression but changing the ``in" index for ``out". One can essentially choose between two conventions, which are clearly represented in Fig. \ref{figDispRelation}: (i) All the frequencies are positive while the normalization of the scattering channels can be positive (plotted in blue) or negative (plotted in red). (ii) All the normalizations are positive but we have to deal with positive and negative frequencies. The first convention is more convenient to perform calculations so we have adopted it throughout this work, as quasiparticle scattering can be viewed as elastic. The second convention permits however a simple physical picture of Hawking radiation: once we have positive and negative frequency outgoing scattering channels, one can expect that, even at zero temperature, two outgoing quasiparticles can be created spontaneously at zero-energy cost. The positive-frequency quasiparticle flows towards the subsonic region, while a negative-energy companion quasiparticle flows towards the supersonic side. Within the first convention, the process of Hawking radiation can be viewed as the vacuum (total absence) of incoming quasiparticles generating outgoing ``particle-antiparticle" pairs (involving the channels $u$ and $d2$). Thus, the Hawking radiation corresponding to the outgoing $u$-phonons in the subsonic region appears to the outside observer as spontaneously created in the horizon region.

The conversion between two normal or two anomalous channels is labeled as a ``normal'' scattering process, while conversion from a ``normal'' to an ``anomalous'' channel (or vice versa) is labeled as an ``anomalous'' scattering process. Hawking radiation can be viewed as the result of anomalous scattering. The Andreev reflection (analog to the Hawking effect but involving $d1$ and $d2$ channels) studied in Ref. \cite{Zapata2009a} is also anomalous. We mainly focus on studying the Hawking effect in this work although many of the results can be straightforwardly translated to the context of Andreev reflection.

Using quantum optics terminology, Hawking radiation can be understood as a non-degenerate parametric amplifier \cite{Walls2008} (see Appendix \ref{app:parametrization} for a explicit demonstration) since the vacuum of the incoming modes can be seen as a squeezed state of the outgoing modes. We will return later to this idea in Sec. \ref{sec:criteriaHR}.

\section{Typical black-hole configurations} \label{sec:typicalbh}

In this section we resume the main theoretical models that have been proposed for studying analog HR in BECs, following Ref. \cite{Larre2012}. Sonic black holes can be produced by several methods: manipulating spatially the value of the coupling constant, introducing potential barriers or accelerating the condensate using a negative potential. All the presented models consider ideal stationary configurations, characterized by the corresponding stationary GP wave function $\Psi_0(x)$ with semi-infinite asymptotic homogeneous regions. The formation of a much more realistic BH configuration, achieved under standard experimental techniques, is studied in Chapter \ref{chapter:MELAFO}. Using the wave function $\Psi_0(x)$ as stationary background, one computes the different BdG scattering states and obtain the corresponding scattering matrix, which characterizes the Hawking processes. We refer the reader to Appendix \ref{app:1DGP} for the specific form of the stationary wave function in the different models and to Appendices \ref{app:BdGsolitonsolutions}, \ref{app:SMatrixBehav} for the computation of the scattering states and the $S$-matrix. We remark that white-hole configurations can be simply obtained from the following black-hole configuration by time-reversal symmetry, i.e., by taking the complex conjugate of $\Psi_0(x)$.

\subsection{Flat profile}\label{subsec:flatprofile}

In this scenario, the GP wave function is everywhere the same plane
wave, $\Psi_{0}(x)=\sqrt{n_{0}}e^{iqx}$, so both velocity and density are uniform, with $v(x)=\hbar q/m$. For achieving this configuration, the 1D constant coupling
$g\left(x\right)$ and the external potential $V\left(x\right)$ depend on position, satisfying the condition
\begin{equation}\label{eq:matchingcondition}
V(x)+g(x)n_{0}=E_b
\end{equation}
with $E_b$ constant. Both quantities can be controlled using standard atomic tools in the laboratory \cite{Pitaevskii2003}. Here, we consider a stepwise constant spatial dependence for $c(x)=\sqrt{g(x)n_0/m}$ in such way that
$c(x<0)=c_1$ and $c(x\geq 0)=c_{2}$, with $c_1>v>c_{2}$ so $x<0$ is subsonic and $x>0$ is supersonic. Then, the black-hole is formed just at $x=0$. Although this configuration is highly idealized from the experimental point of view, it is considered in many works \cite{Balbinot2008,Carusotto2008,Recati2009} because of its appealing simplicity. In particular, we deal with a similar configuration in Chapter \ref{chapter:BHL} in order to study the features of the BH laser.

\subsection{Delta-barrier configuration}\label{subsec:1delta}

This configuration has constant $g$ but it uses a potential of the form $V(x)=Z\delta(x)$, which implies a discontinuity in the derivative of $\Psi_0(x)$ of the form $\Psi'_0(0^{+})-\Psi'_0(0^{-})=2Z\Psi_0(0)$. The local sound speed $c(x)$ will cross the flow velocity $v(x)$ at some point to the left of $x=0$. It was shown in Ref. \cite{Kamchatnov2012} that this kind of configuration could be achieved by projecting a condensate on a localized obstacle (that could be different from a delta potential). It represents a theoretical model of the flow of a condensate through a localized obstacle \cite{Leboeuf2001}. This configuration is schematically depicted in the left panel of Fig. \ref{fig:NonResonantScheme}.

\subsection{Waterfall configuration}\label{subsec:waterfall}

\begin{figure*}[!tb]
\includegraphics[width=1.05\textwidth]{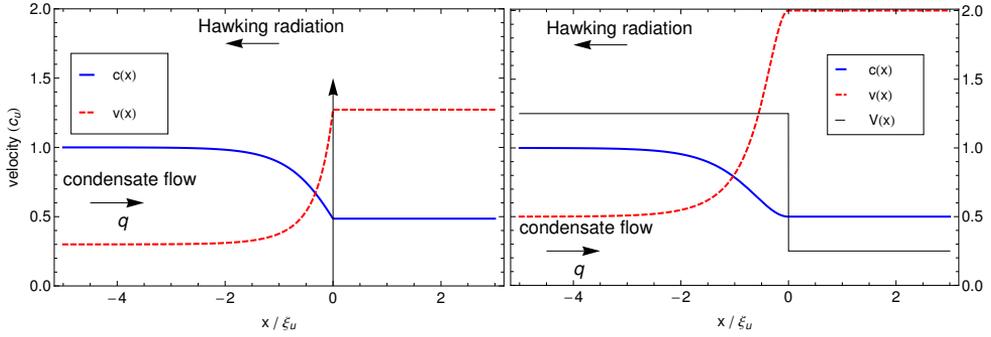}
\caption{Scheme of the delta barrier configuration described in Sec. \ref{subsec:1delta} (left) and the waterfall configuration described in Sec. \ref{subsec:waterfall}
(right).}
\label{fig:NonResonantScheme}
\end{figure*}

This configuration is similar to the previous one but the potential is now given by a negative step, $V(x)=-V_0\theta(x)$ where $\theta(x)$ is the Heaviside function. There are experimental evidences that this stationary configuration can be achieved in the laboratory \cite{Lahav2010}. This configuration is schematically depicted in the right panel of Fig. \ref{fig:NonResonantScheme}.

\subsection{Resonant configurations}\label{subsec:resonant}

We also study BH configurations that present resonant peaks in their Hawking spectra. They can be obtained by combining conveniently the previous configurations as building blocks. In particular, we focus in this work only on two specific resonant configurations, but infinite types of resonant configurations can be considered.

\begin{figure*}[!tb]
\includegraphics[width=1.05\textwidth]{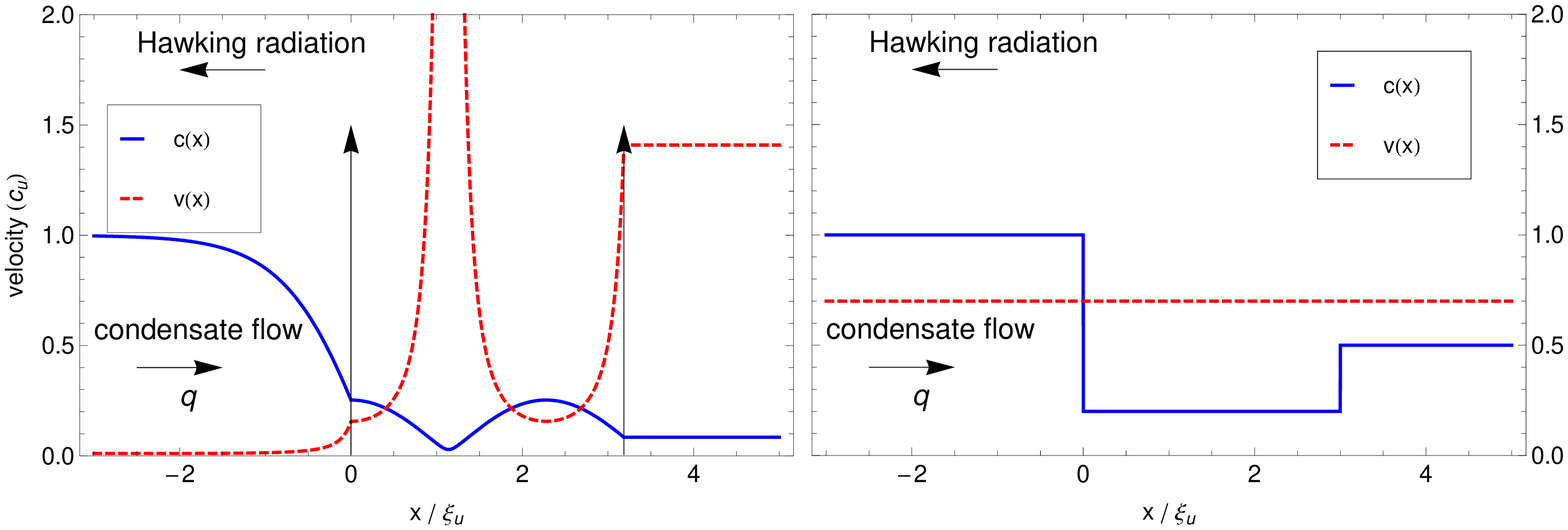}
\caption{Scheme of the double-barrier configuration (left) and the resonant generalization of the flat profile configuration
(right), both described in Sec. \ref{subsec:resonant}.}
\label{fig:ResonantScheme}
\end{figure*}

The first structure that we study is a double-barrier potential represented by two Dirac-deltas $V(x)=Z[\delta(x)+\delta(x-d)]$. Here $d$ is
the distance between the two delta-barriers and $Z$ is their amplitude. This configuration has been already studied in Ref. \cite{Zapata2011}, where it was shown that the HR spectrum can display resonant peaks. Apart from one sonic black hole, depending on the interbarrier distance, several black hole-white hole pairs can appear, as shown in left panel of Fig. \ref{fig:ResonantScheme}.

In the other case, we consider a resonant generalization of the flat profile configuration (see Sec. \ref{subsec:flatprofile}).
We show a plot of this kind of structure in the right panel of Fig. \ref{fig:ResonantScheme}. We take three flat regions with homogeneous value of the sound speed: $c(x<0)=c_1$, $c(0\leq x\leq d)=c_{2}$, $c(x>d)=c_{3}$, with $c_1>v>c_{3}$. The middle region, with speed of sound $c_2$, can be chosen as subsonic or supersonic.

\chapter{Birth of a quasi-stationary black hole in an outcoupled Bose-Einstein condensate}\label{chapter:MELAFO}
\chaptermark{Birth of a black hole in a Bose-Einstein condensate}

\section{Introduction}

As explained at the end of the previous chapter, there are many theoretical models that provide an analog BH scenario in a BEC. However, they are all highly idealized from the experimental point of view as they consider semi-infinite media and perfectly stationary configurations. Moreover, they do not describe the transient regime to such BH configurations. On the other hand, there is a lack of experimental scenarios for BH analogs in BEC; at the present moment, the most relevant experiment was performed by the Technion group \cite{Lahav2010}, where a sonic black hole was created by accelerating a condensate with a negative potential. We numerically explore here an alternative realistic way to produce a quasi-stationary black-hole configuration by outcoupling a large confined condensate so that the coherent outgoing beam is dilute and fast enough to be supersonic.

The main goal of the work presented in this chapter is to explore the actual attainability of the steady-state regime. The hope is that spontaneous HR will be easier to measure in a quasi-stationary BH configuration. Within a mean-field approximation, we investigate the dynamics of an initially confined condensate that begins to leak as the height of one of the confining barriers is driven from an essentially infinite to a finite value that permits a gentle yet appreciable flow of coherently outcoupled atoms.  A similar deconfinement protocol has been explored in Ref. \cite{Wuster2007}. Another theoretical proposal for atom \cite{Kamchatnov2012} and polariton \cite{Gerace2012} condensates is based on the idea of throwing the condensate onto a localized obstacle such as a potential barrier. In the present work, we focus on a finite-sized condensate and on the case where the increasingly transparent potential is formed not by a single \cite{Larre2012} or double \cite{Zapata2011} barrier, but by an extended optical lattice, the main reason being that the latter scenario seems more suitable for the achievement of quasi-stationary flow within this deconfinement scheme, as will be shown later. We will see that close-to-ideal stationary flow within the permitted energy bands is achievable under realistic opening protocols. We also perform a preliminary study of the Hawking radiation spectrum on top of the achieved quasi-stationary black hole. However, more conclusive predictions about the detectability of spontaneous radiation require harder calculations than those presented here and are left for future works.

Besides the motivation of realizing gravitational analogs, the achievement of a quasi-stationary black-hole transport scenarios is of general interest for the investigation of atom quantum transport, in the case of both bosons \cite{Bloch1999,Andersson2002,PhysRevA.76.063605,Guerin2006,PhysRevA.80.041605,PhysRevA.84.043618} and fermions \cite{Brantut2012,Brantut2013}, within the emergent field of atomtronics \cite{Seaman2007,Labouvie2015}.

This chapter is arranged as follows. Section \ref{sec:themodel} presents the model for the gradual reduction of the optical lattice amplitude which we will be investigating. After some preliminary remarks in Section \ref{sec:Prerem}, the main numerical results together with some theoretical arguments (that help to understand the observed trends) are presented in Section \ref{sec:NumericaLOL}. The second part of that section describes the achieved quasi-stationary regime. Section \ref{sec:Gaussian-shaped} addresses the more realistic case of an optical lattice having a Gaussian envelope. Interestingly, we find that the horizon lies at the maximum of the envelope and give a theoretical explanation of that fact. In Sec. \ref{sec:preliminaryBdG} we present some preliminary results for the Hawking spectrum above the quasi-stationary background provided by the black-hole configuration. The main conclusions are summarized in section \ref{sec:OLconclusions}. Appendix \ref{app:confinedconfi} provides a detailed description of the initial configuration of the system as it exists before the deconfinement procedure begins. Appendix \ref{app:nonlinearol} discusses the general features of Bloch waves in the presence of nonlinearities accounting for the interaction. Finally, Appendix \ref{sec:numericalolmethods} describes the numerical method of integration and the use of absorbing boundary conditions, along with the method developed for integrating the resulting quasi-stationary BdG equations.

\section{The model} \label{sec:themodel}

In this chapter we study the outcoupling of a one-dimensional (1D) Bose-Einstein condensate through a finite-size repulsive optical lattice, whose intensity is gradually lowered in such a way that, within a finite time, the periodic barrier shifts from
a regime of practical confinement to one of full transparency. We first focus on the mean-field dynamics, i.e., the evolution of the condensate wave function, which obeys the following effective 1D time-dependent GP equation:
\begin{equation}\label{eq:TDGPOL}
i\hbar\frac{\partial\Psi(x,t)}{\partial t}=\left[-\frac{\hbar^{2}}{2m}\partial_{x}^{2}+V(x,t)+g|\Psi(x,t)|^{2}\right]\Psi(x,t)\, ,
\end{equation}
where $V(x,t)$ is the time-dependent optical lattice potential and $g$ the effective one-dimensional coupling constant $g=2\hbar\omega_{\rm tr}a_{s}$. The previous description is valid whenever the system in the 1D mean-field regime, as explained in Sec. \ref{subsec:1Dmeanfield}. Taking the initial bulk density $n_0$ as a typical value for the density, we can realistically set (see Sec. \ref{sec:NumericaLOL} for typical values of the parameters) $n_0a_s\sim10^{-1}$ and $n_0 a^2_{\rm tr}/a_s\sim10^{3}$, from which we conclude that we are safely in the 1D mean-field regime.

The condensate density is nonzero only for $x>0$ because at all times we assume a sufficiently high barrier at $x=0$, which is simply
implemented via the hard-wall boundary condition $\Psi(0,t)=0$. Initially (at times $t < 0$), we consider an equilibrium condensate made of $N$
atoms occupying the region $0<x\lesssim L$. Thus $n_{0}\simeq N/L$ is the initial atom density, which is defined below more precisely. We also
introduce an optical lattice that spans the region $L \lesssim x \lesssim L+L_{\rm lat}$ and whose initial amplitude $V_0$ is
large enough for particle tunneling through the lattice to be practically forbidden.
The initial wave function is stationary, $\Psi(x,0)=\Psi_0(x)$, with $\Psi_0(x)$ satisfying the 1D time-independent GP equation (\ref{eq:TIGP1D})
\begin{eqnarray}\label{eq:GPinitial}
\left[-\frac{\hbar^{2}}{2m}\partial_{x}^{2}-\mu_{0}+V(x,0)+g|\Psi_0(x)|^{2}\right]\Psi_0(x) & = & 0 \, .
\end{eqnarray}
The initial chemical potential $\mu_0$ is determined by the normalization condition (\ref{eq:normalization}). The initial healing length is defined as $\xi_0\equiv \sqrt{\hbar^{2}/mgn_{0}}$, where $n_{0}\equiv \mu_{0}/g$. Further details on the initial condensate are given in Appendix \ref{app:confinedconfi}. At time $t=0$, the optical lattice intensity starts to decrease and atoms begin to escape towards the region $x \gtrsim L+L_{\rm lat}$, where the potential is assumed to be negligible. On quite general grounds \cite{Leboeuf2001,Zapata2009a,Zapata2011}, the
flow beyond the optical lattice can be expected to be supersonic.

We assume that the optical lattice is
made of two fixed phase lasers of wavelength $\lambda$ and whose wave vectors form an angle $\theta$ \cite{Fabre2011,Blakie2002}.
The time-dependent optical lattice potential is chosen so that in the
lattice region (defined by $L-\frac{d}{2}\leq x\leq L-\frac{d}{2}+L_{\rm lat}$) and
for times $t\geq0$,
\begin{eqnarray}\label{eq:TDPotential}
V(x,t) & = & V(t)\cos^{2}\left[k_L(x-L)\right]\nonumber \\
V(t) & = & V_{\infty}+(V_{0}-V_{\infty})e^{-t/\tau} \, ,
\end{eqnarray}
where $k_L=\pi/d$ and $d=\lambda/\left[2\sin(\theta/2)\right]$ is the lattice period, while $V(x,t)=0$ everywhere else.

\begin{figure}[tb!]
\includegraphics[width=1\columnwidth]{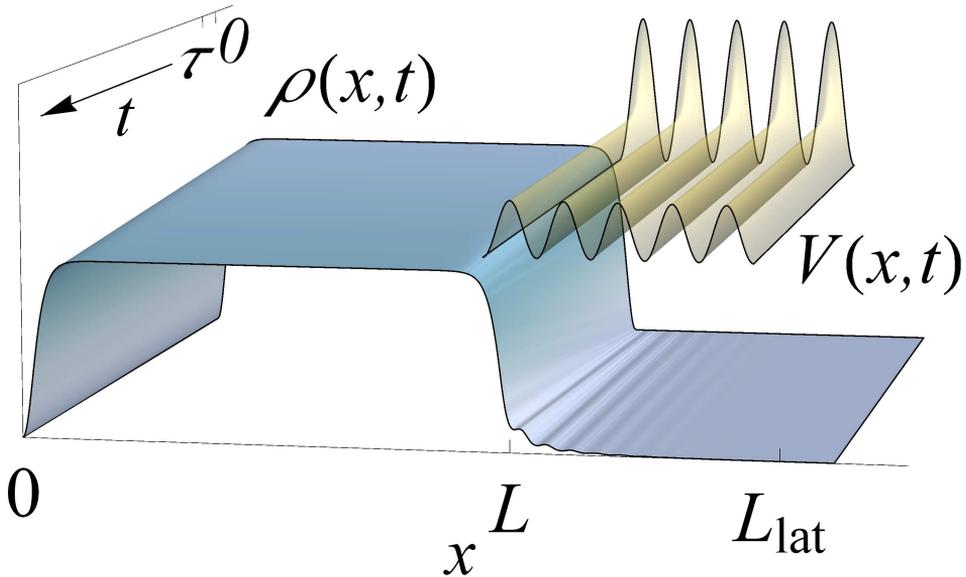}
\caption{Schematic representation of the emitting condensate setup here studied.
Within the ideal lattice scenario, hard-wall boundary conditions are assumed at $x=0$ and
an optical lattice lies in the region $L<x<L_{\rm lat}$ with a time-dependent amplitude such that the potential
$V(x,t)$ (represented by the semi-transparent yellow surface over the $x-t$ plane) evolves from strongly to moderately confining. The resulting time-dependent density profile $n(x,t)$ is represented by the grey-blue surface.
The vertical axis represents the density. The surface $V(x,t)$ is uplifted to provide a better vision of $n(x,t)$. Some parameters defined in the main text are indicated.
The trend towards a long-time quasi-stationary flow regime can be qualitatively observed.}
\label{fig:Scheme}
\end{figure}

The potential profile in Eq. (\ref{eq:TDPotential})
is somewhat idealized. A more realistic
choice should include a Gaussian envelope. For simplicity, we choose to start by considering
a flat-envelope optical lattice, where Bloch's theorem can be invoked with
reasonable confidence. We will see that, remarkably, essentially the same results are obtained when a more realistic Gaussian envelope is used. A sketch of the time-dependent, flat-envelope optical potential and the resulting condensate flow is presented in Fig. \ref{fig:Scheme}.

\section{Preliminary remarks.} \label{sec:Prerem}

The time-dependent amplitude $V(t)$ evolves from $V_{0}\gg\mu_{0}$ at $t \leq 0$ to $V_{\infty}\gtrsim\mu_{0}$ for $t \gg \tau$.
The asymptotic behavior is determined by the initial parameters of the condensate
$(N,g,L)$, the specific form of the final potential ($V_{\infty}$, $d$), and the barrier lowering time scale ($\tau$). The initial potential amplitude, $V_{0}$, plays almost no role since we take it sufficiently large. Here, ``large'' means that $\mu_0$ must lie well below the lowest conduction band of the initial optical lattice potential in order to have an effectively confined condensate. On the other hand, the properties of the final steady state are insensitive to $\tau$ unless $\tau$ is very small (see Subsection \ref{adiabatic}).

The main goal of the present work is to identify the barrier-lowering protocol that best leads to the formation of a quasi-stationary outcoupled black-hole configuration, by which we mean a black-hole flow regime characterized by parameters that vary slowly in time in a sense that will be specified later. As to the space dependence in that regime, we require the density to be as uniform as possible in the upstream region $0<x\lesssim L$.
In the supersonic downstream region ($x \gtrsim  L+L_{\rm lat}$) we also want an uniform
flow profile, even though this will be more difficult to achieve due to the small density. On the other hand, in the optical lattice region, the flow should be as close as possible to that of a propagating Bloch wave. At
the boundary between these two regions, large gradients of the
flow speed and density are likely to occur, but the current density should remain essentially uniform.

Throughout this chapter, when we talk about bands, we will be referring to the Schr\"odinger (non-interacting) bands, unless specified otherwise. Bands in a nonlinear context are discussed in Appendix \ref{app:nonlinearol}. We will see that, in virtually all cases, the relevant properties of the optical lattice are determined by its lowest band since the number of wells of the lattice is sufficiently large. An important result is that, due to the finite size of the subsonic reservoir, the quasi-stationary flow is formed when the space-averaged chemical potential lies slightly above the bottom of the lowest lattice band. This implies that, once in the quasi-stationary regime, the interacting term (and consequently the density) has to be small almost everywhere in the optical lattice region $g|\Psi(x,t)|^{2} \ll (\hbar k_L)^2/m$; see Appendix \ref{app:nonlinearol} for details. In the non-interacting regime, Eq. (\ref{eq:TDGPOL}) becomes the usual Schr\"odinger equation, which for a sinusoidal potential can be transformed into Mathieu's equation \cite{Abramowitz1988}, whose solutions and band structure are well known.

The asymptotic band structure can be characterized by the dimensionless parameter
\begin{equation} \label{small-v}
v \equiv mV_{\infty}/8\hbar^2 k_L^{2} \, .
\end{equation}
The nearly-free
atom regime occurs when $v\ll1$. Then bands are wide while gaps are narrow. By contrast, in the tight-binding regime ($v\gg1$), bands are narrow and widely spaced.
Since $v\propto V_{\infty}d^{2}$, the band structure can be modified by changing the lattice amplitude
or its spacing.

\section{Ideal optical lattice} \label{sec:NumericaLOL}

In this work, the unit length is the initial bulk healing
length $\xi_0$ defined in Section \ref{sec:themodel}.
Accordingly, velocities are measured in units of the sound speed,
$c_{0}\equiv \sqrt{gn_{0}/m}$, and times in units of $t_{0}\equiv \xi_0/c_{0}$. Energies
are expressed in units of the initial chemical potential $\mu_0=mc_{0}^{2}$.
We consider quasi--one-dimensional condensates of $^{87}$Rb, which are typically made of $N\sim10^{4}-10^{7}$ atoms and have mass $m=1.44 \times 10^{-25}$ kg. The typical order of magnitude of the transverse trapping frequency is $\omega_{\rm tr}\sim2\pi\times 10-10^3$ Hz and the confinement length varies in the range
$L\sim10-400 \mu\text{m}$. The scattering length satisfies $a_s\sim100a_0=5.29~\text{nm}$, with $a_0=5.29\times 10^{-11}$ m the Bohr radius. The optical lattice periodicity is bounded
from below, $d>\lambda/2$, for geometrical reasons [see below Eq. (\ref{eq:TDPotential})]. The value of $\lambda$ is chosen to be sufficiently far from the resonance to avoid any spontaneous emission and on the blue-side of the resonance to produce a repulsive potential ($\lambda < 780$ nm for rubidium atoms). The simulations are run up to times $t\sim10^4-10^5t_0$. As $t_0=(2n_0a_s\omega_{\rm tr})^{-1}\sim10^{-4}$ s, then $t\sim1-10$ s for our simulations, which is on the order of the mean lifetime of this type of condensates. Although we focus on the usual case of $^{87}$Rb, the results of the simulations here presented can be extended to arbitrary atoms.

In the simulations we use a numerical scheme based on the Crank-Nicolson
method to integrate the time-dependent GP equation (\ref{eq:TDGPOL}). Hard wall boundary conditions are assumed at $x=0$. At the other end of the finite size computational grid (located at $x=L_{g}$, with $L_{g}$ the total length of the grid), we use absorbing boundary conditions. $L_{g}$ is taken so that the final point of the grid is sufficiently far from the end of the optical lattice for the supersonic region to be clearly observed. Further details on numerical methods are given in Appendix \ref{app:numerical}.

\subsection{Analysis of the simulations\label{sub:AnalysisOfSimulations}}

To characterize the quasi-stationary regime, we define the local chemical
potential as
\begin{equation}\label{eq:LocalChemicalPotential}
\mu(x,t)\equiv -\frac{\hbar^{2}}{2m}\frac{\partial^{2}\Psi(x,t) / \partial x^{2}}{\Psi(x,t)}+V(x,t)+g|\Psi(x,t)|^{2} \, ,
\end{equation}
which, we note, can be complex. For a stationary solution, $\Psi(x,t)=e^{-i\mu t/\hbar}\Psi_0(x)$,
one has $\mu(x,t)=\mu$, real and independent of $(x,t)$. The one-dimensional current, obtained from Eq.(\ref{eq:GPcurrent}):
\begin{equation}\label{eq:1DCurrentFlux}
J(x,t)=-\frac{i\hbar}{2m}\left(\Psi^*(x,t)\frac{\partial\Psi(x,t)}{\partial x}-\Psi(x,t)\frac{\partial\Psi^*(x,t)}{\partial x}\right)
\end{equation}
is also independent of $(x,t)$ for a one dimensional stationary solution, as dictated by the one dimensional version of the continuity equation (\ref{eq:PhaseAmplitude}) [see also Eq. (\ref{eq:1DPhaseamplitude}) and related discussion]. In the quasi-stationary regime the uniformity of $J(x,t)$ is impossible to fulfill strictly, because the current is zero
at $x=0$ while the emitted atoms carry a nonzero current. Thus, there must be a current gradient and,
by the continuity equation, the density has to be time dependent. The hope is however that, in the
quasi-stationary regime, the condensate leak is sufficiently slow for the time dependence to be weak.

On the other hand, we can expect that, in the quasi-stationary regime, $\mu(x,t)$
is a sufficiently uniform function, with small spatial variations around its space-averaged
mean value. To check this expectation in a quantitative manner, we introduce the space-averaged
chemical potential $\bar{\mu}(t)$ together with an appropriate measure of its relative spatial fluctuations $\sigma(t)$:
\begin{eqnarray}
\bar{\mu}(t) & \equiv  & \frac{\int_{0}^{L_{g}}\mathrm{d}x\,n(x,t)\mu(x,t)}{\int_{0}^{L_{g}}\mathrm{d}x\,n(x,t)}\nonumber \\
\sigma(t) & \equiv  & \frac{1}{\bar{\mu}(t)}
\left[
\frac {\int_{0}^{L_{g}}\mathrm{d}x\,n(x,t)|\mu(x,t)-\bar{\mu}
(t)|^{2}} {\int_{0}^{L_{g}}\mathrm{d}x\,n(x,t)}
\right]^{\frac{1}{2}} \, .
\label{eq:AverageChemicalPotential}
\end{eqnarray}
We recall
that $\mu(x,t)$ and $\bar{\mu}(t)$ can be complex. A nonzero imaginary
part of $\mu(x,t)$ reflects a leaking condensate, as revealed by the local relation
\begin{equation}
\frac{\partial n}{\partial t}=\frac{2}{\hbar} n \, \text{Im} \, \mu \, .
\end{equation}
Accordingly, we can define and compute the emission rate per particle  as
\begin{equation}\label{eq:EmissionRate}
\Gamma(t)\equiv \frac{j(L_{g},t)}{\int_{0}^{L_{g}}\mathrm{d}x~n(x,t)}=
-\frac{2}{\hbar} \text{Im}\, \bar{\mu}(t) \, ,
\end{equation}
where the continuity equation has been used.
The spatial average of the time-dependent chemical potential, Eq. (\ref{eq:AverageChemicalPotential}), is mostly determined by the subsonic region, where $\mu(x,t) \simeq g n(x,t)$.

By rescaling the condensate wave function as $\Psi(x,t)\rightarrow\sqrt{n_0}\Psi(x,t)$, we observe clearly the actual degrees of freedom of the system. Given the initial healing length, $\xi_0$, only $L/\xi_0,\, d/\xi_0,\,\tau/t_{0},\, V_{\infty}/mc_{0}^{2}$
and $n_{\rm osc}\equiv L_{\rm lat}/d$ (number of oscillations in the optical lattice) are free parameters of the problem (we have already noted that
$V_{0}$ plays almost no role in the confinement limit $V_{0}\gg\mu_0$).

We find that, for the pertinent experimental ranges, namely, $L\sim10-400\,\mathrm{\mu m}$, variations of $L/\xi_0$ have little effect on the properties of the quasi-stationary regime. Specifically, they have a small influence on the time needed to achieve the desired quasi-stationarity, which grows weakly with the initial size of the condensate as a larger confined condensate needs more time to empty.

A similar point can be made about $n_{\rm osc}$, which becomes unimportant when it lies in the range $n_{\rm osc}\sim10-100$. The (relatively small) effect of increasing $n_{\rm osc}$ even further is that there are more spatial fluctuations in the chemical potential, for two reasons: (a) there are more spatial fluctuations of $\mu(x,t)$ because of multiple atomic reflections across the wells, an effect that is enhanced for larger optical lattices; (b) the larger the lattice, the bigger its contribution to the average chemical potential and its fluctuations.

In summary, except for the above remarks, only the parameters $d/\xi_0$, $V_{\infty}/mc_{0}^{2}$ and $\tau/t_{0}$,
have a noticeable effect on the transport properties of the system. We proceed to study their role on the results of the simulations.

\subsubsection{Role of the final band structure} \label{subsub:bandstructure}

\begin{figure}[tb!]
\includegraphics[width=1\columnwidth]{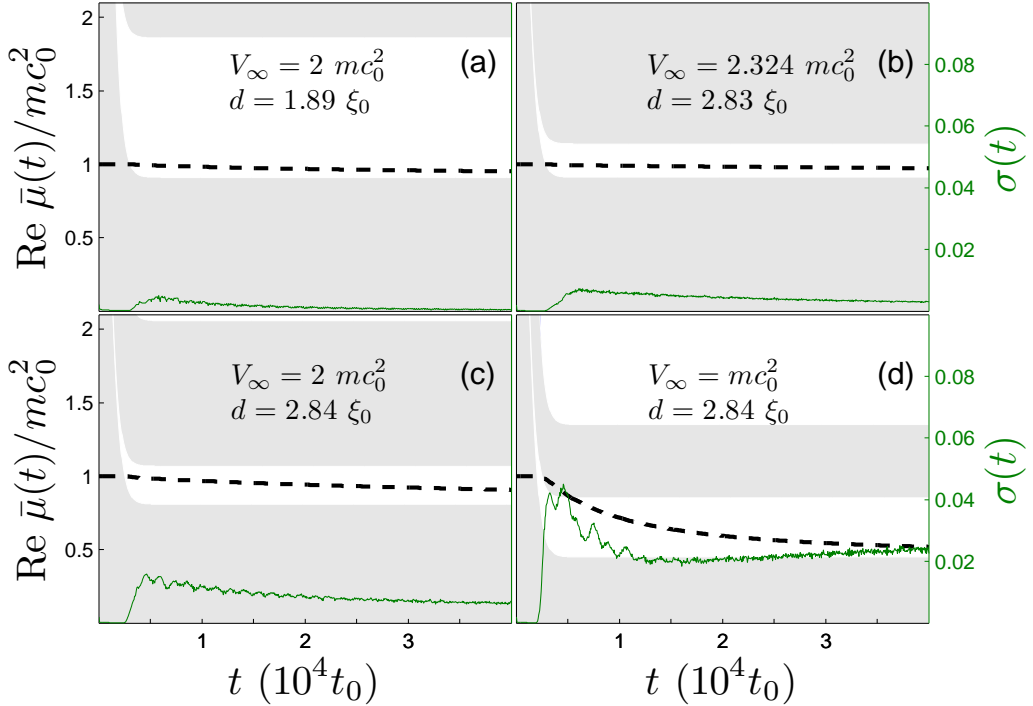} \caption{
Time evolution of the real part of the space-averaged chemical potential $\bar{\mu}(t)$
(black dashed) and its fluctuation spread $\sigma(t)$
(green solid), both defined in Eq. (\ref{eq:AverageChemicalPotential}).
The gap and conduction band of the instantaneous band structure
[computed from the potential Eq. (\ref{eq:TDPotential})] are indicated,
respectively, by grey and white backgrounds.
All graphs are computed with $\tau=500\, t_{0}$,
$L=400\,\mu\text{m}$, $n_{\rm osc}=30$, $N=10^4$, and $\omega_{\rm tr}=2\pi\times4\,\text{kHz}$,
which yields $n_{0}=25\,\mu\text{m}^{-1}$ and $\xi_0=0.3175\, \mu\text{m}$.
The long-time potential amplitude $V_{\infty}$ and the lattice spacing $d$ are indicated
in the graphs. The dimensionless parameter $v$ [see Eq. (\ref{small-v})] takes the values
$(0.0905,0.2350,0.2036,0.1018)$ for graphs (a)-(d).
The setups (a) and (b) are designed to have the same band bottom.
The simulations are run until a time $4\times 10^4\, t_{0}=5.5\,\mathrm{s}$.
Note that the scale of
$\sigma(t)$ is considerably enlarged.
The initial value $\sigma(0)$ (only observed with some magnification)
is spurious, and is related to the discrete approximation to the derivatives
in Eq. (\ref{eq:LocalChemicalPotential}).}
\label{fig: IdealBandComparison}
\end{figure}

As noted before, the combination of $d$ and $V_{\infty}$
fixes the final band structure. Figure \ref{fig: IdealBandComparison} shows the various
scenarios which one may find depending on the long-time width and position of the lowest band with respect to the
initial chemical potential, $\mu_{0}$. The band structure is computed numerically.
The desired steadiness of the long time behavior improves with the width of the band, as the
first row in Fig. \ref{fig: IdealBandComparison} reveals. In Fig. \ref{fig: IdealBandComparison}a, a favorable
case with a wide band is presented and compared with a less favorable case
in Fig. \ref{fig: IdealBandComparison}b, which has the same conduction band minimum but a narrower conduction band.
After a short transient, a comparison of the relative chemical-potential standard deviation  $\sigma(t)$, as defined in Eq. (\ref{eq:AverageChemicalPotential})
and plotted in this graph, shows a clear advantage in the use of wider
bands. For instance, in Fig. \ref{fig: IdealBandComparison}a, $\sigma(t) \sim 10^{-4}$ in the stationary (long time) regime, about 10 times smaller than in Fig. \ref{fig: IdealBandComparison}b.

Besides, after a transition time of order $\sim 5000 t_0$, all the characteristic magnitudes of the system
shown in Fig. \ref{fig: IdealBandComparison}a vary slowly enough in time to properly view the resulting flow regime
as quasi-stationary. In fact, the leak is so slow that other
processes which limit the lifetime of the condensate (such as condensate decay due to inelastic collisions) operate on a shorter
time scale.

If the chemical potential reaches and goes below the bottom of the conducting band  in a relatively short time, then a transition occurs to an essentially confined
situation where the leaking is exponentially small, corresponding to an atom transmission probability $T(L_{\rm lat})\propto \exp(-\kappa L_{\rm lat})$, where $\kappa \propto \sqrt{E_{\rm min}-\mu}$, with $E_{\rm min}$ the bottom of the conducting band.
This situation whereby one soon reaches the regime $\mu < E_{\rm min}$ is not interesting for our purposes because we need some appreciable
flux in order to form a useful black-hole configuration. As a consequence, we typically consider condensates so large that the time needed to reach the bottom of the conduction band is longer than the typical lifetime of the condensate.

A further argument can be invoked in favor of wide bands. In
virtually all the cases we have addressed, ${\rm Re}\,\bar{\mu}(t)$ drops until it almost reaches the bottom of the conduction band, where leaking
is slow. In the resulting regime, the density in the lattice is very small, $g\bar{n}_r\ll \hbar^2k_L^2/m$, where $\bar{n}_r$ is the mean density in the optical lattice, as defined precisely in subsection \ref{qsr}.
It is shown in Appendix \ref{app:nonlinearol} that, in this regime of low interactions, the width of the conduction
band for the linear perturbations (whose evolution is governed by the BdG equations)
is very close to that obtained for the linear Schr\"odinger equation, and the corrections are there computed.
This means that the optical lattice acts like a low-pass
filter, the band width being the equivalent of the cutoff frequency.
The higher the cutoff, the wider is the transmission band of the
lattice. As a consequence, fluctuations on the subsonic side are transmitted away through the optical lattice, which reduces the space fluctuations in
the chemical potential.

Another interesting trend can be observed in the second row of Fig. \ref{fig: IdealBandComparison}. When placing
$\mu_{0}$ slightly below the top of the conduction band or in the
first gap, as Fig. \ref{fig: IdealBandComparison}d illustrates, the leaking occurs faster but the reached regime presents much larger fluctuations than in the other cases shown in Fig. \ref{fig: IdealBandComparison}. For the purpose of keeping $\sigma(t)\ll 1$, a more favorable situation
for the chemical potential is shown in Fig. \ref{fig: IdealBandComparison}c. There, for the
same length $d$ as in  Fig. \ref{fig: IdealBandComparison}d but a higher barrier amplitude $V_{\infty}$, the chemical potential is initially placed in the final conduction band. This case clearly yields smaller fluctuations,
even though the width of the conduction band is smaller. This shows that, besides having wide bands, one also needs that $\mu_0$ be placed close to the bottom of the final conduction band in order to obtain a more favorable quasi-stationary regime.
We discuss this feature in more detail in Sec. (\ref{qsr}).

Within the nearly-free particle approximation ($v\ll 1$), the bottom and top of the first conduction band are given by the relations:
\begin{eqnarray}
E_{\rm min}(v) & =  & 8 E_R (v-v^2+O(v^4)) \nonumber \\
E_{\rm max}(v)& =  & E_R (1+4v-2v^2+O(v^4)) \, ,\label{eq:top-bottom}
\end{eqnarray}
where $E_R \equiv \hbar^2 k_L^2 / 2m$ is the recoil energy of the optical lattice. Given that $k_L=\pi / d$, the condition that the initial chemical potential lies within the final conduction band, i.e.,
\begin{equation}
\label{eq:min-mu-max}
E_{\rm min}(v) < \mu_{0} < E_{\rm max}(v) \, ,
\end{equation}
is guaranteed to be satisfied if
\begin{equation}
\label{eq:m-mu-M}
8E_R v <  \mu_{0} < E_R \, .
\end{equation}
The left inequality is just
\begin{equation}\label{eq:v-min}
\frac{V_{\infty}}{2}<\mu_0~,
\end{equation}
while the right inequality can be rewritten as:
\begin{equation}
\label{eq:d-max}
d < \frac{\pi}{\sqrt{2}}\xi_0 \, .
\end{equation}
Equations (\ref{eq:v-min}), (\ref{eq:d-max}) express a sufficient condition to satisfy Eq. (\ref{eq:min-mu-max}).

%

\subsubsection{Non-adiabatic effects} \label{adiabatic}

\begin{figure}[tb!]
\includegraphics[width=1\columnwidth]{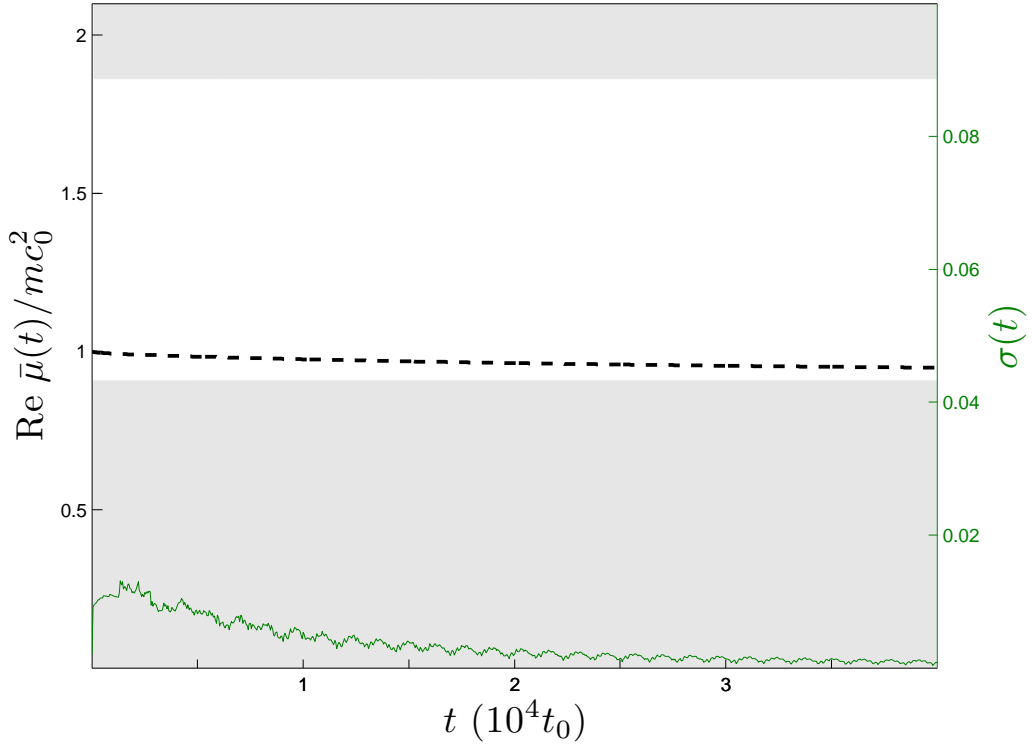} \caption{
Fast barrier lowering.
Same parameters as in Fig. \ref{fig: IdealBandComparison}a except for $\tau=t_{0}$, which is very small to be observed on this scale.}
\label{fig:IdealLambda1200Tau1Vf2}
\end{figure}

In the favorable situation shown in Fig. \ref{fig: IdealBandComparison}a,
the dependence on $\tau$ is not important as long as
$\tau \gg  t_{0}$.
A simulation is presented in Fig. \ref{fig:IdealLambda1200Tau1Vf2}
which shows that, in the fast regime $\tau \sim  t_{0}$, and due to the high-frequency excitations
induced by the short lowering time scale,
$\sigma(t)$ remains higher than in the adiabatic case of Fig. \ref{fig: IdealBandComparison}a, corresponding to a simulation with the same parameters but with $\tau \gg  t_{0}$.

\subsection{Quasi-stationary regime} \label{qsr}

From the discussion in the previous subsection (\ref{sub:AnalysisOfSimulations}),
and particularly through the situation shown in Fig. \ref{fig: IdealBandComparison}a,
we have learned that, in a setup with typical experimental values and that presents wide bands, initial chemical potential close to the bottom of the final conduction band and adiabatic evolution ($\tau\gg t_{0}$), the system evolves towards a quasi-stationary
regime in times shorter than the typical lifetime of a condensate.

The quasi-stationary regime can be defined as that in which
Re$\,\mu(x,t)$ is essentially uniform [$\sigma(t)\ll1$] and its global time variations take place
on a time scale of the order of or greater than the condensate lifetime. We focus our study on the most favorable quasi-stationary scenarios, which here we identify with those satisfying $\sigma(t)\sim 10^{-3}-10^{-4}$.

The achievement of this regime is of general interest as a scenario for the study of atom quantum transport since it provides a quasi-stationary supersonic current. Related directly to the main topic of this work, one may expect spontaneous Hawking radiation to be detectable above a quasi-stationary background with small spatial fluctuations.

In the present section we discuss several features of the condensate
wave function for the quasi-stationary regime. For illustrative purposes,
all the graphs considered in this subsection have been obtained
for a system with the parameters of Fig. \ref{fig: IdealBandComparison}a,
which are sufficiently representative.

We can extend the concept of local sound and flow velocities to time-dependent situations [see Chapter \ref{chapter:Introduction} for the notation]:
\begin{eqnarray}
v(x,t) & \equiv  & \frac{\hbar\partial_{x}\theta(x,t)}{m} \, ,\nonumber \\
c(x,t) & \equiv  & \sqrt{\frac{gn(x,t)}{m}} \, ,\label{eq:SpeedFlowSound}
\end{eqnarray}
The spatial variations of both velocities are small
in the subsonic and supersonic regions, but not in the lattice, where we note that $c(x,t)$ must not be regarded as the actual lattice sound speed; see Appendix \ref{app:nonlinearol} for the computation of the actual sound speed.

\begin{figure}[tb!]
\includegraphics[width=1\columnwidth]{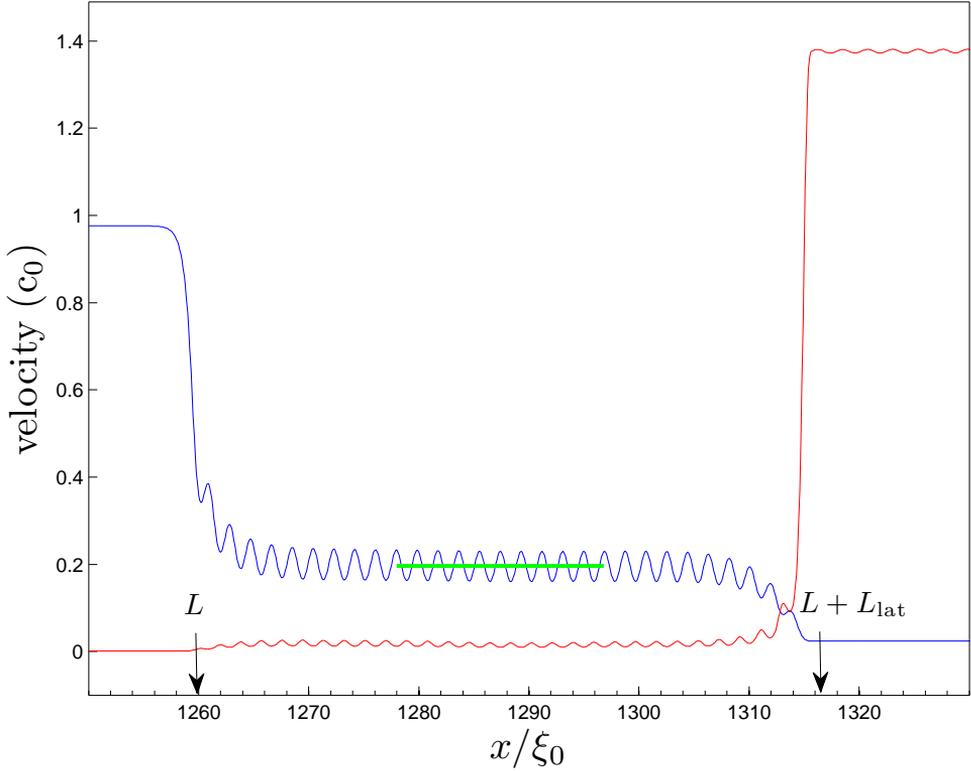} \caption{
Local flow velocity (red) and local speed of sound (blue) at a late time
$t=4\times 10^4\, t_{0}$. The horizontal green segment shows the speed of sound in the optical lattice, computed using (\ref{eq:PerturbativeSound}) with the coefficients there appearing computed numerically. Within this finite lattice the mean density is $\bar{n}_r(t)$, computed by dropping 10 lattice
sites at each end of the lattice.
System parameters are as in Fig. \ref{fig: IdealBandComparison}a.}
\label{fig:IdealCVProfile}
\end{figure}

The profile of both quantities computed at a time $t=4\times 10^4\, t_{0}$ for the simulation of Fig. \ref{fig: IdealBandComparison}a is shown in Fig. \ref{fig:IdealCVProfile}. The subsonic zone shows an essentially flat (uniform) density and flow speed profile in the sense that the spatial fluctuations are on the order of $\sim 10^{-4} n_0$ for the density and $\sim 10^{-3} c_0$ for the flow speed, too small to be observed in Fig. \ref{fig:IdealCVProfile}. In Appendix \ref{app:confinedconfi}, we give an approximate analytical formula [Eq. (\ref{eq:effectivepsi0})] for the wave function of the confined condensate which fits the numerical results within this level of accuracy, but replacing $\xi_0$ by the healing length defined by the quasi-stationary value of the subsonic density $gn_u\simeq \bar{\mu} \simeq E_{\rm min}$. This good agreement reflects the low value of the flow velocity in the upstream region.

\begin{figure}[tb!]
\includegraphics[width=1\columnwidth]{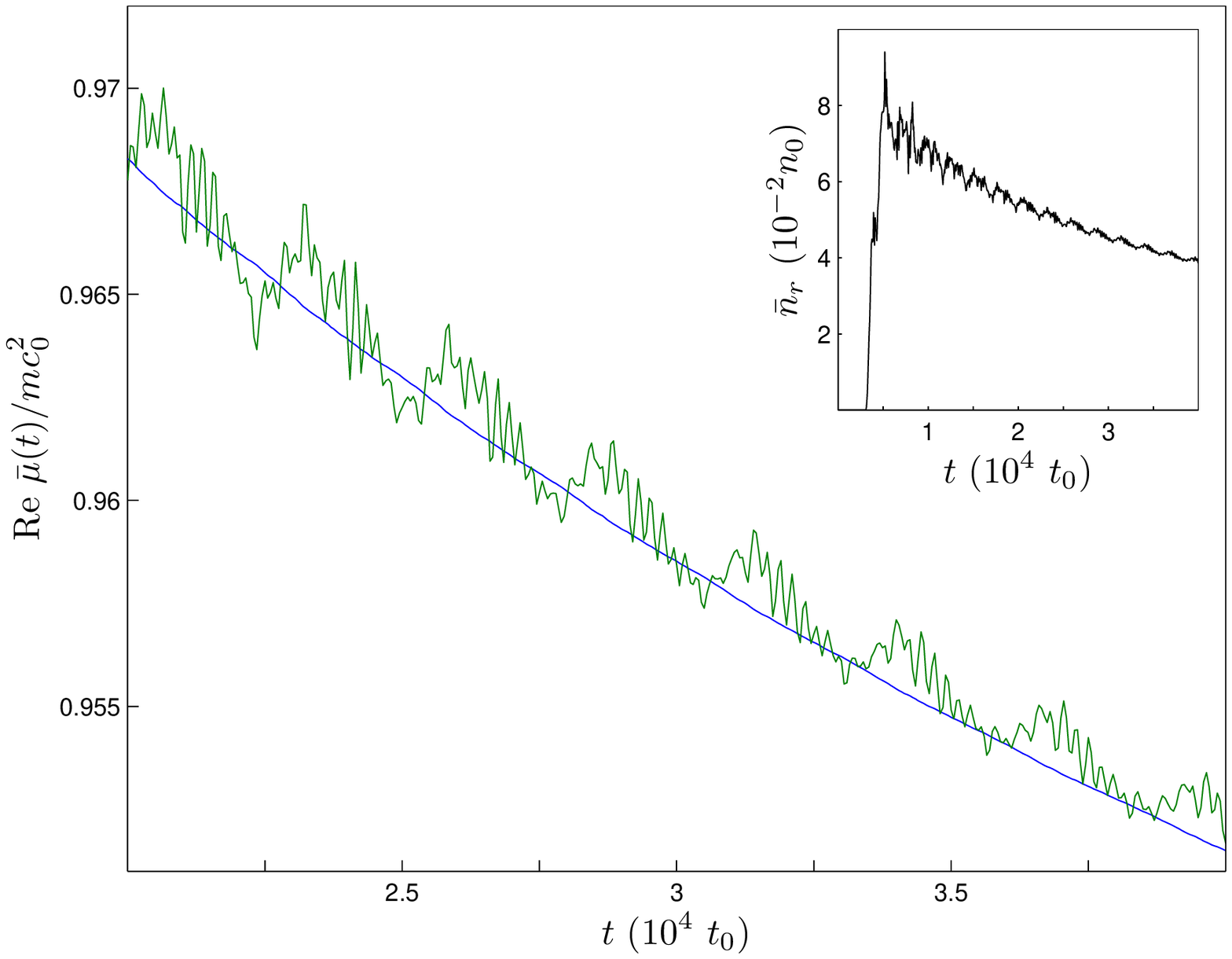}
\caption{System parameters as in Fig.
\ref{fig: IdealBandComparison}a.
Blue: real part of the space-averaged chemical potential $\bar{\mu}(t)$
[see Eq. (\ref{eq:AverageChemicalPotential})]. Green: chemical potential for zero Bloch momentum
computed from the results of Appendix \ref{app:nonlinearol} and using
$\bar{n}_{r}(t)$ as the mean density, which is precisely defined in the previous figure and
plotted in the present inset.
}
\label{fig: LatticeFig}
\end{figure}

On the other hand, in the deep central region of the optical lattice,
Bloch's theorem is satisfied. We introduce the space-averaged density $\bar{n}_{r}(t)$ by averaging $n(x,t)$ over the optical lattice after excluding 10 lattice sites at each end of the lattice.
That average density, combined with the optical lattice parameters, yields an effective sound velocity [see Eq. (\ref{eq:PerturbativeSound})] that is plotted as a horizontal green segment spanning the averaged region in Fig. \ref{fig:IdealCVProfile}.
It can be clearly seen that, within the bulk of the optical lattice, the flow is subsonic, with the horizon on its right edge.
In the quasi-stationary regime, $\bar{n}_{r}(t)$
decreases at a rate comparable to the inverse lifetime of the condensate,
as the inset in Fig. \ref{fig: LatticeFig} shows.

A check on the approximate validity of Bloch's theorem in the presence of non-linear corrections, for the central part of the optical lattice
and in the quasi-stationary regime, is also shown in Fig. \ref{fig: LatticeFig}.
In this graph, the real part of $\bar{\mu}(t)$
is compared with the
time-dependent chemical potential computed using $\bar{n}_{r}(t)$ and assuming zero Bloch momentum, as explained in the first paragraphs of Appendix \ref{app:nonlinearol}. The good agreement between the two curves suggests that the condensate is flowing with very small Bloch momentum.

At the boundaries of the optical lattice there are strong variations of the density due to the
the matching between the vastly different densities found on both the
subsonic and the supersonic sides.

\begin{figure}[tb!]
\includegraphics[width=1\columnwidth]{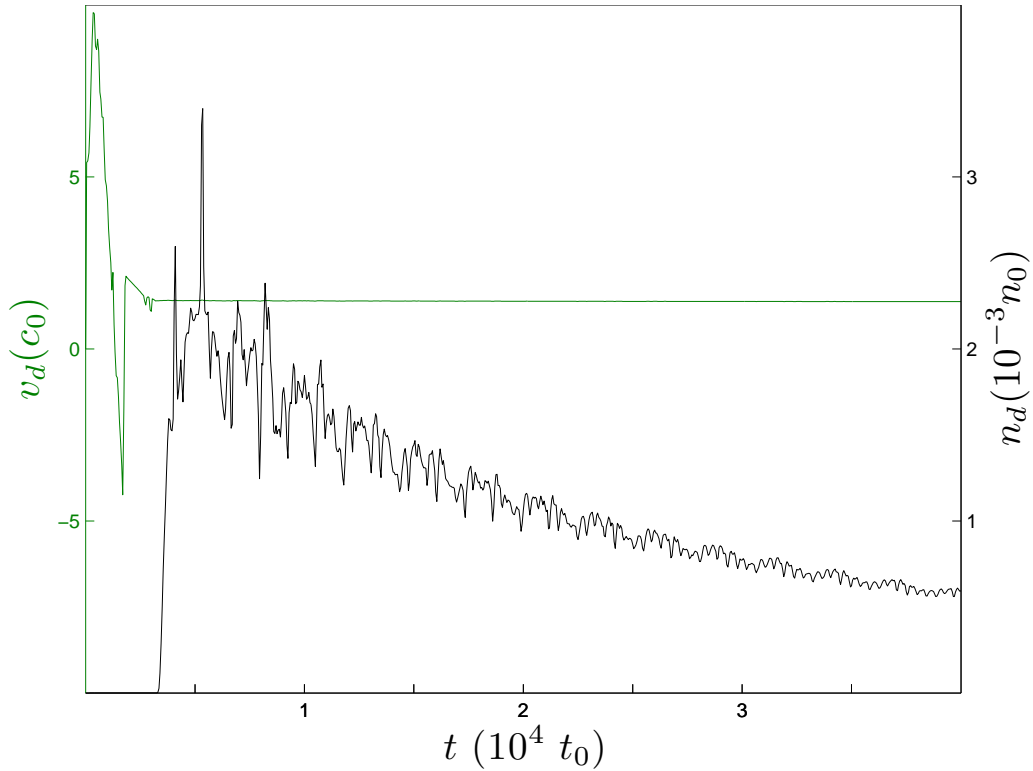} \caption{Time evolution of the mean density and the mean flow
velocity on the supersonic side. The averages are taken in the sense of integrals along the downstream region where the profile is essentially flat. System parameters are as in Fig. \ref{fig: IdealBandComparison}a.}
\label{fig:IdealTimenvD}
\end{figure}

Downstream, once in the quasi-stationary regime, both density and
flow speed profiles are almost uniform. This is hinted at in Fig. \ref{fig:IdealCVProfile} but not shown explicitly. Indeed, part of the non-flat behavior is due to small spurious reflections; see Appendix \ref{app:numerical} for details.
Due to this quasi-homogeneity, we can characterize the density and flow speed in the supersonic region by their average values, $n_d(t)$ and $v_d(t)$, whose time evolution is plotted in Fig. \ref{fig:IdealTimenvD}. The downstream density is much smaller than in the subsonic region so interactions in the supersonic region are negligible. This implies, via conservation of the chemical potential in the quasi-stationary regime, that the supersonic flow speed is constant since the quasi-stationary chemical potential is fully transformed into kinetic energy. On the other hand, the small supersonic density decays with time as the reservoir is depleted, but the process is such that, at each instant, the density profile remains essentially uniform. Even if the density were highly non-homogeneous, this would not pose a problem for computing the BdG solutions since the supersonic Mach number satisfies $v_d/c_d\gg 1$ and then the BdG solutions are the non-interacting plane-wave solutions, corresponding to supersonic scattering channels with vanishing sound speed.

The emission rate per particle $\Gamma(t)$ (not shown), as computed from Eq. (\ref{eq:EmissionRate}), is practically identical to the product $n_d(t) v_d(t)$, except for a numerical factor corresponding to the instantaneous total number of particles, which in the quasi-stationary regime is practically constant. This emission rate gives us the typical time scale for the variation of the number of particles of the system, which is approximately the time scale for the variation of $\mu$. In the quasi-stationary regime considered here, it is $\sim 10^{-6}-10^{-7}~t^{-1}_0$, from which we infer that the typical variation time of the chemical potential is $\sim 10^{6}-10^{7}~t_0$, much longer than the lifetime of the condensate $\sim10^{4}-10^5~t_0$.

We conclude from the results presented along this chapter that the realization of the quasi-stationary regime needs two fundamental ingredients: the existence of a band structure and the presence of interactions. Without a band structure as that provided by the optical lattice, the condensate would continue leaking through the barrier at a fast rate. On the other hand, the presence of interactions (as reflected in the fact that $\partial \mu / \partial n \neq 0$) allows the condensate to stabilize its flow near the bottom of the conduction band. If the interactions in the subsonic region were negligible, the condensate would empty quickly (if $\mu_0$ lied in the final conducting band) or it would remain confined (if $\mu_0$ lied in the final gap).

We have produced some movies to observe such trends, where we consider simulations with the same parameters of Fig. \ref{fig: IdealBandComparison}a but now $L=10 \, \mu\text{m}$ and $N=250$. The qualitative conclusions are similar. In \href{https://www.youtube.com/watch?v=abKDOO5njIo}{Video 1}, we see the time evolution of the density of a condensate confined by an ideal optical lattice of $30$ barriers. We see that the system achieves the desired quasi-stationary regime. On the other hand, in \href{https://www.youtube.com/watch?v=XS1tcpd3aPk}{Video 2}, we introduce the same potential but with just a single barrier, i.e., $n_{\rm osc}=1$. We observe that the fluid leaks faster through the single barrier because the potential does not provide a conduction band that effectively reduces the leaking of the reservoir. This conclusion applies to the scenarios here considered, consisting of an initially confined condensate. In other approaches, such as that of Ref. \cite{Kamchatnov2012}, the condensate is projected onto a potential barrier and a quasistationary black-hole configuration is also eventually reached (the resulting black hole is similar to the delta barrier configuration discussed in Sec. \ref{subsec:1delta}).

The interaction plays the additional role of providing relaxation channels whereby the condensate lowers its energy while some collective modes are excited. The existence of Landau instabilities (see Appendix \ref{app:nonlinearol}) when $\mu_0$ lies well above $E_{\rm min}$ can be clearly observed in the upper right corner of Fig. 5 of Ref. \cite{Wu2003}, whose chosen parameters are similar to those of the present work. The low value of the critical velocity helps to understand the small value of the condensate Bloch momentum which we infer from the numerical results shown in our Fig. \ref{fig: LatticeFig}. The appearance of instabilities can also be viewed as responsible for the fast lowering of the chemical potential after being initially prepared above the final conduction band, as shown in Fig. \ref{fig: IdealBandComparison}d. This interpretation is consistent with the relatively large values found for $\sigma(t)$ when $\mu_0$ is considerably above $E_{\rm min}$.

\section{Gaussian-shaped optical lattice}
\label{sec:Gaussian-shaped}

Here we perform the same analysis as in the previous section
but using a more realistic optical lattice which includes a Gaussian envelope
\cite{Fabre2011,Cheiney2013EPL,Cheiney2013PRA}:
\begin{equation}
V(x,t)=V(t)\cos^{2}\left[k_L(x-L)\right]
\exp\left[-2\left(\frac{x-L}{\tilde{w}}\right)^2\right]
\label{eq:actualpotential}
\end{equation}
where $\tilde{w}=w/\cos(\theta/2)$ (with $w$ the laser beam width and $\theta$ the angle between the laser beams)
plays the role of an effective lattice length, similar to $L_{\rm lat}$ in the ideal optical lattice. The time dependence of $V(t)$ is the same as in Eq. (\ref{eq:TDPotential}). Usually, $\tilde{w}$ varies in a range $10-200~\mu\text{m}$. Here, $L$ is the position of the maximum of the lattice Gaussian envelope. For consistency, we replace the hard wall at $x=0$ by a Gaussian barrier of the type $V_L(x)=U\exp(-2x^2/w_L^2)$ with $w_L=2~\mu\text{m}$ and $U\gg \mu_0$ in order to simulate a more realistic confinement on the left side. This time-independent confining potential must be added to the time-dependent potential of the optical lattice (\ref{eq:actualpotential}). We refer to Appendix \ref{app:confinedconfi} for a detailed description of the initial confinement.

\begin{figure}[tb!]
\includegraphics[width=1\columnwidth]{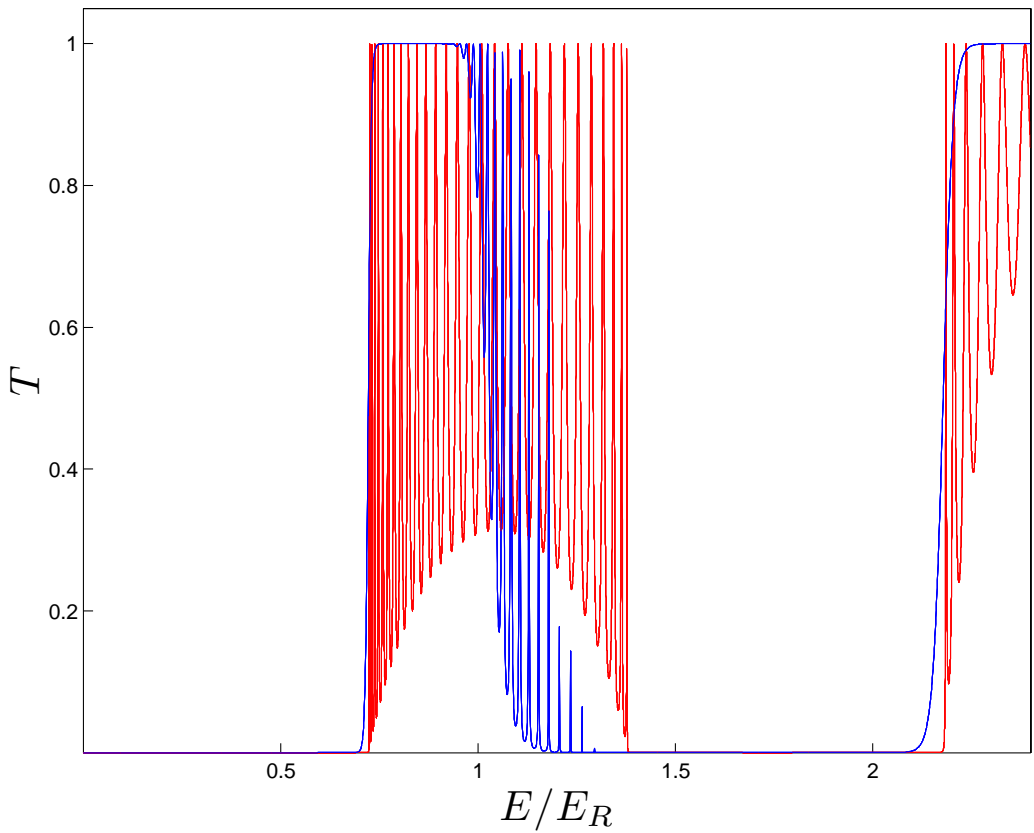}
\caption{
Single atom transmission probability $T(E)$, as a function of energy, for a realistic (Gaussian-shaped) optical lattice
(blue) with the instantaneous value $V(t)=1.6 E_R$ [see Eq. (\ref{eq:actualpotential})] and $\tilde{w}=30 d$, with $E_R$ defined after Eq. (\ref{eq:top-bottom}), and
for an ideal (flat) optical lattice (red) with same amplitude $V(t)$ and $n_{\rm osc}=30$.
}
\label{fig:RealisticIdT}
\end{figure}

\begin{figure}[tb!]
\includegraphics[width=1\columnwidth]{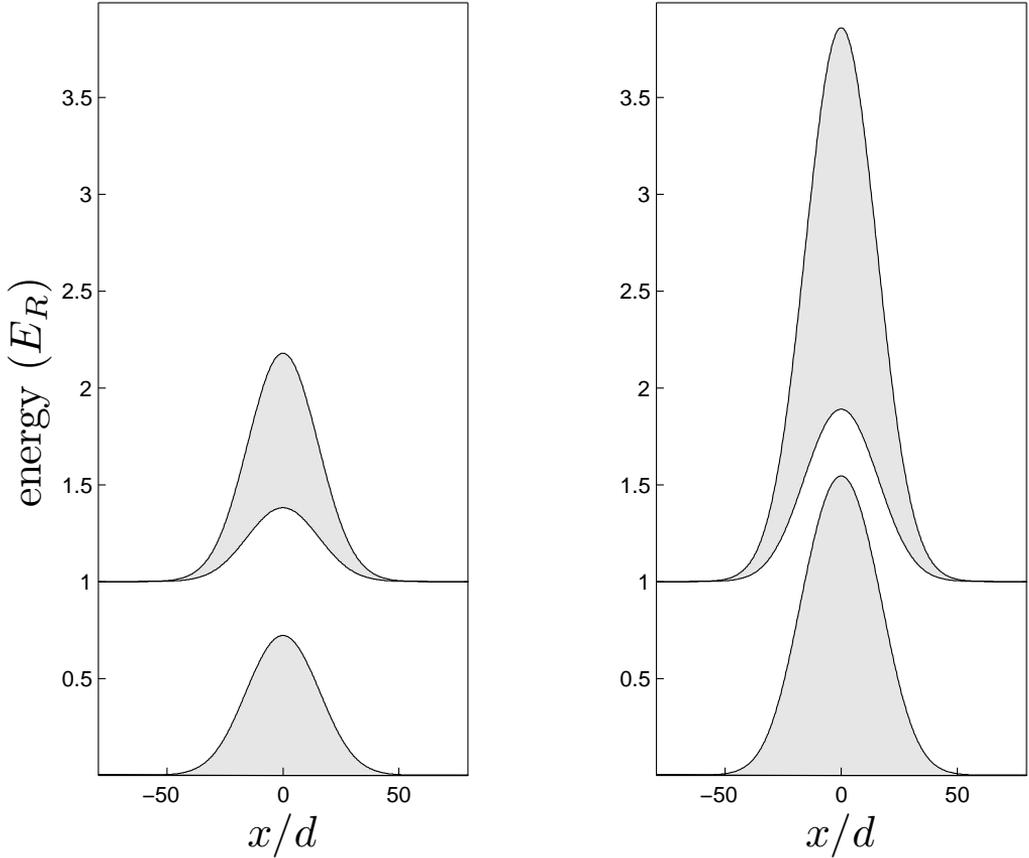}
\caption{Plot of the spatially dependent energy bands for a realistic optical lattice with $\tilde{w}=30 d$. We use the same color criterion as for the band structure of Fig. \ref{fig: IdealBandComparison}. Left panel: the instantaneous value of the amplitude is $V(t)=1.6 E_R$, which corresponds to the case of Fig. \ref{fig:RealisticIdT}. Right panel: the instantaneous value of the amplitude is $V(t)=4 E_R$. In this case, the first conduction band becomes ineffective, as can be expected from the plot, since there is always a transmission gap for the energies of interest.
}
\label{fig:SpaceBands}
\end{figure}

In the ``adiabatic'' regime ($\tilde{w}\gg d$), the solutions of the linear Schr\"odinger equation
for this type of potentials show features similar to those found for an ideal optical
lattice with the same instantaneous amplitude $V(t)$, as can be seen in Fig. \ref{fig:RealisticIdT}, where the transmission bands are compared.
If we focus on the long-time limit ($V(t)=V_{\infty}$), the realistic potential acquires the form
\begin{equation}
V(x)=V_{\infty}(x)\cos^{2}\left[k_L(x-L)\right] \, ,
\end{equation}
where
\begin{equation}\label{V-infty-x}
V_{\infty}(x)=V_{\infty}\exp\left[-2\left(\frac{x-L}{\tilde{w}}\right)^2\right]
\end{equation}
is a slowly varying function. Then, we have a locally ideal optical lattice at each point of the space with amplitude $V_{\infty}(x)$. Bloch's theorem can also be applied locally and the resulting local band structure is plotted as a function of space in Fig. \ref{fig:SpaceBands}. A similar type of reasoning was already used in Refs. \cite{Carusotto2000,Santos1998a,Santos1999}. The left panel presents the setup whose single atom transmission is plotted in Fig. \ref{fig:RealisticIdT}. Since the bottom of the lowest lattice conduction band is an increasing function of the periodic potential amplitude, the bottleneck for transmission across the realistic lattice occurs at the center of its Gaussian envelope. This fact explains the accurate coincidence between the bottom of both conduction bands shown in Fig. \ref{fig:RealisticIdT}. We also see that, for $E>E_R$ [defined after Eq. (\ref{eq:top-bottom})], the particle encounters a gap somewhere along the Gaussian lattice, and this explains why in Fig. \ref{fig:RealisticIdT} the transmission begins to decay for $E>E_R$. For $E_{\rm min}(v)<E<E_R$, the setup shows a plateau of essentially perfect atom transmission. The absence of interference oscillations in this region is due to the adiabatic variation of the lattice envelope. The right panel of Fig. \ref{fig:SpaceBands} presents the qualitatively different case $E_R<E_{\rm min}(v)$. From the previous arguments, we expect not to find a conduction band, as can be numerically confirmed. We conclude that, in order to have a well defined conduction band for the realistic lattice, the condition $E_R>E_{\rm min}(v)$ is required, which implies $V_{\infty}<2.33~E_R$.

Combining all these considerations, we reach the conclusion that a necessary condition for achieving a quasi-stationary regime is $E_{\rm min}(v)<\mu_0$. We also require $\mu_0 < E_R$ to avoid having $\mu_0$ lying too high above $E_{\rm min}(v)$, which, as found for the ideal lattice, tends to generate relatively high values of $\sigma(t)$. This last inequality is equivalent to Eq. (\ref{eq:d-max}).

We can divide the space into three zones as in the ideal optical lattice case. In the quasi-stationary regime, both the subsonic and supersonic zones are located where the Gaussian envelope amplitude is negligible compared to the chemical potential, i.e., where
\begin{equation}
V_{\infty}(x) \ll \, {\rm Re} \, \bar{\mu}(t)
\label{eq: v-inf-ll-mu}
\end{equation}
[see Eqs. (\ref{eq:AverageChemicalPotential}) and (\ref{V-infty-x})]. In order to have a well delimited subsonic side, we require $L\gg \tilde{w}$. The optical lattice region corresponds to the complementary of the subsonic and supersonic zones.

\begin{figure}[tb!]
\includegraphics[width=1\columnwidth]{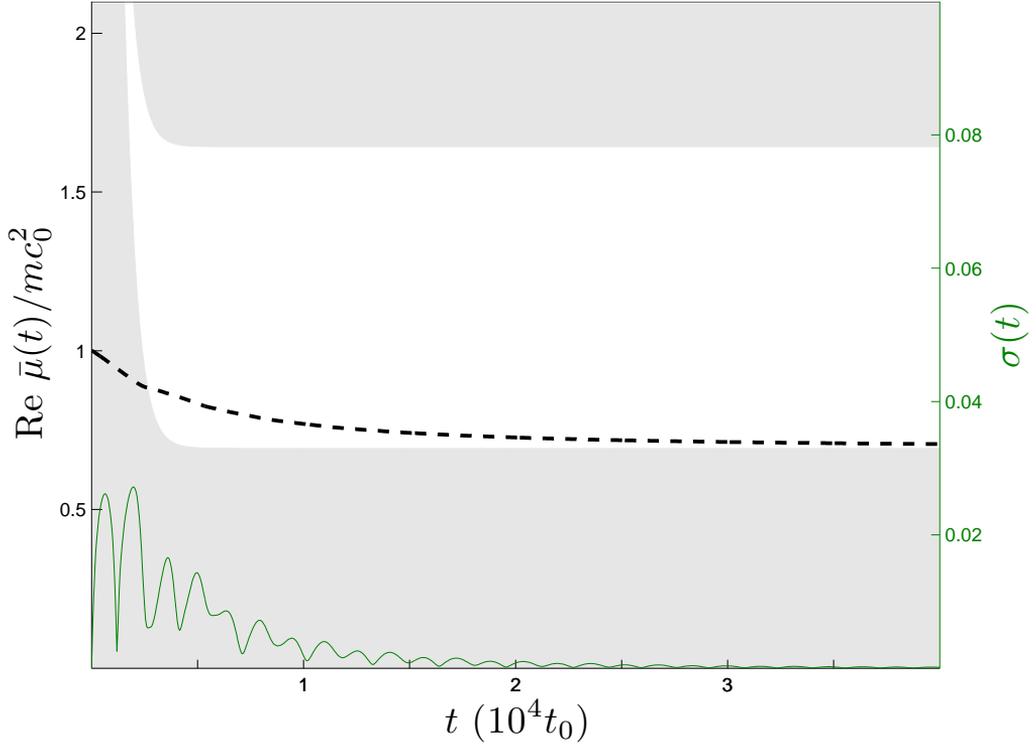} \caption{Time evolution of the real part of the chemical potential and its fluctuation spread in a realistic optical lattice.
The parameters are $\tilde{w}=50~\mu\text{m}$, $d=600~\text{nm}$, $\tau=500~t_{0}$, $V_{\infty}=1.5~ mc_{0}^{2}$ and we have taken $\xi_0=0.3053~\mu\text{m}$. The confinement parameters are $N=9161$, $L=420~\mu\text{m}$, and $\omega_{\rm tr}=2\pi\times 4~\text{kHz}$.}
\label{fig:RealisticBand}
\end{figure}

The requirements of quasi-stationarity are similar to those formulated for the ideal optical lattice. Specifically, the quasi-stationary regime
requires broad conduction bands, an initial chemical potential close to the bottom of the final conduction band,
and a barrier amplitude that evolves not very fast. We also find that the condensate leaks relatively fast until ${\rm Re}~\bar{\mu}(t)$ approaches the bottom of the conduction band. All these features can be observed in Fig. \ref{fig:RealisticBand}, which is the Gaussian-envelope equivalent of Figs. \ref{fig: IdealBandComparison}-\ref{fig:IdealLambda1200Tau1Vf2}. We see that we reach a quasi-stationary state in which $\sigma(t)\sim 10^{-4}$. The bands in Fig. \ref{fig:RealisticBand} are computed as in the ideal case, assuming a uniform barrier amplitude $V(t)$. As noted when discussing Fig. \ref{fig:RealisticIdT}, the positions of the bottom of the ideal and the realistic conduction (or transmission) bands are very similar, so the lower threshold of the transmission band is still a good reference to discuss the evolution of $\bar{\mu}(t)$.

We notice that in the Gaussian case, the condensate apparently leaks from the beginning of the simulation. What is actually happening is that the chemical potential begins to decrease due to the initial expansion of the condensate towards the neighboring, low-amplitude region of the Gaussian optical lattice, even when the actual leaking (towards the right side of the Gaussian envelope) is not yet occurring. This process can be observed in the simulation later presented in \href{https://www.youtube.com/watch?v=b9YA-Efd4F8}{Video 3} at the end of Section \ref{sub:location-sonic-horizon}. The situation contrasts with that shown in Fig. \ref{fig: IdealBandComparison}, where the condensate only begins to leak when the chemical potential is placed within the conduction band due to the sharp envelope of the optical lattice.

\begin{figure}[tb!]
\includegraphics[width=1\columnwidth]{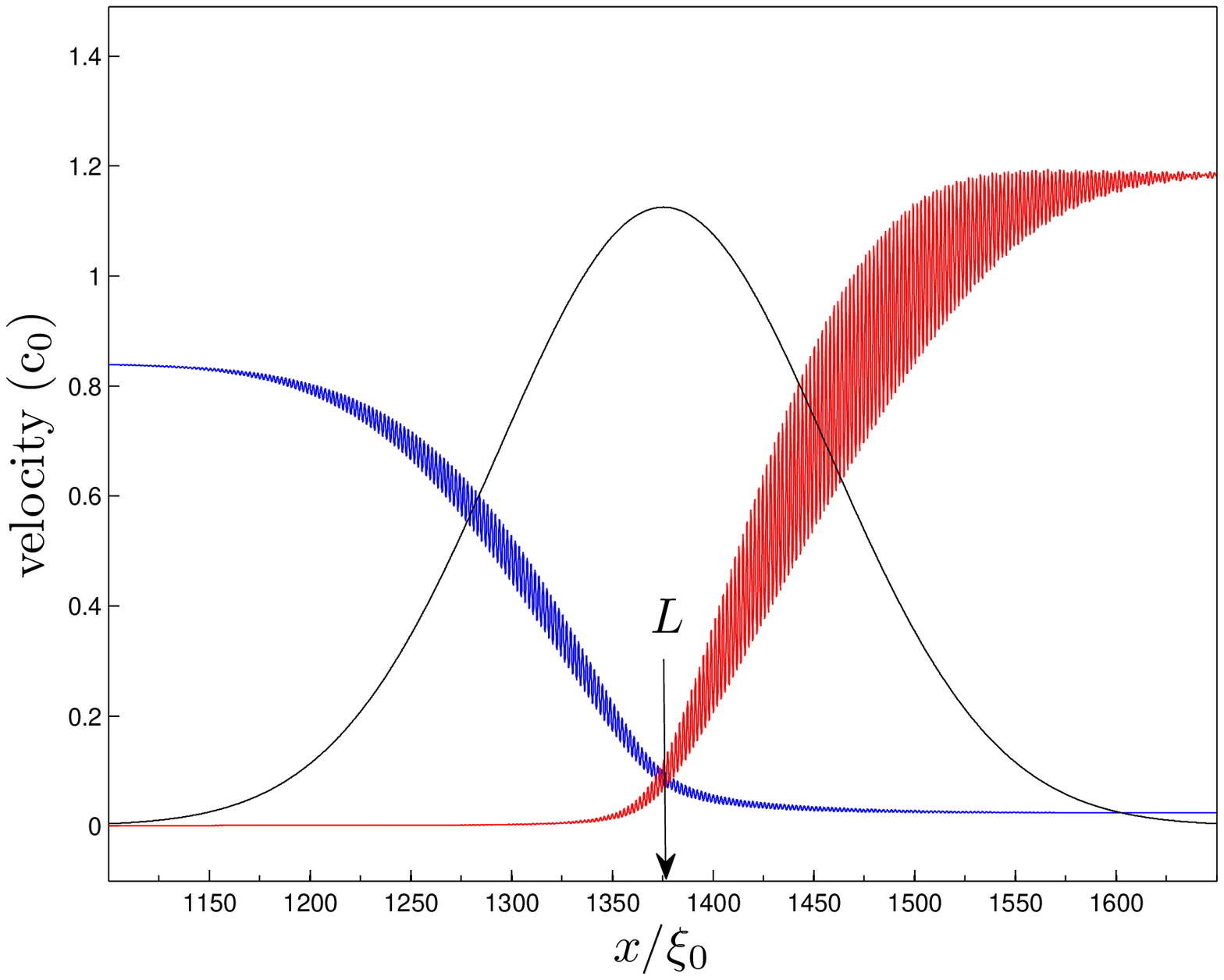} \caption{
Local flow velocity (red) and local speed of sound (blue)
at $t=4\times 10^4\, t_{0}$. System parameters are as in Fig. \ref{fig:RealisticBand}.}
\label{fig:RealisticCVProfile}
\end{figure}

\begin{figure}[tb!]
\includegraphics[width=1\columnwidth]{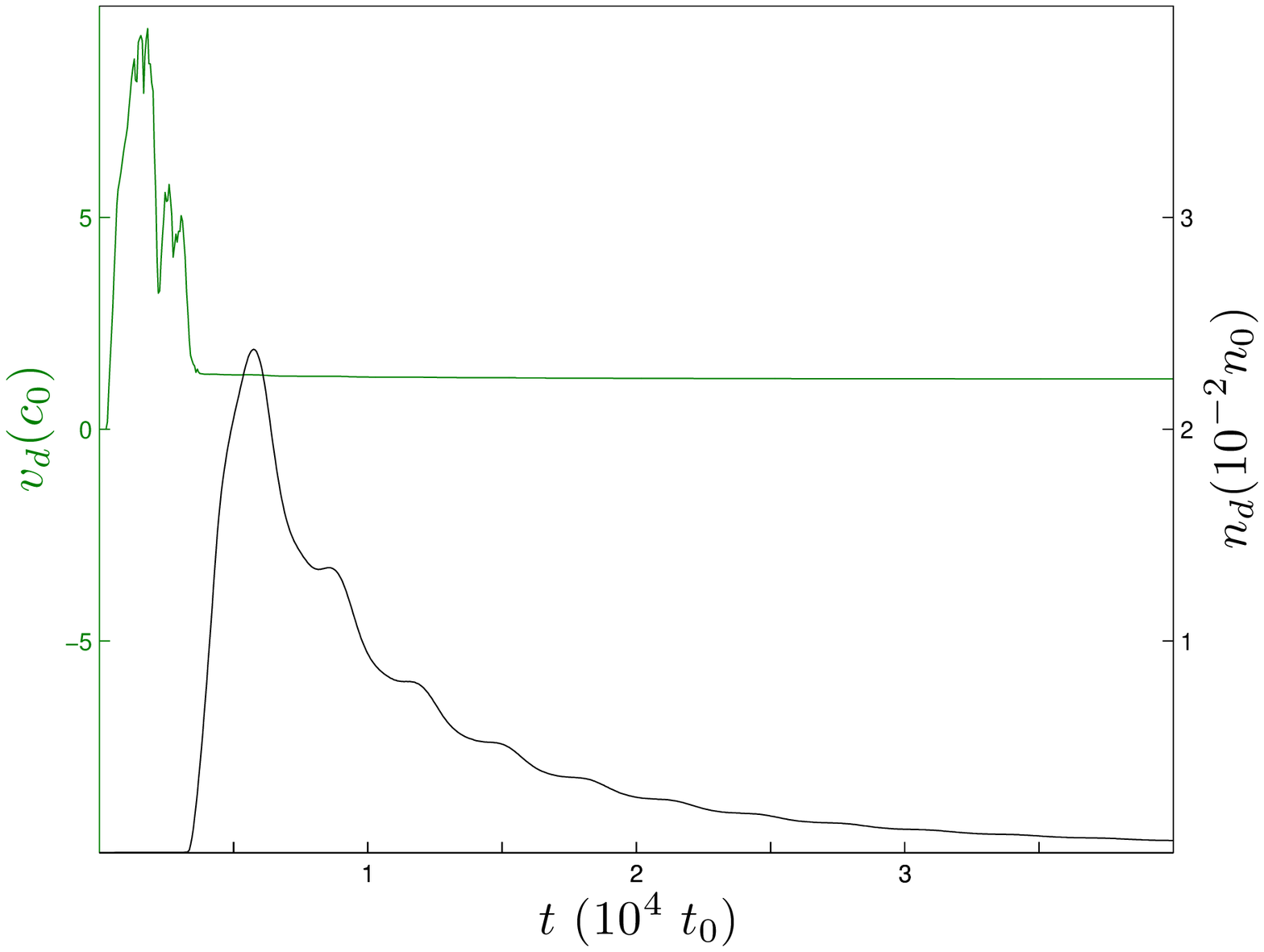} \caption{
Time evolution of the mean density and the mean flow
velocity on the supersonic side for a condensate emitting through a Gaussian-shaped optical lattice. The means are taken over the entire downstream region, see Fig. \ref{fig:IdealTimenvD}. System parameters are as in Fig. \ref{fig:RealisticBand}}
\label{fig:RealisticTimenvD}
\end{figure}

We also study the corresponding quasi-stationary state. For that purpose, we take a snapshot of the configuration at $t=4 \times 10^4~t_0$ for the parameters in Fig. \ref{fig:RealisticBand}. We compare the profiles of $c(x,t)$ and $v(x,t)$ in Fig. \ref{fig:RealisticCVProfile}, which is the realistic equivalent of Fig. \ref{fig:IdealCVProfile}. In the subsonic confined region, the density and velocity are essentially flat, as in the ideal optical lattice region. The apparently sharper oscillations in the optical lattice region, as compared to those in Fig. \ref{fig:IdealCVProfile}, are due to the different horizontal scales used. The larger oscillations of the flow velocity beyond the horizon with respect to those inside the lattice subsonic region can be explained because of the large difference in space-averaged flow velocities. In the supersonic region, we find again essentially flat profiles for the density and flow velocity, with their time evolution shown in Fig. \ref{fig:RealisticTimenvD}. We note, from the results of the latter plot, that the achieved black-hole configuration could be used to produce a quasi-stationary supersonic current with a very well defined velocity. Then, the general features of this quasi-stationary configuration are similar to those of the ideal case although some interesting new features appear.

\subsection{Location of the sonic horizon and related properties \label{sub:location-sonic-horizon}}

We notice in Fig. \ref{fig:RealisticCVProfile} that the horizon seems to be placed at the maximum of the Gaussian envelope.
Actually, this can be explained on quite general grounds by invoking the properties of the quasi-stationary regime and the adiabaticity condition $\tilde{w}\gg d$, which allows us to think in terms of a local band structures, generated by a periodic potential of local amplitude $V_{\infty}(x)$. Then we can use an adiabatically space-dependent version of Eqs. (\ref{eq:oldmu})-(\ref{eq:olhdmu}) (where the sound speed, atom current and chemical potential are obtained for an infinite optical lattice) by making every parameter slowly dependent on $x$. In particular, we take:
\begin{eqnarray}\label{eq:SEQSS}
s(x)&=& \left[\frac{gn_r(x)}{m^*(x)}\alpha_{0}^{(1)}(x)\right]^{\frac{1}{2}}\nonumber \\
j(x)&\simeq&n_r(x)\bar{v}(x) \\
\mu(x)&\simeq&E_{\rm{min}}(x)+\frac{1}{2}m^*(x)\bar{v}^2(x)+m^*(x)s^2(x)\, , \nonumber
\end{eqnarray}
where we neglect the time dependence because the system is assumed to be already in the quasi-stationary regime.
The local averages for $n_r,\bar{v}$ are taken over several lattice periods. In the quasi-stationary regime, the chemical potential is already close to the bottom of the conduction band, so the perturbative treatment considered in Appendix \ref{app:nonlinearol} is valid. By taking spatial derivatives (denoted by $'$) while noting that the chemical potential is almost uniform, $\mu(x)\simeq \bar{\mu}$, and that $\partial_x j(x)$ can be neglected (as implied by the continuity equation and quasi-stationarity), we arrive at:
\begin{equation} \label{eq:derivative-mu-zero}
0=E'_{\rm{min}}+\frac{1}{2}m^{*'}\bar{v}^2+\frac{\alpha^{(1)'}_0}{\alpha^{(1)}_0}m^*s^2+m^*(\bar{v}^2-s^2)\frac{\bar{v}'}{\bar{v}}\, .
\end{equation}
The quantities $E_{\rm{min}}(x),m^*(x),\alpha_{0}^{(1)}(x)$ depend on $x$ through the amplitude of the envelope, $V_{\infty}(x)$, and they increase with its value, provided that the envelope amplitude is always positive [see Eq. (\ref{m-alpha-1})].
Therefore, the first three terms in the r.h.s. of (\ref{eq:derivative-mu-zero}) have the same sign.

Let us assume that we have a horizon ($s=\bar{v}$) somewhere in the optical lattice. We prove next that a necessary implication is that an envelope maximum or minimum exists at that point. As the sum of the first three terms in Eq. (\ref{eq:derivative-mu-zero}) have the same sign, their sum can only be zero whenever the derivative of the amplitude is zero, i.e., when $V'_{\infty}(x)=0$. In our setup, this means that we have an amplitude maximum at the horizon. A proof of a similar result in the case of a single potential barrier, based in a hydrodynamical approximation, was already given in Ref. \cite{Giovanazzi2004}.

Now we consider the inverse implication. Assume we have $V'_{\infty}(x)=0$ (which in our setup is the case at $x=L$). This implies that the first three terms in (\ref{eq:derivative-mu-zero}) are zero. As a consequence, we are left with two possibilities:
\begin{equation} \label{eq:horizon}
s(L)=\bar{v}(L) \,
\end{equation}
(i.e. a horizon) or $\bar{v}'=0$. By the continuity equation, the second option implies a density minimum, which must be ruled out in our current single Gaussian barrier setup. However, it can be a perfectly feasible result in other experimental contexts (see for instance the discussion at the end of this section).

Finally, we note that Eq. (\ref{eq:derivative-mu-zero}) can also be written as
\begin{equation}
0=E'_{\rm{min}}+\frac{1}{2}m^{*'}(\bar{v}^2+2s^2)+m^*(\bar{v}\bar{v}'+2ss')\, ,
\end{equation}
and, as a corollary of the foregoing analysis, we find that, at the horizon, $s'(L)=-\bar{v}'(L)/2$.

We can further exploit the previous results. For example, we can obtain the value of the density and the current at $x=L$ as a function of $\bar{\mu}$:
\begin{eqnarray}\label{eq:njhorizon}
g n(L)&=&\frac{2}{3}\frac{\bar{\mu}-E_{\rm{min}}}{\alpha^{(1)}_0} \, ,\nonumber \\
j(L)&=&n_r(L)\bar{v}(L)=\left(\frac{2}{3}\right)^{\frac{3}{2}}
\frac{\left(\bar{\mu}-E_{\rm{min}}\right)^{\frac{3}{2}}}{g\alpha^{(1)}_0\sqrt{m^*}} \, ,
\end{eqnarray}
which are very good approximations to the actual numerical values. Using (\ref{eq:njhorizon})
we can arrive at a differential equation for the time evolution of $\bar{\mu}$. First we note that, from the continuity equation, we can write:
\begin{equation}\label{eq:Oden}
\frac{dN_L}{dt}=-j(L) \, ,
\end{equation}
where $N_L$ is the number of particles contained between $x=0$ and $x=L$. As the subsonic region is in the Thomas-Fermi regime (see Appendix \ref{app:confinedconfi}), we can take $\bar{\mu}\simeq gN_{\rm{sb}}/L_{\rm{sb}}$, where $N_{\rm{sb}}$ is the number of particles in the subsonic region and $L_{\rm sb}$ its size. As the density in the optical lattice is small, we can assume $N_L\simeq N_{\rm{sb}}$ (which implies $\bar{\mu} \propto N_L$) and rewrite Eq. (\ref{eq:Oden}) as:
\begin{equation}\label{eq:odemu}
\frac{d\bar{\mu}}{dt}\simeq-C\left(\bar{\mu}-E_{\rm{min}}\right)^{\frac{3}{2}} \, ,
\end{equation}
where $C$ is a positive constant independent of $\bar{\mu}$. The solution of this equation is
\begin{equation}\label{eq:mute}
\bar{\mu}(t)=E_{\rm{min}}+\frac{4}{C^2(t-t_1)^2} \, ,
\end{equation}
(with $t_1$ an integration constant), which fits the numerical data of Fig. (\ref{fig:RealisticBand}) reasonably well. As $C$ decreases roughly with the size of the subsonic region, we see that the larger condensate, the slower it decays, as one logically expects.

Finally, we estimate the value of the Hawking temperature [see Eq. (\ref{eq:TH})] as
\begin{equation}\label{eq:THexp}
k_BT_H=\frac{\hbar}{2\pi}\frac{d}{dx}\left[\bar{v}(x)-s(x)\right]_{x=L}
\end{equation}
where we have replaced the local flow and sound velocities by their optical lattice equivalents. If we note that we operate in the nearly-free atom approximation ($v \ll 1$) and in the weak interaction regime ($gn_r \ll E_R$), and derive twice the third equation (\ref{eq:SEQSS}), we obtain
\begin{equation}\label{eq:THfinal}
k_BT_H \simeq \frac{\hbar}{2\pi\tilde{w}}\sqrt{\frac{3V_{\infty}}{m^*}}(1-v) \, ,
\end{equation}
which gives a good estimate of the numerical value of the Hawking temperature. Noting that $V_{\infty}\sim\mu_0,m^*\sim m$, we obtain $k_BT_H\sim\xi_0\mu_0/\tilde{w}\sim10^{-2}\mu_0\ll\mu_0$. The temperature of the condensate is typically of the order of $\mu_0/k_B$, so we conclude $T_H\sim10^{-2}T\ll T$. Similar estimations for the Hawking temperature for single barrier potentials were obtained in a hydrodynamical approximation in Refs. \cite{Giovanazzi2004,Wuster2007}.

To observe the birth of the black hole and to check that the horizon position naturally evolves towards the maximum of the optical lattice envelope, we have created a movie (\href{https://www.youtube.com/watch?v=b9YA-Efd4F8}{Video 3}) that shows the time evolution of the coarse-grained velocities $(\bar{c},\bar{v})$ of the emitting condensate using the setup parameters of Fig. \ref{fig:RealisticBand}. At long times the predicted coincidence between the sonic horizon and the maximum of the Gaussian envelope in the stationary regime can be clearly observed.

When applied to an ideal optical lattice, the above arguments on the position of the horizon yield no preferred point for the location of the horizon because the envelope is uniform. Actually, in the bulk of the lattice, since $V'_{\infty}=0$ everywhere, the natural outcome [from the discussion leading to Eq. (\ref{eq:horizon})] is $\bar{v}'=0$ everywhere, i.e., the mean velocity and (by quasi-stationarity) the mean density are also uniform, as can be observed in Fig. \ref{fig:IdealCVProfile}. This fact only leaves two options: either the lattice bulk is subsonic or it is supersonic. The latter choice is energetically unstable (see Ref. \cite{Wu2003}) and, as a consequence, the subsonic regime is energetically favored in the bulk of the lattice. In the rightmost region, where the potential is not present, the flow has to be supersonic, so the only
possibility for the horizon is to lie at the right extreme of the lattice, as can be seen in Fig. \ref{fig:IdealCVProfile}.

\section{Preliminary results for Hawking spectrum}\label{sec:preliminaryBdG}

We present here a tentative study of the Hawking spectrum in the quasi-stationary black-hole regime previously described. For that purpose, we compute the $S$ matrix treating the quasi-stationary condensate as a true stationary condensate. From Eqs. (\ref{eq:TDGPOL}) (\ref{eq:LocalChemicalPotential}), we see that:
\begin{equation}\label{eq:quasistationarywavefunction}
\Psi(x,t)=\Psi(x,t_i)e^{-i\int_{t_i}^{t}\mathrm{d}t'\frac{\mu(x,t')}{\hbar}}
\end{equation}
for a any given $t=t_i$. In particular, if $t_i$ is chosen such the system is already in the quasi-stationary regime, taking a given snapshot of the condensate at $t=t_i$ as an actual stationary wave function with chemical potential $\bar{\mu}$ provides a good approximation since $\mu(x,t)\simeq \bar{\mu}$. Formally, we consider the time-dependent equation for the field fluctuations (\ref{eq:TDFieldequation}) adapted to our 1D configuration. Once in the quasi-stationary regime, for times $t>t_i$, we define $\Psi_0(x,t)$ through $\Psi(x,t)\equiv\Psi_0(x,t)e^{-i\frac{\bar{\mu}}{\hbar}(t-t_i)}$. Removing the previous time-dependent phase from the field fluctuation operator, $\hat{\varphi}(\mathbf{x},t) \rightarrow \hat{\varphi}(\mathbf{x},t)e^{-i\frac{\bar{\mu}}{\hbar}(t-t_i)}$, we arrive at:
\begin{eqnarray}\label{eq:TDquasistationaryFieldequation}
\nonumber i\hbar\partial_t\hat{\Phi}&=&M(t)\hat{\Phi},\\
M(t)&=&\left[\begin{array}{cc}G(t) & L(t)\\
-L^{*}(t)&-G(t)\end{array}\right]\\
\nonumber G(t)&=&-\frac{\hbar^2}{2m}\frac{\partial^2}{\partial x^2}+V(x,\infty)+2g|\Psi_0(x,t)|^2-\bar{\mu}\\
\nonumber L(t)&=&g\Psi_0^2(x,t)~,
\end{eqnarray}
$V(x,\infty)$ being the asymptotic stationary lattice potential. Now we can make a Bogoliubov expansion for $\hat{\Phi}$ at $t=t_i$ and compute the time evolution of the Bogoliubov modes. The point is that the variation of the wave function $\Psi_0(x,t)$ is sufficiently slow, compared with the time evolution of the modes with frequency $\bar{\mu}\sigma(t)\ll \hbar\omega \lesssim \bar{\mu}$, for us to approximate $\Psi_0(x,t)\simeq \Psi_0(x,t_i)=\Psi(x,t_i)\equiv \Psi_0(x)$. In this way, those modes see the GP wave function as a stationary background and their time evolution is given by a stationary BdG equation:
\begin{equation}\label{eq:effectiveTIBdG}
\hbar\omega z=M_0z,~M_0=M(t_i),~
\end{equation}
In order to compare the results with the theoretical models, we consider the two asymptotic homogeneous regions (subsonic and supersonic) as ideally semi-infinite and we approximate the wave function in both asymptotic regions as a perfect plane wave. From the results of Sec. \ref{qsr}, we see that the condensate in the plateau of the upstream (subsonic) region can be taken as $\Psi_0(x)=\sqrt{n_u}e^{i\theta_u}$ with $\theta_u$ some constant phase. The value of $n_u$ is related to the average chemical potential as $\bar{\mu}=gn_u$, and gives a sound velocity $c_u=\sqrt{gn_u/m}$. We take the momentum of the plane wave as zero since the upstream velocity is negligible. In the supersonic region, the wave function is approximated as $\Psi_0(x)=\sqrt{n_d}e^{iq_dx}e^{i\theta_d}$, where $\hbar q_d=mv_d$, $n_d,v_d$ being the average downstream density and velocity (see Figs. \ref{fig:IdealTimenvD}, \ref{fig:RealisticTimenvD}). As explained in Sec. \ref{qsr}, the Mach number in the supersonic region is very large and the corresponding BdG solutions are very close to free plane waves.
We observe that the current associated to the previous approximation for the GP wave function is non-conserved but, as explained in Sec. \ref{sub:AnalysisOfSimulations}, this condition is impossible to fulfill.

We note that the previous stationary BdG approximation necessarily fails in the limit $\omega\rightarrow 0$ since in that regime the small time-dependence of the wave function cannot be neglected. In particular, the zero mode of Eq. (\ref{eq:zeromode}) is not anymore a solution to this BdG equation as the GP wave function is not strictly stationary. A possible way to overcome this problem is to make use of the relative BdG equations, considered in Ref. \cite{Macher2009a}, instead of the usual BdG equations here considered. However, as will be explained later in Sec. \ref{sec:NumericalCores}, the zero frequency peak is not interesting for the observation of the spontaneous Hawking effect. Thus, for practical purposes, we can focus only on the central frequency region, where the previous stationary BdG approximation is valid. Thus, in the following, we only consider frequencies $\omega > \omega_{\Lambda}$ where $\omega_{\Lambda}$ is an enforced numerical cut-off of order $\omega_{\Lambda}\sim 10^{-2}mc_0^2$. After following these steps, a BH scenario similar to that considered in Sec. \ref{subsec:BHBEC} is reached. We now compute the corresponding scattering states for a given frequency $\omega$ and obtain the elements of the scattering matrix. As usual in one-dimensional scattering theory, this can be done by matching the numerical BdG solutions in the optical lattice with the plane waves in the asymptotic regions, see Sec. \ref{app:SMatrixBehav} for the specific procedure. The previous effective stationary equation also has an associated conserved current $J_z$ which guarantees that the $S$-matrix computed in this way is pseudo-unitary. However, a numerical problem arises because of the solution with exponential increasing behavior in the optical lattice, which comes from one of the Bloch waves with complex wave vector, see discussion at the end of Appendix \ref{app:nonlinearol}. Due to the relatively large size of the optical lattice, this exploding solution gives rise to singular fundamental matrices within the computer's accuracy, spoiling the computation of the scattering matrix. This problem is known as the {\it large fd problem} \cite{Lowe1995} and appears in fields as different as Anderson localization \cite{Slevin2014} or ultrasonic waves in multilayered media \cite{Lowe1995}. We show how to deal with this problem in Appendix \ref{app:numerical}, using the Global Matrix method (considered in ultrasonic wave problems) or the QR decomposition (used in Anderson localization problems).
\begin{figure}[!tb]
\begin{tabular}{@{}cc@{}}
    \includegraphics[width=0.5\columnwidth]{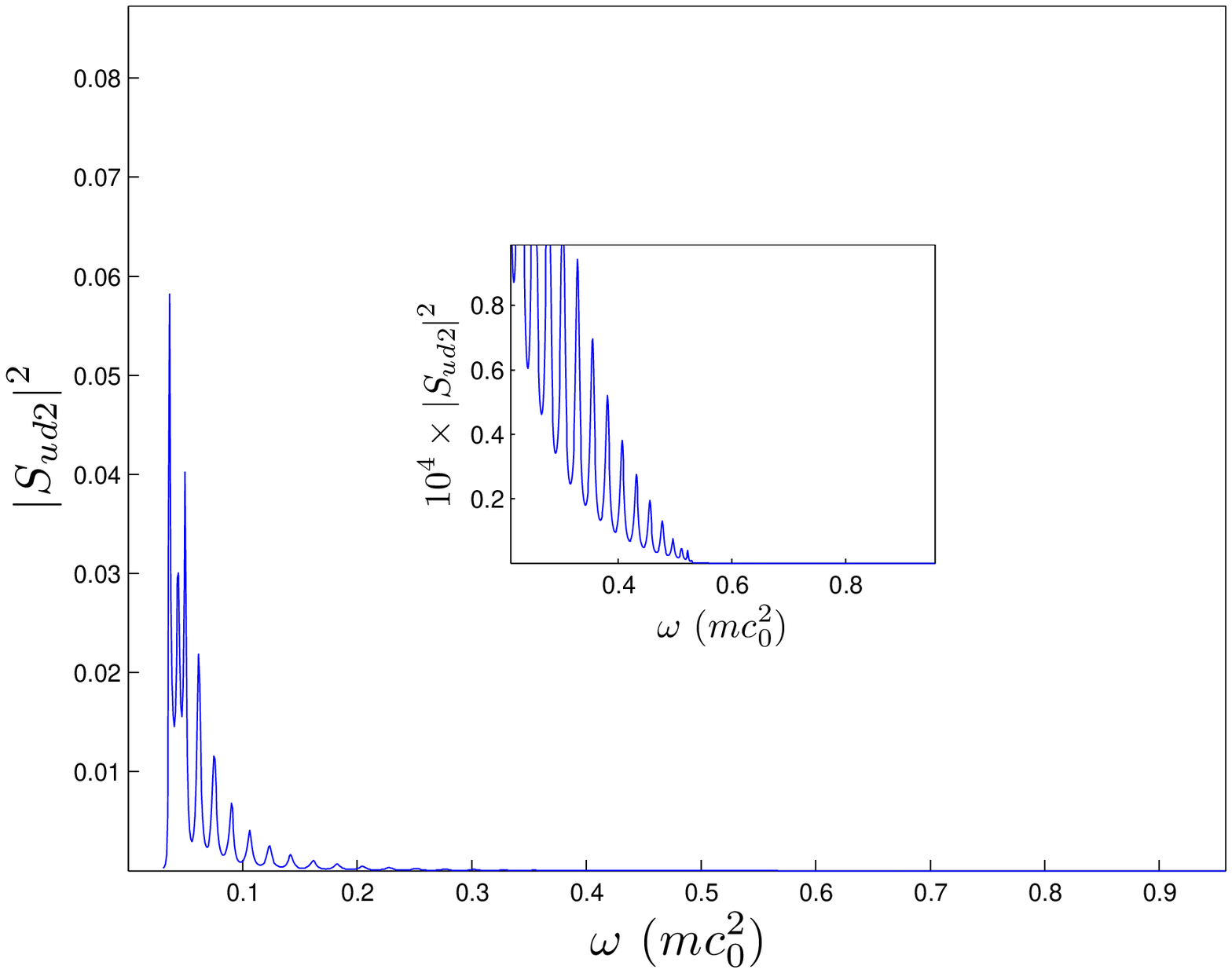} & \includegraphics[width=0.5\columnwidth]{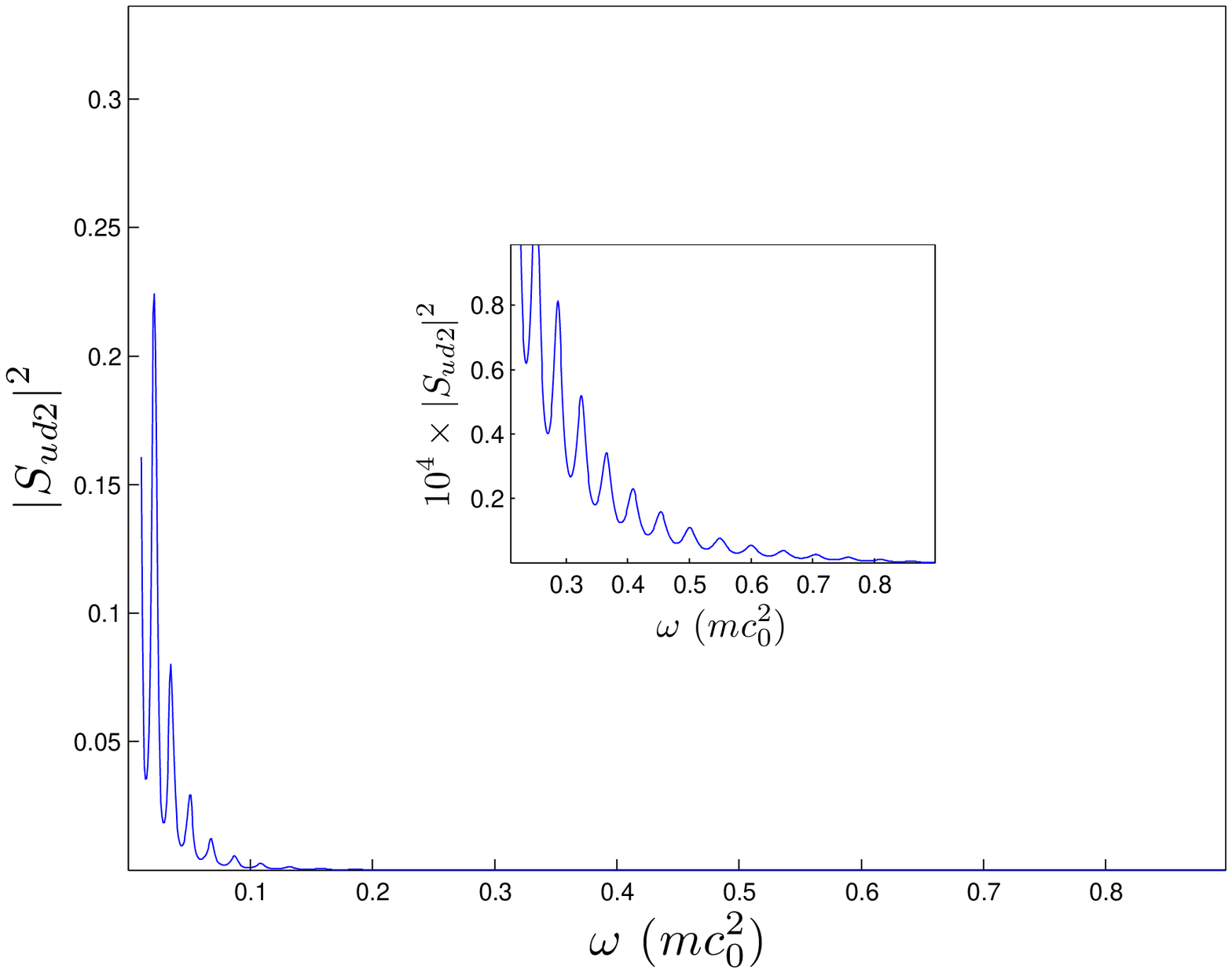} \\
    \includegraphics[width=0.5\columnwidth]{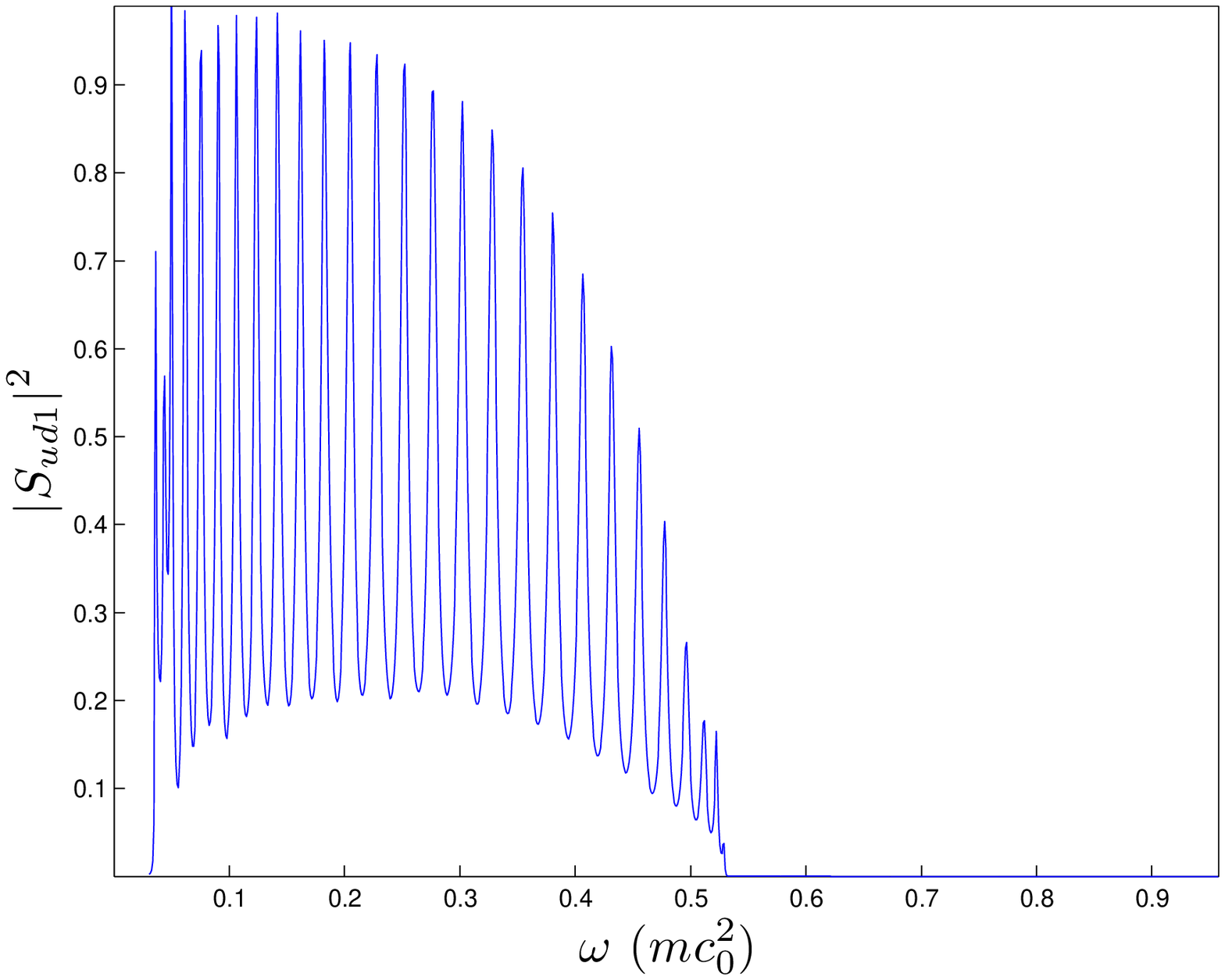} & \includegraphics[width=0.5\columnwidth]{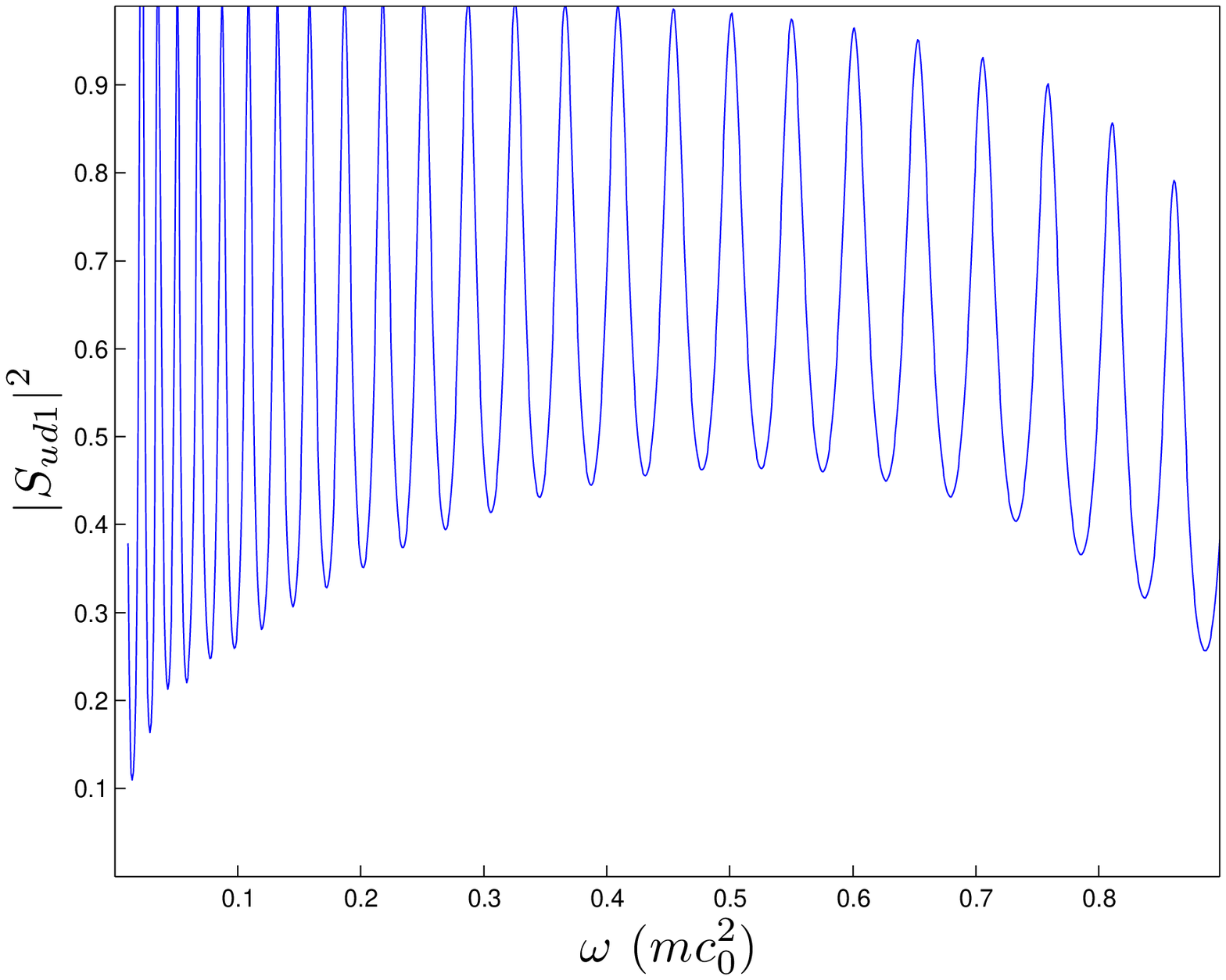}
\end{tabular}
\caption{Hawking spectrum for two different simulations for an ideal optical lattice. The BdG solutions are computed
on top of the quasi-stationary mean-field background provided by the wave function evaluated at a time $4\times 10^4\, t_{0}$.
Both graphs are computed with $\tau=500\, t_{0}$,
$L=400\,\mu\text{m}$, $n_{\rm osc}=30$, $N=10^4$, and $\omega_{\rm tr}=2\pi\times4\,\text{kHz}$,
and the varying parameters are the long-time potential amplitude $V_{\infty}$ and the lattice spacing $d$.
We have $\xi_0=0.3175\,\mu\text{m}$.
Left panel: $V_{\infty}=2.13~mc^2_0$ and $d=2.36\xi_0$, which gives $E_{\rm min}=0.91~mc^2_0$ and $\Delta_c=0.47~mc^2_0$.
The long time parameters of the quasi-stationary flow are $\bar{\mu}\simeq \omega_{\rm max}=0.96~mc^2_0$, $c_u=0.98~c_0$, $v_d=1.39~c_0$ and $c_d=0.02~c_0$.
Right panel: $V_{\infty}=1.86~mc^2_0$ and $d=1.91\xi_0$, which gives $E_{\rm min}=0.85~mc^2_0$ and $\Delta_c=0.95~mc^2_0$. The long time parameters of the quasi-stationary flow are $\bar{\mu}\simeq \omega_{\rm max}=0.90~mc^2_0$, $c_u=0.95~c_0$, $v_d=1.34~c_0$ and $c_d=0.03~c_0$.}
\label{fig:BdGOLIdeal}
\end{figure}

Before we present the numerical results, we make some general statements concerning the quasi-stationary regime. We first note that in this regime the chemical potential is placed near the bottom of the conduction band, so we have $\bar{\mu}\simeq E_{\rm min}$. Moreover, as the Mach number is very large in the supersonic region, $v_d/c_d\gg 1$, we can neglect the sound speed in the dispersion relation and then the Hawking frequency $\omega_{\rm max}$ is given by $\hbar\omega_{\rm max}\simeq mv_d^2/2 \simeq \bar{\mu}$. At the same time, as explained at the end of Appendix \ref{app:nonlinearol}, the width of the conduction band of the BdG solutions, $\Delta^{\rm BdG}_c$, is equal to the width of the Schr\"odinger conduction band $\Delta_{c}$, $\Delta^{\rm BdG}_c \simeq \Delta_{c}$, except for some small corrections due to the non-linearity of the GP wave function. Thus, the whole BdG conduction band is contained within the frequency range of the Hawking spectrum whenever $\Delta_c \lesssim E_{\rm min}$. As shown below, this fact leads to two qualitatively different regimes.

First we study the BdG solutions in the quasi-stationary regime corresponding to the ideal optical lattice of Sec. \ref{qsr}. We compute the scattering matrix using as a background the quasi-stationary wave function $\Psi_0(x)=\Psi(x,t_i)$, with $t_i=4\times 10^4\, t_{0}$. In Fig. \ref{fig:BdGOLIdeal}, two qualitatively different scenarios are displayed. In the left column, we show the results corresponding to a situation in which the BdG conduction band is completely contained inside the HR frequency range $0<\omega<\omega_{\rm max}$ and in the right column, we consider the opposite case, where $\omega_{\rm max}>\Delta_{c}$. In the upper row, we plot the normal-normal transmission element $|S_{ud1}|^2$, corresponding to a normal incident quasiparticle in the scattering channel $d1$ transmitted to the outgoing upstream channel. In the lower row, we show the anomalous-normal transmission, $|S_{ud2}|^2$, which corresponds to the Hawking effect itself.

The Hawking spectrum shows a peaked structure, with a decaying envelope similar to that of non-resonant spectra. When comparing the columns in more detail, we observe that the case of the left column displays a highly non-thermal behavior since the whole conduction band is contained in the Hawking spectrum, creating a sharp cut-off in the transmission band above which the transmission becomes exponentially small, in a similar way to the Schr\"odinger transmission of Fig. \ref{fig:RealisticIdT}.

By comparing both rows, we clearly see that the transmission band corresponds almost entirely to normal-normal transmission while the anomalous-normal transmission plays a secondary role. We can understand this effect by inspecting the form of the BdG solutions in the optical lattice and in the supersonic region. Due to the low value of the interacting term in both regions, for frequencies satisfying  $\hbar\omega \gg g\bar{n}_r, gn_d$, the mixing of the $u$ and $v$ components of the BdG spinors is very small. As the optical lattice is subsonic, the solutions corresponding to propagating Bloch waves have positive normalization and thus, their solutions have a dominant component $u$ and an associated small value of the $v$ component. In the supersonic region, the $d1$ solutions are those with dominant $u$ component while the anomalous $d2$ solutions are those with dominant $v$ component and very small value of the $u$ component. Then, when performing the matching between the solutions of the optical lattice and those of the supersonic region, the normal-normal transmission dominates over the anomalous-normal transmission.

\begin{figure}[!tb]
\begin{tabular}{@{}cc@{}}
    \includegraphics[width=0.5\columnwidth]{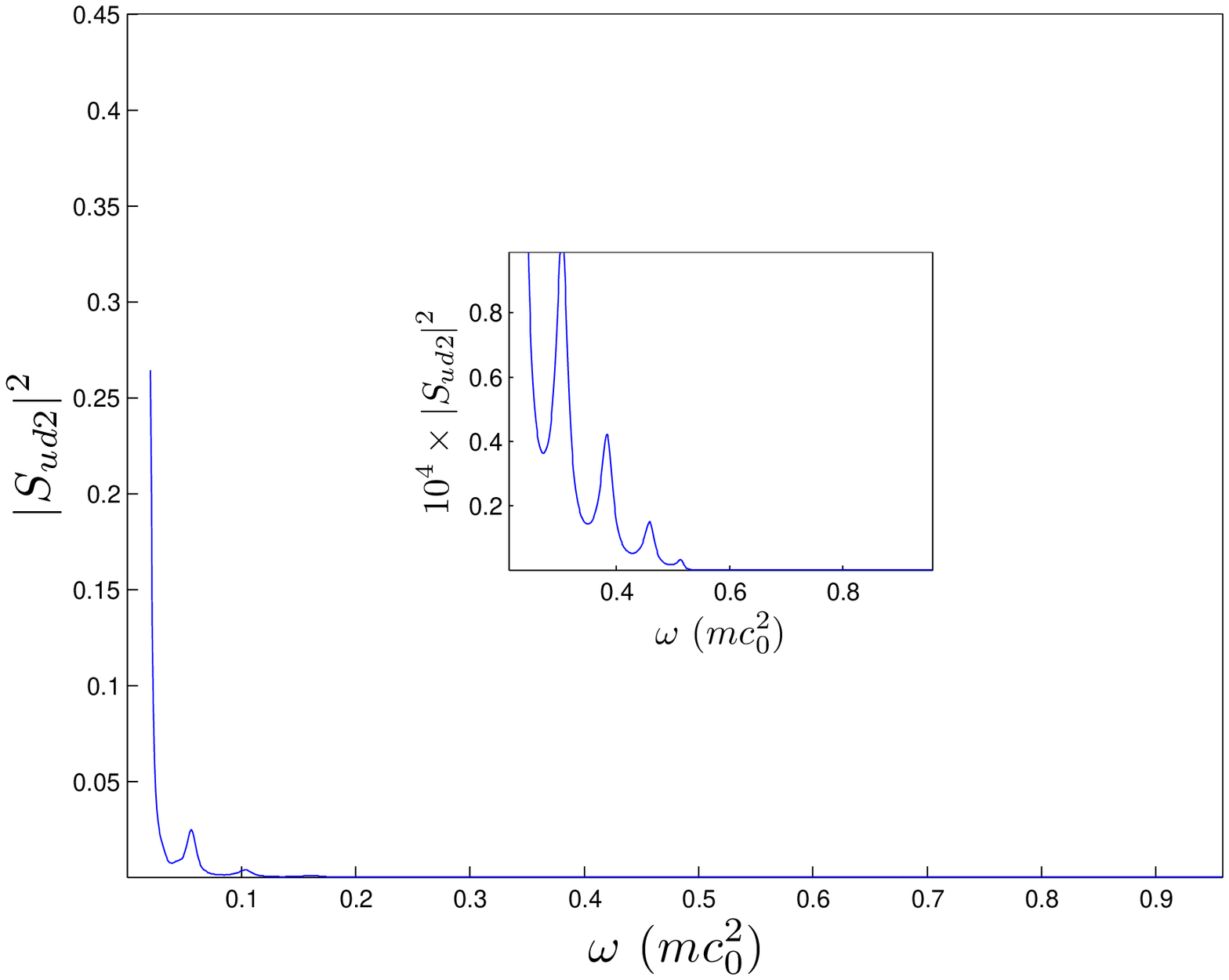} & \includegraphics[width=0.5\columnwidth]{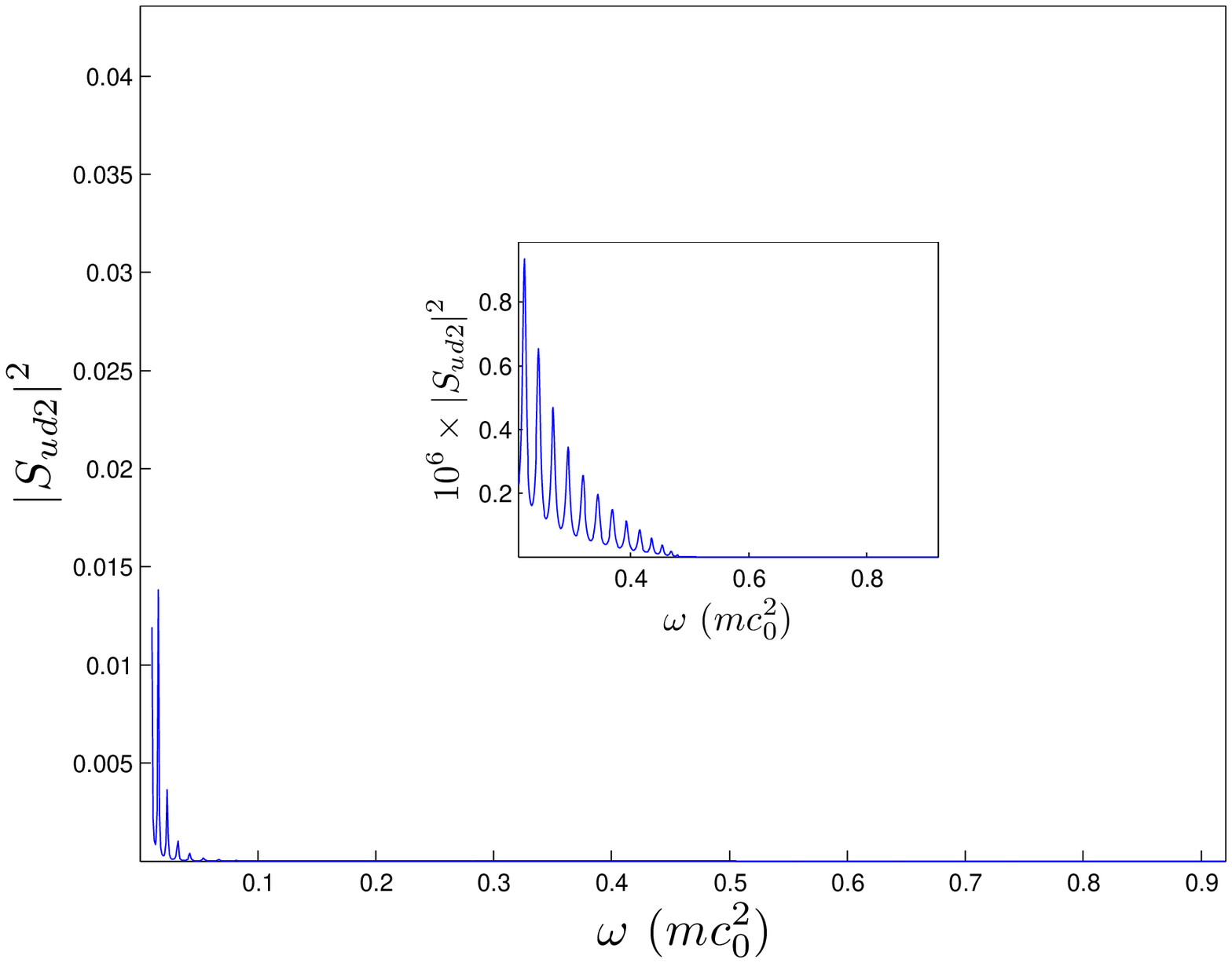}
\end{tabular}
\caption{Same as Fig. \ref{fig:BdGOLIdeal} for two simulations with the same parameters as that of its left panel, but now with a lower number of oscillations $n_{\rm osc}=10$ (left figure) and for a smaller condensate ($N=1000$ and $L=40\,\mu\text{m}$). We note that the value of the supersonic speed of sound in the latter case is as low as $c_d=0.006~c_0$, almost an order of magnitude lower than in Fig. \ref{fig:BdGOLIdeal}.}
\label{fig:BdGOLIdealComparisons}
\end{figure}

We show in Fig. \ref{fig:BdGOLIdealComparisons} the influence of other parameters on the HR spectrum. In the left plot, we show a simulation with the same parameters as those of the left column of Fig. \ref{fig:BdGOLIdeal} but with a reduced number of oscillations in the lattice, $n_{\rm osc}=10$, which results in a less peaked spectrum, as expected from the results of the conventional scattering theory. In the right plot, instead of reducing the number of maxima of the lattice, we reduce the size of the confined condensate, observing a decrease of the intensity of the HR spectrum. This can be understood from the fact that a smaller reservoir provides a lower value of the escaping current and thus, the value of the density is reduced in the supersonic region. By virtue of the arguments given above, a smaller interacting term in the supersonic region reduces the mixing of the $u$ and $v$ components of the BdG spinors, implying a lower anomalous-normal transmission.

\begin{figure}[!t]
\begin{tabular}{@{}cc@{}}
    \includegraphics[width=0.5\columnwidth]{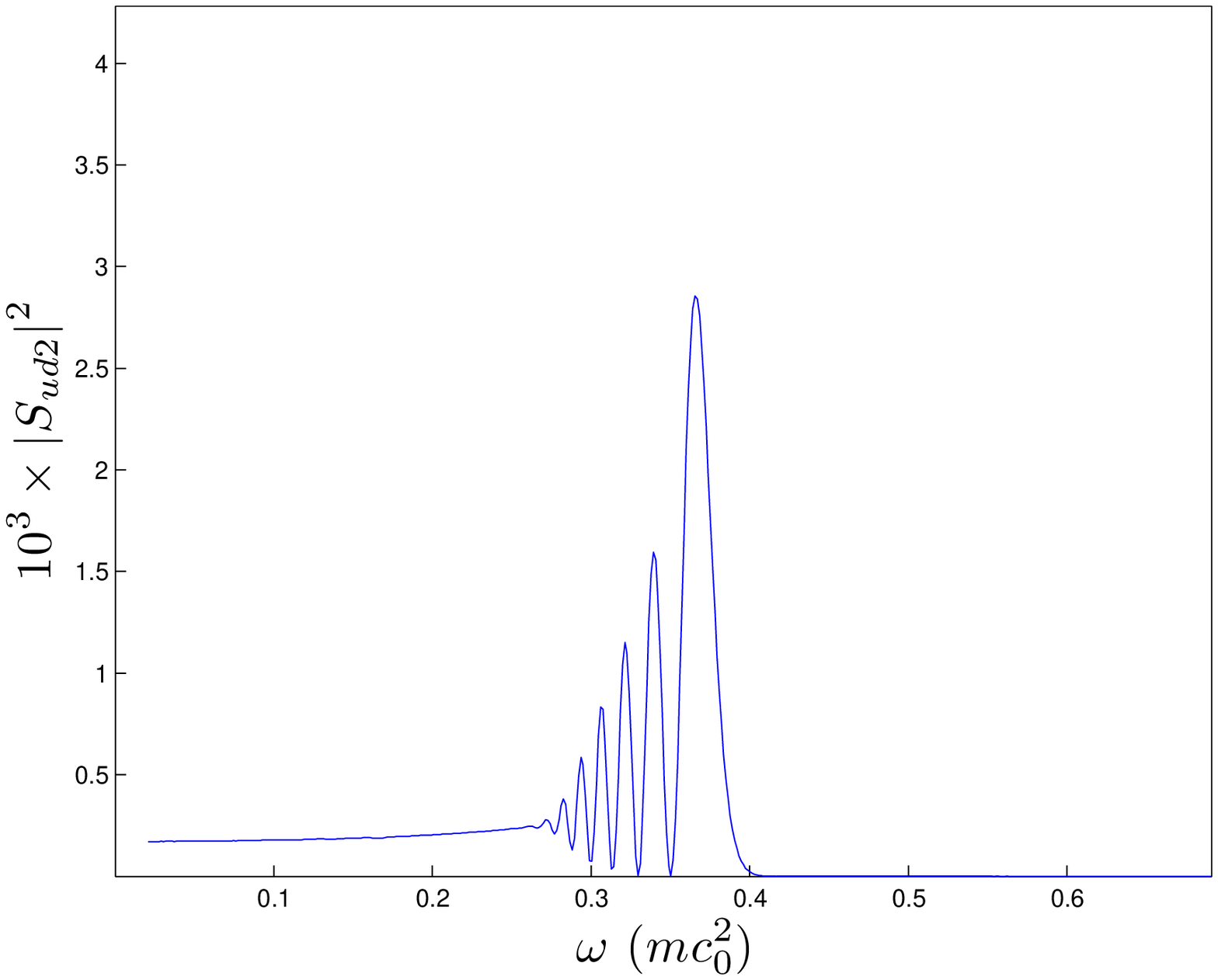} & \includegraphics[width=0.5\columnwidth]{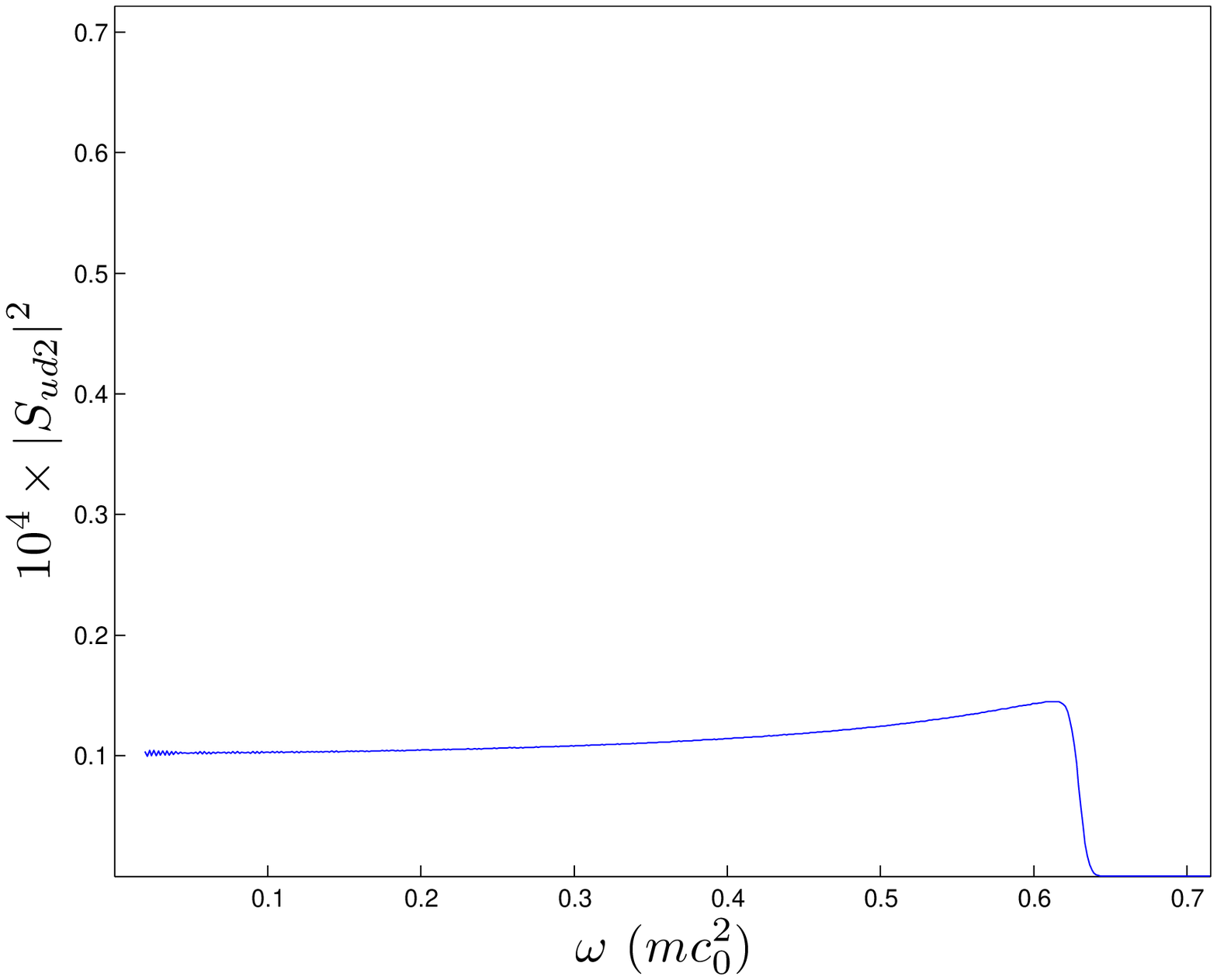}
\end{tabular}
\caption{Hawking spectrum of two different simulations for a realistic optical lattice with a Gaussian envelope. The snapshot of the condensate is taken in both cases at a time $4\times 10^4\, t_{0}$.
Both graphs are computed with $\tau=500\, t_{0}$, $L=470\,\mu\text{m}=1480~\xi_0$, $\tilde{w}=70\,\mu\text{m}=220.5~\xi_0$, $N=8.9\times 10^3$, $\omega_{\rm tr}=2\pi\times4\,\text{kHz}$ and $\xi_0=0.3175\,\mu\text{m}$. They only differ in the values of the long-time potential amplitude $V_{\infty}$ and the lattice spacing $d$. Left panel: $V_{\infty}=1.5~mc^2_0$ and $d=2.21\xi_0$, which gives $E_{\rm min}=0.68~mc^2_0$, $E_R=1.02~mc^2_0$ and $\Delta_c\thickapprox E_R-E_{\rm min}=0.34~mc^2_0$. The long time parameters of the quasi-stationary flow are $\bar{\mu}\simeq \omega_{\rm max}=0.69~mc^2_0$, $c_u=0.83~c_0$, $v_d=1.18~c_0$ and $c_d=0.025~c_0$.
Right panel: $V_{\infty}=1.5~mc^2_0$ and $d=1.73\xi_0$, which gives $E_{\rm min}=0.71~mc^2_0$, $E_R=1.64~mc^2_0$ and $\Delta_c\thickapprox E_R-E_{\rm min}=0.93~mc^2_0$. The long time parameters of the quasi-stationary flow are $\bar{\mu}\simeq \omega_{\rm max}=0.72~mc^2_0$, $c_u=0.85~c_0$, $v_d=1.20~c_0$ and $c_d=0.024~c_0$.}
\label{fig:BdGRealistic}
\end{figure}

After using the ideal optical lattice as a test field, we now shift to the realistic optical lattice considered in Sec. \ref{sec:Gaussian-shaped}. We apply the same procedure as before by taking a snapshot at a given time of the condensate once in the quasi-stationary regime. Most of the previous reasonings still apply in this case. In particular, we again distinguish in Fig. \ref{fig:BdGRealistic} between the case where the conduction band is placed inside the Hawking frequency range (left plot) and the case where it is not (right plot). We see that instead of having a peaked transmission band, as in the ideal optical lattice, the spectrum presents a plateau, in the same fashion as in the corresponding Schr\"odinger spectra of Fig. \ref{fig:RealisticIdT}. Interestingly, the Hawking spectrum displays strong resonant peaks near the top of the conduction band. This contrasts with the case where the end of the conduction band is above $\omega_{\rm max}$, in which the plateau of the transmission band only goes to zero at the end of Hawking spectrum limit without showing any interesting structure.

From these results, we conclude that the optimal scenario for searching the spontaneous Hawking effect requires, apart from the quasi-stationary requirements previously stated, the conduction band to be placed inside the Hawking frequency range and the supersonic density to be as high as possible (in order to increase the signal of the HR spectrum). We note that there must be a trade-off between achieving a very favorable quasi-stationary regime with small fluctuations (which implies having a wide conduction band) and placing the top of the conduction band inside the Hawking spectrum (which implies that the width of the conduction band must be smaller than the Hawking frequency $\Delta_c<\omega_{\rm max}$).

Nevertheless, the calculations presented in this section are preliminary and they only give us an approximated idea of the situation in the quasi-stationary regime. Several points should be addressed in a more complete calculation. First, one has to take into account the large but finite size of the confined subsonic region, which would introduce a hard-wall boundary condition in the BdG equations at $x=0$. Second, a more rigorous model should be introduced for computing the quasi-stationary modes that includes also the lower frequency modes. This could be made, as noted before, by employing the relative BdG equations \cite{Macher2009a} instead of the usual BdG equations. Finally, in order to compute the corresponding expectation values, one should also take into account the initial quantum state of the system at $t=0$, which is expected to be a thermal mixture of the initial quasiparticle confined modes (see Sec. \ref{app:confinedBdG} for more details), and compute its time evolution, which in the Bogoliubov approximation would correspond to a full integration of the time-dependent BdG equations. All these tasks represent harder calculations from both a theoretical and a computational point of view that are left for future works.

\section{Conclusions and outlook} \label{sec:OLconclusions}

Within a mean-field description, we have investigated the process whereby an initially confined atom condensate is coherently outcoupled as the barrier on one side is gradually lowered. The goal has been to identify the barrier-lowering protocol which best leads to a quasi-stationary sonic black hole located at the interface between subsonic and supersonic flow. We find that the use of an optical lattice for the lowered barrier helps to achieve a regime of quasi-stationary flow with a minimal value of the fluctuation spread. First we have focused on an optical lattice of finite length and uniform amplitude. We also find that the long-time band structure of the optical lattice greatly influences the asymptotic behavior of the emitted atom flow. Within this class of setups, the best quasi-stationary flow is achieved when the lowest conduction band is broad and the initial chemical potential lies not much high above the bottom of the final conduction band. In the optimal cases, the relative value of the spatial fluctuations can be as small as $\sigma(t)\sim 10^{-4}$. When we replace the uniform amplitude of the optical lattice by a more realistic Gaussian envelope, we find that the results are similar to those of a uniform lattice with the amplitude of the envelope maximum. Quite interestingly, we argue analytically and check numerically that, in the quasi-stationary regime, the horizon separating the regions of subsonic and supersonic flow is pinned down right at the Gaussian maximum. We also find that the Gaussian envelope is quite efficient in guaranteeing a small deviation from the ideal stationary flow.

The results here obtained are also of relevance in the context of quantum transport. Atom flow through a sonic black hole may also be viewed as one of the paradigms of atom quantum transport through a barrier \cite{Leboeuf2001}. In particular, the setup here presented can be used to provide a quasi-stationary supersonic current, with a well defined velocity (see Figs. \ref{fig:IdealTimenvD}, \ref{fig:RealisticTimenvD}).

After the previous mean-field calculation, we have presented a preliminary study of the Hawking spectrum, in which we assume the condensate to be truly stationary and semi-infinite in order to match with the usual theoretical descriptions. We find that two qualitative regimes can be distinguished depending on whether the top of the BdG conduction band lies below or above the Hawking frequency. In particular, when the top of the conduction band is placed within the frequency range of the Hawking effect, the system displays a highly structured Hawking spectrum which provides a promising scenario for the detection of the Hawking effect. However, we remark that these results are only preliminary. A more detailed future study should confirm the validity of these observations.

\chapter{Quantum signatures of spontaneous Hawking radiation}\label{chapter:CSGPH}

\section{Introduction}

In the previous chapter, we have numerically studied the birth of a quasi-stationary black hole in a Bose-Einstein condensate within a realistic experimental scenario. The question now is how to detect the spontaneous Hawking radiation above the previous quasi-stationary background. The answer is highly non-trivial. On one hand, in a real setup the system is always at a finite non-zero temperature and then, the state of the system is not strictly the vacuum of the incoming scattering modes. In particular, the stimulated signal arising from the thermal occupation of phonon modes can easily overshadow the spontaneous signal due to the genuine quantum Hawking effect.

On the other hand, a clear detection of the Hawking effect should be able to distinguish the quantum origin of the Hawking effect from the small perturbations around the coherent condensate wave function that are impossible to avoid in a realistic scenario, as discussed and quantified in the previous chapter. The point is that the Bogoliubov-de Gennes equations describe both phenomena: linearized collective motion and quantum quasiparticle excitation. The former is fully describable within a mean-field picture; the latter reflects the presence of a many-body ground state (the Bogoliubov vacuum) with strong pair correlations.

Detection schemes based on density-density correlations in real space are not able to distinguish unambiguously the spontaneous Hawking signal \cite{Balbinot2008,Recati2009}. The mere measurement of phonon \cite{Schutzhold2008} or atom \cite{Zapata2011} intensity spectra would not permit those distinctions either.

Hawking radiation is a fundamental quantum phenomenon that results from the impossibility of identifying the vacuum of incoming quasiparticles
with that of outgoing quasiparticles \cite{Birrell1984}. Specifically, the incoming vacuum is a squeezed state of outgoing quasiparticles.
In this respect, it has been long recognized in quantum optical contexts \cite{Loudon1983,Walls2008} that correlation functions characterizing
the electromagnetic radiation satisfy Cauchy-Schwarz (CS) type inequalities, which can, however, be violated in the deep quantum regime, particularly by squeezed light. Thus, the violation of CS inequalities is generally regarded as a conclusive signature of quantum behavior.
We devote part of this chapter to show that the violation of certain CS inequalities is indeed a signal of the presence of spontaneous HR and that the sole presence of linearized collective motion around the stationary wave function or stimulated phonon signal could not provide such violation. As a result, we identify this particular CS violation as an unequivocal smoking gun of the presence of spontaneous HR. The same cannot be said about other CS violations that are always present in a Bogoliubov vacuum; for example, those associated with the paired atom modes $k,-k$ in a condensate at rest. Such CS violations will appear in other correlation functions.

An alternative scheme to identify deep quantum behavior relies on the detection of entanglement between the various outgoing radiation channels from a sonic horizon \cite{Busch2014,Finazzi2013}. A recent work has addressed the possibility of detecting entanglement through density-density correlations in Fourier space \cite{Steinhauer2015}. Approaches based on entanglement detection have been also proposed in analogous contexts such as inflationary cosmology \cite{Campo2006}, astrophysical black holes \cite{Martin-Martinez2010}, general
relativistic quantum fields \cite{Friis2013}, or other black-hole analogs \cite{Horstmann2010,Horstmann2011}. In the present chapter, we establish in great detail the links between the work developed in the present thesis, based on the detection of violation of CS inequalities, and the works of Refs. \cite{Finazzi2013,Busch2014}, based on the detection of entanglement, unifying both approaches.

We also study in this chapter the potential measurability of the previous criteria in specific detection schemes. Importantly, by introducing envelope-modulated Fourier transforms we explicitly take into account the role played by the spatial location of the asymptotic regions. This allows us to show that the violation of only quadratic CS inequalities can be measured. In particular, we prove that the complete implementation of the generalized Peres-Horodecki (GPH) criterion or the quartic CS violation requires the knowledge of parameters that are impossible or very difficult to measure, at least within the class of detection schemes here considered. However, we also show that this does not represent an important limitation in practice, since, under rather broad conditions, all the previous criteria become equivalent.

Although the content developed in this chapter is motivated by the quest for the observation of spontaneous HR, the results here obtained are also of relevance in neighboring fields such as quantum optics \cite{Lopaeva2013,Han2015} and quantum information physics \cite{Clausen2014,Shen2015}, as well as in the broader topic of bosonic condensates \cite{Kheruntsyan2012,Wasak2014}.

We devote Sections \ref{sec:GeneralQuantum}-\ref{sec:CSPH} to review general signatures of quantum behavior and discuss the general properties of CS inequalities and entanglement. We apply these concepts to the specific scenario of analog Hawking radiation in Sec. \ref{sec:criteriaHR} and unify the previous criteria under certain conditions. In Section \ref{sec:Experimental}, we discuss the possibility of experimentally verifying the discussed detection schemes. Finally, in Section \ref{sec:NumericalCores}, we present numerical results on the detection of the CS violation in typical BH setups. The conclusions are presented in Sec. \ref{sec:CSPHConclusions}. We explain in Sec. \ref{app:parametrization} the parametrization of the scattering matrix and of the computation of the correlation functions used in the main text.

\section{General signatures of quantum behavior}\label{sec:GeneralQuantum}

We start by reviewing some general signatures of quantum behavior. For illustrative purposes, we discuss the criteria usually considered in quantum optics \cite{Loudon1983,Walls2008}. We focus on the properties of the normalized second-order correlation function, which for light is defined as
\begin{equation}
g_{ij}^{(2)}(\tau)\equiv\frac{\langle\hat{a}_{i}^{\dagger}(0)\hat{a}_{j}^{\dagger}(\tau)\hat{a}_{j}(\tau)\hat{a}_{i}(0)\rangle}{\langle\hat{a}_{i}^{\dagger}(0)\hat{a}_{i}(0)\rangle\langle\hat{a}_{j}^{\dagger}(0)\hat{a}_{j}(0)\rangle}\,,\label{eqCoherenceDef}
\end{equation}
where $\hat{a}_{i}(t)$ is the Heisenberg operator for photon mode
$i$, and the average is quantum-statistical. The correlation function
for classical light is obtained by removing the quantum average and
replacing the Heisenberg operators $\hat{a}_{i}(t)$ by complex numbers.
The following inequalities can be proven for classical light:
\begin{eqnarray}
1 & \leq & g_{ii}^{(2)}(0),\label{eq:ClassicalIneqa}\\
g_{ii}^{(2)}(\tau) & \leq & g_{ii}^{(2)}(0),\label{eq:ClassicalIneqb}\\
\left[g_{ij}^{(2)}(\tau)\right]^{2} & \leq & g_{ii}^{(2)}(0)g_{jj}^{(2)}(0)\,,\;(i\neq j)\,.\label{eq:ClassicalIneqc}
\end{eqnarray}
These inequalities are satisfied not only classically but also by
quantum thermal states at high temperature. They are also satisfied
by chaotic and coherent states.

The violation of any of the above inequalities is a signature of deep
quantum behavior. States violating (\ref{eq:ClassicalIneqa}) are
said to show sub-Poissonian statistics. Violation of (\ref{eq:ClassicalIneqb})
reflects anti-bunching. Expression (\ref{eq:ClassicalIneqc}) is a
CS inequality. States which violate it at $\tau=0$ are said to exhibit
two-mode sub-Poissonian statistics. In general, the proof of (\ref{eq:ClassicalIneqc})
requires the system to be described by a positive (Glauber-Sudarshan)
$P$ function. Some quantum states such as two-mode squeezed states may
not satisfy this condition and thus can violate (\ref{eq:ClassicalIneqc}).

Another possible signature of quantum behavior is the presence of entanglement \cite{Walls2008}, which also requires the system to be described by non-classical statistics. We define precisely in the next section the concept of entanglement considered throughout this work.

\section{General properties of Cauchy-Schwarz inequalities and Peres-Horodecki criterion} \label{sec:CSPH}
\sectionmark{General properties of quantum signatures}

We now study in more detail the criteria previously discussed. In particular, we focus on two specific signatures: the violation of CS inequalities and the entanglement of the system, which are the only interesting ones for studying HR as we will explain in Sec. \ref{sec:criteriaHR}. First, we review the mathematical CS inequalities for operators in a Hilbert space, which are always fulfilled, see for instance Ref. \cite{Adamek2013}. We define the scalar product associated to a state $\hat{\rho}$ for two operator $\hat{A},\hat{B}$  as:
\begin{equation}\label{eq:scalarproduct}
(\hat{A},\hat{B})\equiv\braket{\hat{A}^{\dagger}\hat{B}}=\text{Tr}(\hat{\rho}\hat{A}^{\dagger}\hat{B}).
\end{equation}
where the expectation value of an operator $\hat{O}$ is taken over the state of the system, described by the density matrix $\hat{\rho}$, $\braket{\hat{O}}=\text{Tr}(\hat{\rho}\hat{O})$. It is easy to check that the previous product satisfies the usual properties of a scalar product. Thus, it satisfies the mathematical Cauchy-Schwarz inequality:
\begin{equation}\label{eq:CSmathematical}
|\braket{\hat{A}^{\dagger}\hat{B}}|^2\leq \braket{\hat{A}^{\dagger}\hat{A}}\braket{\hat{B}^{\dagger}\hat{B}}
\end{equation}
In particular, the previous inequality implies for $B=1$ that
\begin{equation}\label{eq:meansquare}
|\braket{\hat{A}}|^2=|\braket{\hat{A}^{\dagger}}|^2\leq \braket{\hat{A}^{\dagger}\hat{A}}
\end{equation}

For our purposes, we consider, in a general context, correlations between two given quantum modes $i,j$, whose corresponding annihilation operators are $\hat{\gamma}_{i,j}$. The first-order correlation functions are given by
\begin{eqnarray}\label{eq:gdef}
g_{ij}\equiv \braket{\hat{\gamma}_{i}^{\dagger}\hat{\gamma}_{j}},~c_{ij}\equiv \braket{\hat{\gamma}_{i}\hat{\gamma}_{j}} ,
\end{eqnarray}
The mathematical CS inequality associated to the previous correlation functions are obtained by substituting $A=\hat{\gamma}_{i}$ and $B=\hat{\gamma}_{j}$ in Eq. (\ref{eq:CSmathematical})
\begin{equation}\label{eq:CSgijimpossible}
|g_{ij}|^2=|\braket{\hat{\gamma}^{\dagger}_{i}\hat{\gamma}_{j}}|^2
\leq\braket{\hat{\gamma}^{\dagger}_{i}\hat{\gamma}_{i}}\braket{\hat{\gamma}^{\dagger}_{j}\hat{\gamma}_{j}}=g_{ii}g_{jj}
\end{equation}
Thus, the CS inequality $|g_{ij}|^2\leq g_{ii}g_{jj}$ is always satisfied and can never be violated. On the other hand, taking $A=\hat{\gamma}^{\dagger}_{i}$ and $B=\hat{\gamma}_{j}$, we arrive at:
\begin{equation}\label{eq:CScijpos}
|c_{ij}|^2=|\braket{\hat{\gamma}_{i}\hat{\gamma}_{j}}|^2
\leq\braket{\hat{\gamma}_{i}\hat{\gamma}^{\dagger}_{i}}
\braket{\hat{\gamma}^{\dagger}_{j}\hat{\gamma}_{j}}=(g_{ii}+1)g_{jj} \, .
\end{equation}
Interestingly, the mathematical CS inequality for quantum operators is $|c_{ij}|^2\leq(g_{ii}+1)g_{jj}$, instead of the {\it classical} CS inequality $|c_{ij}|^2\leq g_{ii}g_{jj}$, always satisfied in a classical context, in a similar fashion to Eq. (\ref{eq:ClassicalIneqc}). This leaves the possibility of violating the {\it classical} CS inequality, which we can quantify by introducing the following magnitude:
\begin{eqnarray}\label{eq:CSviolation2}
\Delta_{ij}\equiv |c_{ij}|^2-g_{ii}g_{jj}>0 \, .
\end{eqnarray}
The fulfillment of the previous condition reflects the quantum nature of the state of the system. We can also quantify the relative degree of CS violation by introducing
\begin{eqnarray}\label{eq:CSviolationrel2}
\delta_{ij}\equiv \frac{|c_{ij}|^2}{g_{ii}g_{jj}}>1 \, .
\end{eqnarray}
We will refer indistinctly to Eqs. (\ref{eq:CSviolation2}), (\ref{eq:CSviolationrel2}) as the {\it quadratic} CS violation.

We can extend the previous idea to the following second-order correlation function:
\begin{equation}\label{eq:GammaDef}
\Gamma_{ij}\equiv\braket{\hat{\gamma}_{i}^{\dagger}
\hat{\gamma}_{j}^{\dagger}\hat{\gamma}_{j}\hat{\gamma}_{i}}>0\, ,
\end{equation}
A similar reasoning leads to the mathematical CS inequality:
\begin{eqnarray}\label{eq:CSGammaijpos}
\nonumber |\Gamma_{ij}|^2&=&|\braket{\hat{\gamma}^{\dagger}_{i}\hat{\gamma}_{i}\hat{\gamma}^{\dagger}_{j}\hat{\gamma}_{j}}|^2
\leq\braket{\hat{\gamma}^{\dagger}_{i}\hat{\gamma}_{i}\hat{\gamma}^{\dagger}_{i}\hat{\gamma}_{i}}\braket{\hat{\gamma}^{\dagger}_{j}\hat{\gamma}_{j}\hat{\gamma}^{\dagger}_{j}\hat{\gamma}_{j}}\\
&=&(\Gamma_{ii}+g_{ii})(\Gamma_{jj}+g_{jj})
\end{eqnarray}
which also leaves the possibility of violating the {\it classical} CS inequality $|\Gamma_{ij}|^2\leq\Gamma_{ii}\Gamma_{jj}$, as quantified by
\begin{equation}\label{eq:CSviolation4}
\Theta_{ij}\equiv \Gamma_{ij}-\sqrt{\Gamma_{ii}\Gamma_{jj}}>0 \, .
\end{equation}
We can also measure the relative degree of violation using
\begin{equation}\label{eq:CSviolationrel4}
\theta_{ij}\equiv \frac{\Gamma_{ij}}{\sqrt{\Gamma_{ii}\Gamma_{jj}}}>1 \, .
\end{equation}
In analogy to the quadratic CS violation, we will refer to the previous conditions as {\it quartic} CS violation. Both quadratic and quartic violations are signatures of quantum behavior since they cannot be satisfied in a classical context.

Another possible signature of the quantum character of the system is the presence of entanglement between the modes $i,j$. In this work, we follow the convention of defining entanglement as the non-separability of the state of the system. We say that a two-mode state $\hat{\rho}$ is separable when it can be decomposed as:
\begin{equation}\label{eq:separability}
\hat{\rho}=\sum_n p_n\hat{\rho}^{(i)}_n\otimes\hat{\rho}^{(j)}_n
\end{equation}
where $\hat{\rho}^{(i,j)}_n$ are states of the Hilbert subspace associated to the $i,j$ modes, respectively. In order to characterize the entanglement, we make use of the generalized Peres-Horodecki (GPH) criterion \cite{Horodecki1997,Simon2000}. The GPH criterion is based on the fact that, if $\hat{\rho}$ is separable, by taking the partial transpose of $\hat{\rho}$ with respect to only the Hilbert space associated to the mode $j$, we obtain an operator $\hat{\rho}_{t}$ that is also a physical density matrix, satisfying the usual properties of quantum states.

For studying the partial transpose of $\hat{\rho}$, $\hat{\rho}_t$, we first define the vector $\hat{X}\equiv [\hat{q}_i,\hat{p}_i,\hat{q}_j,\hat{p}_j]^{T}$ and the corresponding phase space variables $X\equiv [q_i,p_i,q_j,p_j]^{T}$, where the phase space operators $\hat{q}_k,\hat{p}_k$ are related to the annihilation operator of the modes $k=i,j$ through $\hat{\gamma}_k=(\hat{q}_k+i\hat{p}_k)/\sqrt{2}$. The commutation relations between the components of $\hat{X}$ are $[\hat{X}_{\alpha},\hat{X}_{\beta}]=iL_{\alpha\beta}$, where $L_{\alpha\beta}$ is a $4\times 4$ matrix of the form

\begin{eqnarray}\label{eq:XCommutators}
L=\left[\begin{array}{cc} J & 0\\
0 & J \end{array}\right],~J=\left[\begin{array}{cc}
0 & 1\\
-1 & 0
\end{array}\right]\,
\end{eqnarray}
with $J$ the symplectic matrix in two dimensions. In order to characterize $\hat{\rho}_t$, we study the effect of partial transposition on the Wigner function. The Wigner function of the state $\hat{\rho}$ is given by:
\begin{equation}\label{eq:Wigner}
W(X)\equiv\frac{1}{(2\pi)^2}\int\mathrm{d}^2q'~\braket{q-\frac{q'}{2}|\hat{\rho}|q+\frac{q'}{2}}e^{iq'p}
\end{equation}
with $q=(q_i,q_j)$, $p=(p_i,p_j)$. By transposing the matrix elements of $\hat{\rho}$ with respect to the Hilbert space of the mode $j$, it is straightforward to show from the previous definition that the Wigner distribution associated to $\hat{\rho}_{t}$ is $W_t(q_i,p_i,q_j,p_j)=W(q_i,p_i,q_j,-p_j)$, or, in a more compact form
\begin{equation}\label{eq:Wignertransposed}
W_t(X)=W(\Lambda X),~\Lambda=\rm{diag}[1,1,1,-1].
\end{equation}
We note, from this result, that partial transposition amounts to a local time inversion on the second variable \cite{Simon2000}. The partial transpose has important effects when considering the uncertainties of the phase space operators. If we define the operator
\begin{equation}
\hat{Y}=\sum_{\alpha=1}^4u^\alpha\Delta\hat{X}_\alpha,
\end{equation}
where $\Delta \hat{X}\equiv \hat{X}-\braket{\hat{X}}$, we have that $\braket{\hat{Y}^{\dagger}\hat{Y}}\geq 0$, with $u^\alpha$ the components of an arbitrary four dimensional complex vector. The previous relation can be written compactly in vectorial notation as:
\begin{eqnarray}\label{eq:Uncer}
u^{\dagger}Mu&\geq&0\\
\nonumber M&\equiv& V+i\frac{L}{2}
\end{eqnarray}
$V$ being the covariance matrix with matrix elements $V_{\alpha\beta}=\frac{1}{2}\braket{\{\Delta\hat{X}_{\alpha},\Delta\hat{X}_{\beta}\}}$ and
$\{\Delta\hat{X}_{\alpha},\Delta\hat{X}_{\beta}\}$ the anticommutator. We see that Eq. (\ref{eq:Uncer}) represents an alternative expression of the uncertainty principle. In particular, as it holds for any complex vector $u$, the uncertainty principle simply states that $M$ must be a non-negative matrix, $M\geq 0$. Since $L$ is a constant matrix independent of the state $\hat{\rho}$, we only have to compute the covariance matrix $V$ for obtaining the matrix $M$. As well-known from quantum optics, the symmetric covariances can be computed using the Wigner function \cite{Scully1997}:
\begin{eqnarray}\label{eq:SymmetricCovariances}
\braket{X_\alpha}&=&\int\mathrm{d}^4X~X_\alpha W(X)\\
\nonumber \braket{\{\Delta\hat{X}_{\alpha},\Delta\hat{X}_{\beta}\}}&=&\int\mathrm{d}^4X~\Delta X_\alpha \Delta X_\beta W(X)
\end{eqnarray}

When considering $\hat{\rho}_t$, the state obtained by the partial transposition of $\hat{\rho}$, and following an analog reasoning, we find that the uncertainty principle for $\hat{\rho}_t$ is simply given by the condition $M_t\geq 0$, where:
\begin{eqnarray}\label{eq:Uncertrans}
M_t&\equiv&V_t+i\frac{L}{2}
\end{eqnarray}
Now, $V_t$ is the same covariance matrix as $V$ but with the expectation values evaluated in the state $\hat{\rho}_t$. Taking into account Eqs. \ref{eq:Wignertransposed},\ref{eq:SymmetricCovariances}, we see that $V_t=\Lambda V \Lambda$.

The GPH criterion is based in the fact that if $M_t$ is a non positive-semidefinite matrix, the state is entangled, since if the state $\hat{\rho}$ is separable, $\hat{\rho}_t$ must be a physical state and then, the associated uncertainty principle reads $M_t\geq 0$. Since $M$ is non-negative as $\hat{\rho}$ is a physical density matrix, $M_t$ is non-negative if and only if $\det M_t\geq 0$. The conditions $\det M,M_t\geq 0$ are respectively equivalent to $\mathcal{P}^{\pm}_{ij}\geq0$, where:

\begin{eqnarray}\label{eq:GPHpm}
\mathcal{P}^{\pm}_{ij}&\equiv&\det A_i\det A_j+\left(\frac{1}{4}\mp \det C_{ij}\right)^2\\
\nonumber&-&\text{tr}(JA_iJC_{ij}JA_jJC_{ij}^{T})-\frac{1}{4}(\det A_i+\det A_j) ,
\end{eqnarray}
The matrices $A_i,A_j,C_{ij}$ are $2\times 2$ submatrices of the covariance matrix $V$:
\begin{equation}\label{eq:WBlocks}
V=\left[\begin{array}{cc} A_{i} & C_{ij}\\
C^{T}_{ij} & A_{j} \end{array}\right]
\end{equation}
We note that $\mathcal{P}^{+}_{ij}\geq0$ is always satisfied provided that $\hat{\rho}$ is a physical density matrix. However, when $\mathcal{P}^{-}_{ij}<0$, the state $\hat{\rho}_{t}$ is not a physical state which implies that the state $\hat{\rho}$ is not separable. We can put together both conditions by defining the GPH function $\mathcal{P}_{ij}$ as:
\begin{eqnarray}\label{eq:GPH}
\mathcal{P}_{ij}&\equiv&\det A_i\det A_j+\left(\frac{1}{4}-|\det C_{ij}|\right)^2\\
\nonumber&-&\text{tr}(JA_iJC_{ij}JA_jJC_{ij}^{T})-\frac{1}{4}(\det A_i+\det A_j) ,
\end{eqnarray}
Thus, if $\mathcal{P}_{ij}<0$, the state is automatically entangled. This is the final form of the GPH criterion that we will use in this work. We note that, whenever $\det C_{ij}\geq 0$, the state is separable, as $\mathcal{P}_{ij}=\mathcal{P}^{+}_{ij}\geq0$, so only states with $\det C_{ij}< 0$ can be entangled.

We can compute explicitly the GPH function in terms of the first order correlation functions of Eq. (\ref{eq:gdef}). Hereafter, we focus for simplicity on the usual case where the annihilation and destruction operators have zero mean value [if not, we can simply modify the definition of the correlations in Eq. (\ref{eq:gdef}) by substituting the $\gamma$ operators by their fluctuations]. The matrices $A_{i,j}$ and $C_{ij}$ can be written in terms of the first-order correlation functions as:
\begin{eqnarray}\label{eq:GPHMatrices}
A_{k}&=&\left(g_{kk}+\frac{1}{2}\right)\mathbb{I}_2+\left[\begin{array}{cc}
\text{Re}~c_{kk} & \text{Im}~c_{kk}\\
\text{Im}~c_{kk} & -\text{Re}~c_{kk}
\end{array}\right]\\
C_{ij}&=&
\left[\begin{array}{cc}
\text{Re}(g_{ij}-c_{ij}) & \text{Im}(g_{ij}+c_{ij})\\
\text{Im}(-g_{ij}+c_{ij}) & \text{Re}(g_{ij}-c_{ij})
\end{array}\right]\, ,
\end{eqnarray}
with the index $k=i,j$ and $\mathbb{I}_2$ the $2\times 2$ identity matrix. Then, using Eqs. (\ref{eq:GPH}),(\ref{eq:GPHMatrices}), we obtain:

\begin{eqnarray}\label{eq:GPHcomplete}
\mathcal{P}_{ij}&=&[g_{ii}g_{jj}-|c_{ij}|^2][(g_{ii}+1)(g_{jj}+1)-|c_{ij}|^2]+4\left(g_{ii}+\frac{1}{2}\right)\text{Re}(g_{ij}c^{*}_{jj}c_{ij})\\
\nonumber&+&4\left(g_{jj}+\frac{1}{2}\right)\text{Re}(g_{ij}c_{ii}c^{*}_{ij})-2\text{Re}(c^{2}_{ij}c^{*}_{ii}c^{*}_{jj})-2\text{Re}(g^{2}_{ij}c_{ii}c^{*}_{jj})+|c_{ii}|^2|c_{jj}|^2+|g_{ij}|^4\\
\nonumber &-&|g_{ij}|^2(g_{ii}+g_{jj}+2g_{ii}g_{jj}+2|c_{ij}|^2)-g_{ii}(g_{ii}+1)|c_{jj}|^2-g_{jj}(g_{jj}+1)|c_{ii}|^2 \, .
\end{eqnarray}

This expression applies whenever $|c_{ij}|\geq|g_{ij}|$. For $|c_{ij}|<|g_{ij}|$, we have $\det C_{ij}>0$ and the state is separable so there is no need of using the GPH function. We finally remark that, in the particular case of Gaussian states, the GPH criterion is also a necessary condition for entanglement \cite{Simon2000}. Thus, for Gaussian states, the GPH criterion is equivalent to the entanglement of the system. We can understand this relation from the fact that Gaussian states are completely determined by their second momentums and, precisely, the GPH criterion characterizes entanglement only involving the second momentums of the distribution

\subsection{Quadratic Cauchy-Schwarz violations and the generalized Peres-Horodecki criterion}\label{subsec:quadraticGPH}

On the basis of the previous general results, we prove that the quadratic CS violation implies the fulfillment of the GPH criterion. Suppose that $\mathcal{P}_{ij}\geq0$. In that case, the matrix $M_t$ is non-negative and we can define an associated scalar product for vectors $u,v\in \mathbb{C}^4$ as:
\begin{equation}\label{eq:scalarproduct4}
(u,v)_t\equiv u^{\dagger}M_tv
\end{equation}
which satisfies the associated CS inequality:
\begin{equation}\label{eq:CSscalar4}
|(u,v)_t|^2\leq (u,u)_t(v,v)_t
\end{equation}
By inserting
\begin{equation}\label{eq:CSvectors4}
u=\frac{1}{\sqrt{2}}\left[\begin{array}{c}
0\\ 0\\ 1\\ -i
\end{array}\right],~v=\frac{1}{\sqrt{2}}\left[\begin{array}{c}
1\\ i\\ 0\\ 0
\end{array}\right]
\end{equation}
in Eq. (\ref{eq:CSscalar4}), we obtain the quadratic CS inequality $|c_{ij}|^2\leq g_{ii}g_{jj}$. Thus, if there is quadratic CS violation, the matrix $M_t$ cannot be non-negative, which implies that $\mathcal{P}_{ij}<0$ and the GPH criterion is satisfied. We then conclude that the quadratic CS violation is a sufficient condition for the fulfillment of the GPH criterion.

\subsection{Quartic Cauchy-Schwarz violations and entanglement}\label{subsec:quarticGPH}

We now consider the  relation between quartic CS violations and entanglement. The previous arguments cannot be applied to the quartic CS violation Eq. (\ref{eq:CSviolation4}). As a counterexample, the direct product of two pure number states of the modes $i,j$, $\hat{\rho}=\ket{n}\bra{n}\otimes\ket{n'}\bra{n'}$, which is a manifestly separable state, violates the quartic CS inequality. Even for Gaussian states, the two conditions, quartic CS violation and entanglement, are still independent. For instance, for a general Gaussian state, the quartic CS violation of Eq. (\ref{eq:CSviolation4}) can be expressed, via Wick's theorem, in terms of the quadratic correlations as

\begin{equation}\label{eq:CSviolation4Wick}
\Theta_{ij}=|c_{ij}|^2+|g_{ij}|^2+g_{ii}g_{jj}-\sqrt{2g_{ii}^2+|c_{ii}|^2}\sqrt{2g_{jj}^2+|c_{jj}|^2}
\end{equation}

On the other hand, for Gaussian states, the GPH condition is equivalent to the non-separability of the system. Comparing the expressions for the quartic CS violation (\ref{eq:CSviolation4Wick}) and the GPH criterion (\ref{eq:GPHcomplete}), we clearly see that they represent different conditions for arbitrary Gaussian states. In particular, it is easy to find states violating the quartic CS inequality with $\det C_{ij}>0$, which means that they are separable. Also, we can find entangled states that satisfy the quartic CS inequality.

\section{Criteria for detection of spontaneous Hawking radiation}\label{sec:criteriaHR}

We turn back our attention to analog Hawking radiation in Bose-Einstein condensates; see Sec. \ref{subsec:BHBEC} for checking the notation followed. We study the correlations between the ``out'' modes at the same frequency $\omega$ over the state of the system, $\hat{\rho}$. In particular, we focus on the specific case of states which are Gaussian and incoherent in the ``in'' modes. The latter requirement can be expressed as:
\begin{eqnarray} \label{eq:incoherent}
\nonumber \braket{\hat{\gamma}_{i-\rm{in}}(\omega)\hat{\gamma}_{j-\rm{in}}(\omega')}&=&0\\
\braket{\hat{\gamma}_{i-\rm{in}}^{\dagger}(\omega)\hat{\gamma}_{j-\rm{in}}(\omega')}
&=&
n_i(\omega)\delta_{ij}\delta(\omega-\omega')\, .
\end{eqnarray}
These correlation functions fully characterize the state provided it is Gaussian. It is shown in Appendix \ref{app:parametrization} that, for this class of states, we only need $7$ parameters to describe the CS violations and the GPH function. The case of Gaussian states is important because, within the Bogoliubov approximation, the Hamiltonian is quadratic and hence, the resulting dynamics are Gaussian.

Another interesting feature of Gaussian and incoming incoherent states is that, as we show later, the GPH criterion and the quartic and quadratic CS violations become equivalent criteria. Moreover, a most important particular case of this particular class of states is that considered in many of the main works on analog HR, which involves a thermal distribution of incoming quasi-particles that have thermalized in the comoving reference frame \cite{Macher2009a,Busch2014} so their occupation factor is:
\begin{equation}\label{eq:thermaldistribution}
n_{i}(\omega)=\frac{1}{\exp(\frac{\hbar\Omega_{i}(\omega)}{k_BT})-1}
\end{equation}
with $\Omega_{i}(\omega)$ the commoving frequency of the mode $i$-in at laboratory frequency $\omega$, as given by Eq. (\ref{eq:dispersionrelation}).

For this class of states, neither sub-Poissonian statistics or anti-bunching [violation of Eqs.(\ref{eq:ClassicalIneqa}), (\ref{eq:ClassicalIneqb})] can be observed in the phonon correlations here considered so we focus on studying the criteria discussed in Sec. \ref{sec:CSPH}: CS violations and the GPH criterion.

\subsection{Cauchy-Schwarz inequalities and generalized Peres-Horodecki criterion in analog Hawking radiation}

In order to study CS violations in HR analogs, we evaluate the correlation functions of Eqs. (\ref{eq:gdef}) and (\ref{eq:GammaDef}) for the ``out'' modes at given $\omega$:
\begin{eqnarray}\label{eq:CSGamma}
\nonumber\Gamma_{ij}(\omega)&=&\braket{\hat{\gamma}_{i-\rm{out}}^{\text{\ensuremath{\dagger}}}(\omega)\hat{\gamma}_{j-\rm{out}}^{\dagger}(\omega)\hat{\gamma}_{j-\rm{out}}(\omega)\hat{\gamma}_{i-\rm{out}}(\omega)}\\
g_{ij}(\omega)&\equiv&\braket{\hat{\gamma}_{i-\rm{out}}^{\dagger}(\omega)\hat{\gamma}_{j-\rm{out}}(\omega)}\\
\nonumber c_{ij}(\omega)&\equiv&\braket{\hat{\gamma}_{i-\rm{out}}(\omega)\hat{\gamma}_{j-\rm{out}}(\omega)} \, .
\end{eqnarray}
Hereafter, $i,j=u,d1,d2$. The associated quartic and quadratic CS violations are characterized by the positivity of $\Theta_{ij}(\omega)$ and $\Delta_{ij}(\omega)$ or by the equivalent condition for the relative magnitudes $\theta_{ij}(\omega),\delta_{ij}(\omega)>1$, as defined in Eqs. (\ref{eq:CSviolation4}),(\ref{eq:CSviolation2}). In a similar fashion, we study the GPH function $\mathcal{P}_{ij}(\omega)$ for the $i,j$  ``out'' modes at the same frequency $\omega$. For simplicity, in this section we will obviate the Dirac delta factors appearing in all equal-frequency correlation functions [see Eq. (\ref{eq:incoherent}) for a complete expression].

We define for convenience the complex vector
\begin{eqnarray}\label{eq:defcorrs}
\alpha_i(\omega)&\equiv&\left[\begin{array}{c}
S_{iu}(\omega)\sqrt{n_u(\omega)}\\
S_{id1}(\omega)\sqrt{n_{d1}(\omega)}\\
S_{id2}(\omega)\sqrt{n_{d2}(\omega)+1}
\end{array}\right]\, .
\end{eqnarray}
From the previous definitions, it is easy to show that the only non-zero quadratic correlations for ``out'' modes are:
\begin{eqnarray}\label{eq:quadraticorrs}
\nonumber g_{II'}&=& \alpha_{I}^{\dagger}\cdot\alpha_{I'} \\
\nonumber g_{d2d2}&=&|\alpha_{d2}|^{2}-1\\
c_{Id2}&=&\alpha_{d2}^{\dagger}\cdot\alpha_{I}\, ,
\end{eqnarray}
where the index $I$ stands for a normal, positive-normalization ($u$ or $d1$) mode.
First, we study the correlation between a normal and an anomalous (negative-normalization) mode.
The case $I=u$ corresponds to the proper Hawking effect and the case $I=d1$ corresponds to the bosonic equivalent of the Andreev reflection, both previously discussed in Sec. \ref{subsec:hawkingeffect}.

As we are working with Gaussian states, we can apply Wick's theorem to express the quartic correlations as a function of the quadratic correlations
\begin{eqnarray}\label{eq:CSWick}
\nonumber \Gamma_{II} & = & 2g^2_{II}=2|\alpha_{I}|^{4}\\
\nonumber \Gamma_{d2d2} & = & 2g^2_{d2d2}=2(|\alpha_{d2}|^{2}-1)^{2}\\
\nonumber \Gamma_{Id2} & = & |c_{Id2}|^2+g_{II}g_{d2d2}\\
&=&|\alpha_{d2}^{\dagger}\cdot\alpha_{I}|^{2}+|\alpha_{I}|^{2}(|\alpha_{d2}|^{2}-1)\, .
\end{eqnarray}
Therefore, the condition for the quartic CS violation $\Theta_{Id2}(\omega)>0$ reduces to the simpler quadratic CS violation
\begin{equation}\label{eq:CSnonseparable}
\Theta_{Id2}(\omega)=\Delta_{Id2}(\omega)>0
\end{equation}
which, using Eq. (\ref{eq:quadraticorrs}), can be rewritten as:
\begin{equation}\label{eq:CSnonseparablealphas}
|\alpha_{d2}^{\dagger}\cdot\alpha_{I}|^{2}>|\alpha_{I}|^{2}(|\alpha_{d2}|^{2}-1)
\end{equation}
The $-1$ within the second bracket results from the anomalous character of the $d2$ mode and is responsible for making the violation of the CS inequality possible. In respect to the relative violations, we obtain a similar relation:
\begin{equation}\label{eq:CSrelatives}
\theta_{Id2}(\omega)=\frac{1+\delta_{Id2}(\omega)}{2}>1
\end{equation}
and then, $\theta_{Id2}(\omega)>1$ if and only if $\delta_{Id2}(\omega)>1$. Thus, we have proven that the quartic and the quadratic violations are equivalent for $Id2$, normal-anomalous correlations.

In regard to the correlation between the two normal modes, we find that the quartic CS inequality is equivalent to the condition:
\begin{equation}\label{eq:CSnormal}
|g_{ud1}|^2\leq g_{uu}g_{d1d1} \, ,
\end{equation}
which can never be violated [see Eq. (\ref{eq:CSgijimpossible})]. In particular, the inequality (\ref{eq:CSnormal}) can be rewritten as:
\begin{equation}\label{eq:CSnonviolation}
|\alpha_{u}^{\dagger}\cdot\alpha_{d1}|^2\leq |\alpha_{u}|^2|\alpha_{d1}|^2 \, ,
\end{equation}
which is always satisfied for two complex vectors as it is the usual CS inequality appearing in linear algebra. Thus, there is no CS violation in the correlation between normal modes.

Taking into account the pseudo-unitary character of the scattering matrix, Eq. (\ref{eq:pseudounitarity}), we can transform Eq. (\ref{eq:CSnonseparable}) for $I=u$ into the equivalent relation
\begin{eqnarray}\label{eq:CSequivalent}
 &  & |S_{ud2}|^{2}(1+n_{u}+n_{d1}+n_{d2})>\nonumber \\
 &  & |S_{d1u}|^{2}n_{d1}n_{d2}+|S_{d1d1}|^{2}n_{u}n_{d2}+|S_{d1d2}|^{2}n_{u}n_{d1}\nonumber \\
 &  & +|S_{d2d1}|^{2}n_{u}+|S_{d2u}|^{2}n_{d1}\,.\label{eqFiniteTempCSViol}
\end{eqnarray}
A similar inequality can be derived for $I=d1$ by simply interchanging the $u$-out and $d1$-out labels in the previous equation.

We consider now the CS violation in the zero temperature limit $T=0$, where the state of the system is the incoming vacuum, i.e., all the occupation numbers are zero $n_i(\omega)=0$. We obtain from Eqs. (\ref{eq:defcorrs}-\ref{eq:CSnonseparablealphas}):
\begin{equation}
\theta_{ud2}=\frac{|S_{d2d2}|^{2}-1/2}{|S_{d2d2}|^{2}-1}>1\,,\,\,\Theta_{ud2}=|S_{ud2}|^{2}>0\,,\label{eqZeroTempCSViol}
\end{equation}
Thus, at $T=0$, whenever $S_{ud2}\neq0$, there is CS violation, which further reflects the direct link between CS violation and HR (note that  pseudo-unitarity implies $|S_{d2d2}|>1$ if $S_{ud2}\neq0$).
We note that $\theta_{d1d2}=\theta_{ud2}$
and $\Theta_{d1d2}=|S_{d1d2}|^{2}$ at zero temperature.

A device producing HR works like a nondegenerate parametric amplifier, which is known to generate squeezing from vacuum. However, sources
other than vacuum may also generate squeezing and ultimately CS violation
\cite{Walls2008}. Therefore, the following question arises for a general state different from the vacuum: is the CS violation a conclusive evidence of spontaneous (zero-point) HR radiation?

The answer is yes. Imagine that we neglect the zero-point contributions when computing all the first and second order correlation functions. Then, one arrives at a modified version of (\ref{eq:CSequivalent})
where only the terms quadratic in the $n_{i}$'s of the r.h.s. survive. The resulting
inequality is never satisfied. We conclude that CS violation requires
vacuum fluctuations.

The upshot of this discussion is that the detection of a CS inequality
violation can be regarded as a smoking gun
for the presence of spontaneous HR.
In other words, the spontaneous signal is the
cause of the quantum behavior as reflected in the violation of the classical inequalities.
This contrasts with other schemes in which the spontaneous signal cannot be distinguished unambiguously
from the stimulated (thermal) signal, which has, in this sense, a classical
behavior and thus would never originate the CS violation by itself. The same reasoning applies to the small perturbations around the condensate wave function since they are coherent excitations.

Now we turn our attention to the GPH criterion. From Eq. (\ref{eq:GPHcomplete}) we compute the GPH function for the pair of modes $Id2$:
\begin{eqnarray}\label{eq:GPHanomalous1}
\mathcal{P}_{Id2}(\omega)&=&(g_{II}g_{d2d2}-|c_{Id2}|^2)[(g_{II}+1)(g_{d2d2}+1)-|c_{Id2}|^2]\\
\nonumber &=&\Delta_{Id2}(\omega)[(g_{II}+1)(g_{d2d2}+1)-|c_{Id2}|^2]
 \, .
\end{eqnarray}

As the second (square) bracket in the r.h.s. of Eq. (\ref{eq:GPHanomalous1}) is always positive [see Eq. (\ref{eq:CScijpos})], we conclude that $\mathcal{P}_{Id2}<0$ if and only if Eq. (\ref{eq:CSnonseparable}) is satisfied. As the state is Gaussian, the GPH criterion is equivalent to the entanglement of the state $\hat{\rho}$. Therefore, quadratic and quartic CS violations are equivalent to entanglement. In this way, from the theoretical point of view, we can focus on studying the simpler quadratic CS violation since all the extracted conclusions apply to the GPH criterion and the quartic CS violation.

We note that the equivalence between condition (\ref{eq:CSnonseparable}) and entanglement was already pointed out in Ref. \cite{Busch2014} but the connection with the CS violation criterion was not made explicit. A similar result appeared in Ref. \cite{Busch2014a} on the equivalence between non-separability and CS violation when studying the correlation function at two different times in fluids of light.

When considering the entanglement between the two normal modes, we have $|g_{ud1}|>|c_{ud1}|=0$ and thus there is no entanglement [see Eq. (\ref{eq:GPHcomplete}) and the ensuing discussion].

From the previous results we conclude that, when considering Gaussian and incoherent states [i.e., Gaussian states satisfying Eq. (\ref{eq:incoherent})], all the here considered criteria for characterizing the quantum character of the system become equivalent. This result is important since it unifies the work of Refs. \cite{deNova2014,Finazzi2013,Busch2014}.

\subsection{Final remarks}

\begin{table}[!tb]
\centering
\begin{tabular}[c]{|c|c|c|c|c|}
\hline
Gaussian & Incoherent & $\text{CS4}$ & $\text{CS2}$ & GPH\\
\hline
Yes & Yes &~$\bullet$ $\Leftrightarrow$ GPH &~ $\bullet$ $\Leftrightarrow$ GPH &~ $\bullet$ $\Leftrightarrow$ NS\\
\hline
Yes & No & ~$\bullet$ indep.  NS &~ $\bullet$ $\Rightarrow$ GPH &~ $\bullet$ $\Leftrightarrow$ NS\\
\hline
No & Yes &~$\bullet$ indep.  NS &~ $\bullet$ $\Leftrightarrow$ GPH &~ $\bullet$ $\Rightarrow$ NS\\
\hline
No & No & ~$\bullet$ indep.  NS &~ $\bullet$ $\Rightarrow$ GPH &~ $\bullet$ $\Rightarrow$ NS\\
\hline
\end{tabular}
\caption{Logical relations between the different criteria studied throughout this work in the various physical cases. The three rightmost entries in the upper row indicate the different criteria for the identification of quantum behavior:
$\text{CS2}$, $\text{CS4}$ stand for quadratic and quartic CS violations, respectively, while GPH stands for generalized Peres-Horodecki criterion; NS means non-separability.
The two leftmost columns define
the various physical cases here considered. By ``incoherent'' state, we understand a density matrix which is diagonal in the representation of retarded quasiparticle scattering modes, each characterized by a single incoming channel; see Eq. (\ref{eq:incoherent}). The symbol $\bullet$ stands for the uppermost entry in the corresponding column.
The abbreviation ``indep.'' means that, in the three lower cases, the quartic CS violation is independent of the non-separability of the system.}
\label{table}
\end{table}

We briefly discuss the differences that arise when removing some restrictions. If the state is Gaussian but not incoherent in the incoming channels, the GPH criterion is still equivalent to the non-separability of the system. On the other hand, under the same assumption, the quadratic CS violation is no longer equivalent to the GPH criterion; rather, it is only a sufficient condition for it, as argued in Sec. \ref{subsec:quadraticGPH}. Consequently, the GPH criterion is a more powerful criterion than the quadratic CS violation to detect nonseparability (it can identify nonseparability in cases where the quadratic CS violation would fail). However, when relaxing the requirement of incoherence, the quartic CS violation becomes independent of the GPH criterion (see Sec. \ref{subsec:quarticGPH}). Even when these two conditions are independent, both of them require the system to be described by a non-classical Glauber-Sudarshan function, i.e., a Glauber-Sudarshan function that takes negative values (see discussion in Sec. \ref{sec:GeneralQuantum}).

On the other hand, if the state of the system is not Gaussian but is incoherent in the incoming channels, the quartic CS violation is also independent of the entanglement. The quadratic CS violation and the GPH criterion are still equivalent between them, but no longer equivalent to the entanglement of the system.

For a general state which does not satisfy any of the previous conditions, the quadratic CS violation is only a sufficient condition for the GPH criterion, which in turn is a sufficient condition for the presence of entanglement. We summarize all these logical relations in Table \ref{table}.

Remarkably, the quadratic CS violation reveals at the same time two different aspects of quantum behavior: the violation of a classical inequality and the entanglement of the system.

\section{Experimental detection schemes} \label{sec:Experimental}

In this section, we discuss the possible experimental implementation of the previous criteria. A particular type of CS violation between two colliding condensates was measured using time-of-flight (TOF) detection \cite{Kheruntsyan2012}. Not long ago, a new proposal of detecting spontaneous HR through TOF measurement of the phonon signal has been published \cite{Boiron2015}. On the other hand, the detection of entanglement using the GPH criterion was analyzed in Ref. \cite{Finazzi2013} considering density fluctuations or the optomechanical detection of phonons. Recently, it has been proposed that the entanglement of Hawking radiation in a BEC can be detected experimentally by measuring the density-density correlations through \textit{in situ} imaging \cite{Steinhauer2015}. Following these works, we consider two detection schemes to detect spontaneous HR: TOF experiments and density-density correlations. All the results presented in this section can be straightforwardly translated to study the Andreev reflection involving $d1,d2$ modes. For the calculations, apart from stationarity and ``Gaussianity'', we do not make any particular assumption on the form of the state of the system, $\hat{\rho}$.

\subsection{Correlation between phonon and atomic time-of-flight signals.}\label{subsec:Atom2PhononTOF}

In this specific detection scheme, we aim at detecting CS violation not in the phonon signal but rather in the atomic signal observed in a TOF experiment. We start by considering the atomic wave vectors involved for a given laboratory phonon frequency $\omega>0$, $p_{u}(\omega)\equiv q_{u}+k_{u-{\rm out}}(\omega)$,
$p_{d2}(\omega)\equiv q_{d}-k_{d2-{\rm out}}(\omega)$, with $q_{r}$ the momentum of the wave function in the asymptotic homogeneous regions $r=u,d$. We require the atomic wave vectors from the upstream (downstream)
region to be negative (positive), i.e., we just consider frequencies where
\begin{equation}\label{eq:nocillamix}
p_{u}(\omega)<0
\end{equation}
[note that $p_{d2}(\omega)>0$ is always verified for $\omega<\omega_{\rm max}$ since $k_{d2-\rm{out}}(\omega_{\rm max})<q_{d}$]. If all these conditions are fulfilled, then there is no contribution from the condensate wave function to the atomic signals and the corresponding wave packets travel in opposite directions without mixing. We suppose that the signals from the upstream and downstream regions can be experimentally isolated.

We define the atom destruction operators localized in the asymptotic regions $r$ as
\begin{equation}\label{eq:Fourieroperators}
\hat{c}_{r}\left( p \right)  \equiv\int\mathrm{d}x\, \hat{\Psi}(x)f_{r}^{*}\left(x\right)
e^{-ipx}e^{-i\theta_r}
\end{equation}
where we include the asymptotic phase of the condensate $\theta_r$ in order to get rid of it in the resulting expressions. In the previous definition, the functions $f_{r}\left(x\right)$ are normalized functions ($\int\mathrm{d}x|f_{r}(x)|^{2}=1$) localized only in the corresponding asymptotic regions and
their Fourier transforms are $f_{r}(k)=\frac{1}{\sqrt{2\pi}}\int\mathrm{d}x~e^{-ikx}f_{r}\left(x\right)$. They represent the envelope of the Fourier transform, which has to be introduced to explicitly take into account the fact that the asymptotic subsonic and supersonic regions are placed in different spatial regions in a realistic situation. For simplicity, we
assume that they are of the form
\begin{equation}\label{eq:envelopefunction}
f_{r}(x)=\frac{1}{\sqrt{L_{r}}}f\left(\frac{x-x_{r}}{L_{r}}\right) \, ,
\end{equation}
where $f$ is a symmetric and real dimensionless function. We assume that $L_{r}$ are sufficiently large for $f_{r}(k)$ to be well peaked at zero momentum. In this particular scenario of TOF detection, we can understand $f_{r}(x)$ as the envelopes arising from the finite size, $L_r$, of the asymptotic regions and $x_r$ as the points in which they are centered. In the following, we assume that the size of the scattering region is much smaller than the size of the asymptotic regions. This implies that $x_u<0$ and $x_d>0$ and that:
\begin{equation}\label{eq:centers}
\frac{x_{d}}{L_{d}}\simeq-\frac{x_{u}}{L_{u}}\simeq1/2.
\end{equation}
Using Eq. (\ref{eq:BHFieldOperator}) and following Eq. (\ref{eq:scatteringchannelstate}), we obtain the atomic operators:
\begin{eqnarray}\label{eq:cks}
\nonumber\hat{c}_{u}\left(p_{u}\left(\omega\right)\right) & \simeq & \int\mathrm{d}\omega'~f^{*}_u(k_{u-\rm{out}}(\omega')-k_{u-\rm{out}}(\omega))\frac{u_{u-\rm{out}}(\omega')}
{\left|w_{u-\rm{out}}\left(\omega'\right)\right|^{1/2}}
\hat{\gamma}_{u-\rm{out}}(\omega')\\
& + & f^{*}_u(-k_{u-\rm{in}}(\omega')-k_{u-\rm{out}}(\omega))
\frac{v_{u-\rm{in}}^{*}(\omega')}
{|w_{u-\rm{in}}\left(\omega'\right)|^{1/2}}
\hat{\gamma}_{u-\rm{in}}^{\dag}(\omega')\\
\nonumber\hat{c}_{d}\left(p_{d2}\left(\omega\right)\right) & \simeq & \int_{0}^{\omega_{\rm max}}\mathrm{d}\omega'~f^{*}_d(k_{d2-\rm{out}}(\omega)-k_{d2-\rm{out}}(\omega'))
\frac{v_{d2-\rm{out}}^{*}(\omega')}
{|w_{d2-\rm{out}}\left(\omega'\right)|^{1/2}}
\hat{\gamma}_{d2-\rm{out}}(\omega')\\
\nonumber & + & \int~\mathrm{d}\omega'~f^{*}_d(k_{d2-\rm{out}}(\omega)-k_{d1-\rm{out}}(\omega'))
 \frac{v_{d1-\rm{out}}^{*}(\omega')}
 {|w_{d1-\rm{out}}\left(\omega'\right)|^{1/2}}
 \hat{\gamma}_{d1-\rm{out}}^{\dag}(\omega'),
\end{eqnarray}

\begin{figure*}[!tb]
\begin{tabular}{@{}cc@{}}
    \includegraphics[width=0.5\columnwidth]{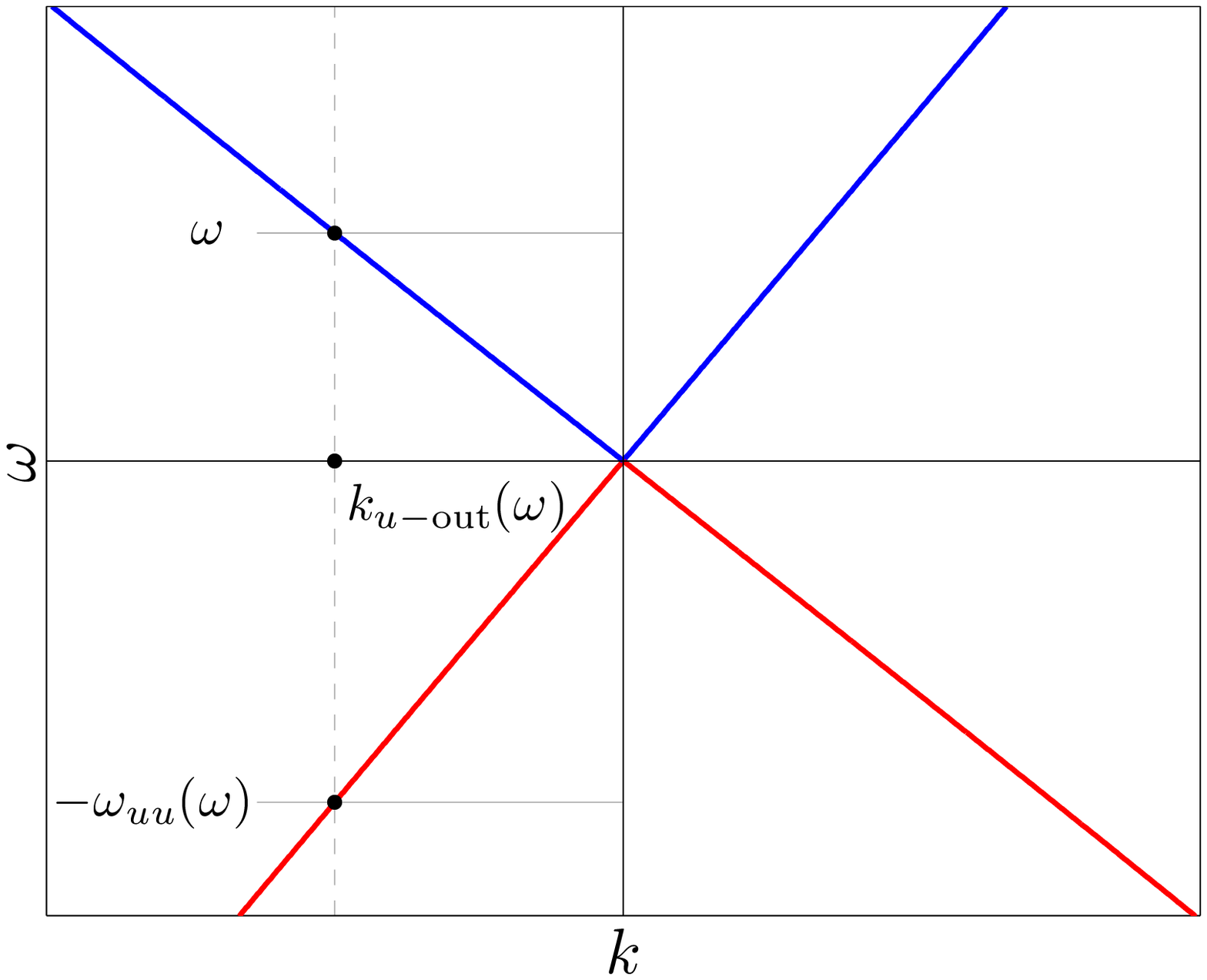} &
    \includegraphics[width=0.482\columnwidth]{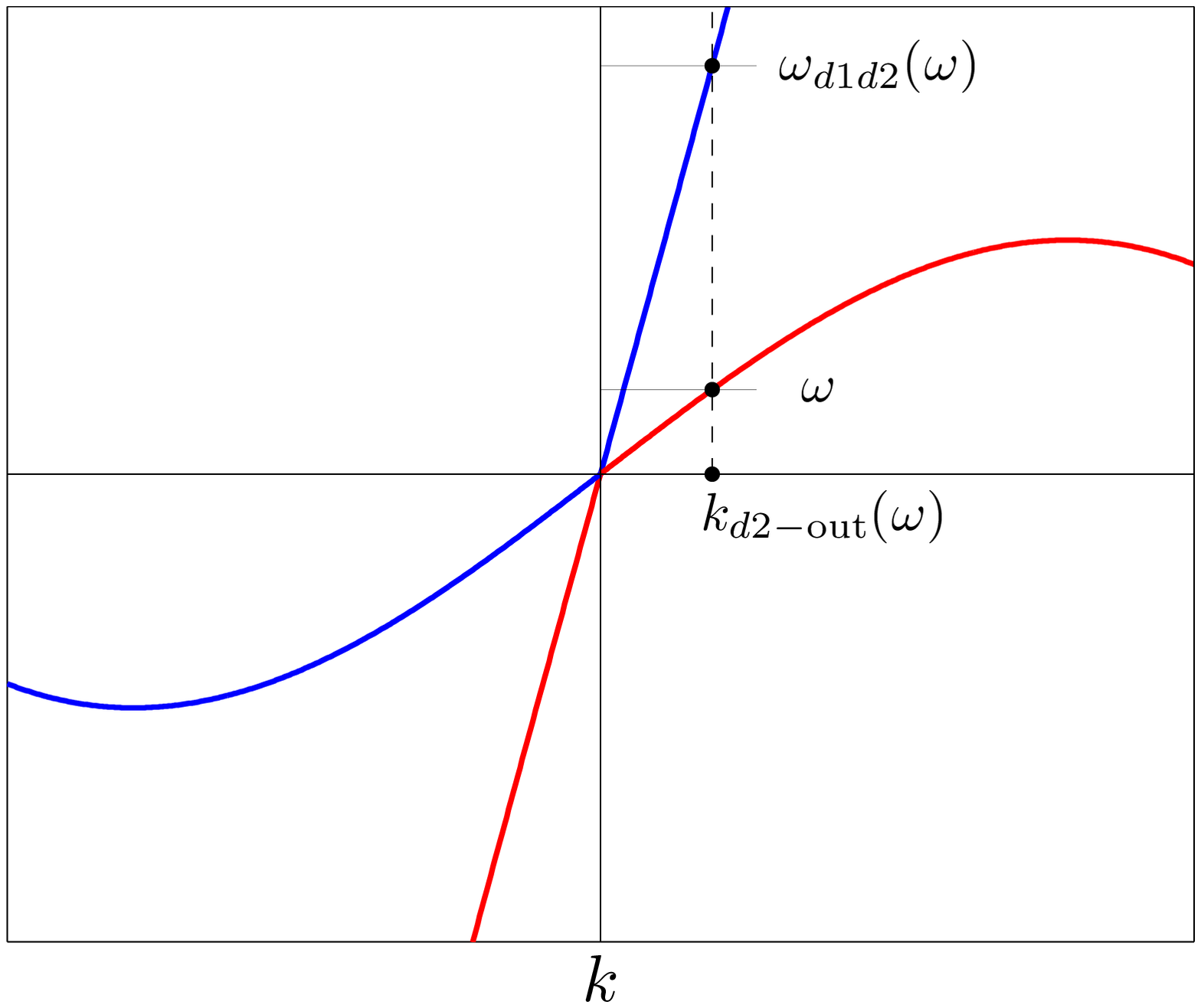} \\
\end{tabular}
\caption{Scheme that shows graphically the values of $\omega_{d1d2}(\omega)$ and $-\omega_{uu}(\omega)$.
See Fig. \ref{figDispRelation} for mode notation.}
\label{fig:DepletionOmegas}
\end{figure*}

We can now compute the corresponding atomic first-order correlation functions, namely,
\begin{eqnarray}\label{eq:atomicfirstorder}
\braket{\hat{c}_{u}^{\dagger}\left(p_{u}\left(\omega\right)\right)\hat{c}_{u}\left(p_{u}\left(\omega\right)\right)} & = & |u_{u-\rm{out}}(\omega)|^{2}|\alpha_{u}(\omega)|^{2}
+ |v_{u-\rm{in}}(\omega_{uu}(\omega))|^{2}[1+n_{u}(\omega_{uu}(\omega))]\nonumber \\
\braket{\hat{c}_{d}^{\dagger}\left(p_{d2}\left(\omega\right)\right)\hat{c}_{d}\left(p_{d2}\left(\omega\right)\right)} & = & |v_{d2-\rm{out}}(\omega)|^{2}[|\alpha_{d2}(\omega)|^{2}-1]\\
\nonumber &+&|v_{d1-\rm{out}}(\omega_{d1d2}(\omega))|^{2}[|\alpha_{d1}(\omega_{d1d2}(\omega))|^{2}+1] \\ \nonumber\braket{\hat{c}_{d}\left(p_{d2}\left(\omega\right)\right)\hat{c}_{u}\left(p_{u}\left(\omega\right)\right)} & = & [\alpha_{d2}^{\dag}(\omega)\cdot \alpha_{u}(\omega)] u_{u-\rm{out}}(\omega)v_{d2-\rm{out}}^{*}(\omega)F(\omega),
\end{eqnarray}
where $\omega_{d1d2}(\omega)=\omega_{d1}(k_{d2-\rm{out}}(\omega))$ and
$\omega_{uu}(\omega)=\omega_{u}(-k_{u-\rm{out}}(\omega))$, with $\omega_{i}(k)$
the dispersion relation of scattering channel $i$ (see Fig. \ref{fig:DepletionOmegas}). We see that the terms related to the direct phonon signal are weighted with the dominant component of the BdG spinor (remember that for the $d2$ modes, $v_{d2-\rm{out}}>u_{d2-\rm{out}}$ as they are anomalous) while the contribution related to the depletion cloud is weighted by the minor component of the BdG spinor.

The overlap function in the last line of Eq. (\ref{eq:atomicfirstorder}) is:
\begin{eqnarray}\label{eq:overlap}
&&F(\omega) = \int\mathrm{d}k~f_{u}^{*}\left(\frac{k}{\zeta(\omega)}\right)f_{d}^{*}\left(k\zeta(\omega)\right)
\\ \nonumber
&=& \frac{1}{\sqrt{L_uL_d}}
\int\mathrm{d}x\, f^{*}\left(\frac{\zeta\left(\omega\right)x+x_{u}}{L_{u}}\right)f^{*}\left(\frac{\zeta\left(\omega\right)^{-1}x-x_{d}}{L_{d}}\right)\nonumber \, ,
\end{eqnarray}
with
\begin{equation}\label{eq:zetaomega}
\zeta(\omega) \equiv\left|\frac{w_{u-\rm{out}}(\omega)}{w_{d2-\rm{out}}(\omega)}\right|^{1/2} \, .
\end{equation}

From a mathematical point of view, the integral $F\left(\omega\right)$ can be regarded as
the scalar product of two normalized functions, hence $F(\omega)\leq1$.
Taking into account Eq. (\ref{eq:centers}) and that the $f$ function is normalized, it easy to show that this inequality saturates for $\omega_0$ such that
\begin{equation}\label{eq:TOFcriterion}
\zeta(\omega_0)=\sqrt{\frac{L_{u}}{L_{d}}}
\end{equation}
The previous condition has a straightforward physical interpretation. The phonon operators
are evaluated in the frequency domain. The frequency resolution near a given frequency $\omega_0$ for both wave packets is given by
\begin{equation}\label{eq:resolution}
\Delta\omega_{u-\rm{out}/d2-\rm{out}}(\omega_0)=|w_{u-\rm{out}/d2-\rm{out}}\left(\omega_0\right)|\Delta k_{u/d}.
\end{equation}
where the resolution in momentum space is $\Delta k_{r}\sim1/L_{r}$. As we are correlating the phonon operators of $u-\rm{out}$ and $d2-\rm{out}$
modes in the vicinity of a given frequency, the maximum correlation is achieved whenever $\Delta\omega_{u-\rm{out}}(\omega_0)=\Delta\omega_{d2-\rm{out}}(\omega_0)$ and
this implies the condition (\ref{eq:TOFcriterion}). For an arbitrary frequency $\omega$, we obtain, supposing that the $f$ function is the characteristic function of the interval $[0,1]$, that
\begin{equation}\label{eq:TOFintegral}
F(\omega)=\frac{1+\min\left[\upsilon(\omega),\upsilon^{-1}(\omega)\right]}{2},~\upsilon(\omega)=\zeta(\omega)\sqrt{\frac{L_{d}}{L_{u}}}
\end{equation}

We define the following second-order correlation functions for the atomic operators, which can be experimentally measured in a TOF experiment:
\begin{equation}\label{eq:TOFatomcorrelations}
G^{\rm{TOF}}_{ij}(\omega)\equiv\braket{\hat{c}_{i}^{\dagger}\left(p_{i}\right)\hat{c}^{\dagger}_{j}\left(p_{j}\right)\hat{c}_{j}\left(p_{j}\right)\hat{c}_{i}\left(p_{i}\right)}
\end{equation}
where $\hat{c}_k$ for $k=d1,d2$ must be interpreted as $\hat{c}_d$. Their associated absolute and relative violations are given (for $i\neq j$) by:
\begin{eqnarray}\label{eq:atomicviolation}
Z_{ij}(\omega)&\equiv&G^{\rm{TOF}}_{ij}-\sqrt{G^{\rm{TOF}}_{ii}}\sqrt{G^{\rm{TOF}}_{jj}}\\
\nonumber z_{ij}(\omega)&\equiv&\frac{G^{\rm{TOF}}_{ij}}{\sqrt{G^{\rm{TOF}}_{ii}}\sqrt{G^{\rm{TOF}}_{jj}}}
\end{eqnarray}
which are also a signal of quantum behavior by themselves. Taking into account the assumption that the state is Gaussian, the second-order correlation functions can be expressed, using Wick's theorem, in terms of the first-order correlation functions as
\begin{eqnarray}\label{eq:atomicGWick}
G^{\rm{TOF}}_{uu}(\omega) & = & \braket{\hat{c}_{u}^{\dagger}\left(p_{u}\right)\hat{c}^{\dagger}_{u}\left(p_{u}\right)\hat{c}_{u}\left(p_{u}\right)\hat{c}_{u}\left(p_{u}\right)}\nonumber\\&=&2|\braket{\hat{c}_{u}^{\dagger}\left(p_{u}\right)\hat{c}_{u}\left(p_{u}\right)}|^{2}\nonumber \\
G^{\rm{TOF}}_{d2d2}(\omega) & = & \braket{\hat{c}_{d}^{\dagger}\left(p_{d2}\right)\hat{c}_{d}^{\dagger}\left(p_{d2}\right)\hat{c}_{d}\left(p_{d2}\right)\hat{c}_{d}\left(p_{d2}\right)}\nonumber\\&=&2|\braket{\hat{c}_{d}^{\dagger}\left(p_{d2}\right)\hat{c}_{d}\left(p_{d2}\right)}|^{2}\\
\nonumber G^{\rm{TOF}}_{ud2}(\omega) & = & \braket{\hat{c}_{d}^{\dagger}\left(p_{d2}\right)\hat{c}_{u}^{\dagger}\left(p_{u}\right)\hat{c}_{d}\left(p_{d2}\right)\hat{c}_{u}\left(p_{u}\right)}\nonumber\\&=&\braket{\hat{c}_{d}^{\dagger}\left(p_{d2}\right)\hat{c}_{u}^{\dagger}\left(p_{u}\right)}
\braket{\hat{c}_{d}\left(p_{d2}\right)\hat{c}_{u}\left(p_{u}\right)}\nonumber\\ &+&\braket{\hat{c}_{d}^{\dagger}\left(p_{d2}\right)\hat{c}_{d}\left(p_{d2}\right)}\braket{\hat{c}_{u}^{\dagger}\left(p_{u}\right)\hat{c}_{u}\left(p_{u}\right)}\nonumber \, .
\end{eqnarray}

By comparing Eqs. (\ref{eq:atomicfirstorder}) and (\ref{eq:atomicGWick}) with Eq. (\ref{eq:quadraticorrs}), one can infer that $z_{ud2}>1$ implies $\delta_{ud2}>1$, i.e., that the quartic {\it atom} CS violation is a sufficient condition for the quadratic CS violation in the phonon signal.

We see that in Eq. (\ref{eq:atomicGWick}) the crossed correlations of the type $\braket{\hat{c}_{u}\left(p_{u}\left(\omega\right)\right)\hat{c}_{u}\left(p_{u}\left(\omega\right)\right)}$, $\braket{\hat{c}_{d}\left(p_{d2}\right)\hat{c}_{d}\left(p_{d2}\right)}$, $\braket{\hat{c}_{u}^{\dagger}\left(p_{u}\left(\omega\right)\right)\hat{c}_{d}\left(p_{d2}\right)}$ and so on vanish. This is of course true for an incoming incoherent state since in that case they are identically zero. Interestingly, this result also holds for a general state but for a very different reason: for those correlations, the appearing overlap integrals, similar to $F(\omega)$, give zero. We refer to the end of Sec. \ref{subsec:densitycorrelations} for the complete explanation of this phenomenon.

\subsection{Density-density correlations}\label{subsec:densitycorrelations}

The role of the spatial density correlation function in analog models has been extensively studied in several works \cite{Balbinot2008,Recati2009,Larre2012}. In the BEC context, the measurement of the spatial density correlations has been experimentally used to characterize the BH laser \cite{Steinhauer2014}. The density-density correlation function also plays an important role in the polariton analog \cite{Gerace2012,Nguyen2015}.

We start our analysis by considering the density in the BdG approximation of Sec. \ref{sec:physmodel} and expanding it up to first order in the fluctuations of the field operator
\begin{equation}\label{eq:BdGdensity}
\hat{n}(x)=\hat{\Psi}^{\dagger}(x)\hat{\Psi}(x)\simeq n(x)+\delta\hat{n}(x),~\delta\hat{n}(x)=\Psi^*_0(x)\hat{\varphi}(x)+\Psi_0(x)\hat{\varphi}^{\dagger}(x)
\end{equation}
and $n(x)=|\Psi_0(x)|^2$ the usual mean-field density. The expression for the density fluctuations is formally similar to that of $\hat{\varphi}(x)$, see Eq. (\ref{eq:BHFieldOperator}). Importantly, the density fluctuation operator $\delta\hat{n}(x)$ in the BdG approximation is linear in the destruction operators, rather than quadratic. Thus, we can extract the quadratic correlations of Sec. \ref{sec:CSPH} by measuring density-density correlations \cite{Finazzi2013,Steinhauer2015}. For that purpose, we restrict ourselves to the asymptotic (subsonic and supersonic) regions. By taking the Fourier transform of the density in the asymptotic regions at $k\neq 0$, we can get rid of the condensate signal and extract the phonon signal. Analogously to Eq. (\ref{eq:Fourieroperators}), we define:
\begin{equation}\label{eq:Fourierdensity}
\hat{n}_{r}\left(k\right)\equiv\int\mathrm{d}x\, \hat{n}(x)f_{r}^{*}\left(x\right)e^{-ikx}\, ,
\end{equation}
with $f_{r}\left(x\right)$ as given in Eq. (\ref{eq:envelopefunction}). However, in this case, the envelope has a different physical explanation. In a TOF experiment, we only have access to the signal in momentum space and thus, the envelopes arise from the finite size of the asymptotic regions. Using \textit{in situ} imaging \cite{Schley2013,Steinhauer2015}, we have direct access to the density in real space so the envelopes $f_{r}\left(x\right)$ can be appropriately selected to extract the corresponding Fourier transforms. In particular, this implies that we can choose freely the values of $x_{u,d},L_{u,d}$, in contrast to the situation of TOF detection, where they were fixed by the size of the asymptotic regions, see Eq. (\ref{eq:centers}) and related discussion.

Taking into account all the previous considerations, we obtain:
\begin{eqnarray}\label{eq:density}
\nonumber\hat{n}_{u}\left(k_{u-\rm{out}}(\omega)\right) & \simeq & \sqrt{n_u}\left[\int\mathrm{d}\omega'~f^{*}_u(k_{u-\rm{out}}(\omega')-k_{u-\rm{out}}(\omega))\frac{r_{u-\rm{out}}(\omega')}
{\left|w_{u-\rm{out}}\left(\omega'\right)\right|^{1/2}}
\hat{\gamma}_{u-\rm{out}}(\omega')\right.\\
& + & \left. f^{*}_u(-k_{u-\rm{in}}(\omega')-k_{u-\rm{out}}(\omega))
\frac{r_{u-\rm{in}}(\omega')}
{|w_{u-\rm{in}}\left(\omega'\right)|^{1/2}}
\hat{\gamma}_{u-\rm{in}}^{\dag}(\omega')\right]\\
\nonumber\hat{n}_{d}\left(-k_{d2-\rm{out}}(\omega)\right) & \simeq & \sqrt{n_d}\int_{0}^{\omega_{\rm max}}\mathrm{d}\omega'~f^{*}_d(k_{d2-\rm{out}}(\omega)-k_{d2-\rm{out}}(\omega'))
\frac{r_{d2-\rm{out}}(\omega')}
{|w_{d2-\rm{out}}\left(\omega'\right)|^{1/2}}
\hat{\gamma}_{d2-\rm{out}}(\omega')\\
\nonumber & + & \sqrt{n_d}\int~\mathrm{d}\omega'~f^{*}_d(k_{d2-\rm{out}}(\omega)-k_{d1-\rm{out}}(\omega'))
\frac{r_{d1-\rm{out}}(\omega')}
 {|w_{d1-\rm{out}}\left(\omega'\right)|^{1/2}}
 \hat{\gamma}_{d1-\rm{out}}^{\dag}(\omega'),
\end{eqnarray}
where $r_a(\omega)$ is [see Eq. (\ref{eq:PlaneWaveSpinors})]
\begin{equation} \label{r-a}
r_{a}(\omega)=u_{a}(\omega)+v_{a}(\omega)=\sqrt{\frac{k^2_a}{2\Omega_a(\omega)}},
\end{equation}
and we have used $|\omega-v_rk_a(\omega)|=\Omega_a(\omega)=\sqrt{c_r^2k^2+\frac{k^4}{4}}$. It is worth noting that, if we leave freedom to choose the values of $x_{u,d},L_{u,d}$ in the envelopes of the atomic operators of Eq. (\ref{eq:Fourieroperators}), we find:
\begin{equation} \label{densityatomoperators}
\hat{n}_r(k)=\sqrt{n_r}[\hat{c}_r(q_r+k)+\hat{c}^{\dagger}_r(-q_r-k)]
\end{equation}
where $n_r$ is the density in the asymptotic region $r$. Now, we can connect the correlations studied in Sec. \ref{sec:CSPH} with the density correlations by taking into account that $\hat{n}^{\dagger}_{r}(k)=\hat{n}_{r}(-k)$. We obtain:

\begin{eqnarray}\label{eq:measuredcorrelations}
\nonumber G_{uu}(\omega)\equiv\braket{\hat{n}_{u}\left(-k_{u-\rm{out}}\left(\omega\right)\right)\hat{n}_{u}\left(k_{u-\rm{out}}\left(\omega\right)\right)}&=& n_ur^2_{u-\rm{out}}(\omega)[g_{uu}(\omega)+n_{u}(\omega_{uu}(\omega))+1]\\
\nonumber G_{d2d2}(\omega)\equiv\braket{\hat{n}_{d}\left(k_{d2-\rm{out}}(\omega)\right)\hat{n}_{d}\left(-k_{d2-\rm{out}}(\omega)\right)} & = & n_dr^2_{d2-\rm{out}}(\omega)[g_{d2d2}(\omega)+g_{d1d1}(\omega_{d1d2}(\omega))+1] \\
G_{ud2}(\omega)\equiv\braket{\hat{n}_{d}\left(-k_{d2-\rm{out}}(\omega)\right)\hat{n}_{u}\left(k_{u-\rm{out}}\left(\omega\right)\right)} & = & F(\omega)\sqrt{n_un_d}r_{u-\rm{out}}r_{d2-\rm{out}}c_{ud2}(\omega) \, ,
\end{eqnarray}
with $\omega_{d1d2}(\omega)$ and $\omega_{uu}(\omega)$ defined after Eq. (\ref{eq:atomicfirstorder}) and $F(\omega)$ given by Eq. (\ref{eq:overlap}). We note the strong analogy of the previous equation with its TOF equivalent (\ref{eq:atomicfirstorder}), which results from the formal analogy between the expressions for the density fluctuations and the field operator fluctuations, compare for instance Eqs. (\ref{eq:cks}) and (\ref{eq:density}).

Similar to the TOF case, it is easy to prove that $F(\omega)=1$ when
\begin{equation}\label{eq:densitylines}
\zeta(\omega)=\sqrt{\frac{L_{u}}{L_{d}}}=\sqrt{\left|\frac{x_{u}}{x_{d}}\right|},
\end{equation}
We have extended the condition (\ref{eq:TOFcriterion}) to all frequencies $\omega$ since $L_{u,d2},x_{u,d}$ are now free parameters and we can vary them to fulfill the previous relation for each value of the frequency. This condition has been interpreted in the case of TOF detection as the condition for the wave-packets to have equal width in frequency space. However, resulting from the well-known properties of the spatial density correlations, Eq. (\ref{eq:densitylines}) can be also interpreted as expressing that the wave packets in the subsonic and supersonic regions have to be placed along the correlation lines that maximize the spatial density correlation function, see Refs. \cite{Balbinot2008,Recati2009,Larre2012}. In the following, we take $F(\omega)=1$.

It is worth noting that the correlation functions $G_{ij}(\omega)$ can never violate the CS inequality by themselves since $G_{ud2}<\sqrt{G_{uu}G_{d2d2}}$. This is due to the fact that now the depletion cloud contribution is weighted with the same coefficient as the direct phonon signal, in contrast to the atomic correlations of the TOF case, see discussion after Eq. (\ref{eq:atomicfirstorder}).

In order to get over this problem, we try to relate these functions, which can be measured, to the phonon correlations. For that purpose, we introduce the normalized density-density correlations:
\begin{equation}\label{eq:normalizedcor}
\mathcal{G}_{ij}(\omega) \equiv \frac{G_{ij}(\omega)}{\sqrt{n_in_j}r_{i-\rm{out}}(\omega)r_{j-\rm{out}}} \, ,
\end{equation}
From Eq. (\ref{eq:measuredcorrelations}), we see that $\mathcal{G}_{ud2}(\omega)=c_{ud2}(\omega)$. For extracting the other correlations, in a similar way to Ref. \cite{Steinhauer2015} we define the magnitudes:
\begin{eqnarray}\label{eq:measuredcorrelations-1}
\nonumber \tilde{g}_{uu}(\omega)&\equiv&g_{uu}(\omega)+n_{u}(\omega_{uu}(\omega))=\mathcal{G}_{uu}(\omega)-1\\
\nonumber \tilde{g}_{d2d2}(\omega)&\equiv&g_{d2d2}(\omega)+g_{d1d1}(\omega_{d1d2}(\omega))=\mathcal{G}_{d2d2}(\omega)-1 \\
\end{eqnarray}
The $-1$ appearing in the definitions of $\tilde{g}_{uu}(\omega),\tilde{g}_{d2d2}(\omega)$ reflects the subtraction of the atomic depletion contribution.

We see that the functions $\tilde{g}_{uu},\tilde{g}_{d2d2}$ are over-estimations of correlations $g_{uu},g_{d2d2}$, since $n_{u},g_{d1d1}$ are always positive. From this, we can define the quantity:
\begin{equation}\label{eq:Delta-ud2}
\tilde{\Delta}_{ud2}\equiv|c_{ud2}|^2-\tilde{g}_{uu}\tilde{g}_{d2d2} \, , \\
\end{equation}
noting that $\Delta_{ud2}>\tilde{\Delta}_{ud2}$. Thus, measuring $\tilde{\Delta}_{ud2}>0$ amounts to experimentally observing the quadratic CS violation, in a close analogy to case of TOF detection, where the measurement of CS violation in the atomic signal implies the quadratic CS violation. We remark that, for these calculations, we have only assumed that the state of the system is stationary and here, we do not even have assumed Gaussian behavior as it was done in the TOF scheme.

The upshot of this discussion is that we can observe the violation of a quadratic CS inequality through the measurement of the function $\tilde{\Delta}_{ud2}$. Similar claims can be made about the measurement of the quadratic CS inequality involving the $d1,d2$ modes, where we can repeat the same strategy of above to obtain the correlation functions $\tilde{g}_{d1d1},c_{d1d2}$. However, it is not possible to obtain the correlation functions $c_{uu},c_{d1d1},c_{d2d2},g_{ud2},g_{d1d2}$ from this scheme. This is due to the vanishing overlap integrals that appear when trying to obtain the corresponding correlations. For instance, if we compute $\braket{\hat{n}_{u}\left(k_{u-\rm{out}}\left(\omega\right)\right)\hat{n}_{u}\left(k_{u-\rm{out}}\left(\omega\right)\right)}$, we face an overlap integral of the type
\begin{equation}\label{eq:overlap}
\int\mathrm{d}k~f_{u}^{*}\left(k\right)f_{u}^{*}\left(k\right)=\int\mathrm{d}x\, f^*_u\left(x\right)f^*_u\left(-x\right)=0,
\end{equation}
because the $f_{u}$ function is, by construction, well localized in the subsonic region, far from the scattering region around $x=0$, see Eq. (\ref{eq:Fourierdensity}) and the discussion below. Thus, we cannot obtain $c_{uu}$ by this method. A similar reasoning applies to other correlations.

It is important to note the crucial role played by the spatial location of the asymptotic regions. If we do not take into account the spatial location by introducing the envelopes in Eq. (\ref{eq:Fourierdensity}) and we take the Fourier transforms as ideally infinite, we would obtain:

\begin{eqnarray}\label{eq:densitynaive}
&~&\hat{n}_{u}\left(k_{u-\rm{out}}(\omega)\right) \simeq \sqrt{n_u}\times\\
\nonumber &~& \times \left(\frac{r_{u-\rm{out}}(\omega)}{\left|w_{u-\rm{out}}\left(\omega\right)\right|^{1/2}}\hat{\gamma}_{u-\rm{out}}(\omega)+
\frac{r_{u-\rm{in}}(\omega_{uu}(\omega))}
{|w_{u-\rm{in}}\left(\omega_{uu}(\omega)\right)|^{1/2}}
\hat{\gamma}_{u-\rm{in}}^{\dag}(\omega_{uu}(\omega))\right)\\
\nonumber &~&\hat{n}_{d}\left(-k_{d2-\rm{out}}(\omega)\right) \simeq \sqrt{n_d}\times\\
\nonumber &~& \times \left(
\frac{r_{d2-\rm{out}}(\omega)}
{|w_{d2-\rm{out}}\left(\omega\right)|^{1/2}}
\hat{\gamma}_{d2-\rm{out}}(\omega)+\frac{r_{d1-\rm{out}}(\omega_{d1d2}(\omega))}
{|w_{d1-\rm{out}}\left(\omega_{d1d2}(\omega)\right)|^{1/2}}
\hat{\gamma}_{d1-\rm{out}}^{\dag}(\omega_{d1d2}(\omega))\right),
\end{eqnarray}
This assumption would lead to contradictory results. For instance, the commutator
\begin{equation}\label{eq:contradiction}
[\hat{n}_{u}\left(k_{u-\rm{out}}(\omega)\right),\hat{n}_{u}\left(-k_{u-\rm{in}}(\omega)\right)]=n_u\frac{r_{u-\rm{out}}(\omega)}{\left|w_{u-\rm{out}}\left(\omega\right)\right|^{1/2}}
\frac{r_{u-\rm{in}}(\omega)}
{|w_{u-\rm{in}}\left(\omega\right)|^{1/2}}S_{uu}(\omega)\neq 0
\end{equation}
is non-vanishing but it should be zero as $[\hat{n}(x),\hat{n}(x')]=0$ for every $x,x'$. However, when using the full expressions of Eq. (\ref{eq:density}) we arrive at the correct result
\begin{eqnarray}\label{eq:consistent}
[\hat{n}_{u}\left(k_{u-\rm{out}}(\omega)\right),\hat{n}_{u}\left(-k_{u-\rm{in}}(\omega)\right)]&=&n_u\frac{r_{u-\rm{out}}(\omega)}{\left|w_{u-\rm{out}}\left(\omega\right)\right|^{1/2}}
\frac{r_{u-\rm{in}}(\omega)}
{|w_{u-\rm{in}}\left(\omega\right)|^{1/2}}\times\\
\nonumber &\times& S_{uu}(\omega)\int\mathrm{d}\omega'~f_{u}^{*}\left(-\frac{\omega'-\omega}{|w_{u-\rm{in}}\left(\omega\right)|}\right)f_{u}^{*}\left(-\frac{\omega'-\omega}{|w_{u-\rm{out}}\left(\omega\right)|}\right)
\end{eqnarray}

The integral of the r.h.s, after turning back to real space, is similar to that considered in Eq. (\ref{eq:overlap}) and gives zero.

We arrive at the same conclusion when considering the density-density correlations in real space. As noted in Ref. \cite{Recati2009}, only the terms with a stationary phase should be kept when computing $\braket{\hat{n}(x)\hat{n}(x')}$, which yields the condition (\ref{eq:densitylines}). In particular, only the ``proper'' correlations between out-out modes can be extracted $c_{ud2},c_{d1d2},g_{ud1}$. The other correlations $c_{uu},c_{d1d1},c_{d2d2},g_{ud2},g_{d1d2},c_{ud1}$, even when they are non-zero, cannot be obtained because the associated exponential terms do not present a stationary phase when integrating over frequencies.

Thus, an important consequence of taking into account the spatial location of the subsonic and supersonic regions is the conclusion that only a limited number of correlations can be experimentally observed. In particular, this implies that we cannot measure all the correlation functions appearing in the GPH criterion (\ref{eq:GPH}). Moreover, and for the same reason, if the state is Gaussian, we cannot obtain all the correlation functions appearing in the quartic CS violation, see Eq. (\ref{eq:CSviolation4Wick}). The results here presented show that we can only aim at observing a quadratic CS violation within this kind of detection schemes. A similar claim can be maid about the TOF detection scheme here presented, where we can only observe indirectly the quadratic CS violation and we can never extract the full GPH criterion from the measurements. We can expect that this phenomenon is not exclusive of these specific procedures. Any realistic attempt to obtain the correlation functions between phonons has to take necessarily into account the spatial location of the asymptotic regions and thus, similar reasonings would arise.

Nevertheless, the previous considerations do not pose a problem for the detection of the quantum Hawking effect for two reasons: (a) if the state of the system is incoherent over the incoming channels, we have proven that the GPH criterion and the quartic CS violation are equivalent to the quadratic CS violation and (b) even for an arbitrary state in which the previous relations do not hold, the quadratic CS violation is still a signature of the presence of the entanglement; in particular, it is a sufficient condition for the fulfillment of the GPH criterion. More than that, as remarked at the end of Sec. \ref{sec:CSPH}, the quadratic CS violation is also by itself a clear indication of the quantum nature of the system, as CS inequalities are always satisfied in a classical system.

\section{Numerical results} \label{sec:NumericalCores}

Here, we compute numerically the different correlation functions and the associated phonon CS violation. We also compute the experimental magnitudes corresponding to the different detection schemes. For definiteness, we consider the usual case where the state of the system is given by a thermal distribution over the incoming modes, Eq. (\ref{eq:thermaldistribution}). For this class of states, all the discussed criteria become equivalent, as argued before. Since we only aim at the identification of some physical trends and not at the study of the whole parameter space, characterized by $7$ variables (see Appendix \ref{app:parametrization}), we study the CS violation using the BH configurations presented in Sec. \ref{sec:typicalbh}. For simplifying the results, we set units such $\hbar=m=c_u=k_B=1$, $c_u$ being the upstream sound speed.

First, we make some general theoretical remarks before showing the numerical results. In the conventional ($\omega$ = 0) peak of the Hawking spectrum $|S_{ud2}(\omega)|^{2}$, the scattering matrix elements diverge as $|S_{ij}(\omega)|\sim1/\sqrt{\omega}$
with $i$ arbitrary and $j=d1,d2$ . By contrast, $S_{iu}(\omega)$
in the same limit ($\omega\rightarrow0^+$) saturates to a nonzero constant
(the asymptotic behavior of the $S-$matrix coefficients is discussed in Appendix
\ref{app:SMatrixBehav}). On the other hand, the only occupation factor
which diverges is $n_{u}(\omega)\sim1/\omega$, because $\Omega_{u}(\omega)$
is the only comoving frequency that vanishes for small $\omega$.
From pseudo-unitarity it follows that $|S_{ud2}|^{2}-|S_{d2d1}|^{2}=|S_{d2u}|^{2}-|S_{d1d2}|^{2}$.
We then conclude from Eq. (\ref{eq:CSequivalent}) that there is no CS violation in
the $\omega\rightarrow0^+$ region. This argument relies solely on the
presence of a uniform condensate flow connecting the subsonic and supersonic
asymptotic regions, and not on other details of the scattering structure.

In the opposite frequency region, $\omega\rightarrow\omega^-_{\rm max}$, it is easy to show from the results of Appendix
\ref{app:SMatrixBehav} that $\theta_{ud2}(\omega)=1/2+O\left(\sqrt{\omega_{\rm max}-\omega}\right)$.
This leaves us with only the central frequency region
in the quest for CS violation and motivates the study of the resonant structures discussed in Sec. \ref{subsec:resonant}, which are able to display peaks in that central region.

\subsection{Time-of-flight detection}

We consider the detection of CS violation within the TOF detection scheme of Sec. \ref{subsec:Atom2PhononTOF}. In this case, we compute the quartic CS violation, instead of the quadratic CS violation, in order to compare it with the quartic {\it atom} CS violation. However, as we deal with thermal distributions, quadratic and quartic violations are equivalent so we can speak about CS violation in general. At sufficiently high temperatures, it is expected that the system behaves classically and there is no CS violation because the high occupation numbers crush the spontaneous contribution which makes possible the violation. Thus, we may define the violation temperature $T_{v}(\omega)$ as the highest temperature at which there is CS violation in the phonon signal ($\Delta_{ud2}(\omega)>0$ or $\delta_{ud2}(\omega)>1$) for given $\omega$ and try to identify some trends in its behavior.

\begin{figure*}[!tb]
\begin{tabular}{@{}cc@{}}
    \includegraphics[width=0.5\columnwidth]{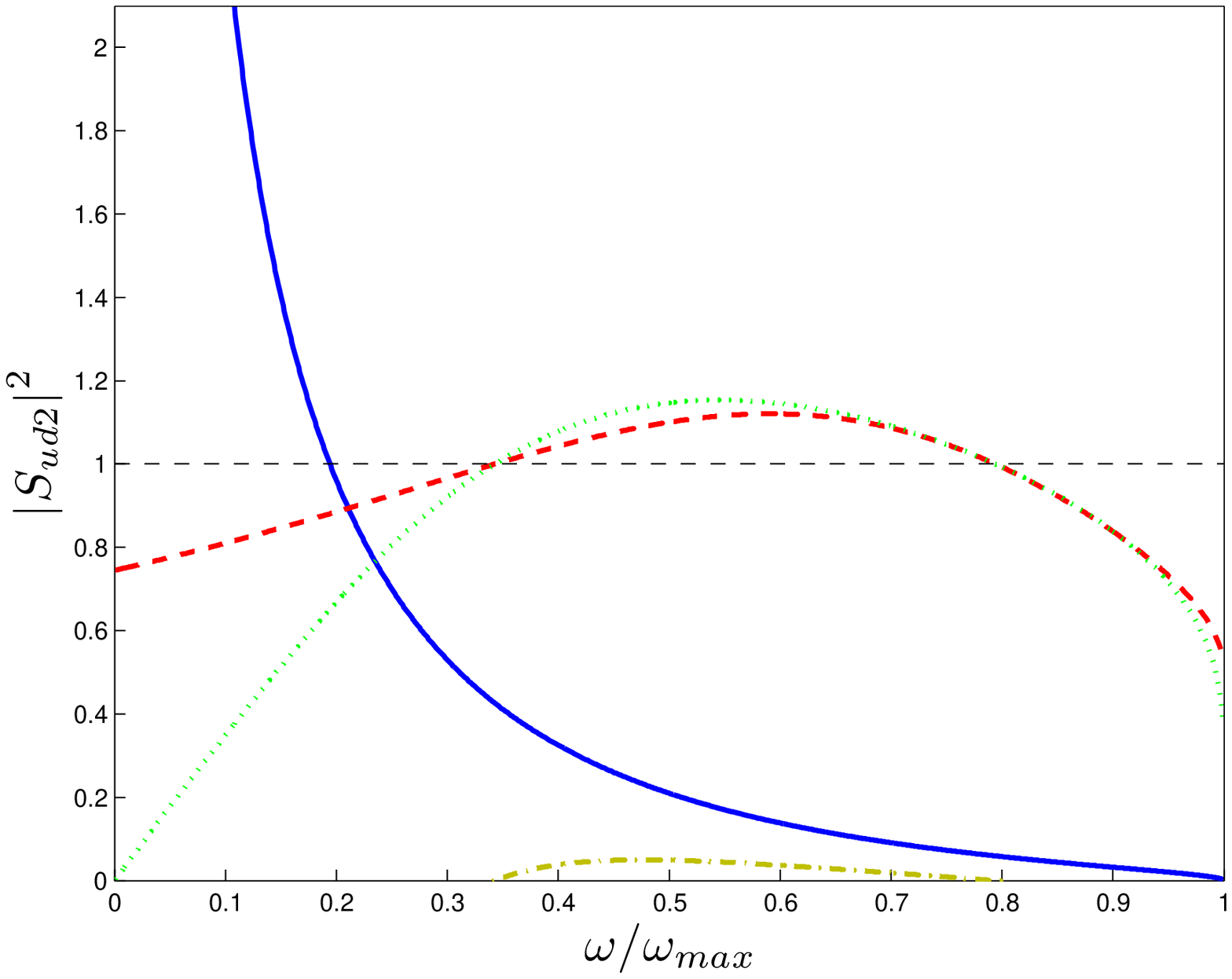} &
    \includegraphics[width=0.5\columnwidth]{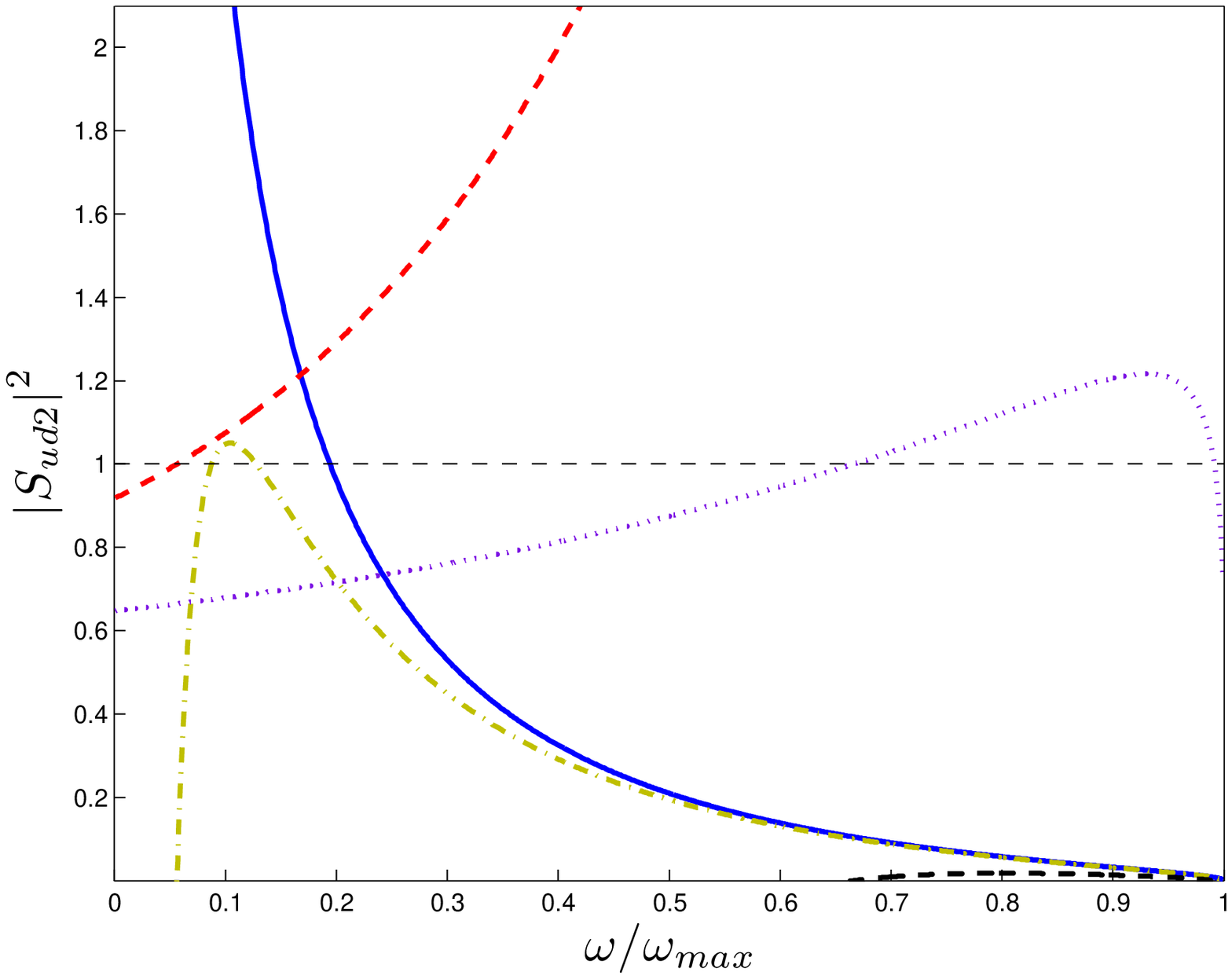} \\
\end{tabular}
\caption{Hawking radiation for the delta barrier
structure shown in left Fig. \ref{fig:NonResonantScheme}. The amplitude of the barrier is $Z=0.62$, with subsonic flow velocity $v_{u}=0.3$ and $\omega_{{\rm max}}=0.6$. In both plots, the solid
blue line represents the zero-temperature HR spectrum $|S_{ud2}(\omega)|^{2}$. Left panel: the dashed red curve corresponds to $\theta_{ud2}(\omega)$ at temperature $T=1$, the dotted green curve to the maximum violation temperature $T_{v}(\omega)$ and the dashed-dotted golden brown to $\Theta_{ud2}(\omega)$ also at $T=1$. Right plot: same as left plot but now at $T=0.2$. We also include the relative atom violation, $z_{ud2}$, in dotted purple and the absolute atom violation, $Z_{ud2}$, in thick black dashed line. }
\label{fig:Delta1CS}
\end{figure*}

\begin{figure*}[!tb]
\begin{tabular}{@{}cc@{}}
    \includegraphics[width=0.5\columnwidth]{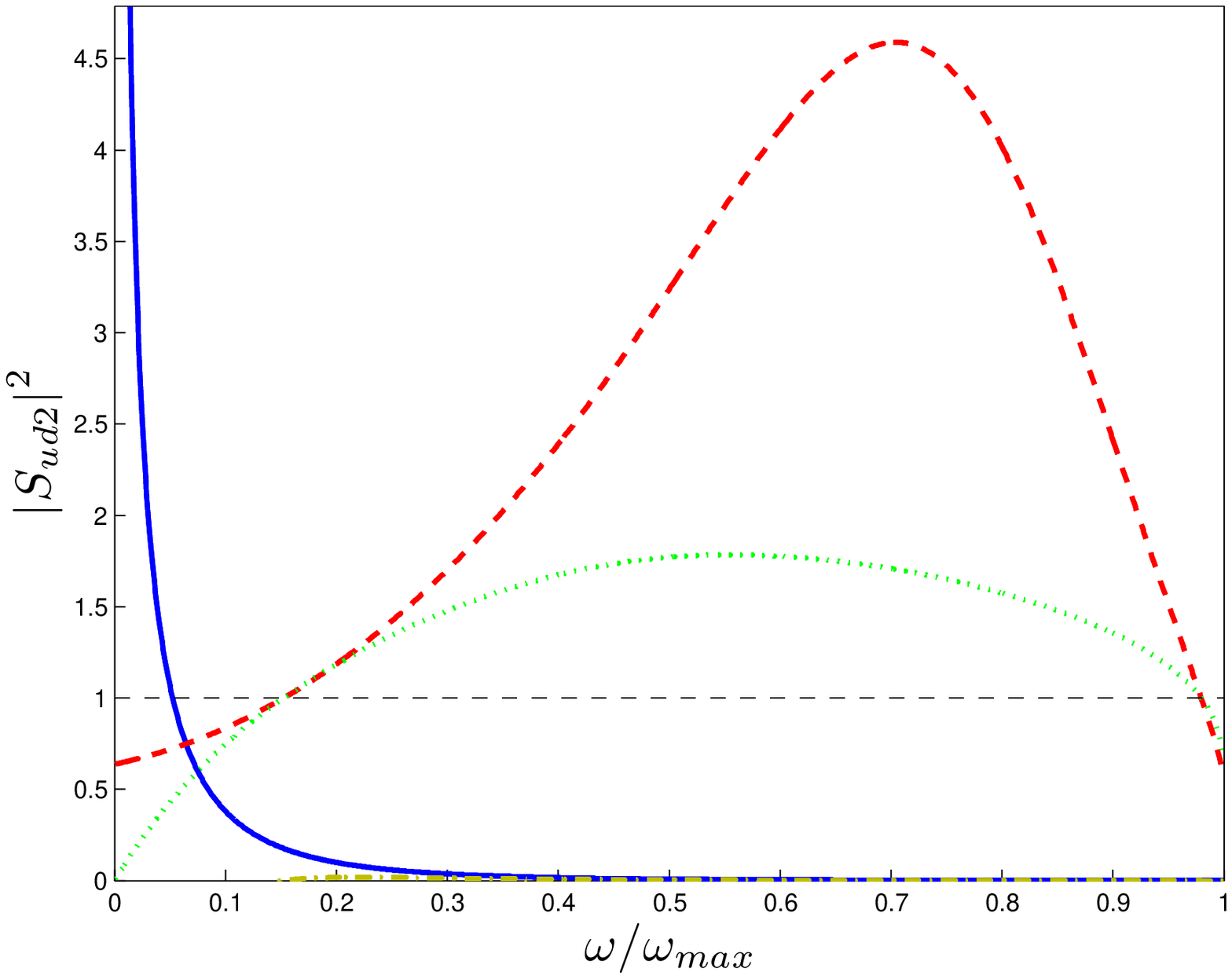} &
    \includegraphics[width=0.5\columnwidth]{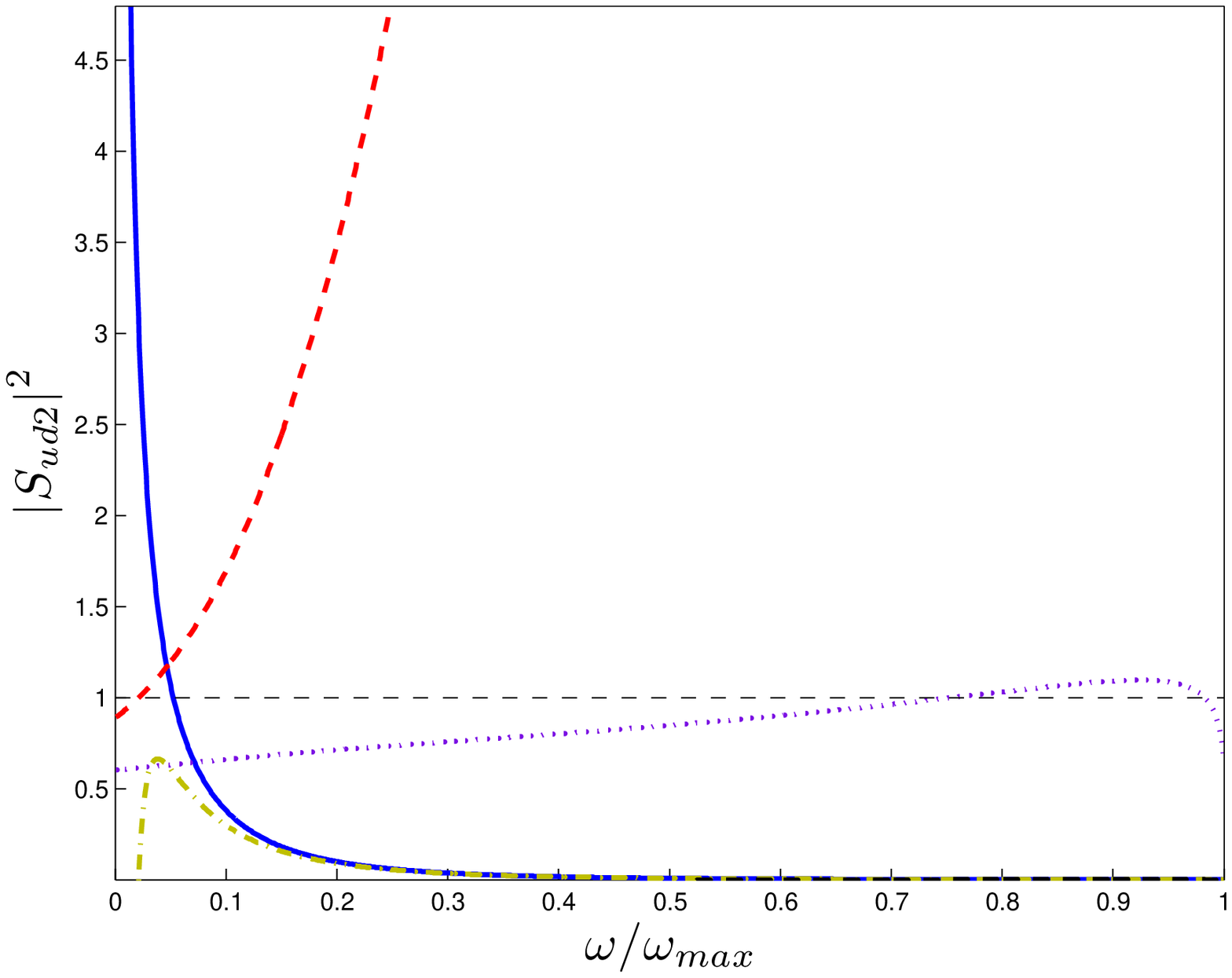} \\
\end{tabular}
\caption{Same as Fig. \ref{fig:Delta1CS} but for a waterfall structure (see right Fig. \ref{fig:NonResonantScheme}) with $v_{u}=0.5$ and $\omega_{{\rm max}}=0.6$. The negative amplitude of the potential is $V_0=1.125$.}
\label{fig:WFCS}
\end{figure*}

We first study CS violation for non-resonant configurations in Figs. \ref{fig:Delta1CS}, \ref{fig:WFCS}. In Fig. \ref{fig:Delta1CS} we consider the delta-barrier configuration of Sec. \ref{subsec:1delta}. In the left plot, we represent the spontaneous HR spectrum $|S_{ud2}(\omega)|^{2}$, the relative degree of quartic CS violation $\theta_{ud2}(\omega)$ (at $T=1$), the violation
temperature $T_{v}(\omega)$ and the absolute amount of quartic CS violation $\Theta_{ud2}(\omega)$. As the HR spectrum is non-resonant, we only have the universal peak at $\omega=0$. We see that CS violation is present in the phonon signal at typical experimental temperatures, $T=1$.
There is no CS violation in the atom signal at this high temperature.

In the right plot, we consider the same setup of the left one but at $T=0.2$, which is still of the order of magnitude of typical experimental setups, where it can be as low as $T\sim0.1$ \cite{Steinhauer2014}. We represent the spontaneous HR spectrum and the relative and absolute phonon and atom signals, $\theta_{ud2},\Theta_{ud2},z_{ud2},Z_{ud2}$. As we only want to discuss the physical features of the atom signal, for simplicity we take the overlap integral as $F(\omega)=1$. The absolute phonon signal is now larger than before as the temperature is lower. In particular, for high $\omega$, both $\Delta_{ud2},\tilde{\Delta}_{ud2}$ converge to $|S_{ud2}|^2$, since in that limit the occupation numbers are negligible and thus, we recover the zero-temperature limit, for which $\Theta_{ud2}(T=0)=|S_{ud2}|^2$. It is also observed that, while the relative phonon violation at this temperature is large in a wide range of frequencies, only a small atom relative violation is found in the large frequency regime, where the number of occupation and depletion contributions are sufficiently small. Indeed, the absolute atom violation (thick black dashed line) is so small that it is hard to observe within this scale.

A similar behavior is found for the waterfall configuration, analyzed in Fig. \ref{fig:WFCS}, where we represent the same magnitudes as before. The only particularity of the waterfall configuration is that, for low values of the subsonic flow velocity $v_u$, it can be proven that $\omega_{\rm max}\simeq 1/2v_u^2\gg 1$ and then for $\omega\lesssim\omega_{\rm max}$ all the occupation numbers satisfy $n_i(\omega)\ll1$ even for $T\sim 1$. This implies that the total phonon signal is almost equal to the spontaneous signal for typical experimental temperatures. Nevertheless, the absolute amount of CS violation is still quite small to be clearly observed in the plots.

\begin{figure}[!htb]
\includegraphics[width=1\columnwidth]{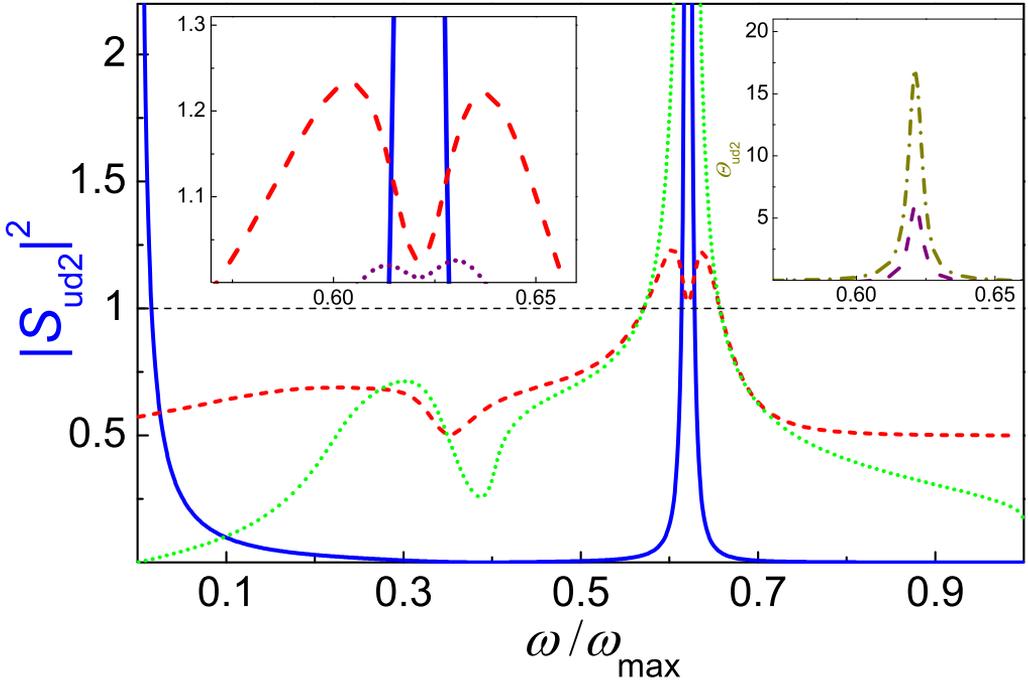}
\caption{Hawking radiation for a condensate leaking through a double barrier
structure. The
strength of both delta barriers is $Z=2.2$. They
are separated by a distance $d=3.62$. The flow is such that
$v_{u}=0.01$ and $\omega_{{\rm max}}=0.99$. Solid
blue: zero-temperature HR spectrum $|S_{ud2}(\omega)|^{2}$.
Dashed red: $\theta_{ud2}(\omega)$ at temperature $T=1$,
where $\mu$ is the comoving chemical potential on the subsonic side.
Dotted green: maximum violation temperature $T_{v}(\omega)$. Not shown in the figure, $T_{v}(\omega)$ rises up
to $T_{v}(\omega_{0})\simeq21$ at the peak, where $|S_{ud2}(\omega_{0})|^{2}\simeq8$.
Left inset: Zoom of the peak region, with $T_{v}(\omega)$ removed
and $z_{ud2}(\omega)$ (relative atom CS violation) added (dotted
purple). Right inset: Same as left inset; it shows $\Theta_{ud2}(\omega)$
(dashed-dotted brown) and $Z_{ud2}(\omega)$ (dashed purple), which
measure the absolute amount of CS violation in the phonon and atom
signals.}
\label{fig:GraphCSViol_2DeltaV1}
\end{figure}

We switch now to resonant configurations. In Fig. \ref{fig:GraphCSViol_2DeltaV1} we plot the same magnitudes as in Figs. \ref{fig:Delta1CS}, \ref{fig:WFCS} but for a double barrier structure. The left inset magnifies the peak region
and includes $z_{ud2}(\omega)$, which measures the relative amount
of CS violation in the atomic signal. The right inset shows $\Theta_{ud2}(\omega)$
and $Z_{ud2}(\omega)$, i.e., the absolute amount of CS violation
by the phonon and atom signals. All the magnitudes are computed at $T=1$.

From Eq. (\ref{eqZeroTempCSViol}) and the pseudo-unitary relation $|S_{ud2}|^{2}+|S_{d1d2}|^{2}+1=|S_{d2d2}|^{2}$,
it may appear that a shortcoming of a large peak is its small relative
degree of CS violation, as reflected in $\theta_{ud2}$ lying slightly above unity (see left inset),
which can be generally inferred from the bound
\begin{equation}
\theta_{ud2}-1\leq\frac{1}{2(|\alpha_{d2}|^2-1)} \, .
\end{equation}
However, this does not imply that the experimental signal is necessarily small.
Quite the opposite, the absolute amount of violation (as measured
by $\Theta_{ud2}$) can be quite large, as Eq. (\ref{eqZeroTempCSViol})
directly reveals. We can check this observation in the right inset, where we see that now we have a much greater absolute phonon signal than in the non-resonant case. Also, we are able to detect atomic CS violation, even at a high temperature as $T=1$. This can be understood from the fact that, when $|S_{ud2}(\omega_{0})|^{2}\!\gg\!1$, the main contribution to the atomic operators near the peak
frequency comes from their respective phononic
counterparts, i.e., $\hat{c}_{u}\left(p_{u}\left(\omega_{0}\right)\right)\sim\hat{\gamma}_{u-\rm{out}}(\omega_{0})$
and $\hat{c}_{d}\left(p_{d2}\left(\omega_{0}\right)\right)\sim\hat{\gamma}_{d2-\rm{out}}(\omega_{0})$, so the relative atomic signal is equal to the relative phonon signal with small corrections near the peak.

\begin{figure}[!htb]
\includegraphics[width=1\columnwidth]{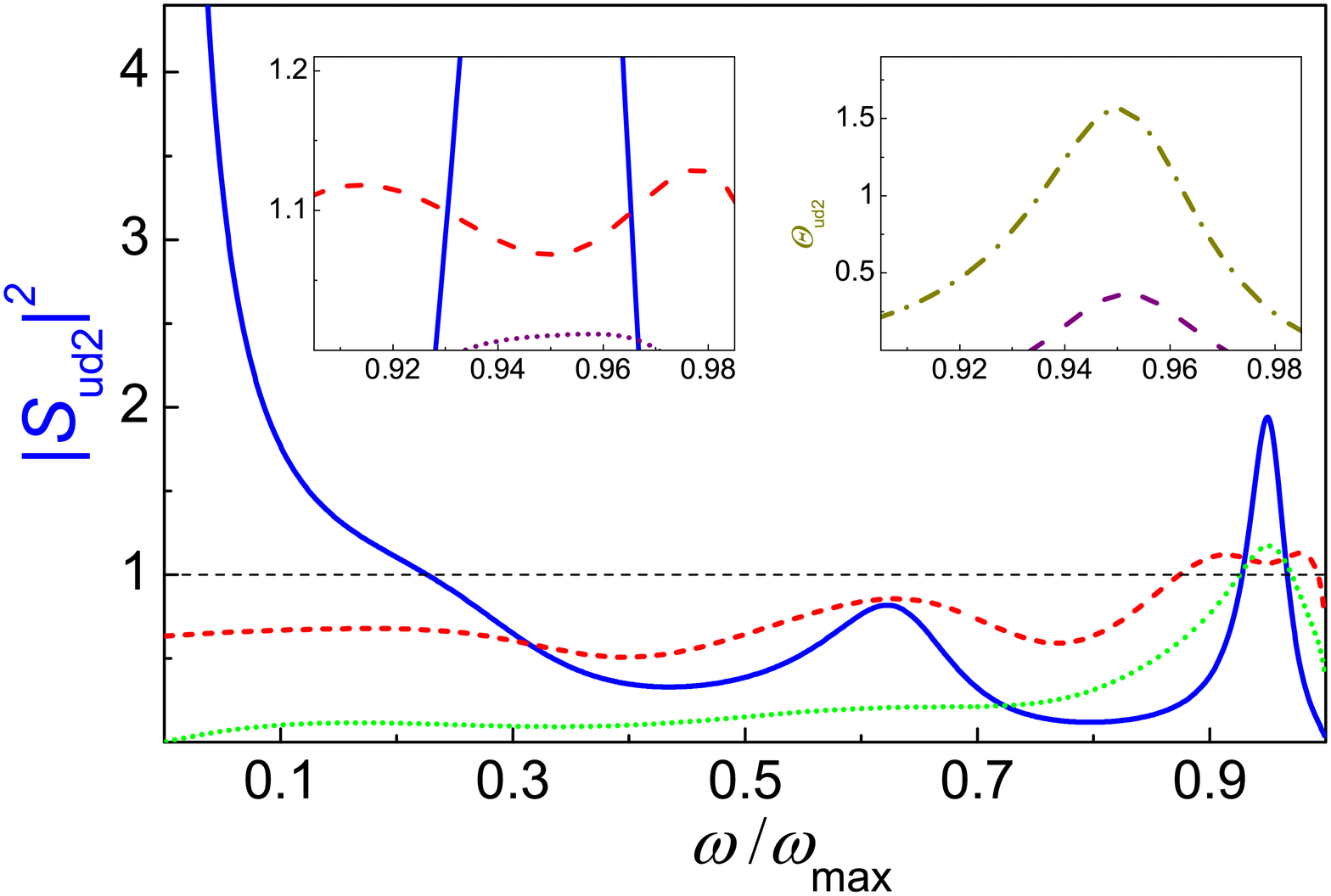}
\caption{Same as Fig. \ref{fig:GraphCSViol_2DeltaV1} but for $T=0.6$ and
for a resonant flat-profile structure with two sharp variations in the local speed of sound,
which takes the successive values $1,0.43,0.6$,
as depicted in the right Fig. \ref{fig:ResonantScheme}. The
intermediate region has a length $L=26$. The flow is such that
$q_{u}=0.95$ and $\omega_{{\rm max}}=0.18$.}
\label{fig:CSResFlatProfile}
\end{figure}

Finally, for completeness, Fig. \ref{fig:CSResFlatProfile} shows the same
curves as those of Fig. \ref{fig:GraphCSViol_2DeltaV1} but for a setup with a resonant flat-profile configuration such as that depicted in the right Fig. \ref{fig:ResonantScheme} with temperature $T=0.6$. These graphs indicate that resonant structures are a promising scenario for the unambiguous detection of spontaneous HR using a TOF experiment since their Hawking signal is much stronger than that of non-resonant structures and is able to provide a clear CS violation directly in the atom signal at relatively high temperatures.

\subsection{Density-density correlations}

We now consider the possibility of observing CS violation through the measurement of density-density correlations. In this case, we focus on studying the quadratic CS violation $\Delta_{ud2}$ in order to compare its value with the measurable quantity $\tilde{\Delta}_{ud2}$, defined in Eq. (\ref{eq:Delta-ud2}). As noted in the previous subsection, the quadratic and quartic CS violations are equivalent and then the results previously presented for the pure phonon CS violation are still valid; indeed, the violation temperature for the phonon signal $T_v(\omega)$ is the same and $\Delta_{ud2}(\omega)=\Theta_{ud2}(\omega)$.

\begin{figure*}[t!]
\begin{tabular}{@{}cc@{}}
    \includegraphics[width=0.5\columnwidth]{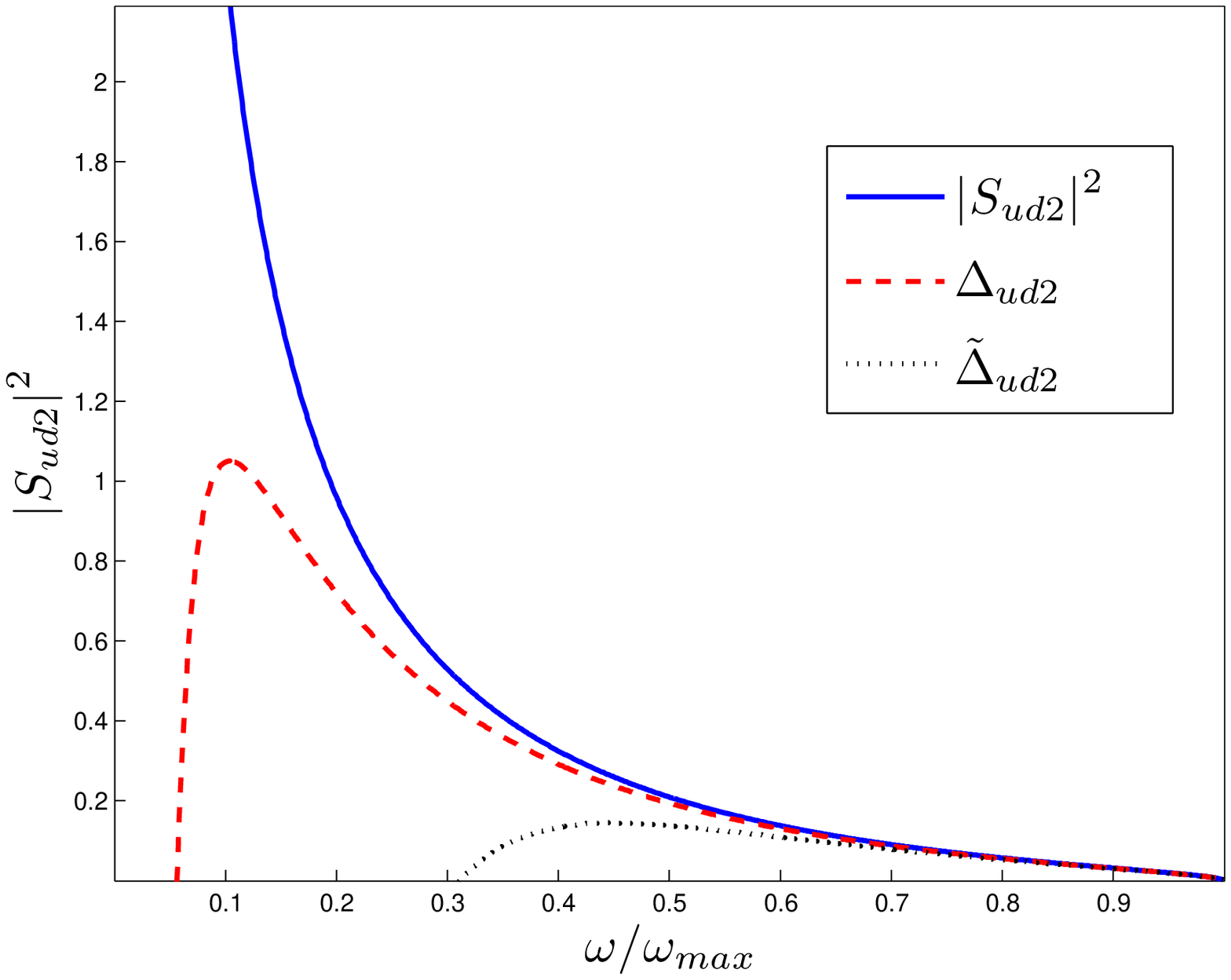} &
    \includegraphics[width=0.5\columnwidth]{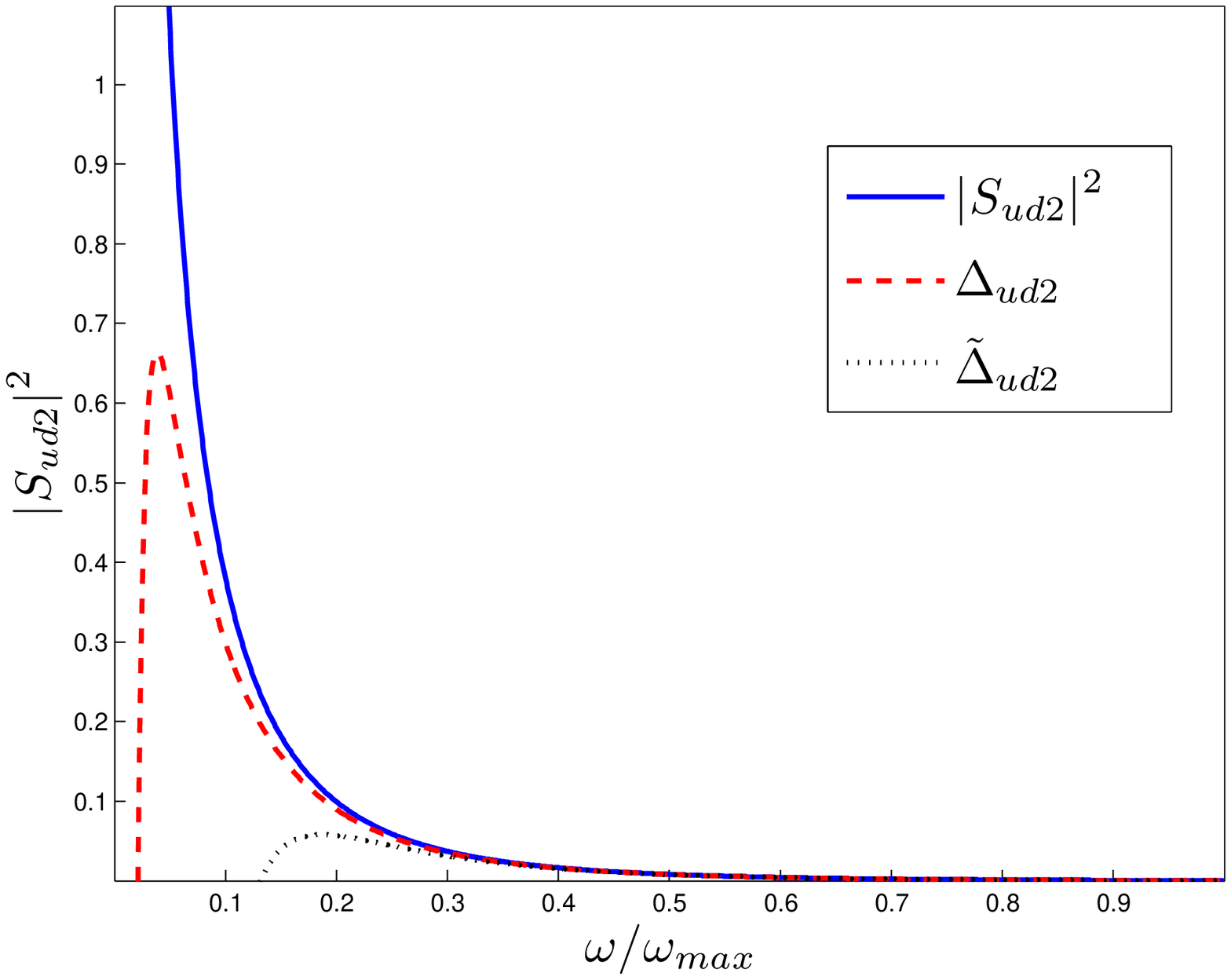} \\
\end{tabular}
\caption{Plot of the Hawking radiation spectrum and the functions $\Delta_{ud2},\tilde{\Delta}_{ud2}$ for a delta barrier configuration (left panel) and a waterfall configuration (right panel), at temperature $T=0.2$. For the delta barrier, the subsonic flow speed is $v_u=0.3$ and the barrier strength is $Z=0.62$. For the waterfall setup, the subsonic flow speed is $v_u=0.5$ and the potential depth is $V_0=1.125$.}
\label{fig:NonResonantViolations}
\end{figure*}

In Fig. \ref{fig:NonResonantViolations}, we represent the Hawking spectrum, $|S_{ud2}|^2$ and the functions $\Delta_{ud2},\tilde{\Delta}_{ud2}$ for the same delta barrier and the waterfall setups of the right panels of Figs. \ref{fig:Delta1CS}, \ref{fig:WFCS}, both at temperature $T=0.2$. As noted previously, $\tilde{\Delta}_{ud2}>0$ is a more restrictive condition than the bare quadratic CS violation $\Delta_{ud2}>0$. However, we see that for sufficiently large $\omega$, both $\Delta_{ud2},\tilde{\Delta}_{ud2}$ converge to $|S_{ud2}|^2$, since for high $\omega$ we have that the occupation numbers are negligible and thus, as discussed before, we recover the zero-temperature result $\Delta_{ud2}(T=0)=|S_{ud2}|^2$. Moreover, also at high $\omega$, $|S_{d1d2}(\omega_{d1d2}(\omega))|^2=0$, because for $\omega$ such that $\omega_{d1d2}(\omega)>\omega_{\rm max}$ the anomalous scattering channel disappears, and then we have $\tilde{\Delta}_{ud2}\simeq \Delta_{ud2}(T=0)=|S_{ud2}|^2$. Thus, we see that taking $\tilde{\Delta}_{ud2}\simeq \Delta_{ud2}(T=0)$, as made in Ref. \cite{Steinhauer2015}, is a very good approximation at high frequencies, even for finite temperatures.

\begin{figure}[htb!]
\includegraphics[width=1\columnwidth]{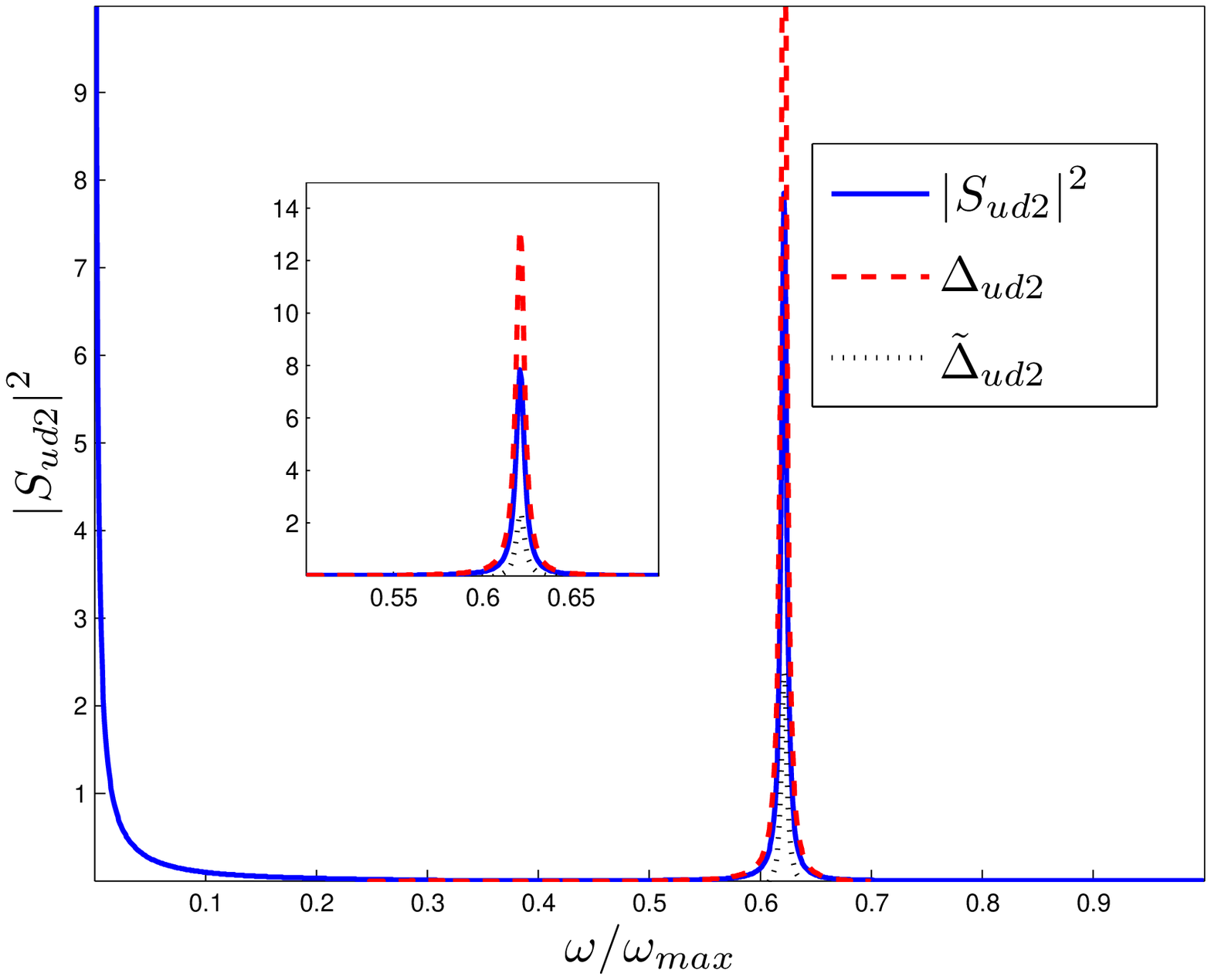}
\caption{Same as Fig. \ref{fig:NonResonantViolations}, but now for a double delta barrier configuration with parameters $Z=2.2$, $d=3.62$ and $v_u=0.01$. The temperature of the system is $T=0.7$. The inset zooms into the peak region.}
\label{fig:2Deltaud2}
\end{figure}

We consider now this scheme detection for resonant configurations. As expected from the previous results, resonant spectra present a strong signal of CS violation. This trend can be observed in Fig. \ref{fig:2Deltaud2}, where we represent the same magnitudes as in Fig. \ref{fig:NonResonantViolations} but now for a higher temperature $T=0.7$. We see that, even for this higher temperature, the experimental signal $\tilde{\Delta}_{ud2}$ is, at the resonance frequency, substantially larger than in the non-resonant case.

\section{Conclusions and outlook} \label{sec:CSPHConclusions}

In this chapter, we have proposed a criterion for detecting the spontaneous Hawking effect based on the violation of CS inequalities. We have shown that it represents an unambiguous signal of the presence of spontaneous Hawking radiation. After that, we have compared the proposed criterion with other one often considered in the literature, the GPH criterion. In particular, we have shown that, under certain physical conditions (Gaussian processes and, simultaneously, incoherent incoming channels), all the considered criteria are equivalent.

We have also investigated the possible measurement of the different criteria in experimental detection schemes. By taking into account the different spatial location of the subsonic and supersonic regions, we have shown that only certain correlation functions can be obtained. For simplicity we have focused on two specific detection schemes, but we expect similar problems to arise in other kind of measurements, since in any realistic situation the supersonic and subsonic regions are necessarily placed in different regions. However, our work also shows that this limitation is not a major problem, as we can often measure the quadratic CS violation, which is also a sufficient condition for the entanglement of the system. Finally, our numerical results of Sec. \ref{sec:NumericalCores} show that, in typical analog configurations, the CS violation can be detected in an achievable range of temperatures.

\chapter{Time-dependent study of a black-hole laser in a flowing atomic condensate}\label{chapter:BHL}
\chaptermark{Time evolution of the black-hole laser}

\section{Introduction}

So far, we have focused on the emission of Hawking radiation by a single sonic black-hole. We now switch to a slightly different gravitational analog system: the so-called black-hole laser, formed by a pair of neighboring horizons and first discussed in the gravitational context in Ref. \cite{Corley1999}. Within this analog scenario, for a quantum Bose field with a superluminal dispersion, the negative energy partner of the Hawking emission by the outer horizon can bounce on the inner horizon and travel back to the outer one, so to stimulate further Hawking emission. In suitable conditions, this process may give rise to a dynamical instability (similar to that arising in a laser system) and then to the exponential growth of a self-amplifying coherent Hawking emission. This coherent emission contrasts with the spontaneous emission Hawking radiation from a single horizon, with unambiguously quantum features, as discussed in the previous chapter. The first experimental evidence of the BH laser instability has been recently reported in Ref. \cite{Steinhauer2014} by looking at the exponential growth in time of a complex density modulation pattern between the horizons.

While the first works on BH laser configurations in astrophysics \cite{Corley1999} and in analog models \cite{leonhardt2007a,Coutant2010,Finazzi2010,Steinhauer2014} have focused on the linear dynamics at early times after the onset of the instability, quite interesting physics also occurs at late times when the exponentially growing BH laser emission has become strong enough to exert a measurable back-reaction effect onto the horizons \cite{Michel2013,Michel2015}. Although there are strong conceptual differences between the two scenarios, one can reasonably expect that fully understanding this simpler case will help to understand the back-reaction effect by the spontaneous Hawking radiation that is responsible for the (extremely slow and hardly observable) evaporation of astrophysical black-holes.

In this chapter, we present a campaign of numerical simulations of the long-time dynamics of a Bose-Einstein condensate based on the integration of the 1D GP equation starting from a BH laser configuration with a black hole-white hole pair of horizons. They provide a numerical confirmation of the different behaviors anticipated in Refs. \cite{Michel2013,Michel2015}, where it was observed the quick relaxation of the BH laser configuration towards a horizonless sub-sonic flow by ``evaporating'' away the horizons as well as it was pointed the possibility of spontaneously oscillating behaviors. In addition, we perform here a detailed characterization of these oscillating regimes where the classical dynamics of the system quickly forgets its initial condition and tends at late times to a limit cycle attractor. As a consequence of these oscillations, soliton waves keep being emitted for indefinite times in a continuous and periodic way.

While the possibility of such oscillating behaviors in astrophysics has never been proposed to the best of our knowledge, a physical comparison to optical laser devices suggests that this self-oscillating regime is the closest gravitational counterpart of the monochromatic emission of radiation by continuous-wave laser devices \cite{QuantumNoise,Walls2008,Petruccione,MandelWolf}: in both cases, a highly coherent emission follows from the late-time oscillations along a limit cycle attractor in the classical dynamics.

In a similar fashion to the work presented in Chapter \ref{chapter:MELAFO}, the results of this work are also of interest in the context of quantum transport and atomtronics since we show that a BH laser configuration is able to reach a regime of continuous emission of solitons, providing a hydrodynamic analog of an optical laser.

The scheme of the chapter is the following. In Sec. \ref{sec:BHLaserintro} we briefly revisit the basic theory of black-hole lasers in Bose-Einstein condensates. In Sec. \ref{sec:numresults} we present our numerical results and we classify the different regimes according to the number of linearly unstable modes in the initial condition (Secs. \ref{subsec:n=0}, \ref{subsec:n=1}). The continuous soliton emission is then discussed in Sec. \ref{subsec:CES}. Comparison between this mechanism of emission of solitons with the operation of optical laser devices is given in Sec. \ref{sec:LaserComparison}. Conclusions and outlook are finally given in Sec. \ref{sec:BHLconclusions}. Technical details about the method used for the numerical calculations are presented in Appendix \ref{sec:numericalbhl}. We discuss in detail the computation of the dynamical instabilities and of the non-linear Gross-Pitaevskii solutions of the BH laser in Appendix \ref{app:technicalBH}. A summary of the movies and of the parameters used in each of them is given in Appendix \ref{app:Movies}.

\section{Black-hole laser configurations}\label{sec:BHLaserintro}

We start by revising the basic concepts of the theory of black-hole lasers in Bose-Einstein condensates. For further details, we refer the reader to the recent literature on the topic \cite{leonhardt2007a,Coutant2010,Finazzi2010,Michel2013,Michel2015}.

Following the same theoretical model presented in Chapter \ref{chapter:Introduction}, we consider the flow of an atomic one-dimensional condensate near $T=0$ in the 1D mean-field regime. In particular, we focus on a system in which the time evolution of the system is governed by the following 1D time-dependent GP equation:
\begin{equation}\label{eq:TDGPBH}
i\hbar\frac{\partial\Psi(x,t)}{\partial t} = \left[-\frac{\hbar^{2}}{2m}\frac{\partial^2}{\partial x^2} +V(x)+g(x)|\Psi(x,t)|^{2}\right]\Psi(x,t)\, ,
\end{equation}
where we have allowed for a space dependence of the external potential $V(x)$ and of the coupling constant $g(x)$, in the same fashion as the flat profile configuration of Sec. \ref{subsec:flatprofile}. As there explained, if $g(x),V(x)$ satisfy the condition (\ref{eq:matchingcondition}), $g(x)n_0+V(x)=E_b$, then exists a stationary homogeneous plane wave GP solution given by
\begin{equation}\label{eq:stationaryhomogeneousplanewave}
\Psi(x,t)=\sqrt{n_0}e^{iqx}e^{-i\frac{\mu}{\hbar}t},
\end{equation}
Specifically, we take here a piecewise coupling constant such $g(x)=g_1$ for $|x|>X/2$ and $g(x)=g_2$ for $|x|<X/2$ with the corresponding sound speed $c(x)=c_{1,2}$ for $|x|\gtrless X/2$. A BH laser configuration appears when the external regions are subsonic and the internal region of size $X$ is supersonic, i.e., $c_2<v<c_1$, $v=\hbar q/m$ being the (constant) velocity of the condensate. As a result, two sonic sonic horizons are formed, the upstream one located at $x=-X/2$ has a black-hole nature, while the downstream one at $x=X/2$ has a white-hole nature.

\begin{figure}[t!]
\centering
\begin{tabular}{@{}c@{}}
    \includegraphics[width=0.73\columnwidth]{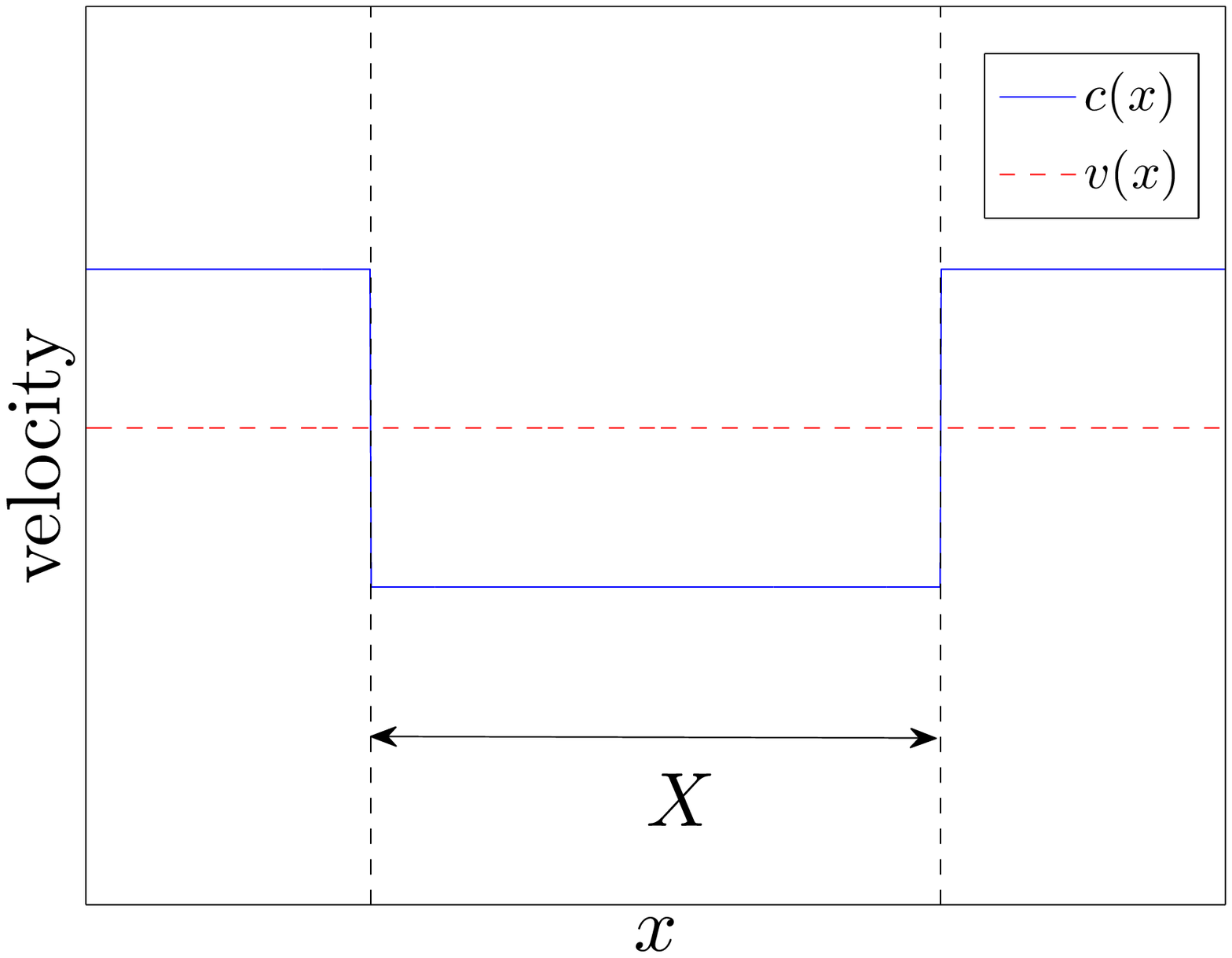} \\
    ~~~\includegraphics[width=0.7\columnwidth]{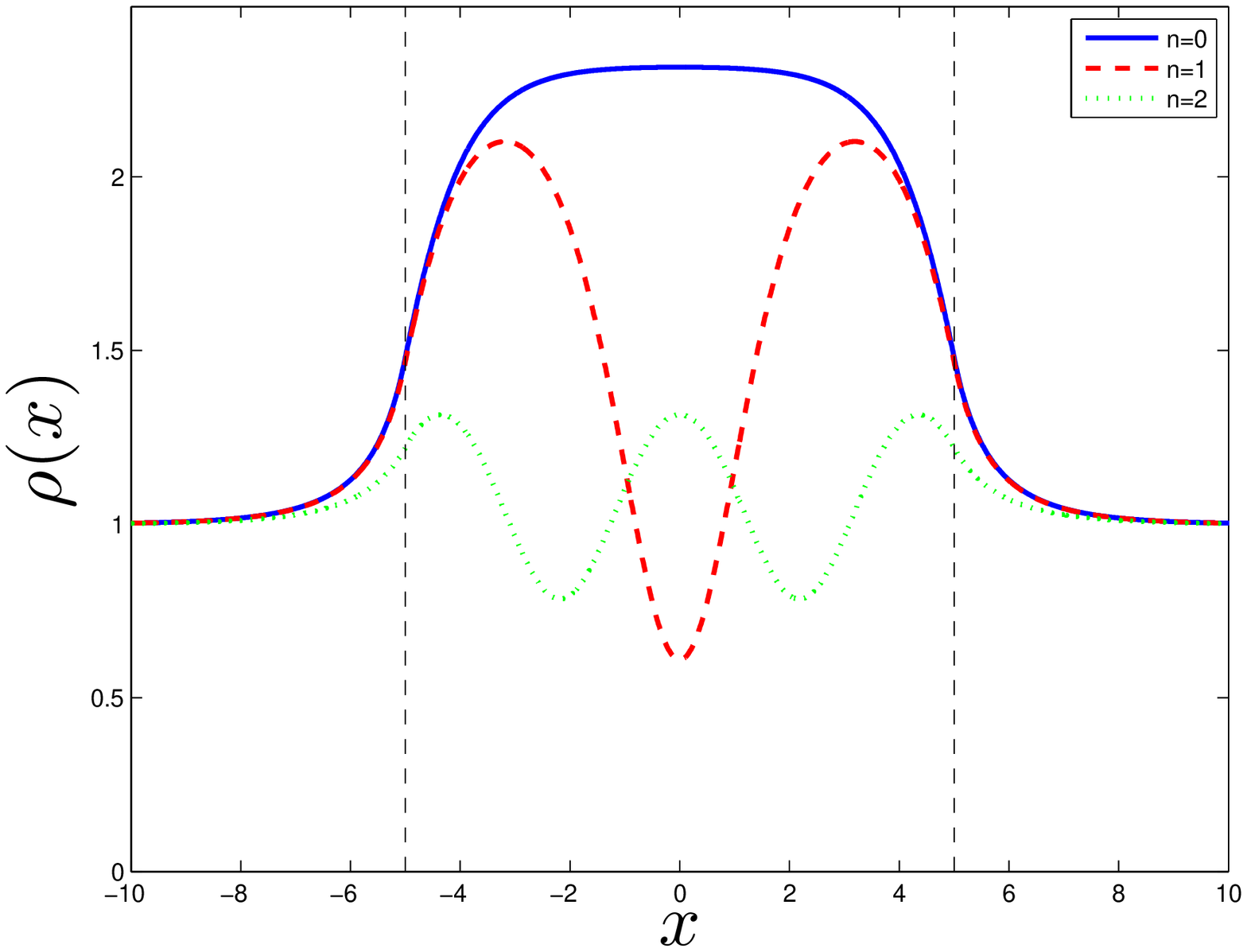}
\end{tabular}
\caption{Upper panel: Scheme of the BH laser configuration discussed in this work. The local speed of sound, $c(x)$, is depicted (solid blue line) along with the (here constant) flow velocity, $v(x)=v>0$ (red dashed line). Lower panel: Plot of the $n=0,1,2$ nonlinear stationary solutions of the GP equation, Eq. (\ref{eq:groundnonlinearstates}), for parameters $v=0.9$, $c_2=0.5$ and $X=10$. Dashed vertical black lines represent the limits of the internal region, $|x|<X/2$.}
\label{fig:BHSchemes}
\end{figure}

This configuration is schematically illustrated in the upper panel of Fig. \ref{fig:BHSchemes}. For computational convenience, hereafter we rescale the wave-function $\Psi(x,t)\rightarrow\sqrt{n_0}\Psi(x,t)$ and set $\hbar=m=c_1=1$. In these units, the initial condensate density is $1$, the healing length in the external sub-sonic regions is initially $1$, the wavevector is $q=v$ and the system is completely characterized by the three values of $(c_2,v,X)$. Also, within this chapter, we relabel the density as $\rho(x)\equiv n(x)=|\Psi_0(x)|^2$ in order to avoid confusions with the quantum number $n$ of the stationary non-linear solutions (see Sec. \ref{subsec:nonlinearsolutions}).

\subsection{Unstable modes}\label{subsec:unstablemodes}

The linear perturbations of the wave-function around the stationary solution, $\Psi(x,t)=\left[1+\delta\Psi(x,t)\right]e^{ivx}e^{-i\mu t}$, are described in terms of the Bogoliubov-de Gennes (BdG) equations as discussed in Sec. \ref{sec:physmodel}. In our system, the solutions of the BdG equations for the perturbation $\delta \Psi(x,t)$ can be written within each subsonic or supersonic region as linear combinations of waves satisfying the dispersion relation of Eq. (\ref{eq:dispersionrelation}):
\begin{equation}\label{eq:dispersionrelationBHL}
(\omega-vk)^2=c^2_ik^2+\frac{k^4}{4},~i=1,2
\end{equation}
For a given frequency $\omega$, we obtain the global solutions to the BdG equations by matching the plane-wave solutions at the boundaries between the external and internal regions, in a similar way to scattering in BH configurations. However, the system here considered is much more interesting since for sufficiently large value of $X$, apart from the usual scattering states with real frequency, a discrete number of dynamical instabilities can also appear, labeled by the quantum number $n=0,1,2\ldots$ and characterized by a complex eigenfrequency $\gamma_n=\omega_n+i\Gamma_n$, with growth rate $\Gamma_n>0$. As $X$ increases further, more and more unstable modes appear. In particular, the critical length at which a new dynamical instability appears for given $(v,c_2)$ is \cite{Michel2013,Michel2015}:
\begin{equation}\label{Eq:UnstableLength}
X_n=X_0+n\lambda_0
\end{equation}
where
\begin{equation}
X_0=\frac{\arctan\sqrt{\frac{1-v^2}{v^2-c^2_2}}}{\sqrt{{v^2-c^2_2}}} \textrm{\;\;and\;\;} \lambda_0=\frac{\pi}{\sqrt{{v^2-c^2_2}}}.
\end{equation}
The intermediate values $X=X_{n+1/2}$ separate qualitatively different unstable behaviors: while for $X_n<X<X_{n+1/2}$ the complex frequency $\gamma_n$ is purely imaginary, for $X_{n}<X<X_{n+1/2}$ it acquires a finite oscillation frequency $\omega_n\neq 0$. Remarkably, the instability rate satisfy $\Gamma_n=0$ right at the transition point $X=X_{n+1/2}$ \cite{Michel2013,Michel2015}. We refer the reader to Appendix \ref{app:technicalBH} for the technical details on the obtention of the unstable modes.

\subsection{Non-linear stationary solutions}\label{subsec:nonlinearsolutions}

Another step of crucial importance to understand the long-time behavior of the system after the onset of an instability is to identify and characterize the non-linear stationary solutions of the GP equation $\Psi_0(x)$ that asymptotically match the homogeneous plane wave solution $\Psi_0(x)=e^{ivx}$ far in the external sub-sonic regions at $x=\pm \infty$. The quest of such solutions is motivated by the fact that the total energy and number of particles are conserved during the growth of the instabilities and only these solutions present a finite energy and particle number difference with respect to the initial plane wave \cite{Michel2013}.

As a criterion to choose among the many possible solutions of these equations, it is reasonable to expect that the system, when dynamically unstable, will evolve towards a new stationary solution which minimizes its grand-canonical energy $K$ since, as explained in Sec. \ref{subsec:timeindependent}, energetic stability implies dynamical stability. The expression for the functional $K[\Psi]$ of Eq. (\ref{eq:Kfunctional}) in our system of units is:
\begin{equation}\label{eq:grandcanonicalenergy}
K[\Psi]=n_0\int\mathrm{d}x~\frac{1}{2}\partial_x\Psi^*\partial_x\Psi+(V(x)-\mu)|\Psi|^2+\frac{g(x)}{2}|\Psi|^4
\end{equation}
When $\Psi=\Psi_0$ is a stationary solution, the energy difference with respect to the homogeneous plane wave solution (with energy $K_h$) can be computed from Eq. (\ref{eq:meanfieldhamiltonian}) as
\begin{equation}\label{eq:grandcanonicaldiff}
\Delta K=K[\Psi_0]-K_h=-n_0\int\mathrm{d}x~\frac{g(x)}{2}(|\Psi_0(x)|^4-1)~.
\end{equation}
We see that the system can reduce its energy by increasing the amplitude of the wave function. As the amplitude of the wave function is asymptotically fixed in the external subsonic regions, an increase of the amplitude within the internal $|x|<X/2$ region leads to an overall reduction of the grand-canonical energy. This increase of the amplitude implies an increase of the density that carries a finite increment in the number of particles:
\begin{equation}\label{eq:deltaN}
\Delta N=\int\mathrm{d}x~(|\Psi_0(x)|^4-1)
\end{equation}
Following this reasoning, it can be shown that the family of solutions with lower grand-canonical energy are the so-called sh-sh solutions \cite{Michel2013,Michel2015}. They consist of a shadow-soliton solution in the external regions and of elliptic functions in the internal region (see Appendix \ref{app:technicalBH} for their specific expression) and they are also characterized by a discrete quantum number $n=0,1,2...$, in analogy to the discrete number of unstable modes. This is not a random coincidence; in fact, using the results there presented, it can be easily shown that the minimum length at which the $n$ non-linear solution appears coincides with the onset of the dynamical instability at $X=X_n$ as given by Eq. (\ref{Eq:UnstableLength}). Hence, there is a remarkable correspondence between dynamically unstable modes and non-linear solutions of the GP~\cite{Michel2013,Michel2015}.

The first few non-linear solutions of Eq. (\ref{eq:groundnonlinearstates}) for a given configuration are plotted in the lower panel of Fig. \ref{fig:BHSchemes}. The grand-canonical energies of these solutions satisfy $\Delta K_0<\Delta K_1<\ldots<0$. As it was first anticipated in Ref.\cite{Michel2013,Michel2015} by studying linearized fluctuations around the stationary state, only the $n=0$ solution is dynamically stable while all other solutions are dynamically unstable. A time-dependent integration of the GP equation fully confirms this fact.

A qualitative argument justifying the fact that the $n=0$ solution is the most stable solution can be put forward as follows: the density profile of this solution, displayed in Fig. \ref{fig:BHSchemes}, shows an accumulation of atoms in the central region at $|x|<X/2$, so that the local sound velocity increases and, according to the continuity equation, the flow velocity is reduced. As a result, the supersonic unstable character of the central region is lost and the flow becomes everywhere subsonic. This restores full dynamical stability of the configuration.

The situation is of course completely different for the higher $n>0$ nonlinear solutions shown in the same plot, where the density minima that are present in these solutions in the central $|x|<X/2$ region correspond to a locally supersonic flow. As usual in hydrodynamics~\cite{Frisch:PRL1992,Hakim1997,Pavloff:PRA2002}, the presence of such localized super-sonic flow regions in combination with spatial inhomogeneities is responsible for the appearance of dynamical instabilities of the flow.

\section{Numerical results} \label{sec:numresults}

We proceed to present our numerical results for the time evolution of the BH laser. For numerical convenience, our simulations are based on a simplified form of the GP equation
\begin{equation}\label{eq:GPNumerical}
i\frac{\partial \Psi(x,t)}{\partial t}=-\frac{1}{2}\frac{\partial^2 \Psi(x,t)}{\partial x^2}+g(x)\left[|\Psi(x,t)|^2-1\right]\Psi(x,t)
\end{equation}
where the constant $E_b$, $E_{b}=g(x)+V(x)$ in our units, has been subtracted to Eq. (\ref{eq:TDGPBH}). The size of the numerical grid is taken sufficiently large to avoid finite size effects, $L_g\sim10^3\gg X$. Spurious reflections from the boundaries of the integration box are further suppressed by implementing a diffusive term at the edges of the grid. This allows us to extend the numerical simulations up to long times $t\sim 10^4$ without suffering from numerical artifacts. A more detailed description of the numerical integration scheme and of the implementation of the diffusive term is given in Appendix \ref{app:numerical}.

As the initial condition at $t=0$, we take a plane wave of the form $e^{ivx}$ supplemented by a weak random noise on top of it; this noise is essential to trigger the dynamical instabilities. We restrict ourselves to the case where $g(x)$ and $V(x)$ has the same form as in Sec. \ref{subsec:unstablemodes}. In the next subsections, the main features of the time evolution of the system will be characterized as a function of the relevant parameters $(c_2,v,X)$.

\subsection{Long-time stationary state}
\label{subsec:longtimestat}

Three regimes appear depending on whether $X<X_0(v,c_2)$ or $X_0(v,c_2)<X<X_1(v,c_2)$ or $X_1(v,c_2)<X$ with $X_{0,1}$ defined in (\ref{Eq:UnstableLength}).

The first case $X<X_0(v,c_2)$ is trivial as there are no instabilities and the time evolution reduces to the dynamics of the (very weak) noise on top of the stationary solution. In the following we therefore focus our attention on the other two cases that show the most interesting physics.

Before entering into the details, some preliminary remarks are in order. First, we note that the limit $c_2\rightarrow0$ is ill-defined, as in this limit the $n=0$ non-linear solution presents an infinite accumulation of particles between the two horizons, which cannot be achieved starting from a finite condensate as that considered in our numerical simulations. We then restrict our simulations to finite values of $c_2$ for which the particle accumulation $\Delta N$ is much smaller than the total number of particles in the system, $\Delta N\ll N$.

For improving the presentation of the results, we have prepared some movies to show qualitative trends of the system. All the technical details about the simulations presented in those movies are collected in Appendix \ref{app:Movies}.

\subsubsection{Single unstable mode $X_0<X<X_1$}\label{subsec:n=0}

In the range $X_0(v,c_2)<X<X_1(v,c_2)$, only one unstable mode is present, associated with the appearance of the lowest energy $n=0$ nonlinear solution described by Eq. (\ref{eq:groundnonlinearstates}), whose energy lies below that of the homogeneous solution. After some transient governed by the unstable mode, we can expect that the system will expel the extra energy in the form of waves propagating to $x\to \pm\infty$ and eventually relax towards this non-linear stationary solution.

Whilst this behavior was indeed observed in Ref. \cite{Michel2015}, we have noticed that it is restricted only to a certain region in parameter space, corresponding to sufficiently low values of $v$ and high values of $c_2$. Outside this region, the instability of the supersonic region becomes too severe and a smooth convergence towards the $n=0$ stationary solution is replaced by a time-dependent regime of continuous emission of solitons (CES). Importantly, we have numerically observed that the choice between the two behaviors only depends on the system parameters $(c_2,v,X)$ and not on the specific initial configuration of the noise used to trigger the instability. In this subsection, we focus our attention on the first behavior while the detailed characterization of the CES regime is postponed to Sec.\ref{subsec:CES}. The phase diagrams summarizing the parameter regions corresponding to each of these behaviors is shown and discussed in Fig.\ref{fig:CESregion}.

For given values of $(v,c_2)$ outside the CES region in the parameter space, two cases $X_0(v,c_2)<X<X_{1/2}(v,c_2)$ and $X_{1/2}(v,c_2)<X<X_{1}(v,c_2)$ can be further distinguished. In the first one, $X_0(v,c_2)<X<X_{1/2}(v,c_2)$, the frequency of the unstable mode is purely imaginary and the mode does not oscillate [see Eq. (\ref{Eq:UnstableLength}) and paragraph below]. As a result, the density shows for times $t\gtrsim \Gamma^{-1}_0$ a monotonic exponential evolution of the form
\begin{equation}\label{eq:densityzero}
\rho(x,t)\approx1+\alpha_0\delta\rho_0(x)e^{\Gamma_0t}~,
\end{equation}
where $\delta\rho_0(x)$ is the linear density perturbation corresponding to the unstable mode and $\alpha_0$ its initial amplitude. Depending on the sign of the real number $\alpha_0$, the density will either grow or decrease: as the exponential evolution is triggered by a random noise on top of the stationary solution, the probabilities of the occurrence of both behaviors are equal to $1/2$. This fact has been numerically checked.

When the density initially increases, the system smoothly reaches the $n=0$ ground state solution. The increase of the density in the central region is associated with a small emission of waves and a small soliton to the upstream region ($x\rightarrow-\infty$) in order to conserve the total number of particles $N$. We can observe this behavior in \href{https://www.youtube.com/watch?v=JyLDGYGep-I}{Movie 1}, which shows the result of a simulation for a choice of parameters $v=0.75$, $c_2=0.3$ and $X=2$, satisfying $X_0(v,c_2)<X<X_{1/2}(v,c_2)$. In each frame, the time-evolving spatial density profile is shown as a thin blue line and the $n=0$ stationary solution is plotted as a thick black line.

On the other hand, when the density initially decreases, the system has to emit a larger soliton to the downstream region ($x\rightarrow\infty$) in order to compensate the initial decrease in the particle density. Once the soliton has been emitted, the system is again free to evolve to the $n=0$ solution by locally increasing the density in the central region. This scenario is shown in \href{https://www.youtube.com/watch?v=OBHGV2xQ9Fo}{Movie 2}, generated with the same parameters $(c_2,v,X)$ used in Movie 1 but a different configuration of initial noise.

For $X_{1/2}(v,c_2)<X<X_{1}(v,c_2)$, the system also evolves towards the $n=0$ stationary solution. The only difference is the transient: as the instability keeps oscillating while growing, the dependence on the initial noise condition is no longer relevant and no distinction between initially increasing and decreasing cases can be made any longer. We can observe this behavior in \href{https://www.youtube.com/watch?v=F681IomByYw}{Movie 3}.

Our numerical results then confirm that in this stable regime the (very different) transient dynamics does not play any significant role on the long time behavior of the system, which is only determined by the values of $(c_2,v,X)$ and in all cases eventually reaches the ground-state $n=0$ solution.

\subsubsection{Several unstable modes $X_1<X$}\label{subsec:n=1}

For larger sizes $X$ of the central supersonic region, several unstable modes are present and, because of that, the dynamics is potentially much more complicated. From Eq. (\ref{Eq:UnstableLength}), we infer that the minimum length at which a new unstable mode appears is given by:
\begin{equation}\label{eq:minimum10length}
X_n^{min}=X_n(v=1,c_2=0)=n\pi
\end{equation}
Also in this case, two strongly different behaviors are observed depending on the degree of instability: for sufficiently slow speeds $v$ and high values of $c_2$, the long-time evolution after a (sometimes complex) transient tends to the $n=0$ stationary state. On the other hand, for sufficiently high speeds $v$ and low values of $c_2$, the mentioned CES regime appears. Once more, we refer to the next subsection for the discussion about the CES. A phase diagram summarizing the regions of parameters where each of these behaviors is observed can be found in the lower panels of Fig.\ref{fig:CESregion}.

It was shown in Ref. \cite{Michel2013} that the unstable mode with largest value of $n$ is typically the dominant one in the early evolution as it has the largest growth rate $\Gamma_n$. This feature has been qualitatively confirmed in our simulations by looking at the spatial profile of the density modulation at intermediate times and its oscillating/non-oscillating temporal dependence.

For what concerns the later dynamics, the situation is quite similar to that of the previous section; the main difference is that several stationary solutions with $\Delta K<0$ are now available. While on long times the system typically tries to relax to the lowest energy $n=0$ solution (which is the only stable one) by emitting the extra energy and particles in the form of solitons and small waves, at intermediate times it may or may not approach a higher $n>0$ stationary solution and remain for a quite macroscopic time ($t\ll 1$) in its vicinity. Of course, as such solutions are dynamically unstable, the system eventually departs from them and finally converges to the stable $n=0$ solution after another stage of soliton and wave emission.

When $X_n(v,c_2)<X<X_{n+1/2}(v,c_2)$, some useful information about whether the system does or does not spend time in the vicinity of a higher $n>0$ stationary solution can be obtained along the lines of the discussion around Eq. \ref{eq:densityzero}. Two different behaviors can in fact be identified depending on the initial noise configuration: depending on the sign of its projection onto the $n$ dominant unstable mode, the system can smoothly evolve towards the $n$ non-linear solution or it can start evolving in the opposite direction. In this latter case, more solitons and waves need emitting to conserve the total number of particles and the system never approaches the corresponding $n$ non-linear stationary solution. These two behaviors can be neatly observed in \href{https://www.youtube.com/watch?v=SsC5OlM1dcc}{Movie 4} and in \href{https://www.youtube.com/watch?v=QO1RRSHeEWQ}{Movie 5}, respectively. In both cases, after a (possibly very long) transient, the simulation ends up in the lowest energy $n=0$ stationary solution.

This dramatic dependence on the initial conditions is a simplest example of the chaotic behavior of the system dynamics for $X_1<X$, when many unstable modes are present. In more complicate cases, it may even happen that the system intercepts some other nonlinear stationary solution and the $n=0$ solution is not reached for very long times on the order of $t=10^4$. An example of such a behavior is shown in \href{https://www.youtube.com/watch?v=2_TyxbXALxQ}{Movie 6}. Unless some CES behavior sets in at late times, as all other $n>0$ stationary solutions are dynamically unstable, we can reasonably expect that the system will eventually tend to the $n=0$ ground state solution.

\subsection{Continuous emission of solitons}\label{subsec:CES}

This subsection expands the discussion of the CES mechanism briefly mentioned in Secs. \ref{subsec:n=0} and \ref{subsec:n=1} and reports the main results of this work. Similar scenarios of emission of trains of solitons have been predicted in Refs. \cite{Frisch:PRL1992,Hakim1997,Pavloff:PRA2002} and experimentally observed in~\cite{engels2007}: in these works the soliton emission process takes place in the vicinity of a localized defect potential where the condensate density is locally depleted and a localized supersonic flow appears. Related, but somehow different soliton emission mechanisms were discussed in~\cite{Kamchatnov2002,Kamchatnov2012}.

\begin{figure*}[t!]
\begin{tabular}{@{}cc@{}}
    \includegraphics[width=0.5\columnwidth]{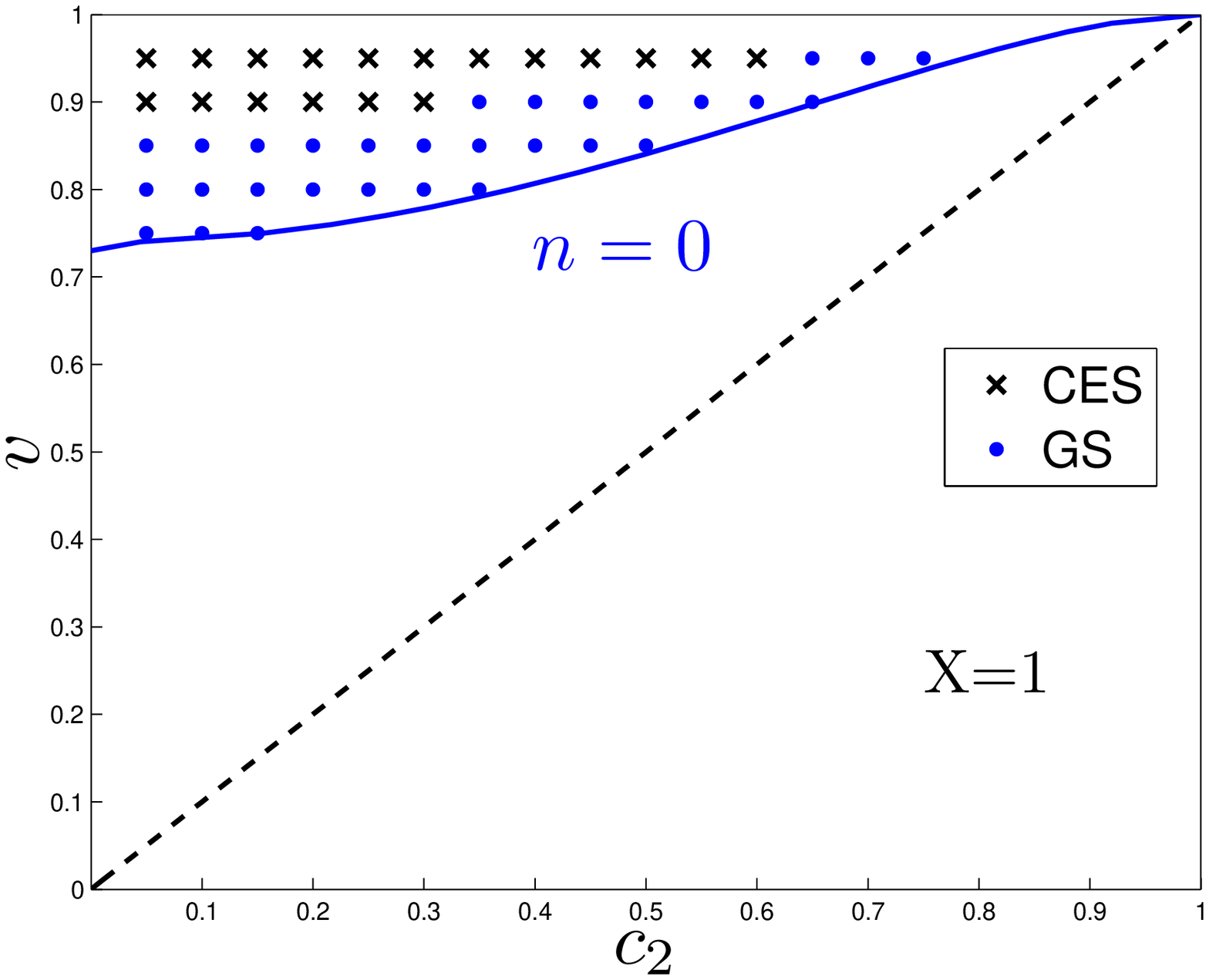} &
    \includegraphics[width=0.5\columnwidth]{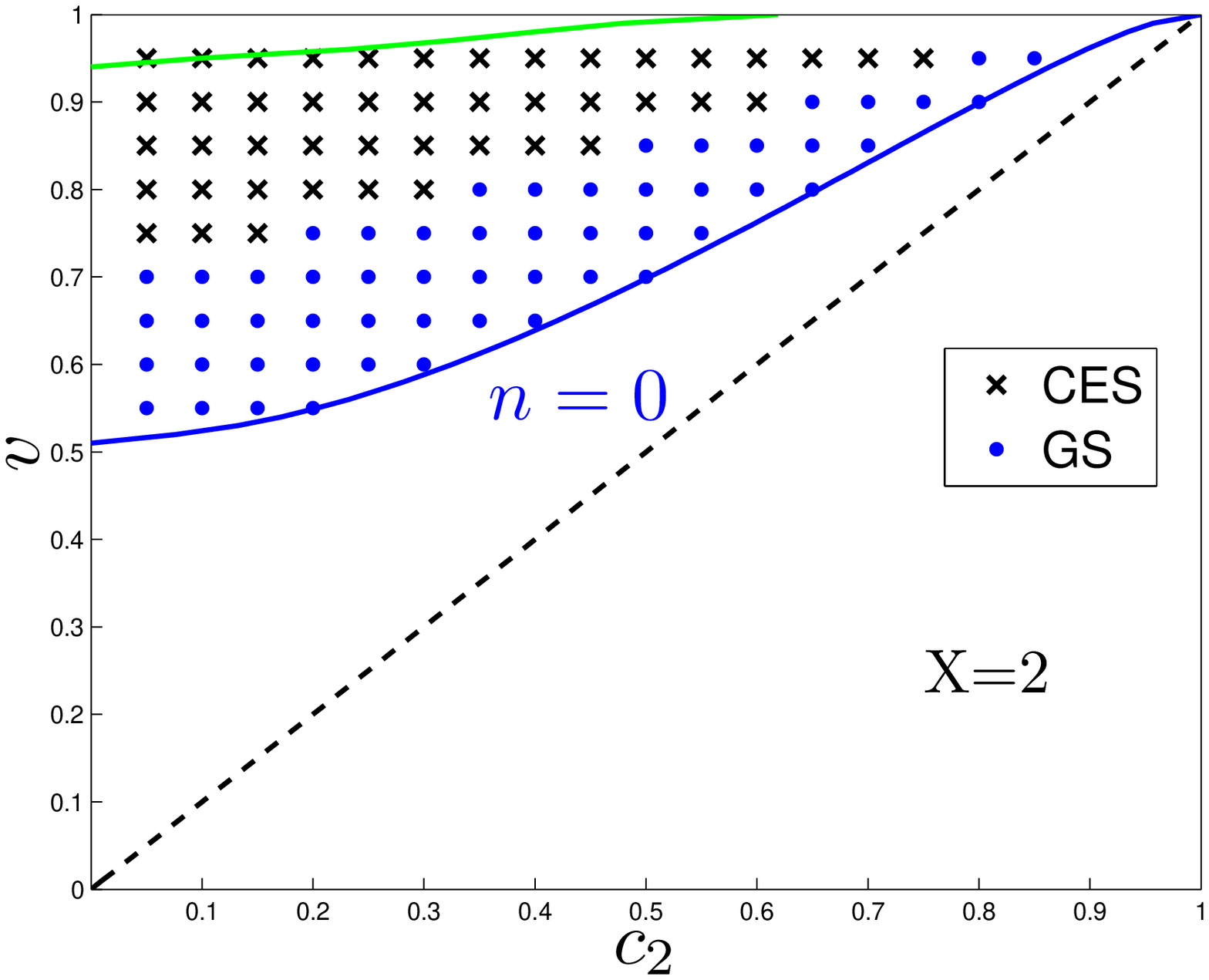} \\
    \includegraphics[width=0.5\columnwidth]{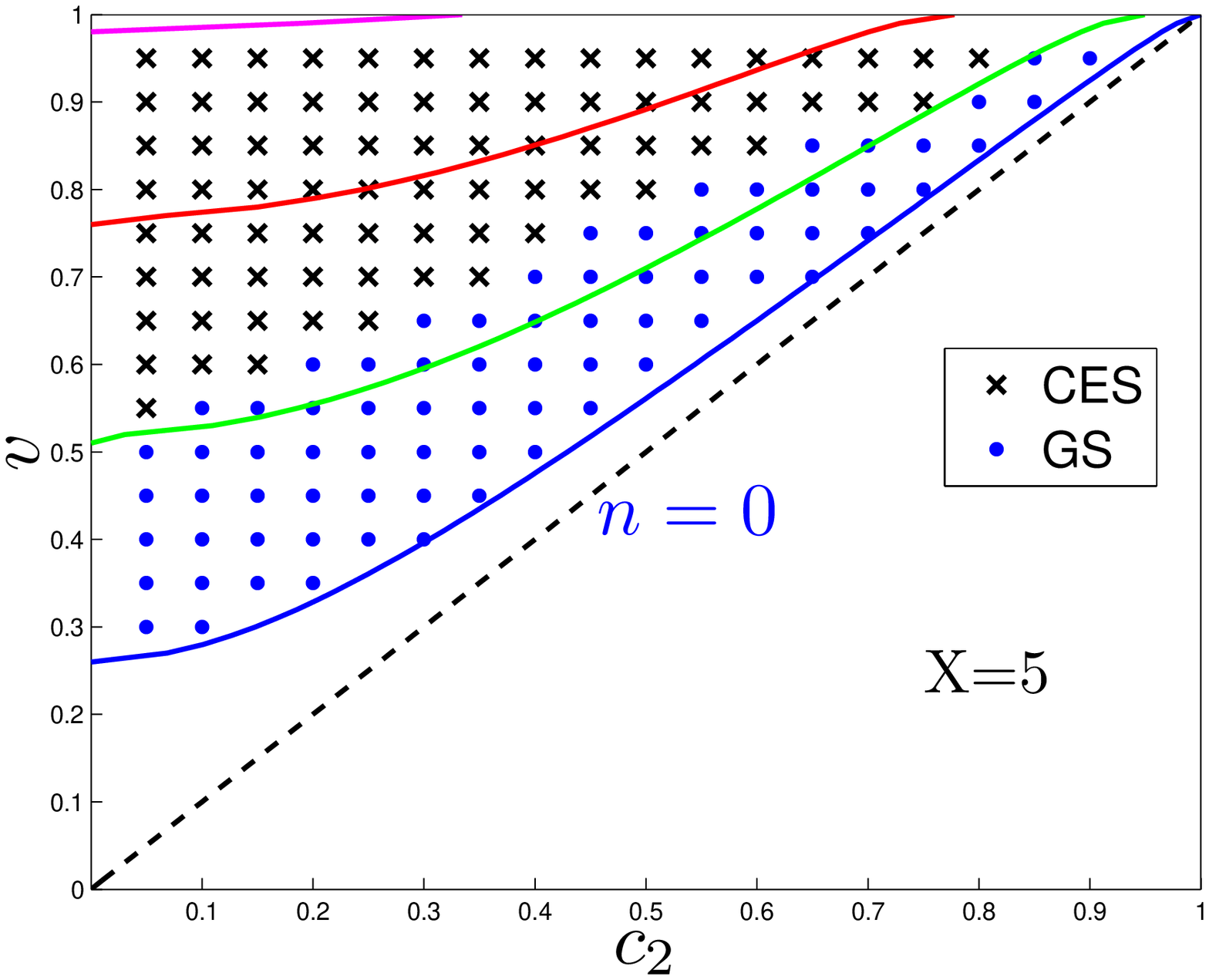} &
    \includegraphics[width=0.5\columnwidth]{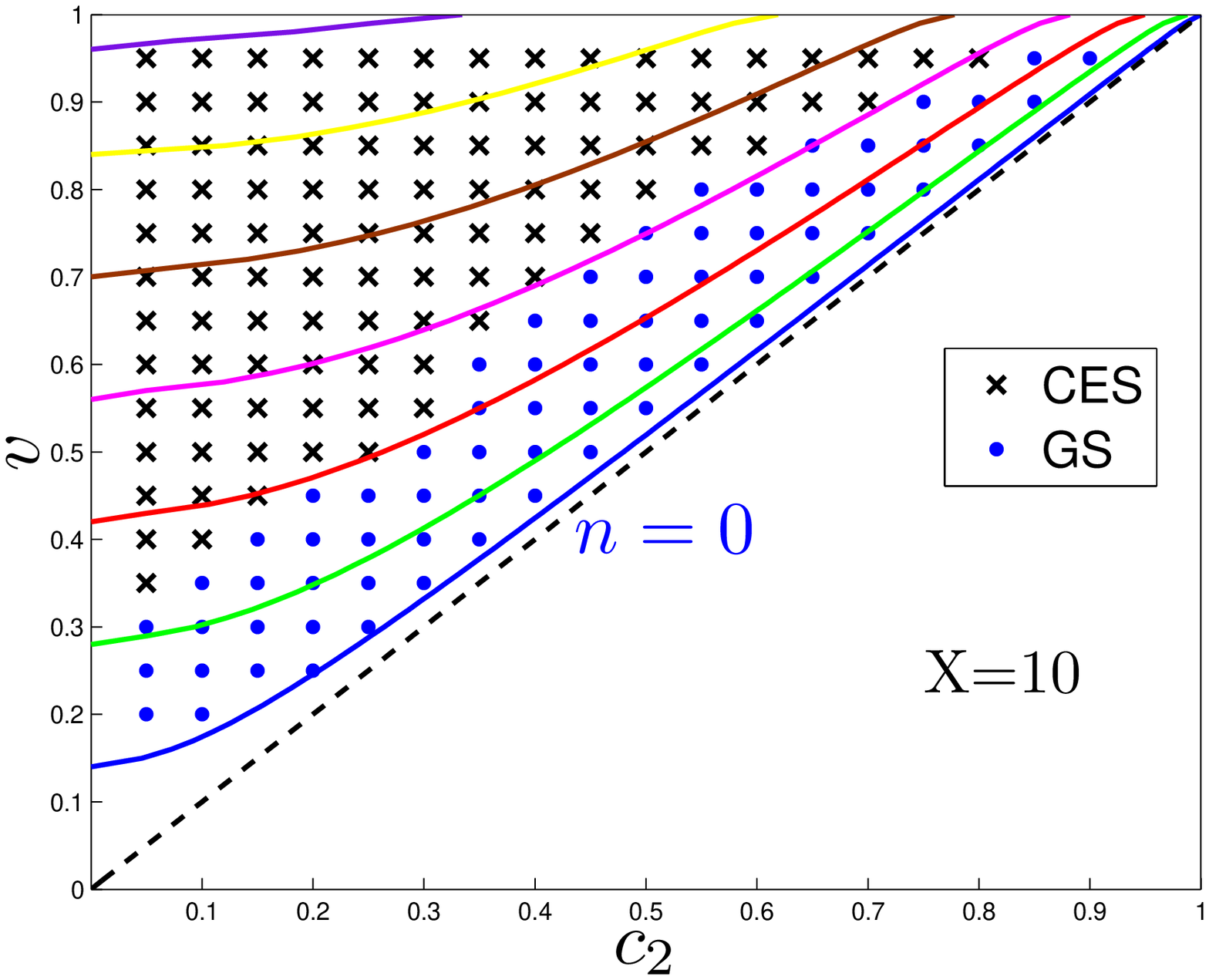} \\
\end{tabular}
\caption{Phase diagram in the $(c_2,v)$ plane for different values of $X$. The blue dots represents numerical simulations that end in the $n=0$ ground state (labeled as GS in the legend). The black crosses represents simulations in which CES has been observed. The solid curves from bottom to top represent the lower boundary of the linear instability regions defined in Eq. \ref{Eq:UnstableLength} with $n=0$ (blue), $1/2$ (green), $1$ (red), $3/2$ (pink), etc. The dashed oblique straight line is the upper boundary of the region where the flow is everywhere subsonic.}
\label{fig:CESregion}
\end{figure*}

The first step for characterizing the CES regime is to identify the CES regions in the $(X,c_2,v)$ parameter space: cuts of such phase diagram along the $(c_2,v)$ plane are shown in the different panels of Fig. \ref{fig:CESregion} for growing values of $X$. The region where CES happens is indicated by black crosses and grows with $X$ from the upper-left corner (the region with high $v$ and low $c_2$). This result can be explained as the CES regime is qualitatively related to the degree of instability of the central supersonic region, and this is favored by a large flow speed $v$ and a low $c_2$. Although a chaotic CES can be observed in strongly unstable systems with many unstable modes, we do not focus our attention on this chaotic regime and we restrict the use of the CES expression to periodic soliton emission processes.

\begin{figure*}[t!]
\begin{tabular}{@{}cccc@{}}
    \includegraphics[width=0.25\columnwidth]{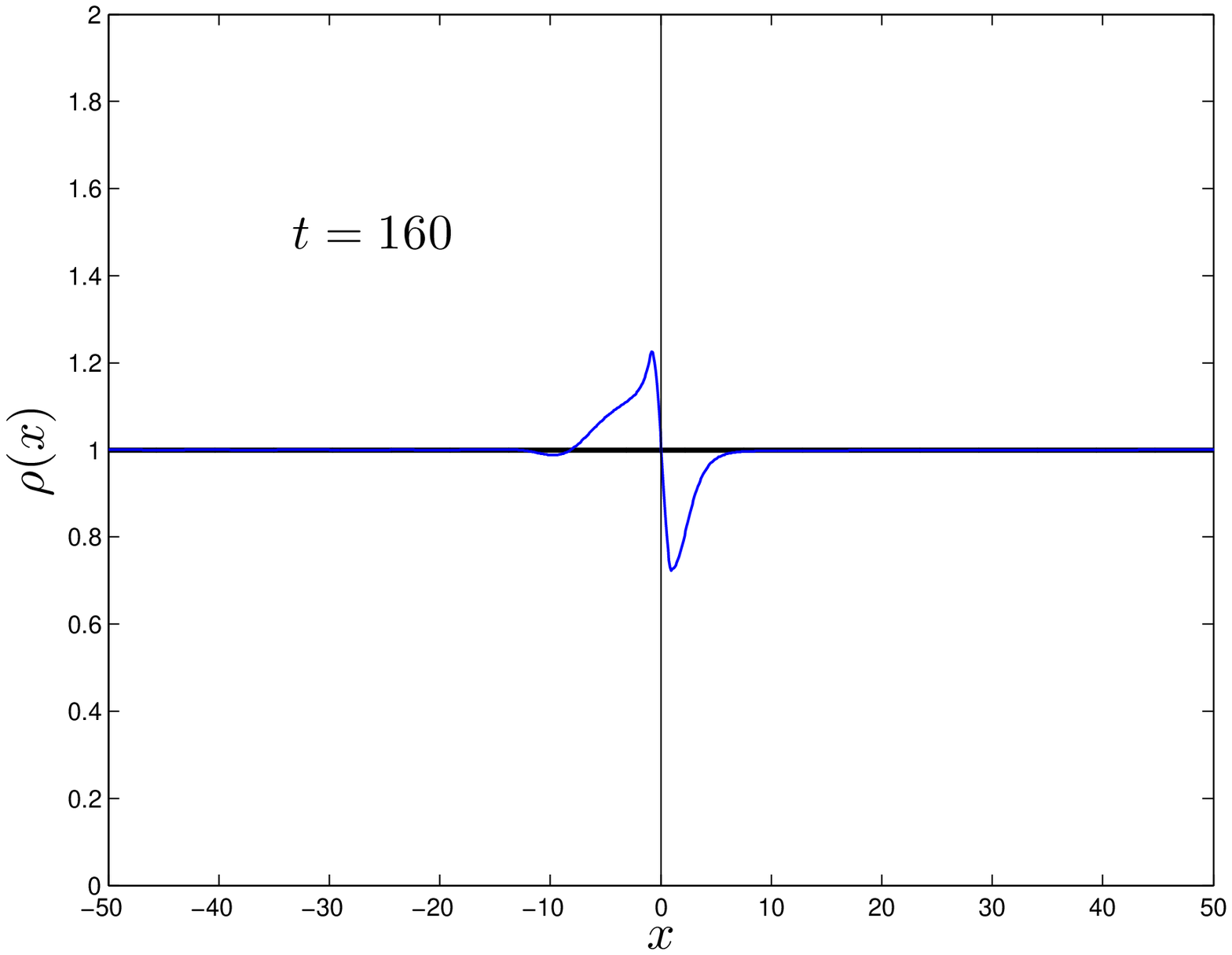} &
    \includegraphics[width=0.25\columnwidth]{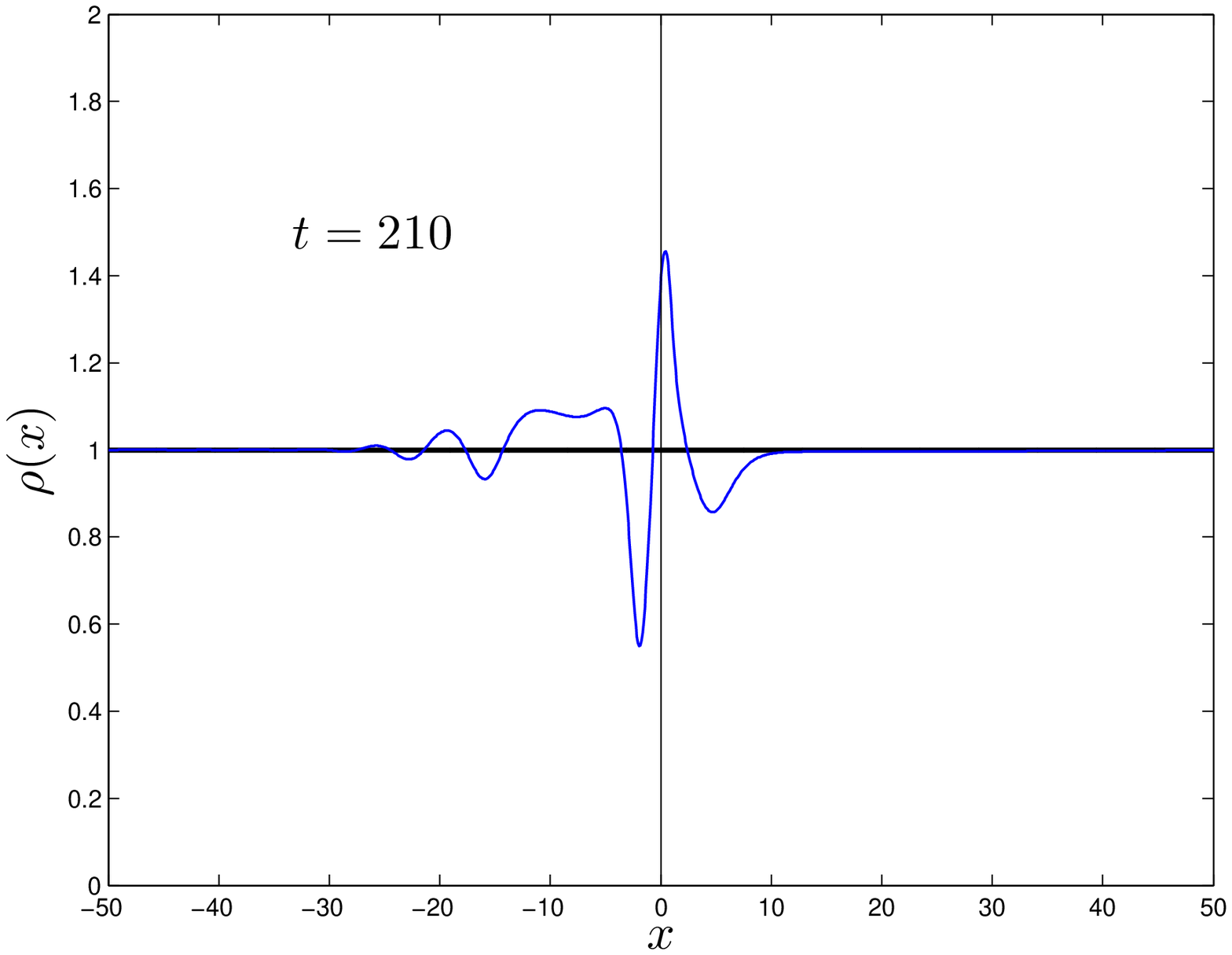} &
    \includegraphics[width=0.25\columnwidth]{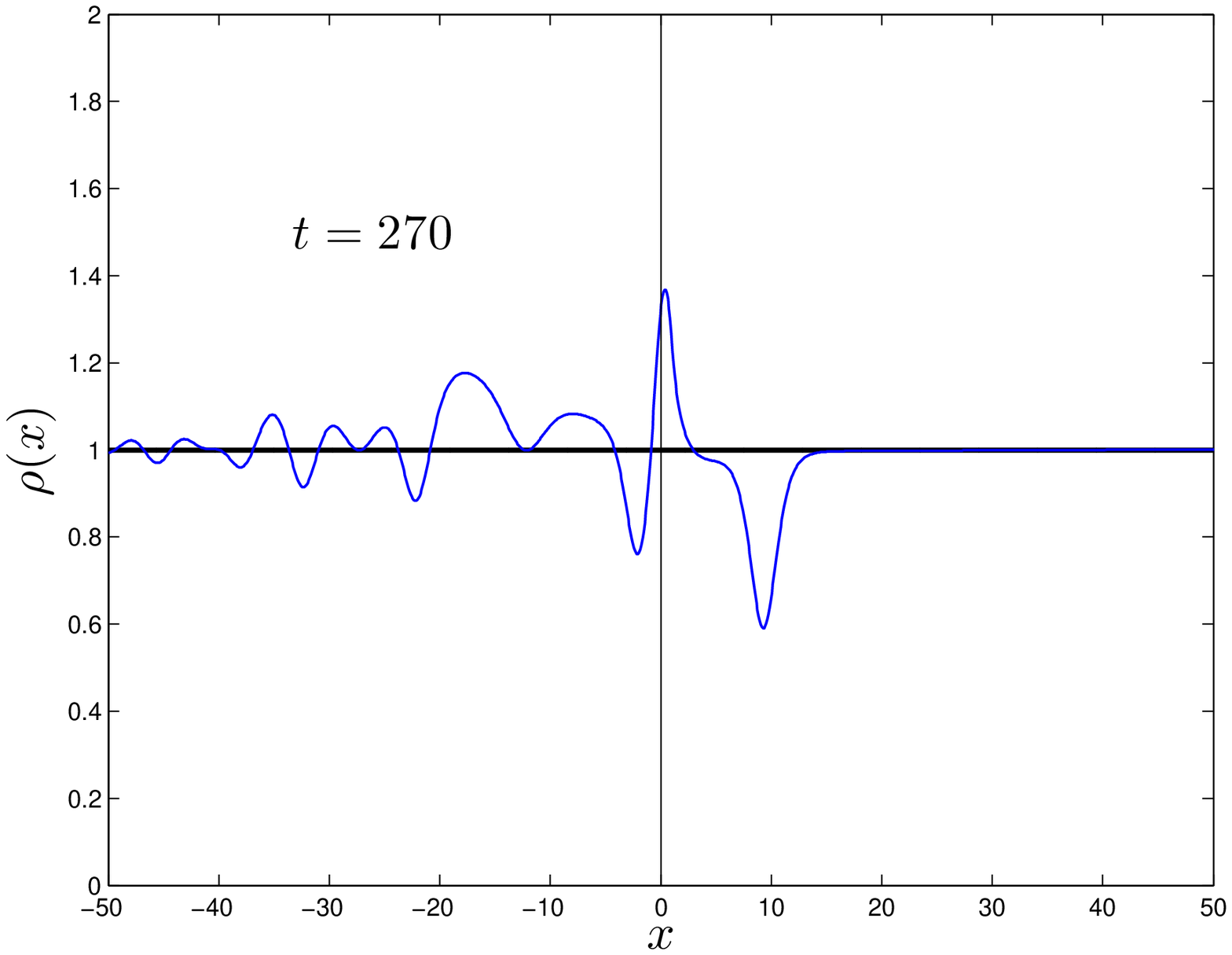} &
    \includegraphics[width=0.25\columnwidth]{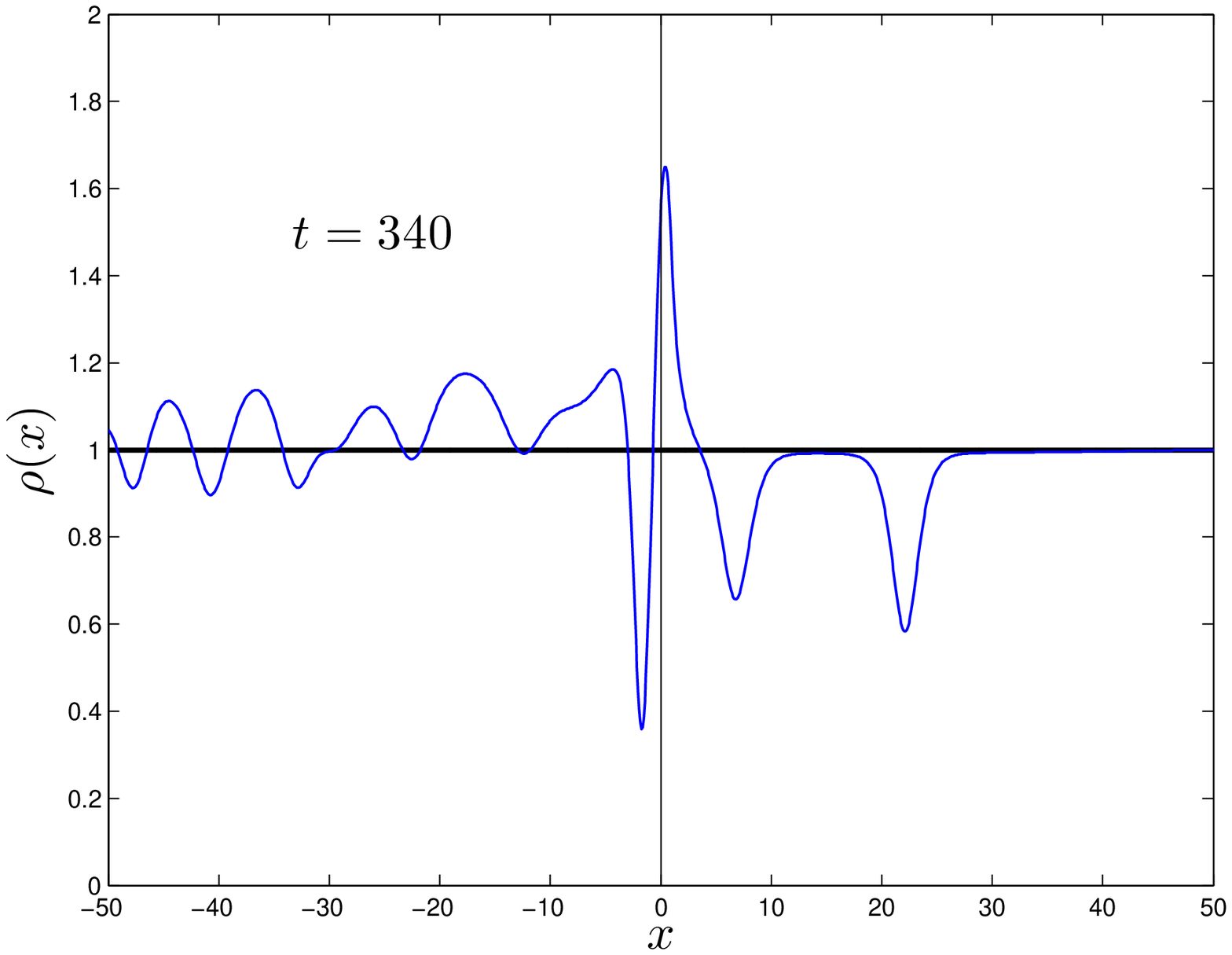} \\
    \includegraphics[width=0.25\columnwidth]{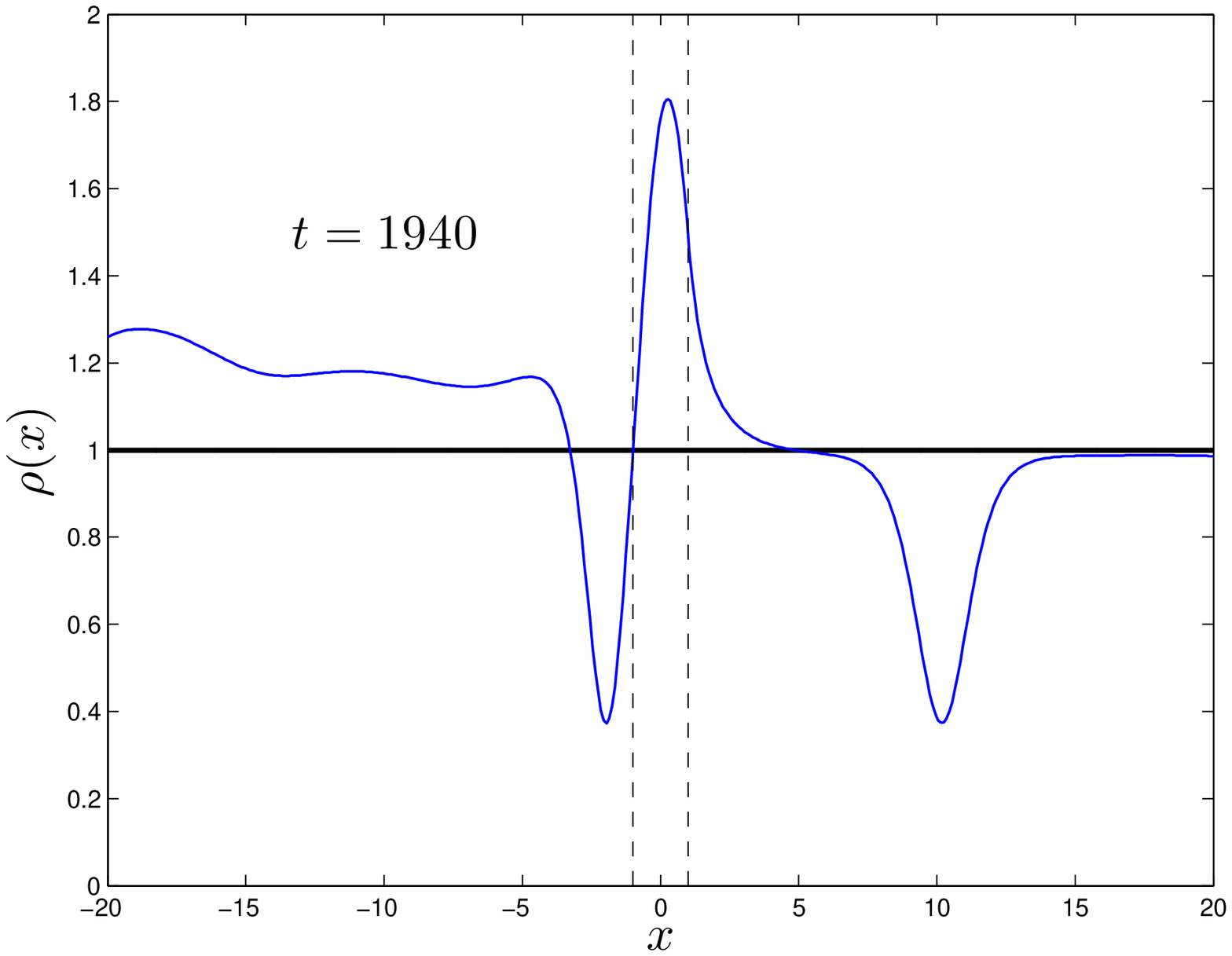} &
    \includegraphics[width=0.25\columnwidth]{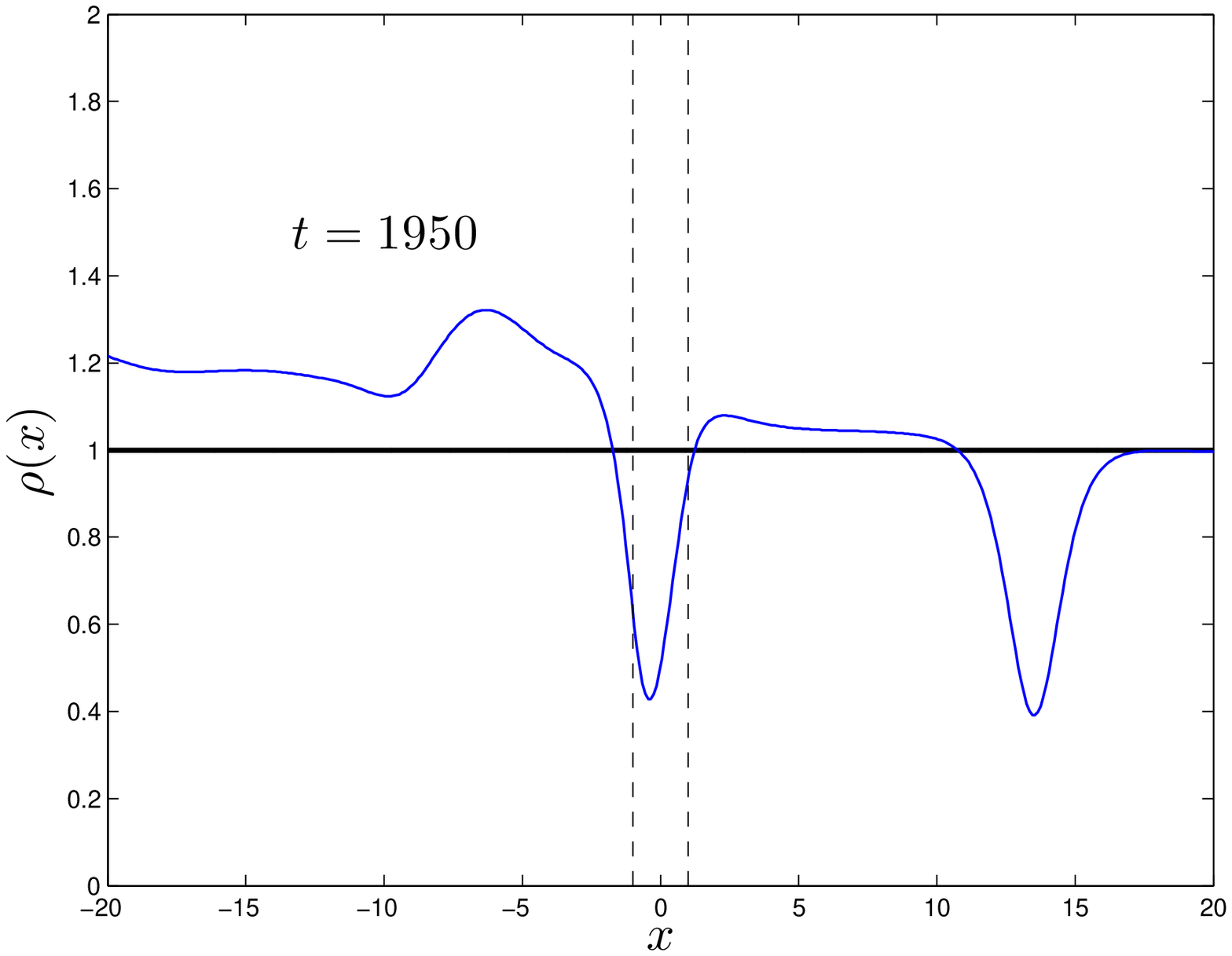} &
    \includegraphics[width=0.25\columnwidth]{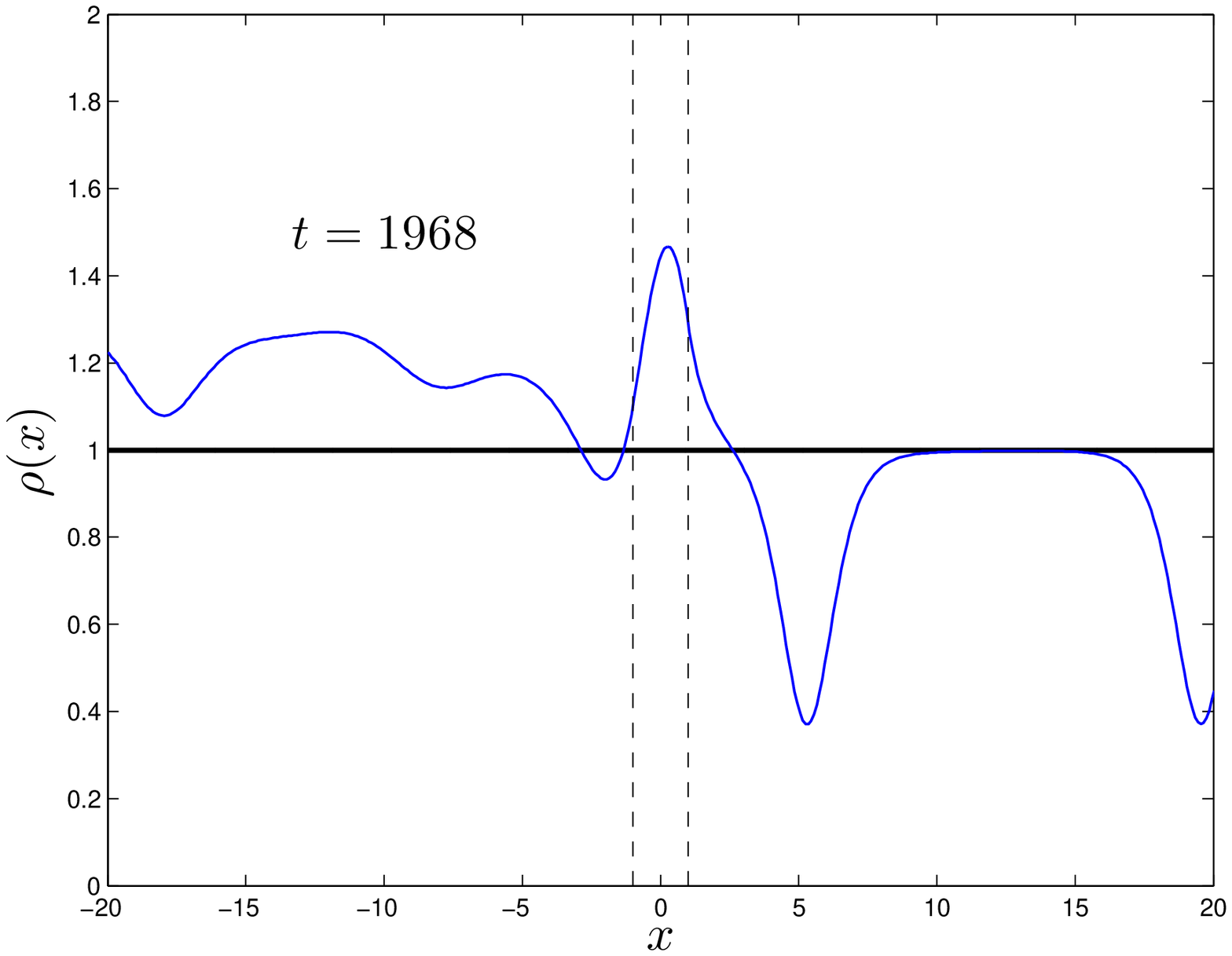} &
    \includegraphics[width=0.25\columnwidth]{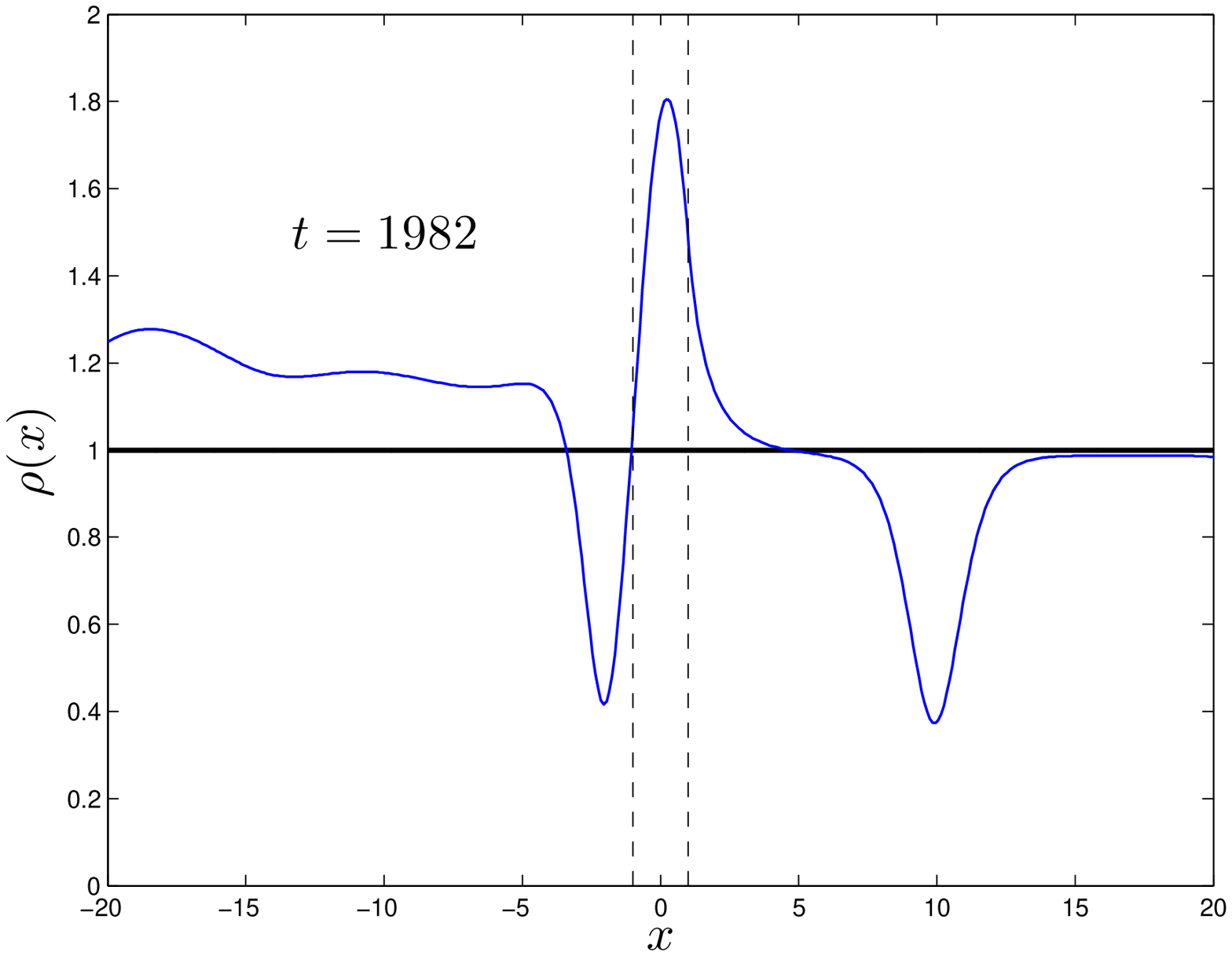}
\end{tabular}
\caption{Snapshots of the condensate density (solid blue line) at different evolution times (indicated in the panels) for a configuration with $v=0.95$, $c_2=0.4$ and $X=2$ that eventually reaches the CES regime. The horizontal black line represent the initial homogeneous condensate density. After some transient, the continuous emission of solitons begins. Upper row: initial evolution of the system when the initial linear instability sets in. The vertical black line marks the $x=0$ position. Lower row: periodic evolution at late times after the system has reached the CES regime. The vertical dashed lines marks the boundaries of the supersonic region. At $t=1940$, a soliton emerges near the black hole (subsonic-supersonic) interface at $x<0$. This soliton cannot travel upstream and then it bounces back, traveling to the downstream region and finally crossing the supersonic region, see plot at $t=1950$. After this soliton has left around $t=1968$, the density grows again and a new soliton starts growing near the interface. Finally, at $t=1982$, the system recovers the same state as at $t=1940$ and the process repeats.}
\label{fig:CESmechanism}
\end{figure*}

As done in the previous section, in order to understand the physics underlying the CES process we focus on the case where only the $n=0$ nonlinear stationary solution is present and, correspondingly, there is only one unstable mode. Several snapshots of the corresponding time evolution are shown in Fig. \ref{fig:CESmechanism} for a parameter choice that ends in the CES regime. In the upper row, we represent the initial evolution of the system. We see that, after the onset of the initially instability, the emission of solitons begins. In the lower row, we analyze in detail the CES mechanism, which persists indefinitely for arbitrarily long times. The soliton that tries to be emitted in the upstream direction is dragged by the flowing condensate and bounces back. Eventually, it ends up in the downstream region and travels towards $x\to +\infty$. After the soliton has gone, the density modulation in the supersonic region begins to grow again until a new soliton is generated. Continuous periodic repetition of this process leads to the emission of a train of solitons into the downstream region.

\begin{figure}[t]
\centering
\includegraphics[width=0.7\columnwidth]{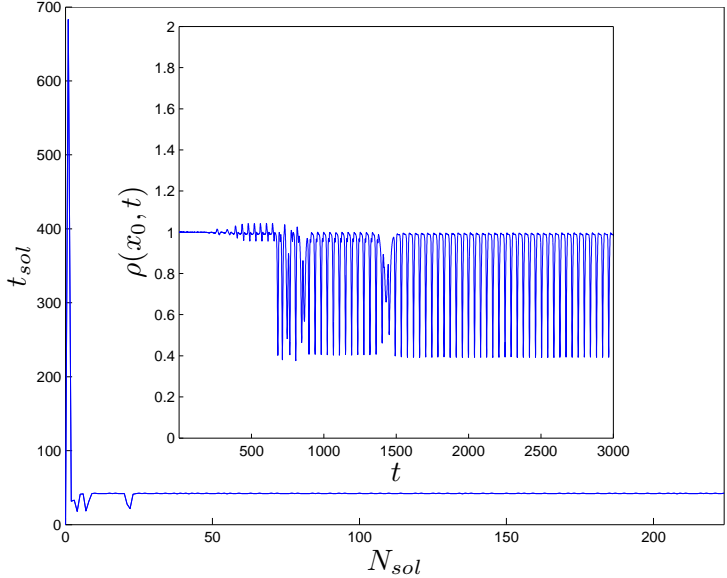}
\caption{Main panel: plot of $t_{sol}$, the time lapse between the arrival of consecutive solitons, as a function of the soliton number: after the emission of some solitons, the system reaches a regime of periodic soliton emission of period. Inset: time evolution of the density at a given point, $\rho(x_0,t)$ with $x_0=100$. The density minima correspond to the periodic passage of solitons. System parameters: $v=0.95$, $c_2=0.4$ and $X=2$.}
\label{fig:regularplot}
\end{figure}

In order to check quantitatively the periodicity of the soliton emission process, we introduce the quantity $t_{sol}(N)$, which is defined as the time lapse between the emission of the consecutive $(N-1)$th soliton and $N$th soliton. If the emission of solitons is periodic, $t_{sol}$ should be constant and equal to the period. For counting solitons, we monitor the density at a generic point $x_0\gg X/2$ in the downstream region as a function of time, as shown in the inset of Fig. \ref{fig:regularplot}. The minima of the density correspond to the passage of a soliton. In the main panel we represent the time-evolution of the time lapse $t_{sol}(N)$. After some irregular transient, we see that this quantity approaches a constant value $T$, meaning that the soliton emission process becomes an almost perfectly periodic one.

\begin{figure}[h]
\includegraphics[width=1\columnwidth]{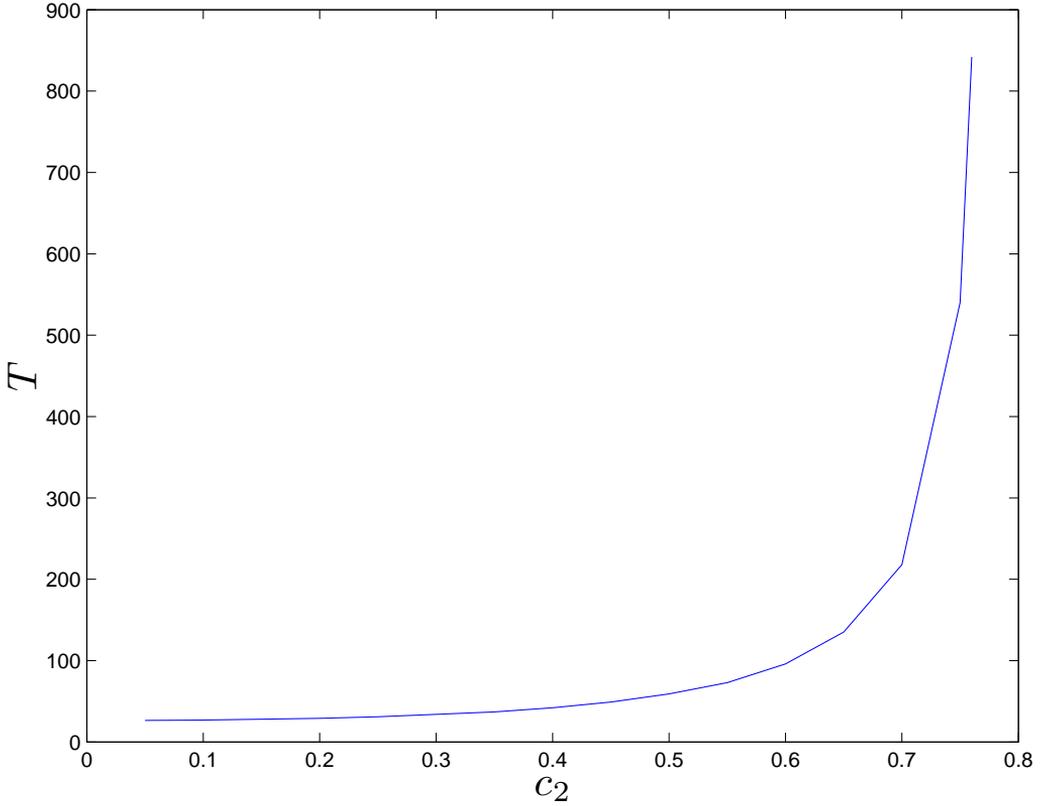} \caption{Soliton emission period $T$ as a function of $c_2$ for fixed $v=0.95$ and $X=2$.}
\label{fig:SolPeriod}
\end{figure}

Interesting light on the soliton emission process can be obtained by studying the dependence of the emission period on the system parameters. As an example, in Fig. \ref{fig:SolPeriod} we plot the CES period $T$ as a function of $c_2$ for fixed $v=0.95$ and $X=2$. The plot stops at $c_2=0.77$ since for larger values the CES dissapears and a sole soliton is emitted to the upstream region while the system relaxes to the $n=0$ nonlinear solution, in the same line of Sec. \ref{subsec:n=0}. For $c_2$ below the maximum value $c_2=0.77$, the period increases with $c_2$ as expected as for larger $c_2$ the instability of the the supersonic region becomes weaker. The validity of this physical interpretation has been further confirmed by numerically checking that the period also increases when either $v$ or $X$ are decreased.

As a last ingredient before proposing a qualitative explanation of the CES mechanism, it is useful to remind a few crucial features of solitons in condensates (check Sec. (\ref{app:1DGP} for a detailed discussion on 1D GP solutions). As we have seen in the previous sections, solitons are often emitted into the upstream and/or downstream external regions in order to compensate the increase of the density in the internal region. In the external region, the coupling constant is homogeneous $g(x)=1$ and the condensate asymptotically flows at a constant speed $v$. Hence, in the external regions, solitons are described by solutions of the form
\begin{equation}\label{eq:soliton}
\Psi(x)=e^{ivx}\left[ik+\sqrt{1-k^2}\tanh(\sqrt{1-k^2}\,x)\right]
\end{equation}
which corresponds to a Galilean transformation of Eq. (\ref{eq:solitonwavefunction}). As $v$ and the asymptotic density are fixed, the previous soliton solution is parametrized by a single real degree of freedom $-1\leq k \leq 1$. The corresponding density profile [given by Eq. (\ref{eq:solitondensity})] shows a minimum with value $k^2$. In respect to its dynamical evolution, apart from picking some irrelevant global time-dependent phase, the previous soliton rigidly moves at a $k$-dependent velocity
\begin{equation}\label{eq:solitonvelocity}
v_s=v+k
\end{equation}
which results from the Galilean combination of an overall drag at the condensate speed $v$ and a relative motion with speed $k$ equal to the local value of the speed of sound at the density minimum. Thus, the deeper the soliton, the smaller its relative velocity with respect to the surrounding fluid.

A soliton is typically emitted into the upstream direction in order to conserve the total number of particles. Hence, we can expect that its minimum density $k^2$ is a roughly decreasing function of the amplitude of the non-linear stationary solution $n=0$. As the soliton moves in the upstream direction, the sign of $k$ is fixed to $k<0$. In particular, we have numerically observed that the $k$ parameter of the emitted soliton depends very weakly on the initial condition so we can consider it to be a function of the system parameters, $k=k(v,c_2,X)$. As the $k$ parameter fixes the relative velocity of the soliton with respect to the underlying fluid, a qualitative change of behavior is expected to occur around $v=|k(v,c_2,X)|$. For $v<|k(v,c_2,X)|$, the soliton can freely travel in the upstream direction. In the opposite case, $v>|k(v,c_2,X)|$, the flow speed $v$ is larger than the relative soliton velocity $|k|$, so the soliton cannot travel to the upstream region and is dragged by the condensate flow into the downstream direction, originating the CES mechanism previously described.

Movies 7,8 and 9 illustrate this transition to the CES regime by reporting the result of numerical simulations for fixed $v=0.8,c_2=0.4$ and growing $X=2.2$ (\href{https://www.youtube.com/watch?v=olTkoSW5eR4}{Movie 7}), $X=2.4$ (\href{https://www.youtube.com/watch?v=CX-A-Rxy7MU}{Movie 8}), and $X=2.5$ (\href{https://www.youtube.com/watch?v=CNPdeuaL0jk}{Movie 9}). As $X$ is increased, the soliton has to carry away more and more particles to allow the system to relax to the $n=0$ nonlinear stationary state, which implies a reduction of $k$. As a result, the soliton becomes slower and slower: while for $X=2.2$ and $X=2.4$ the soliton is able to escape to the upstream direction, compensating the higher density of the $n=0$ nonlinear stationary solutions, this no longer happens for $X=2.5$, where we have in fact entered the $v>k$ regime and the soliton is forced to bounce back to the central region. As a result, it finally escapes in the downstream direction leaving the system again in an unstable configuration, so that the CES process keeps periodically repeating for indefinite times.

Although this qualitative observation is in good agreement with the numerical results, it is far from offering a complete theoretical picture. In particular, finding a quantitative relation between the minimum amplitude of the soliton emitted upstream and the parameters of the system, i.e., the function $k(v,c_2,X)$, is still a theoretically unsettled problem. Obtaining and understanding this relation is an open question that should be addressed in future works.

The CES regime when several unstable modes are present can be much more complex due to the appearance of other $n>0$ nonlinear stationary solutions. Typically, the system emits several solitons before reaching the periodic CES regime described previously. We show this scenario in \href{https://www.youtube.com/watch?v=7_H6jmMCfaI}{Movie 10}, which summarizes the result of a simulation for $v=0.65, c_2=0.3, X=8$. In other situations, the system first evolves towards a $n>0$ solution, remains in the vicinity of such solution for quite some time, then departs again from it and, after some transient, reaches the CES regime. We show an example of this latter case in \href{https://www.youtube.com/watch?v=1InNpUDHNVc}{Movie 11}. In this simulation, the system stays in the vicinity of the dynamically unstable $n=1$ solution for quite some time before entering in the CES regime. Finally, for a large number of unstable modes, the periodicity of the soliton emission may completely disappear, with the system strongly oscillating around different solutions with $n>0$. This chaotic scenario is illustrated in \href{https://www.youtube.com/watch?v=6W3smK7QZcY}{Movie 12}, where the system shows a very complicated aperiodic emission of solitons.

\section{Discussion and comparison with optical lasers}\label{sec:LaserComparison}

On the basis of the results presented in the previous Sections, we are
now in a position to critically discuss the actual physical content of
the widely used expression {\em black-hole laser}.

Historically, when this expression was first introduced by Corley and
Jacobson~\cite{Corley1999}, the authors had in mind the linear dynamical instability of the
configuration with a pair of neighbouring black- and white-hole
horizons. In terms of laser devices in quantum optics, this dynamical
instability corresponds to the linear instability of the electromagnetic vacuum state of the
cavity when the (unsaturated) gain exceeds losses. In this case, any weak signal due to,
e.g., spontaneous emission, is able to trigger the instability and quickly gets coherently amplified by stimulated
emission.

The actual operation of a continuous-wave laser device~\cite{QuantumNoise,Walls2008,Petruccione,MandelWolf} takes
place in a very different regime, where nonlinear gain saturation brings
the system to a nonlinear stationary state where (saturated) gain and losses
exactly compensate each other and the field keeps oscillating at a well-defined frequency with a well-defined amplitude.
This clean and coherent laser emission is attained at long times after switch-on once all (possibly complex) transient phenomena are gone: from the point of view of the field dynamics, the long-time limit is a limit cycle in the classical dynamics, and the only observable quantum effect are small fluctuations in the oscillation frequency, leading to a very long but finite coherence time of the emission.

Even though from a quantum optics perspective the expression {\em black-hole laser} suggests a spontaneously oscillating behavior with no memory of the initial state and strongly determined by nonlinear effects, the complex nonlinear dynamics of analog models in the two-horizon configurations has began being investigated only very recently in Refs. \cite{Michel2013,Michel2015}. Building on top of these works, we have seen in the previous sections that the long-time behavior of such configurations is much richer than that of standard optical lasers.

In many cases (Sec.~\ref{subsec:longtimestat}), the system is in fact able to quickly find a new time-independent stationary state where the instability is suppressed by a suitable redistribution of the density which eliminates the supersonic character of the central region. During this rapid ``evaporation'' of the horizons, the extra energy and particle density are compensated by the emission of a short sequence of sound and solitons to the outer regions.

On the other hand, regimes (Sec. \ref{subsec:CES}) where the density keeps oscillating in a periodic way in time can be regarded as a hydrodynamic counterpart of a continuous-wave monochromatic optical laser oscillation, leading to a continuous and periodic emission of trains of solitons. The excellent periodicity and insensitivity to initial conditions of this soliton train guarantees a very high degree of coherence of the emission, closely analogous to that of the output beam of an optical laser device. On this basis, a rigorous reader may then argue that the {\em black-hole lasing} expression should only be used to refer to this latter case.

As a final remark, it is worth noting that in contrast to standard single-mode laser devices which emit light of a single frequency, the solitons emitted by the BH laser device are intrinsically nonlinear objects containing a number of frequency components. Differently from pulsed lasers where the pulse-generating mechanism serves to modulate a much faster carrier frequency and creates tightly spaced sidebands, here the nonlinearity provides new frequencies at integer multiples of the fundamental one. In a qualitative picture, these components can be understood as resulting from a phase-locking of the different modes of the central, super-sonic region that are simultaneously oscillating. The energy for these oscillations is provided by the macroscopic flow of the condensate, and the mechanism responsible for the phase-locking of the different components originates from the hydrodynamic nonlinearities at large density modulations.

\section{Conclusions and outlook} \label{sec:BHLconclusions}

In this chapter, we have presented an extensive campaign of numerical simulations of the GP equation describing the time-evolution of a Bose-Einstein condensate in a BH laser scenario. In agreement with existing theoretical results about linear instabilities and nonlinear stationary solutions, our simulations provide useful physical insight on the system behavior as a function of the initial system parameters. In some cases, the central supersonic region is eventually evaporated away and the system at late times approaches a stationary stable configuration; in other cases, the system approaches a novel regime of continuous and periodic emission of solitons (CES) for indefinite times.

For different reasons, both regimes have a potentially strong interdisciplinary interest. On the one hand, the horizon evaporation process studied in our numerical simulations is a classical counterpart of black-hole evaporation under the effect of the spontaneous Hawking emission. From this point of view, a full understanding of the classical dynamics of the black-hole laser under the back-reaction effect of the classical emission is a natural first step towards a quantitative modeling of the back-reaction of quantum fluctuations on the curved space-time metric of an astrophysical black hole \cite{fabbri2005modeling}.

On the other hand, interesting analogies and differences between the continuous soliton emission in black-hole lasers and the operation of an optical laser device can be drawn from our simulations. On this basis, a more restrictive use of the {\em black-hole laser} expression is proposed. While all our discussion has focussed on the classical black-hole laser dynamics, future work will address the effect of quantum fluctuations on these systems: taking again inspiration from quantum optics and laser theory, we may legitimately expect that quantum effects should reduce to weak fluctuations in the emission frequency and, therefore, to a slow decoherence of the soliton emission analogous to the long, but finite coherence time of optical laser devices \cite{QuantumNoise,Walls2008,Petruccione,MandelWolf}. A more deep understanding of the CES regime should be also provided by forthcoming theoretical works.

Apart from the gravitational considerations, the CES regime here described can also represent an interesting scenario for the fields of quantum transport and atomtronics since it provides the soliton analog of an optical laser in a Bose-Einstein condensate.

From the experimental point of view, we can reasonably expect that an upgraded version of the ultracold atom experiment in \cite{Steinhauer2014} will soon be able to investigate the nonlinear dynamics of the black-hole laser at late times after the onset of the dynamical instability and hopefully characterize the rich phenomenology discussed in this work. Even though all the discussion in this work has been carried out having an atomic implementation in mind, we expect that most conclusions can be directly transferred to analog models based on fluids of light in the so-called propagating geometry, for which theoretical works on black hole configurations have recently appeared \cite{FleurovEPL2011,CarusottoPRSA2014} as well as first experimental evidences of superfluid behaviors \cite{Vocke_Optica2015}.

\part{Thermal clouds}

\chapter{Thermal decay in a trapped gas}\label{chapter:thermaldecay}

\section{Introduction}

In the first part of this thesis, we have studied the implementation of gravitational analogues in condensed-matter systems. In particular, we have focused on Bose gases near $T=0$, where the mean-field Gross-Pitaevskii equation is valid. In this chapter, we switch to the opposite limit: a gas of bosons above the critical temperature $T_c$. In particular, we analyze the response of a thermal cloud (without any condensed fraction) to the application of a short Bragg pulse. As one may expect, this process induces oscillations in the density profile that eventually decay after removing the lattice potential. Following the standard textbooks on the field \cite{Pitaevskii2003,Pethick2008}, one would typically attribute the damping mechanism to the collisional relaxation of the collective modes above $T_c$, described using the classical Boltzmann distribution function. However, we show in this chapter that the decay of the oscillations is actually due to the thermal disorder of the particles and shows a characteristic time scale that only depends on the temperature $T$, the mass of the atoms $m$ and the wave vector of the Bragg lattice $k$. This result is found when using both quantum and classical formalisms. Indeed, we show that this thermal decay is a general feature of the introduction of an arbitrary spatially periodic pulse. Moreover, under some general assumptions, the results here presented can be applied to other systems such a gas of bosons below $T_c$ (with some condensed fraction present) or a gas of fermions at high temperatures.

We start the chapter by describing the physical setup in Sec. \ref{sec:physexpsetup}. After that, under certain specific assumptions, we provide a theoretical explanation for the phenomenon, using both classical (Sec. \ref{sec:classical}) and quantum (Sec. \ref{sec:quantum}) formalisms. We discuss the validity and physical meaning of the approximations used and unify the quantum and classical results in Sec. \ref{sec:approxs}. The obtained results are generalized to broader scenarios in Sec. \ref{sec:generalcase}. We finally compare the results from the calculations with actual experimental data in Sec. \ref{sec:experimentaldata}. The conclusions are drawn in Sec. \ref{sec:decayconclusions}.

\section{Physical setup}\label{sec:physexpsetup}

We consider an initially confined cloud of $N$ bosonic atoms at thermal equilibrium with temperature $T>T_{c}$, so there is no condensate. The atomic cloud is confined by the harmonic trap created by the potential
\begin{equation}\label{eq:harmonictrap}
V_{\rm {trap}}(\mathbf{x})=\sum_i \frac{1}{2}mw^2_ix_i^2
\end{equation}
with $i=1,2,3$ labeling the coordinates $x,y,z$. Then, at $t=0$, a short Bragg pulse is switched on during a time $\tau$, inducing a sinusoidal potential in the $z$ direction of the form
\begin{equation}\label{eq:potentialswitch}
V(z,t)=-\chi\left(\frac{t}{\tau}\right)V_0\cos(kz)
\end{equation}
where $\chi$ is the characteristic function of the interval $[0,1]$.

This Bragg pulse is created by two incident laser beams with same wavelength and forming some angle between them, in the same way as the optical lattice considered in Chapter \ref{chapter:MELAFO}. After time $\tau$, the pulse is switched off and the atom cloud is left to evolve within the trap. As we will show in this work, the Fourier components of the formed density pattern decay with a Gaussian behavior and the density returns to the initial equilibrium configuration. The introduction of a short Bragg pulse was also applied below $T_c$ for measuring the phonon dispersion relation \cite{Shammass2012}.

For studying the dynamics of the previous scenario, we consider both quantum and classical descriptions. In the two models we neglect interaction between particles, see Sec. \ref{subsec:interactingrole} for a detailed discussion on the role of interactions.

\section{Classical description}\label{sec:classical}

As a first step, we study the problem in a classical context. For that purpose, we use the Boltzmann equation, which governs the time evolution of the Boltzmann distribution function $f(\mathbf{x},\mathbf{p},t)$ that gives the density in both real and momentum space \cite{Pitaevskii2003,Pethick2008}. The Boltzmann equation, for non-interacting particles, can be written as:
\begin{equation}\label{eq:boltzmanneq}
\frac{\partial f}{\partial t}+\frac{\mathbf{p}}{m}\nabla f+\mathbf{F}_{ext} \nabla_{p} f=0
\end{equation}
where $\nabla_{p}$ denotes the gradient with respect the momentum components and $\mathbf{F}_{ext}$ is the force that acts on the particles due to the external field, $\mathbf{F}_{ext}=-\nabla H$, with $H$ the total Hamiltonian. The general solution of Eq. (\ref{eq:boltzmanneq}) can be computed using the method of characteristics:
\begin{equation}\label{eq:boltzmannsolution}
f(\mathbf{x},\mathbf{p},t)=f_0(\mathbf{x}_0(\mathbf{x},\mathbf{p},t),\mathbf{p}_0(\mathbf{x},\mathbf{p},t))
\end{equation}
where $f_0(\mathbf{x},\mathbf{p})$ is the initial Boltzmann distribution function at $t=0$. The vectors $\mathbf{x}_0(\mathbf{x},\mathbf{p},t)$, $\mathbf{p}_0(\mathbf{x},\mathbf{p},t)$ are initial conditions at $t=0$ such their time evolution through the classical equations of motion will carry them to $\mathbf{x},\mathbf{p}$ in a time $t$ or, from another point of view, they are the result of integrating the corresponding time-reversed equations of motion with initial conditions $\mathbf{x},\mathbf{p}$. In our case, as the system is in equilibrium, the initial condition is the thermal Boltzmann distribution
\begin{equation}\label{eq:boltzmannthermal}
f_0(\mathbf{x},\mathbf{p})=N\frac{e^{-\beta H_0}}{Z},~Z=\int\mathrm{d}^3\mathbf{x}\mathrm{d}^3\mathbf{p}~e^{-\beta H_0}
\end{equation}
with $H_0$ being the trapped equilibrium Hamiltonian
\begin{equation}\label{eq:equilibriumHam}
H_0=\frac{\mathbf{p}^2}{2m}+V_{\rm {trap}}(\mathbf{x})
\end{equation}
and $\beta=1/k_BT$. The previous Hamiltonian is separable as it corresponds to a harmonic oscillator (HO) in each spatial direction,
\begin{equation}\label{eq:separability}
H_0=\sum_i H_{0i},~H_{0i}=\frac{p^2_i}{2m}+\frac{1}{2}mw^2_ix_i^2
\end{equation}
As a consequence, $f_0$ factorizes in the form:
\begin{equation}\label{eq:boltzmannfactor}
f_0(\mathbf{x},\mathbf{p})=N\prod_i \frac{\beta \omega_i}{2\pi}e^{-\frac{\beta}{2m}p_i^2}e^{-\frac{\beta}{2}mw^2_ix_i^2}
\end{equation}
We see from this expression that the typical momentum of the particles is of order $p_T=\sqrt{2m/\beta}$ and the width of the trapped cloud in the direction $i$ is $R_i=(\beta m \omega_i^2)^{-\frac{1}{2}}$.

We now proceed to integrate the Boltzmann equation for our problem. The time evolution of the transverse degrees of freedom is trivial as the potential applied, Eq. (\ref{eq:potentialswitch}), only depends on the $z$ coordinate. Thus, the only dynamical part is that related with the $z$ coordinate and the solution is given, according to Eq. (\ref{eq:boltzmannsolution}), by:

\begin{equation}\label{eq:boltzmanngeneralsolution}
f(\mathbf{x},\mathbf{p},t)=N\left(\frac{\beta\tilde{\omega}}{2\pi}\right)^3e^{-\beta H_{0x}}e^{-\beta H_{0y}}e^{-\frac{\beta}{2m}p^2_{z0}}e^{-\frac{\beta}{2}mw^2_zz_0^2}
\end{equation}
where $\tilde{\omega}=\sqrt[3]{\omega_x\omega_y\omega_z}$. Therefore, we only need to solve the one-dimensional equations of motion in the $z$ direction for a given initial condition $z_0,p_{z0}$ in order to obtain $f(\mathbf{x},\mathbf{p},t)$.

During the pulse, the particles move according to
\begin{eqnarray}\label{eq:Hamiltontau}
m\dot{z}&=&p_z\\
\nonumber \dot{p}_z&=&-V_0k\sin kz-m\omega^2_z z
\end{eqnarray}
while for $t>\tau$ the time evolution is governed by the well-known HO solutions:
\begin{equation}\label{eq:zclassho}
\left[\begin{array}{c}
z\\
p_z
\end{array}\right]=\left[\begin{array}{cc}
\cos[\omega_z (t-\tau)] & \frac{\sin[\omega_z (t-\tau)]}{m\omega_z}\\
-m\omega_z \sin[\omega_z (t-\tau)] & \cos[\omega_z (t-\tau)]
\end{array}\right]\left[\begin{array}{c}
z(\tau)\\
p_z(\tau)
\end{array}\right]
\end{equation}
In order to obtain analytical results, we make use of some approximations. First, we suppose that $t$ satisfies $\omega_zt\ll 1$. In that case, we can approximate Eq. (\ref{eq:zclassho}) as:
\begin{equation}\label{eq:zclassapprox}
z(\tau)\simeq z-p_z(t-\tau),~p_z(\tau)\simeq p_z
\end{equation}
In the previous approximation, we have neglected the variation of momentum resulting from the term $m\omega_z z \sin[\omega_z (t-\tau)]$ as it represents a small correction to the thermal momentum of the particles, $m\omega_z z \sin[\omega_z (t-\tau)] \sim m\omega_zR_z \omega_z (t-\tau)\sim p_T\omega_z (t-\tau)\ll p_T$. The same argument can be applied to the harmonic oscillator part of Eq. (\ref{eq:Hamiltontau}) and then it is reduced to $\dot{p}_z\simeq-V_0k\sin kz$. This equation can be integrated in terms of elliptic functions, as it corresponds to a classical pendulum. In this way, we express $z_0$ in terms of $z(\tau)$ as:
\begin{eqnarray}\label{eq:zelliptic}
\nonumber z_0&=&\frac{2}{k}\arcsin\left(\nu~\text{sn}\left[F\left(\arcsin\left(\frac{\frac{kz(\tau)}{2}}{\nu}\right),\nu\right)-\sqrt{\frac{V_0}{m}}k\tau,~\nu\right]\right),~\nu=\frac{E+V_0}{2V_0},~E<V_0\\
z_0&=&\frac{2}{k}\arcsin~\text{sn}\left[F\left(\frac{kz(\tau)}{2},\nu\right)-\sqrt{\frac{E+V_0}{2m}}k\tau,~\nu\right],~\nu=\frac{2V_0}{E+V_0},~E>V_0
\end{eqnarray}
with $E$ the conserved energy during the presence of the Bragg pulse:
\begin{equation}
E=\frac{p^2_z(\tau)}{2m}-U\cos\left[kz(\tau)\right]=\frac{p^2_{z0}}{2m}-U\cos\left[kz_0\right]
\end{equation}
from which we can also obtain the momentum $p_{z0}$. For the notation of elliptic functions, see Eq. (\ref{eq:ellipticdensity}) and associated discussion.

If the increase in the momentum due to the force of the pulse satisfy $V_0k\tau/p_T\lesssim1$, then $p_T$ is still the typical momentum of the particles after the pulse. This implies that, as the total displacement of the $z$ coordinate is of order $z\sim z_0+v_Tt$, with $v_{T}=p_T/m$ the thermal velocity of the atoms, $\beta m\omega_z^2z_0^2\simeq\beta m\omega_z^2z^2$ with corrections $O(\omega_zt)$. Hence, we set $z_0=z$ in Eq. (\ref{eq:boltzmanngeneralsolution}):
\begin{equation}\label{eq:boltzmannLDA}
f(\mathbf{x},\mathbf{p},t)\simeq N\left(\frac{\beta\tilde{\omega}}{2\pi}\right)^3e^{-\beta V_{\rm {trap}}(\mathbf{x})} e^{-\frac{\beta}{2m}p_x^2} e^{-\frac{\beta}{2m}p_y^2}e^{-\frac{\beta}{2m}p^2_{z0}}
\end{equation}

We see that the previous approximations amount to consider the gas as locally homogeneous, with an envelope factor $e^{-\beta V_{\rm {trap}}(\mathbf{x})}$, similar to the local density approximation that we will consider in Sec. \ref{subsec:staLDA}. The only remaining task is to compute $p_{z0}$. Although it can be expressed in terms of elliptic functions from the results of Eq. (\ref{eq:zelliptic}), in order to obtain explicit formulas for the density we assume that the pulse duration $\tau$ is sufficiently small for a Runge-Kutta second-order approximation to be valid. In that case, we compute the time evolution during the pulse as:
\begin{equation}\label{eq:RK2}
p_{z0}\simeq p_z+V_0k\tau \sin kz\left(\frac{\tau}{2}\right),~z\left(\frac{\tau}{2}\right)\simeq z(\tau)-\frac{p_z\tau}{2m}=z-\frac{p_z}{m}\bar{t}
\end{equation}
with $\bar{t}\equiv t-\tau/2$ and we have used that $p_z=p_z(\tau)$, following Eq. (\ref{eq:zclassapprox}). The first corrections to the previous equation are $O(\tau^3)$. We refer to Sec. \ref{subsec:LTA} for a detailed analysis of the validity of the previous approximation.

The density is obtained after integrating over the momentum variables of $f(\mathbf{x},\mathbf{p},t)$ in Eq. (\ref{eq:boltzmannLDA}):
\begin{equation}\label{eq:boltzmanndensity}
n(\mathbf{x},t)=\int\mathrm{d}^3\mathbf{p}~f(\mathbf{x},\mathbf{p},t)= N\left(\frac{\beta m\tilde{\omega}^2}{2\pi}\right)^{\frac{3}{2}}e^{-\beta V_{\rm {trap}}(\mathbf{x})}G(z,t)
\end{equation}
where the dimensionless function $G(z,t)$ takes into account the dynamics in the $z$ coordinate
\begin{equation}\label{eq:integralBdimension}
G(z,t)=\sqrt{\frac{\beta}{2\pi m}}\int\mathrm{d}p_z~e^{-\frac{\beta}{2m}[p_z+V_0k\tau\sin\left(kz-k\frac{p_z}{m}\bar{t}\right)]^2}
\end{equation}
The previous integral can be computed analytically from the identity
\begin{equation}\label{eq:integralB}
I(a,b,\phi)\equiv \frac{1}{\sqrt{\pi}}\int_{-\infty}^\infty\mathrm{d}x~e^{-[x+a\sin\left(\phi-b x\right)]^2}=\sum_{n=-\infty}^\infty J_n\left(nab\right)e^{-\frac{n^2b^2}{4}}e^{in\phi}
\end{equation}
where the Bessel functions are defined as
\begin{equation}\label{eq:BesselDef}
J_n(x)=\sum_{k=0}^\infty\frac{(-1)^k}{(n+k)!k!}\left(\frac{x}{2}\right)^{n+2k},~n\geq 0
\end{equation}
and $J_{-n}(x)=(-1)^nJ_n(x)$.

In order to arrive at Eq. (\ref{eq:integralB}), we have followed these steps. First, we expand the square in the exponential of the integral and consider the generating function of the Hermite polynomials:
\begin{equation}\label{eq:generatingfunction}
e^{2xt-t^2}=\sum_{n=0}^\infty\frac{H_n(x)}{n!}t^n
\end{equation}
The previous relation can be easily proven from the definition of the Hermite polynomials
\begin{equation}\label{eq:hermitepolynomialsdef}
H_n(x)=(-1)^n e^{x^2}\frac{d^n}{dx^n}e^{-x^2}
\end{equation}
After that, we insert the binomial expansion of the powers of the sine function:
\begin{equation}\label{eq:sinbinomial}
\sin^n(x)=\sum_{k=0}^n\frac{n!}{k!(n-k)!}(-1)^{k}\frac{e^{i(n-2k)x}}{(2i)^n}
\end{equation}
and finally, we take into account the following Fourier transform:
\begin{equation}
\frac{1}{\sqrt{\pi}}\int_{-\infty}^\infty\mathrm{d}x~H_n(x)e^{-x^2}e^{-ikx}=(-ik)^ne^{-\frac{k^2}{4}}
\end{equation}
Performing all these operations and rearranging the sum conveniently, we obtain Eq. (\ref{eq:integralB}). Finally, applying this identity to Eq. (\ref{eq:integralBdimension}), we get:
\begin{equation}\label{eq:MBfunctionG}
G(z,t)=\sum_{n=-\infty}^\infty J_n\left(n\frac{V_0 k^2\tau}{m}~\bar{t}\right)e^{-n^2\frac{k^2}{2m\beta}\bar{t}^2}e^{inkz}
\end{equation}
We see that the density presents decaying oscillations with the same spatial periodicity as the potential, $d=2\pi/k$. In particular, each discrete Fourier component $n$ decays with a characteristic time:
\begin{equation}\label{eq:decaytime}
\tau^d_{n}=\frac{\tau^d}{n},~\tau^d=\frac{\sqrt{2m\beta}}{k}
\end{equation}
In this way, the dominant harmonics are the $n=\pm 1$ Fourier components, that can be extracted by performing the Fourier transform of the density in Eq. (\ref{eq:boltzmanndensity}):
\begin{equation}\label{eq:Fouriercomponent}
n_k(t)=\int\mathrm{d}^3\mathbf{x}~n(\mathbf{x},t)e^{-ikz}=NJ_1\left(\frac{V_0 k^2\tau}{m}~\bar{t}\right)e^{-\left(\frac{\bar{t}}{\tau^d}\right)^2}
\end{equation}
We have neglected the mixing of the discrete Fourier components due to the envelope $e^{-\beta V_{\rm {trap}}(\mathbf{x})}$ by supposing that the wavelength of the pulse is much lower than the size of the trap, $\exp(-k^2/\beta m \omega_z^2)\ll1$.

\begin{figure}[!htb]
\centering
\includegraphics[width=0.7\columnwidth]{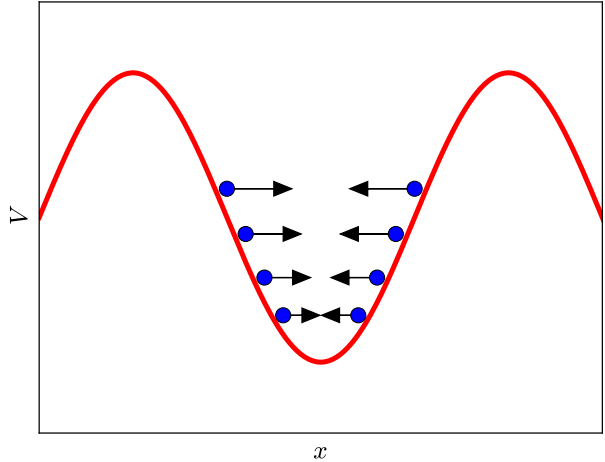}
\caption{Illustrative scheme of the situation. The red line represents the sinusoidal potential and the black arrows represent the force on the atoms.}
\label{fig:Focus}
\end{figure}

The physical explanation of this phenomenon is indeed quite simple. During the pulse, the atoms are focused onto the minima of the potential originating the initial periodic density pattern, with the same periodicity of the Bragg potential $d$. This situation is represented in Fig. \ref{fig:Focus}. After the pulse, the atoms evolve locally as free particles according to their instantaneous momenta. Then, the fringes decay because of disordering due to thermal motion, where the decay time associated to each Fourier component can be estimated as
\begin{equation}
v_{T}\tau^d_{n}\sim \frac{d}{n} \sim \frac{1}{nk}
\end{equation}
which is just the time that takes the particles, with typical velocity $v_T$, to travel the spatial period of the $n$-th Fourier harmonic.

\section{Quantum description}\label{sec:quantum}

We now consider a quantum description, using the second-quantization formalism to compute all the desired magnitudes. We first present a general review of the initial equilibrium configuration, where many of the results can be consulted in the standard literature of the field \cite{Pitaevskii2003,Pethick2008}. After that, following the techniques developed for the equilibrium study, we compute the dynamical evolution of the density. Throughout this chapter, when dealing with second-quantization operators, we write them with ``hats'' in order to distinguish them from their first-quantization equivalents.

\subsection{Equilibrium properties}\label{subsec:equilibriumproperties}

In equilibrium, the system is described by the density matrix
\begin{equation}\label{eq:equilibriumstate}
\hat{\rho}=\frac{e^{-\beta\hat{K}}}{Q},~Q=\text{Tr}\left(e^{-\beta \hat{K}}\right)
\end{equation}
where $\hat{K}=\hat{H}_0-\mu\hat{N}$ is the grand-canonical Hamiltonian and the operator $\hat{H}_0$ is the second-quantized version of the trap Hamiltonian (\ref{eq:equilibriumHam}). The equilibrium density can be computed from:
\begin{equation}\label{eq:equilibriumdensity}
n_0(\mathbf{x})=\braket{\hat{\Psi}^\dagger(\mathbf{x})\hat{\Psi}(\mathbf{x})}=\text{Tr}\left(\hat{\Psi}^\dagger(\mathbf{x},t)\hat{\Psi}(\mathbf{x},t)\hat{\rho}\right)
\end{equation}
where the stationary field operator is given by:
\begin{equation}\label{eq:equilibriumspfieldoperator}
\hat{\Psi}(\mathbf{x})=\sum_n \Psi_n(\mathbf{x})\hat{a}_n
\end{equation}
$\Psi_n(\mathbf{x})$ being the single-particle eigenfunctions of the (first-quantized) equilibrium Hamiltonian $H_0$, with eigenenergy $\varepsilon_n$, and $\hat{a}_n$ their associated second-quantized annihilation operators. Here, $n$ labels any possible quantum number of the eigenfunctions.

From the previous results, it is straightforward to prove that the density is given by:
\begin{equation}\label{eq:equilibriumstaticdensity}
n_0(\mathbf{x})=\sum_n |\Psi_n(\mathbf{x})|^2\frac{1}{e^{\beta(\varepsilon_n-\mu)}-1}
\end{equation}
which simply shows that the total density is just the sum of the Bose-Einstein occupation number of every state multiplied by its associated spatial probability density. Integrating over all the space, the total number of particles is:
\begin{equation}\label{eq:equilibriumN}
N=\int\mathrm{d}^3\mathbf{x}~n_0(\mathbf{x})=\sum_n \frac{1}{e^{\beta(\varepsilon_n-\mu)}-1}
\end{equation}
These are the usual results found in the textbooks for a Bose gas. However, for further calculations, it is useful to rewrite the previous identities in a more convenient way. Taking into account the identity
\begin{equation}\label{eq:harmonicserie1}
\frac{x}{1-x}=x\sum_{j=0}^{\infty}x^j=\sum_{j=1}^{\infty}x^j
\end{equation}
the density can be expressed as:
\begin{eqnarray}\label{eq:equilibriumtotaldensityBE}
\nonumber n_0(\mathbf{x})&=&\sum_{j=1}^{\infty}n_j(\mathbf{x})\\
n_j(\mathbf{x})&=&\eta^j\sum_n |\Psi_n(\mathbf{x})|^2e^{-j\beta\varepsilon_n}=\eta^jF(j\beta,\mathbf{x}),
\end{eqnarray}
$\eta=e^{\beta\mu}$ being the fugacity. The term $j=1$ in the previous sum corresponds to the Maxwell-Boltzmann approximation ($\eta\ll 1$) and the higher order terms arise from corrections of the actual Bose-Einstein occupation factor. Thus, we can obtain the total density by just computing the density in the Maxwell-Boltzmann approximation, i.e., by computing the function $F(\beta,\mathbf{x})$, which can be rewritten as:
\begin{equation}\label{eq:equilibriumMBfunctionF}
F(\beta,\mathbf{x})=\sum_n |\Psi_n(\mathbf{x})|^2e^{-\beta\varepsilon_n}=\braket{\mathbf{x}|e^{-\beta H_0}|\mathbf{x}},
\end{equation}
As in the classical description, the previous function factorizes since the total equilibrium Hamiltonian is separable:
\begin{equation}\label{eq:equilibriumdensityfactorization}
F(\beta,\mathbf{x})=F_x(\beta,x)F_y(\beta,y)F_z(\beta,z)=\prod_iF_i(\beta,x_i)
\end{equation}
with
\begin{equation}\label{eq:equilibrium1Dfactor}
F_i(\beta,x_i)=\braket{x_i|e^{-\beta H_{0i}}|x_i}
\end{equation}
$H_{0i}$ being the sub-Hamiltonian associated to the $i$-direction, see Eq. (\ref{eq:separability}). Hence, we can restrict ourselves to the corresponding one-dimensional problem in order to calculate the requested magnitudes. It is worth noting that the function $F(\beta,\mathbf{x})$ represents the diagonal elements of the single-particle propagator of $H_0$ for imaginary time $i\hbar\beta$, so the formalism developed here can also be used for computing the propagators.

\subsection{Local density approximation}\label{subsec:staLDA}

We first compute $F(\beta,\mathbf{x})$ in a local density approximation (LDA), which consists in considering the gas as locally uniform, with a space-dependent chemical potential $\mu(\mathbf{x})=\mu-V_{\rm {trap}}(\mathbf{x})$ that gives a space-dependent fugacity of the form $\eta(\mathbf{x})=\eta e^{-\beta V_{\rm {trap}}(\mathbf{x})}$ \cite{Pethick2008}.

The previous approximation amounts to:
\begin{equation}\label{eq:equilibriumLDAfactorization}
F(\beta,\mathbf{x})=e^{-\beta V_{\rm {trap}}(\mathbf{x})}F^0(\beta,\mathbf{x})
\end{equation}
where $F^0(\beta,\mathbf{x})$ is the function arising from setting $H_0=\mathbf{P}^2/2m$ in Eq. (\ref{eq:equilibriumMBfunctionF}):
\begin{equation}\label{eq:LDAtransversex}
F^0(\beta,\mathbf{x})=\int\mathrm{d}^3\mathbf{p}~|\braket{\mathbf{x}|\mathbf{p}}|^2e^{-\frac{\beta \mathbf{p}^2}{2m} }=\left(\frac{m}{2\pi\hbar^2\beta}\right)^{\frac{3}{2}}\equiv\frac{1}{\Lambda^3_T}~,
\end{equation}
$\Lambda_T$ being the thermal de Broglie wavelength and $\braket{\mathbf{x}|\mathbf{p}}$ the wave function of the plane wave with momentum $\mathbf{p}$
\begin{equation}\label{eq:planewavefunction}
\braket{\mathbf{x}|\mathbf{p}}=\frac{e^{i\frac{\mathbf{p}}{\hbar}\mathbf{x}}}{(2\pi\hbar)^{\frac{3}{2}}}
\end{equation}
Then, in the LDA,
\begin{equation}\label{eq:equilibriumLDAF}
F(\beta,\mathbf{x})=\frac{e^{-\beta V_{\rm {trap}}(\mathbf{x})}}{\Lambda^3_T}
\end{equation}
and the total density reads
\begin{equation}\label{eq:equilibriumLDADensity}
n_0(\mathbf{x})=\frac{g_{\frac{3}{2}}[\eta(\mathbf{x})]}{\Lambda^3_T},~g_{s}(z)\equiv\sum_{j=1}^{\infty}\frac{z^{j}}{j^{s}}
\end{equation}
which is the usual relation for a homogeneous ideal gas but replacing $\eta$ by its space-dependent version $\eta(\mathbf{x})$, as previously explained. Integrating, we find:
\begin{equation}\label{eq:equilibriumLDAN}
N=\frac{g_{3}(\eta)}{(\beta\hbar\tilde{\omega})^3}
\end{equation}
which allows us to write
\begin{equation}\label{eq:equilibriumLDADensityclassical}
n_0(\mathbf{x})=N\left(\frac{\beta m \tilde{\omega}^2}{2\pi}\right)^{\frac{3}{2}}\frac{g_{\frac{3}{2}}[\eta(\mathbf{x})]}{g_3(\eta)}
\end{equation}
that gives the classical expression for $\eta\ll 1$. In the opposite limit, a phase transition to a Bose-Einstein condensate occurs at $T$ such that $\eta=1$. As $g_{s}(1)=\zeta(s)$, with $\zeta(s)$ the Riemann zeta function, we find that the {\it ideal} critical temperature $T^{0}_c$ is given by:
\begin{equation}\label{eq:criticaltemperature}
k_BT^{0}_c=\left(\frac{N}{\zeta(3)}\right)^{\frac{1}{3}}\hbar\tilde{\omega}
\end{equation}

As explained later, the LDA corresponds to the {\it ideal} thermodynamic limit of an infinite number of particles.

\subsection{Harmonic oscillator}\label{subsec:staHO}

On the other hand, we note that for a HO-type Hamiltonian the previous magnitudes can be also computed analytically. Since the full Hamiltonian $H_0$ is separable, we have a collection of 1D HOs so we first review the general properties of the 1D HO Hamiltonian:
\begin{eqnarray}\label{eq:1DHO}
\nonumber H&=&\frac{P^2}{2m}+\frac{1}{2}m\omega^2X^2=\hbar\omega\left(a^{\dagger}a+\frac{1}{2}\right)\\
a&=&\frac{\frac{X}{l}+i\frac{P}{\hbar}l}{\sqrt{2}},~l=\sqrt{\frac{\hbar}{m\omega}}
\end{eqnarray}
In the previous expression, $X,P$ are the position and momentum operators and $l$ the harmonic oscillator length. The annihilation operator $a$ satisfies the well-known commutation relation $[a,a^{\dagger}]=1$. The eigenvalues and eigenstates of the HO Hamiltonian are the well-known number states:
\begin{eqnarray}\label{eq:numberstates}
H\ket{n}&=&\hbar\omega\left(n+\frac{1}{2}\right)\ket{n},~\ket{n}=\frac{(a^{\dagger})^n}{\sqrt{n!}}\ket{0},~n=0,1,2\ldots\\
\nonumber \braket{x|n}&\equiv&\frac{1}{\sqrt{2^nn!\sqrt{\pi}l}}H_n(\frac{x}{l})e^{-\frac{x^2}{2l^2}}
\end{eqnarray}
We also introduce the coherent states $\ket{\alpha}$, where $\alpha$ is a complex number, as they will be very useful for our calculations. They are the eigenstates of the annihilation operator $a\ket{\alpha}=\alpha\ket{\alpha}$,
\begin{equation}\label{eq:coherentstates}
\ket{\alpha}=e^{-\frac{|\alpha|^2}{2}}\sum_{n=0}^{\infty}\frac{\alpha^n}{\sqrt{n!}}\ket{n}=e^{\alpha a^{\dagger}-\alpha^*a}\ket{0}
\end{equation}
where the rightmost relation follows from Eq. (\ref{eq:formalsumexp}). It is straightforward to show from this expression that, for any complex number $z$,
\begin{equation}\label{eq:coherentexponential}
e^{za^{\dagger}a}\ket{\alpha}=e^{-\frac{(1-|e^{2z}|)|\alpha|^2}{2}}\ket{e^z\alpha}
\end{equation}
The spatial representation of the coherent states reads
\begin{eqnarray}\label{eq:coherentstatesposition}
\nonumber \braket{x|\alpha}&=&\frac{e^{-\frac{(x-\sqrt{2}\alpha_1l)^2}{2l^2}}}{\pi^{\frac{1}{4}}l^{\frac{1}{2}}}e^{i\sqrt{2}\alpha_2\frac{x}{l}}e^{-i\alpha_1\alpha_2}=\frac{e^{-\frac{x^2}{2l^2}}}{\pi^{\frac{1}{4}}l^{\frac{1}{2}}}e^{\sqrt{2}\alpha \frac{x}{l}}e^{-\alpha_1^2}e^{-i\alpha_1\alpha_2}\\
&=&\frac{e^{-\frac{x^2}{2l^2}}}{\pi^{\frac{1}{4}}l^{\frac{1}{2}}}e^{\sqrt{2}\alpha \frac{x}{l}}e^{-\frac{|\alpha|^2}{2}}e^{-\frac{\alpha^2}{2}}
\end{eqnarray}
where $\alpha_{1,2}$ are the real and imaginary parts of $\alpha$, $\alpha=\alpha_1+i\alpha_2$. The previous identity can be obtained by expanding $a$ in terms of the position and momentum operators $X,P$ in Eq. (\ref{eq:coherentstates}) or well by using the generating function of Eq. (\ref{eq:generatingfunction}) and the spatial representation of $\ket{n}$. It can also be proven, from Eq. (\ref{eq:coherentstates}), that the coherent states form an overcomplete set of states of the Hilbert state:
\begin{equation}\label{eq:coherentcomplete}
1=\frac{1}{\pi}\int\mathrm{d}^2\alpha~\ket{\alpha}\bra{\alpha},~\mathrm{d}^2\alpha=\mathrm{d}\alpha_1\mathrm{d}\alpha_2
\end{equation}
Another useful identity follows from Eq. (\ref{eq:coherentstatesposition})
\begin{equation}\label{eq:coherentwavemult}
e^{\gamma \frac{x}{l}}\braket{x|\alpha}=e^{\frac{\gamma^2}{4}}e^{\frac{|\gamma|^2}{4}}e^{\frac{(\alpha+\alpha^*)\gamma+(\gamma+\gamma^*)\alpha}{2\sqrt{2}}}\braket{x|\alpha+i\frac{\gamma}{\sqrt{2}}}~,
\end{equation}
satisfied for any complex number $\gamma$.

The above results allow us to rewrite in a more convenient way the operator $e^{-\beta H}$:
\begin{eqnarray}\label{eq:thermalcoherentoperator}
\nonumber e^{-\beta H}&=&e^{-\beta H}\frac{1}{\pi}\int\mathrm{d}^2\alpha~\ket{\alpha}\bra{\alpha}=\frac{e^{- \frac{\beta\hbar\omega}{2}}}{\pi}\int\mathrm{d}^2\alpha ~e^{-\frac{(1-e^{-2\beta\hbar\omega})|\alpha|^2}{2}}\ket{e^{-\beta\hbar\omega}\alpha}\bra{\alpha}\\
&=&\frac{e^{\frac{\beta\hbar\omega}{2}}}{\pi}\int\mathrm{d}^2\alpha~e^{-\sinh(\beta\hbar\omega)|\alpha|^2}\ket{e^{- \frac{\beta\hbar\omega}{2}}\alpha}\bra{e^{\frac{\beta\hbar\omega}{2}}\alpha}
\end{eqnarray}
We are now in a position to compute the equilibrium density since it is expressed in terms of the spatial diagonal matrix elements of $e^{-\beta H}$:
\begin{eqnarray}\label{eq:HOspaceexp}
\braket{x|e^{-\beta H}|x}&=&\frac{e^{\frac{\beta\hbar\omega}{2}}e^{-\frac{x^2}{l^2}}}{l\pi^{\frac{3}{2}}}\int\mathrm{d}^2\alpha~e^{-\frac{1}{2}\mathbf{\alpha}^T M \mathbf{\alpha}+\mathbf{v}^T\mathbf{\alpha}}\\
\nonumber M&=&\left[\begin{array}{cc}
4\cosh(\beta\hbar\omega)+2\sinh(\beta\hbar\omega) & -2i\sinh(\beta\hbar\omega)\\
-2i\sinh(\beta\hbar\omega) & 2\sinh(\beta\hbar\omega)
\end{array}\right]\\
\nonumber \mathbf{\alpha}&=&\left[\begin{array}{c}
\alpha_1\\
\alpha_2
\end{array}\right]\\
\nonumber \mathbf{v}&=&\sqrt{2}\frac{x}{l}\left(e^{-\frac{\beta\hbar\omega}{2}}\left[\begin{array}{c}
1\\
i
\end{array}\right]+e^{\frac{\beta\hbar\omega}{2}}\left[\begin{array}{c}
1\\
-i
\end{array}\right]\right)
\end{eqnarray}
We have introduced some vectorial magnitudes in order to simplify the notation. Note that the matrix $M$ is a symmetric complex matrix. If we write $M=A+iB$, with $A,B$ real symmetric matrices, the previous integral only converges for $A>0$. In that case, $M$ can be reduced to a diagonal matrix by the transformation $U^TMU=D$, with $U=OD_A^{-1/2}\tilde{O}D_A^{1/2}$, being $O$ the orthogonal matrix that diagonalizes $A$, $D_A=O^TAO$, and $\tilde{O}$ the orthogonal matrix that diagonalizes $D_A^{-1/2}O^TBOD_A^{-1/2}$. As $U$ is a real matrix and $|\det U|=1$, we arrive at the usual result for Gaussian integrals for definite positive real matrices:
\begin{equation}\label{eq:Gaussiancomplexintegral}
\int\mathrm{d}^2\alpha~e^{-\frac{1}{2}\mathbf{\alpha}^T M \mathbf{\alpha}+\mathbf{v}^T\mathbf{\alpha}}=\frac{2\pi}{\sqrt{\det M}} e^{\frac{1}{2}\mathbf{v}^T M^{-1} \mathbf{v}}
\end{equation}
The previous formula trivially generalizes to $n$ dimensions by replacing $2\pi$ by $(2\pi)^{n/2}$. Applying this result to Eq. (\ref{eq:HOspaceexp}), we find:
\begin{eqnarray}\label{eq:HOspaceexpdef}
\braket{x|e^{-\beta H}|x}=\sqrt{\frac{m\omega}{2\pi\hbar\sinh(\beta\hbar \omega)}}e^{-\tanh\left(\frac{\beta\hbar \omega}{2}\right)\frac{m\omega}{\hbar}~x^2}
\end{eqnarray}

Turning back to our physical problem, the last equation gives directly the values of $F_i$ in Eq. (\ref{eq:equilibrium1Dfactor}) by just changing $\omega$ by $\omega_i$:
\begin{eqnarray}\label{eq:equilibriumHOFactor}
F_i(\beta,x_i)=\sqrt{\frac{m\omega_i}{2\pi\hbar\sinh(\beta\hbar \omega_i)}}e^{-\tanh\left(\frac{\beta\hbar \omega_i}{2}\right)\frac{m\omega_i}{\hbar}~x_i^2}
\end{eqnarray}
From here, after integrating and performing some straightforward manipulations, the total number of particles can be expressed as:
\begin{equation}\label{eq:equilibriumHON}
N=\sum_{j=1}^{\infty}\frac{\eta^j}{8\sinh\left(j\frac{\beta\hbar \omega_x}{2}\right)\sinh\left(j\frac{\beta\hbar \omega_y}{2}\right)\sinh\left(j\frac{\beta\hbar \omega_z}{2}\right)}
\end{equation}
We note that, for the full HO Hamiltonian, the phase transition occurs when the chemical potential is equal to the zero-point energy
\begin{equation}\label{eq:HOphasetransition}
\mu=\varepsilon_0=\frac{3\hbar\omega_m}{2},~\omega_m=\frac{\omega_x+\omega_y+\omega_z}{3}~,
\end{equation}
$\omega_m$ being the mean frequency. Absorbing the zero-point energy in the definition of the chemical potential, $\bar{\eta}\equiv\eta e^{-\beta\varepsilon_0}$, the previous condition is equivalent to $\bar{\eta}=1$. After this redefinition, the expression for the total number of particles reads:
\begin{equation}\label{eq:equilibriumHONRedefined}
N=\sum_{j=1}^{\infty}\frac{\bar{\eta}^je^{j\beta\varepsilon_0}}{8\sinh\left(j\frac{\beta\hbar \omega_x}{2}\right)\sinh\left(j\frac{\beta\hbar \omega_y}{2}\right)\sinh\left(j\frac{\beta\hbar \omega_z}{2}\right)}
\end{equation}
In the limit $\beta\hbar\omega_m\ll1$ (which implies $\beta\hbar\omega_i\ll1$) we recover the result of Eq. (\ref{eq:equilibriumLDAN}). Moreover, we can also compute the first correction to the LDA, given by:
\begin{equation}\label{eq:equilibriumNfinitesized}
N=\frac{g_{3}(\bar{\eta})}{(\beta\hbar\tilde{\omega})^3}+\frac{3}{2}\frac{\omega_m}{\tilde{\omega}}\frac{g_2(\bar{\eta})}{(\beta\hbar\tilde{\omega})^2}
\end{equation}
This correction shifts the critical temperature by an amount $\delta T_c\equiv T_c-T^0_c$, with $T^0_c$ the ideal critical temperature obtained in the LDA, which is perturbatively obtained from the previous equation:
\begin{equation}\label{eq:criticalfinitesize}
\frac{\delta T_c}{T^0_c}=-\beta_c\hbar\omega_m\frac{\zeta(2)}{2\zeta(3)}=-\frac{\omega_m}{2\tilde{\omega}}\frac{\zeta(2)}{\zeta^{\frac{2}{3}}(3)}N^{-\frac{1}{3}}\thickapprox-0.73\frac{\omega_m}{\tilde{\omega}}N^{-\frac{1}{3}}
\end{equation}
The rightmost expression tells us that the corrections to the LDA arise from the actual finite number of particles. For sufficiently large number of particles $N$, however, these corrections are negligible. Therefore, the LDA corresponds to the thermodynamic limit $N\rightarrow \infty$. On the other hand, the previous relation implies that, if finite-size effects are small, $\beta_c\hbar\omega_m\ll1$ and then, as we are dealing with temperatures $T\gtrsim T_c$, $\beta\hbar\omega_m\ll1$. Thus, the LDA is valid whenever the previous finite-size correction is small. We discuss in more detail these conditions in Sec. \ref{subsec:LDAjust}.

Interestingly, the previous results can also be obtained directly from the sum of the series of Eq. (\ref{eq:equilibriumN})
\begin{equation}
N=\sum_{\mathbf{n}} \frac{1}{e^{\beta(\varepsilon_{\mathbf{n}}-\mu)}-1},~\varepsilon_{\mathbf{n}}=\varepsilon_{0}+\sum_in_i\hbar\omega_i
\end{equation}
where $\mathbf{n}=(n_x,n_y,n_z)$ are the quantum numbers of the oscillators in each direction. After absorbing the zero-point energy in the chemical potential, expanding the previous expression using Eq. (\ref{eq:harmonicserie1}) and performing first the summations over the $n_i$ quantum numbers, which are just the usual harmonic series, we arrive at Eq. (\ref{eq:equilibriumHON}).

\subsection{Dynamical evolution}\label{subsec:dynamicalevolution}

After reviewing the thermodynamic properties of the initial state of the trapped cloud, we return to our dynamical problem. Similar to Eq. (\ref{eq:equilibriumdensity}), the time-dependent density reads:
\begin{equation}
n(\mathbf{x},t)=\braket{\hat{\Psi}^\dagger(\mathbf{x},t)\hat{\Psi}(\mathbf{x},t)}=\text{Tr}\left(\hat{\Psi}^\dagger(\mathbf{x},t)\hat{\Psi}(\mathbf{x},t)\hat{\rho}\right)
\end{equation}
where $\hat{\Psi}(\mathbf{x},t)$ is now the time evolution of the field operator $\hat{\Psi}(\mathbf{x})$ of Eq. (\ref{eq:equilibriumspfieldoperator}) in the Heisenberg picture while $\hat{\rho}$ is the same equilibrium state of Eq. (\ref{eq:equilibriumstate}). As we are neglecting interactions, the time-evolution of the field operator is:
\begin{equation}\label{eq:spfieldoperator}
\hat{\Psi}(\mathbf{x},t)=\sum_n \Psi_n(\mathbf{x},t)\hat{a}_n
\end{equation}
with $\Psi_n(\mathbf{x},t)$ the time-evolution of the eigenfunction $\Psi_n(\mathbf{x})$ under the total first-quantization Hamiltonian
\begin{equation}\label{eq:firstquantizationhamiltonian}
H=H_0+V(z,t)
\end{equation}
with $V(z,t)$ being the potential of the perturbation of Eq. (\ref{eq:potentialswitch}). Following the same reasonings leading to Eq. (\ref{eq:equilibriumtotaldensityBE}), we can write the expression for the density as
\begin{eqnarray}\label{eq:totaldensityBE}
\nonumber n(\mathbf{x},t)&=&\sum_{j=1}^{\infty}n_j(\mathbf{x},t)\\
n_j(\mathbf{x},t)&=&\eta^jF(j\beta,\mathbf{x},t),
\end{eqnarray}
and $F(\beta,\mathbf{x},t)$ is now
\begin{equation}\label{eq:MBfunctionF}
F(\beta,\mathbf{x},t)=\sum_n |\Psi_n(\mathbf{x},t)|^2e^{-\beta\varepsilon_n}=\braket{\mathbf{x}|U(t)e^{-\beta H_0}U^{\dagger}(t)|\mathbf{x}},
\end{equation}
with $U(t)$ being the time-evolution operator associated to the Hamiltonian (\ref{eq:firstquantizationhamiltonian}). Following the same reasoning as in the classical calculation, only the $z$-coordinate has a non-trivial time evolution and then the function $F(\beta,\mathbf{x},t)$ factorizes in the form:
\begin{equation}\label{eq:densityfactorization}
F(\beta,\mathbf{x},t)=F_x(\beta,x)F_y(\beta,y)F_z(\beta,z,t)
\end{equation}
where the functions associated to the transverse coordinates are those of the equilibrium state and the dynamical part is:
\begin{equation}\label{eq:dynamicalzfactor}
F_z(\beta,z,t)=\braket{z|U_z(t)e^{-\beta H_{0z}}U_z^{\dagger}(t)|z}
\end{equation}
with $U_z(t)$ the time evolution operator in the longitudinal direction. For times $t>\tau$, $U_z(t)$ is given by:
\begin{eqnarray}\label{eq:zevolution}
U_z(t)&=&e^{-i\frac{H_{0z}}{\hbar}(t-\tau)}U_{\tau}\\
\label{eq:tauevolution}U_{\tau}&=&e^{-i\frac{H_{0z}}{\hbar}\tau+iu\cos(kz)}\\
\label{eq:impulseparameter}u&\equiv&\frac{V_0\tau}{\hbar}
\end{eqnarray}
For the computation of the function $F(\beta,\mathbf{x},t)$, in the same fashion as in the equilibrium state, we first consider the local density approximation and after that we perform the complete calculation using the full Hamiltonian.

\subsubsection{Local density approximation}\label{subsec:dynLDA}

As previously explained, the LDA consists in considering the gas as locally homogeneous with an space-dependent chemical potential. Following the same reasonings as in Sec. \ref{subsec:staLDA}, we take:
\begin{equation}\label{eq:LDAfactorization}
F(\beta,\mathbf{x},t)=e^{-\beta V_{\rm {trap}}(\mathbf{x})}F^0(\beta,\mathbf{x},t)
\end{equation}
where $F^0(\beta,\mathbf{x},t)$ is the result from taking $H_0=\mathbf{P}^2/2m$ in Eq. (\ref{eq:MBfunctionF}). For the transverse degrees of freedom we recover the equilibrium expressions:
\begin{equation}\label{eq:LDAtransversex}
F_i(\beta,x_i)=\frac{e^{-\frac{\beta}{2}mw^2_ix_i^2} }{\Lambda_T},~i=1,2
\end{equation}
while for the longitudinal part we get:
\begin{equation}\label{eq:LDAdynamical}
F_z(\beta,z,t)=e^{-\frac{\beta}{2}mw^2_zz^2}\int\mathrm{d}p~ e^{-\frac{\beta p^2}{2m}} |\braket{z|U_z(t)|p}|^2
\end{equation}
with $\braket{z|U_z(t)|p}$ the time evolution of the 1D plane wave $\braket{z|p}=\frac{e^{i\frac{p}{\hbar}z}}{(2\pi\hbar)^{\frac{1}{2}}}$

Now, in order to obtain explicit expressions for the density, we approximate the time evolution operator during the pulse, $U_{\tau}$, using a split-step approximation (as that used considered for the numerical integration of the GP equation considered in Chapter \ref{chapter:BHL}; see Sec. \ref{sec:numericalbhl}):
\begin{equation}\label{eq:impulsesplitstepapproximation}
U_{\tau}\simeq e^{-i\frac{H_{0z}}{\hbar}\frac{\tau}{2}}e^{iu\cos(kz)}e^{-i\frac{H_{0z}}{\hbar}\frac{\tau}{2}}
\end{equation}
Since the time evolution due to the rightmost term of Eq. (\ref{eq:impulsesplitstepapproximation}) is trivial as it commutes with $e^{-\beta H_{0z}}$, we take $U_z(t)$ in the following as:
\begin{equation}\label{eq:zevolutionapprox}
U_z(t)=e^{-i\frac{H_{0z}}{\hbar}\bar{t}}e^{iu\cos(kz)}
\end{equation}
The Fourier expansion of $e^{iu\cos(kz)}$ is
\begin{equation}\label{eq:Fourierseries}
e^{iu\cos(kz)}=\sum_{n=-\infty}^\infty i^nJ_n(u)e^{inz}
\end{equation}
which is obtained from the integral representation of the Bessel function $J_n(x)$:
\begin{equation}\label{eq:Besselintegral}
J_n(x)=\frac{1}{2\pi}\int_0^{2\pi}\mathrm{d}\varphi ~e^{ix\sin\varphi}e^{-in\varphi}
\end{equation}
The previous identity is proven by using the exponential series of $e^{ix\sin\varphi}$, the binomial expansion Eq. (\ref{eq:sinbinomial}) and the definition of $J_n(x)$, Eq. (\ref{eq:BesselDef}). Then, after making use of Eq. (\ref{eq:zevolutionapprox}) for the time operator, we find:
\begin{equation}\label{eq:planewavetimeevolution}
\braket{z|U_z(t)|p}=\frac{e^{i\frac{p}{\hbar}z}}{(2\pi\hbar)^{\frac{1}{2}}}e^{-i\frac{p^2}{2m\hbar}\bar{t}}\sum_{n=-\infty}^\infty i^nJ_n(u)e^{inkz}e^{-in\frac{pk}{m}\bar{t}}e^{-in^2\frac{\hbar k^2}{2m}\bar{t}}
\end{equation}
Inserting the previous relation in Eq. (\ref{eq:LDAdynamical}) yields:
\begin{equation}\label{eq:LDAz0}
F_z(\beta,z,t)=\frac{e^{-\frac{1}{2}\beta mw^2_zz^2}}{\Lambda_T}\sum_{n=-\infty}^\infty\sum_{n'=-\infty}^\infty i^{(n-n')}J_n(u)J_{n'}(u)e^{-i(n^2-n'^2)\frac{\hbar k^2}{2m}\bar{t}}e^{-(n-n')^2\frac{k^2}{2m\beta}\bar{t}^2}e^{i(n-n')kz}
\end{equation}
Now, we conveniently rearrange the previous sum as
\begin{equation}\label{eq:LDAz1}
F_z(\beta,z,t)=\frac{e^{-\frac{\beta}{2}mw^2_zz^2}}{\Lambda_T}\sum_{n=-\infty}^\infty i^n\left(\sum_{n'=-\infty}^\infty J_{n'}(u)J_{n'-n}(u)e^{-i(2nn'-n^2)\frac{\hbar k^2}{2m}\bar{t}}\right)e^{-n^2\frac{k^2}{2m\beta}\bar{t}^2}e^{inkz}
\end{equation}
The series between brackets can be computed analytically. For that purpose, we consider the following Fourier series:
\begin{eqnarray}
\nonumber e^{iA\sin(\varphi+\alpha)}&=&\sum_{n'=-\infty}^\infty J_{n'}(u)e^{in'\alpha}e^{in'\varphi}\\
e^{iA\sin(\varphi-\alpha)}e^{in\varphi}&=&\sum_{n'=-\infty}^\infty J_{n'-n}(u)e^{-i(n'-n)\alpha}e^{in'\varphi}
\end{eqnarray}
and then we find
\begin{equation}
\sum_{n'=-\infty}^\infty J_{n'}(u)J_{n'-n}(u)e^{-i\alpha(2n'-n)}=\frac{1}{2\pi}\int_0^{2\pi}\mathrm{d}\varphi ~e^{-2iA\sin\alpha\cos\varphi}e^{in\varphi}=(-i)^nJ_n(2u\sin\alpha)
\end{equation}
By setting $\alpha=n\frac{\hbar k^2}{2m}~\bar{t}$ in the previous identity, Eq. (\ref{eq:LDAz1}) is rewritten as:
\begin{equation}\label{eq:LDAz2}
F_z(\beta,z,t)=\frac{e^{-\frac{\beta}{2}mw^2_zz^2}}{\Lambda_T}\sum_{n=-\infty}^\infty J_n\left[2u\sin\left(n\frac{\hbar k^2}{2m}~\bar{t}\right)\right]e^{-n^2\frac{k^2}{2m\beta}\bar{t}^2}e^{inkz}
\end{equation}
Putting all together, we get:
\begin{equation}\label{eq:LDAF}
F(\beta,\mathbf{x},t)=\frac{e^{-\beta V_{\rm {trap}}(\mathbf{x})}}{\Lambda^3_T}\sum_{n=-\infty}^\infty J_n\left[2u\sin\left(n\frac{\hbar k^2}{2m}~\bar{t}\right)\right]e^{-n^2\frac{k^2}{2m\beta}\bar{t}^2}e^{inkz}
\end{equation}
This equation is quite similar to the classical result of Eq. (\ref{eq:MBfunctionG}) and presents the same decaying oscillations. In particular, in the Maxwell-Boltzmann approximation, the density is:
\begin{equation}\label{eq:quantumMBdensity}
n(\mathbf{x},t)\simeq n_1(\mathbf{x},t)=N\left(\frac{\beta m\tilde{\omega}^2}{2\pi}\right)^{\frac{3}{2}}e^{-\beta V_{\rm {trap}}(\mathbf{x})}\sum_{n=-\infty}^\infty J_n\left[2u\sin\left(n\frac{\hbar k^2}{2m}~\bar{t}\right)\right]e^{-n^2\frac{k^2}{2m\beta}\bar{t}^2}e^{inkz}
\end{equation}
where we have replaced the fugacity by its Maxwell-Boltzmann value [obtained by setting $g_3(\eta)=\eta$ in Eq. (\ref{eq:equilibriumLDAN})]. This expression is the same as Eq. (\ref{eq:boltzmanndensity}) except for the argument of the Bessel functions but, whenever
\begin{equation}\label{eq:quantumclassical}
\frac{\hbar k^2}{2m}~\bar{t}\ll1~,
\end{equation}
we can approximate the sine's argument to first order and then both expressions give exactly the same result. The condition (\ref{eq:quantumclassical}) can be simply interpreted as that the total phase acquired due to the recoil energy must be small. Remarkably, the physics of $\hbar$ only enters through the corrections to the first order approximation of the sine. As we are dealing with times such that $\bar{t}\sim \tau^d$, we find that the previous condition reads
\begin{equation}\label{eq:quantumclassicalphoton}
\frac{\hbar k^2}{2m}~\tau^d\sim \frac{\hbar k}{p_T}\ll1~,
\end{equation}
which means that the momentum of the photon must be much smaller than the thermal momentum of the particles. We will return later to the relation between classical and quantum physics in Sec. \ref{subsec:QuantumClassical}.

In the opposite limit, $T$ close to $T_c$, $\eta\lesssim 1$ and then higher corrections to the Maxwell-Boltzmann density $n_1$ appear in Eq. (\ref{eq:totaldensityBE}) because of the accumulation of particles in the lower energy states, characteristic of Bose statistics. Specifically, the total expression for the Fourier transform of the density reads:
\begin{equation}\label{eq:Fouriercomponentquantum}
n_k(t)=NJ_1\left[2u\sin\left(\frac{\hbar k^2}{2m}\bar{t}\right)\right]\frac{\sum_{j=1}^\infty\frac{\eta^j e^{-\frac{\bar{t}^2}{j (\tau^d)^2}}}{j^3}}{g_3(\eta)}
\end{equation}
where we have taken into account the LDA result (\ref{eq:equilibriumLDAN}). Interestingly, the higher-order corrections decay more slowly than the first-order Maxwell-Boltzmann term.

\subsubsection{Harmonic oscillator}\label{subsec:dynHO}

We now address the complete problem by using the full Hamiltonian of the trap $H_0$. Once more, when computing $F(\beta,\mathbf{x},t)$, the transverse part is the same as that of the equilibrium state, Eq. (\ref{eq:equilibriumHOFactor}), while the dynamical part, corresponding to the $z$ direction, is [see Eq. (\ref{eq:thermalcoherentoperator})]
\begin{eqnarray}\label{eq:coherentdyn}
F_z(\beta,z,t)&=&\frac{e^{\frac{\beta\hbar\omega}{2}}}{\pi}\int\mathrm{d}^2\alpha~e^{-\sinh(\beta\hbar\omega)|\alpha|^2}\bra{z}U_z(t)\ket{e^{- \frac{\beta\hbar\omega}{2}}\alpha}\bra{e^{\frac{\beta\hbar\omega}{2}}\alpha}U_z^{\dagger}(t)\ket{z}~ ~~~~~~~  ~
\end{eqnarray}
Using the same split approximation for the time evolution during the pulse as in the previous subsection, which results in the expression for $U_z(t)$ of Eq. (\ref{eq:zevolutionapprox}), and taking into account the properties (\ref{eq:coherentexponential}), (\ref{eq:Fourierseries}) and (\ref{eq:coherentwavemult}) of coherent states, we can write
\begin{eqnarray}\label{eq:coherenttimeevolution}
&~&\bra{z}U_z(t)\ket{\alpha}=e^{-i\frac{\omega_z}{2}\bar{t}}\sum_{n=-\infty}^\infty i^nJ_n(u)e^{in\alpha_1\frac{kl_z}{\sqrt{2}}}\braket{z|e^{-i\omega_z\bar{t}}\left(\alpha+in\frac{kl_z}{\sqrt{2}}\right)}\\
\nonumber &~&=e^{-i\frac{\omega_z}{2}\bar{t}}\sum_{n=-\infty}^\infty i^nJ_n(u)e^{-n^2\frac{k^2l^2_z}{4}\left(1-e^{-i2\omega_z\bar{t}}\right)}e^{-\sqrt{2}\alpha e^{-i\omega_z\bar{t}} nkl_z\sin(\omega_zt)}e^{inke^{-i\omega_z\bar{t}}z}\braket{z|\alpha e^{-i\omega_z\bar{t}}}
\end{eqnarray}
Evaluating the previous expression for $\alpha e^{\pm\frac{\beta\hbar\omega}{2}}$, making the change of variable $\alpha e^{-i\omega_z\bar{t}}\rightarrow\alpha$ (which is just a rotation in the complex plane) and performing the resulting Gaussian integral in Eq. (\ref{eq:coherentdyn}), we finally arrive at:
\begin{eqnarray}\label{eq:HOdynamical}
\nonumber &~&F_z(\beta,z,t)=\sqrt{\frac{m\omega_z}{2\pi\hbar\sinh(\beta\hbar \omega_z)}}e^{-\tanh\left(\frac{\beta\hbar \omega_z}{2}\right)\frac{m\omega_z}{\hbar}~z^2}\\ \nonumber &~&\sum_{n=-\infty}^\infty\sum_{n'=-\infty}^\infty i^{(n-n')}J_n(u)J_{n'}(u)e^{\tanh\left(\frac{\beta\hbar \omega_z}{2}\right)(n+n')kz\sin(\omega_z\bar{t})}e^{-(n^2+n'^2)\tanh\left(\frac{\beta\hbar \omega_z}{2}\right)\frac{\hbar k^2}{2m\omega_z}\sin^2(\omega_z\bar{t})}\\
&~&e^{-i(n^2-n'^2)\frac{\hbar k^2}{2m\omega_z}\sin(\omega_z\bar{t})\cos(\omega_z\bar{t})}e^{-(n-n')^2\frac{k^2}{2m\beta \omega^2_z}\sin^2(\omega_z\bar{t})}e^{i(n-n')kz\cos(\omega_z\bar{t})}
\end{eqnarray}
We see that this equation reduces to the LDA expression (\ref{eq:LDAz2}) whenever $\beta\hbar\omega_z\ll1$ and $\omega_z\bar{t}\ll1$. However, for sufficiently high times, $\omega_zt\sim 2\pi$, the density oscillates periodically in this non-interacting picture as it is characteristic of a harmonic oscillator. In the next section, we discuss in detail the physical meaning of these conditions.

\section{Discussion of the approximations}\label{sec:approxs}

In the theoretical calculations of the previous section, several approximations have been made in order to get analytical results. We discuss here in detail the validity and physical meaning of the previous conditions and with usual values of the magnitudes.

As explained in Sec. \ref{sec:NumericaLOL}, the range of variation of the confinement frequencies is $\omega_{i}\sim2\pi\times 10-10^3$ Hz, which gives typical HO lengths $l_i\sim 10^{-6}$ m. In respect to the lattice period, usual values are $d\lesssim1~\mu\text{m}$ and then $k\sim 10^6$ m. The clouds are formed by a sufficiently large number of atoms, $N\sim 10^4-10^7$, for the finite-size effects to be small. The {\it ideal} critical temperature reads in terms of these magnitudes as
\begin{equation}\label{eq:estcritical}
T^0_c=4.51\times10^{-9}\times\left(\frac{N}{10^6}\right)^{\frac{1}{3}}\times\left(\frac{\tilde{\omega}[\text{Hz}]}{2\pi}\right)~\text{K}
\end{equation}
which also provides an estimation of the values for the temperature $T$ as we are dealing with $T\gtrsim T_c$. From the considerations of Sec. \ref{subsec:staLDA}, the expression for the de Broglie thermal wavelength can be written as:
\begin{equation}\label{eq:estdebroglie}
\Lambda_T=(2\pi\zeta^{\frac{1}{3}}(3))^{\frac{1}{2}}\sqrt{\frac{T^0_c}{T}}\frac{\tilde{l}}{N^{\frac{1}{6}}}=2.585\times\sqrt{\frac{T^0_c}{T}}\times\frac{\tilde{l}}{N^{\frac{1}{6}}},~\tilde{l}=\sqrt{\frac{\hbar}{m\tilde{\omega}}}
\end{equation}
with $\tilde{l}$ being the HO length associated to $\tilde{\omega}$.

On the other hand, the decay time satisfies
\begin{equation}\label{eq:estdecay}
\tau^d=3.81\times10^{-4}\times\frac{10^6}{k[\text{m}]} \times\sqrt{\frac{m[\text{kg}]}{10^{-25}}}\times \sqrt{\frac{10^{-7}}{T[\text{K}]}}~\text{s}\sim 10^{-4}~\text{s}
\end{equation}
In principle, there are no tight bounds for the values of the amplitude and duration of the pulse $U,\tau$. Bearing these values in mind, we analyze the approximations used in this work.

\subsection{Local density approximation}\label{subsec:LDAjust}

We first study the validity of the LDA. For the equilibrium state, there is a {\it static} LDA, valid whenever
\begin{equation}\label{eq:LDAstacond}
\beta\hbar\omega_m\ll 1~,
\end{equation}
This relation is fulfilled for $T\gtrsim T_c$ if the relative finite-size shift of the critical temperature is small; see Eq. (\ref{eq:criticalfinitesize}) and accompanying discussion. The previous condition can be rewritten as
\begin{equation}\label{eq:LDAmeaning}
\beta\hbar\omega_m=\frac{\frac{\beta\hbar^2}{m}}{\frac{\hbar}{m\omega_m}}\sim\frac{\Lambda^2_T}{l^2_m}\ll1
\end{equation}
with $l_m$ the HO length associated to the mean frequency. This relation reveals more neatly the physical meaning of the LDA: the thermal de Broglie wavelength must be much smaller than the length scales of the trapping potential. It can also be rewritten as $\beta\hbar\omega_i=\beta m\omega_i^2l_i^2\ll1$ which just means that the size of the thermal cloud is much greater than the HO length in each direction. These conditions imply that the gas can be regarded as locally homogeneous. We can try to justify this last statement in a more rigorous way. Let's consider the equilibrium density as given by Eq. (\ref{eq:equilibriumtotaldensityBE}) and in particular the expression for the function $F(\beta,\mathbf{x})$ of Eq. (\ref{eq:equilibriumMBfunctionF}). The total Hamiltonian is the sum of the free Hamiltonian $\mathbf{P}^2/2m$ and the potential of the trap $V_{\rm {trap}}(\mathbf{x})$, so the exponent of $e^{-\beta H_0}$ is the sum of two terms. For two operators, $A,B$, it can be proven the following relation:
\begin{equation}\label{eq:formalsumexp}
e^{A+B}=e^Ae^Be^{-\frac{[A,B]}{2}}e^{R(A,B)}
\end{equation}
where the function $R(A,B)$ contains higher order commutators of the operators $A,B$. Following the basis of the split-step method, we ``split'' the exponential as
\begin{equation}\label{eq:formalsplitstep}
e^{A+B}=e^{\frac{A+B}{2}}e^{\frac{A+B}{2}}\simeq e^{\frac{A}{2}}e^Be^{\frac{A}{2}}
\end{equation}
with third-order corrections $O([A,[A,B]],[B,[A,B]])$. The point is that if they are small, the previous approximation is valid. Formally, we define the function:
\begin{equation}
G(t)\equiv e^{-\frac{A}{2}t}e^{(A+B)t}e^{-\frac{A}{2}t}
\end{equation}
and inverting the relation we obtain
\begin{equation}
e^{(A+B)t}=e^{\frac{A}{2}t}G(t)e^{\frac{A}{2}t}
\end{equation}
which for $t=1$ gives the desired exponential. Using standard results, $G(t)$ can be written as
\begin{equation}\label{eq:formalmathsplit}
G(t)=e^{Bt+R(t)},~R(t)=\frac{[A,[A,B]]}{24}t^3+\frac{[B,[A,B]]}{12}t^3+\ldots
\end{equation}
where the first corrections to $R(t)$ are of fourth-order in the operators $A,B$. Thus, whenever $R(1)$ is small, the approximation (\ref{eq:formalsplitstep}) is valid. We note the large values of the denominators in the expression of $R(1)$, which make stronger the previous approximation. Going back to our problem, we consider the 1D HO Hamiltonian of Eq. (\ref{eq:1DHO}) and we find that the third-order corrections satisfy:
\begin{eqnarray}\label{eq:3rdordercommutators}
\nonumber [A,[A,B]]&=&-(\beta\hbar\omega)^2\frac{\beta m\omega^2X^2}{4}\sim (\beta\hbar\omega)^2\\
\nonumber [B,[A,B]]&=&-(\beta\hbar\omega)^2 \frac{\beta P^2}{2m}\sim (\beta\hbar\omega)^2\\
A&=&-\frac{1}{2}\beta m\omega^2X^2,~B=-\frac{\beta P^2}{2m}
\end{eqnarray}
Thus, whenever $\beta\hbar\omega\ll1$,
\begin{equation}\label{eq:1DLDAstatic}
e^{-\beta H}\simeq e^{-\frac{1}{4}\beta m\omega^2X^2}e^{-\frac{\beta P^2}{2m}}e^{-\frac{1}{4}\beta m\omega^2X^2}
\end{equation}
with corrections $O\left([\beta\hbar\omega]^2\right)$. Extending this reasoning to the HO in each direction of the trap, we conclude that, whenever $\beta\hbar\omega_i\ll1$,
\begin{equation}\label{eq:LDAstatic}
e^{-\beta H_0}\simeq e^{-\frac{\beta V_{\rm {trap}}(\mathbf{x})}{2}}e^{-\frac{\beta\mathbf{P}^2}{2m} }e^{-\frac{\beta V_{\rm {trap}}(\mathbf{x})}{2}}
\end{equation}
and then, relation (\ref{eq:equilibriumLDAfactorization}) is fulfilled, which in the end amounts to the mentioned ``trick'' of replacing $\eta$ by $\eta(\mathbf{x})=\eta e^{-\beta V_{\rm {trap}}(\mathbf{x})}$. We note that this discussion only appears when considering the quantum description, since in the classical regime the Boltzmann function already factorizes in the form of the LDA.

In addition to the previous {\it static} LDA, in our problem there is also a {\it dynamical} LDA for the time-dependent computation and its condition of validity is
\begin{equation}\label{eq:dynamicLDA}
\omega_zt\ll 1
\end{equation}
appearing both in the classical and quantum regimes, see discussions after Eqs. (\ref{eq:zclassho}) and (\ref{eq:HOdynamical}). In fact, when one performs a more rigorous dynamical calculation within the LDA using Eq. (\ref{eq:LDAstatic}) from the beginning for the computation of $F(\beta,\mathbf{x},t)$ instead of directly Eq. (\ref{eq:LDAfactorization}), one arrives at the same final result (\ref{eq:LDAF}) but with corrections in $\omega_z\bar{t}$.

Condition (\ref{eq:dynamicLDA}) just tells us that the LDA description is only valid for times $t$ much lower than the period of harmonic oscillation in the $z$ direction $2\pi/\omega_z$. Even when the {\it static} LDA is valid, for times $\omega_zt \gtrsim 1$ the particles are able to ``see'' the confinement of the trap and then we can no longer regard the gas as locally homogeneous. Indeed, the condition for the {\it dynamic} LDA is the same as that of {\it static} LDA when replacing $\beta$ by $t/\hbar$, in the same spirit of the discussion after Eq. (\ref{eq:equilibrium1Dfactor}); they both represent the validity of using the splitting approximation (\ref{eq:formalsplitstep}).

In order to evaluate $\omega_zt$, as we focus on the observation of the thermal decay of the Fourier components, we restrict to times $t$ such that $(t-\tau)\sim \tau^d$. Thus, Eq. (\ref{eq:dynamicLDA}) is fulfilled whenever:
\begin{equation}\label{eq:dynamicLDAtimescales}
\omega_z\tau\ll 1,~\omega_z\tau^d\ll1
\end{equation}
and the latter condition reads
\begin{equation}
\omega_z\tau^d\sim \sqrt{\frac{\beta m \omega_z^2}{k^2}}=\sqrt{\frac{\beta \hbar \omega_z}{(kl_z)^2}}\ll1
\end{equation}
The physical meaning of this relation is that the size of the cloud must be much larger than the spatial period of the pulse (note that this condition appeared when computing the Fourier transform of the density at the end of Sec \ref{sec:classical}). Since $kl_z\sim 1$, $\omega_z\tau^d\sim \sqrt{\beta \hbar \omega_z}$, and thus $\omega_z\tau^d\ll 1$ reduces to the condition of validity for the {\it static} LDA in the $z$-direction. As a consequence, the {\it dynamic} LDA only requires $\omega_z\tau\ll 1$ provided that the {\it static} LDA is valid.

Although we have focused on the case of harmonic trapping for practical purposes, the analysis here presented for the validity of the LDA can be extended to arbitrary trap potentials.

\subsection{Role of interactions}\label{subsec:interactingrole}

So far, we have neglected interaction between particles. Now we take them into account by using the same contact pseudo-potential considered in the BH analogs; see Eq. (\ref{eq:2ndHamiltonian}) and discussion below. First, we consider interactions in the equilibrium state. In the classical regime, interactions in dilute gases add a collisional term at the r.h.s of the Boltzmann equation (\ref{eq:boltzmanneq}):
\begin{equation}\label{eq:boltzmanneqcol}
\frac{\partial f}{\partial t}+\frac{\mathbf{p}}{m}\nabla f+\mathbf{F}_{ext} \nabla_{p} f=\left(\frac{\partial f}{\partial t}\right)_{col}
\end{equation}
However, it can be shown that $\left(\frac{\partial f}{\partial t}\right)_{col}$ vanishes when $f\propto e^{-\frac{\beta\mathbf{p}^2}{2m}}$ and then collisions do not modify the initial thermal Boltzmann distribution \cite{Huang2001}.

On the other hand, in the quantum problem, we can estimate the magnitude of interactions as $g_{3D}n_0(0)$, where $n_0(0)$ is the density at the center of the trap, as given by Eq. (\ref{eq:equilibriumLDADensityclassical}). Comparing with the thermal energy $k_BT$
\begin{equation}
\frac{g_{3D}n_0(0)}{k_BT}\sim \frac{a_{s}}{\Lambda_T}
\end{equation}
and considering that we are always dealing with temperatures of the order $T^0_c$, we find that interactions represent a small correction whenever
\begin{equation}\label{eq:estinteractions}
N^{\frac{1}{6}}\frac{a_{s}}{\tilde{l}}\ll1
\end{equation}
Quantitatively, interactions can be taken into account to lowest order using a mean-field Hartree-Fock-Bogoliubov (HFB) approximation \cite{Griffin1996,Giorgini1996}. In that case, the trap Hamiltonian $H_0$ is replaced by an effective Hamiltonian of the form:
\begin{equation}\label{eq:effinteractingHamiltonian}
H_{eff}=H_0+V_{HFB}(\mathbf{x}),~V_{HFB}(\mathbf{x})=2g_{3D}n_0(\mathbf{x})
\end{equation}
As interactions represent a small correction, we can work at first order in $g_{3D}$ and use a non-self-consistent HFB approximation by setting $n_0(\mathbf{x})=n^{(0)}_0(\mathbf{x})$, $n^{(0)}_0$ being the non-interacting equilibrium density. Moreover, as the length scale of the effective mean-field potential is that of the density, we can treat $V_{HFB}$ also in a LDA, which amounts to replace $V_{\rm {trap}}(\mathbf{x})$ in Eq. (\ref{eq:LDAstatic}) by $V_{eff}(\mathbf{x})=V_{\rm {trap}}(\mathbf{x})+V_{HFB}(\mathbf{x})$. Due to this effective potential, the chemical potential suffers a shift $V_{HFB}(0)$ at the phase transition, similar to the case of finite size effects [see Eq. (\ref{eq:HOphasetransition}) and ensuing discussion]. Indeed, taking into account both effects, we define $\bar{\eta}\equiv\eta e^{-\beta\varepsilon_0}e^{-\beta V_{HFB}(0)}$ and, considering them as small corrections, we find that Eq. (\ref{eq:equilibriumNfinitesized}) is extended to
\begin{eqnarray}\label{eq:HFBfinitesizeN}
\nonumber N&=&\frac{g_{3}(\bar{\eta})}{(\beta\hbar\tilde{\omega})^3}+\frac{3}{2}\frac{\omega_m}{\tilde{\omega}}\frac{g_2(\bar{\eta})}{(\beta\hbar\tilde{\omega})^2}+\frac{4a_s}{\Lambda_T}\frac{\Phi(\bar{\eta})}{(\beta\hbar\tilde{\omega})^3}\\
\Phi(z)&\equiv&g_{\frac{3}{2}}(z)g_{2}(z)-f(z),~f(z)=\sum_{n=1}^{\infty}\sum_{m=1}^{\infty}\frac{z^{n+m}}{n^{\frac{3}{2}}n^{'\frac{1}{2}}(n+n')^{\frac{3}{2}}}
\end{eqnarray}
In particular, the total shift to the critical temperature can be obtained from the previous expression in a similar way to Eq. (\ref{eq:criticalfinitesize}). Numerically, we find that $f(1)\thickapprox1.208$ and then, $\Phi(1)=\zeta(\frac{3}{2})\zeta(2)-f(1)\thickapprox3.805$, hence
\begin{equation}\label{eq:criticalfinitesizeinteraction}
\frac{\delta T_c}{T^0_c}=-\frac{\omega_m}{2\tilde{\omega}}\frac{\zeta(2)}{\zeta^{\frac{2}{3}}(3)}N^{-\frac{1}{3}}-\frac{4\Phi(1)}{3\sqrt{2\pi}\zeta^{\frac{7}{6}}(3)}N^{\frac{1}{6}}\frac{a_{s}}{\tilde{l}}\thickapprox-0.73\frac{\omega_m}{\tilde{\omega}}N^{-\frac{1}{3}}-1.32N^{\frac{1}{6}}\frac{a_{s}}{\tilde{l}}
\end{equation}
From the rightmost term, we conclude that if the relative shift of the critical temperature due to interactions is small, Eq. (\ref{eq:estinteractions}) is fulfilled. This mean-field scheme is only accurate to first order in the scattering length. Higher-order corrections in the scattering length arise from critical fluctuations \cite{Arnold2001}.

We switch now to the dynamical problem. In the classical context, the typical rate of collisions can be estimated as
\begin{equation}\label{eq:classcol}
n(0)v_T\sigma_0\sim \frac{1}{\tau^{col}}
\end{equation}
with $n(0)$ the density at the center of the trap, $n(0)\sim N(\beta m \tilde{\omega})^{3/2}$ and $\sigma_0=8\pi a^2_s$ the $s$-wave cross section for bosonic atoms \cite{Pethick2008}. Thus, for temperatures $T\gtrsim T^0_c$, we can neglect interactions during the time evolution if the considered times are much lower than $\tau^{col}$. Similar to the discussion leading to Eq. \ref{eq:dynamicLDAtimescales}, this is achieved whenever $\tau,\tau^d\ll \tau^{col}$. The corresponding ratios read:
\begin{eqnarray}\label{eq:collisionlessregime}
\frac{\tau}{\tau^{col}}&\sim& \left(\frac{N^{\frac{1}{3}}a_s}{\tilde{l}}\right)^2\tilde{\omega}\tau\ll 1\\
\nonumber \frac{\tau^d}{\tau^{col}}&\sim& \frac{a^2_s}{\Lambda^2_T}\frac{d}{\Lambda_T}\ll 1
\end{eqnarray}
It is worth noting that the collisional mean life $\tau^{col}$ indeed {\it increases} with temperature, since $\tau^{col}\sim (n(0)v_T\sigma_0)^{-1}\sim T$, in contrast to the thermal decay time $\tau^d$, that decreases as $T^{-1/2}$.

In the quantum problem, we can use a time-dependent HFB approximation to describe the dynamics, where the mean-field potential of the effective Hamiltonian of Eq. (\ref{eq:effinteractingHamiltonian}) is now $V_{HFB}(\mathbf{x},t)=2g_{3D}n(\mathbf{x},t)$. Once more, a non-self-consistent scheme is valid up to first order in interactions and then we set $n(\mathbf{x},t)=n^{(0)}(\mathbf{x},t)$, $n^{(0)}(\mathbf{x},t)$ being the non-interacting time-dependent density computed in Sec. \ref{subsec:dynamicalevolution}. In fact, from Eqs. (\ref{eq:LDAF}),(\ref{eq:HOdynamical}), we see that $n^{(0)}(\mathbf{x},t)=n^{(0)}_0(\mathbf{x})[1+\delta(\mathbf{x},t)]$ where the dominant terms in the relative corrections to the equilibrium density are the first Fourier components. As they go like $\delta(\mathbf{x},t)\propto J_1\left[2u\sin\left(\frac{\hbar k^2\bar{t}}{2m}\right)\right]$, if we assume that the value of the Bessel function is small, then these relative corrections are small and, as the interacting mean-field potential itself is in turn a perturbative correction, we conclude that to leading order $V_{HFB}(\mathbf{x},t)\simeq V_{HFB}(\mathbf{x})$ which means that the interactions of the particles with the cloud amount to a static trapping potential in this mean-field picture. Besides, if the dynamic LDA condition (\ref{eq:dynamicLDA}) is satisfied, the trapping potential does not play any dynamic role as explained in the previous section so the effect of interactions reduces, also in the dynamical problem, to change $V_{\rm {trap}}(\mathbf{x})$ by $V_{eff}(\mathbf{x})$. Mean-field approaches do not give rise to damping processes \cite{Pitaevskii2003} but, for temperatures $T\gtrsim T^0_c$, the rate of collisional relaxation can be also estimated through Eq. (\ref{eq:classcol}) \cite{Pethick2008}. Thus, as long as the conditions of Eq. (\ref{eq:collisionlessregime}) are satisfied, interactions do not play a significant role in our problem.

\subsection{Pulse evolution}\label{subsec:LTA}

For the time evolution during the pulse, we have also made some approximations in order to obtain analytical results. In the classical problem, we used a second order Runge-Kutta (RK2) approximation [see Eq. (\ref{eq:RK2})] to integrate the equation for $p_z$ during the pulse instead of the exact expression in terms of elliptic functions. The RK2 approximation is extracted from the associated Taylor series
\begin{eqnarray}\label{eq:classicalcorrections}
\nonumber z\left(\frac{\tau}{2}\right)&=&z(\tau)-\frac{p_z(\tau)}{m}\frac{\tau}{2}-\frac{V_0k}{m}\sin\left[k z(\tau)\right]\frac{\tau^2}{8}+O(\tau^3)\\
p_{z0}&=&p_z(\tau)-\dot{p}_z\left(\frac{\tau}{2}\right)\tau-\frac{\dddot{p}_z\left(\frac{\tau}{2}\right)}{24}\tau^3+O(\tau^5)
\end{eqnarray}
In order to quantify the corrections to the adopted Runge-Kutta scheme, we define the dimensionless parameters:
\begin{eqnarray}\label{eq:perturbativeparameter}
\nonumber \lambda &\equiv&\frac{V_0k^2\tau^2}{m}=u\frac{\hbar k^2\tau}{m}\\
\theta &\equiv&\frac{p_Tk\tau}{m}=2\frac{\tau}{\tau^d}
\end{eqnarray}
$\lambda$ represents the ratio between the distance traveled by the particles due to the momentum proportionated by the force, $V_0k\tau$, and the lattice period; $\theta$ is the same but rather considering the thermal momentum $p_T$. Taking into account that $p_z\sim p_T$, the first derivative and the third-order corrections of the momentum satisfy:
\begin{eqnarray}\label{eq:RK2corrections}
\nonumber \dot{p}_z\left(\frac{\tau}{2}\right)\tau&=&-V_0k\tau\sin\left[k z\left(\frac{\tau}{2}\right)\right]\\
kz\left(\frac{\tau}{2}\right)&=&k\left[z(\tau)-\frac{p_z(\tau)}{m}\frac{\tau}{2}\right]+O(\lambda)\\
\frac{\dddot{p}_z\left(\frac{\tau}{2}\right)}{24}\tau^3&\sim& \frac{V_0k\tau}{24}[O(\lambda)+O(\theta^2)]
\end{eqnarray}
Gathering all the pieces, we arrive at:
\begin{equation}\label{eq:classcorrdimensionless}
p_{z0}\simeq p_z(\tau)+V_0k\tau\left[\sin\left(kz(\tau)-\frac{p_z(\tau)k\tau}{2m}\right)+\frac{O(\lambda)+O(\theta^2)}{24}\right]
\end{equation}
We note the important role played by the $24$ of the denominator since, even for $\theta,\lambda\sim 1$, the first corrections to the RK2 approximation can be still regarded as small corrections.

In the quantum case, we perform an analog approximation to the RK2 method used in the classical problem for the time evolution during the pulse. Specifically, we consider the splitting of Eq. (\ref{eq:impulsesplitstepapproximation}) which is precisely the same split-step method analyzed in Sec. \ref{subsec:LDAjust}. Following the same path between Eqs. (\ref{eq:formalsumexp})-(\ref{eq:formalmathsplit}), we write:
\begin{equation}\label{eq:formaltimesplitstep}
U_{\tau}=e^{-i\frac{H_{0z}}{\hbar}\tau+iu\cos(kz)}=e^{-i\frac{H_{0z}}{\hbar}\frac{\tau}{2}}G_{\tau}e^{-i\frac{H_{0z}}{\hbar}\frac{\tau}{2}}
\end{equation}
and now
\begin{equation}\label{eq:formaltimekernel}
G_{\tau}=e^{iu\cos(k z)+R_{\tau}},~R_{\tau}=\left(\frac{[H_{0z},[H_{0z},V(z)]]}{24}+\frac{[V(z),[H_{0z},V(z)]]}{12}\right)\left(\frac{\tau}{\hbar}\right)^3+O(\tau^4)
\end{equation}
with $V(z)=-V_0\cos(k z)$ the potential of the Bragg pulse. In the same fashion of Eq. (\ref{eq:3rdordercommutators}), we compute the third-order commutators
\begin{eqnarray}\label{eq:3rdordertemporalcommutators}
\left[H_{0z},[H_{0z},V(z)]\right]\left(\frac{\tau}{\hbar}\right)^3 & \sim & u\theta^2\left[1+O\left(\frac{\hbar k}{p_T}\right)\right]\\
\nonumber \left[V(z),[H_{0z},V(z)]\right]\left(\frac{\tau}{\hbar}\right)^3 & \sim & u\lambda
\end{eqnarray}
where we have taken into account that the typical momentum is $P_z\sim p_T$ and we have neglected the contributions from the trapping potential to $[H_{0z},[H_{0z},V(z)]]$, since they give rise to terms $O(\omega_z^2\tau^2)$. We also have supposed that $\hbar k\lesssim p_T$ (see discussion at the end of the next subsection). Thus, we have:
\begin{equation}\label{eq:formaltimekernel}
G_{\tau}\simeq e^{iu\left[\cos(k z)+\frac{O(\lambda)+O(\theta^2)}{24}\right]}\simeq e^{iu\cos(kz)}
\end{equation}
which is a similar result to Eq. (\ref{eq:classcorrdimensionless}). The physical meaning of the previous approximation is further studied using time-dependent perturbation theory in Sec. \ref{sec:generalcase}.

We note that there is a strong analogy between the classical (RK2 method) and quantum (split-step) approximations here considered. This relation results from the fact that both methods are accurate to second-order, with corrections $O(\tau^3)$. Indeed, this resemblance is only one aspect of the more general connection between both formalisms, analyzed in detail in the next section.

\subsection{Classical-quantum correspondence}\label{subsec:QuantumClassical}

Along this chapter we have seen that there is a close relation between the classical and the quantum problem. This connection is rigorously established through the Wigner function \cite{Schleich2001}. The Wigner function, which has already appeared when studying the GPH criterion in Chapter \ref{chapter:CSGPH}, is defined for our present problem as:
\begin{eqnarray}\label{eq:Wignerfunctionspace}
\nonumber W(\mathbf{x},\mathbf{p},t)&=&\int\frac{\mathrm{d}^3\Delta\mathbf{x}}{(2\pi\hbar)^3}~e^{i\frac{\mathbf{p}\Delta\mathbf{x}}{\hbar}}n(\mathbf{x}+\frac{\Delta\mathbf{x}}{2},\mathbf{x}-\frac{\Delta\mathbf{x}}{2},t)\\
n(\mathbf{x},\mathbf{x}',t)&=&\braket{\hat{\Psi}^{\dagger}\left(\mathbf{x},t\right)\hat{\Psi}\left(\mathbf{x}',t\right)}
\end{eqnarray}
where $n(\mathbf{x},\mathbf{x}',t)$ is the so-called one-body density matrix \cite{Pitaevskii2003} and gives the off-diagonal elements of the density matrix. In particular, the density of particles is $n(\mathbf{x},t)=n(\mathbf{x},\mathbf{x},t)$. A similar expression can be written in term of momentum correlations
\begin{eqnarray}\label{eq:Wignerfunctionmomentum}
\nonumber W(\mathbf{x},\mathbf{p},t)&=&\int\frac{\mathrm{d}^3\Delta\mathbf{p}}{(2\pi\hbar)^3}~e^{i\frac{\mathbf{x}\Delta\mathbf{p}}{\hbar}}n(\mathbf{p}-\frac{\Delta\mathbf{p}}{2},\mathbf{p}+\frac{\Delta\mathbf{p}}{2},t)\\
n(\mathbf{p},\mathbf{p}',t)&=&\braket{\hat{c}^{\dagger}\left(\mathbf{p},t\right)\hat{c}\left(\mathbf{p}',t\right)}
\end{eqnarray}
where $\hat{c}\left(\mathbf{p}\right)$ is the annihilation operator of the plane wave state with momentum $\mathbf{p}$. The Wigner function plays an analog role to the Boltzmann distribution function since it gives the density and momentum distributions,
\begin{eqnarray}\label{eq:Wignerfunctionmomentum}
\nonumber \int\mathrm{d}^3\mathbf{p}~W(\mathbf{x},\mathbf{p},t)&=&\braket{\hat{\Psi}^{\dagger}\left(\mathbf{x},t\right)\hat{\Psi}\left(\mathbf{x},t\right)}=n(\mathbf{x},t)\\
\int\mathrm{d}^3\mathbf{x}~W(\mathbf{x},\mathbf{p},t)&=&\braket{\hat{c}^{\dagger}\left(\mathbf{p},t\right)\hat{c}\left(\mathbf{p},t\right)}
\end{eqnarray}
In the first place, we compute the Wigner function in the equilibrium state. Following the same considerations that lead to Eq. (\ref{eq:equilibriumtotaldensityBE}), we find that we can write the one-body density matrix as
\begin{eqnarray}\label{eq:equilibriumcorrelations}
n(\mathbf{x},\mathbf{x}')&=&\sum_{j=1}^{\infty}\eta^jF(j\beta,\mathbf{x}',\mathbf{x})\\
F(\beta,\mathbf{x}',\mathbf{x})&=&\braket{\mathbf{x}'|e^{-\beta H_0}|\mathbf{x}},
\end{eqnarray}
which gives a similar decomposition for the Wigner function
\begin{eqnarray}\label{eq:Wignerequilibrium}
\nonumber W(\mathbf{x},\mathbf{p})&=&\sum_{j=1}^{\infty}\eta^jw(j\beta,\mathbf{x},\mathbf{p})\\
w(\beta,\mathbf{x},\mathbf{p})&=&\int\frac{\mathrm{d}^3\Delta\mathbf{r}}{(2\pi\hbar)^3}~e^{i\frac{\mathbf{p}\Delta\mathbf{x}}{\hbar}}F(\beta,\mathbf{x}-\frac{\Delta\mathbf{x}}{2},\mathbf{x}+\frac{\Delta\mathbf{x}}{2})
\end{eqnarray}
As before, keeping just the term $j=1$ corresponds to the Maxwell-Boltzmann approximation. We focus as usual on the 1D HO problem of Eq. (\ref{eq:1DHO}) and then extend the results to each direction. Considering the 1D version of the LDA of Eq. (\ref{eq:1DLDAstatic}), the corresponding one-body density matrix reads, in terms of the center of mass and relative position, as:
\begin{eqnarray}\label{eq:WignerLDAF}
\nonumber F(\beta,x-\frac{\Delta x}{2},x+\frac{\Delta x}{2})&\equiv&F(\beta,x,\Delta x)=\frac{e^{-\frac{1}{2}\beta m\omega^2x^2}}{\Lambda_T}e^{-\frac{\pi}{\Lambda^2_T}\left[1+\frac{(\beta\hbar\omega)^2}{4}\right](\Delta x)^2}\\
&\simeq&\frac{e^{-\frac{1}{2}\beta m\omega^2x^2}}{\Lambda_T}e^{-\frac{\pi}{\Lambda^2_T}(\Delta x)^2}
\end{eqnarray}
where we consistently neglect the corrections $O(\left[\beta\hbar\omega\right]^2)$. The associated function $w(\beta,x,p)$ reads
\begin{eqnarray}\label{eq:WignerLDA}
w(\beta,x,p)=\frac{e^{-\beta\left(\frac{p^2}{2m}+\frac{1}{2}m\omega^2x^2\right)}}{2\pi\hbar}=\frac{e^{-\beta H}}{2\pi\hbar}
\end{eqnarray}
which is the same form of the equilibrium Boltzmann distribution. Extending this result to the complete 3D problem gives:
\begin{equation}\label{eq:WignerLDA3D}
w(\beta,\mathbf{x},\mathbf{p})=\frac{e^{-\beta H_0}}{(2\pi\hbar)^3}
\end{equation}
The series for the total Wigner function of Eq. (\ref{eq:Wignerequilibrium}) can be explicitly computed from here, obtaining the {\it semi-classical} distribution function \cite{Pethick2008}:
\begin{equation}\label{eq:Wignersemiclass}
W(\mathbf{x},\mathbf{p})=\frac{1}{(2\pi\hbar)^3}\frac{1}{e^{\beta (H_0-\mu)}-1}=\frac{1}{(2\pi\hbar)^3}\frac{1}{e^{\beta\left[\frac{\mathbf{p}^2}{2m}-\mu(\mathbf{x})\right]}-1}
\end{equation}
with $\mu(\mathbf{x})=\mu-V_{\rm {trap}}(\mathbf{x})$. The previous expression corresponds to the homogeneous Wigner distribution but replacing the chemical potential by its space-dependent version, in the spirit of the LDA.

For completeness, we also give the results for the total HO Hamiltonian (\ref{eq:1DHO})
\begin{eqnarray}\label{eq:WignerHO}
\nonumber F(\beta,x,\Delta x)&=&\sqrt{\frac{m\omega}{2\pi\hbar\sinh(\beta\hbar \omega)}}e^{-\tanh\left(\frac{\beta\hbar \omega}{2}\right)\frac{m\omega}{\hbar}~x^2}e^{-\coth\left(\frac{\beta\hbar \omega}{2}\right)\frac{m\omega}{\hbar}~\frac{(\Delta x)^2}{4}}\\
w(\beta,x,p)&=&\frac{1}{2\pi\hbar\cosh\left(\frac{\beta\hbar \omega}{2}\right)}e^{-\tanh\left(\frac{\beta\hbar \omega}{2}\right)\left(\frac{m\omega}{\hbar}~x^2+\frac{p^2}{m\hbar\omega}\right)}
\end{eqnarray}
which, in the limit $\beta\hbar\omega\ll1$ reduce to the LDA expressions, as expected.

We now address the dynamical problem. In the same fashion, we compute the one-body density matrix as
\begin{eqnarray}\label{eq:Wignerdynamicalfactor}
n(\mathbf{x},\mathbf{x}',t)&=&\sum_{j=1}^{\infty}\eta^jF(j\beta,\mathbf{x}',\mathbf{x},t)\\
F(\beta,\mathbf{x}',\mathbf{x},t)&=&\braket{\mathbf{x}'|U(t)e^{-\beta H_0}U^{\dagger}(t)|\mathbf{x}},
\end{eqnarray}
and the Wigner function can be obtained from the Fourier transform of the previous function. However, the point of considering the Wigner function is that its time evolution follows the differential equation
\begin{eqnarray}\label{eq:Wignertimeevolution}
\nonumber \frac{\partial W}{\partial t}+\frac{\mathbf{p}}{m}\nabla W-\nabla V ~\nabla_{p}W&=&\sum_{n=1}^{\infty}\sum_{i_1,i_2\ldots i_{2n+1}}\frac{(-1)^n\hbar^{2n}}{2^{2n}(2n+1)!}\nabla_{i_{1}i_{2}\ldots i_{2n+1}}V\nabla^{i_{1}i_{2}\ldots i_{2n+1}}_{p}W\\
\nabla^{i_{1}i_{2}\ldots i_{2n+1}}_{p}W&\equiv&\frac{\partial^{2n+1}W}{\partial p_{i_1}\partial p_{i_2}\ldots \partial p_{i_{2n+1}}}
\end{eqnarray}
with $V$ the total potential acting on the particles. We note that the l.h.s. of the previous equation is precisely the non-interacting Boltzmann equation of Eq. (\ref{eq:boltzmanneq}) and the r.h.s. introduces the quantum physics of $\hbar$. Interestingly, the time evolution of the Wigner function in a harmonic potential satisfies exactly the Boltzmann equation as the r.h.s. vanishes. As a consequence, the only possible quantum corrections in the dynamics arise from the Bragg pulse. The relative magnitude of the quantum corrections is $O(\hbar^2k^2/p^2_T)$ so they become negligible whenever Eq. (\ref{eq:quantumclassicalphoton}) is fulfilled. Precisely, this condition appeared in Sec. \ref{subsec:dynLDA} as the condition for the smallness of the quantum corrections in $\hbar$ to the classical result, in agreement with the discussion here presented. Its physical meaning is that the momentum of the photon has to be much smaller than the typical momentum of the particles, $\hbar k\ll p_T$. We can further evaluate this condition using the experimental values presented at the beginning of this section. Since $k\tilde{l}\sim 1$, Eq. (\ref{eq:quantumclassicalphoton}) gives
\begin{equation}\label{eq:estquantumclassical}
\frac{\hbar k}{p_T}\sim k\Lambda_T\sim N^{-\frac{1}{6}}\ll 1
\end{equation}
as we are dealing with large numbers of particles $N\sim 10^4-10^7$ and $T\sim T^0_c$. The above equation provides a similar interpretation to the LDA: the scale of variation of the lattice has to be much larger than the associated thermal de Broglie wavelength.

From all the above considerations, we conclude that, as long as interactions are negligible, if the cloud has a sufficiently large number of particles, the Wigner function and the Boltzmann distribution function evolve with the same (classical) differential equation. Moreover, a large number of particles also implies that the LDA is valid, so one can use the {\it semi-classical} distribution function of Eq. (\ref{eq:Wignersemiclass}). In this way, quantum features enter only through the Bose statistics but not through the time-evolution of the particles, as their distribution obeys the {\it classical} Boltzmann equation.

\section{Extension of the model}\label{sec:generalcase}

The theoretical model presented in the previous sections has shown that, whenever the considered approximations are valid, the introduction of a short Bragg pulse in a trapped cloud induces decaying oscillations in the spatial density profile. However, as we show now, this result is in fact more general than the previous scenario. In order to address the problem, we use the quantum description and keep assuming that the LDA is valid and interactions do not play any dynamical role, see Secs. \ref{subsec:LDAjust}-\ref{subsec:interactingrole}. In this non-interacting picture, as the trapping potential does not enter into the dynamics, the problem is reduced to compute the time-evolution of a plane wave along the $z$-axis as given by Eq. (\ref{eq:LDAdynamical}). Since in this regime the motion after the pulse is the free one, we only need to compute the time evolution operator during the pulse, $U_{\tau}$, given by Eq. (\ref{eq:tauevolution}). However, instead of making use of the split-step approximation, we study more general situations. In particular, we first use the Dirac (or interaction) picture to compute the perturbative series for $U_{\tau}$ and after that we consider its formal expression for the general case.

\subsection{Dirac picture}\label{subsec:Diracpicture}

During the pulse, the motion along the $z$ coordinate is governed (in the LDA) by the Hamiltonian
\begin{equation}\label{eq:BraggHamiltonian}
H_z=H_{0z}+V(z),~H_{0z}=\frac{P^2_z}{2m},~
\end{equation}
with $V(z)$ the potential created by the pulse. In order to solve the corresponding Schr\"odinger equation, we make use of the so-called Dirac picture, consisting in an unitary transformation of the operators $A^D\equiv e^{i\frac{H_{0z}}{\hbar}t}Ae^{-i\frac{H_{0z}}{\hbar}t}$, with $A$ an arbitrary quantum operator. This implies that the time-evolution operator in the Dirac picture satisfies the differential equation:
\begin{equation}\label{eq:Diracpictureequation}
V^D(t)U_z^D(t)=i\hbar\frac{dU_z^D(t)}{dt}
\end{equation}
which can be simply integrated iteratively
\begin{equation}\label{eq:Diraciterative}
U_z^D(t)=1-\frac{i}{\hbar}\int_0^t\mathrm{d}t'~V^D(t')U_z^D(t')=\sum_{n=0}^{\infty}\left(\frac{-i}{\hbar}\right)^n\int_0^t\mathrm{d}t_1\int_0^{t_1}\mathrm{d}t_2\ldots\int_0^{t_{n-1}}\mathrm{d}t_{n}~V^D(t_1)\ldots V^D(t_n)
\end{equation}
In particular, for $t=\tau$, we obtain $U_{\tau}=e^{-i\frac{H_{0z}}{\hbar}\tau}U^D_z(\tau)$. Comparing with Eq. (\ref{eq:formaltimesplitstep}), we see that $G_{\tau}$ is an unitary transformation of $U^D_z(\tau)$ and is given by a similar expression
\begin{equation}\label{eq:Dirackernelperturbative}
G_{\tau}=e^{-i\frac{H_{0z}}{\hbar}\frac{\tau}{2}}U^D_z(\tau)e^{i\frac{H_{0z}}{\hbar}\frac{\tau}{2}}=\sum_{n=0}^{\infty}\left(-\frac{i}{\hbar}\right)^n\int_{-\frac{\tau}{2}}^{\frac{\tau}{2}}\mathrm{d}t_1\int_{-\frac{\tau}{2}}^{t_1}\mathrm{d}t_2\ldots\int_{-\frac{\tau}{2}}^{t_{n-1}}\mathrm{d}t_{n}~V^D(t_1)\ldots V^D(t_n)
\end{equation}
As a consequence, in analogy with Eq. (\ref{eq:zevolutionapprox}), we rewrite the total evolution operator as
\begin{equation}\label{eq:Diraczevolutionexact}
U_z(t)=e^{-i\frac{H_{0z}}{\hbar}\bar{t}}G_{\tau}
\end{equation}
which we note that it is an exact relation for $U_z(t)$ (apart from the trivial evolution factor $e^{-i\frac{H_{0z}}{\hbar}\frac{\tau}{2}}$ that we neglect as it does not enter into the dynamics). As argued in Sec. \ref{subsec:LTA}, for a short pulse, $G_{\tau}\simeq e^{iu\cos(k z)}$ with corrections $O(\tau^3)$. This result is easily recovered within this picture by supposing that the time dependence of $V^D(t)$ is very weak and hence $V^D(t)\simeq V(z)$. Under this assumption, we get
\begin{equation}\label{eq:DirackernelRaman}
G_{\tau}\simeq \sum_{n=0}^{\infty}\frac{\left(\frac{-iV(z)\tau}{\hbar}\right)^n}{n!}=e^{-i\frac{V(z)\tau}{\hbar}}
\end{equation}
We can try to understand this approximation by computing explicitly the matrix elements of the first and second order terms of the previous perturbative series. After replacing $V(z)$ by its specific expression $V(z)=-V_0\cos(k z)$ and noting that, as it is spatially periodic, it only connects $\ket{p}$ with states with momentum differing an integer multiple of $\hbar k$, we find
\begin{eqnarray}\label{eq:Dirackernelfirstsecondorders}
\bra{p,n}G^{(0)}_{\tau}\ket{p}&=&\braket{p+n\hbar k| p}=\delta_{n,0}\\
\nonumber \bra{p,n}G^{(1)}_{\tau}\ket{p}&=&\frac{iu}{2}\text{sinc}\left(\frac{\Omega^p_{n0}\tau}{2}\right)\delta_{n,\pm 1}\\
\nonumber \bra{p,n}G^{(2)}_{\tau}\ket{p}&=&-\frac{u^2}{8}\left(\left(i\frac{\Omega^p_{10}\tau}{2}\right)^{-1}\left[\text{sinc}\left(\frac{\Omega^p_{20}\tau}{2}\right)-e^{-i\frac{\Omega^p_{10}\tau}{2}}\text{sinc}\left(\frac{\Omega^p_{21}\tau}{2}\right)\right]\delta_{n,2}\right.\\
\nonumber &+&\left[\left(i\frac{\Omega^p_{10}\tau}{2}\right)^{-1}\left[1-e^{-i\frac{\Omega^p_{10}\tau}{2}}\text{sinc}\left(\frac{\Omega^p_{10}\tau}{2}\right)\right]\right.\\
\nonumber &+&\left.\left(i\frac{\Omega^p_{-10}\tau}{2}\right)^{-1}\left[1-e^{-i\frac{\Omega^p_{-10}\tau}{2}}\text{sinc}\left(\frac{\Omega^p_{-10}\tau}{2}\right)\right]\right]\delta_{n,0}\\
\nonumber &+&\left.\left(i\frac{\Omega^p_{-10}\tau}{2}\right)^{-1}\left[\text{sinc}\left(\frac{\Omega^p_{-20}\tau}{2}\right)-e^{-i\frac{\Omega^p_{-10}\tau}{2}}\text{sinc}\left(\frac{\Omega^p_{-2-1}\tau}{2}\right)\right]\delta_{n,-2}\right)
\end{eqnarray}
with $\ket{p,n}\equiv\ket{p+n\hbar k}$, $n$ an integer number and
\begin{eqnarray}\label{eq:sincfunctions}
\text{sinc}(x)&\equiv&\frac{\sin(x)}{x}=\sum_{n=0}^\infty\frac{(-1)^n}{(2n+1)!}x^{2n}\\
\nonumber \hbar\Omega^p_{nn'}&=&\frac{(p+n\hbar k)^2}{2m}-\frac{(p+n'\hbar k)^2}{2m}=(n-n')\frac{\hbar p k}{m}+(n^2-n'^2)\frac{\hbar^2k^2}{2m}~,
\end{eqnarray}
$\hbar\Omega^p_{nn'}$ being the kinetic-energy difference between $\ket{p,n}$ and $\ket{p,n'}$. By comparing this result with the lowest order terms in $u$ from the Fourier series of $e^{iu\cos(k z)}$ in Eq. (\ref{eq:Fourierseries}), we conclude that Eq. (\ref{eq:DirackernelRaman}) provides a good approximation whenever the phase differences $\Omega^p_{nn'}\tau$ acquired during the presence of the pulse satisfy $\Omega^p_{nn'}\tau\ll 1$. In particular, this condition implies
\begin{eqnarray}\label{eq:Ramanphases}
\frac{\hbar k^2}{2m}\tau&\ll&1\\
\label{eq:Ramanphasesthermal}\frac{p k \tau}{m}&\ll&1
\end{eqnarray}
Note that the latter equation reduces to $\theta\ll1$ [see Eq. (\ref{eq:perturbativeparameter})] after taking into account that we are dealing with momenta $p\sim p_T$. Indeed, in the classical regime $\hbar k\ll p_T$, Eq. (\ref{eq:Ramanphasesthermal}) automatically implies Eq. (\ref{eq:Ramanphases}). The leading order corrections to Eq. (\ref{eq:DirackernelRaman}) can be explicitly computed from Eq. (\ref{eq:Dirackernelfirstsecondorders}), obtaining:
\begin{eqnarray}\label{eq:Diracestimations}
\nonumber \bra{p}G_{\tau}\ket{p}&=&J_0(u)+iu^2\frac{\hbar k^2 \tau}{24m}+\ldots\simeq J_0(u)+i\frac{u\lambda}{24}\\
\bra{p,1}G_{\tau}\ket{p}&=&iJ_1(u)-i\frac{u}{48}\left(\frac{pk\tau}{m}\right)^2+\ldots\sim iJ_1(u)+\frac{u\theta^2}{48} \\
\nonumber \bra{p,2}G_{\tau}\ket{p}&=&-J_2(u)-iu^2\frac{\hbar k^2 \tau}{48m}+\ldots\simeq -J_2(u)-i\frac{u\lambda}{48}
\end{eqnarray}
in agreement with the estimations of Eq. (\ref{eq:3rdordertemporalcommutators}). All these reasonings can be straightforwardly translated to arbitrary pulses. We also note that the approximation here presented for the time evolution during the pulse corresponds to an improved Raman-Nath approximation \cite{Berman2010}.

However, the above considerations necessarily fail at sufficiently high temperature since then $\theta\gg 1$. In this limit, we treat perturbatively the potential by using the series of Eq. (\ref{eq:Dirackernelperturbative}). Supposing that $V_0\ll\hbar \Omega^p_{10}$, that Eq. (\ref{eq:Ramanphases}) is fulfilled and keeping only the first term of the perturbative series, we get
\begin{equation}\label{eq:DiracHTlimit}
G_{\tau}\ket{p}\simeq\ket{p}+\frac{iu}{2}\text{sinc}\left(\frac{pk\tau}{2m}\right)\ket{p+k}+\frac{iu}{2}\text{sinc}\left(\frac{pk\tau}{2m}\right)\ket{p-k}
\end{equation}
Using Eq. (\ref{eq:Diraczevolutionexact}) for the total time evolution operator, inserting this result in Eq. (\ref{eq:LDAdynamical}) and keeping consistently only first-order terms yields
\begin{eqnarray}\label{eq:HTDynFactor}
F_z(\beta,z,t)&=&\frac{e^{-\frac{1}{2}\beta mw^2_zz^2}}{\Lambda_T}\left[1+F_1(\beta,\bar{t})e^{ikz}+F^*_{1}(\beta,\bar{t})e^{-ikz}\right]\\
\nonumber F_1(\beta,\bar{t})&=&u\sin\left(\frac{\hbar k^2\bar{t}}{2m}\right)\sqrt{\frac{\beta}{2\pi m}}\int_{-\infty}^{\infty}\mathrm{d}p~ \text{sinc}\left(\frac{pk\tau}{2m}\right)e^{-i\frac{pk}{m}\bar{t}}e^{-\frac{\beta p^2}{2m}}
\end{eqnarray}
In order to deal with the previous integral, we use Parseval's theorem and the Fourier transform of the sinc function
\begin{equation}
\frac{1}{\sqrt{2\pi}}\int_{-\infty}^\infty\mathrm{d}x~\text{sinc}(ax)e^{-ikx}=\sqrt{\frac{\pi}{2}}\frac{\chi\left(\frac{1}{2}+\frac{k}{2a}\right)}{a}~,
\end{equation}
obtaining:
\begin{eqnarray}\label{eq:Percival}
\nonumber &~&\int_{-\infty}^{\infty}\mathrm{d}p~ \text{sinc}\left(\frac{pk\tau}{2m}\right)e^{-i\frac{pk}{m}\bar{t}}e^{-\frac{\beta p^2}{2m}}=\frac{1}{2}\int_{-1}^1\mathrm{d}x~e^{-\frac{k^2}{2m\beta}\bar{t}^2\left(1-\frac{\tau}{2\bar{t}}x\right)^2}\\
&~&=\frac{\sqrt{\pi}\tau^d}{2\tau}\left[\text{erf}\left(\frac{\bar{t}+\frac{\tau}{2}}{\tau^d}\right)-\text{erf}\left(\frac{\bar{t}-\frac{\tau}{2}}{\tau^d}\right)\right]\\
\nonumber &~&\text{erf}(x)\equiv\frac{2}{\sqrt{\pi}}\int_{0}^x\mathrm{d}t~e^{-t^2}~,
\end{eqnarray}
$\text{erf}(x)$ being the error function. For sufficiently large times $\bar{t}\gg \tau$, the asymptotic expansion of the previous integral is $e^{-\frac{k^2}{2m\beta}\bar{t}^2}$ and then
\begin{eqnarray}\label{eq:HTexponentialdecay}
F_1(\beta,\bar{t})\simeq u\sin\left(\frac{\hbar k^2\bar{t}}{2m}\right)e^{-\frac{k^2}{2m\beta}\bar{t}^2}
\end{eqnarray}
If we insert explicitly this results in Eq. (\ref{eq:HTDynFactor}), we find that
\begin{equation}\label{eq:HTDynFactor}
F_z(\beta,z,t)=\frac{e^{-\frac{1}{2}\beta mw^2_zz^2}}{\Lambda_T}\left[1+2u\sin\left(\frac{\hbar k^2\bar{t}}{2m}\right)e^{-\frac{k^2}{2m\beta}\bar{t}^2}\cos(kz)\right]
\end{equation}
which corresponds to keep only the first order term in $u$ in Eq. (\ref{eq:LDAz2}). We then see that also out of the split-step approximation, the system displays the same Gaussian decay as before. This behavior makes us thinking that this property is something intrinsic to the system.

\subsection{General case}

In light of the previous discussion, one may wonder how general is the described thermal decay. For that purpose, we consider an arbitrary pulse that creates during time $\tau$ a periodic potential $V(z)$ along the $z$-axis (although we note that the particular direction of the potential does not matter provided that the LDA is valid). The Hamiltonian during the pulse is that of Eq. (\ref{eq:BraggHamiltonian}), which corresponds to a one-dimensional periodic Hamiltonian and whose eigenfunctions are Bloch waves (a discussion of Bloch solutions in an optical lattice is presented in Appendix \ref{app:nonlinearol}). The time evolution operators reads, in terms of these eigenfunctions, as:
\begin{equation}
U_{\tau}=\sum_{n_b}\int_{-\frac{k}{2}}^{\frac{k}{2}}\mathrm{d}q~e^{-i\frac{E_{q,n_b}}{\hbar}\tau}\ket{q,n_b}\bra{q,n_b}
\end{equation}
where $\ket{q,n_b}$ is the Bloch wave with pseudomomentum $q$ that belongs to the band labeled by the discrete index $n_b$. As Bloch waves only involve Fourier components differing by an integer multiple of $k$, when acting on a plane wave $\ket{p}$ with $U_{\tau}$ we get
\begin{equation}\label{eq:GENERALBlochwavepulse}
\ket{\Psi_p}\equiv U_{\tau}\ket{p}=\sum_{n=-\infty}^{\infty}c_{n}(p)\ket{p+n\hbar k}
\end{equation}
The total time evolution of an arbitrary plane wave follows straightforwardly from the previous relation
\begin{equation}\label{eq:GENERALplanewavetimeevolution}
\braket{z|U_z(t)|p}=\bra{z}e^{-i\frac{H_{0z}}{\hbar}\Delta t}\ket{\Psi_p}=
\frac{e^{i\frac{p}{\hbar}z}}{(2\pi\hbar)^{\frac{1}{2}}}e^{-i\frac{p^2}{2m\hbar}\Delta t}\sum_{n=-\infty}^\infty c_n(p)e^{inkz}e^{-in\frac{pk}{m}\Delta t}e^{-in^2\frac{\hbar k^2}{2m}\Delta t}
\end{equation}
where we have defined $\Delta t\equiv t-\tau$ for convenience. Turning back to Eq. (\ref{eq:LDAdynamical}), one obtains, after some straightforward manipulations:
\begin{equation}\label{eq:GENERALFINALexpression}
F_z(\beta,z,t)=\frac{e^{-\frac{\beta}{2}mw^2_zz^2}}{\Lambda_T}\sum_{n=-\infty}^\infty F_n(\beta,\Delta t)e^{inkz}
\end{equation}
with
\begin{eqnarray}
\nonumber F_n(\beta,\Delta t)&=&\sqrt{\frac{\beta}{2\pi m}}\int_{-\infty}^{\infty}\mathrm{d}p~ f_n(p,\Delta t)e^{-in\frac{pk}{m}\Delta t}e^{-\frac{\beta p^2}{2m}}\\
f_n(p,\Delta t)&\equiv& \sum_{n'=-\infty}^\infty c_{n'}(p)c^*_{n'-n}(p)e^{-i(2nn'-n^2)\frac{\hbar k^2}{2m}\Delta t}
\end{eqnarray}
Operating in the same fashion of Eq. (\ref{eq:Percival})
\begin{eqnarray}
\nonumber F_n(\beta,\Delta t)&=&\frac{1}{\sqrt{2\pi}}\int_{-\infty}^\infty\mathrm{d}x~\tilde{f}_n(x,\Delta t)e^{-\frac{n^2k^2}{2m\beta}\Delta {t}^2\left(1-\frac{m}{k\Delta{t}}x\right)^2}\\
\tilde{f}_n(x,\Delta t)&=&\frac{1}{\sqrt{2\pi}}\int_{-\infty}^\infty\mathrm{d}p~f_n(p,\Delta t)e^{-ipx}
\end{eqnarray}
with $\tilde{f}_n$ the Fourier transform of $f(p,\Delta t)$.
Once more, for sufficiently large times, the previous integral gives
\begin{equation}
F_n(\beta,\Delta t)\simeq f_n(0,\Delta t) e^{-\frac{n^2k^2}{2m\beta}\Delta{t}^2}
\end{equation}
which is the same asymptotic decaying behavior for the Fourier components of the main theoretical model. Thus, this thermal decay is a general feature of a trapped gas of bosons within the approximated regime here considered (LDA and non-interacting dynamics).

The previous theoretical model can be further extended to the thermal component for temperatures below $T_c$ since in the regime $T_{TF}<T<T_c$ the excitations of the uncondensed cloud are free particles to a good approximation \cite{Pitaevskii2003,Pethick2008}. The temperature $T_{TF}$ is the temperature associated to the Thomas-Fermi chemical potential $\mu_{TF}=k_BT_{TF}$
\begin{equation}
\mu_{TF}=\frac{\hbar\tilde{\omega}}{2}\left(\frac{15Na_s}{\tilde{l}}\right)^{\frac{2}{5}}
\end{equation}
Moreover, under the same general assumptions, the previous calculations can also be translated to a fermion gas at sufficiently high temperature, where $\eta\lesssim 1$.

\section{Experimental data}\label{sec:experimentaldata}

\begin{figure}[!htb]\includegraphics[width=\columnwidth]{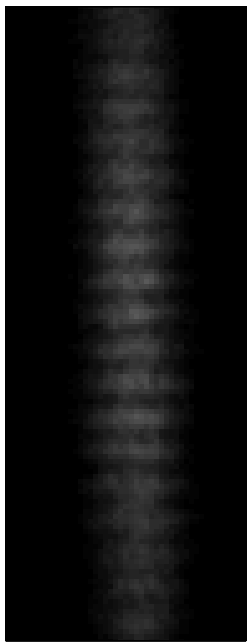}
\caption{Experimental {\it in situ} image of a thermal trapped cloud $0.07$ ms after the pulse. This plot has been provided by Technion group.}
\label{fig:ExperimentalImage}
\end{figure}

Here, we compare the previous theoretical calculations with the experimental data provided by the Technion group for a setup similar to that of our model. It consists of a confined thermal cloud of atoms of $N$ $^{87}Rb$ atoms (the same element of the simulations of Chapter \ref{chapter:MELAFO}) with $N\sim 10^5-10^6\gg 1$ and trap frequencies $\omega_x=\omega_y=\omega_r=2\pi\times224~\text{Hz}$ and $\omega_z=2\pi\times26~\text{Hz}$. The mass of the atoms is the same as in Sec. \ref{sec:NumericaLOL}, $m=1.44 \times 10^{-25}$ kg. Suddenly, it is introduced a short Bragg pulse of the same form of that of Eq. (\ref{eq:potentialswitch}) with parameters $V_0=2.85 \times10^{-30} \text{J}$, $\tau=1.10\times 10^{-4} \text{s}$ and $k=1.43 \times 10^6~\text{m}^{-1}$. The pulse induces a periodic pattern in the density (which can be observed in Fig. \ref{fig:ExperimentalImage}) that eventually decays to the equilibrium configuration.

Before of applying our theoretical model, we evaluate the relevant parameters of the system in order to check the considered approximations. The HO lengths are
\begin{eqnarray}\label{eq:experimentalHOlengths}
\nonumber l_r&=&\sqrt{\frac{\hbar}{m\omega_r}}=7.21\times10^{-7}~\text{m}\\
l_z&=&\sqrt{\frac{\hbar}{m\omega_z}}=2.11\times10^{-6}~\text{m}\\
\nonumber \tilde{l}&=&\sqrt{\frac{\hbar}{m\tilde{\omega}}}=1.03\times10^{-6}~\text{m}
\end{eqnarray}
while we find, from Eqs. (\ref{eq:estcritical})-(\ref{eq:estdecay}), that
\begin{eqnarray}\label{eq:experimentalestimations}
\nonumber T^0_c&=&4.93\times10^{-7}\times\left(\frac{N}{10^6}\right)^{\frac{1}{3}}~\text{K}\sim 10^{-7}~\text{K}\\
\Lambda_T&=&5.92\times 10^{-7}\sqrt{\frac{10^{-7}}{T[\text{K}]}}~\text{m}\sim 10^{-7}~\text{m}\\
\nonumber \tau^d&\sim& 10^{-4}~\text{s}\sim \tau
\end{eqnarray}

We first analyze the validity of the LDA. Following the discussion of Sec. \ref{subsec:LDAjust}, the {\it static} LDA is valid when the relative finite-size shift of the critical temperature is small. This condition is fulfilled in our configuration as
\begin{equation}\label{eq:criticalexpfinitesize}
\frac{\delta T_c}{T^0_c}=-0.73\frac{\omega_m}{\tilde{\omega}}N^{-\frac{1}{3}}\sim 0.01\ll 1
\end{equation}
On the other hand, the {\it dynamic} LDA is also a good approximation since $\tau^d\sim \tau$ and, as estimated after Eq. (\ref{eq:dynamicLDAtimescales}), $\omega_z\tau^{d}\sim \sqrt{\beta\hbar\omega_z} \sim 0.02\ll1 $. In order to quantify interactions between particles, we use the value of the scattering length for $^{87}Rb$, given in Sec. \ref{sec:NumericaLOL}, $a_s\sim100a_0=5.29~\text{nm}$. Interactions in the equilibrium configuration originate a shift in the critical temperature given by Eq. (\ref{eq:criticalfinitesizeinteraction})
\begin{equation}\label{eq:expinteractionshift}
\frac{\delta T_c}{T^0_c}=-1.32N^{\frac{1}{6}}\frac{a_{s}}{\tilde{l}}\sim 0.07
\end{equation}
which represents a small correction and consequently interactions can be treated to first order in a HFB approach, as done in Sec. \ref{subsec:interactingrole}. Also, as the LDA is valid, the mean-field HFB potential does not enter into the dynamics and only acts as an envelope. In respect to the collisional rate, we find from Eq. (\ref{eq:collisionlessregime}) that for $T\sim T^0_c$
\begin{equation}\label{eq:expcollisionlessregime}
\frac{\tau^{d}}{\tau^{col}}\sim \frac{a^2_s}{\Lambda^2_T}\frac{d}{\Lambda_T}\sim 5\times 10^{-3}\ll 1
\end{equation}
Thus, we are safely in the non-interacting picture considered in the theory. In respect to the approximations for the motion during the pulse of Sec. \ref{subsec:LTA}, the dimensionless impulse parameter $u$ of Eq. (\ref{eq:impulseparameter}) is $u=2.97$. The phase acquired during the pulse due to the recoil energy is:
\begin{equation}\label{eq:exprecoilphase}
\frac{\hbar k^2}{2m}\tau\simeq 0.08\ll 1
\end{equation}
On the other hand, the dimensionless parameters of Eq. (\ref{eq:perturbativeparameter}) are $\lambda=0.5$ and $\theta\gtrsim 1$ which are not strictly small. Nevertheless, the associated corrections (\ref{eq:Diracestimations}) are indeed small:
\begin{equation}\label{eq:expdiracestimations}
\frac{u\lambda}{24}=0.06,~\frac{u\theta^2}{48}\gtrsim 0.1
\end{equation}
Finally, we check that the classical picture for the motion of the particles is valid since
\begin{equation}\label{eq:expquantumclassical}
\frac{\hbar k}{p_T}\sim 0.1
\end{equation}
and hence both classical and quantum descriptions are valid. Therefore, all the approximations of our model are good ones as they are fulfilled in a real situation.

\begin{figure}[!htb]
\centering
\begin{tabular}{@{}ll@{}}
\includegraphics[width=0.55\columnwidth]{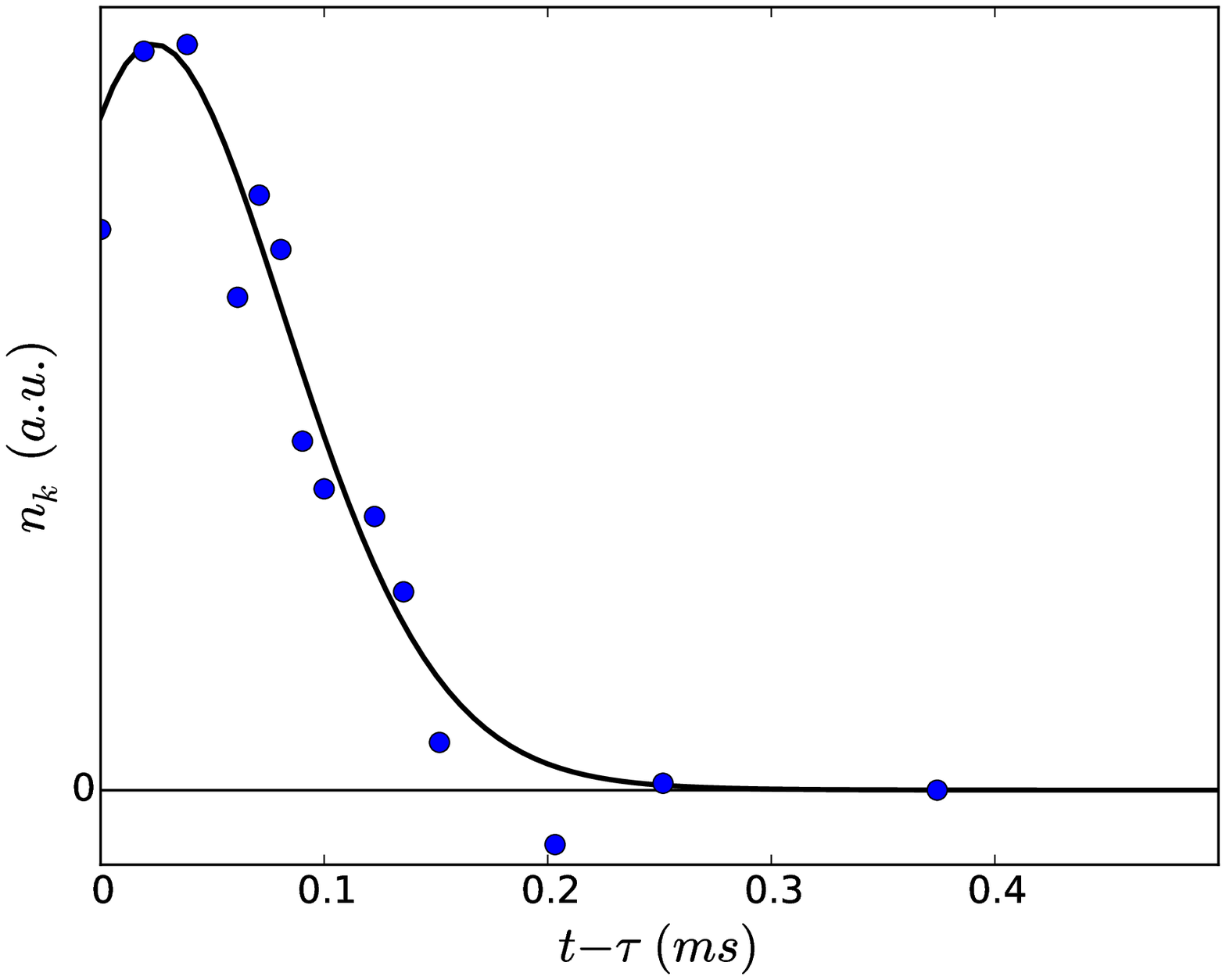} & \includegraphics[width=0.55\columnwidth]{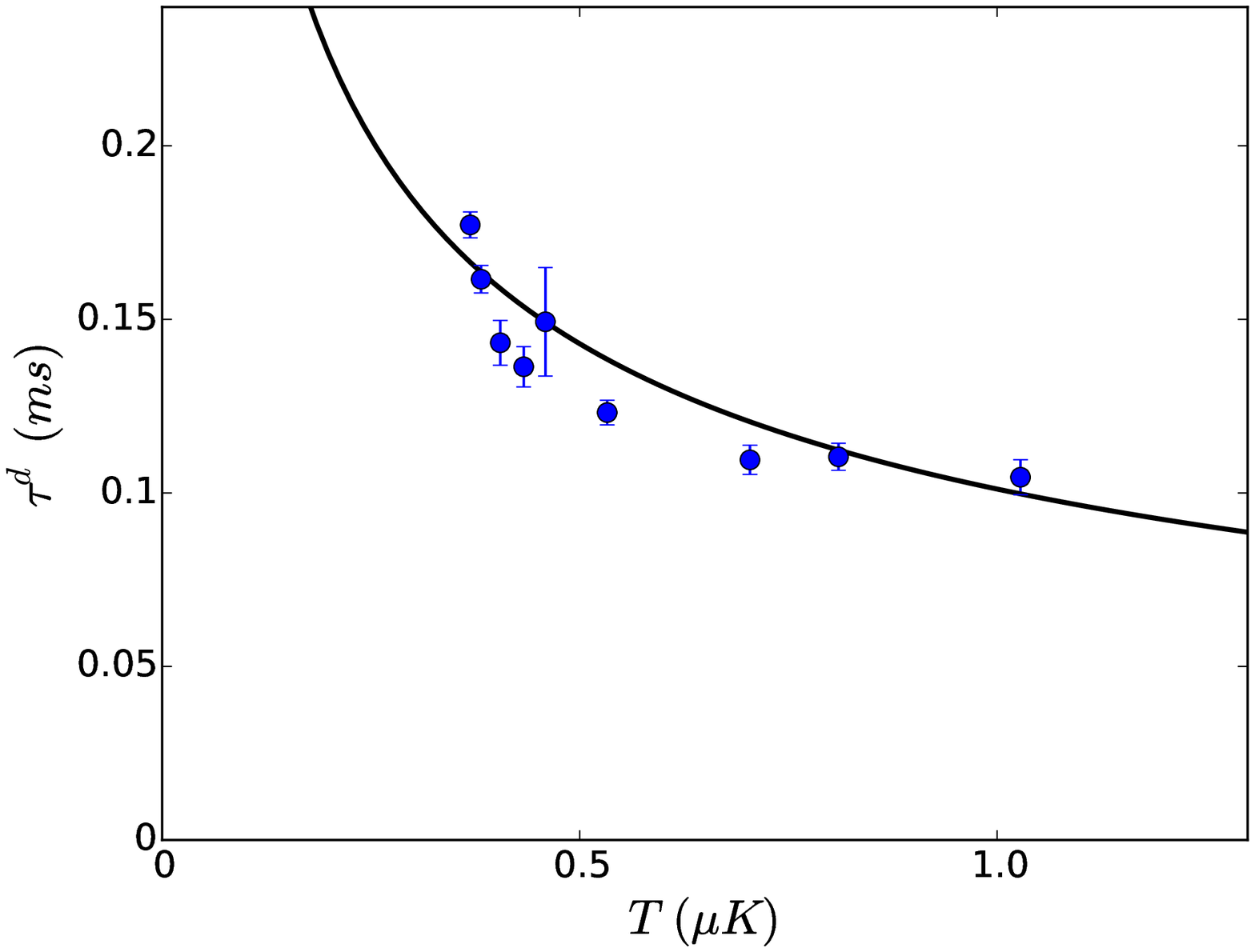}
\end{tabular}
\caption{Left panel: comparison for $t>\tau$ between the Fourier transform of the imaged density (blue dots) and a fit to Eq. (\ref{eq:Fouriercomponentquantum}) with the total amplitude, $\eta$ and $\tau^d$ as degrees of freedom (solid black line). The atom cloud is at temperature $T=810$ nK and contains $N=7.55\times10^5$ atoms. Right panel: plot of the decay time $\tau^d$ vs. temperature. The blue dots are the values extracted from the fit and the black line is the theoretical value of Eq. (\ref{eq:decaytime}). The error bars only include the statistical error of the fit and not the experimental one.}
\label{fig:ExperimentalData}
\end{figure}

We proceed to compare the experimental data with our theoretical model by fitting the measured Fourier signal after the pulse to Eq. (\ref{eq:Fouriercomponentquantum}) with free parameters $\eta$, $\tau^d$ and the amplitude of the signal. In the left panel of Fig. \ref{fig:ExperimentalData} we plot the time dependence Fourier transform of the experimental data (blue dots) and the corresponding fitting (black solid line) for an experiment at $T=8.1\times 10^{-7}$ K and with $N=7.55\times10^5$ atoms. In the right panel we represent the values of $\tau^d$ extracted from the previous fit (blue dots) and the theoretical prediction (black solid line). The measured decay time decreases with temperature, in agreement with the behavior predicted for $\tau^{d}$ by Eq. (\ref{eq:decaytime}). If the decay of the oscillations was due to collisions between particles, the estimation of Eq. (\ref{eq:classcol}) gives that the characteristic time of damping increases with the temperature as $T$. Note also that the time dependence of the exponential decay is quadratic rather than linear, as one would expect from usual damping mechanisms, described by an imaginary frequency. Therefore, the thermal disorder predicted by our model is the mechanism responsible for the decay of the oscillations. We remark that the error bars of the right Fig. \ref{fig:ExperimentalData} only take into account the statistical error arising from the fit and do not include any experimental uncertainty. Even though, there is a quite good agreement between theory and experiment, as it should be!

\section{Conclusions and outlook}\label{sec:decayconclusions}

We have proven that the introduction of a short Bragg pulse in a trapped thermal cloud induces a periodic density pattern that, after the removal of the pulse, decays with a Gaussian behavior. In particular, we have shown that this behavior is explained by both classical and quantum descriptions of the problem and we have established the link between the two formalisms. We have also checked that the approximations made for the model are satisfied in typical setups. The physical origin of this decay is indeed a single-particle phenomenon, arising from the thermal disorder of the particles, in contrast to the usual collisional damping mechanism of the collective modes. Finally, we have compared the theoretical predictions with actual experimental data, finding a very good agreement.

As argued in the main text, the obtained results can be extended to more general situations. In particular, as long as the local density approximation is valid and the role of interactions can be neglected, the same behavior appears for an arbitrary spatially periodic pulse.
{\it Mutatis mutandi}, the theoretical model developed is translatable to other similar configurations such as the thermal component of a Bose gas at temperature $T_{TF}<T<T_c$ (with $T_{TF}$ the temperature of the Thomas-Fermi chemical potential) or a Fermi gas at sufficiently high temperatures. A future computation could try to apply these ideas to the phonons above the condensate in the regime $0<T<T_{TF}$.

\part{Quantum Hall effect on graphene}

\chapter{Collective modes of the phase diagram of the $\nu=0$ quantum Hall state}\label{chapter:QHFM}
\chaptermark{Collective modes of the $\nu=0$ quantum Hall state}

\section{Introduction}

We switch in the last part of the thesis to a very different system: electrons in graphene under the effect of a strong magnetic field. One of the main contrasts with the former chapters arises from the fact that electrons are fermions. This implies that they obey the Pauli exclusion principle so creation/annihilation operators anti-commute instead of commute. Besides, graphene is a two-dimensional system. We also have to deal with the extra degrees of freedom corresponding to spin. Moreover, in graphene there are other extra degrees of freedom corresponding to valley and sublattice, originating a rich structure in the spectrum of excitations, as will be discussed.

We focus here on the $\nu=0$ quantum Hall (QH) state of bilayer graphene (BLG), in which the zero-energy Landau level (LL) is half-filled. Several works have been devoted to study its phase diagram, see for instance Refs. \cite{Gorbar2010,Kim2011,Kharitonov2012PRL}. In particular, the latter reference \cite{Kharitonov2012PRL} shows a complete study of the phase diagram by taking into account the most general spin symmetric interactions (including short-range valley/sublattice asymmetric interactions) and the introduction of a voltage between the two layers. The identified phases are ferromagnetic (F), canted anti-ferromagnetic (CAF), fully layer-polarized (FLP) and partially layer-polarized (PLP) \cite{Kharitonov2012,Kharitonov2012PRL}. The resulting phase diagram is analog to that of monolayer graphene (MLG) \cite{Kharitonov2012} since the form of the short-range valley/sublattice asymmetric interactions is the same. On the other hand, some works have computed the collective modes of the $\nu=0$ QH state for both monolayer and bilayer graphene \cite{Iyengar2007,Toke2011,Sari87,Lambert2013,Wu2014}; however, to our best knowledge, a computation of the collective modes of the complete phase diagram of Ref. \cite{Kharitonov2012} using the full short-range Hamiltonian has not been performed yet.

Thus, in this chapter we use the time-dependent Hartree-Fock approximation (TDHFA) to compute, neglecting LL mixing, the collective modes of the $\nu=0$ QH phases of Ref. \cite{Kharitonov2012PRL}.  We find that the short-range valley/sublattice interactions play a crucial role in order to understand the stability of the different phases since for zero-momentum the lowest energy collective modes are independent of the Coulomb interaction strength. In particular, we prove that the boundary between the F and FLP phases presents a gapless mode that arises from a residual $U(1)$ symmetry, which can be regarded as a remanent of the full $SO(5)$ symmetry existing for the same boundary at zero Zeeman energy in MLG \cite{Wu2014}. A similar gapless mode appears in the boundary between the CAF and PLP phases. Besides, we show that the CAF and PLP phases can present dynamical instabilities and discuss possible experimental scenarios for their observation. Due to the strong analogy between the $\nu=0$ QH states in bilayer and monolayer graphene, most of these results are straightforwardly translated to MLG. In this way, the work presented here provides a natural continuation of those of Refs. \cite{Kharitonov2012PRL,Toke2011,Sari87,Lambert2013,Wu2014}.

We first introduce the physical model for low energies of bilayer graphene in Sec. \ref{sec:basicmodel}. We present in Sec. \ref{subsec:projection} the effective Hamiltonian considered in this work and re-derive in Sec. \ref{subsec:HF} the phase diagram of Ref. \cite{Kharitonov2012PRL} within a Hartree-Fock mean-field scheme. The formalism of the TDHFA is presented in Sec. \ref{subsec:dyson} and the dispersion relation of the different collective modes is computed in Sec. \ref{subsec:results}. We translate the same calculations to monolayer graphene in Sec. \ref{sec:MLG}. We consider the effects of LL mixing in Sec. \ref{sec:renorm}. Possible scenarios for the observation of the dynamical instabilities are discussed in Sec. \ref{sec:remarksexps}. The conclusions are given in Sec. \ref{sec:QHFMConclusions}. Technical details about the diagonalization of the Hartree-Fock equations and the TDHFA are given in Appendices \ref{app:magneticFF}-\ref{app:analyticalTDHFA}.

\section{Physical model of bilayer graphene} \label{sec:basicmodel}

In the first place, we briefly review the main details of the effective model of bilayer graphene at low energies. For more details on the Hamiltonian of monolayer and bilayer graphene in the presence of a magnetic field, see for instance Refs. \cite{McCann2006,Kharitonov2012}.

In this approximation, the field operator for the electrons has $8$ components and reads:
\begin{eqnarray}\label{eq:BLGfieldoperator}
\nonumber\hat{\psi}(\mathbf{x})&=&\left[\begin{array}{c}
\hat{\psi}_{+}(\mathbf{x})\\
\hat{\psi}_{-}(\mathbf{x})
\end{array}\right]\\
\hat{\psi}_{\sigma}(\mathbf{x})&=&\left[\begin{array}{c}
\hat{\psi}_{KA\sigma}(\mathbf{x})\\
\hat{\psi}_{K\tilde{B}\sigma}(\mathbf{x})\\ \hat{\psi}_{K'\tilde{B}\sigma}(\mathbf{x})\\ \hat{\psi}_{K'A\sigma}(\mathbf{x})
\end{array}\right]\equiv\left[\begin{array}{c}
\hat{\psi}_{K\bar{A}\sigma}(\mathbf{x})\\
\hat{\psi}_{K\bar{B}\sigma}(\mathbf{x})\\ \hat{\psi}_{K'\bar{A}\sigma}(\mathbf{x})\\\hat{\psi}_{K'\bar{B}\sigma}(\mathbf{x})
\end{array}\right],~\sigma=\pm
\end{eqnarray}
where $A,\tilde{B}$ represent the most far apart sublattices, $K,K'$ the two valleys and $\sigma$ the spin polarization. We note that the two sublattices are interchanged in the $K'$ valley so, as usually done, we will refer to the corresponding subspace as $\bar{A}\bar{B}$ in order to avoid confusions. The $8$ components of the field operator then correspond to the total subspace $KK'\otimes\bar{A}\bar{B} \otimes s$, $s$ being the spin subspace. As electrons are fermions, the components $\mu,\nu$ of the field operator obey the canonical {\it anti-commutation} rules:
\begin{equation}\label{eq:anticommie}
\left\{\hat{\psi}^{\mu}(\mathbf{x}),\hat{\psi}^{\nu}(\mathbf{x'})\right\}=\delta_{\mu\nu}\delta(\mathbf{x}-\mathbf{x'})
\end{equation}
in contrast to the canonical {\it commutation} rules for bosons, discussed in Sec. \ref{subsec:timeindependent}.
Ignoring trigonal warping effects (they are negligible for sufficiently high magnetic field in the $\nu=0$ quantum Hall state \cite{McCann2006,Sari87,Toke2013}, the main object of study of our work), we can write the effective Hamiltonian considered in this work as $\hat{H}=\hat{H}_0+\hat{H}_C+\hat{H}_{sr}$. The single particle Hamiltonian, $\hat{H}_0$, is given by:
\begin{equation}\label{eq:spHamiltonian}
\hat{H}_0 = \int\mathrm{d}^2\mathbf{x}~\hat{\psi}^{\dagger}(\mathbf{x})H(p)\hat{\psi}(\mathbf{x}),~H(p) = -\frac{1}{2m}\left[\begin{array}{cc}
0 & (P_x-i P_y)^2\\
(P_x+i P_y)^2 & 0
\end{array}\right]
\end{equation}
where the matrix $H$ acts in the $\bar{A}\bar{B}$ subspace and $P_i=-i\hbar \partial_i$, $i=1,2$, are the components of the momentum operator in the two-dimensional plane of the graphene sample. The effective mass $m$ takes the value $m=0.028m_e$, with $m_e$ the electron mass \cite{Mayorov2011}. This term represents the kinetic energy contribution. In the second place, $\hat{H}_C$ represents the long range Coulomb interaction:
\begin{equation}\label{eq:coulombHamiltonian}
\hat{H}_C=\frac{1}{2}\int\mathrm{d}^2\mathbf{x}~\mathrm{d}^2\mathbf{x'}:[\hat{\psi}^{\dagger}(\mathbf{x})\hat{\psi}(\mathbf{x})]V_0(\mathbf{x}-\mathbf{x'})[\hat{\psi}^{\dagger}(\mathbf{x'})\hat{\psi}(\mathbf{x'})]:
\end{equation}
Here, $:$ denotes normal ordering of the field operators and $V_0(\mathbf{x})=e^2_c/\kappa|\mathbf{x}|$ is the Coulomb potential, with $e^2_c\equiv e^2/4\pi\epsilon_0$ and $\kappa$ the dielectric constant of the environment. Finally, for the short-range interaction Hamiltonian, $\hat{H}_{sr}$, we consider the most general expression compatible with all the symmetries of the problem \cite{Kharitonov2012}
\begin{equation}\label{eq:srHamiltonian}
\hat{H}_{sr}=\sideset{}{'}\sum_{i,j}\frac{1}{2}\frac{4\pi\hbar^2}{m}\int\mathrm{d}^2\mathbf{x}\ g_{ij}:[\hat{\psi}^{\dagger}(\mathbf{x})T_{ij}\hat{\psi}(\mathbf{x})]^2:
\end{equation}
where $T_{ij}=\tau_i^{KK'}\otimes\tau_j^{\bar{A}\bar{B}}\otimes\hat{\mathbf{1}}^s$ and $i,j=0,x,y,z$ with $\tau_i$, $i=x,y,z$, the usual Pauli matrices and $\tau_0=\hat{\mathbf{1}}$; the $'$ in the sum denotes that we exclude the symmetric term $i=j=0$, already accounted by the  long-range Coulomb interaction. Note that, within this convention, the coupling constants $g_{ij}$ are dimensionless. These interactions are asymmetric in the valley and sublattice subspaces. The origin of these short-range interactions are the Coulomb interaction between sublattice/valley subspaces and also the electron-phonon interactions, which we also treat as short-ranged \cite{Kharitonov2012}. Except for the form of the quadratic kinetic energy term (\ref{eq:spHamiltonian}), this Hamiltonian is formally similar to that of monolayer graphene, see Sec. \ref{sec:MLG}.

We now consider a more general situation in which we add a magnetic field $B$ and a voltage between the two layers $Ea_z$, with $E$ the perpendicular electric field and $a_z\approx 0.35~\text{nm}$ the separation between the layers. As a consequence, the single-particle Hamiltonian (\ref{eq:spHamiltonian}) is modified and reads
\begin{equation}\label{eq:spBHamiltonian}
\hat{H}_{0}=\int\mathrm{d}^2\mathbf{x}~\hat{\psi}^{\dagger}(\mathbf{x})\left[H(\pi)+\epsilon_VT_{zz}-\epsilon_Z\sigma_z\right]\hat{\psi}(\mathbf{x}),~\epsilon_V=\frac{Ea_z}{2}
\end{equation}
where $\sigma_z$ is the corresponding Pauli matrix in the spin space. In the following, the Pauli matrices in valley or sublattice space are denoted using the letter $\tau$ and the Pauli matrices in spin space are denoted using the letter $\sigma$. The first term is still the kinetic energy contribution, but replacing $P_i$ by
\begin{equation}\label{eq:peierlssub}
\pi_i=P_i+eA_i
\end{equation}
The second term arises from the voltage difference between the two layers. The third term takes into account the Zeeman effect with $\epsilon_Z=\mu_BB, B=\sqrt{B_{\parallel}^2+B_{\perp}^2}$. Here, we assume that the magnetic field is not necessarily in the $z$ direction, i.e., it can present a parallel component to the graphene plane $\mathbf{B}=[B_{\parallel}\cos\phi_B,B_{\parallel}\sin\phi_B,B_{\perp}]$. The polarizations $\sigma=\pm$ correspond to the spin components that are antiparallel (parallel) to the total magnetic field $\mathbf{B}$, respectively. We may define the operator
\begin{equation}\label{eq:aniquilacion}
a_B=\frac{l_B}{\hbar}\frac{\pi_y+i\pi_x}{\sqrt{2}},~l_B=\sqrt{\frac{\hbar}{eB_{\perp}}}=\frac{25.7}{\sqrt{B_{\perp}[\text{T}]}}~\text{nm},
\end{equation}
$l_B$ being the magnetic length. This operator is an annihilation-type operator that satisfies the usual commutation rule $[a_B,a_B^{\dagger}]=1$, similar to that of the harmonic oscillator Hamiltonian discussed in Sec. \ref{subsec:staHO}. The kinetic energy term of Eq. (\ref{eq:spBHamiltonian}) is then rewritten in terms of $a_B,a_B^{\dagger}$ as
\begin{equation}\label{eq:BilayerdestructionHO}
H(\pi)=\hbar\omega_B\left[\begin{array}{cc}
0 & a_B^2\\
(a_B^{\dagger})^2 & 0
\end{array}\right],~\omega_B=\frac{eB_{\perp}}{m}=1.76\times 10^{11}\frac{m_e}{m}B_{\perp}[\text{T}]~\text{Hz}=6.28\times 10^{12} B_{\perp}~\text{Hz}
\end{equation}
Taking into account that only the perpendicular magnetic field affects the orbital motion and using the Landau gauge, we can write the potential vector in Eq. (\ref{eq:peierlssub}) as $\mathbf{A}(\mathbf{x})=[0,B_{\perp}x,0]$. In this particular gauge, the eigenstates the operator (\ref{eq:BilayerdestructionHO}) are characterized by the following quantum numbers: the magnetic index $n$, which is an integer number and characterizes the energy of the corresponding Landau level, $\epsilon_n$, the momentum in the $y$-direction $k$ and the polarization in the $KK'\otimes s$ subspace $\alpha$. Specifically, they are given by $\Psi^{0}_{n,k,\alpha}(\mathbf{x})=\Psi^{0}_{n,k}(\mathbf{x})\chi_{\alpha}$, where $\chi_{\alpha}$ is an arbitrary 4-component spinor in the valley-spin subspace and the orbital wave function with components in the subspace $\bar{A}\bar{B}$ is
\begin{equation}\label{eq:Landaueigenfunctions}
\Psi^{0}_{n,k}(\mathbf{x})=\frac{e^{iky}}{\sqrt{L_y}}\frac{1}{\sqrt{2}}\left[\begin{array}{c}
\textrm{sgn}\ n\ \phi_{|n|-2}(x+kl^2_B)\\ \phi_{|n|}(x+kl^2_B)
\end{array}\right],~\epsilon_n=\textrm{sgn}\ n\ \sqrt{|n|(|n|-1)}\hbar\omega_B
\end{equation}
for $|n|\neq 0,1$ and
\begin{equation}\label{eq:ZLL}
\Psi^{0}_{n,k}(\mathbf{x})=\frac{e^{iky}}{\sqrt{L_y}}\left[\begin{array}{c}
0\\
\phi_{|n|}(x+kl^2_B)
\end{array}\right],~\epsilon_n=0
\end{equation}
for the degenerate levels $|n|=0,1$ with zero energy (note that $n=\pm 1$ are indeed the same state). Hereafter, we refer to this manifold of states as the ZLL (zero Landau level). In the previous equations, $L_y$ is the length of the system in the $y$ direction and $\phi_n(x)$ is the usual harmonic oscillator wave function, see Eq. \ref{eq:oscillatorwavefunctions} for its explicit expression. The kinetic energy does not depend on the polarization $\chi_{\alpha}$ as it is degenerate in that subspace. Interestingly, the wave functions in the ZLL only have non-vanishing components in the subspace $KK'\otimes\bar{B}\otimes s$. The field operator is decomposed in terms of the previous eigenfunctions as
\begin{equation}\label{eq:fieldoperatorLL}
\hat{\psi}(\mathbf{x})=\sideset{}{'}\sum_{n=-\infty}^{\infty}\sum_{k,\alpha}\Psi^0_{n,k,\alpha}(\mathbf{x})\hat{c}_{n,k,\alpha}
\end{equation}
where $'$ means that $n$ takes every integer value except $n=-1$.

\section{Effective model and mean-field phase diagram}\label{sec:effmod}
\sectionmark{Effective model and mean-field phase diagram}

\subsection{Projection onto the lowest Landau Level}\label{subsec:projection}
In this section, we develop an effective model in order to study the low-energy dynamics of the $\nu=0$ quantum Hall state, which corresponds to a half-filling of the ZLL and complete filling of all the LLs with $n\leq -2$. First, we make an estimation of the order of magnitude of the different terms in the Hamiltonian compared to the typical energy  between LLs $\hbar\omega_B$. For the Zeeman term, we have
\begin{equation}
\frac{\epsilon_Z}{\hbar\omega_B}=0.014\frac{B}{B_{\perp}} \ll 1
\end{equation}
for typical values of the ratio $B/B_{\perp}$. For the Coulomb interaction,
\begin{equation}\label{eq:coulombfactor}
F_c=\frac{e^2_c}{\kappa l_B\hbar\omega_B}=\frac{13.58}{\kappa\sqrt{B_{\perp}[\text{T}]}}
\end{equation}
Usual values for the magnetic field are $B\gtrsim 1~\text{T}$. The highest available value of continuous magnetic field in the laboratory is $\lesssim 45~\text{T}$, which means that the dimensionless strength of the Coulomb interaction verifies $F_c\gtrsim1$ when the environment is the vacuum ($\kappa=1$).

Finally, a dimensional analysis of the short-range terms gives an order of magnitude estimate $E_{sr}\sim e^2_ca_L/\kappa l^2_B$ with $a_L\approx0.14~\text{nm}$ the nearest-neighbor constant. Then:
\begin{equation}\label{eq:shortrangeestimation}
\frac{E_{sr}}{\hbar\omega_B}\sim F_c \frac{a_L}{l_B}=\frac{0.028\times a_L}{\kappa a_0}\simeq\frac{0.1}{\kappa} \ll 1
\end{equation}
with $a_0$ the Bohr radius. We see that the only term that could not be small compared to the separation between LLs is that related with the Coulomb interaction.

At the present moment, we treat $F_c$ as a small parameter by taking $F_c\lesssim 1$. For instance, by supposing a typical value $\kappa\sim5$ and magnetic field $B\sim~20~T$, we can achieve $F_c\sim 0.5$, which can be regarded as sufficiently small to treat it perturbatively. In Sec. \ref{sec:renorm}, we address the usual situation $F_c\gtrsim1$ and explain how to deal with it.

Taking into account the previous considerations, we neglect LL mixing and restrict ourselves to the ZLL, by projecting the full Hamiltonian into that subspace \cite{Kharitonov2012,Kharitonov2012PRL}. We remark, see Eq. (\ref{eq:ZLL}) and ensuing discussion, that the states in the ZLL belong to the $KK'\otimes\bar{B}\otimes s$ subspace, which means that they are localized, for each valley, in one sublattice or the other and correspondingly, in one layer or the other; see Eq. (\ref{eq:BLGfieldoperator}) and related discussion. Thus, within the ZLL, the sublattice degree of freedom becomes equivalent to the valley degree of freedom. The resulting effective Hamiltonian for the ZLL is:
\begin{eqnarray}\label{eq:EffectiveHamiltonian}
\nonumber\hat{H}^{(0)}&=&\int\mathrm{d}^2\mathbf{x}~\hat{\psi}^{\dagger}(\mathbf{x})\left[-\epsilon_VT_z-\epsilon_Z\sigma_z\right]\hat{\psi}(\mathbf{x})+\frac{1}{2}\int\mathrm{d}^2\mathbf{x}~\mathrm{d}^2\mathbf{x'}:[\hat{\psi}^{\dagger}(\mathbf{x})\hat{\psi}(\mathbf{x})]V_0(\mathbf{x}-\mathbf{x'})[\hat{\psi}^{\dagger}(\mathbf{x'})\hat{\psi}(\mathbf{x'})]:\\
&+&\sum_{i}\frac{1}{2}\frac{4\pi\hbar^2}{m}\int\mathrm{d}^2\mathbf{x}\ g_{i}:[\hat{\psi}^{\dagger}(\mathbf{x})T_{i}\hat{\psi}(\mathbf{x})]^2:+\int\mathrm{d}^2\mathbf{x}~\mathrm{d}^2\mathbf{x'}\hat{\psi}^{\dagger}(\mathbf{x})V_{DS}(\mathbf{x},\mathbf{x'})\hat{\psi}(\mathbf{x'})
\end{eqnarray}
where $T_{i}\equiv\tau^{KK'}_i$ and $g_{i}=g_{i0}+g_{iz}, i=1,2,3$. The potential $V_{DS}(\mathbf{x},\mathbf{x'})$ represents the mean-field interaction of the ZLL with the Dirac sea compound by all the occupied states with $n\leq -2$ \cite{Shizuya2012,Shizuya2013,Toke2013}. The explicit expression of $V_{DS}(\mathbf{x},\mathbf{x'})$ is given in Eq. (\ref{eq:HFprojected}); check Appendix \ref{app:magneticFF} for more details about the obtention of the previous effective Hamiltonian. As we stay in a single LL, the kinetic term is quenched. We have also neglected the symmetric short-range interaction, arising from the coupling $g_{0z}$, due to its smallness compared to the symmetric Coulomb interaction \cite{Kharitonov2012}. Using symmetry considerations, it can be proven that $g_x=g_y\equiv g_{\perp}$ \cite{Aleiner2007,Kharitonov2012}, so there are only two independent coupling constants $g_{\perp},g_z$. The projection into the ZLL leads to a field operator of the form
\begin{equation}\label{eq:fieldoperatorZLL}
\hat{\psi}(\mathbf{x})=\sum_{n=0,1}\sum_{k,\alpha}\Psi^0_{n,k,\alpha}(\mathbf{x})\hat{c}_{n,k,\alpha}
\end{equation}

\subsection{Hartree-Fock approximation and phase diagram}\label{subsec:HF}

In order to obtain the zero temperature mean-field phase diagram, we use the Hartree-Fock (HF) approximation for the self-consistent single-particle wave functions, which we denote as $\Psi_{n,k,\alpha}$. For a complete discussion about the Hartree-Fock approximation, see for instance Ref. \cite{Fetter2003}. Note that this kind of mean-field scheme is similar to the Hartree-Fock-Bogoliubov approximation considered in Sec. \ref{subsec:interactingrole} to study interactions between particles. The corresponding HF equations for the Hamiltonian (\ref{eq:EffectiveHamiltonian}) read
\begin{eqnarray}\label{eq:HFeqs}
\epsilon_{n,\alpha}\Psi_{n,k,\alpha}(\mathbf{x})&=&\int\mathrm{d}^2\mathbf{x'}~V_{DS}(\mathbf{x},\mathbf{x'})\Psi_{n,k,\alpha}(\mathbf{x'})\\
\nonumber&-&\sum_{m,p,\beta}\int\mathrm{d}^2\mathbf{x'}~V_0(\mathbf{x}-\mathbf{x'})\nu_{m,\beta}\Psi_{m,p,\beta}(\mathbf{x})\Psi_{m,p,\beta}^{\dagger}(\mathbf{x'})\Psi_{n,k,\alpha}(\mathbf{x'})\\
\nonumber&+&\frac{4\pi\hbar^2}{m}\sum_{i}\sum_{m,p,\beta}g_{i}\nu_{m,\beta}\left([\Psi_{m,p,\beta}^{\dagger}(\mathbf{x})T_{i}\Psi_{m,p,\beta}(\mathbf{x})]T_{i}\Psi_{n,k,\alpha}(\mathbf{x})\right.\\
\nonumber&-&\left. T_{i}\Psi_{m,p,\beta}(\mathbf{x})\Psi_{m,p,\beta}^{\dagger}(\mathbf{x})T_{i}\Psi_{n,k,\alpha}(\mathbf{x})\right)-\epsilon_VT_z\Psi_{n,k,\alpha}(\mathbf{x})-\epsilon_Z\sigma_z\Psi_{n,k,\alpha}(\mathbf{x})
\end{eqnarray}
where the indices $n,m$ label the magnetic levels, $k,p$ are the momenta in the $y$-direction and $\alpha,\beta$ represent the polarization in the $KK'\otimes s$ subspace. We consider homogeneous solutions in which every orbital $p$ is filled in the same way so the occupation number $\nu_{n,p,\alpha}$ of every state solely depends on $n$ and $\alpha$, $\nu_{n,p,\alpha}=\nu_{n,\alpha}$. In our projected model, the indices $n,m$ take the values $0,1$. As usual, the direct (Hartree) term for the Coulomb interaction is suppressed by the positive charge background. An important result is that the orbital part of the self-consistent HF wave functions are equal to that of the non-interacting wave functions, given in Eq. (\ref{eq:ZLL}); see Appendix \ref{app:magneticFF} for a detailed discussion on the diagonalization of the HF equations.

Thus, the only remaining task is to specify the spinors $\chi_{\alpha}$ in the $KK'\otimes s$ subspace. In the $\nu=0$ QH state, only half of the ZLL is filled. Thus, for each value of the $y$-momentum $k$, only four states of the eightfold degenerate space, formed by the $0,1$ magnetic states and the $KK'\otimes s$ subspace, are occupied. In order to minimize the Coulomb interaction, the electrons occupy in the same way the valley-spin subspace for the two magnetic levels, i.e., $\nu_{0,a}=\nu_{0,b}=\nu_{1,a}=\nu_{1,b}=1$, with $\chi_{a,b}$ being two orthogonal spinors \cite{Abanin2009,Kharitonov2012PRL}. The four remaining unoccupied states of the ZLL are characterized by the orthogonal spinors $\chi_{c,d}$. In this way, the polarization index $\alpha$ takes the values $\alpha=a,b,c,d$ corresponding to an orthonormal base of the $KK'\otimes s$ subspace.

\begin{figure}[tb]
\includegraphics[width=1\columnwidth]{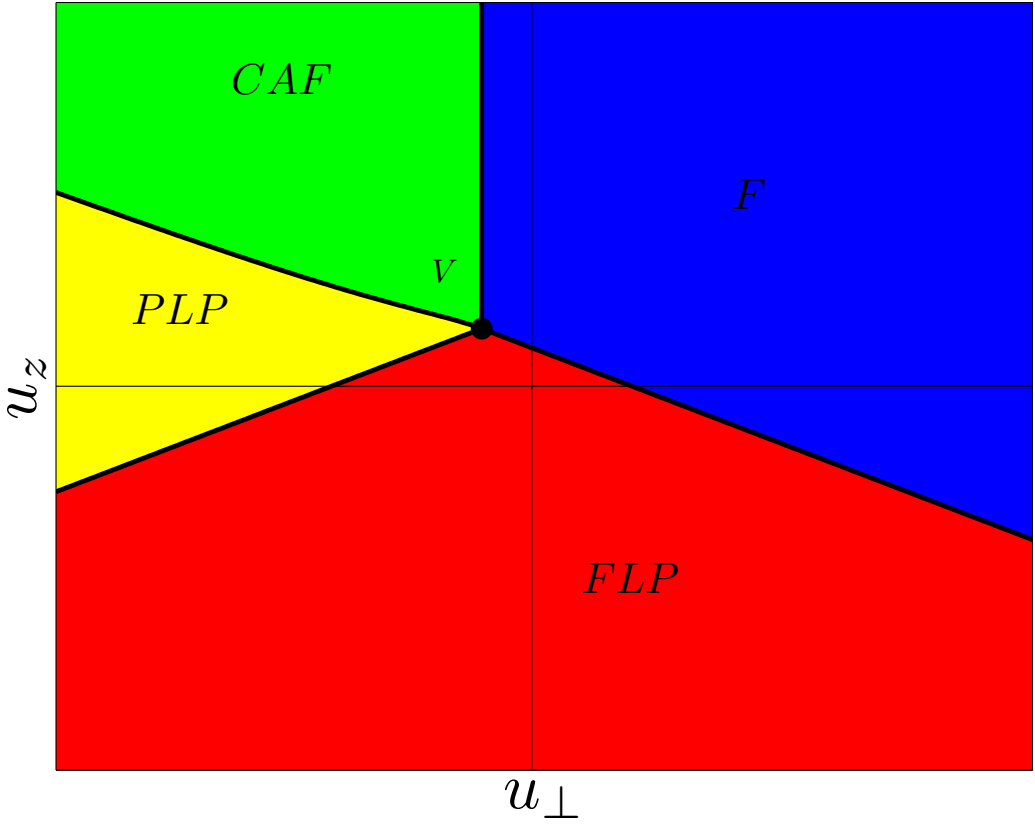}
\caption{Phase diagram of the $\nu=0$ QH state in the parameter space $u_{\perp},u_{z}$. The point $V$ is the intersection of all phase boundaries.}
\label{fig:PhaseDiagram}
\end{figure}

By projecting into the orbital part of the wave functions, we obtain closed algebraic equations for the spinorial part (check Appendix \ref{app:magneticFF} for more details):
\begin{equation}\label{eq:HFenergy}
\epsilon_{n,\alpha}\chi_{\alpha}=\frac{F_n}{2}\chi_{\alpha}-F_nP\chi_{\alpha}+\sum_{i}u_{i}\left([\text{tr}(PT_{i})]T_{i}-T_{i}PT_{i}\right)\chi_{\alpha}-\epsilon_VT_z\chi_{\alpha}-\epsilon_Z\sigma_z\chi_{\alpha}
\end{equation}
where $u_{i}=4\hbar \omega_B g_{i}$ and $F_n=F_{n0}+F_{n1}$; see Eqs. (\ref{eq:HFCoulombEnergies}), (\ref{eq:Fockeigenvalues}) for their explicit expressions. The term $F_{n}/2$ is the analog of the Lamb shift, arising from the interaction with the Dirac sea \cite{Shizuya2012,Shizuya2013}. The term $-F_nP$ arises from the exchange Coulomb interaction, where the matrix $P$ is the projector onto the subspace formed by $\chi_{a,b}$, $P=\chi_{a}\chi^{\dagger}_{a}+\chi_{b}\chi^{\dagger}_{b}$ and hence $P\chi_{a,b}=\chi_{a,b},~P\chi_{c,d}=0$. The remaining terms are those related with the short-range interactions and the single-particle Hamiltonian. We see that the sole dependence on the magnetic level is through the Coulomb and Lamb-shift terms, while the other contributions to the energy only depend on the spinor $\chi_{\alpha}$. Following the previous reasonings, we write the HF energies as:
\begin{equation}\label{eq:HFenergystructure}
\epsilon_{n,(a,b)}=-\frac{F_n}{2}+\epsilon_{(a,b)},~\epsilon_{n,(c,d)}=\frac{F_n}{2}+\epsilon_{(c,d)}
\end{equation}
with $\epsilon_{\alpha}$ depending only on the polarization $\alpha$. The energy of the ground state (per wave vector orbital) is:
\begin{eqnarray}\label{eq:meanfieldenergy}
\nonumber E_{HF}&=&-\frac{(F_0+F_1)}{2}+2E(P)\\
E(P)&=&\frac{1}{2}\sum_{i}u_{i}\left\{[\text{tr}(PT_{i})]^2-\text{tr}(T_{i}PT_{i}P)\right\}-\epsilon_V\text{tr}(PT_{z})-\epsilon_Z\text{tr}(P\sigma_z)
\end{eqnarray}
This result is analog to that obtained in Refs. \cite{Kharitonov2012,Kharitonov2012PRL}. The contribution from the Coulomb interaction to the total energy turns to be degenerate and does not depend on the specific form of the occupied spinors $\chi_{a,b}$. Then, the actual ground state is determined by comparing the energies corresponding to all possible solutions to the HF equations and selecting that with lower energy $E(P)$. The corresponding mean-field phase diagram for the $\nu=0$ QH state is represented in Fig. \ref{fig:PhaseDiagram}. The different possible phases are ferromagnetic (F), canted anti-ferromagnetic (CAF), fully layer-polarized (FLP) and partially layer-polarized (PLP) \cite{Kharitonov2012,Kharitonov2012PRL}.

\subsubsection{Ferromagnetic phase}\label{subsec:Fphase}

In the F phase, all the electrons have the spin aligned with the magnetic field. The complete set of solutions involves the following 4 spinors in the $KK'\otimes s$ subspace with eigenvalues:
\begin{eqnarray}\label{eq:Fphase}
\nonumber \chi_{a}=\ket{n_z}\otimes\ket{s_z},~\chi_{b}&=&\ket{-n_z}\otimes
\ket{s_z},~\chi_{c}=\ket{n_z}\otimes\ket{-s_z},~\chi_{d}=\ket{-n_z}\otimes\ket{-s_z}\\
\nonumber\epsilon_{a,b}&=&-2u_{\perp}-u_z-\epsilon_Z\mp\epsilon_V,~\epsilon_{c,d}=\epsilon_Z\mp\epsilon_V\\
E(P)&=&-(2u_{\perp}+u_z)-2\epsilon_Z
\end{eqnarray}
Here, $\ket{\pm s_z}$ denotes the state with spin polarization $\sigma=\pm$ [see discussion after Eq. (\ref{eq:peierlssub})] and $\ket{n_z}=\ket{K},\ket{-n_z}=\ket{K'}$. The F phase is always a solution of the HF equations although it does not always correspond to the actual ground-state. It has a $Z_2$ symmetry corresponding to a flip in the valley $KK'$ space.

\subsubsection{Full layer-polarized phase}\label{subsec:FLPphase}

The FLP phase is the equivalent of the F phase but in the $KK'$ subspace, which means that all the electrons are concentrated on one layer (due to the equivalency of valley-sublattice-layer of the ZLL in BLG). The corresponding spinors are now:
\begin{eqnarray}\label{eq:FLPphase}
\nonumber \chi_{a}=\ket{n_z}\otimes\ket{s_z},~\chi_{b}&=&\ket{n_z}\otimes\ket{-s_z},~\chi_{c}=\ket{-n_z}\otimes\ket{s_z},~\chi_{d}=\ket{-n_z}\otimes
\ket{-s_z}\\
\nonumber\epsilon_{a,b}&=&u_z\mp\epsilon_Z-\epsilon_V,~\epsilon_{c,d}=-2u_{\perp}-2u_z\mp\epsilon_Z+\epsilon_V\\
E(P)&=&u_z-2\epsilon_V
\end{eqnarray}
In analogy to the F phase, it is always a solution to the HF equations but not necessarily the ground-state. It also has a $Z_2$ symmetry corresponding to flip the spins. By comparing the energies of the two phases, we can obtain the boundary between the F and FLP phases:

\begin{equation}\label{eq:FFLPborder}
u_{\perp}+u_z=\epsilon_V-\epsilon_Z
\end{equation}

\subsubsection{Canted anti-ferromagnetic phase}\label{subsec:CAFphase}

The previous phases would be the only possible phases if there were not short-range interactions, i.e., $u_{\perp}=u_z=0$. However, when taking into account these interactions, the system can exhibit canted anti-ferromagnetism or partially layer-polarization in order to minimize the interaction energy. In the CAF phase, we have that:

\begin{eqnarray}\label{eq:CAFphase}
\nonumber \chi_{a}=\ket{n_z}\otimes
\ket{s_a},~\chi_{b}&=&\ket{-n_z}\otimes\ket{s_b},~\chi_{c}=\ket{n_z}\otimes\ket{-s_a},~\chi_{d}=\ket{-n_z}\otimes\ket{-s_b}\\
\nonumber\epsilon_{a,b}&=&-u_z-2u_{\perp}\cos^2\theta_s-\epsilon_Z\cos\theta_s \mp\epsilon_V=-u_z\mp\epsilon_V \\
\nonumber \epsilon_{c,d}&=&-2u_{\perp}\sin^2\theta_s+\epsilon_Z\cos\theta_s \mp \epsilon_V=-2u_{\perp}\mp \epsilon_V\\
E(P)&=&-u_z-\epsilon_Z\cos\theta_s
\end{eqnarray}
where $s_{a,b}=[\pm\sin\theta_s\cos\phi_s,\pm\sin\theta_s\sin\phi_s,\cos\theta_s]$ and the tilting angle is $\cos\theta_s=-\epsilon_Z/2u_{\perp}$. As the azimut $\phi_s$ is a free parameter, we see that CAF phase has U(1) symmetry. When this solutions exists, it always has lower energy than the regular F phase, hence the condition for the presence of the CAF phase is just
\begin{equation}\label{eq:FCAFborder}
\cos\theta_s<1\Rightarrow u_{\perp}<-\frac{\epsilon_Z}{2}
\end{equation}

\subsubsection{Partially layer-polarized phase}\label{subsec:PLPphase}

In analogy with the relation between the FLP and F phases, the PLP is similar to the CAF phase but in the valley subspace:

\begin{eqnarray}\label{eq:PLPphase}
\nonumber \chi_{a}=\ket{\mathbf{n}}\otimes\ket{s_z},~\chi_{b}&=&\ket{\mathbf{n}}\otimes\ket{-s_z},~\chi_{c}=\ket{-\mathbf{n}}\otimes\ket{s_z},~\chi_{d}=\ket{-\mathbf{n}}\otimes\ket{-s_z}\\
\nonumber\epsilon_{a,b}&=&u_{\perp}\sin^2\theta_v+u_z\cos^2\theta_v\mp\epsilon_Z-\epsilon_V\cos\theta_v=u_{\perp}\mp\epsilon_Z \\
\nonumber \epsilon_{c,d}&=&-2u_{\perp}-u_z-u_{\perp}\sin^2\theta_v-u_z\cos^2\theta_v+\epsilon_V\cos\theta_v\mp \epsilon_Z=-3u_{\perp}-u_z\mp \epsilon_Z\\
E(P)&=&u_{\perp}\sin^2\theta_v+u_z\cos^2\theta_v-2\epsilon_V\cos\theta_v=u_{\perp}-\epsilon_V\cos\theta_v
\end{eqnarray}
with $\mathbf{n}=[\sin\theta_v\cos\phi_v,\sin\theta_v\sin\phi_v,\cos\theta_v]$ and $\cos\theta_v=\epsilon_v/(u_z-u_{\perp})$. The PLP phase also presents an $U(1)$ symmetry. In analogy with the F-CAF phase transition, whenever this solution exists, it has lower energy than the FLP phase. The existence condition for the PLP phase is
\begin{equation}\label{eq:FLPPLPborder}
\cos\theta_v<1\Rightarrow u_z-u_{\perp}>\epsilon_V
\end{equation}
On the other hand, the boundary between the FLP and the CAF phases is placed at
\begin{equation}\label{eq:PLPCAFborder}
u_z+u_{\perp}=\frac{\epsilon^2_V}{u_z-u_{\perp}}+\frac{\epsilon^2_Z}{2u_{\perp}}
\end{equation}

\subsubsection{Critical point}\label{subsec:Criticalpoint}

The resulting phase diagram from this HF scheme, depicted in Fig. \ref{fig:PhaseDiagram}, is the same as that of Ref. \cite{Kharitonov2012PRL}. It is straightforward to show that all phase boundaries intersect at the critical point
\begin{equation}\label{eq:criticalpoint}
V=(u^*_{\perp},u^*_z)=(-\epsilon_Z/2,\epsilon_V-\epsilon_Z/2).
\end{equation}

As discussed in the previous section, we only need two parameters to characterize the asymmetric interaction, $u_{\perp},u_z$. The expected phase for the $\nu=0$ QH state of bilayer graphene is the CAF phase at $\epsilon_V=0$ and zero in-plane component of the magnetic field \cite{Kharitonov2012a,Maher2013}, which implies that $u_z>-u_{\perp}>0$. The exact values of these short-range energies remain unknown but their order of magnitude is $u_z,|u_{\perp}|\sim 0.1 \hbar \omega_B$ \cite{Kharitonov2012b,Wu2014,Young2014}. In the following, we treat them as phenomenological inputs for the theory. We note that, since $u_{\perp}<0$, all the phases can be explored because, as Eq. \ref{eq:criticalpoint} reveals, the critical $V$ point can be placed anywhere in the whole semi-plane $u_{\perp}<0$ by just changing the in-plane component of the magnetic field or the layer voltage.

\section{Collective modes}\label{sec:colmod}

\subsection{Correlation functions and the time-dependent Hartree-Fock approximation}\label{subsec:dyson}

The collective mode frequencies are given by the poles of the set of correlation functions:
\begin{equation}\label{eq:corrfunc}
\Pi_A(x,x')=-i\braket{T\{\Delta[\hat{\psi}^{\dagger}(x)\theta_A^{\dagger}\hat{\psi}(x)]\Delta[\hat{\psi}^{\dagger}(x')\theta_A\hat{\psi}(x')]\}}
\end{equation}
where $x=(\mathbf{x},t)$, $T$ denotes time-ordering, all the expectation value are evaluated in the ground state of the system and $\theta_A$ is an arbitrary operator in the valley-spin subspace. As in Sec. \ref{sec:CSPH}, $\Delta \hat{O}=\hat{O}-\braket{\hat{O}}$ denotes the fluctuations of an operator around its mean value.

\begin{figure}[t]
\begin{fmffile}{Diagram}
\begin{eqnarray*}\nonumber
G=~\parbox{20mm}{\begin{fmfgraph}(50,50)
\fmfleft{i1} \fmfright{o1}
\fmf{dbl_plain_arrow}{i1,o1}
\end{fmfgraph}}
&=& \parbox{20mm}{\begin{fmfgraph}(50,50)
\fmfleft{i2} \fmfright{o2}
\fmf{plain_arrow}{i2,o2}
\end{fmfgraph}}
+~~\parbox{35mm}{\begin{fmfgraph}(100,100)
\fmfleft{i2} \fmfright{o2}
\fmf{dbl_plain_arrow}{i2,v2}
\fmf{dbl_plain_arrow}{v2,w2}
\fmf{plain_arrow}{w2,o2}
\fmf{boson,tension=0,left}{v2,w2}
\fmfdot{v2,w2}
\end{fmfgraph}}
~~~~+~~~\parbox{35mm}{\begin{fmfgraph}(80,40)
	\fmfleft{i2} \fmfright{o2} \fmftop{w2}
	\fmf{dbl_plain_arrow,tension=100}{i2,v2}
	\fmf{plain_arrow,tension=100}{v2,o2}
	\fmf{dbl_plain_arrow,tension=0.5}{w2,w2}
	\fmf{dashes}{v2,w2}
	\fmfdot{v2,w2}
\end{fmfgraph}}\\
&+&\nonumber \parbox{20mm}{\begin{fmfgraph}(100,100)
	\fmfleft{i2} \fmfright{o2}
	\fmf{dbl_plain_arrow}{i2,v2}
	\fmf{dbl_plain_arrow}{v2,w2}
	\fmf{plain_arrow}{w2,o2}
	\fmf{dashes,tension=0,left}{v2,w2}
	\fmfdot{v2,w2}
\end{fmfgraph}}
\end{eqnarray*}
\end{fmffile}
\caption{Diagrammatic representation of the HF equations. The double (single) lines represent the dressed (bare) Green's function. The wiggly line represents the Coulomb interaction and the dashed line the short-range interactions. The direct (Hartree) term of the Coulomb interaction is suppressed by the uniform positive charge background.}\label{fig:HFDiagram}
\end{figure}
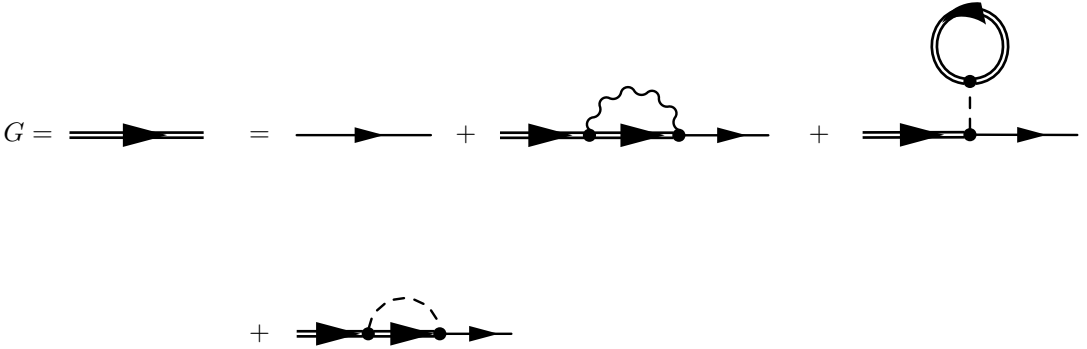

The previous correlation functions are computed within the TDHFA. We now briefly present the diagrammatic formalism considered, following closely Refs. \cite{Kallin1984,Wang2002}. The HF eigenfunctions and eigenvalues of Eq. (\ref{eq:HFeqs}) lead to the self-consistent Green's functions:
\begin{eqnarray}\label{eq:GreenZLLBare}
G^{\mu\nu}(\mathbf{x},\mathbf{x'},\omega)&=&\sum_{n,p,\alpha}\Psi^\mu_{n,p,\alpha}(\mathbf{x})\Psi^{\nu*}_{n,p,\alpha}(\mathbf{x'})G_{n,\alpha}(\omega)\\
\nonumber G_{n,\alpha}(\omega)&=&\frac{1-\nu_{n,\alpha}}{\omega-\omega_{n,\alpha}+i\eta}+\frac{\nu_{n,\alpha}}{\omega-\omega_{n,\alpha}-i\eta}
\end{eqnarray}
with $\omega_{n,\alpha}=\epsilon_{n,\alpha}/\hbar$, $\eta=0^{+}$ and $\mu,\nu$ labeling the components of the wave functions in the total subspace $KK'\otimes\bar{A}\bar{B} \otimes s$ (although we remember that the projection into the ZLL that we consider is restricted to the subspace $KK'\otimes\bar{B} \otimes s$, see discussion at the end of Sec. \ref{sec:basicmodel}). In the following, in order to simplify the notation, we use the dummy index $k$ to label all the quantum numbers $(n_k,p_k,\alpha_k)$ at the same time, with $n_k=0,1$, $p_k$ the $y$-momentum and $\alpha_k=a,b,c,d$ the polarization in the $KK'\otimes s$, $\chi_{a,b}$ being the spinors of the occupied states and $\chi_{c,d}$ those of the empty ones. For instance, the previous expression reads in this notation as
\begin{eqnarray}\label{eq:GreenZLL}
G^{\mu\nu}(\mathbf{x},\mathbf{x'},\omega)&=&\sum_{k}\Psi^\mu_{k}(\mathbf{x})\Psi^{\nu*}_{k}(\mathbf{x'})G_{k}(\omega)\\
\nonumber G_k(\omega)&=&\frac{1-\nu_{k}}{\omega-\omega_{k}+i\eta}+\frac{\nu_{k}}{\omega-\omega_{k}-i\eta}
\end{eqnarray}
The matrix elements of the total interaction potential
\begin{equation}\label{eq:totalpotential}
V_{\mu\nu,\lambda\sigma}(\mathbf{x}-\mathbf{x'})=V_0(\mathbf{x}-\mathbf{x'})\delta_{\mu\nu}\delta_{\lambda\sigma}+\sum_i \frac{4\pi\hbar^2}{m} g_i~\delta(\mathbf{x}-\mathbf{x'}) (T_i)_{\mu\nu}(T_i)_{\lambda\sigma}
\end{equation}
are given by
\begin{equation}\label{eq:potentialmatrixelements}
V_{lk,jm}\equiv\int\mathrm{d}^2\mathbf{x}\mathrm{d}^2\mathbf{x'}~\Psi^{\mu^*}_{l}(\mathbf{x})\Psi^{\nu}_{k}(\mathbf{x})V_{\mu\nu,\lambda\sigma}(\mathbf{x}-\mathbf{x'})\Psi^{\lambda^*}_{j}(\mathbf{x'})\Psi^{\sigma}_{m}(\mathbf{x'})
\end{equation}

In the previous identity, we have momentarily used Einstein summation convention for simplicity. The complete expression for these matrix elements is given in Eq. (\ref{eq:matrixelementsexplicit}). Using this notation, we can rewrite the HF equations using the Green's function and the interaction potentials as given by the diagrammatic representation of Fig. \ref{fig:HFDiagram}. For further details on the Green's function, the correlation function and the diagrammatic formalism, we refer the reader to Ref. \cite{Fetter2003}.

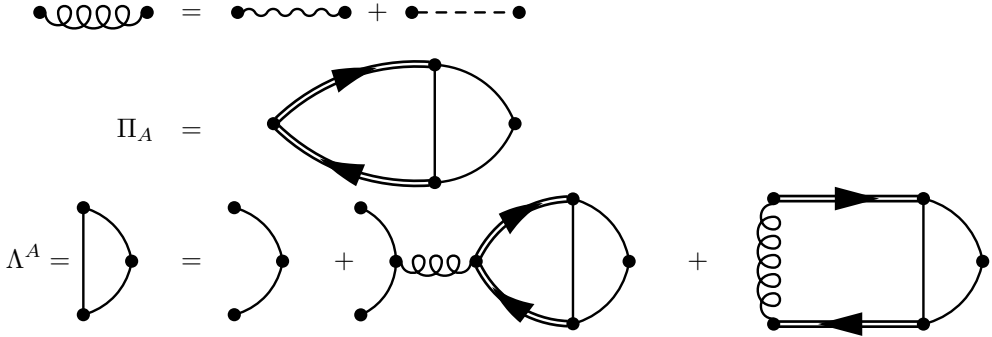
\begin{figure*}
		\begin{fmffile}{TDHFA}
			\begin{eqnarray*}
                    \nonumber
					\parbox{15mm}{\begin{fmfgraph}(40,40)
					\fmfleft{v1} \fmfright{v2}
					\fmf{curly}{v1,v2}
					\fmfdot{v1,v2}
					\end{fmfgraph}}&=&~
					\parbox{15mm}{\begin{fmfgraph}(40,40)
						\fmfleft{v1} \fmfright{v2}
						\fmf{boson}{v1,v2}
						\fmfdot{v1,v2}
						\end{fmfgraph}}~+~~
						\parbox{15mm}{\begin{fmfgraph}(40,40)
							\fmfleft{v1} \fmfright{v2}
							\fmf{dashes}{v1,v2}
							\fmfdot{v1,v2}
							\end{fmfgraph}}\\
				\nonumber
				\Pi_A&=&~~~~~\parbox{30mm}{\begin{fmfgraph*}(80,40)
				\fmfleft{i1} \fmfright{o1} \fmftopn{w}{10} \fmfbottomn{y}{10}
				\fmflabel{$\bar{A}$}{i1}
				\fmflabel{$A$}{o1}
				\fmf{heavy,left=0.225}{i1,w7}
				\fmf{plain,left=0.3,tension=1}{w7,o1}
				\fmf{plain,left=0.3,tension=1}{o1,y7}
				\fmf{heavy,left=0.225}{y7,i1}
				\fmf{plain}{w7,y7}
				\fmfdot{i1,o1,w7,y7}
				\end{fmfgraph*}}\\ \nonumber
			\Lambda^A=\parbox{10mm}{\begin{fmfgraph*}(20,40)
				\fmfright{o1}  \fmfleft{v2,v1}
				\fmflabel{$A$}{o1}
				\fmf{plain,left=0.3,tension=1}{v1,o1}
				\fmf{plain,left=0.3,tension=1}{o1,v2}
				\fmf{plain}{v1,v2}
				\fmfdot{o1,v1,v2}
				\end{fmfgraph*}}&=&
			\parbox{13mm}{\begin{fmfgraph*}(20,40)
				\fmfright{o1}  \fmfleft{v2,v1}
				\fmflabel{$A$}{o1}
				\fmf{plain,left=0.3,tension=1}{v1,o1}
				\fmf{plain,left=0.3,tension=1}{o1,v2}
				\fmfdot{o1,v1,v2}
				\end{fmfgraph*}}+
			\parbox{35mm}{\begin{fmfgraph*}(100,50)
				\fmfright{o1} \fmftopn{w}{20} \fmfbottomn{y}{20}
				\fmf{plain,left=0.3,tension=1}{w1,i1}
				\fmf{plain,left=0.3,tension=1}{i1,y1}
				\fmf{phantom}{w6,i1,y6}
				\fmffreeze
				\fmf{phantom}{w12,i2,y12}
				\fmf{curly}{i1,i2}
				\fmffreeze
				\fmflabel{$A$}{o1}
				\fmf{heavy,left=0.3}{i2,w16}
				\fmf{heavy,left=0.3}{y16,i2}
				\fmf{plain,left=0.3,tension=1}{w16,o1}
				\fmf{plain,left=0.3,tension=1}{o1,y16}
				\fmf{plain}{w16,y16}
				\fmfdot{w1,y1,i1,i2,w16,y16,o1}
				\end{fmfgraph*}}
			~~~~~~+
			\parbox{35mm}{\begin{fmfgraph*}(100,50)
				\fmfright{o1} \fmftopn{w}{10} \fmfbottomn{y}{10}
				\fmflabel{$A$}{o1}
				\fmf{heavy,tension=0}{w3,w8}
				\fmf{plain,left=0.3,tension=1}{w8,o1}
				\fmf{plain,left=0.3,tension=1}{o1,y8}
				\fmf{heavy}{y8,y3}
				\fmf{plain}{w8,y8}
				\fmf{curly}{w3,y3}
				\fmfdot{o1,w3,y3,w8,y8}
				\end{fmfgraph*}}
			\end{eqnarray*}
		\end{fmffile}
		\caption{Diagrammatic representation of the total potential (top), the correlation function $\Pi_A$, Eq. (\ref{eq:corrfunc}), (middle) and the vertex equation in the TDHFA, Eq. (\ref{eq:vertexequation}), (bottom).}\label{fig:TDHFADiagram}
\end{figure*}

The Fourier transform of the correlation function gives:
\begin{equation}\label{eq:FTcorrfunc}
\Pi_A(\mathbf{k},\omega)=\int\mathrm{d}^3x~e^{-ikx}\Pi_A(x,0),~\mathrm{d}^3x\equiv \mathrm{d}^2\mathbf{x}\mathrm{d}t,~kx=\mathbf{k}\mathbf{x}-\omega t
\end{equation}
where we are taking into account that the correlation function is invariant under time translations (due to the fact that the Hamiltonian is time independent) and also under spatial translations (this last property is shown explicitly later).

After introducing the dressed vertex function $\Lambda^A_{kl}(\mathbf{k'},\omega)$, Eq. (\ref{eq:FTcorrfunc}) reads:
\begin{equation}\label{eq:FTcorrvertex}
\Pi_A(k)=\Pi_A(\mathbf{k},\omega)=-\frac{i}{\hbar}\sum_{k,l}\int\frac{d\omega'}{2\pi}~\frac{1}{S}\sum_{\mathbf{k'}}\bar{A}_{kl}(\mathbf{k})G_k(\omega')G_l(\omega+\omega')\Lambda^A_{kl}(\mathbf{k'},\omega)
\end{equation}
In the above equation, we have made use of the identity $\sum_{\mathbf{k'}}e^{i\mathbf{k'}\mathbf{x'}}=S\delta(\mathbf{x'})$, $S$ being the total area. We remark that the index $k$ in Eq. (\ref{eq:FTcorrvertex}) labels the quantum numbers $(n_k,p_k,\alpha_k)$ and not momentum. The non-interacting vertex $\bar{A}_{kl}$ can be expressed using Dirac notation as $\bar{A}_{kl}=\braket{k|e^{-i\mathbf{k}\mathbf{x}}\theta_A^{\dagger}|l}$. The corresponding Dyson's equation for the vertex function is given, in the TDHFA, by:
\begin{eqnarray}\label{eq:vertexequation}
\nonumber\Lambda^A_{kl}(\mathbf{k'},\omega)&=&\Lambda^{A(0)}_{kl}(\mathbf{k'},\omega)+\frac{i}{\hbar}\sum_{j,m}\left[V_{jk,lm}-V_{lk,jm}\right]\int\frac{d\omega'}{2\pi}~G_j(\omega')G_m(\omega+\omega')\Lambda^A_{jm}(\mathbf{k'},\omega)\\
\Lambda^{A(0)}_{kl}(\mathbf{k'},\omega)&=&\braket{l|e^{i\mathbf{k'}\mathbf{x}}\theta_A|k}=\bar{A}^*_{kl}(\mathbf{k'})
\end{eqnarray}

The diagrams for the total bare interaction, the correlation function and the vertex equation in the TDHFA are shown in Fig. \ref{fig:TDHFADiagram}. Within this scheme, the dressed vertex function is the bare vertex (first term) plus the series corresponding to ladder (second term) and bubble (third term) diagrams. By defining the two-particle propagator
\begin{equation}\label{eq:electronholefunction}
i D_{kl}(\omega)\equiv \int\frac{d\omega'}{2\pi}~G_k(\omega')G_l(\omega+\omega')=i\left[\frac{\nu_k(1-\nu_l)}{\omega+\omega_k-\omega_l+i\eta}-\frac{\nu_l(1-\nu_k)}{\omega+\omega_k-\omega_l-i\eta}\right]~,
\end{equation}
we can rewrite Eqs. (\ref{eq:FTcorrvertex}),(\ref{eq:vertexequation}) as:

\begin{eqnarray}\label{eq:FTvertex}
\hbar\Pi_A(\mathbf{k},\omega)&=&\frac{1}{S}\sum_{k,l}\sum_{\mathbf{k'}}\bar{A}_{kl}(\mathbf{k})D_{kl}(\omega)\Lambda^A_{kl}(\mathbf{k'},\omega)\\
\nonumber \Lambda^A_{kl}(\mathbf{k'},\omega)&=&\Lambda^{A(0)}_{kl}(\mathbf{k'},\omega)+\frac{1}{\hbar}\sum_{j,m}\left[V_{lk,jm}-V_{jk,lm}\right]D_{jm}(\omega)\Lambda^A_{jm}(\mathbf{k'},\omega)
\end{eqnarray}

We note that the two-particle propagator $D_{kl}(\omega)$ does not depend on the momenta $p_k,p_l$ and it is only non-zero whenever the pair index $kl$ represents one filled level and one empty. The non-interacting vertex $\bar{A}_{kl}(\mathbf{k})$ is:
\begin{equation}
\bar{A}_{kl}(\mathbf{k})=\braket{k|e^{-i\mathbf{k}\mathbf{x}}\theta_A^{\dagger}|l}=e^{ik_x\frac{p_k+p_l}{2}l^2_B}A_{n_kn_l}(\mathbf{k})\delta_{p_l-p_k-k_y}\bar{\theta}_{kl}
\end{equation}
with $\bar{\theta}_{\alpha_k\alpha_l}=\chi^{\dagger}_{\alpha_k}\theta_A^{\dagger}\chi_{\alpha_l}$ and $A_{n_kn_l}(\mathbf{k})$ the usual magnetic form factor given in Eq. (\ref{eq:defmagneticFF}). We now simplify the correlation function as
\begin{eqnarray}\label{eq:Pidiagonalization}
\hbar\Pi_A(\mathbf{k},\omega)&=&\sum_{\substack{n_k,n_l\\ \alpha_k \alpha_l}}\sum_{\mathbf{k'}}~\bar{\theta}_{\alpha_k\alpha_l}A_{n_kn_l}(\mathbf{k})D_{kl}(\omega)L^A_{kl}(\mathbf{k},\mathbf{k'},\omega)\\
\label{eq:vertexdiagonalization}
L^A_{kl}(\mathbf{k},\mathbf{k'},\omega)&\equiv&\frac{1}{S}\sum_{p_{lk},\Delta p_{lk}}e^{ip_{lk}k_xl^2_B} \delta_{\Delta p_{lk},k_y}\Lambda^A_{kl}(\mathbf{k'},\omega)
\end{eqnarray}
In the last line, we have changed the sum in $p_k,p_l$ to the variables $p_{lk}=(p_k+p_l)/2$ and $\Delta p_{lk}=p_l-p_k$. The function $L^A_{kl}$ only depends in $k,l$ through the magnetic level and the valley-spin polarization as we are summing over all the momenta $p_k,p_l$. Then, thanks to these manipulations, we get rid of the momentum indices and simplify the vertex equation. For instance, for the bare vertex, $\Lambda^{A(0)}_{kl}(\mathbf{k'},\omega)$, we find:
\begin{equation}\label{eq:zeroL}
L^{(0)A}_{kl}(\mathbf{k},\mathbf{k'},\omega)=\frac{1}{S}\sum_{p_{lk},\Delta p_{lk}}e^{ip_{lk}k_xl^2_B}\delta_{\Delta p_{lk},k_y} \bar{A}^*_{kl}(\mathbf{k})=\frac{1}{2\pi l^2_B}\delta_{\mathbf{k},\mathbf{k'}}A^*_{n_kn_l}(\mathbf{k})\bar{\theta}^*_{\alpha_k\alpha_l}
\end{equation}
and we have used that the number of states $p_k$ per magnetic level $n_k$ is $N_B=S/2\pi l^2_B$. For the ladder and bubble diagrams, we obtain, after using the expression for the matrix elements given in Eq. (\ref{eq:matrixelementsexplicit}):
\begin{eqnarray}\label{eq:calculationladderRPA}
&~&\frac{1}{S}\frac{1}{\hbar}\sum_{p_{lk},\Delta p_{lk}}\sum_{j,m}e^{ip_{lk}k_xl^2_B}\delta_{\Delta p_{lk},k_y}
\left[V_{lk,jm}-V_{jk,lm}\right]D_{jm}(\omega)\Lambda^A_{jm}(\mathbf{k'},\omega)\\
\nonumber &~&=
\frac{1}{\hbar}\sum_{\substack{n_j,n_m\\ \alpha_j \alpha_m}}W_{lk,jm}(\mathbf{k})D_{jm}(\omega)L^A_{jm}(\mathbf{k},\mathbf{k'},\omega)
\end{eqnarray}
with
\begin{eqnarray}\label{eq:ladderRPA}
W_{lk,jm}(\mathbf{k})&=&U^{RPA}_{lk,jm}(\mathbf{k})-U^{LAD}_{jk,lm}(\mathbf{k})\\
\nonumber U^{RPA}_{lk,jm}(\mathbf{k})&=&\frac{1}{2\pi l^2_B}A_{n_ln_k}(-\mathbf{k})A_{n_jn_m}(\mathbf{k})U_{\alpha_l\alpha_k,\alpha_j\alpha_m}(\mathbf{k})\\
\nonumber U^{LAD}_{jk,lm}(\mathbf{k})&=&
\int\frac{\mathrm{d}^2\mathbf{q}}{(2\pi)^2}~e^{i(q_xk_y-q_yk_x)l^2_B}A_{n_jn_k}(-\mathbf{q})A_{n_ln_m}(\mathbf{q})U_{\alpha_j\alpha_k,\alpha_l\alpha_m}(\mathbf{q})\\
\nonumber U_{\alpha_l\alpha_k,\alpha_j\alpha_m}(\mathbf{k})&=&
V_0(\mathbf{k})\delta_{\alpha_l\alpha_k}\delta_{\alpha_j\alpha_m}+\frac{4\pi\hbar^2}{m}\sum_i g_i (T_i)_{\alpha_l\alpha_k}(T_i)_{\alpha_j\alpha_m}
\end{eqnarray}
$V_0(\mathbf{k})=2\pi e^2_c/|\mathbf{k}|$ being the Fourier transform of the Coulomb interaction and $(T_i)_{\alpha_l\alpha_k}=\chi^{\dagger}_{\alpha_l}T_i\chi_{\alpha_k}$. The quantities $U^{RPA},U^{LAD}$ take into account the energy contribution from bubble and ladder diagrams, respectively. Since we have performed the summation in all the momentum dummy variables, from now on we use the index $k$ to label just the quantum numbers $(n_k,\alpha_k)$. After inserting the expression for $\Lambda^A_{kl}(\mathbf{k},\omega)$ of Eq. (\ref{eq:FTvertex}) in Eq. (\ref{eq:vertexdiagonalization}) and using Eqs. (\ref{eq:zeroL})-(\ref{eq:ladderRPA}), one realizes that $L^A_{kl}(\mathbf{k},\mathbf{k'},\omega)$ is diagonal in $(\mathbf{k},\mathbf{k'})$. Hence, if we define a new dressed vertex function:
\begin{equation}\label{eq:newvertex}
L^A_{kl}(\mathbf{k},\mathbf{k'},\omega)\equiv \frac{1}{2\pi l^2_B}\delta_{\mathbf{k},\mathbf{k'}}\tilde{\Lambda}^{A}_{kl}(\mathbf{k},\omega)
\end{equation}
we have managed to transform the original integral equation (\ref{eq:vertexequation}) for the dressed vertex into a discrete matrix equation, given by:
\begin{eqnarray}\label{eq:diagonalvertexequation}
\tilde{\Lambda}^{A}_{kl}(\mathbf{k},\omega)&=&A^*_{n_kn_l}(\mathbf{k})\bar{\theta}^*_{\alpha_k\alpha_l}+\frac{1}{\hbar}\sum_{jm}W_{lk,jm}(\mathbf{k})D_{jm}(\omega)\tilde{\Lambda}^{A}_{jm}(\mathbf{k},\omega)
\end{eqnarray}
where we insist that the labels $l,k,j,m$ now only represent the pair index corresponding to the magnetic level $n$ and the valley-spin polarization $\alpha$. The correlation function then reads:
\begin{equation}\label{eq:PiAfinal}
\hbar\Pi_A(\mathbf{k},\omega)=\frac{1}{2\pi l^2_B}\sum_{kl}~\bar{\theta}_{\alpha_k\alpha_l}A_{n_kn_l}(\mathbf{k})D_{kl}(\omega)\tilde{\Lambda}^{A}_{kl}(\mathbf{k},\omega)
\end{equation}
The $\delta_{\mathbf{k},\mathbf{k'}}$ factor appearing in Eq. (\ref{eq:newvertex}) ensures the translational invariance of the correlation function [see Eq. (\ref{eq:FTcorrfunc}) and associated discussion]. In order to solve the matrix equation (\ref{eq:diagonalvertexequation}) we define:
\begin{eqnarray}
\label{eq:Polvector}\Pi^A_{kl}(\mathbf{k},\omega)&\equiv&\frac{1}{\hbar}D_{kl}(\omega)\tilde{\Lambda}^{A}_{kl}(\mathbf{k},\omega)\\\label{eq:dispersionmatrix}
X_{kl,jm}(\mathbf{k},\omega)&\equiv&W_{lk,jm}(\mathbf{k})-\delta_{kj}\delta_{lm} \hbar D^{-1}_{kl}(\omega)
\end{eqnarray}
The reader is advised to pay attention to the interchange of the indices $kl$ in the element $W_{lk,jm}$ for the definition of $X_{kl,jm}$. From these results, we find:
\begin{eqnarray}
\nonumber \sum_{jm} X_{kl,jm}(\mathbf{k},\omega)\Pi^A_{jm}(\mathbf{k},\omega)&=&-B^{A*}_{kl}(\mathbf{k})\\
B^{A}_{kl}(\mathbf{k})&=&A_{n_kn_l}(\mathbf{k})\bar{\theta}_{\alpha_k\alpha_l}
\end{eqnarray}
The matrix $X_{kl,jm}$ is referred as dispersion matrix in Ref. \cite{Wang2002}. The dispersion relation of the collective modes is computed from the equation
\begin{equation}\label{eq:detcollmod}
\det~[X(\mathbf{k},\omega)]=0
\end{equation}
as Eq. (\ref{eq:PiAfinal}) can be rewritten as:
\begin{equation}\label{eq:Pi}
\Pi_A(\mathbf{k},\omega)=-\frac{1}{2\pi l^2_B}\sum_{kljm}~B^{A}_{kl}(\mathbf{k})X^{-1}_{kl,jm}B^{A*}_{jm}(\mathbf{k})
\end{equation}
We clearly see from this relation that the poles of the correlation function, which give the collective-mode energies, are given by the values at which the matrix $X_{kl,jm}$ is singular. The matrices $X_{kl,jm}(\mathbf{k},\omega), W_{lk,jm}(\mathbf{k})$ present the following useful properties:
\begin{eqnarray}
\label{eq:hermiticw}W_{lk,jm}(\mathbf{k},\omega)&=&W^{*}_{mj,kl}(\mathbf{k})\\
\label{eq:symmetryminusomega} X_{kl,jm}(\mathbf{k},\omega)&=&X^{*}_{lk,mj}(-\mathbf{k},-\omega^*)
\end{eqnarray}
According to Eqs. (\ref{eq:electronholefunction}),(\ref{eq:Polvector}) and (\ref{eq:dispersionmatrix}), the only valid matrix elements of $X_{kl,jm}(\mathbf{k},\omega)$ are those where the index pair $kl$ correspond to one level occupied and the other level empty (the same goes for $jm$). The occupation number only depend on the valley-spin index, $\nu_k=\nu_{\alpha_k}$, as argued in Section \ref{subsec:HF}, and then the level unoccupied has necessarily different valley-spin polarization from the occupied level. This fact implies that for $\theta_A=1$, i.e., the case where $\Pi_A$ is the density-density correlation function, $\Pi_A(\mathbf{k},\omega)=0$. This is just an artifact of our model, as we restrict ourselves to the ZLL. When we allow arbitrary transitions between LLs a non-vanishing density correlation function is recovered (see Sec. \ref{sec:renorm}).

\subsection{Dispersion relation}\label{subsec:results}

\subsubsection{Preliminary remarks}\label{subsec:preliminar}
In this subsection, we compute the collective mode energies using the results derived in Sec. \ref{subsec:dyson}. As a first step, we calculate $D^{-1}_{kl}(\omega)$ from Eq. (\ref{eq:electronholefunction}):
\begin{equation}\label{eq:inverseD}
D^{-1}_{kl}(\omega)=\frac{\omega-(\omega_l-\omega_k)}{\nu_k-\nu_l}
\end{equation}
We remark that we follow the notation introduced after Eq. (\ref{eq:ladderRPA}) and $k$ labels at the same time the magnetic index and valley-spin polarization, $(n_k,\alpha_k)$. Thus, each index $k,l$ takes $8$ values, corresponding to the dimension of the space composed of the magnetic levels and the valley-spin subspace, $\ket{01}\otimes KK' \otimes s $. As mentioned in the discussion at the end of Sec. \ref{subsec:dyson}, the occupation numbers of the pair $kl$ must correspond to one filled level and one empty level, so we have $\nu_k-\nu_l=\pm 1$. We remind that the occupation numbers only depend on the valley-spin component $\alpha_k$ and $\alpha_l$. Specifically, if we denote the index pair $kl$ as normal (anomalous) when $\nu_{\alpha_k}-\nu_{\alpha_l}=1$ ($\nu_{\alpha_k}-\nu_{\alpha_l}=-1$),
we have that $kl$ is normal for $\alpha_k\alpha_l=ac,ad,bc,bd$ and anomalous when $\alpha_k\alpha_l=ca,da,cb,db$. Therefore, the collective modes here computed correspond to intra-LL valley-spin waves \cite{Iyengar2007}. A direct consequence of this fact is that the Coulomb contribution to the RPA energy, $U^{RPA}$, vanishes; see Eq. (\ref{eq:ladderRPA}). In respect to the orbital structure of the collective modes, it is well known that they are described in terms of the so-called magnetoexciton wave-function \cite{Kallin1984}.

The considerations of the previous paragraph induce a very special structure in the total matrix $X_{kl,jm}(\mathbf{k},\omega)$, which can be expressed using the $4\times4$ matrices $Y^{\alpha_k\alpha_l,\alpha_j\alpha_m}(\mathbf{k},\omega)$ as building blocks. These matrices result from considering all the possible values in the magnetic level subspace, $n_kn_l=00,01,10,11$, for a fixed set of values of the valley-spin components $(\alpha_k\alpha_l,\alpha_j\alpha_m)$, i.e.
\begin{eqnarray}\label{eq:Ymagneticstructure}
&~&Y^{\alpha_k\alpha_l,\alpha_j\alpha_m}(\mathbf{k},\omega)=\\
\nonumber &~&=\left[\begin{array}{cccc}
Y^{\alpha_k\alpha_l,\alpha_j\alpha_m}_{00,00}(\mathbf{k},\omega)&Y^{\alpha_k\alpha_l,\alpha_j\alpha_m}_{00,01}(\mathbf{k},\omega)&Y^{\alpha_k\alpha_l,\alpha_j\alpha_m}_{00,10}(\mathbf{k},\omega)&Y^{\alpha_k\alpha_l,\alpha_j\alpha_m}_{00,11}(\mathbf{k},\omega)\\
Y^{\alpha_k\alpha_l,\alpha_j\alpha_m}_{01,00}(\mathbf{k},\omega)&Y^{\alpha_k\alpha_l,\alpha_j\alpha_m}_{01,01}(\mathbf{k},\omega)&Y^{\alpha_k\alpha_l,\alpha_j\alpha_m}_{01,10}(\mathbf{k},\omega)&Y^{\alpha_k\alpha_l,\alpha_j\alpha_m}_{01,11}(\mathbf{k},\omega)\\
Y^{\alpha_k\alpha_l,\alpha_j\alpha_m}_{10,00}(\mathbf{k},\omega)&Y^{\alpha_k\alpha_l,\alpha_j\alpha_m}_{10,01}(\mathbf{k},\omega)&Y^{\alpha_k\alpha_l,\alpha_j\alpha_m}_{10,10}(\mathbf{k},\omega)&Y^{\alpha_k\alpha_l,\alpha_j\alpha_m}_{10,11}(\mathbf{k},\omega)\\
Y^{\alpha_k\alpha_l,\alpha_j\alpha_m}_{11,00}(\mathbf{k},\omega)&Y^{\alpha_k\alpha_l,\alpha_j\alpha_m}_{11,01}(\mathbf{k},\omega)&Y^{\alpha_k\alpha_l,\alpha_j\alpha_m}_{11,10}(\mathbf{k},\omega)&Y^{\alpha_k\alpha_l,\alpha_j\alpha_m}_{11,11}(\mathbf{k},\omega)\\
\end{array}\right]
\end{eqnarray}
with
\begin{eqnarray}\label{eq:Ydef}
Y^{\alpha_k\alpha_l,\alpha_j\alpha_m}_{n_kn_l,n_jn_m}(\mathbf{k},\omega)&\equiv& X_{kl,jm}(\mathbf{k},\omega)
\end{eqnarray}

The explicit form of $Y^{\alpha_k\alpha_l,\alpha_j\alpha_m}(\mathbf{k},\omega)$ is computed from Eqs. (\ref{eq:ladderRPA}), (\ref{eq:dispersionmatrix}) and reads:
\begin{eqnarray}\label{eq:pseudospinmatrices}
\nonumber Y^{\alpha_k\alpha_l,\alpha_j\alpha_m}(\mathbf{k},\omega)&=&[(\nu_{\alpha_k}-\nu_{\alpha_l})(E^{\alpha_k\alpha_l}-\hbar\omega I)-C(\mathbf{k})]\delta_{\alpha_k\alpha_j}\delta_{\alpha_l\alpha_m}+R(\mathbf{k})M^{\alpha_k\alpha_l,\alpha_j\alpha_m}\\
M^{\alpha_k\alpha_l,\alpha_j\alpha_m}&\equiv&\sum_i u_i[(T_i)_{\alpha_l\alpha_k}(T_i)_{\alpha_j\alpha_m}
-(T_i)_{\alpha_j\alpha_k}(T_i)_{\alpha_l\alpha_m}]~,
\end{eqnarray}
$I$ being the identity $4\times 4$ matrix. The matrix $C$ takes into account the (ladder) contribution from the Coulomb interaction while short-range interactions give rise to the $R$ matrix, which is the same for both ladder and bubble contributions. The expressions for these matrices are given in Appendix \ref{app:analyticalTDHFA}. The matrix $E^{\alpha_k\alpha_l}$ is diagonal and arises from the term $D^{-1}_{kl}(\omega)$; its elements are given by $E^{\alpha_k\alpha_l}_{n_kn_l,n_jn_m}=(\epsilon_{n_l,\alpha_l}-\epsilon_{n_k,\alpha_k})\delta_{n_kn_j}\delta_{n_ln_m}$.

On the other hand, as explained before, we only have to take into account indices $\alpha_k\alpha_l$ in such way that they correspond to one filled and one empty level. A simple count then reveals that the total matrix $X_{kl,jm}(\mathbf{k},\omega)$ is $32\times32$. We also see from Eq. (\ref{eq:pseudospinmatrices}) that the matrix $Y^{\alpha_k\alpha_l,\alpha_j\alpha_m}(\mathbf{k},\omega)$ can be separated as
\begin{equation}\label{eq:Ystructure}
Y^{\alpha_k\alpha_l,\alpha_j\alpha_m}(\mathbf{k},\omega)=Y^{\alpha_k\alpha_l,\alpha_j\alpha_m}(\mathbf{k})-(\nu_{\alpha_k}-\nu_{\alpha_l})\hbar\omega I,~Y^{\alpha_k\alpha_l,\alpha_j\alpha_m}(\mathbf{k})\equiv Y^{\alpha_k\alpha_l,\alpha_j\alpha_m}(\mathbf{k},\omega=0)
\end{equation}
This relation tells us that Eq. (\ref{eq:detcollmod}) is a generalized eigenvalue equation that reduces to the usual eigenvalue equation for $\omega$ ($-\omega$) for normal (anomalous) indices [this fact is at the origin of the symmetry (\ref{eq:symmetryminusomega})]. The conclusion reached is that the computation of the collective-mode frequencies amounts to the obtention of the eigenvalues of the following $32\times32$ matrix:
\begin{eqnarray}\label{eq:XYgen32}
&~&\tilde{X}_{kl,jm}(\mathbf{k})\equiv\tilde{\sigma}_zX_{kl,jm}(\mathbf{k},0)\\
\nonumber &~&=\left[\begin{array}{rrrr|rrrr}
Y^{ac,ac} & Y^{ac,ad} & Y^{ac,bc} & Y^{ac,bd} & Y^{ac,ca} & Y^{ac,da} & Y^{ac,cb} & Y^{ac,db}\\
Y^{ad,ac} & Y^{ad,ad} & Y^{ad,bc} & Y^{ad,bd} & Y^{ad,ca} & Y^{ad,da} & Y^{ad,cb} & Y^{ad,db}\\
Y^{bc,ac} & Y^{bc,ad} & Y^{bc,bc} & Y^{bc,bd} & Y^{bc,ca} & Y^{bc,da} & Y^{bc,cb} & Y^{bc,db}\\
Y^{bd,ac} & Y^{bd,ad} & Y^{bd,bc} & Y^{bd,bd} & Y^{bd,ca} & Y^{bd,da} & Y^{bd,cb} & Y^{bd,db}\\
\hline
- Y^{ca,ac} & - Y^{ca,ad} & - Y^{ca,bc} & - Y^{ca,bd} & - Y^{ca,ca} & - Y^{ca,da} & - Y^{ca,cb} & - Y^{ca,db}\\
- Y^{da,ac} & - Y^{da,ad} & - Y^{da,bc} & - Y^{da,bd} & - Y^{da,ca} & - Y^{da,da} & - Y^{da,cb} & - Y^{da,db}\\
- Y^{cb,ac} & - Y^{cb,ad} & - Y^{cb,bc} & - Y^{cb,bd} & - Y^{cb,ca} & - Y^{cb,da} & - Y^{cb,cb} & - Y^{cb,db}\\
- Y^{db,ac} & - Y^{db,ad} & - Y^{db,bc} & - Y^{db,bd} & - Y^{db,ca} & - Y^{db,da} & - Y^{db,cb} & - Y^{db,db}
\end{array}\right]
\end{eqnarray}
with $\tilde{\sigma}_z$ a $32\times 32$ diagonal matrix of the form $\tilde{\sigma}_z=\text{diag}[I,I,I,I,-I,-I,-I,-I]$. The horizontal and vertical lines in Eq. (\ref{eq:XYgen32}) separate the normal and anomalous sectors, yielding an analog structure to that of the bosonic BdG equations discussed along the first chapters. Indeed, the BdG equations can be regarded as generalized RPA equations \cite{Negele2008}.

Because the term between square brackets in Eq. (\ref{eq:pseudospinmatrices}) (which contains the electron-hole energy difference and the Coulomb interaction contributions) is multiplied by a diagonal tensor with respect to the valley-spin indices, the different valley-spin sectors of $X$ are only connected through the term arising from short-range interactions $M^{\alpha_k\alpha_l,\alpha_j\alpha_m}$, that is, for $\alpha_k\alpha_l\neq \alpha_j\alpha_m$
\begin{equation}\label{eq:pseudospinmatricesnondiagonal}
Y^{\alpha_k\alpha_l,\alpha_j\alpha_m}(\mathbf{k},0)=R(\mathbf{k})M^{\alpha_k\alpha_l,\alpha_j\alpha_m}
\end{equation}
The coefficients $M^{\alpha_k\alpha_l,\alpha_j\alpha_m}$, given in Eq. (\ref{eq:pseudospinmatrices}), present the following symmetries
\begin{equation}\label{eq:Melements}
M^{\alpha_k\alpha_l,\alpha_j\alpha_m}=(M^{\alpha_j\alpha_m,\alpha_k\alpha_l})^*=(M^{\alpha_l\alpha_k,\alpha_m\alpha_j})^*=-M^{\alpha_k\alpha_j,\alpha_l\alpha_m}
\end{equation}
The point of writing the matrix $X$ in this form is that the previous properties simplify notably the calculations since, in practice, many of the coefficients $M^{\alpha_k\alpha_l,\alpha_j\alpha_m}$ vanish and consequently we do not need to take into account the full $32\times 32$ matrix in order to compute the collective modes. In particular, whenever $M^{\alpha_k\alpha_l,\alpha_j\alpha_m}$ does not mix normal and anomalous sectors of the matrix $X_{kl,jm}$ [i.e., the off-diagonal boxes in Eq. (\ref{eq:XYgen32}) vanish], the computation of the dispersion relation is reduced to the computation of the eigenvalues of $\pm X_{kl,jm}(\mathbf{k},0)$ evaluated in the normal (anomalous) sector. If this is the case, the property (\ref{eq:hermiticw}) automatically guarantees that $\omega$ is real. In Appendix \ref{app:analyticalTDHFA}, we show that the dispersion relation is isotropic and hence only depends on $k=|\mathbf{k}|$ (not to be confused with the index $k$ in the matrix elements).

Interestingly, at $\mathbf{k}=0$, the dispersion relation can be obtained analytically after making a symmetric and antisymmetric combination of the magnetic indices of the form
\begin{equation}\label{eq:basechange}
S,A=\frac{00\pm 11}{\sqrt{2}}~,
\end{equation}
that yields a simplified matrix
\begin{eqnarray}\label{eq:zerofrequencyequation}
Y^{\alpha_k\alpha_l,\alpha_j\alpha_m}(\mathbf{k}=0)&=&\left(F+\Delta^{\alpha_k\alpha_l}I\right)\delta_{\alpha_k\alpha_j}\delta_{\alpha_l\alpha_m}+Q M^{\alpha_k\alpha_l,\alpha_j\alpha_m}\\
\nonumber F&=&\text{diag}[0,\frac{F_{00}+F_{11}}{2},\frac{F_{00}+F_{11}}{2},2F_{01}],~Q=\text{diag}[1,0,0,0]
\end{eqnarray}
where the order of the magnetic indices for rows and columns is now $S,01,10,A$. The quantity $\Delta^{\alpha_k\alpha_l}=|\epsilon_{\alpha_l}-\epsilon_{\alpha_k}|$ is the energy difference related only to the valley-spin part [see Eq. (\ref{eq:HFenergystructure})]. Thus, all matrices in Eq. (\ref{eq:zerofrequencyequation}) are already diagonal in the magnetic subspace. In particular, the only magnetic matrix element which is connected to other valley-spin sectors is that associated to the symmetric combination $S$ of Eq. (\ref{eq:basechange}). For the other magnetic matrix elements, the collective-mode frequencies are immediately obtained from the diagonal of $F+\Delta^{\alpha_k\alpha_l}I$. Hence, at $k=0$, the symmetric modes $S$ represent the lowest-frequency modes because they do not depend on the strength of the Coulomb interaction. We note that these modes are the same as the ``even'' modes of Ref. \cite{Toke2011} and they arise due to the exact conservation of the quantum number $l_z\equiv |n_l|-|n_k|$ at $\mathbf{k}=0$, that corresponds to the angular momentum of the magnetoexciton at zero momentum \cite{Iyengar2007}. The prolongations for $k\neq 0$ of the symmetric modes frequencies, $\omega_S(0)$, increase with $k$. Therefore, we conclude that the symmetric modes $S$ characterize the phase transitions as they represent the lowest energy excitations of the system and, remarkably, their calculation is always analytical. On the opposite limit, $kl_B\gtrsim 1$, only Coulomb interactions are relevant since $C(\mathbf{k})$ decays as $k^{-1}$ while $R(\mathbf{k})$ does it as
\begin{equation}\label{eq:exponentialdecayR}
R(\mathbf{k})\sim e^{-\frac{(kl_B)^2}{2}}~,
\end{equation}
see Appendix \ref{app:analyticalTDHFA} for the proof. We then expect to recover in this limit the results of Refs. \cite{Toke2011,Lambert2013} for the dispersion relation. For $kl_B\gg 1$, the collective modes approach to the mean-field excitations $\omega\simeq \omega_{n_l,\alpha_l}-\omega_{n_k,\alpha_k}$. For arbitrary momentum, the dispersion relation is computed numerically.

\subsubsection{Collective modes: numerical and analytical results}\label{subsec:numericalresults}

In this section, we obtain the dispersion relation for the different ground states obtained in Sec. \ref{subsec:HF}, in which the occupied (unoccupied) valley-spin polarizations were labeled as $a,b~(c,d)$, see Eqs. (\ref{eq:Fphase}),(\ref{eq:FLPphase}),(\ref{eq:CAFphase}),(\ref{eq:PLPphase}) to check the mean-field solutions and the notation. For all the phases, we give the values of the valley-spin energies $\Delta^{\alpha_k\alpha_l}$ and of the independent non-vanishing coefficients $M^{\alpha_k\alpha_l,\alpha_j\alpha_m}$ and obtain analytically the symmetric frequencies $\omega_S(0)$ that represent the lowest-energy excitations. In the rest of this section, we only use $k$ to denote the norm of the wave vector, $k=|\mathbf{k}|$.

\paragraph{Ferromagnetic phase:}\label{subsec:Fcol}

\begin{eqnarray}\label{eq:characterizationF}
\Delta^{ac}&=&\Delta^{bd}=2\epsilon_Z+2u_{\perp}+u_z,~\Delta^{ad,bc}=\Delta^{ac}\pm 2\epsilon_V\\
\nonumber M^{ac,ac}&=&M^{bd,bd}=-u_z,~M^{ac,bd}=-2u_{\perp},~M^{bc,bc}=M^{ad,ad}=u_z
\end{eqnarray}
As the unoccupied levels have orthogonal spin polarizations to the occupied levels, there is no mixing between normal and anomalous indices and the RPA contribution vanishes. The computation of the collective modes frequency then reduces to solve an eigenvalue equation that gives always real frequencies, see discussion below Eq. (\ref{eq:pseudospinmatricesnondiagonal}). Consequently, we focus on computing the dispersion relation $\omega(k)$ on the normal sector, and the dispersion relation in the anomalous sector is just $-\omega(k)$.

We consider two kind of excitations: spin-flip excitations and full-flip excitations, where both valley and spin indices are flipped at the same time. The spin-flip excitations involve transitions between two different spin polarizations, keeping the same valley index. They correspond to indices $(kl,jm)=(\{ac,bd\},\{ac,bd\})$ and hence we restrict to the $8\times 8$ submatrix $\tilde{X}^{sf}$ of the dispersion matrix $\tilde{X}$ in order to compute the spin-flip excitations:
\begin{equation}\label{eq:Fspinflipmatrix}
\tilde{X}^{sf}(\mathbf{k})=\left[\begin{array}{cc}
Y^{ac,ac}(\mathbf{k}) & Y^{ac,bd}(\mathbf{k})\\
Y^{bd,ac}(\mathbf{k}) & Y^{bd,bd}(\mathbf{k})
\end{array}\right]
\end{equation}
As $Y^{ac,ac}(\mathbf{k})=Y^{bd,bd}(\mathbf{k})$ and $Y^{ac,bd}(\mathbf{k})=Y^{bd,ac}(\mathbf{k})$, by making the transformation in the valley-spin index pairs $(ac\pm bd)/\sqrt{2}$, we separate the matrix $X^{sf}$ into two disconnected $4\times4$ matrices
\begin{equation}\label{eq:spinorialtransformation}
Y^{\pm}(\mathbf{k})=Y^{ac,ac}(\mathbf{k})\pm Y^{ac,bd}(\mathbf{k})
\end{equation}
and thus, for computing the collective modes frequency, we only have to obtain the eigenvalues of $Y^{\pm}(\mathbf{k})$. In particular, it is immediate to calculate $\omega_S(0)$
\begin{equation}\label{eq:FCAFinstability}
\hbar\omega^{\pm}_S(0)=2\epsilon_Z+2u_{\perp}\mp2u_{\perp}
\end{equation}
The $+$ branch gives $\hbar\omega^{+}_S(0)=2\epsilon_Z>0$ but the $-$ branch gives $\hbar\omega^{-}_S(0)=2\epsilon_Z+4u_{\perp}$. We see that  $\omega^{-}_S(0)=0$ is precisely the condition for the transition between the F and the CAF phase, Eq.(\ref{eq:FCAFborder}). Indeed, in the region where the CAF phase is present, the system is energetically unstable, $\omega^{-}_S(0)<0$.

Now, we turn our attention to full-flip excitations. They are related to $(kl,jm)=(ad,ad),(bc,bc)$. These subspaces are not connected between them as $M^{ad,bc}=0$, so we only have to compute the eigenvalues of $4\times 4$ matrices corresponding to $Y^{ad,ad}(\mathbf{k})$ and $Y^{bc,bc}(\mathbf{k})$. Moreove, it is straightforward to show from Eq. (\ref{eq:characterizationF}) that $Y^{ad,ad}(\mathbf{k})=Y^{bc,bc}(\mathbf{k})+4\epsilon_VI$, so the frequencies corresponding to the $ad$ sector satisfy $\hbar\omega^{ad}(k)=\hbar\omega^{bc}(k)+4\epsilon_V$. Thus, we only need to obtain the dispersion relation for the $bc$ indices, $\omega^{bc}(k)$. Specifically, we find for the symmetric mode that
\begin{equation}\label{eq:FFLPinstability}
\hbar\omega^{bc}_S(0)=2(u_{z}+u_{\perp}+\epsilon_Z-\epsilon_V)
\end{equation}
which is proportional to the difference between the mean-field energies of the FLP state and the F state, see Sec. \ref{subsec:HF}. In analogy to the case of spin-flip excitations, when we are in the region where the ground state is the FLP phase, $\omega^{bc}_S(0)<0$ and $\omega^{bc}_S(0)=0$ is the boundary between the two phases. The existence of a gapless mode can be understood from the appearance of a residual degeneracy at the boundary between the FLP and the F phase. Specifically, when the condition (\ref{eq:FFLPborder}) is satisfied, the HF equations present a degenerate ground state of the form:
\begin{eqnarray}\label{eq:Residualphase}
\nonumber \chi_{a}&=&\ket{n_z}\otimes\ket{s_z},~\chi_{b}=\cos{\varphi}\ket{n_z}\otimes\ket{-s_z}+\sin{\varphi}\ket{-n_z}\otimes\ket{s_z}\\
E(P)&=&-u_{\perp}-\epsilon_Z-\epsilon_V,
\end{eqnarray}
see Eqs. (\ref{eq:meanfieldenergy}),(\ref{eq:Fphase}) and (\ref{eq:FLPphase}). As we see, the energy does not depend on $\varphi$ and an associated $U(1)$ symmetry arises from this fact. This symmetry can be regarded as a remanent of the full exact $SO(5)$ symmetry displayed by the short-range interaction Hamiltonian for $g_{\perp}=-g_z$ and $\epsilon_Z=\epsilon_V=0$ \cite{Wu2014}.

\begin{figure*}[tb!]
\begin{tabular}{@{}cc@{}}
    \includegraphics[width=0.5\columnwidth]{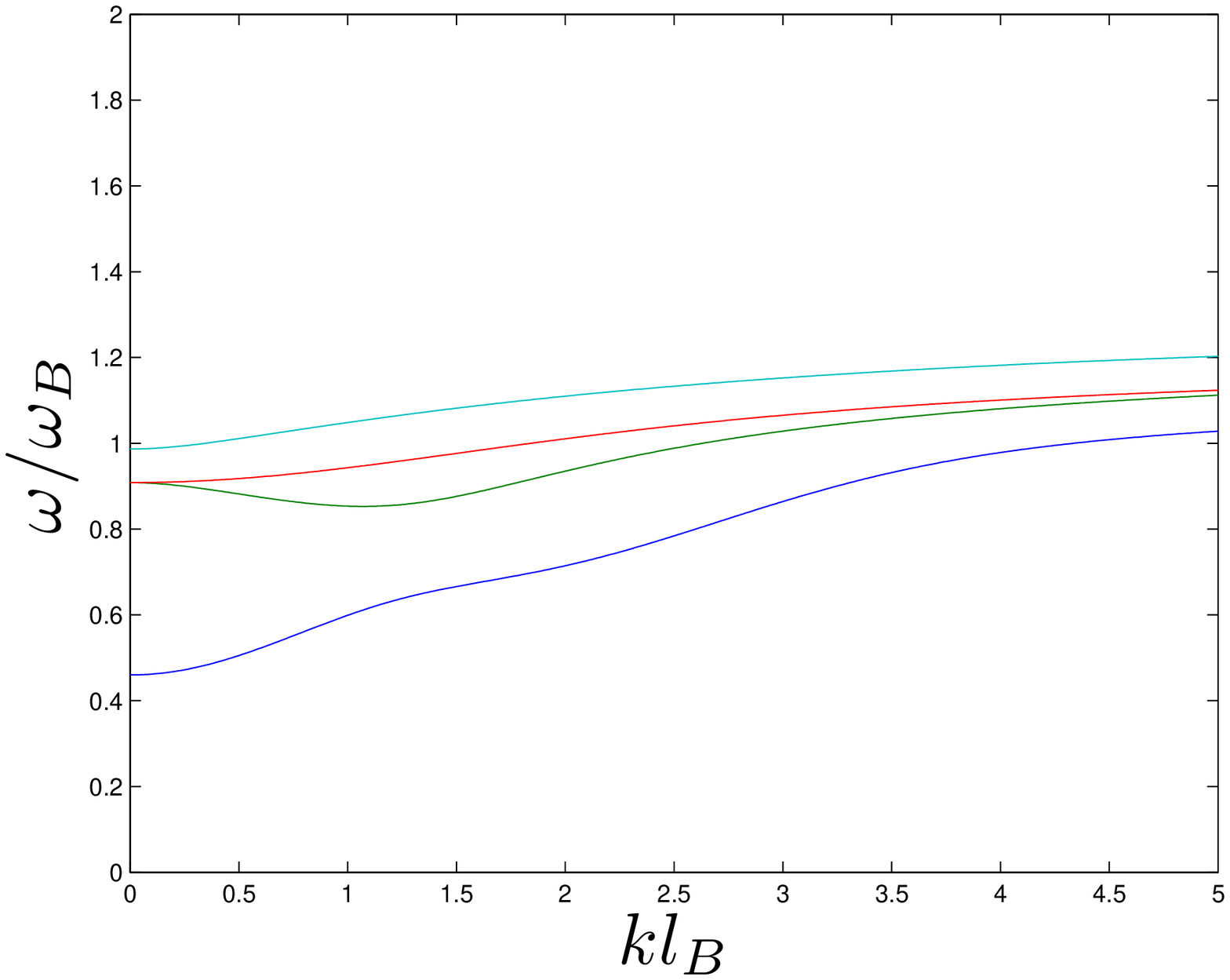} &
    \includegraphics[width=0.5\columnwidth]{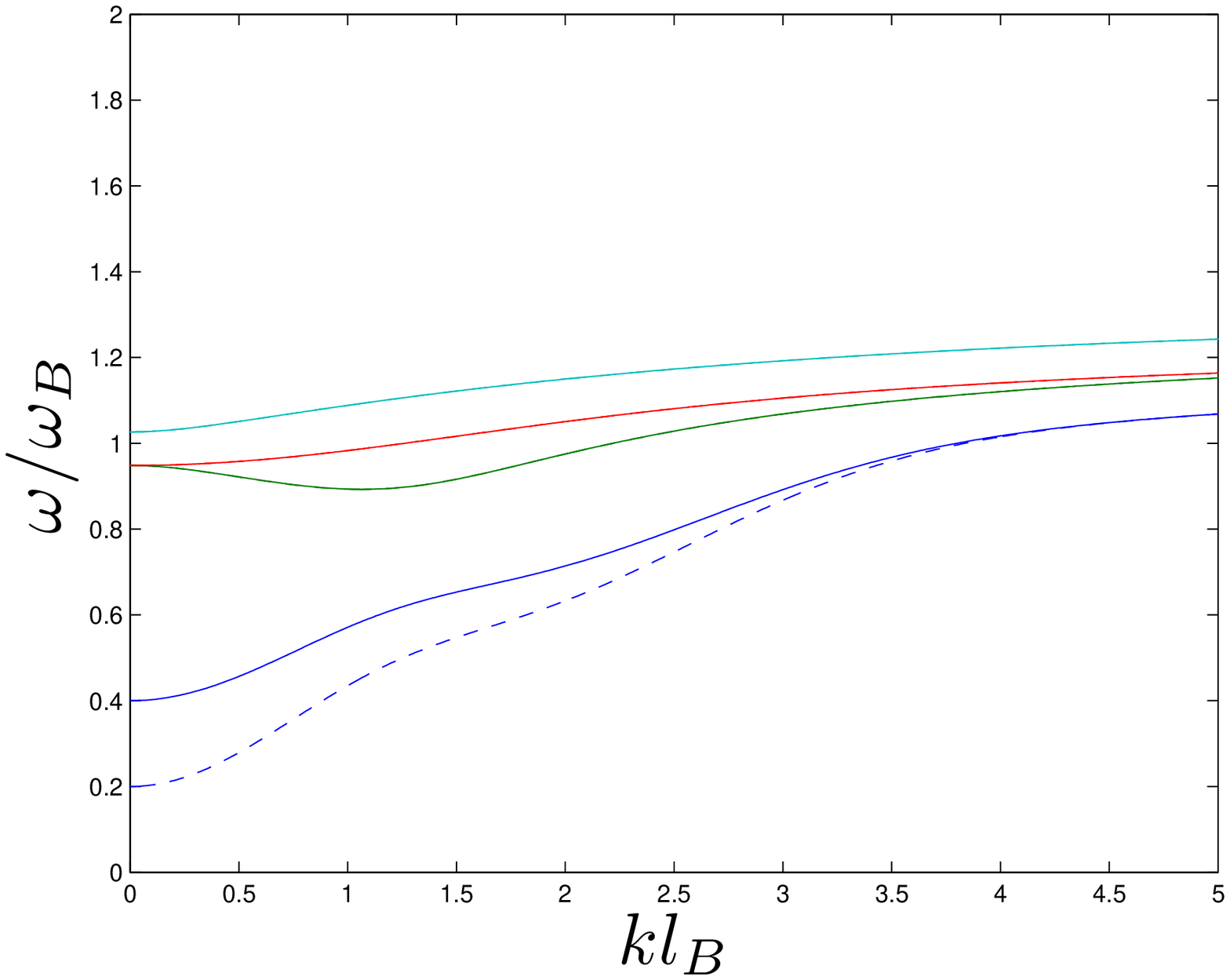} \\
\end{tabular}
\caption{Collective modes of the F phase with $F_c=0.5$, $u_z=0.1 \hbar \omega_B$, $u_{\perp}=-0.05 \hbar \omega_B$, $\epsilon_V=0.02 \hbar \omega_B$ and $\epsilon_Z=0.2 \hbar \omega_B$. Left panel: plot of the $4$ eigenvalues corresponding to full-flip excitations, $\omega^{bc}(k)$. Right panel: plot of the eigenvalues corresponding to the spin-flip excitations. The $+(-)$ branches are plotted in solid (dashed) line. We note that the largest eigenvalues of the $\pm$ branches are so close to each other that it is difficult to distinguish them.}
\label{fig:FColMod}
\end{figure*}

In Fig. \ref{fig:FColMod} we plot the dispersion relation of the full-flip (left panel) and spin-flip (right panel) excitations, obtained by computing the eigenvalues of $Y^{bc,bc}(\mathbf{k})$ and $Y^{\pm}(\mathbf{k})$. In the case of spin-flip excitations, the $+(-)$ branches are plotted in solid (dashed) line. For these modes, as $\Delta^{ac}=\Delta^{bd}$, only the symmetric modes $\omega^{\pm}_S$, corresponding to the lowest energy modes, are not degenerate for $k=0$, see Eqs. (\ref{eq:zerofrequencyequation}), (\ref{eq:spinorialtransformation}). On the other hand, for $kl_B\gg 1$, all the modes of the $\pm$ branches are degenerate due to the exponential decay of $R(\mathbf{k})$ as given by Eq. (\ref{eq:exponentialdecayR}). This explains why the curves corresponding to the larger-energy modes of the $\pm$ branches are so close to each other and why only those corresponding to the prolongations for $k\neq 0$ of the symmetric modes are clearly distinguished.

When comparing the plots of spin-flip and full-flip excitations, we see the curves have the same qualitative form, very close to those of Refs. \cite{Toke2011,Lambert2013}. Again, this can be explained by invoking the dominant character of Coulomb interactions. The main differences are observed once more for the lowest energy modes since for low $k$ their dependence on the strength of the Coulomb interaction is weak.

Finally, we mention that varying $\epsilon_Z,\epsilon_V$ only changes the origin of the curves at $k=0$ as they only appear in the constant energy shifts $\Delta^{ac},\Delta^{bc}$.

\paragraph{Full layer-polarized phase:}\label{subsec:FLPcol}

\begin{eqnarray}\label{eq:characterizationFLP}
\Delta^{ac}&=&\Delta^{bd}=2\epsilon_V-2u_{\perp}-3u_z,~\Delta^{ad,bc}=\Delta^{ac}\pm 2\epsilon_Z\\
\nonumber M^{ac,ac}&=&M^{bd,bd}=2u_{\perp}+u_z,~M^{ac,bd}=2u_{\perp},~M^{bc,bc}=M^{ad,ad}=u_z
\end{eqnarray}

This state is the analog of the F phase in the valley subspace. Here, we consider valley-flip excitations and full-flip excitations. Again, it can be shown that there is no mixing between normal and anomalous sectors, so the computation can be carried out in a similar fashion to the F state. The computation for the valley-flip excitations is formally equal to the F case, see Eq. (\ref{eq:spinorialtransformation}) and related discussion. However, a peculiarity arises for the FLP phase. It can be shown that $\hbar\omega^{-}(k)=\hbar\omega^{bc}(k)+2\epsilon_Z$, where $\omega^{bc}(k)$ is the frequency corresponding to the full-flip excitations, as explained later. Thus, half of the modes corresponding to valley-flip excitations are indeed full-flip excitations. The origin of this fact can be trace back to the spin invariance of the interaction. As a consequence, we only need to compute $\omega^{+}(k)$ for characterizing the valley-flip excitations. At $k=0$, the symmetric mode is given by:
\begin{equation}\label{eq:FLPPLPinstability}
\hbar\omega^{+}_S(0)=2u_{\perp}-2u_{z}+2\epsilon_V
\end{equation}
In analogy to the case of the F-CAF transition [see Eq. (\ref{eq:FCAFinstability}) and related discussion], $\omega^{+}_S(0)$ becomes negative in the region where the PLP phase is present. Precisely, $\omega^{+}_S(0)=0$ is the boundary between the two phases. Thus, the FLP state becomes energetically unstable in the region where the PLP phase is present, as expected.

On the other hand, as for the F phase, the full-flip excitations describe the transition between the F and FLP phase. The calculation is formally analog to that of the F phase and we also find a similar relation, $\hbar\omega^{ad}(k)=\hbar\omega^{bc}(k)+4\epsilon_V$. For studying the phase transition, we obtain the symmetric mode
\begin{equation}
\hbar\omega^{bc}_S(0)=2(\epsilon_V-\epsilon_Z-u_{\perp}-u_{z})
\end{equation}
which is just $-\omega^{bc}_S(0)$ for the F state, see Eq. (\ref{eq:FFLPinstability}). Reasoning as before, the FLP phase becomes unstable in the region where the F phase is present and the appearance of the zero-frequency mode at the boundary between the two regions results from the accidental $U(1)$ symmetry, see Eqs. (\ref{eq:FFLPinstability}),(\ref{eq:Residualphase}). 

\begin{figure*}[bt]
\begin{tabular}{@{}cc@{}}
    \includegraphics[width=0.5\columnwidth]{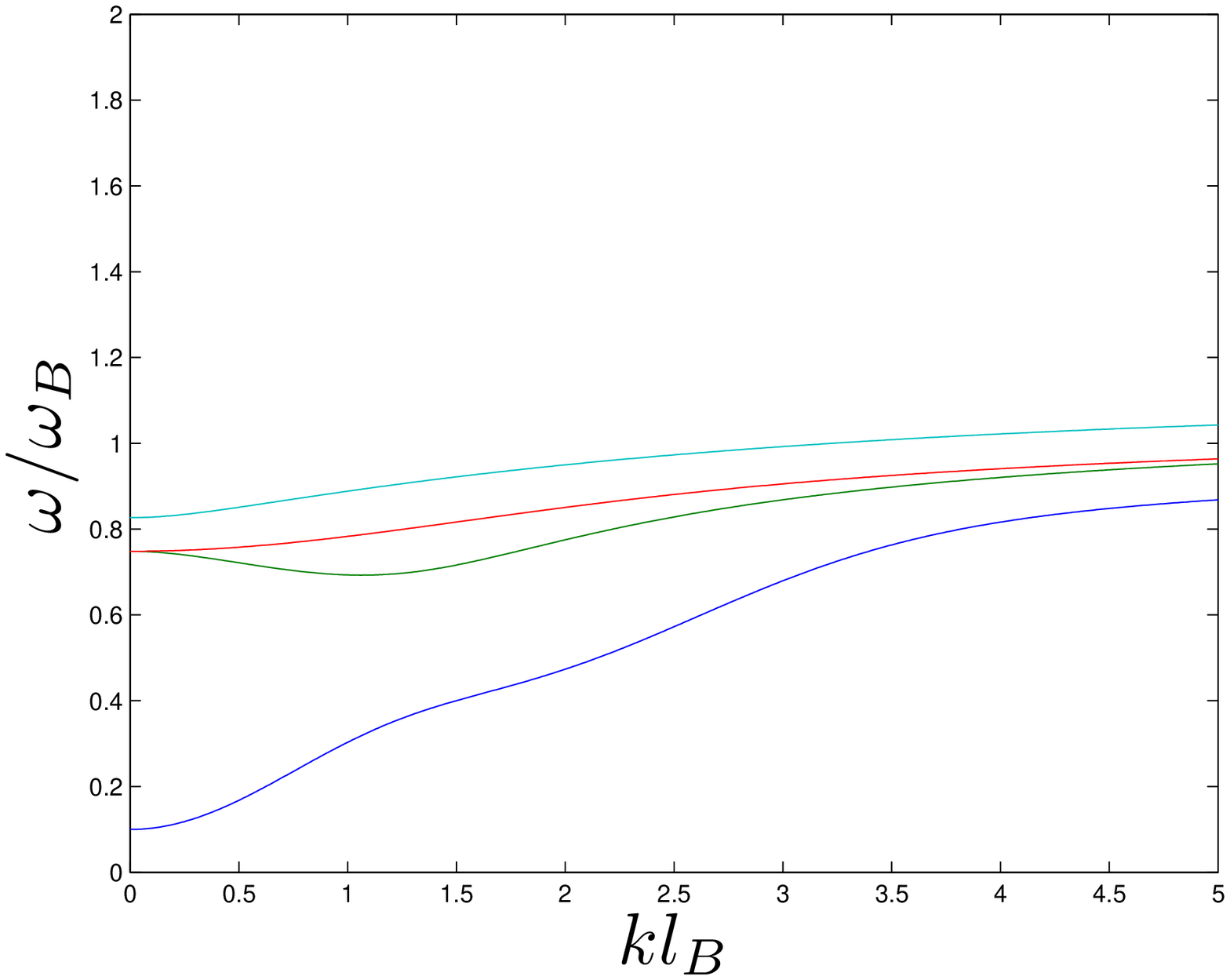} &
    \includegraphics[width=0.5\columnwidth]{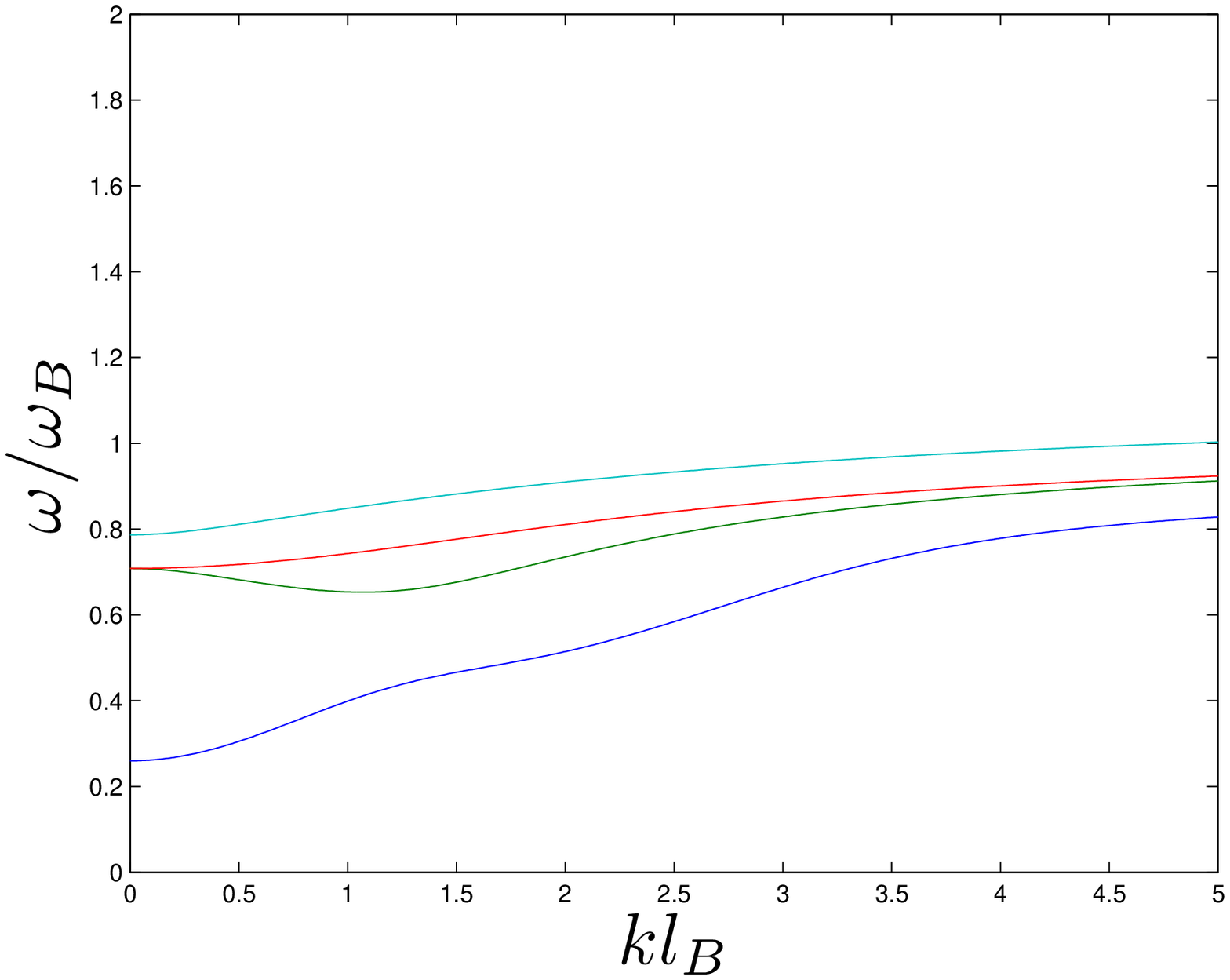} \\
\end{tabular}
\caption{Collective modes of the FLP phase with $F_c=0.5$, $u_z=0.1 \hbar \omega_B$, $u_{\perp}=-0.05 \hbar \omega_B$, $\epsilon_V=0.2 \hbar \omega_B$ and $\epsilon_Z=0.02 \hbar \omega_B$. Left panel: plot of the frequencies corresponding to full-flip excitations, $\omega^{bc}(k)$. Right panel: plot of the frequencies corresponding to the proper valley-flip excitations $\omega^{+}(k)$.}
\label{fig:FLPColMod}
\end{figure*}

In Fig. \ref{fig:FLPColMod}, we represent the dispersion relation of the full-flip (left panel) and proper valley-flip (right panel) excitations. The qualitative trends and properties of the curves are similar to those of the F state.

\paragraph{Canted anti-ferromagnetic phase:}\label{subsec:CAFcol}

\begin{eqnarray}\label{eq:characterizationCAF}
\Delta^{ac}&=&\Delta^{bd}=u_z-2u_{\perp},~\Delta^{ad,bc}=\Delta^{ac}\pm 2\epsilon_V\\
\nonumber M^{ac,ac}&=&M^{bd,bd}=-u_z,~M^{ac,bd}=-2u_{\perp}\cos^2 \theta_s\\
\nonumber M^{bc,bc}&=&M^{ad,ad}=u_z+2u_{\perp}\sin^2 \theta_s,~M^{ac,db}=-2u_{\perp}\sin^2 \theta_s
\end{eqnarray}
In this case, the coefficients $M^{\alpha_k\alpha_l,\alpha_j\alpha_m}$ mix normal and anomalous indices. In particular, for studying the equivalent of the spin-flip modes, we have to compute the eigenvalues of the $16\times 16$ matrix $\tilde{X}^{sf}(\mathbf{k})$:

\begin{equation}\label{eq:CAFspinflipmatrix}
\tilde{X}^{sf}(\mathbf{k})=\left[\begin{array}{cccc}
Y^{ac,ac}(\mathbf{k}) &  Y^{ac,bd}(\mathbf{k})  & 0 & Y^{ac,db}(\mathbf{k})\\
Y^{bd,ac}(\mathbf{k}) & Y^{bd,bd}(\mathbf{k})   &  Y^{bd,ca}(\mathbf{k}) & 0 \\
0 & -Y^{ca,bd}(\mathbf{k}) & -Y^{ca,ca}(\mathbf{k})  &-Y^{ca,db}(\mathbf{k})  \\
-Y^{db,ac}(\mathbf{k}) & 0  & -Y^{db,ca}(\mathbf{k})  &  -Y^{db,db}(\mathbf{k})\\
\end{array}\right]
\end{equation}
The $-$ signs in the previous expression arise from the contribution of the anomalous sector, in which the $\omega$ term changes sign, see Eq. (\ref{eq:XYgen32}). Because of this fact, the matrix $\tilde{X}^{sf}(\mathbf{k})$ is non-hermitian and can present complex eigenvalues. By making the same transformation of Eq. (\ref{eq:spinorialtransformation}), we separate the matrix $\tilde{X}^{sf}$ into two $8\times 8$ blocks, $\tilde{X}^{sf,\pm}(\mathbf{k})$:
\begin{equation}\label{eq:CAFspinflipreduction}
\tilde{X}^{sf,\pm}(\mathbf{k})=\left[\begin{array}{cc}
Y^{\pm}(\mathbf{k})  & Y^{ac,db}(\mathbf{k}) \\
-Y^{ac,db}(\mathbf{k}) & -Y^{\pm}(\mathbf{k})   \\
\end{array}\right]
\end{equation}
It is worth noting that $Y^{ac,db}(\mathbf{k})=Y^{db,ac}(\mathbf{k})=Y^{bd,ac}(\mathbf{k})=Y^{ca,bd}(\mathbf{k})$.
After some straightforward algebra, it can be shown that if $\omega$ is an eigenvalue of $\tilde{X}^{sf}(\mathbf{k})$, $(-\omega,\omega^*)$ are also eigenvalues. This relation implies that, if a complex-frequency mode is present, it must satisfy $\text{Re}(\omega)=0$. It also implies that we can focus on computing only the positive frequency modes and the other half will be given by just changing the sign.

In particular, we compute the symmetric modes frequencies, finding:

\begin{equation}\label{eq:CAFspmodessimplified}
\hbar\omega^{sf,\pm}_S(0)=|2u_{\perp}|\sqrt{(1\pm\cos^2 \theta_s)^2 - \sin^4 \theta_s}
\end{equation}

The previous equation gives $\hbar\omega^{sf,+}_S(0)=|4u_{\perp}|\cos \theta_s=2\epsilon_Z>0$ and $\omega^{sf,-}_S(0)=0$, where we take into account that $u_{\perp}<0$ is a necessary condition for being in the CAF state. The appearance of this zero-energy mode is expected because the CAF phase presents an $U(1)$ symmetry, as mentioned in Sec. \ref{subsec:HF}. Also, we see that, at $k=0$, there are no complex-frequency modes. As $R(\mathbf{k})$ decays exponentially, one can expect that there are not complex-frequency modes also for $k\neq0$, see Eqs. (\ref{eq:pseudospinmatrices}),(\ref{eq:exponentialdecayR}) and (\ref{eq:CAFspinflipmatrix}). This observation has been numerically checked.

The other kind of excitations are the analog of the full-flip excitations of the F phase, Eq. (\ref{eq:FFLPinstability}) and related comments. In a general CAF state, the full-flip excitations involve the normal and anomalous sectors $ad,cb$ and $bc,da$ separately, so they are given by $8\times 8$ matrices. In particular, the $\omega^{ad,cb}$ dispersion relation satisfy $\hbar\omega^{ad,cb}(k)=\hbar\omega^{bc,da}(k)+4\epsilon_V$. Thus, it is only necessary to compute $\omega^{bc,da}(k)$, which can be done by obtaining the eigenvalues of the matrix $\tilde{X}^{ff}(\mathbf{k})$, with
\begin{equation}\label{eq:CAFpseudospinflip}
\tilde{X}^{ff}(\mathbf{k})=\left[\begin{array}{cc}
Y^{bc,bc}(\mathbf{k}) & Y^{bc,da}(\mathbf{k})\\
-Y^{da,bc}(\mathbf{k}) & -Y^{da,da}(\mathbf{k})
\end{array}\right]
\end{equation}
It can be shown that if $\omega$ is an eigenvalue of $\tilde{X}^{ff}(\mathbf{k})$, $(-4\epsilon_V/\hbar-\omega,\omega^*)$ are also eigenvalues and then, only $4$ eigenvalues are independent. Another consequence resulting from the previous relation is that the complex-frequency modes satisfy $\hbar\text{Re}(\omega)=-2\epsilon_V$. The lower-energy symmetric mode frequency is:
\begin{equation}\label{eq:CAFpszeromode}
\hbar\omega^{bc,da}_S(0)=-2\epsilon_V + \sqrt{(2u_z-2u_{\perp}\cos^2 \theta_s)^2-4u^2_{\perp}\sin^4 \theta_s}
\end{equation}
Interestingly, there are situations in which the quantity in the square root is negative and then a dynamical instability appears, in analogy to the BH laser scenario of Chapter \ref{chapter:BHL}. It is straightforward to show that $\omega^{bc,da}_S(0)$ is purely real whenever
\begin{equation}\label{eq:CAFdynin}
u_{\perp}+u_{z}>\frac{\epsilon^2_Z}{2u_{\perp}}
\end{equation}
is fulfilled. Also, whenever $u_z-u_{\perp}<0$, $\omega^{bc,da}_S(0)$ is real, but this condition is incompatible with the CAF phase and it is not expected to be satisfied in an actual experiment (see comments at the end of Sec. \ref{subsec:HF}). On the other hand, $\omega^{bc,da,+}_S(0)=0$ at the boundary between the PLP and CAF phases, i.e, whenever Eq. (\ref{eq:PLPCAFborder}) is satisfied. Once more, this gapless mode can be traced back to the broken $SO(5)$ symmetry; indeed, its appearance was already hinted in Ref. \cite{Wu2014}. A summary of the above statements is given by
\begin{eqnarray}\label{eq:CAFinstabilitycondition}
\nonumber u_{\perp}+u_{z}&>&\frac{\epsilon^2_Z}{2u_{\perp}}+\frac{\epsilon^2_V}{u_z-u_{\perp}}>\frac{\epsilon^2_Z}{2u_{\perp}},~\omega^{bc,da}_S(0)>0\\
\frac{\epsilon^2_Z}{2u_{\perp}}+\frac{\epsilon^2_V}{u_z-u_{\perp}}&>&u_{\perp}+u_{z}>\frac{\epsilon^2_Z}{2u_{\perp}} ,~\omega^{bc,da}_S(0)<0\\
\nonumber \frac{\epsilon^2_Z}{2u_{\perp}}+\frac{\epsilon^2_V}{u_z-u_{\perp}}&>&\frac{\epsilon^2_Z}{2u_{\perp}}>u_{\perp}+u_{z} ,~\text{Im}~\omega^{bc,da}_S(0)\neq0
\end{eqnarray}

We remark that $\omega^{bc,da}_S(0)>0$ as long as we stay in the space parameter of the CAF phase, see Fig. \ref{fig:PhaseDiagram}. In the region of the parameters where the PLP state is the ground state, there is a small region in which $\omega^{bc,da}_S(0)<0$ but real and when (\ref{eq:CAFdynin}) is not satisfied, a dynamical instability appears. Thus, the system can only be dynamically unstable whenever is energetically unstable, in close analogy with the BdG equations; see discussion at the end of Sec. \ref{subsec:timeindependent}. We remark that the possible appearance of the dynamical instability arises from the presence of short-range interactions, that mix normal and anomalous sectors of the matrix $\tilde{X}$.

\begin{figure*}[bt]
\begin{tabular}{@{}cc@{}}
    \includegraphics[width=0.5\columnwidth]{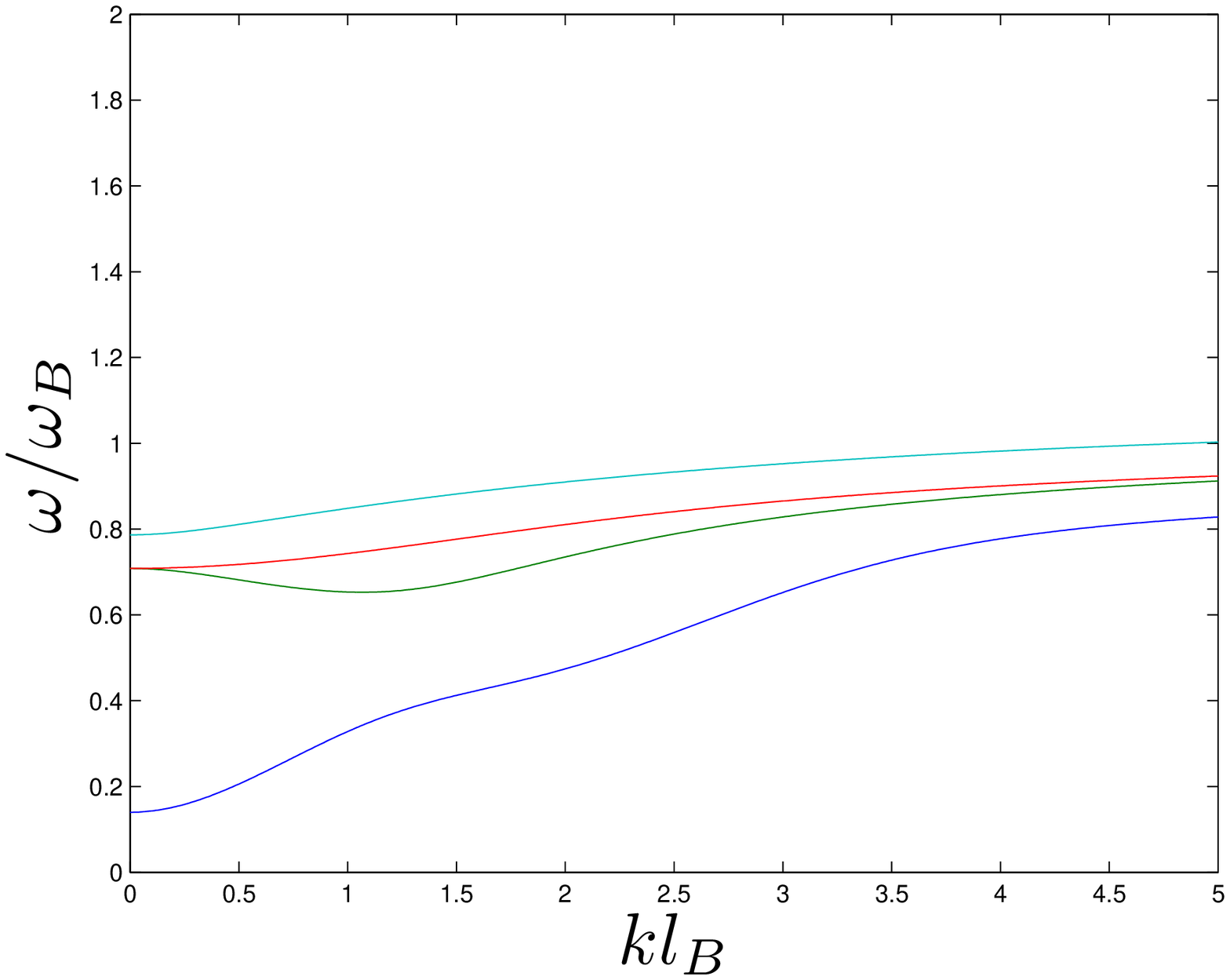} &
    \includegraphics[width=0.5\columnwidth]{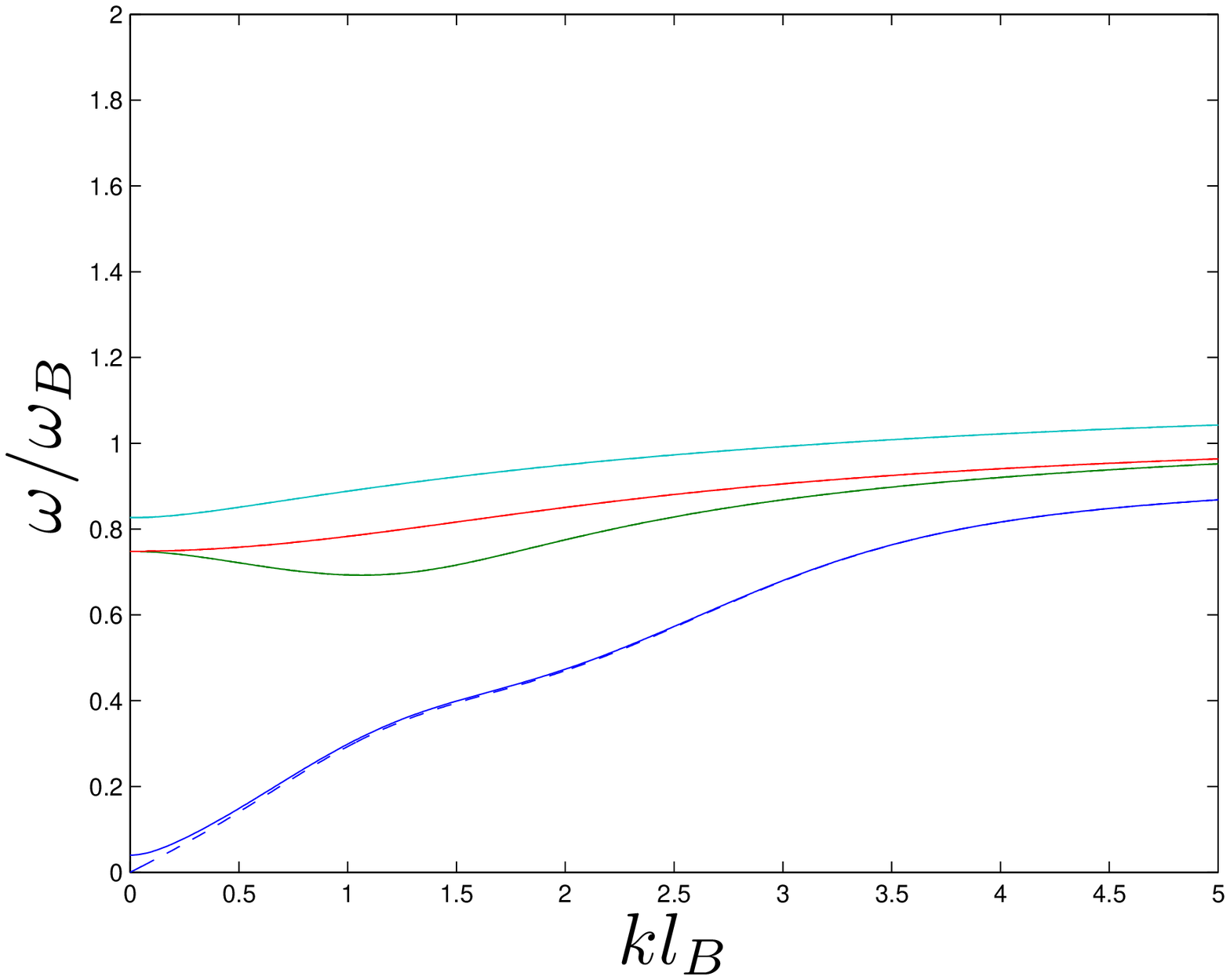} \\
\end{tabular}
\caption{Collective modes of the CAF phase for $F_c=0.5$, $u_z=0.1 \hbar \omega_B$, $u_{\perp}=-0.05 \hbar \omega_B$, $\epsilon_V=0.02 \hbar \omega_B$ and $\epsilon_Z=0.02 \hbar \omega_B$. Left panel: plot of the $4$ frequencies corresponding to full-flip excitations, $\omega^{bc,ad}(k)$. Right panel: plot of the frequencies corresponding to the spin-flip excitations. The $+(-)$ branches are plotted in solid (dashed) line. We note that the largest eigenvalues of the $\pm$ branches are so close to each other that it is difficult to distinguish them.}
\label{fig:CAFColMod}
\end{figure*}

In Fig. \ref{fig:CAFColMod}, we plot the dispersion relation of the full-flip (left panel) and spin-flip (right panel) excitations. We see that, for the spin-flip excitations, we always have a mode with zero-energy for $k=0$, corresponding to the Goldstone mode associated to the $U(1)$ symmetry of the CAF state. Apart from this consideration, the qualitative form of the curves is similar to the other previous cases.

\begin{figure}[bt]
\includegraphics[width=\columnwidth]{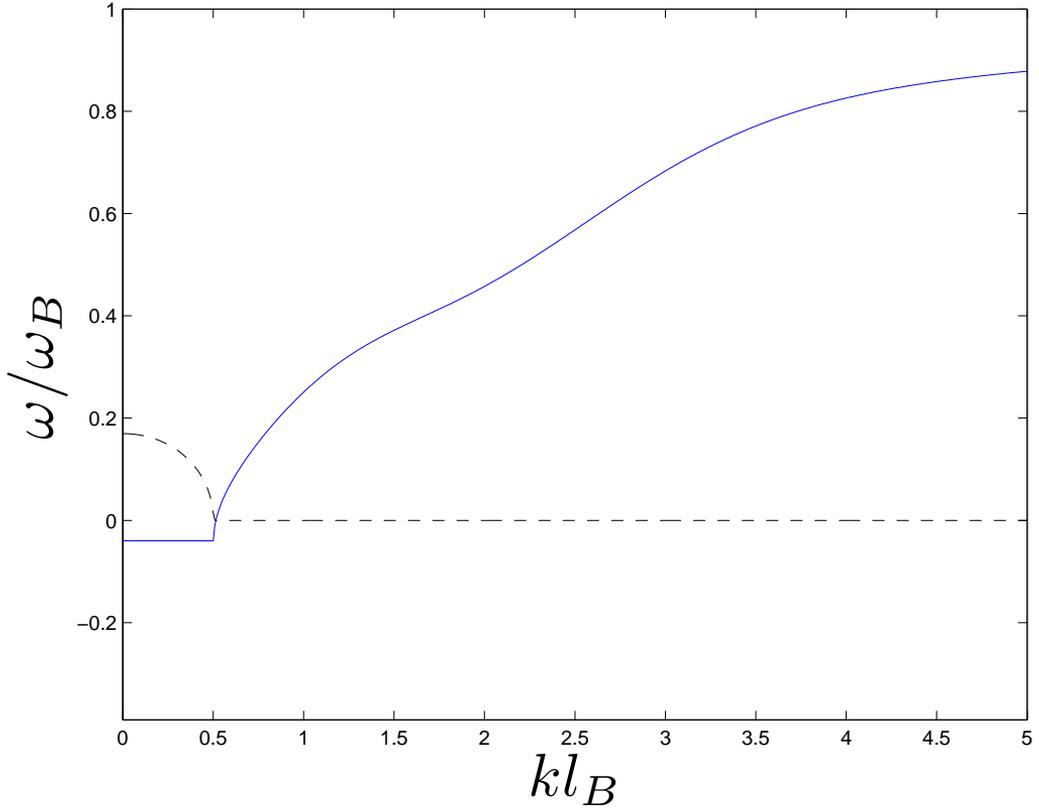}
\caption{Full-flip complex-frequency mode $\omega^{bc,ad}(k)$ of the CAF state for $F_c=0.5$, $u_z=0.05 \hbar \omega_B$, $u_{\perp}=-0.1 \hbar \omega_B$, $\epsilon_V=0.02 \hbar \omega_B$ and $\epsilon_Z=0.02 \hbar \omega_B$. The real (imaginary) part of the complex-frequency mode is plotted in solid (dashed) line.}
\label{fig:CAFexpmode}
\end{figure}

In Fig. \ref{fig:CAFexpmode}, we study a case where full-flip complex-frequency modes appear. We have not observed complex modes for spin-flip excitations. The plot shows the real (solid curve) and the imaginary (dashed curve) parts of the unstable mode. The decreasing of the imaginary part of the frequency is a consequence of Eq. (\ref{eq:exponentialdecayR}), as the matrix $R(\mathbf{k})$ connects the normal and anomalous sectors. We see that as long as $\text{Im}[\omega^{bc,ad}_S(0)]>0$, $\hbar\text{Re}[\omega^{bc,ad}_S(0)]=-2\epsilon_V$, as previously shown. At high values of $k$, the complex frequency becomes purely real and only contains the Coulomb contribution, in the same fashion of the other plots.

\paragraph{Partially layer-polarized phase:}\label{subsec:PLPcol}

\begin{eqnarray}\label{eq:characterizationPLP}
\Delta^{ac}&=&\Delta^{bd}=-u_z-4u_{\perp},~\Delta^{ad,bc}=\Delta^{ac}\pm 2\epsilon_Z\\
\nonumber M^{ac,ac}&=&M^{bd,bd}=u_z+2u_{\perp},~M^{ac,bd}=2u_{\perp}+(u_z-u_{\perp})\sin^2\theta_v\\
\nonumber M^{bc,bc}&=&M^{ad,ad}=u_z-(u_z-u_{\perp})\sin^2\theta_v,~M^{ac,db}=(u_z-u_{\perp})\sin^2\theta_v
\end{eqnarray}

The computation of the collective-mode frequencies for the PLP state is formally equal to that of the CAF phase and most of the conclusions also apply here. The equivalent of the valley-flip modes of the FLP phase are now computed by obtaining the eigenvalues of a $16\times 16$ matrix $\tilde{X}^{vf}$ whose formal expression is the same of $\tilde{X}^{sf}$ in Eq. (\ref{eq:CAFspinflipmatrix}). As in the FLP phase, half of the valley-flip modes are indeed full-flip modes, $\hbar\omega^{-}(k)=\hbar\omega^{bc,ad}(k)+2\epsilon_Z$, where $\omega^{bc,ad}(k)$ is the frequency of the full-flip excitations (see discussion below). Thus, in order to characterize the proper valley-flip excitations, we compute $\omega^{vf,+}(k)$. At $k=0$, the symmetric mode frequency is null,
\begin{equation}
\omega^{vf,+}_S(0)=0~,
\end{equation}
as it is the Goldstone mode associated to the $U(1)$ symmetry of the PLP state.

In respect to the analog of the full-flip modes of the FLP phase, the computation of these modes is also formally equal to that of the CAF phase, see the discussion associated to Eqs. (\ref{eq:CAFpseudospinflip}), (\ref{eq:CAFpszeromode}). The symmetric mode frequency now reads:
\begin{equation}\label{eq:PLPpszeromode}
\hbar\omega^{bc,da}_S(0)=-2\epsilon_Z+\sqrt{(4u_{\perp}+A)^2-A^2},~A=(u_z-u_{\perp})\sin^2\theta_v
\end{equation}
It can be proven that $\omega^{bc,da}_S(0)$ is real if:
\begin{equation}\label{eq:PLPdynin}
u_{\perp}+u_{z}<\frac{\epsilon^2_V}{u_z-u_{\perp}}
\end{equation}
as $A>0$ in the PLP state. Also $u_{\perp}>0$ guarantees that $\omega^{vf}_S(0)>0$, but this condition is incompatible with the PLP state. A zero-frequency mode $\omega^{bc,da}_S(0)=0$ appears at the boundary between the PLP and CAF phases, as expected from the analog result of the CAF phase. Putting all together as in the CAF state:
\begin{eqnarray}\label{eq:PLPinstabilitycondition}
\nonumber u_{\perp}+u_{z}&<&\frac{\epsilon^2_Z}{2u_{\perp}}+\frac{\epsilon^2_V}{u_z-u_{\perp}}<\frac{\epsilon^2_V}{u_z-u_{\perp}},~\omega^{bc,da}_S(0)>0\\
\frac{\epsilon^2_Z}{2u_{\perp}}+\frac{\epsilon^2_V}{u_z-u_{\perp}}&<&u_{\perp}+u_{z}<\frac{\epsilon^2_V}{u_z-u_{\perp}} ,~\omega^{bc,da}_S(0)<0\\
\nonumber \frac{\epsilon^2_Z}{2u_{\perp}}+\frac{\epsilon^2_V}{u_z-u_{\perp}}&<&\frac{\epsilon^2_V}{u_z-u_{\perp}}<u_{\perp}+u_{z} ,~\text{Im}~\omega^{bc,da}_S(0)\neq0
\end{eqnarray}

\begin{figure*}[bt]
\begin{tabular}{@{}cc@{}}
    \includegraphics[width=0.5\columnwidth]{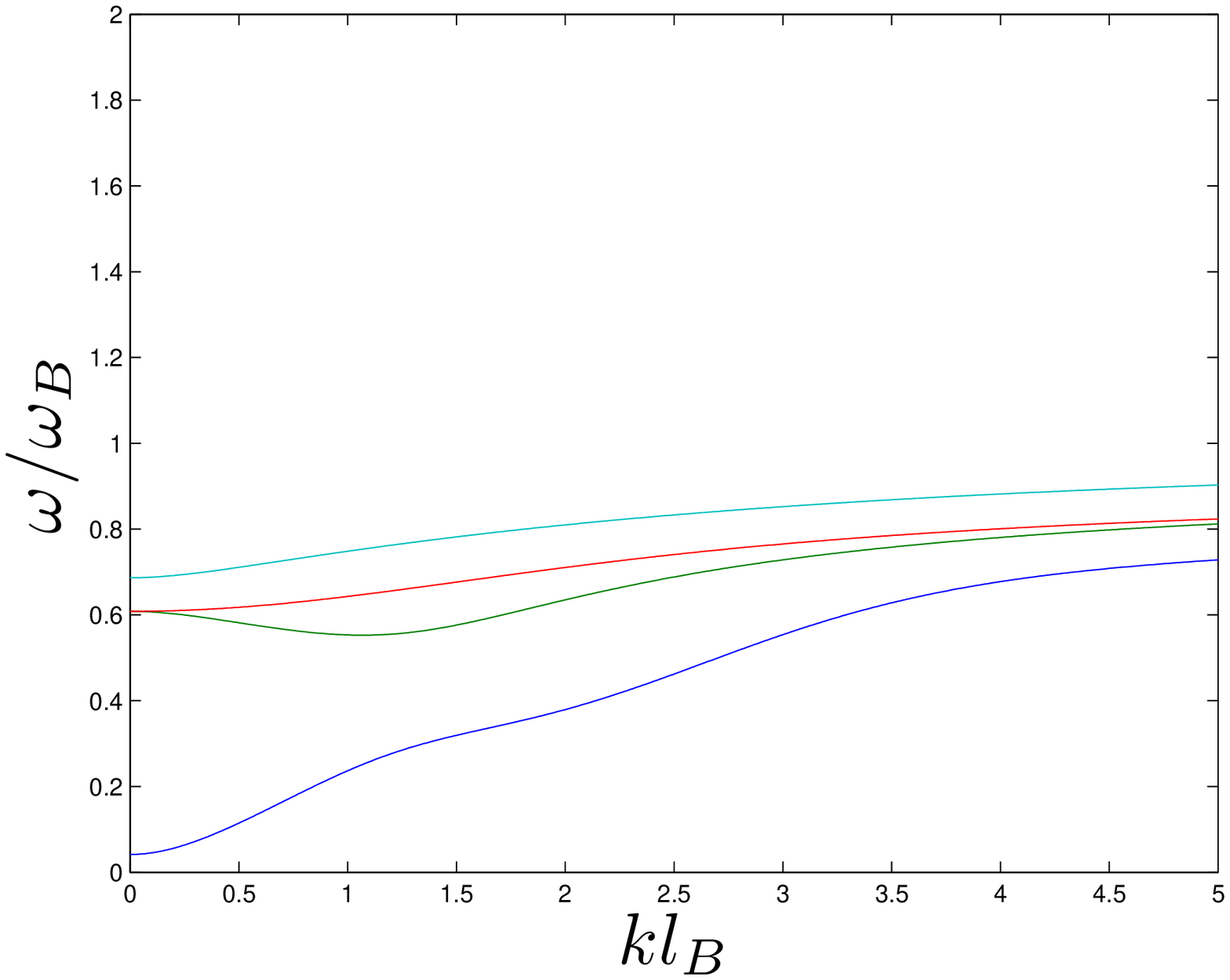} &
    \includegraphics[width=0.5\columnwidth]{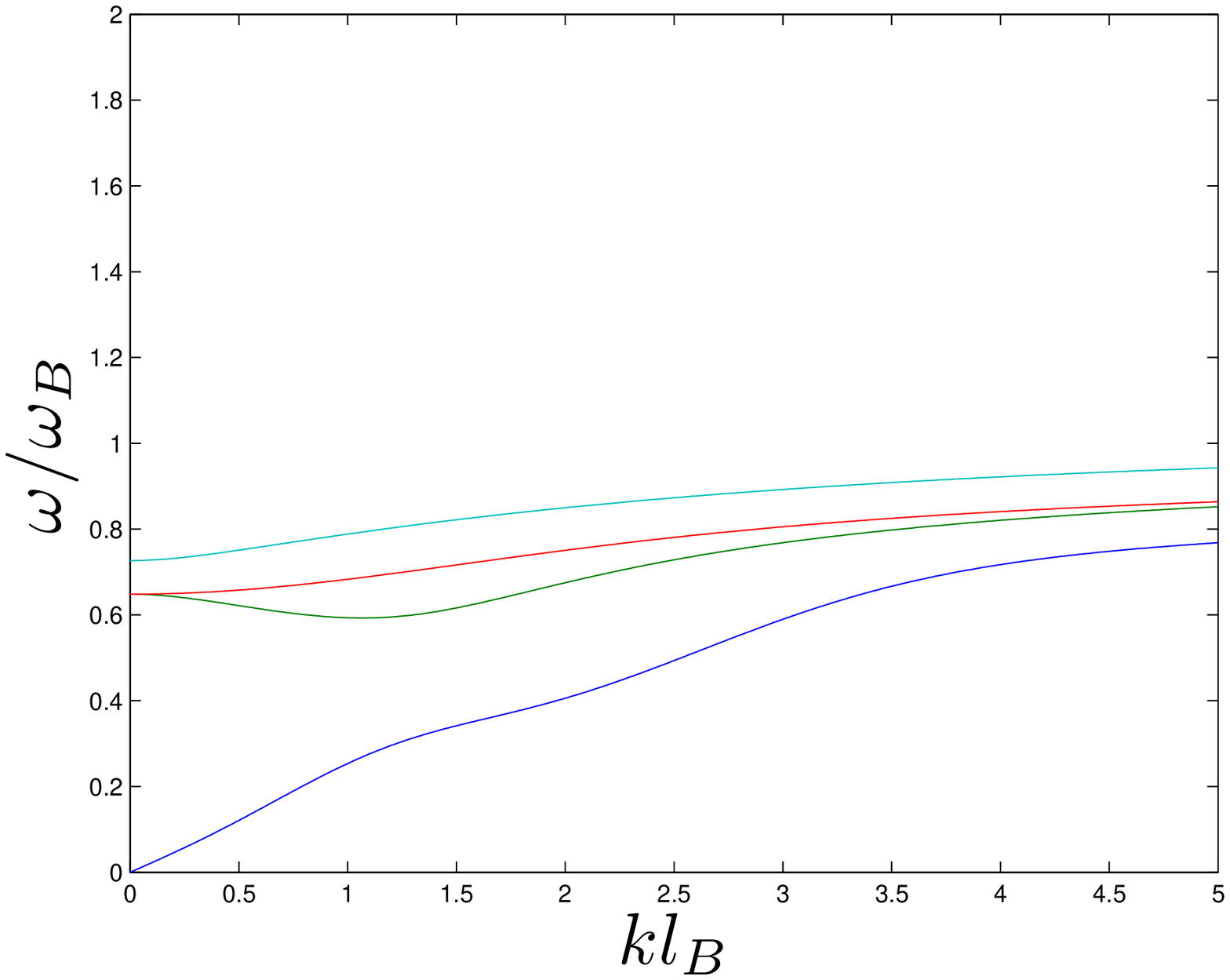} \\
\end{tabular}
\caption{Collective modes of the PLP phase with $F_c=0.5$, $u_z=0.1 \hbar \omega_B$, $u_{\perp}=-0.05 \hbar \omega_B$, $\epsilon_V=0.5 \hbar \omega_B$ and $\epsilon_Z=0.02 \hbar \omega_B$. Left panel: plot of the frequencies corresponding to full-flip excitations, $\omega^{bc,ad}(k)$. Right panel: plot of the frequencies corresponding to the proper valley-flip excitations $\omega^{+}(k)$.}
\label{fig:PLPColMod}
\end{figure*}

\begin{figure}[bt]
\includegraphics[width=\columnwidth]{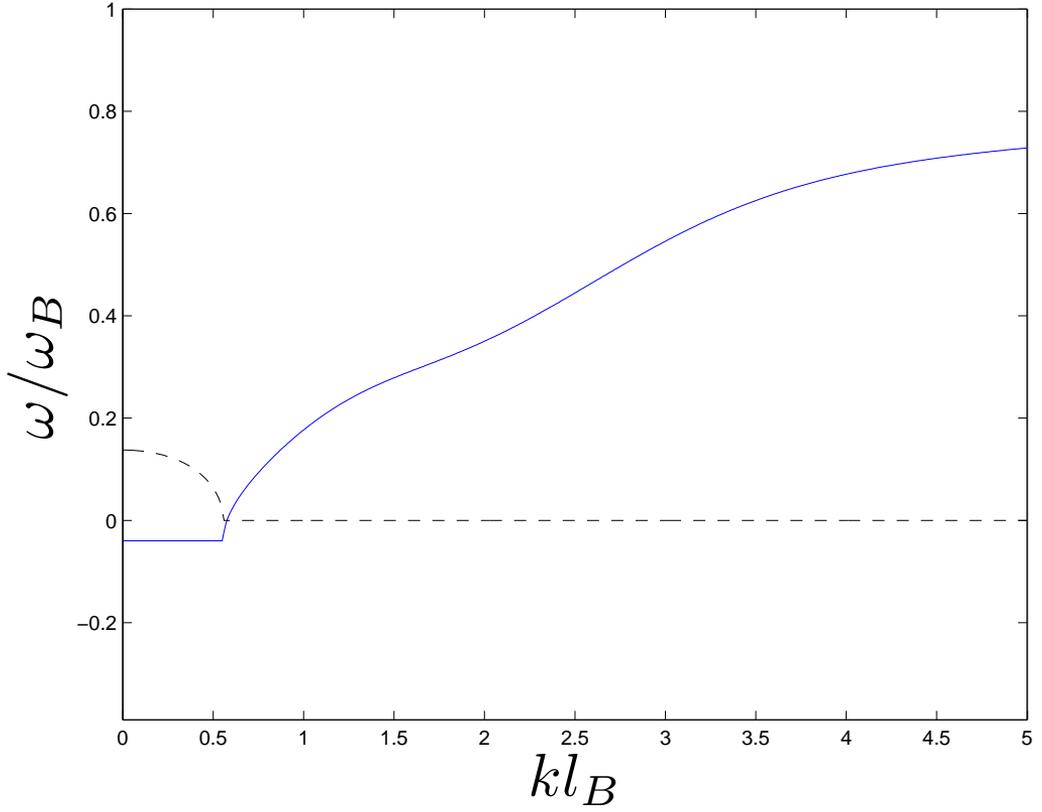}
\caption{Full-flip complex-frequency mode $\omega^{bc,ad}(k)$ of the PLP state for $F_c=0.5$, $u_z=0.4 \hbar \omega_B$, $u_{\perp}=-0.2 \hbar \omega_B$, $\epsilon_V=0.02 \hbar \omega_B$ and $\epsilon_Z=0.02 \hbar \omega_B$. The real (imaginary) part of the complex-frequency mode is plotted in solid (dashed) line.}
\label{fig:PLPexpmode}
\end{figure}

In Fig. \ref{fig:PLPColMod}, we plot the dispersion relation of the full-flip (left panel) and spin-flip (right panel) excitations. We see that, for the valley-flip excitations, there is always a mode with zero-energy for $k=0$, resulting from the Goldstone mode of the $U(1)$ symmetry of the PLP state. The rest of the trends are the same as in the previous cases.

In Fig. \ref{fig:PLPexpmode}, we study a case where full-flip complex-frequency modes appear. The plot shows the real (solid curve) and the imaginary (dashed curve) parts of the unstable mode. The qualitative behavior is similar to that of the unstable mode appearing in the CAF state, see the discussion related to Fig. \ref{fig:CAFexpmode}

\section{Monolayer graphene}\label{sec:MLG}

Although the previous formalism was developed in order to study the collective modes of the $\nu=0$ QH state of bilayer graphene, it can be straightforwardly adapted to the $\nu=0$ QH state of monolayer graphene (MLG). We briefly revisit the content of Secs. \ref{sec:basicmodel}-\ref{sec:colmod}, adjusting the results to the present case.

\subsection{Effective Hamiltonian}

We start by writing the effective Hamiltonian of MLG (see Ref. \cite{Kharitonov2012} for a more complete review). The corresponding field operator is given by:
\begin{eqnarray}\label{eq:MLGfieldoperator}
\nonumber\hat{\psi}(\mathbf{x})&=&\left[\begin{array}{c}
\hat{\psi}_{+}(\mathbf{x})\\
\hat{\psi}_{-}(\mathbf{x})
\end{array}\right]\\
\hat{\psi}_{\sigma}(\mathbf{x})&=&\left[\begin{array}{c}
\hat{\psi}_{KA\sigma}(\mathbf{x})\\
\hat{\psi}_{KB\sigma}(\mathbf{x})\\ \hat{\psi}_{K'B\sigma}(\mathbf{x})\\-\hat{\psi}_{K'A\sigma}(\mathbf{x})
\end{array}\right]\equiv\left[\begin{array}{c}
\hat{\psi}_{K\bar{A}\sigma}(\mathbf{x})\\
\hat{\psi}_{K\bar{B}\sigma}(\mathbf{x})\\ \hat{\psi}_{K'\bar{A}\sigma}(\mathbf{x})\\\hat{\psi}_{K'\bar{B}\sigma}(\mathbf{x})
\end{array}\right],~\sigma=\pm
\end{eqnarray}
where now $A,B$ are the two sublattices of the same graphene layer. Once more, the two sublattices are interchanged in the $K'$ valley so we denote the corresponding subspace as $\bar{A}\bar{B}$. The $8$ components of the field operator belong to the total subspace $KK'\otimes\bar{A}\bar{B} \otimes s$. The MLG Hamiltonian is decomposed in the same fashion of the bilayer case, $\hat{H}=\hat{H}_0+\hat{H}_C+\hat{H}_{sr}$, but the kinetic operator of the single-particle Hamiltonian $\hat{H}_0$ now reads
\begin{equation}\label{eq:MLGspHamiltonian}
H(p)=v_F[T_{0x}P_x+T_{0y}P_y]
\end{equation}
with $v_F\thickapprox 10^6$ m/s the Fermi velocity. In the presence of a magnetic field $B$, the previous single-particle Hamiltonian (\ref{eq:spHamiltonian}) changes to
\begin{equation}\label{eq:MLGspBHamiltonian}
\hat{H}_{0}=\int\mathrm{d}^2\mathbf{x}~\hat{\psi}^{\dagger}(\mathbf{x})\left[H(\pi)-\epsilon_Z\sigma_z\right]\hat{\psi}(\mathbf{x})
\end{equation}
(note that now there are no voltage between the layers). After an appropriated unitary transformation in the sublattice subspace and following the same reasonings leading to Eq. (\ref{eq:BilayerdestructionHO}), we write $H(\pi)$ in terms of $a_B$ as:
\begin{equation}\label{eq:MLGdestructionHO}
H(\pi)=\hbar\omega_B\left[\begin{array}{cc}
0 & a_B\\
a_B^{\dagger} & 0
\end{array}\right],~\omega_B=\frac{\sqrt{2}v_F}{l_B}\simeq 5.51\times 10^{13}\sqrt{B[\text{T}]}~\text{Hz}
\end{equation}
The corresponding eigenfunctions and eigenvalues are analog to the bilayer scenario, $\Psi^{0}_{n,k,\alpha}(\mathbf{x})=\Psi^{0}_{n,k}(\mathbf{x})\chi_{\alpha}$, with
\begin{equation}\label{eq:MLGLandaueigenfunctions}
\Psi^{0}_{n,k}(\mathbf{x})=\frac{e^{iky}}{\sqrt{L_y}}\frac{1}{\sqrt{2}}\left[\begin{array}{c}
\textrm{sgn}\ n\ \phi_{|n|-1}(x+kl^2_B)\\ \phi_{|n|}(x+kl^2_B)
\end{array}\right],~\epsilon_n=\textrm{sgn}\ n\ \sqrt{|n|}\hbar\omega_B
\end{equation}
for $|n|\neq 0$ and
\begin{equation}\label{eq:MLGZLL}
\Psi^{0}_{n,k}(\mathbf{x})=\frac{e^{iky}}{\sqrt{L_y}}\left[\begin{array}{c}
0\\
\phi_{n}(x+kl^2_B)
\end{array}\right]
\end{equation}
for the ZLL, that now corresponds to just the magnetic index $n=0$.

While the Zeeman effect is the same as before, an order of magnitude analysis of the Coulomb interaction gives the dimensionless strength
\begin{equation}\label{eq:MLGcoulombfactor}
F_c=\frac{e^2_c}{\kappa l_B\hbar\omega_B}=\frac{e^2_c}{\kappa v_F}\thickapprox\frac{2.2}{\kappa}~,
\end{equation}
Note that it does not depend on the value of the magnetic field, in contrast to the case of BLG. Once more, for $\kappa=1$, $F_c$ is not a small parameter. However, by taking $\kappa\sim 5$ we can get $F_c\sim 0.4$, which can be regarded as a small value. The considerations about short-range interactions are similar to the bilayer case.

Under this assumption, we neglect LL mixing and project the Hamiltonian into the ZLL, by projecting the full Hamiltonian into that subspace \cite{Kharitonov2012,Kharitonov2012PRL}. The states for the ZLL of MLG are also restricted to the $KK'\otimes\bar{B}\otimes s$ subspace, which means that they are localized, for each valley, in one sublattice or the other hence the sublattice degree of freedom becomes equivalent to the valley degree of freedom. Reasoning in the same fashion as in BLG, the resulting effective Hamiltonian is formally equal to that of Eq. (\ref{eq:EffectiveHamiltonian}) but with $\epsilon_V=0$:
\begin{eqnarray}\label{eq:MLGEffectiveHamiltonian}
\nonumber\hat{H}^{(0)}&=&\int\mathrm{d}^2\mathbf{x}~-\epsilon_Z\hat{\psi}^{\dagger}(\mathbf{x})\sigma_z\hat{\psi}(\mathbf{x})+\frac{1}{2}\int\mathrm{d}^2\mathbf{x}~\mathrm{d}^2\mathbf{x'}:[\hat{\psi}^{\dagger}(\mathbf{x})\hat{\psi}(\mathbf{x})]V_0(\mathbf{x}-\mathbf{x'})[\hat{\psi}^{\dagger}(\mathbf{x'})\hat{\psi}(\mathbf{x'})]:\\
\nonumber &+&\sum_{i}\frac{1}{2}4\pi\hbar \omega_B l^2_B\int\mathrm{d}^2\mathbf{x}\ g_{i}:[\hat{\psi}^{\dagger}(\mathbf{x})T_{i}\hat{\psi}(\mathbf{x})]^2:+\int\mathrm{d}^2\mathbf{x}~\mathrm{d}^2\mathbf{x'}\hat{\psi}^{\dagger}(\mathbf{x})V_{DS}(\mathbf{x},\mathbf{x'})\hat{\psi}(\mathbf{x'})\\
\end{eqnarray}
We have defined the coupling constants of the short-range interactions in such way that their expressions are formally equal to the case of BLG in the presence of a magnetic field, see Eq. (\ref{eq:EffectiveHamiltonian}). The Dirac sea that creates the mean-field potential $V_{DS}(\mathbf{x},\mathbf{x'})$ is composed now by all the occupied states with $n\leq -1$.

\subsection{Mean-field phase diagram}

The HF equations for MLG in the ZLL have the same form of Eq. (\ref{eq:HFeqs}). As the $\nu=0$ QH state corresponds to half-filling of the ZLL, the electrons occupy in the same way two orthogonal spinors $\chi_{a,b}$ in valley-spin space and leave empty the remaining orthogonal spinors $\chi_{c,d}$. Then, after projecting the HF equations into the orbital part of the wave functions, we get the algebraic equation
\begin{equation}\label{eq:MLGHFenergy}
\epsilon_{0,\alpha}\chi_{\alpha}=\frac{F_{00}}{2}\chi_{\alpha}-F_{00}P\chi_{\alpha}+\sum_{i}u_{i}\left([\text{tr}(PT_{i})]T_{i}-T_{i}PT_{i}\right)\chi_{\alpha}-\epsilon_VT_z\chi_{\alpha}-\epsilon_Z\sigma_z\chi_{\alpha}
\end{equation}
where now $u_{i}=2\hbar \omega_B g_{i}$. The previous equation presents the same valley-spin structure of Eq. (\ref{eq:HFenergy}). As a consequence, the spinorial solutions $\chi_{\alpha}$ to the HF equations are identical to those of BLG and their mean-field energies are:
\begin{equation}\label{eq:MLGHFenergystructure}
\epsilon_{0,(a,b)}=-\frac{F_{00}}{2}+\epsilon_{(a,b)},~\epsilon_{0,(c,d)}=\frac{F_{00}}{2}+\epsilon_{(c,d)}
\end{equation}
Moreover, the total energy of the ground state (per wave vector state) is:
\begin{equation}\label{eq:MLGmeanfieldenergy}
E_{HF}=-\frac{F_{00}}{2}+E(P)\\
\end{equation}
with $E(P)$ given by Eq. (\ref{eq:meanfieldenergy}). As the Coulomb contribution is degenerate, the true ground state is that which minimizes $E(P)$. Therefore, we conclude that the mean-field phase diagram for the $\nu=0$ QH state in MLG is the same of BLG for $\epsilon_V=0$. In that case, the PLP phase of Sec. \ref{subsec:PLPphase} changes to a fully interlayer coherent (ILC) phase since then $\theta_V=\pi/2$ and the vector $\mathbf{n}$ pointing the polarization in the valley space is fully contained in the plane, $\mathbf{n}=[\cos\phi_v,\sin\phi_v,0]$. As the valley degree of freedom is equivalent to the sublattice in the ZLL, this state corresponds to a coherent mixture of the two sublattices.

There are experimental evidences that the phase for the $\nu=0$ QH state of monolayer graphene for zero in-plane component of the magnetic field is also the CAF phase \cite{Young2014}, so $u_z>-u_{\perp}>0$ as in the bilayer case.

\subsection{Collective modes}\label{sec:MLGcolmod}

The TDHFA developed for computing the collective modes of BLG is also valid for MLG. In particular, the dispersion relation is still obtained from Eq. (\ref{eq:detcollmod}) but now there is only one possible value for the magnetic index, $n_k=0$. Thus, as the structure in the valley-spin subspace is the same of the bilayer problem, the building blocks of the dispersion matrix $X$, $Y^{\alpha_k\alpha_l,\alpha_j\alpha_m}(\mathbf{k},\omega)$, are now scalars instead of $4\times 4$ matrices:
\begin{equation}\label{eq:MLGYstructure}
Y^{\alpha_k\alpha_l,\alpha_j\alpha_m}(\mathbf{k},\omega)=Y^{\alpha_k\alpha_l,\alpha_j\alpha_m}(k)-(\nu_{\alpha_k}-\nu_{\alpha_l})\hbar\omega
\end{equation}
with
\begin{eqnarray}\label{eq:MLGpseudospinmatrices}
Y^{\alpha_k\alpha_l,\alpha_j\alpha_m}(k)&=&\left[F(k)+\Delta^{\alpha_k\alpha_l}\right]\delta_{\alpha_k\alpha_j}\delta_{\alpha_l\alpha_m}+Q(k)M^{\alpha_k\alpha_l,\alpha_j\alpha_m}\\
\nonumber F(k)&\equiv& F_{00}-C_{00,00}(k)=F_{00}\left[1-I_0\left(\frac{(kl_B)^2}{4}\right)e^{-\frac{(kl_B)^2}{4}}\right]\\
\nonumber Q(k)&\equiv& 2R_{00,00}(k)=e^{-\frac{(kl_B)^2}{2}}
\end{eqnarray}
and we have made explicit the isotropy in the wave vector dependence of $Y^{\alpha_k\alpha_l,\alpha_j\alpha_m}(k)$. The expression of the function $C_{00,00}(k)$ can be obtained from the results of Appendix \ref{app:analyticalTDHFA}. By the same arguments of Sec. \ref{subsec:preliminar}, the collective-mode frequencies are the eigenvalues of the matrix $\tilde{X}(k)$ in Eq. (\ref{eq:XYgen32}), which now is reduced to a $8\times 8$ matrix as $Y^{\alpha_k\alpha_l,\alpha_j\alpha_m}(k)$ are scalars. Since the ground states in MLG have the same valley-spin structure that in BLG, all the coefficients $\Delta^{\alpha_k\alpha_l}$, $M^{\alpha_k\alpha_l,\alpha_j\alpha_m}$ remain identical for every phase. Thus, Eqs. (\ref{eq:characterizationF}), (\ref{eq:characterizationFLP}), (\ref{eq:characterizationCAF}) and (\ref{eq:characterizationPLP}) can be still used for computing the eigenvalues of $\tilde{X}(k)$. Moreover, as $F(0)=0$ and $Q(0)=1$, we recover at zero momentum the analytical results previously obtained for the symmetric modes $\omega_S$ [see discussion accompanying Eq. (\ref{eq:zerofrequencyequation}]. In fact, for MLG, the collective modes within the ZLL can be computed explicitly at arbitrary momentum $k$.

\subsubsection{Ferromagnetic phase}

Following Sec. \ref{subsec:Fcol}, there are spin flip excitations and full-flip excitations. {\it Mutatis mutandi}, the frequencies of the
two branches of the spin-flip modes are [see Eq. (\ref{eq:Fspinflipmatrix}) and related text]:
\begin{equation}\label{eq:MLGspinflip}
\hbar\omega^{\pm}(k)=Y^{\pm}(k)=Y^{ac,ac}(k)\pm Y^{ac,bd}(k)=F(k)+2\epsilon_Z+2u_{\perp}+u_z+Q(k)\left[-u_z\mp2u_{\perp}\right]
\end{equation}
On the other hand, for the full-flip modes, we get:
\begin{equation}\label{eq:MLGFfullflip}
\hbar\omega^{bc}(k)=Y^{bc,bc}(k)=F(k)+2\epsilon_Z+2u_{\perp}+u_z+Q(k)u_z
\end{equation}
As $\epsilon_V=0$, $\omega^{bc}(k)=\omega^{ad}(k)$.

\subsubsection{Full layer-polarized phase}\label{subsec:MLGFLP}

In this phase, there are valley-flip and full-flip modes, see Sec. \ref{subsec:FLPcol}. The valley-flip modes are
\begin{equation}\label{eq:MLGvalleyflip}
\hbar\omega^{\pm}(k)=Y^{\pm}(k)=Y^{ac,ac}(k)\pm Y^{ac,bd}(k)=F(k)-2u_{\perp}-3u_z+Q(k)\left[2u_{\perp}+u_z\pm2u_{\perp}\right]
\end{equation}
while the frequency of the full-flip modes is
\begin{equation}\label{eq:MLGFLPfullflip}
\hbar\omega^{bc}(k)=Y^{bc,bc}(k)=F(k)-2\epsilon_Z-2u_{\perp}-3u_z+Q(k)u_z=Y^{-}(k)-2\epsilon_Z
\end{equation}
and $\omega^{ad}(k)=\omega^{bc}(k)+4\epsilon_Z$. In correspondence with the results of Sec. \ref{subsec:FLPcol}, one of the valley-flip modes is indeed a spin-flip mode, $\omega^{bc}(k)=\hbar\omega^{-}(k)-2\epsilon_Z$.

\subsubsection{Canted anti-ferromagnetic phase}\label{subsec:MLGCAF}

The equivalent of the spin-flip modes of the CAF phase, Sec. \ref{subsec:CAFcol}. Following the same steps leading to the matrix of Eq. (\ref{eq:CAFspinflipreduction}), we arrive at the corresponding $2\times 2$ matrix
\begin{equation}\label{eq:MLGCAFspinflip}
\tilde{X}^{sf,\pm}(\mathbf{k})=\left[\begin{array}{cc}
Y^{\pm}(k)  & Y^{ac,db}(k) \\
-Y^{ac,db}(k) & -Y^{\pm}(k)   \\
\end{array}\right]
\end{equation}
The dispersion relation is obtained from its eigenmodes:
\begin{eqnarray}\label{eq:MLGCAFsp}
\hbar\omega^{sf,\pm}(k)&=&\sqrt{\left[Y^{\pm}(k)\right]^2 - 4u^2_{\perp}\sin^4 \theta_s Q^2(k)}\\
\nonumber Y^{\pm}(k)&=&F(k)+u_z-2u_{\perp}+Q(k)\left[-u_z\mp2u_{\perp}\cos^2 \theta_s\right]
\end{eqnarray}

On the other hand, for the full-flip modes of Eq. (\ref{eq:CAFpseudospinflip}), one gets the analog matrix
\begin{equation}\label{eq:MLGCAFff}
\tilde{X}^{ff}(k)=\left[\begin{array}{cc}
Y^{bc,bc}(k) & Y^{bc,da}(k)\\
-Y^{da,bc}(k) & -Y^{da,da}(k)
\end{array}\right]
\end{equation}
that gives:
\begin{eqnarray}\label{eq:MLGCAFfffinal}
\hbar\omega^{bc,da}(k)&=&\sqrt{\left[Y^{bc,bc}(k)\right]^2-4u^2_{\perp}\sin^4 \theta_s Q^2(k)}\\
\nonumber Y^{bc,bc}(k)&=&F(k)+u_z-2u_{\perp}+Q(k)\left[u_z+2u_{\perp}\sin^2 \theta_s\right]
\end{eqnarray}
The remaining mode satisfy $\omega^{ad,cb}(k)=\omega^{bc,da}(k)$ as $\epsilon_V=0$.

\subsubsection{Inter-layer coherent phase}\label{subsec:MLGILC}

For the ILC phase, we set $\epsilon_V=0$ and, consistently, $\theta_V=\pi/2$ in Eq. (\ref{eq:characterizationPLP}). Following the lines of Sec. \ref{subsec:PLPcol}, an analog computation to the spin-flip modes of the CAF phase gives the equivalent of the valley-flip modes. In particular, their dispersion relation is:
\begin{eqnarray}
\hbar\omega^{vf,\pm}(k)&=&\sqrt{\left[Y^{\pm}(k)\right]^2-(u_z-u_{\perp})^2Q^2(k)}\\
\nonumber Y^{\pm}(k)&=&F(k)-u_z-4u_{\perp}+Q(k)\left[u_z+2u_{\perp}\pm \left(u_{\perp}+u_z\right)\right]
\end{eqnarray}
Finally, for the full-flip excitations, we get
\begin{equation}
\hbar\omega^{bc,da}(k)=-2\epsilon_Z+\sqrt{\left[Y^{-}(k)\right]^2-(u_z-u_{\perp})^2Q^2(k)}=\hbar\omega^{vf,-}(k)-2\epsilon_Z
\end{equation}
As in the FLP phase, the valley-flip branch $-$ is indeed a full-flip mode. The remaining full-flip mode is $\hbar\omega^{ad,bc}(k)=\hbar\omega^{vf,-}(k)+2\epsilon_Z$.

\subsubsection{Numerical results}

\begin{figure*}[tb!]\label{fig:MLGFFLP}
\begin{tabular}{@{}cc@{}}
    \includegraphics[width=0.5\columnwidth]{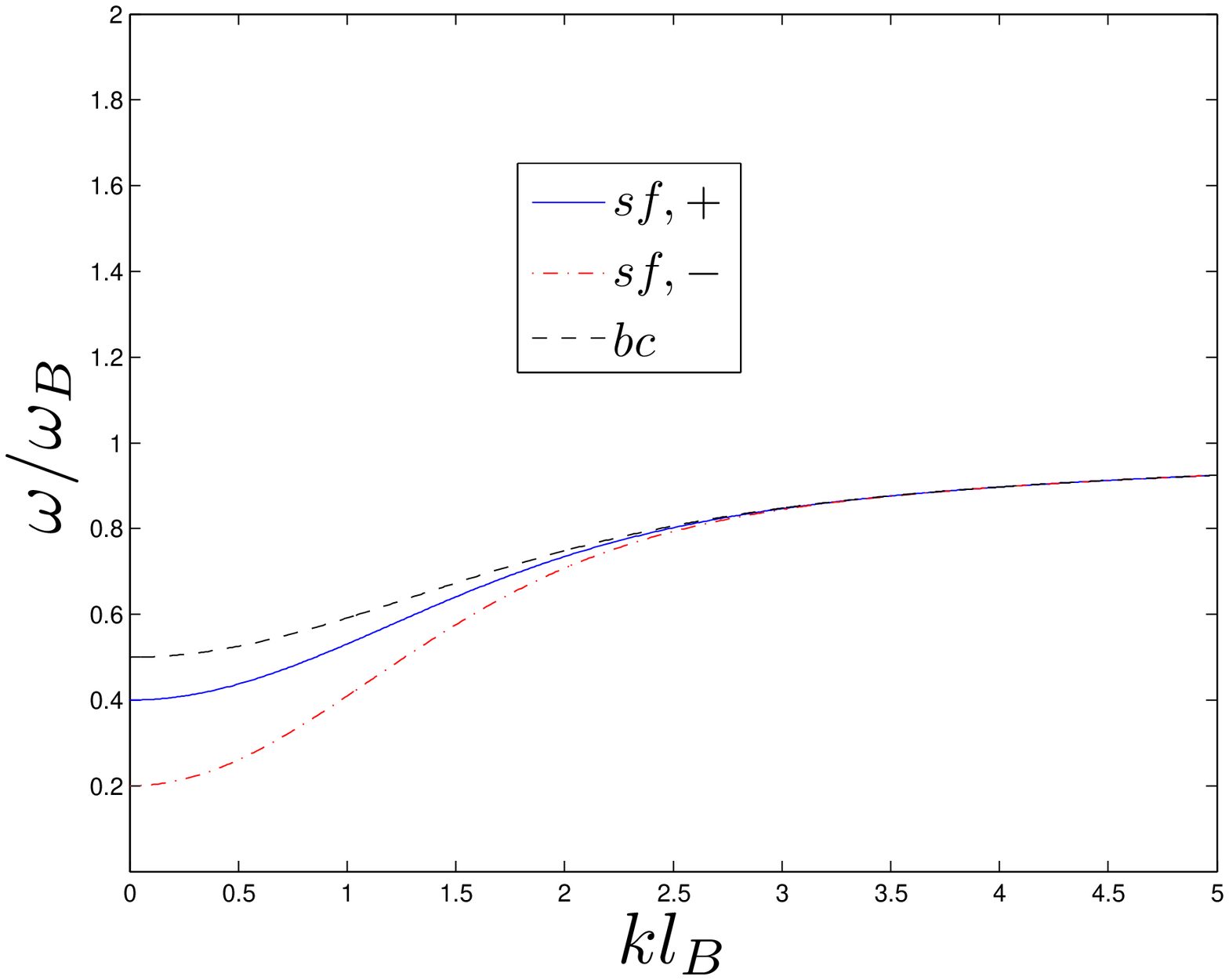} &
    \includegraphics[width=0.5\columnwidth]{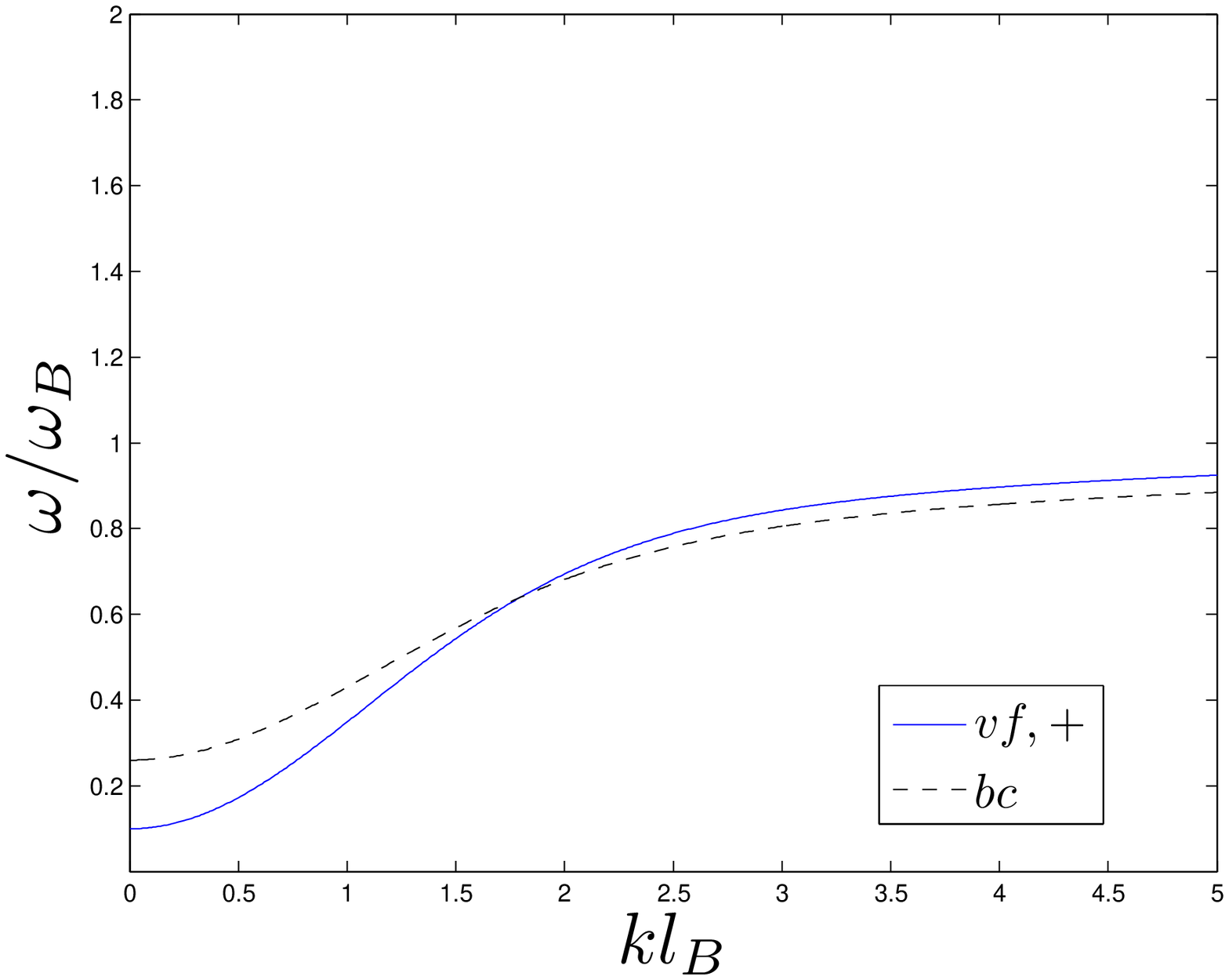} \\
\end{tabular}
\caption{Plot of the collective modes of the $\nu=0$ QH state in monolayer graphene for the F and FLP phases. Left: dispersion relation of the F phase with parameters $F_c=0.5$, $u_z=0.1$, $u_{\perp}=-0.05$ and $\epsilon_Z=0.2$. The two spin-flip modes are plotted in solid blue ($+$ branch) and in dot-dashed red ($-$ branch). The full-flip mode frequency $\omega^{bc}(k)$ is plotted is dashed black. Right: dispersion relation of the FLP phase with parameters $F_c=0.5$, $u_z=-0.1$, $u_{\perp}=-0.05$ and $\epsilon_Z=0.02$. The solid blue line is the true valley flip mode (labeled as $+$ in Sec. \ref{subsec:MLGFLP}) and the black dashed line is the full-flip mode $\omega^{bc}(k)$}.
\end{figure*}

\begin{figure*}[tb!]\label{fig:MLGCAFPLP}
\begin{tabular}{@{}cc@{}}
    \includegraphics[width=0.5\columnwidth]{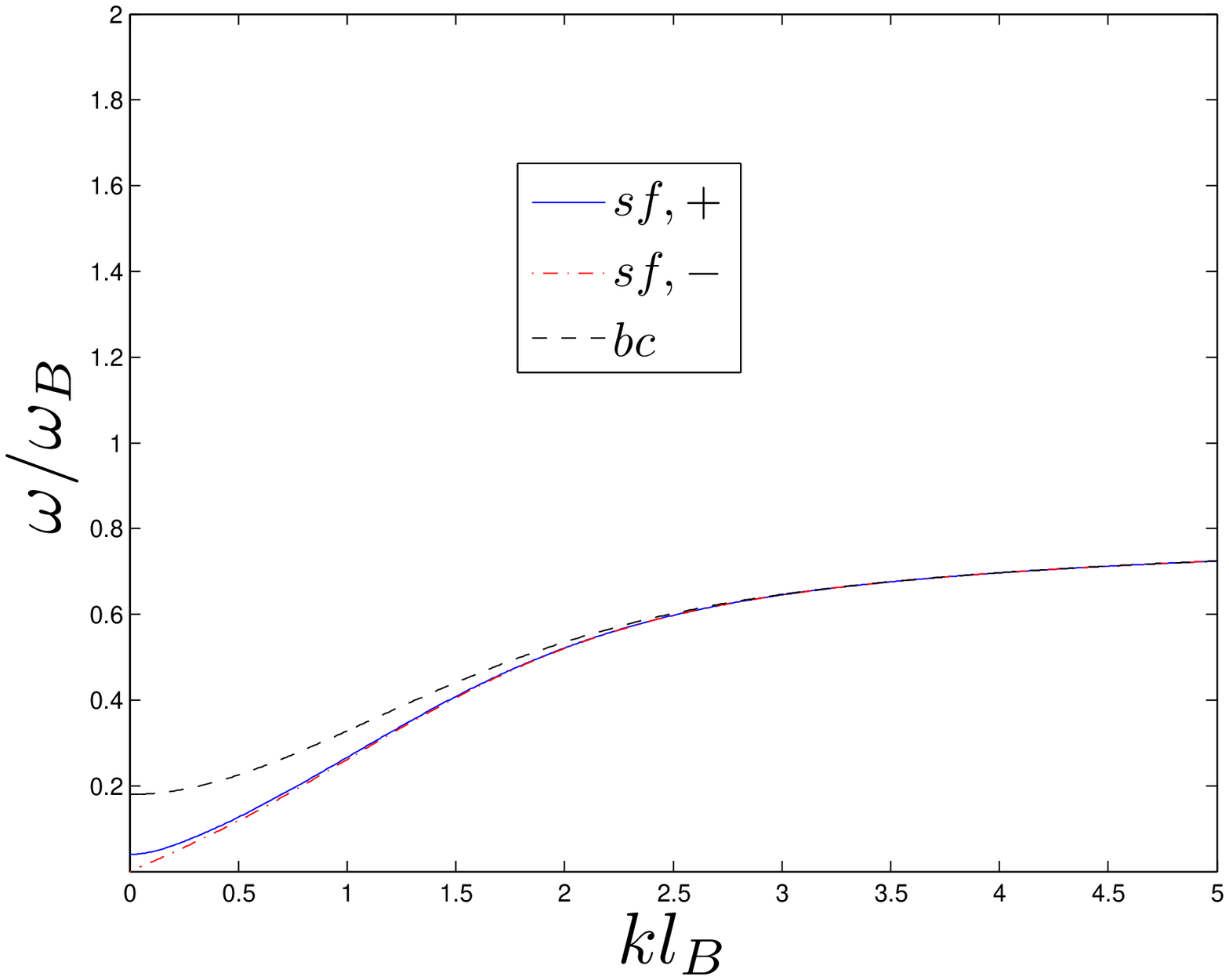} &
    \includegraphics[width=0.5\columnwidth]{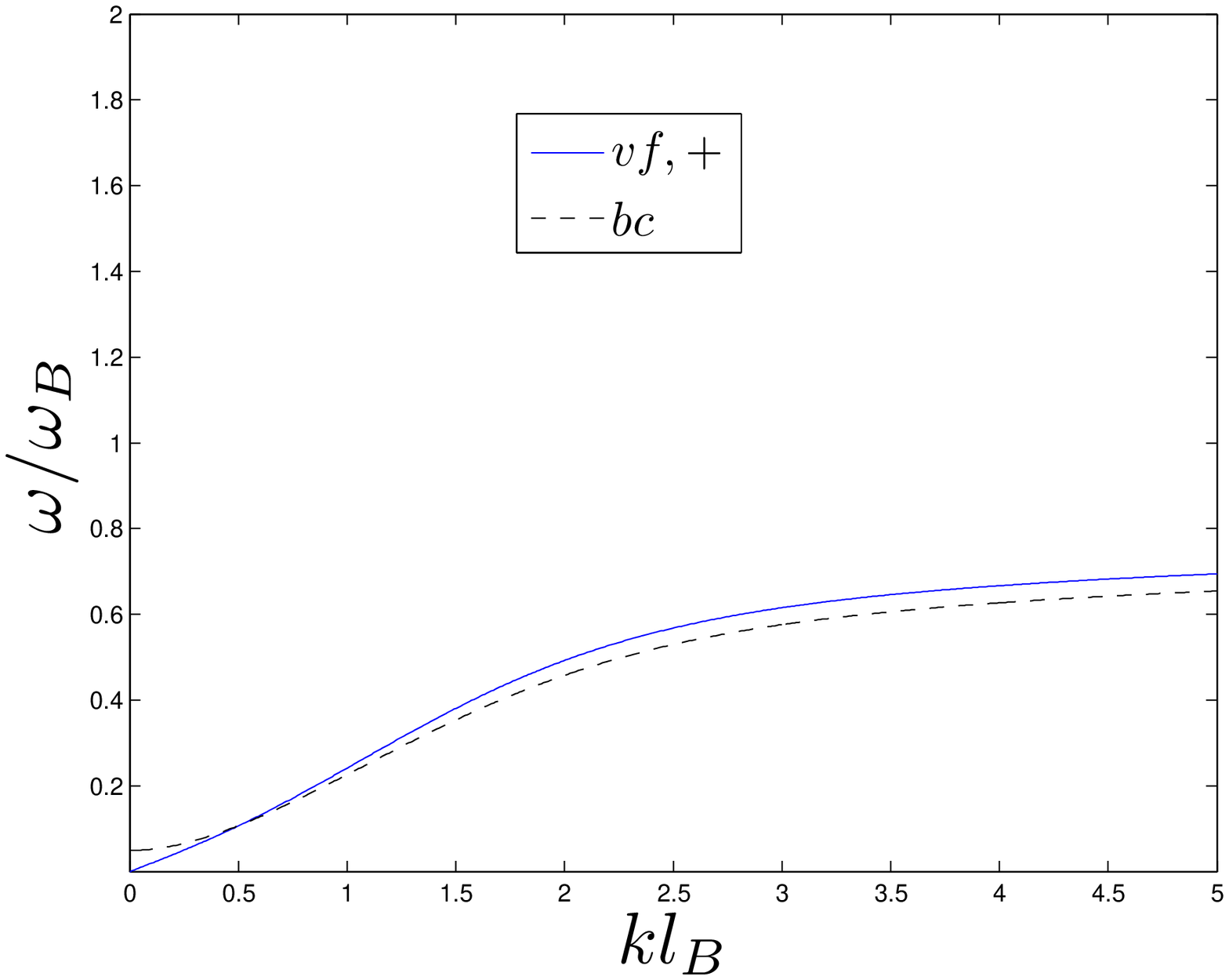} \\
\end{tabular}
\caption{Plot of the collective modes of the $\nu=0$ QH state in monolayer graphene for the CAF and ILC phases. Left: dispersion relation of the CAF phase with parameters $F_c=0.5$, $u_z=0.1$, $u_{\perp}=-0.05$ and $\epsilon_Z=0.02$. The two spin-flip modes are plotted in solid blue ($+$ branch) and in dot-dashed red ($-$ branch). The full-flip mode frequency $\omega^{bc,da}(k)$ is plotted is dashed black. Right: dispersion relation of the ILC phase with parameters $F_c=0.5$, $u_z=0.03$, $u_{\perp}=-0.05$ and $\epsilon_Z=0.02$. The solid blue line is the true valley flip mode (labeled as $vf,+$ in Sec. \ref{subsec:MLGILC}) and the black dashed line is the full-flip mode $\omega^{bc,da}(k)$.}
\end{figure*}

We display here the dispersion relations for the different phases. The results for the F (left panel) and FLP (right panel) phases are shown in Fig. \ref{fig:MLGFFLP} while the results for the CAF (left) and ILC (right) are shown in Fig. \ref{fig:MLGCAFPLP}. Their qualitative trends are similar to the symmetric modes $S$ of bilayer graphene; indeed, at $k=0$, they have the same expression. The dispersion relation of the dynamical instabilities of the CAF and ILC phases, when exist, is quite similar to the case of bilayer graphene (not shown).

\section{Effects of Landau-level mixing}\label{sec:renorm}

In this section, we consider the case of suspended samples of bilayer graphene, where $\kappa=1$. In this case, $F_c\gtrsim 1$ for any realistic value of the magnetic field, so we cannot neglect LL mixing. One possible way to take into account the effects of the LL mixing is the screening of the Coulomb interaction in the large-$N$ approximation \cite{Foster2008,Basko2008,Aleiner2007,Lemonik2010,Kharitonov2012}, which leads to an RPA-type screening given by:

\begin{equation}\label{eq:RPACoulomb}
\bar{V}(k)=\frac{V_0(k)}{1-\Pi^{0}(k,0)V_0(k)}
\end{equation}
where
\begin{equation}\label{eq:freepolarization}
\hbar\Pi^{0}(\mathbf{k},\omega)=\sideset{}{'}\sum_{\substack{n_k,n_l\\ \alpha_k \alpha_l}}~\frac{\delta_{\alpha_k,\alpha_l}}{2\pi l^2_B}D^{0}_{kl}(\omega)|A^{(2)}_{n_kn_l}(\mathbf{k})|^2
\end{equation}
is the non-interacting polarization for the $\nu=0$ QH state and $n_k,n_l$ take values for all integers except $-1$, as in Eq. (\ref{eq:fieldoperatorLL}). In the same fashion, $D^{0}_{kl}(\omega)$ is the non-interacting two-particle propagator, with the non-interacting values for the energies and occupation numbers. Other approaches allow for LL mixing in the Hartree-Fock formalism (see for instance Ref. \cite{Toke2013}); however, they are not expected to describe correctly the dispersion relation near $\mathbf{k}=0$. The polarization $\Pi^{0}(\mathbf{k},\omega)$ is obtained from the non-interacting density-density correlation function, following an analog calculation to that leading to Eq. (\ref{eq:PiAfinal}) but using the bare vertex and allowing for LL mixing. Here, $A^{(2)}_{nn'}(\mathbf{k})$ is the bilayer graphene magnetic form factor:
\begin{eqnarray}\label{eq:magneticFFBilayer}
\nonumber A^{(2)}_{nn'}(\mathbf{k})&=&\frac{1}{2}\left[a^{-}_{n,n'}A_{|n|-2,|n'|-2}(\mathbf{k})\textrm{sgn}\ n~\textrm{sgn}\ n'+a^{+}_{n,n'}A_{nn'}(\mathbf{k})\right]\\
a^{\pm}_{n,n'}&=&\sqrt{1\pm(\delta_{n0}+\delta_{n1})}\sqrt{1\pm(\delta_{n'0}+\delta_{n'1})}
\end{eqnarray}
with $A_{nn'}(\mathbf{k})$ the usual magnetic form factor, given by Eq. (\ref{eq:magneticFF}). Neglecting the small corrections due to the Zeeman effect and the layer voltage, the free static polarization reads:
\begin{eqnarray}\label{eq:polarizationseries}
\Pi^{0}(k,0)&=&-\frac{N}{2\pi l^2_B}\frac{f(kl_B)}{\hbar\omega_B}\\
\nonumber f(kl_B)&=&\sum^{\infty}_{n_k=0}\sum^{\infty}_{n_l=2}~\frac{2|A^{(2)}_{n_k,-n_l}(\mathbf{k})|^2}{\sqrt{n_k(n_k-1)}+\sqrt{n_l(n_l-1)}}
\end{eqnarray}
where $N=4$ is the number of the valley-spin components of the field, which is expected to be a sufficiently large value to provide a good approximation \cite{Basko2008,Lemonik2010,Kharitonov2012}. The function $f(x)$ is dimensionless as $|A^{(2)}_{n_kn_l}(\mathbf{k})|^2$ only depends on the value $kl_B$, with $k=|\mathbf{k}|$. The effective screened Coulomb interaction can be then rewritten as:
\begin{equation}\label{eq:Screened}
\bar{V}(k)=\frac{V_0(k)}{1+NF_c\frac{f(kl_B)}{kl_B}}
\end{equation}

\begin{figure}[bt]\label{fig:dimensionlessfunction}
\includegraphics[width=\columnwidth]{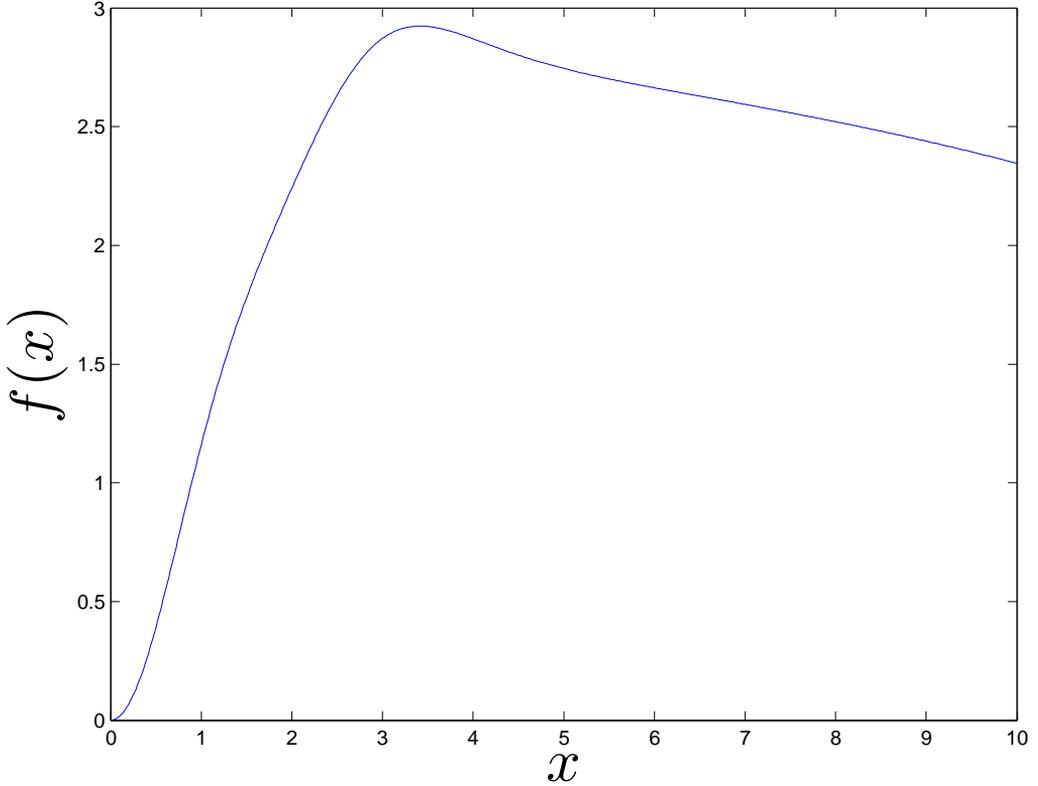}
\caption{Plot of the function $f(x)$ defined in Eq. (\ref{eq:polarizationseries}), where we have imposed a finite cutoff in the series $N_C=150$ for the computation.}
\end{figure}

The dimensionless function $f(x)$ is plotted in Fig. \ref{fig:dimensionlessfunction}. As $f(x)\sim1$ for $x\gtrsim 1$, for relevant momentum $kl_B\sim 1$ the screened potential satisfies:
\begin{equation}\label{eq:ScreenedShortRange}
\bar{V}(k)\sim \frac{4\pi \hbar^2}{m}\frac{1}{N}
\end{equation}
The dimensionless strength of the screened Coulomb interaction is then of order $N^{-1}=0.25\ll 1$ and thus, we can now safely neglect LL mixing and restrict once more ourselves to the ZLL.

The other effect that results from considering LL mixing is the renormalization of the coupling constants \cite{Kharitonov2012,Kharitonov2012PRL}, that amounts to replace the bare coupling constants by their renormalized values, $\bar{g}_{\perp},\bar{g}_{z}$. Taking into account the previous observations, we consider the following effective Hamiltonian
\begin{eqnarray}\label{eq:EffectiveHamiltonianScreened}
\nonumber\hat{H}^{(0)}&=&\int\mathrm{d}^2\mathbf{x}~\hat{\psi}^{\dagger}(\mathbf{x})(-\epsilon_VT_z-\epsilon_Z\sigma_z)\hat{\psi}(\mathbf{x})+
\frac{1}{2}\int\mathrm{d}^2\mathbf{x}~\mathrm{d}^2\mathbf{x'}:[\hat{\psi}^{\dagger}(\mathbf{x})\hat{\psi}(\mathbf{x})]\bar{V}(\mathbf{x}-\mathbf{x'})[\hat{\psi}^{\dagger}(\mathbf{x'})\hat{\psi}(\mathbf{x'})]:\\
&+&\sum_{i}\frac{1}{2}\frac{4\pi\hbar^2}{m}\int\mathrm{d}^2\mathbf{x}~\bar{g}_{i}:[\hat{\psi}^{\dagger}(\mathbf{x})T_{i}\hat{\psi}(\mathbf{x})]^2:
\end{eqnarray}
instead of that of Eq. (\ref{eq:EffectiveHamiltonian}). We see that the result of considering LL mixing is reduced to
replacing the Coulomb interaction and the coupling constants by their effective values $\bar{V}$ and $\bar{g}_{i}$ in Eq. (\ref{eq:EffectiveHamiltonian}), respectively. The solutions to the corresponding HF equations are the same as long as the screened Coulomb interaction dominates over the short-range interactions. Under this assumption, the phase diagram is exactly the same of Fig. \ref{fig:PhaseDiagram} but replacing $u_{i}$ by $\bar{u}_i$. The same holds for the computation of the collective modes, where we arrive at the renormalized version of the dispersion matrix, $\bar{X}$. As the ground states are the same, the valley-spin sector is identical so all the collective modes preserve the structure discussed in Sec. \ref{subsec:results}. Specifically, the symmetric modes still describe the phase transition and their frequencies are given by the same expressions but replacing $u_{i}$ by $\bar{u}_i$. Thus, only the quantitative values change but not the qualitative features and the resulting physical conclusions. In particular, the structure of the Goldstone modes and of the dynamical instabilities is preserved.

This kind of effective renormalized Hamiltonian has already been used for studying edge excitations in the $\nu=0$ state or the fractional QH effect in graphene in the ZLL \cite{Kharitonov2012b,Abanin2013}. It can been also applied to study the $\nu=0$ state for MLG \cite{Kharitonov2012,Kharitonov2012b}.


\section{Experimental remarks}\label{sec:remarksexps}

We make here some experimental remarks, following the results quoted in Sec. \ref{subsec:Criticalpoint}. The value of the parameters $u_{\perp},u_z$ is such that $u_z>-u_{\perp}>0$ so the state is in the CAF phase for $\epsilon_Z=\epsilon_V=0$. All the phases of Fig. \ref{fig:PhaseDiagram} can be explored by varying the in-plane component of the magnetic field or the layer voltage. On the other hand, we see from Eqs. (\ref{eq:CAFinstabilitycondition}), (\ref{eq:PLPinstabilitycondition}) that the conditions for the appearance of the unstable modes for both CAF and PLP phases are
\begin{eqnarray}\label{eq:grapheneinstability}
\frac{\epsilon^2_Z}{2u_{\perp}}&>&u_{\perp}+u_{z},~\text{(CAF)}\\
\nonumber u^2_{z}&>&u^2_{\perp}+\epsilon^2_Z,~\text{(PLP)}
\end{eqnarray}
Interestingly, the saturation of the two previous inequalities correspond to the boundary between the PLP and CAF phases for $\epsilon_V=0$ ($\epsilon_Z=0$). The first condition is unlikely to be achieved from the arguments displayed above. The second condition, however, is compatible with the values expected the coupling constants. Let's imagine that the in-plane magnetic field and the layer voltage are such that the system is initially in the PLP phase. If we now reduce suddenly the layer voltage sufficiently enough for the instability condition of the PLP phase to be fulfilled, then we would be able to observe this dynamical instability, in a similar way to the BH laser effect described in Chapter \ref{chapter:BHL}. Future work should address this issue more deeply, exploring the possible experimental consequences of the appearance of the dynamical instability.

On the other hand, another possible interesting feature is that the energy gap of the symmetric modes is easily manipulable by tuning the in-plane component of the magnetic field or the layer voltage as they do not depend on the strength of the Coulomb interaction.

The above comments also apply to monolayer graphene except that, unfortunately, there is not an equivalent of the layer voltage so we cannot reproduce the mentioned protocol for observing the dynamical instability.

\section{Conclusions and outlook} \label{sec:QHFMConclusions}

In this chapter, we have studied the $\nu=0$ QH state within a mean-field Hartree-Fock approach restricted to the zero-energy Landau level.
We have reproduced the mean-field phase diagram of the $\nu=0$ QH state of Ref. \cite{Kharitonov2012} by solving the self-consistent Hartree-Fock equations. Within the time-dependent Hartree-Fock approximation, we have computed the intra-LL collective modes, that in the limit of zero momentum can be obtained analytically. In particular, the lowest-energy modes that describe the phase transition are the symmetric modes, corresponding to a symmetric combination of excitations conserving the magnetic number. We have seen that, at the boundary between the ferromagnetic and the fully layer-polarized phases, there is a gapless mode, resulting from a residual symmetry [see Eq. (\ref{eq:Residualphase})] that can be regarded as a remanent of the broken $SO(5)$ symmetry described in Ref. \cite{Wu2014}. In respect to the boundary between the canted anti-ferromagnetic and the partially layer-polarized phases, a similar gapless mode appears. Interestingly, we have seen that these phases can present dynamical instabilities. Although the dynamical instability of the CAF phase is unlikely to be observed since it requires $u_{\perp}+u_z<0$ and this condition is incompatible with the current experimental observations, we have discussed in Sec. \ref{sec:remarksexps} a possible protocol for the observation of the dynamical instability associated to the PLP phase. Importantly, we remark that the possible appearance of such unstable modes results from the short-range valley/sublattice interactions. Another interesting feature is that the gap of the symmetric modes is easily tunable by varying the in-plane magnetic field or the layer voltage.

The performed calculations are straightforwardly translated to monolayer graphene, due to the formal analogy of the short-range interactions. We have found that most of the conclusions of the previous paragraph still hold in the monolayer scenario. We also have analyzed the effects of LL mixing and we have accounted them by screening the long-range Coulomb interaction and renormalizing the coupling constants of the short-range interactions, following Ref. \cite{Kharitonov2012}. The resulting effective Hamiltonian is formally similar to that previously considered for the calculations and hence, the phase diagram and the collective modes present the same structure as that discussed above.

\part{Global conclusions}

\chapter{Global conclusions and future perspectives}\label{chapter:conclusions}

In this section we briefly summarize the content of this thesis, remarking the most important results and discussing future perspectives. In the first part of the thesis, we have explored the implementation of gravitational analogs in Bose-Einstein condensates. After an introductory chapter on the matter, we have reported numerical simulations of the formation of a black-hole configuration in a realistic setup. Specifically, we have studied the process whereby an initially confined condensate starts leaking after the lowering of the optical lattice that acts as an initial confining agent. We have first considered the problem in a slightly idealized model in order to understand the main features of the system. We have seen that, under certain conditions (broad conduction bands, adiabatic evolution and chemical potential near the bottom of the conduction band) a quasi-stationary black-hole configuration is achieved. After this training, we have switched to a realistic case where we have confirmed the previous observations. In addition, for the realistic optical lattice with a Gaussian envelope, we have observed and theoretically explained that the sonic horizon is precisely formed at the maximum of the envelope. Above the reached black-hole configuration, we have made some preliminary calculations of the Hawking spectrum. We have seen that, whenever the top of the conduction band is below the Hawking frequency, a highly non-thermal peaked structure appears near the top of the conduction band, which could be a promising scenario for an eventual detection of the Hawking effect.

Thus, we conclude from the results of the simulations that the considered protocol is a good candidate to produce a quasi-stationary black-hole configuration in achievable experimental setups. However, we remark the preliminary character of the results for the Hawking spectrum. A more detailed work should address the problem in the future, taking into account the (weak) time-dependence of the background and the finite size of the subsonic region. Also, an adapted calculation should be performed for this setup as the quantum state of the system in the quasi-stationary regime is the time evolution of the initial thermal equilibrium state, which could probably lead to an integration of the time-dependent BdG equations, requiring much harder numerical computations than those presented here. Apart from the gravitational analogy, the quasi-stationary black hole here studied could also be interesting in quantum transport scenarios since it provides a quasi-stationary supersonic current with a very well defined velocity.

In the third chapter of the thesis we have addressed the existence of criteria for the detection of the spontaneous Hawking radiation. Borrowing some ideas from quantum optics, we have proposed the violation of Cauchy-Schwarz type inequalities as a signature of quantum behavior. In particular, we have argued that the violation of the Cauchy-Schwarz inequality involving the outgoing channels associated to the Hawking effect is a smoking gun of the presence of the spontaneous Hawking signal since stimulated Hawking radiation or linearized coherent motion above the condensate wave function are not able to produce such violation just by themselves. We have compared the proposed criterion with the generalized Peres-Horodecki criterion; we have found that, under typical assumptions, both criteria become equivalent. After that, we have presented a discussion on the possible experimental implementation. Importantly, we have found that only the violation of quadratic Cauchy-Schwarz violations can be detected due to the different spatial location of the asymptotic regions. The numerical results show, however, that this does not represent a major limitation in practice for the detection of the Hawking effect. Also, as noticed before, all the above criteria are equivalent under general conditions and, besides, the quadratic Cauchy-Schwarz violation is indeed a signal of quantum behavior by itself. The results of this chapter of great importance since they provide the theoretical ground for the implementation of experimental detection schemes. Apart from the direct implications for the detection of the Hawking effect, the content of this chapter can also be of interest for other fields such as quantum optics, quantum information physics or in the broader topic of bosonic condensates.

The last chapter of the gravitational analog part is devoted to the study of the so-called black-hole laser, formed by a black hole-white hole pair. By integrating the corresponding time-dependent Gross-Pitaevskii equation, we have analyzed the long-time behavior of the system, where the instability has grown up to saturation. On the one hand, we have confirmed the theoretical predictions for the unstable behavior at short times and for the long-time stationary regime, in which the system reaches the non-linear ground state solution. On the other hand, we have extensively characterized a new regime of continuous emission of solitons, where the system is sufficiently unstable to keep oscillating in such a way that emits perfectly periodic trains of solitons into the upstream region. A comparison with standard laser devices suggests that the periodic emission of solitons is the {\it actual} black-hole laser effect. The results here presented can be interesting in several fields. For instance, the ``evaporation'' of the horizons caused by the evolution towards the non-linear ground state solution can help to understand the much complex case of the evaporation in an actual cosmological black hole due to the spontaneous Hawking effect. In respect to the continuous emission of solitons, they represent a non-linear analog of an optical laser device, with potential interest in quantum transport or atomtronics scenarios. In the future, new theoretical works should provide a better understanding of the regime of continuous emission of solitons. The effect of quantum fluctuations above the coherent soliton laser regime is an open subject that should also be addressed. In respect to the experimental part, it is expected that forthcoming experiments will be able to characterize the long-time non-linear behavior of the black-hole laser. We remark that most of the results presented in this chapter can be directly translated to other analog scenarios such as quantum fluids of light, for instance.

In the fifth chapter, we have switched, within the same context of boson gases, to the opposite limit of a thermal cloud above the critical temperature. We have studied the response induced in the system by the introduction of a temporal Bragg pulse. We have shown using both classical and quantum descriptions that, whenever the local-density approximation is valid, interactions are negligible and the pulse is sufficiently short, the formed periodic density pattern decays to the equilibrium configuration. However, this damping mechanism is not due to the collisional relaxation of the collective modes usually described in the standard literature but rather to the thermal disorder of the particles. In particular, the decay is given by a Gaussian function with a characteristic time that only depends on the temperature, the mass of the atoms and the wave vector of the lattice potential. We have discussed thoroughly the validity of the approximations made and the link between the classical and quantum calculations. We have also shown that, indeed, this decay is a very general phenomenon that appears under quite general grounds for any spatially periodic pulse. A very good agreement is found between the theoretical predictions and the experimental data provided by the Technion group. The developed formalism can be translated to other systems such as Fermi gases at sufficiently high temperature or the thermal component of a Bose gas below the critical temperature whenever the picture of non-interacting atoms holds. A future work could try to extend this formalism to the phonons above the condensate.

Finally, we have devoted the content of the last chapter to a very different system: quantum Hall effect on graphene. In particular, we have studied the $\nu=0$ quantum Hall state of bilayer graphene. Using a mean-field Hartree-Fock approach, we have re-derived the corresponding mean-field phase diagram. After that, we have computed the collective modes within the zero Landau level trough the time-dependent Hartree-Fock approximation. Among the most remarkable results, we have found that at the boundary between the full layer-polarized and the ferromagnetic phases a gapless mode appears resulting from an accidental symmetry that can be regarded as a remanent of a broken $SO(5)$ symmetry. Interestingly, we have seen that the canted anti-ferromagnetic and partially layer-polarized phases can present dynamical instabilities. We emphasize the crucial role played by the valley/sublattice short-range interactions in the appearance of such instabilities. The theoretical calculations previously developed are easily applied to monolayer graphene. Following previous works on the field, we have discussed the effects of LL mixing and we have taken them into account, arriving at an effective Hamiltonian formally similar to that previously considered, which preserves the structure of the phase diagram and of the collective modes and hence, the resulting physical conclusions are the same. One interesting feature is that the gap of the symmetric modes does not depend on the strength of the Coulomb interaction and it can be easily manipulated with the help of the layer voltage or the in-plane component of the magnetic field. We have also discussed a possible protocol to observe the dynamical instability of the PLP phase; this issue could be further investigated in the future.

\appendix

\part{Appendices}

\chapter{1D solutions of the Gross-Pitaevskii equation}\label{app:1DGP}

We review in this Appendix the solutions to the 1D stationary GP equation, in similar terms to Refs. \cite{Zapata2011,Larre2012,Michel2013}. Using the same amplitude-phase decomposition that leads to Eq. (\ref{eq:PhaseAmplitude}), we can write the 1D time-independent GP equation as:

\begin{eqnarray}\label{eq:1DPhaseamplitude}
\frac{dJ}{dx}&=&0\\
\nonumber \mu A(x)&=&-\frac{\hbar^2}{2m}\frac{d^2A}{dx^2}+\frac{1}{2}mv^2(x)A+V(x)A+gA^3\, ,
\end{eqnarray}
where $v,J$ are the 1D velocity and current. The first equation tells us that the current is constant. Specifically, $A^2(x)v(x)=J$ and thus we can rewrite the equation for the amplitude as:
\begin{equation}\label{eq:1DGPamplitude}
\mu A(x)=-\frac{\hbar^2}{2m}A''+\frac{mJ^2}{2A^3}+V(x)A+gA^3\, ,
\end{equation}
where $'$ denotes spatial derivative. We note that, if $J\neq 0$, neither the amplitude or the flow velocity vanish and hence we can safely choose $J>0$ so $v(x)>0$ at every point.

We only consider in this Appendix the case in which the external potential is zero or constant, where in the latter case we can absorb it in the definition of the chemical potential. The importance of this particular case is based on that it is the most considered scenario in analog condensed systems due to its analytical simplicity. First, we look for homogeneous solutions corresponding to plane waves, $A(x)=A_0$, which are given by the zeros of:
\begin{equation}\label{eq:1Dhomogeneous}
gn_0^3-\mu n_0^2+\frac{mJ^2}{2}=0,~n_0=A^2_0
\end{equation}
This equation only has real solutions whenever:
\begin{equation}\label{eq:currentcondition}
J\leq \sqrt{\frac{8\mu^3}{27mg^2}}
\end{equation}
Suppose now that we find one solution to the previous equation, let's say $n_0=n_b$. Then, after rescaling all the previous magnitudes we obtain:
\begin{eqnarray}\label{eq:1Dhomogeneous}
0&=&X^3-\tilde{\mu}X^2+\frac{\tilde{v}_b^2}{2}=(X-1)\left(X^2-\frac{\tilde{v}_b^2}{2}X-\frac{\tilde{v}_b^2}{2}\right)\\
X&=&\frac{n_0}{n_b}=\frac{c_0^2}{c_b^2},~\tilde{v}_b=\frac{J}{n_bc_b},~\tilde{\mu}=1+\frac{\tilde{v}_b^2}{2}
\end{eqnarray}
where $c_{0,b}$ is the sound speed associated to the density $n_{0,b}$, $c_{0,b}=\sqrt{gn_{0,b}/m}$. It is immediate from here to obtain the other two roots for Eq. (\ref{eq:1Dhomogeneous}):
\begin{equation}\label{eq:1Dhomogeneousroots}
X=\frac{\tilde{v}_b^2\pm\sqrt{\tilde{v}_b^4+8\tilde{v}_b^2}}{4}
\end{equation}
The previous equation has two solutions for $n_0$, one positive and one negative, the latter unphysical so we only consider the positive one, which we label as $n_p$.

As explained in the main text, a solution is subsonic whenever $c>v$ and supersonic whenever $c<v$, where $c$ is the sound speed and $v$ the flow velocity. Taking into account that $n_bv_b=J$, we find that $\tilde{v}_b=\frac{v_b}{c_b}$ is the Mach number for the particular solution $n_b$. For $\tilde{v}_b\gtrless1$ we have a supersonic (subsonic) solution. For the other one, taking into account that $n_bv_b=J=n_pv_p$, we find that its Mach number is:
\begin{equation}\label{eq:1Dsubsupsols}
\frac{v_p}{c_p}=\frac{z}{(1+\sqrt{1+z})^{\frac{3}{2}}}
\end{equation}
where we have introduced the dimensionless quantity $z\equiv 8/\tilde{v}_b^2$. As the function $f(z)=z^{2/3}-1-\sqrt{1+z}$ increases monotonically for $z>0$, it only has one zero, at $z=8$ or, equivalently, $\tilde{v}_b=1$. Then, if $\tilde{v}_b\lessgtr1$, $v_p/c_p\gtrless1$ and thus, we always have one solution subsonic and one supersonic. The limit case $\tilde{v}_b=1$ presents a doubly degenerated solution. In the following, we take $n_b$ as the subsonic solution and $n_p$ as the supersonic one, satisfying $n_b>n_p$.

\begin{figure}[!htb]
\begin{tabular}{@{}cc@{}}
    \includegraphics[width=0.5\columnwidth]{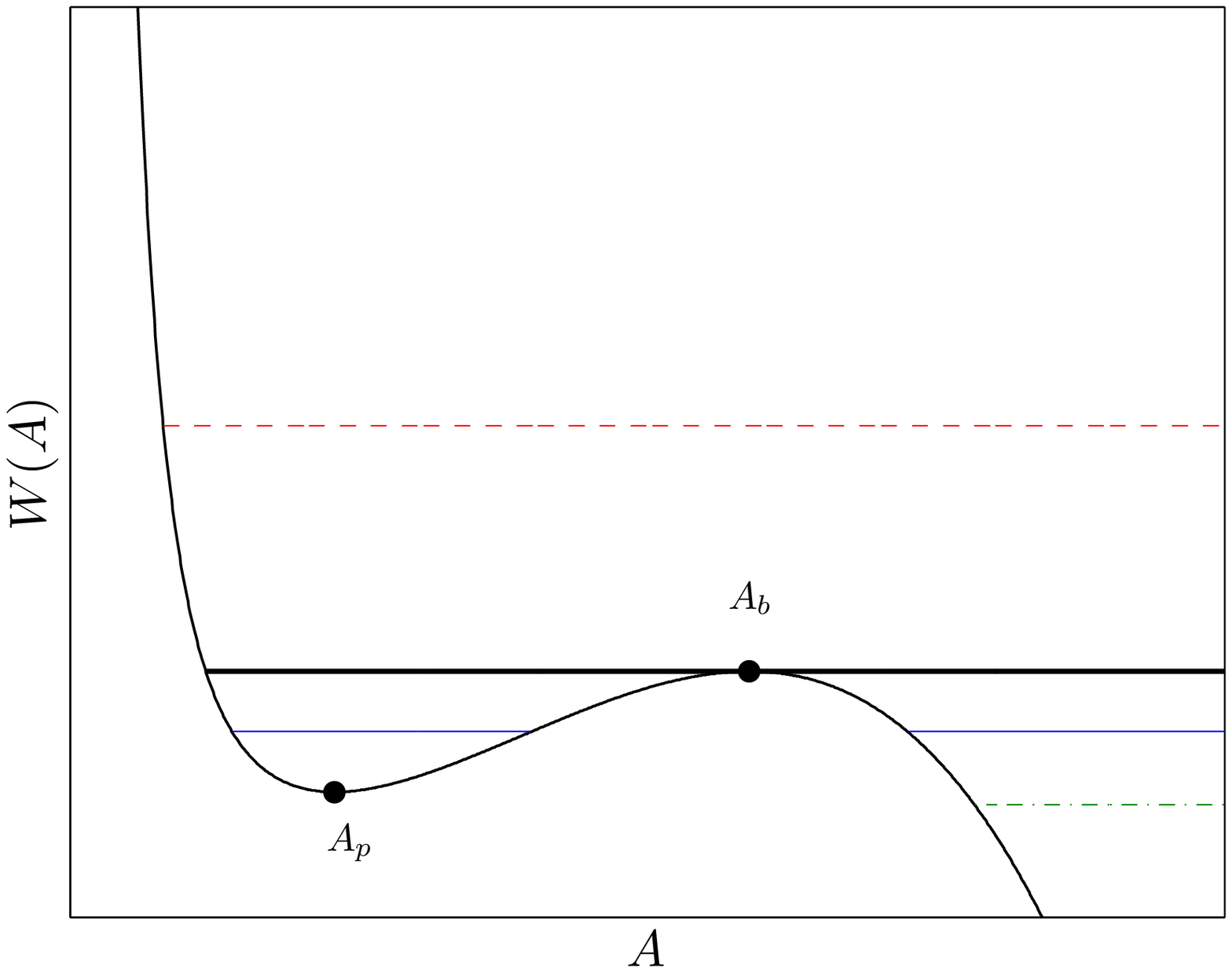} & \includegraphics[width=0.5\columnwidth]{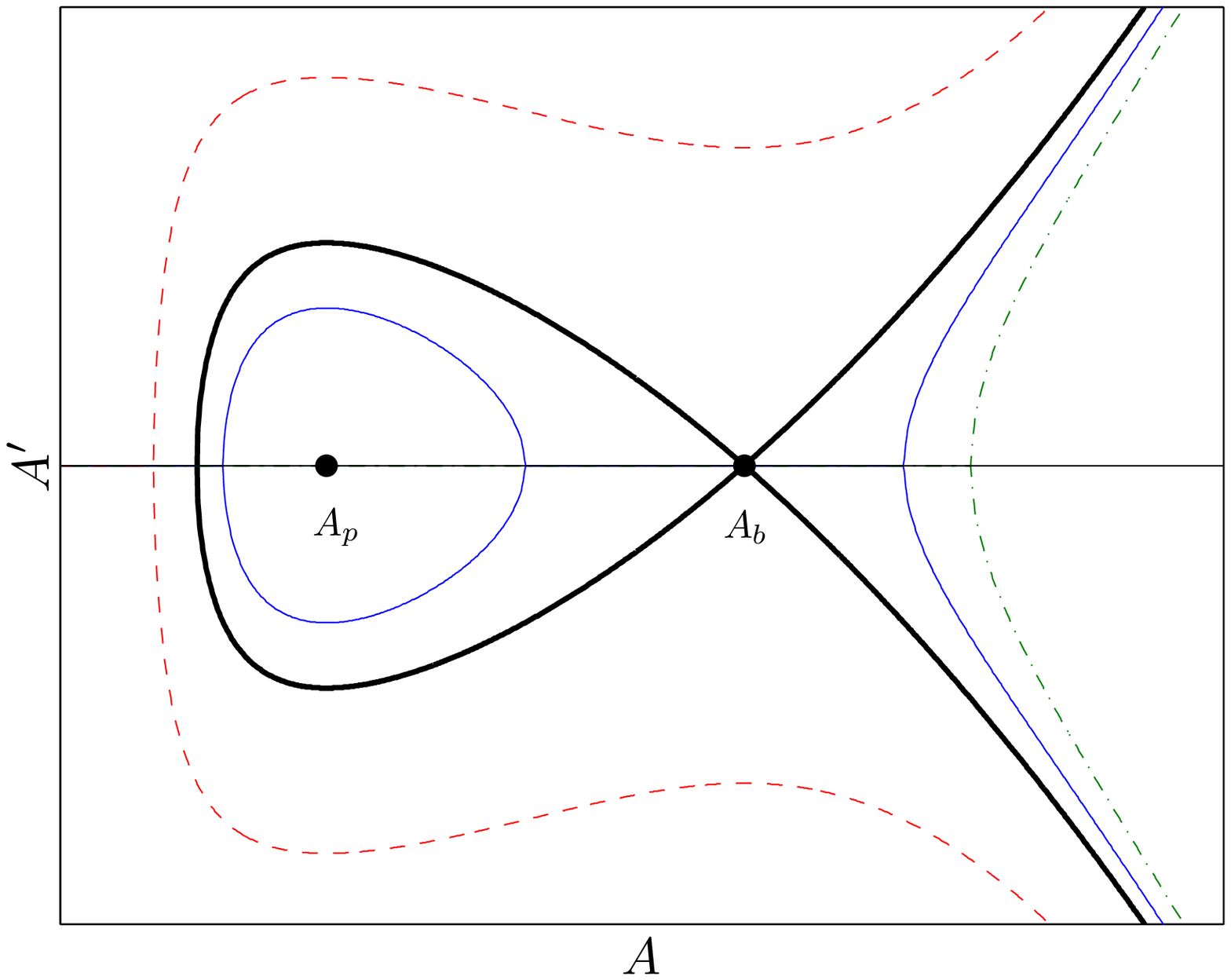}
\end{tabular}
\caption{Left plot: representation of the amplitude potential $W(A)$ (black solid line). We depict some values of the amplitude energy $E_A$ corresponding to qualitatively different solutions; $E=W(A_b)$ is plotted in black thick line, $W(A_p) < E_A < W(A_b)$ is plotted in solid blue line, $E_A>W(A_b)$ is plotted in red dashed line and $E_A<W(A_p)$ is plotted in green dashed-dotted line. The homogeneous solutions $A=A_b$ and $A=A_p$ are marked with a black dot. Right panel: orbits in the space $(A,A')$ associated to the values of the amplitude energy shown in the left panel, which can be obtained from the identity $A'^2=2[E_A-W(A)]$. The horizontal black line marks $A'=0$.}
\label{fig:AmplitudePotential}
\end{figure}

Non-homogeneous solutions can also appear. In order to compute them, we note that Eq. (\ref{eq:1DGPamplitude}) can be regarded as the classical motion equation of a particle in a certain potential $W(A)$. Integrating the equation, we obtain:
\begin{eqnarray} \label{eq:GP-potential}
\frac{1}{2}A'^{2}+W(A) &=& E_A \\
\nonumber W(A)&=&\frac{m}{\hbar^{2}}\left(\frac{mJ^2}{2A^2}+\mu A^{2}-\frac{g}{2}A^{4}\right)
\end{eqnarray}
where $E_A$ is the analogous to the mechanical energy of the particle and we denote it as the amplitude energy. We plot the potential $W(A)$ in Fig. \ref{fig:AmplitudePotential}. The local minimum corresponds to the supersonic homogeneous solution with amplitude $A_p=\sqrt{n_p}$ (which means that it is an stable fixed point) and the local maximum to the subsonic homogeneous solution with amplitude $A_b=\sqrt{n_b}$ (which means that it is an unstable fixed point).

We can further rewrite the previous identity in terms of the density as:
\begin{equation}\label{eq:ellipticdensityenergy}
n'^2=4\tilde{g}(n-n_1)(n-n_2)(n-n_3),~\tilde{g}\equiv \frac{gm}{\hbar^{2}}
\end{equation}
where $n_i,~i=1,2,3$ are the roots of the polynomial
\begin{equation}\label{eq:roots}
gn^3-2\mu n^2+2\frac{\hbar^2}{m}E_An-mJ^2=g(n-n_1)(n-n_2)(n-n_3)=0
\end{equation}
which is equivalent to find the zeros of the equation $W(A)=E_A$. We distinguish several cases according to the value of the amplitude energy $E_A$.

For $W(A_p) < E_A < W(A_b)$, the three roots are real (blue line of left Fig. \ref{fig:AmplitudePotential}) and we take them such $0<n_1 < n_2 < n_3$. We first consider the oscillating solution $n_1<n(x)<n_2$, corresponding to the closed blue line of right Fig. \ref{fig:AmplitudePotential}. Integrating Eq. (\ref{eq:ellipticdensityenergy}), we find:
\begin{equation}\label{eq:ellipticintegration}
\int~\frac{\mathrm{d}n}{2\sqrt{(n-n_1)(n_2-n)(n_3-n)}}=\pm\sqrt{\tilde{g}}(x-x_0)
\end{equation}
with $x_0$ some integration constant. The solution of the previous indefinite integral is given in term of elliptical functions \cite{Abramowitz1988,Byrd1971}. The phase of the wave function can be obtained by integrating the relation $\hbar\theta'(x)/m=v(x)=J/n(x)$. Putting all together, we get:
\begin{eqnarray}\label{eq:ellipticdensity}
\Psi_0(x)&=&A(x)e^{i\theta(x)}\\
\nonumber n(x)&=&n_1+(n_2-n_1)\text{sn}^2(\sqrt{\tilde{g}(n_3-n_1)}(x-x_0),\nu),~\nu=\frac{n_2-n_1}{n_3-n_1}\\
\nonumber \theta(x)&=&\theta(0)+\frac{m J}{\hbar n_1\sqrt{\tilde{g}(n_3-n_1)}}\\
\nonumber ~&\times&\Theta\left(\sqrt{\tilde{g}(n_3-n_1)}(x-x_0),-\sqrt{\tilde{g}(n_3-n_1)}x_0,n_1,n_2,n_3,\nu\right)\\
\nonumber\Theta\left(u_2,u_1,n_1,n_2,n_3,\nu\right)&=&\Pi\left(\text{am}(u_2,\nu),1-\frac{n_2}{n_1},\nu\right)-\Pi\left(\text{am}(u_1,\nu),1-\frac{n_2}{n_1},\nu\right)
\end{eqnarray}
For understanding better the notation in the previous equation, we briefly review some concepts about elliptic functions. The function $\text{sn}(u,\nu)$ is a Jacobi elliptic function. It is defined as $\text{sn}(u,\nu)=\sin\left(\text{am}(u,\nu)\right)$, where $\text{am}(u,\nu)$ is the inverse function of $F(\phi,\nu)$,
\begin{equation}\label{eq:incompleteelliptic}
F(\phi,\nu)\equiv\int_0^\phi\frac{\mathrm{d}\varphi}{\sqrt{1-\nu\sin^2\varphi}}
\end{equation}
so $u=F(\text{am}(u,\nu),\nu)$. The function $F(\phi,\nu)$ is known as the incomplete elliptic integral of the first kind. As a consequence of the previous relations, $\text{sn}(u,\nu)$ is a periodic function with period $4K(\nu)$, $K(\nu)=F(\frac{\pi}{2},\nu)$ being the complete elliptic integral of the first kind. The function $\Pi(\phi,m,\nu)$ is the incomplete elliptic integral of the third kind:
\begin{equation}\label{eq:incompleteellipticthird}
\Pi(\phi,m,\nu)\equiv\int_0^\phi\frac{\mathrm{d}\varphi}{(1-m\sin^2\varphi)\sqrt{1-\nu\sin^2\varphi}}
\end{equation}

On the other hand, for $W(A_p) < E_A < W(A_b)$ and $n(x)>n_3$, the solution grows indefinitely, see the blue orbit for $A>A_b$ in right plot in Fig. \ref{fig:AmplitudePotential}). Moreover, for high values of $n$, we have $n'\propto n^{3/2}$ which implies that the solution blows up at some finite value $x_{bu}$ as $n(x)\sim (x-x_{bu})^{-2}$. The same reasoning holds for $E_A<W(A_p)$ (green dashed-dotted line of Fig. \ref{fig:AmplitudePotential}) or $E_A>W(A_b)$ (red dashed line of Fig. \ref{fig:AmplitudePotential}). We do not care about these exploding solutions since they do not appear in this work.

Finally, we consider the degenerate cases $E_A=W(A_p)$ and $E_A=W(A_b)$. For $E_A=W(A_p)$, we have $n_1=n_2=n_p$. One possible solution corresponds to the stable fixed point of the homogeneous supersonic solution $n(x)=n_p$:
\begin{equation}\label{eq:supersonicplanewave}
\Psi_0(x)=\sqrt{n_p}e^{iq_px},~q_p=\frac{mv_p}{\hbar}
\end{equation}
with $q_p$ the supersonic momentum. The other possible solution corresponds to $n(x)>n_3$, which blows up as explained.

The other degenerate case is $E_A=W(A_b)$, where the roots satisfy $n_2=n_3=n_b$ and $n_1=n_b \tilde{v}_b^2$. For $n(x)=n_b$, the solution is the subsonic plane wave
\begin{equation}\label{eq:subsonicplanewave}
\Psi_0(x)=\sqrt{n_b}e^{iq_bx},~q_b=\frac{mv_b}{\hbar}
\end{equation}
For $n(x)\neq n_b$, we get from Eq. (\ref{eq:ellipticintegration})
\begin{eqnarray}\label{eq:solitonintegration}
\int~\frac{\mathrm{d}X}{2\sqrt{(X-1)^2(X-\tilde{v}^2_b)}}&=&\pm\frac{x-x_0}{\xi_b}\\
\nonumber X&=&\frac{n}{n_b},~\xi_b=\frac{1}{\sqrt{\tilde{g}n_b}}
\end{eqnarray}
$\xi_b$ being the healing length associated to the subsonic density. For $X<1$, i.e., $n(x)<n_b$, we obtain:
\begin{equation}\label{eq:solitondensity}
n(x)=n_b\left[\tilde{v}^2_b+(1-\tilde{v}^2_b)\tanh^2\left(\sqrt{1-\tilde{v}^2_b}\frac{x-x_0}{\xi_b}\right)\right]
\end{equation}
which is just the same of Eq. (\ref{eq:ellipticdensity}) taking into account that $\text{sn}(u,1)=\tanh(u)$. The phase of the wave function can be obtained analytically in a simple form and we can write the total wave function as:
\begin{equation}\label{eq:solitonwavefunction}
\Psi_0(x)=n_be^{iq_bx}e^{i\theta_0}\left[\sqrt{1-\tilde{v}^2_b}\tanh\left(\sqrt{1-\tilde{v}^2_b}\frac{x-x_0}{\xi_b}\right)-i\tilde{v}_b\right]
\end{equation}
being $\theta_0$ some constant phase. This solution represents a soliton solution with zero velocity \cite{Pethick2008}. We note that $\tilde{v}_b<1$ since it is the Mach number of a subsonic solution.

On the other hand, taking $X>1$ in Eq. (\ref{eq:solitonintegration}), we obtain the so-called shadow soliton solution \cite{Michel2013}, which gives:
\begin{equation}\label{eq:shadowsolitonwavefunction}
\Psi_0(x)=n_be^{iq_bx}e^{i\theta_0}\left[\sqrt{1-\tilde{v}^2_b}\text{cotanh}\left(\sqrt{1-\tilde{v}^2_b}\frac{x-x_0}{\xi_b}\right)-i\tilde{v}_b\right]
\end{equation}
Although, as previously discussed, this solution blows up at a finite value of $x$, it appears when studying the BH lasers of Chapter \ref{chapter:BHL}.

From all the previous solutions, the most relevant ones for BH configurations are the supersonic wave function (\ref{eq:supersonicplanewave}), which appears in the supersonic region of a BH; the subsonic soliton (\ref{eq:solitonwavefunction}), which typically appears in the subsonic region and the oscillating solution of Eq. (\ref{eq:ellipticdensity}) that can appear in the scattering region between the two asymptotic regions.

\section{Black-hole type solutions}\label{subsec:BHanalyticsolutions}

We now compute the GP wave functions for the different BH configurations of Sec. \ref{sec:typicalbh}. In the flat-profile type configurations, the wave function is trivial since it is a plane wave. We focus on the more interesting cases of non-homogeneous wave functions. In the following, we assume that the condition (\ref{eq:currentcondition}) is satisfied in both asymptotic regions. In order to simplify the notation, we rescale the wave function to the density of the subsonic region $\Psi_0\rightarrow \sqrt{n_u}\Psi_0$ and set units such $\hbar=m=c_u=1$, with $c_u$ the subsonic sound speed. In these units, we have that the amplitude potential $W(A)$ is determined only by the value of the subsonic flow velocity $v_u$, which we denote in the following as simply $v$, since in these units the current is just $J=v$ and the chemical potential is $\mu=1+v^2/2$:
\begin{equation} \label{eq:GPBH-potential}
W(A)=\frac{v^2}{2A^2}+\left(1+\frac{v^2}{2}\right) A^{2}-\frac{A^{4}}{2}
\end{equation}
The amplitude of the subsonic homogeneous solution now corresponds to $A=1$. As it represents an unstable fixed point, it can only be reached at $x\rightarrow -\infty$. On the other hand, the stable point corresponding to the supersonic homogeneous solution can be extracted from Eq. (\ref{eq:1Dhomogeneousroots}):
\begin{equation}\label{eq:1Dsupersonicamplitude}
A_{p}=\sqrt{\frac{v^2+\sqrt{v^4+8v^2}}{4}}<1
\end{equation}
since $v<1$ is also the subsonic Mach number. In all the cases considered in the rest of this section, the GP wave function in the subsonic region is the soliton solution of Eq. (\ref{eq:solitonwavefunction}), which in our units can be rewritten as:
\begin{eqnarray}\label{eq:solitonGP}
\nonumber\Psi_0(x)&=&\left(\gamma(x)+iv\right)e^{iqx}e^{i\theta_0},\\
\gamma(x)&=&\sqrt{1-v^2}\tanh\left[\sqrt{1-v^2}(x_0-x)\right],
\end{eqnarray}
where we have conveniently multiplied the original wave function by $-1$. The amplitude energy of Eq. (\ref{eq:GP-potential}) for the soliton subsonic solution is simply given by:
\begin{equation}\label{eq:subsonicfictionenergy}
E_A=W(1)=\frac{1}{2}+v^2
\end{equation}

If we look at the $x$ coordinate as a time coordinate, our GP solution starts at $x=-\infty$ in the subsonic solution (\ref{eq:solitonGP}) and ends at $x=\infty$ in the supersonic plane wave solution. As the amplitude energy $E_A$ is a conserved quantity for homogeneous problems, we need some inhomogeneity in order to switch from one asymptotic solution to the other one. This can be achieved by introducing a scattering structure such as a potential or an inhomogeneous coupling constant.

We note that, for these models, the Hawking temperature \cite{Macher2009a,Busch2014} can be obtained analytically. The expression for the Hawking temperature, after momentarily restoring dimensions, is given by:
\begin{equation}\label{eq:TH}
k_BT_H=\frac{\hbar}{2\pi}\frac{d}{dx}\left[v(x)-c(x)\right]_{x=x_H}
\end{equation}
where $x=x_H$ is the location of the horizon, $v(x_H)=c(x_H)$. Going back to our dimensionless system of units, we have that $c(x)=A(x)$ and then the continuity equation gives $c^2(x)v(x)=v$. Thus, the local Mach number is
\begin{equation}
M(x)\equiv \frac{v(x)}{c(x)}=\frac{v}{c^3(x)}
\end{equation}
and then we find that $M(x_H)=1$ implies $A(x_H)=c(x_H)=v^{\frac{1}{3}}$, which gives the location of the horizon. From all the previous results, we can rewrite the equation for the Hawking temperature as simply:
\begin{equation}\label{eq:THsimply}
T_H=-\frac{3}{2\pi}c'(x=x_H)=\frac{3}{2\pi}\sqrt{2\left[E_A-W(v^{\frac{1}{3}})\right]}
\end{equation}
where $T_H$ is in units of $mc^2_u/k_B$. In particular, if the horizon is placed in the region where the GP solution is given by the soliton solution of Eq. (\ref{eq:solitonGP}), we find:
\begin{equation}\label{eq:THsimply}
T_H=\frac{3}{2\pi}\left(1-v^{\frac{2}{3}}\right)\sqrt{1-v^{\frac{4}{3}}}<\frac{3}{2\pi}<\frac{1}{2}
\end{equation}
The importance of the Hawking temperature is that, in most non-resonant configurations, the Hawking spectrum is described reasonably well by a Planck distribution in frequencies with effective temperature $T_H$, see Refs. \cite{Macher2009a,Busch2014} for further details.

\subsection{Delta-barrier configuration}\label{subsec:GPdelta}

By introducing a potential of the form $V(x)=Z\delta(x)$, we create a discontinuity in the derivative of $\Psi_0(0)$, $\Psi'_0(0^{+})-\Psi'_0(0^{-})=2Z\Psi_0(0)$, which amounts to $A'_0(0^{+})-A'_0(0^{-})=2ZA_0(0)$ in terms of the amplitudes. In the rest of the space we have the same homogeneous problem previously discussed. In this way, for $x<0$, the wave function is the same soliton solution of Eq. (\ref{eq:solitonGP}). For $x>0$, the solution is the supersonic plane wave, whose amplitude is given by $A_{p}$, see Eq. (\ref{eq:1Dsupersonicamplitude}). Joining both solutions gives
\begin{eqnarray}\label{eq:1deltaGP}
\nonumber\Psi_0(x)&=&\left[\gamma(x)+iv\right]e^{iqx}e^{i\theta_0},~x<0\\
\Psi_0(x)&=&A_{p}e^{iv_{p}x},~v_{p}=v/A^2_{p},~x>0
\end{eqnarray}
We take the phase $\theta_0$ such that $\Psi_0(0)$ is real and $x_0$ such that the amplitude of the wave function is continuous at $x=0$,
\begin{eqnarray}\label{eq:zerophase}
e^{i\theta_0}&=&\frac{\gamma(0)-iv}{\sqrt{\gamma^2(0)+v^2}} \, ,
x_0=\frac{1}{\sqrt{1-v^2}}\textrm{arctanh}\left(\sqrt{\frac{A^2_{p}-v^2}{1-v^2}}\right)
\end{eqnarray}

\subsection{Waterfall configuration}\label{subsec:GPwaterfall}

The potential is now a negative step function with amplitude $-V_0<0$. For $x<0$, we have the same type of soliton solution. We note that, for $x>0$, the amplitude potential $W(A)$ is different to that of the subsonic region and so its supersonic amplitude. As the step potential does not introduce a discontinuity in the amplitude or the derivatives of the wave function, $W(A)$ must be such that its supersonic solution matches with the soliton solution at $x=0$. Then, $A'(0)=0$, which implies that the amplitude of the wave function at zero must be the minimum of the soliton amplitude $A(0)=v$, see Eq.(\ref{eq:solitonGP}). The value of $V_0$ must be such the supersonic amplitude for $x>0$ is $A_{p}=v$ and by conservation of the current $J$ we find that $v_d=1/v$, $v_d$ being the value of the supersonic velocity. From these considerations, we obtain that the total wave function can be written as:

\begin{eqnarray}\label{eq:WFGP}
\nonumber \Psi_0(x)&=&\left[v+i\sqrt{1-v^2}\tanh\left(\sqrt{1-v^2}x\right)\right]e^{ivx},~x<0\\
\Psi_0(x)&=&ve^{ix/v},~x\geq 0
\end{eqnarray}

\subsection{Resonant double delta-barrier configuration}\label{subsec:GPresonant}

This case is quite similar to the $\delta$ barrier previously considered but now with a double barrier $V(x)=Z[\delta(x)+\delta(x-d)]$. We divide the problem in three regions: region 1, for $x<0$; region 2, for $0<x<d$ and region 3, for $x>d$. At the boundaries between the regions, the derivative of the amplitude of the wave function is discontinuous due to the $\delta$ barriers. In each region, the GP equation reduces to the homogeneous problem given of Eq. (\ref{eq:GP-potential}), with three different values of the constant $E_A$, labeled as $E_i$, $i=1,2,3$. For $x<0$, the solution is again a soliton wave and for $x>d$, the solution is the homogeneous supersonic plane wave of Eq. (\ref{eq:1deltaGP}) so $E_1=W(1)$ and $E_3=W(A_{p})$. The value of $E_2$ can be extracted by the discontinuity $A'(d^+)-A'(d^-)=2ZA(d)$. Since at $x=d^+$ we have the supersonic plane wave solution, we find $E_2=2Z^2A^2_{p}+E_3$. The value of $E_2$ must satisfy $E_2<E_1$ because if not, the solution in the region 2 cannot be matched with the soliton in the region 1 as we infer from Fig. \ref{fig:AmplitudePotential}. For obtaining an equation for the amplitude at $x=0$, $A(0)$, we compute the difference between Eq. (\ref{eq:GP-potential}) in the regions 1 and 2 and taking into account the discontinuity in the derivative, $A'(0^+)-A'(0^-)=2ZA(0)$, we arrive at:

\begin{eqnarray}\label{eq:doubledeltaA0}
\nonumber n'(0^-)&=&\frac{\Delta E}{Z}-2Zn(0),~\Delta E=E_2-E_1\\
n'^2(0^-)&=&4[n(0)-1]^2[n(0)-v^2]
\end{eqnarray}
where we have rewritten the equation for $A(0)$ and $A'(0^-)$ in terms of the density $n(0)=A^2(0)$ and its derivative at $x=0^-$, $n'(0^-)=2A(0)A'(0^-)$. The previous system of equations has two solutions for $n(0)$ provided that $Z$ takes values in the range $Z\in (Z_{\rm{min}}(v),Z_{\rm{max}}(v))$, where the limits of the range depend only on the value of the subsonic flow speed $v$. In particular, the value of $Z_{\rm{max}}(v)$ is the maximum value that $Z$ can take in such way that the solution is not a exploding one, i.e., $E_2\leq E_1$:
\begin{equation}
Z_{\rm{max}}(v))=\sqrt{\frac{E_1-E_3}{2A^2_p}}
\end{equation}
The solution in the region 2 is given in terms of elliptic functions, see Eq. (\ref{eq:ellipticdensity}). Specifically, the inter-barrier solution can be written as:
\begin{eqnarray}\label{eq:interbarrierwavefunction}
\Psi^{\text{int}}_0(x)&=&A(x)e^{i\theta(x)}\\
\nonumber n(x)&=&n_1+(n_2-n_1)\text{sn}^{2}(\sqrt{n_3-n_1}x+u_0,\nu),~\nu=\frac{n_2-n_1}{n_3-n_1}\\
\nonumber u_0&=&\textrm{sgn}[n'(0^+)]~\text{sn}^{-1}\left(\sqrt{\frac{n(0)-n_1}{n_2-n_1}},\nu\right)\\
\nonumber \theta(x)&=&\frac{v}{n_1\sqrt{n_3-n_1}}\times\\
\nonumber ~&\times&\left[\Pi\left(\text{am}(\sqrt{n_3-n_1}x+u_0,\nu),1-\frac{n_2}{n_1},\nu\right)-\Pi\left(\text{am}(u_0,\nu),1-\frac{n_2}{n_1},\nu\right)\right]
\end{eqnarray}
where we have chosen the phase such the wave function is real at $x=0$, as in the single barrier case. Here, $\text{sn}^{-1}(u,\nu)$ is the inverse function of $\text{sn}(u,\nu)$, given by:
\begin{equation}\label{eq:sninverse}
\text{sn}^{-1}(u,\nu)=F(\arcsin(u),\nu)
\end{equation}
By imposing $n(d)=A^2_{p}$, we find that for given values of $v$ and $Z$, only a discrete number of solutions exist for the value of the distance between barriers $d$:
\begin{eqnarray}
\nonumber d_n&=&\frac{\left[2n+1+\text{sgn}(u_1+u_0)\right]K(\nu)-u_1-u_0}{\sqrt{n_3-n_1}}\\
u_1&=&\text{sn}^{-1}\left(\sqrt{\frac{A^2_{p}-n_1}{n_2-n_1}},\nu\right)
\end{eqnarray}
In the previous expression, $n=0,1,2\ldots$ is the number of periods of the density between the barriers.
Gathering the solutions of all regions:
\begin{eqnarray}\label{eq:2deltaGP}
\Psi_0(x)=\left\{
\begin{array}{cc}
               \left[\gamma(x)+iv\right]e^{iqx}e^{i\theta_0}, & x<0\\
               \Psi^{\text{int}}_0(x), & 0<x<d\\
               A_{p}e^{i[v_{p}x+\theta(d)]},~v_{p}=v/A^2_{p}, & x>d
             \end{array} \right.
\end{eqnarray}
where $\gamma(x),\theta_0$ are chosen such the function is continuous and real at $x=0$ and $\theta(d)$ is the phase of Eq. (\ref{eq:interbarrierwavefunction}) evaluated at $x=d$.

\chapter{Non-homogeneous Bogoliubov-de Gennes equations: Lax pair of equations.}\label{app:BdGsolitonsolutions}
\chaptermark{Lax pair of equations}

We devote this Appendix to compute the stationary BdG solutions in 1D whenever the wave function $\Psi_0$ is inhomogeneous [when $\Psi_0$ is a plane wave, the BdG solutions are trivial plane waves, given by Eq. (\ref{eq:PlaneWaveSpinors})]. Following Refs. \cite{Chen1998,Faddeev2007,Zapata2011}, we obtain the solutions to the 1D time-dependent BdG equations from the following system of equations, known as the Lax pair of equations of the Zhakarov-Shabat problem:

\begin{eqnarray}\label{eq:Laxpair}
\frac{\partial}{\partial x}\left[\begin{array}{c} w_1(x,t) \\ w_2(x,t) \end{array}\right]&=&\left[\begin{array}{cc}-i\lambda & \Psi^*(x,t)\\
\Psi(x,t)& i\lambda \end{array}\right]\left[\begin{array}{c} w_1(x,t) \\ w_2(x,t) \end{array}\right]\\
\nonumber \frac{\partial}{\partial t}\left[\begin{array}{c} w_1(x,t) \\ w_2(x,t) \end{array}\right]&=&\left[\begin{array}{cc}i\lambda^2+i\frac{|\Psi(x,t)|^2}{2} & \left(-\frac{i}{2}\frac{\partial }{\partial x}-\lambda\right)\Psi^*(x,t)\\
\left(\frac{i}{2}\frac{\partial }{\partial x}-\lambda\right)\Psi(x,t)&-i\lambda^2-i\frac{|\Psi(x,t)|^2}{2}\end{array}\right]\left[\begin{array}{c} w_1(x,t) \\ w_2(x,t) \end{array}\right]
\end{eqnarray}
where $\lambda$ is some constant parameter and $\Psi(x,t)$ a function that depends on space and time. The previous system of equations is compatible (i.e., the mixed derivatives satisfy the Schwarz lemma) whenever $\Psi(x,t)$ satisfy:
\begin{equation}\label{eq:TDGPFreedimensionless}
i\frac{\partial \Psi(x,t)}{\partial t}=-\frac{1}{2}\frac{\partial^2 \Psi(x,t)}{\partial x^2}+|\Psi(x,t)|^2\Psi(x,t)
\end{equation}
which means that $\Psi(x,t)$ is a solution of the time-dependent GP equation, expressed in a dimensionless system of units. Moreover, it can be shown that the solutions to the Lax pair of equations (\ref{eq:Laxpair}) satisfy:
\begin{eqnarray}\label{eq:LaxBdG}
\nonumber i\frac{\partial}{\partial t}z(x,t)&=&
\left[\begin{array}{cc} -\frac{1}{2}\frac{\partial^2}{\partial x^2}+2|\Psi(x,t)|^2 & \Psi^2(x,t)\\
\Psi^{*2}(x,t)&\frac{1}{2}\frac{\partial^2}{\partial x^2}-2|\Psi(x,t)|^2\end{array}\right]z(x,t)\\
z(x,t)&=&\left[\begin{array}{c} w^2_2(x,t) \\ w^2_1(x,t) \end{array}\right]
\end{eqnarray}
Thus, we can identify the components of the BdG spinor with the solutions of the Lax pair of equations as $u=w^2_2$ and $v=w^2_1$.

We consider now stationary solutions of the GP equation $\Psi(x,t)=\Psi_0(x)e^{-i\mu t}$. In that case, we look for stationary solutions to the BdG equations, which take the following form in terms of the Lax spinors:
\begin{eqnarray}\label{eq:Laxansatzstationary}
\left[\begin{array}{c} w_1(x,t) \\ w_2(x,t) \end{array}\right]=e^{-i\frac{\omega}{2}t}\left[\begin{array}{c} e^{i\frac{\mu}{2}t}w_1(x) \\ e^{-i\frac{\mu}{2}t}w_2(x) \end{array}\right]
\end{eqnarray}
where $\omega$ is the corresponding energy. The Lax pair of equations takes then the form:

\begin{eqnarray}\label{eq:Laxstationary}
\nonumber 0&=&\left[\begin{array}{cc}i\left(\lambda^2+\frac{|\Psi_0(x)|^2}{2}-\frac{\mu}{2}\right)+i\frac{\omega}{2} & \left(-\frac{i}{2}\frac{\partial }{\partial x}-\lambda\right)\Psi_0^*(x)\\
\left(\frac{i}{2}\frac{\partial }{\partial x}-\lambda\right)\Psi_0(x)&-i\left(\lambda^2+\frac{|\Psi_0(x)|^2}{2}-\frac{\mu}{2}\right)+i\frac{\omega}{2}\end{array}\right]\left[\begin{array}{c} w_1(x) \\ w_2(x) \end{array}\right]\\
\frac{\partial}{\partial x}\left[\begin{array}{c} w_1(x) \\ w_2(x) \end{array}\right]&=&\left[\begin{array}{cc}-i\lambda & \Psi^*_0(x)\\
\Psi_0(x)& i\lambda \end{array}\right]\left[\begin{array}{c} w_1(x) \\ w_2(x) \end{array}\right]
\end{eqnarray}

We see that obtaining the eigenmodes of the BdG equations reduces to integrating a first order system of ODEs, which is easier than the original second order equation corresponding to the BdG equations. Due to this fact, in some situations we can obtain analytical solutions. Specifically, we focus on the often considered case where the stationary GP wave function is a soliton wave, Eq. (\ref{eq:solitonwavefunction}). For matching the dimensionless notation of the Lax equations, we rescale the wave function as $\Psi_{0}(x)\rightarrow \sqrt{n_b}\Psi_{0}(x)$ and set units such $\hbar=m=\xi_b$. In order to unify the results with BH solutions, we relabel $v=\tilde{v}_b$ and multiply the wave function by $-1$, obtaining the same formal expression of Eq. (\ref{eq:solitonGP}). We then look for asymptotic plane wave solutions of the form:
\begin{eqnarray}\label{eq:Laxansatz}
\left[\begin{array}{c} w_1(x) \\ w_2(x) \end{array}\right]=e^{i\frac{k}{2}x}\left[\begin{array}{c} e^{-i\frac{(vx+\theta_0)}{2}}\tilde{w}_1(x) \\ e^{i\frac{(vx+\theta_0)}{2}}\tilde{w}_2(x) \end{array}\right]
\end{eqnarray}
where $\tilde{w}'_i(x)\rightarrow 0$ for $x\rightarrow \pm\infty$ in order to asymptotically match the corresponding plane wave solutions. We can identify $k$ as the total momentum of the final BdG solution. Inserting the previous ansatz in the spatial Lax equation of (\ref{eq:Laxstationary}) gives
\begin{eqnarray}\label{eq:Laxspatialsoliton}
\frac{\partial}{\partial x}\left[\begin{array}{c} \tilde{w}_1(x) \\ \tilde{w}_2(x) \end{array}\right]&=&\left[\begin{array}{cc}-i\left(\lambda-\frac{v}{2}\right)-i\frac{k}{2} & \gamma(x)-iv\\
\gamma(x)+iv& i\left(\lambda-\frac{v}{2}\right)-i\frac{k}{2} \end{array}\right]\left[\begin{array}{c} \tilde{w}_1(x) \\ \tilde{w}_2(x) \end{array}\right]
\end{eqnarray}
In order to find non-trivial solutions, for $x\rightarrow \pm \infty$, the determinant of the previous matrix has to be zero, which gives the condition:
\begin{eqnarray}\label{eq:Laxlambda}
\lambda=\frac{v}{2}\pm\sqrt{1+\frac{k^2}{4}}
\end{eqnarray}
Applying the same reasoning to the time Lax equation of (\ref{eq:Laxstationary}), we find
\begin{equation}\label{eq:Laxdispersion}
\omega=k(\lambda+v/2)
\end{equation}
which is nothing else than the usual dispersion relation of Eq. (\ref{eq:dispersionrelation}). By making the following change of variables
\begin{eqnarray}\label{eq:Laxcombination}
\tilde{w}_{\pm}(x)=w_1(x)\pm w_2(x)
\end{eqnarray}
and deriving to arrive at a second order differential equation, we obtain:
\begin{eqnarray}\label{eq:Lax2order}
\tilde{w}_{\pm}''&=&-ik\tilde{w}_{\pm}'+\left(\pm\gamma'(x)-(1-v^2)+\gamma^2(x)\right)\tilde{w}_{\pm}
\end{eqnarray}
As our soliton solution satisfy $\gamma'(x)=-(1-v^2)+\gamma^2(x)$ and $w_{-}'(-\infty)=0$, it follows $w_{-}'(x)=0$ and then, $w_{-}(x)=R$, with $R$ some constant. Going back to the Lax spatial equation, we find:
\begin{equation}
w_{+}(x)=i\frac{\gamma(x)+i\frac{k}{2}}{\lambda+\frac{v}{2}}R
\end{equation}
and inverting the relation
\begin{equation}
\tilde{w}_{1,2}(x)=\frac{R}{2}\left(1\pm\frac{\gamma(x)+i\frac{k}{2}}{\lambda+\frac{v}{2}}\right)
\end{equation}
Finally, taking the constant $R$ such the previous solutions are properly normalized plane waves for $x\rightarrow\pm\infty$, the $u_{a,\omega}(x),v_{a,\omega}(x)$ components of the BdG solutions that asymptotically match the plane wave with wave vector $k_a$ are given by:
\begin{eqnarray}\label{eq:solitonspinors}
\nonumber s_{a,\omega}=\left[\begin{array}{c}
u_{a,\omega}(x)\\
v_{a,\omega}(x)
\end{array}\right] &=&\frac{C_ae^{ik_{a}x}}{\sqrt{2\pi|w_{a}|}}\left[\begin{array}{c}
e^{i(vx+\theta_0)}\left[1+\frac{k_a}{\omega}\left(\frac{k_a}{2}-i\gamma(x)\right)\right]^2\\
e^{-i(vx+\theta_0)}\left[1-\frac{k_a}{\omega}\left(\frac{k_a}{2}-i\gamma(x)\right)\right]^2
\end{array}\right]\\
C_a&=&\frac{\omega}{\sqrt{8k^2_a|\omega-vk_a|}}
\end{eqnarray}
As mentioned, it is easy to check that these solutions reduce to the corresponding scattering channel $a$ for $x\rightarrow\pm\infty$ (apart from some trivial $k$-dependent phase).

When considering stationary GP solutions involving elliptic functions such Eq. (\ref{eq:ellipticdensity}), we are not able to find the explicit solution from this method and we have to use numerical integration to find the corresponding BdG solutions, which are Bloch type waves since the density is periodic. We do not need to compute the BdG solutions for exploding GP solutions as we never use them.

\chapter{Computation of the scattering matrix}\label{app:SMatrixBehav}

\section{General case and asymptotic behavior}

We explain in this Appendix how to compute the $S$-matrix elements. For that purpose, we study the matching of the BdG solutions in the asymptotic subsonic and supersonic regions. When considering Eq. (\ref{eq:BdGequation}) in 1D, we see that we can rewrite the eigenvalue problem of the BdG equations as a four-dimensional linear system  of ODEs:
\begin{eqnarray}\label{eq:4BdG}
\frac{d\sigma}{dx}&=&D(x,\omega)\sigma,\\
\sigma(x)&=&\left[ u(x), v(x),u'(x),v'(x)\right]^{T}
\end{eqnarray}
where $D(x,\omega)$ is a $4\times4$ matrix of the form:
\begin{eqnarray}
D(x,\omega)&=&\left[\begin{array}{cc}
0 & \mathbb{I}_2\\
M(x,\omega) & 0
\end{array}\right]\\
\nonumber M(x,\omega)&=&\frac{2m}{\hbar^2}\left[\begin{array}{cc}
V(x)+2g|\Psi_0(x)|^2-\mu-\hbar\omega & L(x)\\
L^*(x) & V(x)+2g|\Psi_0(x)|^2-\mu+\hbar\omega
\end{array}\right]\\
\nonumber L(x)&=&g\Psi^2_0(x)
\end{eqnarray}
and $\mathbb{I}_2$ is the identity matrix in two dimensions.
We divide the system in three regions: the subsonic region ($x\rightarrow-\infty$), the supersonic region ($x\rightarrow\infty$) and the scattering region (that between the subsonic and supersonic regions). In both asymptotic regions, the solutions are combinations of the different scattering channels, as explained in Sec. \ref{subsec:BHBEC}. Writing the matching equations at the points $x_u\rightarrow-\infty$ and $x_d\rightarrow\infty$ for $\omega<\omega_{\rm{max}}$:

\begin{eqnarray}\label{eq:udmatching}
\nonumber b_{u-\rm{in}}\sigma_{u-\rm{in},\omega}\left(x_u\right)&+&b_{u-\rm{out}}\sigma_{u-\rm{out},\omega}\left(x_u\right)
+b_{\rm{ev}}\sigma_{\rm{ev},\omega}\left(x_u\right)=\sum_{j=1}^{4}b_{j}\sigma_{i,\omega}\left(x_u\right)\\
\sum_{j=1}^{4}b_{j}\sigma_{i,\omega}\left(x_d\right)&=&b_{d2-\rm{in}}\sigma_{d2-\rm{in},\omega}\left(x_d\right)+b_{d1-\rm{in}}\sigma_{d1-\rm{in},\omega}\left(x_d\right)\\
\nonumber &+&b_{d1-\rm{out}}\sigma_{d1-\rm{out},\omega}\left(x_d\right)+b_{d2-\rm{out}}\sigma_{d2-\rm{out},\omega}\left(x_d\right)
\end{eqnarray}
where $\sigma_{a,\omega}$ is a four-component vector of the form:
\begin{equation}
\sigma_{a,\omega}(x)=\left[\begin{array}{c}
s_{a,\omega}(x)\\
s'_{a,\omega}(x)
\end{array}\right]
\end{equation}

In Eq. (\ref{eq:udmatching}), $\sigma_{i,\omega}\left(x\right)$ are four arbitrary independent solutions of the BdG equations in the scattering region and the index "ev" labels the real exponential evanescent wave that exists in the subsonic region (the other real exponential solution is exploding and then unphysical). Thus, there are $11$ amplitudes and $8$ restrictions so we only have $3$ degrees of freedom. In particular, we can choose these independent amplitudes as the ``in" scattering channel amplitudes. Writing Eq. (\ref{eq:udmatching}) as a linear system of $8$ equations:
\begin{eqnarray}\label{eq:scatteringlinearsystem}
\nonumber d&=&A\cdot b\\
b&=&\left[b_{u-\rm{out}},b_{\rm{ev}},b_{1},b_{2},b_{3},b_{4},b_{d1-\rm{out}},b_{d2-\rm{out}}\right]^{T}\\
\nonumber d&=&b_{u-\rm{in}}\left[\begin{array}{c}\sigma_{u-\rm{in},\omega}\left(x_u\right)\\ 0 \end{array}\right]+b_{d1-\rm{in}}\left[\begin{array}{c}0\\ \sigma_{d1-\rm{in},\omega}\left(x_d\right)\end{array}\right]+b_{d2-\rm{in}}\left[\begin{array}{c}0\\ \sigma_{d2-\rm{in},\omega}\left(x_d\right)\end{array}\right],\\
\nonumber A&=&\left[\begin{array}{ccccc}-\sigma_{u-\rm{out},\omega}\left(x_u\right)&-\sigma_{\rm{ev},\omega}\left(x_u\right)&F(x_u)&0&0\\ 0&0&F(x_d)&-\sigma_{d1-\rm{out},\omega}\left(x_d\right)&-\sigma_{d2-\rm{out},\omega}\left(x_d\right)\end{array}\right]
\end{eqnarray}
where $F(x)$ is a $4\times 4$ matrix whose columns are the four linearly independent solutions $\sigma_i(x)$ in the scattering region. The matrix $F$ is often denoted in the literature as the fundamental matrix of the system of equations.

In practice, we can solve the previous matching equations by choosing the points $x_u,x_d$ sufficiently deep in the regions where the wave function tends to a plane wave. Specifically, in the cases considered in the present work, when the wave function is not directly the homogeneous subsonic plane wave as in the flat profile type configurations, the wave function tends to a plane wave with exponentially small corrections in the subsonic region [see the soliton wave function of Eq. (\ref{eq:solitonGP})], and it is a perfect plane wave in the supersonic region for a finite value of $x_d$. The fundamental matrix $F$, if it cannot obtained analytically, is computed numerically by integrating the $4\times 4$ system of equations (\ref{eq:4BdG}) for four independent initial conditions. We note that the fundamental matrices of two points $x_1,x_2$ are connected through the so-called transfer matrix $F(x_2)=T(x_2,x_1)F(x_1)$. From the previous expression, we obtain:
\begin{equation}\label{eq:transfermatrix}
T(x_2,x_1)=F(x_2)F^{-1}(x_1)
\end{equation}
In particular, in order to numerically compute $T(x_2,x_1)$, we can select the four independent solutions such the initial condition is $F(x_1)=1$ and then we simply have $T(x_2,x_1)=F(x_2)$. An interesting property of the transfer matrix for any second order system is that
\begin{equation}\label{eq:wronskian}
\det T(x_2,x_1)=1
\end{equation}
since the Wronskian $W(x)\equiv\det F(x)$ is a conserved quantity.

We now consider Cramer's rule to solve the system of equations (\ref{eq:scatteringlinearsystem}):
\begin{equation}\label{eq:Cramer}
b_l=\frac{\textrm{det} ~A_l}{\textrm{det} ~A}, ~l=1,2...8 \, ,
\end{equation}
$b_l$ being the $l$ component of the vector $b$ and $A_l$ the matrix $A$ but with the column $l$ changed by the solution vector $d$. Although in the actual numerical computation of the $S$-matrix we do not use Cramer's rule, we consider this tool for theoretical purposes. For example, using Cramer's rule, it is straightforward to check that every amplitude is a linear function in the amplitudes of the ``in" modes. In particular, the amplitude of the ``out" modes is related to the amplitude of the ``in" modes through Eq. (\ref{eq:inoutmodesrelation}) by changing $\hat{\gamma}_{u,d1-\rm{in/out}}\rightarrow b_{u,d1-\rm{in/out}}$ and $\hat{\gamma}^{\dagger}_{d2-\rm{in/out}}\rightarrow b_{d2-\rm{in/out}}$. Then, in order to compute the elements of the column $a=u,d1,d2-$in of the $S$-matrix, we only have to set the corresponding amplitude $b_{a-\rm{in}}=1$ and the other ``in" amplitudes to zero in the vector $d$ of Eq. (\ref{eq:scatteringlinearsystem}). The solution of the system gives us the corresponding $S$-matrix elements.

Interestingly, using Cramer's rule we can also obtain the asymptotic behavior of the $S$-matrix in both limits of the HR spectrum, $\omega\rightarrow0^+$ and $\omega\rightarrow\omega^-_{\rm max}$. First, we consider the lower limit of the spectrum, $\omega\rightarrow0^+$. In this case, the incoming $u-$in and the three outgoing modes behave as $k_{a}\propto\omega$ (as can be seen in Fig. \ref{figDispRelation}), $k_a$ being their respective wave vector.
From (\ref{eq:PlaneWaveSpinors}), we can prove that the four-component vectors corresponding to these modes can be written in this limit as
$\propto \sigma_{0}\left(x\right)/\sqrt{\omega}$ with corrections $O(\sqrt{\omega})$, where
\begin{equation}\label{eq:zeromodelimit}
\sigma_0(x) \equiv \left[ \Psi_0(x), -\Psi_0^*(x),\Psi'_0(x),-\Psi_0^{'*}(x) \right]^{T} \, ,
\end{equation}
is the vector associated to the zero-mode of the BdG equations (\ref{eq:zeromode}), which is solution of Eq. (\ref{eq:4BdG}) for $\omega=0$. The wave vectors corresponding to the remaining modes tend to
a non-zero value in this limit (see again Fig. \ref{figDispRelation}). From these considerations, it can be proven, using Cramer's rule,
that $\left|S_{ij}\left(\omega\right)\right|\propto1/\sqrt{\omega}$ for $i,j=d1,d2$.

Let us considerate, as an example, the computation of $S_{d1d1}$ using Cramer's rule. At least, the $d2$-out mode has the behavior described in Eq. (\ref{eq:zeromodelimit}). We first consider the determinant of $A$ present in the denominator of Eq. (\ref{eq:Cramer}). As we can perform linear combinations between columns in a determinant, by an appropriated subtraction between the $d2$-out and $d1$-out columns [last two columns of the matrix $A$ of Eq. (\ref{eq:scatteringlinearsystem})], the $d1$-out column gives a term $O(\sqrt{\omega})$. In the matrix of the numerator, after substituting the column vector of the $d1$-out mode for that of the $d1$-in mode, which presents a regular behavior near $\omega=0$, we can no longer perform the previous linear combination. Thus, we have in the denominator a similar determinant to that of the numerator but with a factor $\propto O(\sqrt{\omega})$. We then conclude that the matrix element behaves as $\left|S_{d1d1}\left(\omega\right)\right|\propto1/\sqrt{\omega}$. A very similar proof can be given for the other elements $S_{ij}$ for $i,j=d1,d2$.

For studying the remaining $S$-matrix elements in the limit $\omega\rightarrow0^+$, we have to use the fact that $\sigma_{0}\left(x\right)$ is a zero-frequency solution and also that, in the scattering region, there exists (at least) one solution that tends to $\sigma_0$ with corrections $O(\omega)$. This last proposition can be proven using a theorem (also called Poincare's theorem) on the analytic dependence of the solutions on the parameters of the differential equation \cite{Arnold1992,Kaltchev2013}. We can apply safely this theorem in our case because $\omega$ enters as a non-singular parameter, which means that it is not multiplying the highest order derivative of the ODE. In particular, this result is true for homogeneous flows (no matter if subsonic or supersonic) since two modes are proportional to the zero-mode vector $\sigma_0(x)$ for $\omega\rightarrow0^+$ (the labeled as ``out'' modes for supersonic flows and the two propagating modes for subsonic flows). Taking these results into account and proceeding in a similar way as before, it can be proven that $\left|S_{ui}\left(\omega\right)\right|\propto1/\sqrt{\omega}$ for $i=d1,d2$ and that $S_{iu}$, $i=d1,d2$, tend to a constant.

In the opposite limit of the spectrum, $\omega\rightarrow\omega^-_{\rm max}$, the only irregular behavior is present in the $d2$ modes because in this limit their wave vectors tend to the same value $k_{d2-\rm{in/out}}\left(\omega\right)=k_{d2}(\omega_{\rm max})\pm O(\sqrt{\omega_{\rm max}-\omega})$. Also, the group velocity tends to zero as $\sim\sqrt{\omega_{\rm max}-\omega}$. Then, the corresponding four-component vectors are in this limit $\sigma_{d2-{\rm in/out},\omega}(x) \propto \sigma_{\omega_{\rm max}}\left(x\right) \left(\omega_{\rm max}-\omega\right)^{-1/4}+O\left(\omega_{\rm max}-\omega\right)^{1/4}$
where $\sigma_{\omega_{\rm max}}\left(x\right)$ is the (not normalized) vector that solves the system for $\omega_{\rm max}$. The divergent factor $\left(\omega_{\rm max}-\omega\right)^{-1/4}$ comes from the group velocity that ensures the proper normalization for the modes. Now, operating in a similar fashion as in the other limit, it is straightforward to show that $S_{ud2}\left(\omega\right),S_{d1d2}\left(\omega\right),S_{d2u}\left(\omega\right),S_{d2d1}\left(\omega\right)\propto\left(\omega_{\rm max}-\omega\right)^{1/4}$ and $S_{d2d2}\left(\omega\right)=-1+O(\sqrt{\omega_{\rm max}-\omega})$.

\section{Explicit computation of the scattering matrix for the different black-hole configurations}\label{subsec:BHanalyticsolutions}
\sectionmark{Explicit computation}
We make now explicit the computation of the $S$ matrix coefficients for the different BH configurations considered in this work. We also verify explicitly the behavior of the $S$ matrix near $\omega=0$, since in most cases we have analytical formulas for the BdG solutions and we do not need to make use of Poincare's theorem.

\subsection{Flat profile configurations}\label{subsec:scatflatprofile}

For the flat profile configuration of Sec. (\ref{subsec:flatprofile}), the solutions are the corresponding scattering channels for each homogeneous region (subsonic and supersonic), given by Eq. (\ref{eq:PlaneWaveSpinors}). We only need to match the solutions at one point, $x=0$, so we arrive at a reduced system of $4$ equations of the form:
\begin{eqnarray}\label{eq:scatteringreducedlinearsystem}
\nonumber d&=&A\cdot b\\
b&=&\left[b_{u-\rm{out}},b_{\rm{ev}},b_{d1-\rm{out}},b_{d2-\rm{out}}\right]^{T}\\
\nonumber d&=&b_{u-\rm{in}}\sigma_{u-\rm{in},\omega}(0)-b_{d1-\rm{in}}\sigma_{d1-\rm{in},\omega}(0)-b_{d2-\rm{in}}\sigma_{d2-\rm{in},\omega}(0),\\
\nonumber A&=&\left[-\sigma_{u-\rm{out},\omega}(0)~-\sigma_{\rm{ev},\omega}(0)~\sigma_{d1-\rm{out},\omega}(0)~\sigma_{d2-\rm{out},\omega}(0)\right]
\end{eqnarray}
resulting from taking $x_u=x_d=0$ in Eq. (\ref{eq:scatteringlinearsystem}). As explained before, by setting $b_{a-\rm{in}}=1$ and the other ``in" amplitudes to zero in the column vector $d$, we obtain the column $a$ of the $S$-matrix from the solution of the previous system of equations.

Since all the solutions are plane wave spinors, the verification of the previous results on the limiting behavior of the scattering matrix is straightforward, by applying directly Cramer's rule to the previous $4\times 4$ system in the same fashion as before and hence there is no need of using Poincare's theorem.

\subsection{Delta-barrier configuration}\label{subsec:scat1delta}
From Eq. (\ref{eq:1deltaGP}), we see that for $x<0$ the wave function is a soliton wave, so the BdG solutions are now obtained from Eq. (\ref{eq:solitonspinors}). For $x>0$, the GP solution is a supersonic plane wave and then the solutions are the usual scattering channels associated to the corresponding supersonic flow. The match between subsonic and supersonic BdG solutions is made at $x=0$, as in the previous case. However, we note that the delta barrier potential $V(x)=Z\delta(x)$ also creates a discontinuity in the derivative of the spinors $z$ of the BdG equations $z'(0^+)=z'(0^-)+2Zz(0)$, which amounts to a discontinuity in the last two components of the vector $\sigma(\omega)$. Then, for the matching equations, we define the vector
\begin{equation}\label{eq:matchspinors}
\tilde{\sigma}(x)\equiv \sigma +2Z\left[\begin{array}{c} 0 \\ 0 \\ u(x) \\ v(x) \end{array}\right]
\end{equation}

We arrive at a similar system of equations to that of Eq. (\ref{eq:scatteringreducedlinearsystem}) but now
\begin{eqnarray}\label{eq:matching1delta}
d&=&b_{u-\rm{in}}\tilde{\sigma}_{u-\rm{in},\omega}(0)-b_{d1-\rm{in}}\sigma_{d1-\rm{in},\omega}(0)-b_{d2-\rm{in}}\sigma_{d2-\rm{in},\omega}(0),\\
\nonumber A&=&\left[-\tilde{\sigma}_{u-\rm{out},\omega}(0)~-\tilde{\sigma}_{\rm{ev},\omega}(0)~\sigma_{d1-\rm{out},\omega}(0)~\sigma_{d2-\rm{out},\omega}(0)\right]
\end{eqnarray}
where we remark again that the subsonic solutions are the corresponding ones to Eq. (\ref{eq:solitonspinors}).

As in the flat profile configuration, we do not need Poincare's theorem for proving the limiting behavior of the scattering matrix near $\omega=0$. In this limit, it is easy to prove that the subsonic solutions satisfy:
\begin{eqnarray}\label{eq:outevcombination}
\nonumber &&a_{u-\rm{out}}(x)\sigma_{u-\rm{out},\omega}(x)+a_{\rm{ev}}(x)\sigma_{\rm{ev},\omega}\left(x\right)\\
&&=\sigma_0(x)+O(\omega)
\end{eqnarray}
for every point $x$, where $a_{u-\rm{out}}(x),a_{\rm{ev}}(x)$ are some space dependent coefficients. In particular, this holds for $x=0$. Also, it can be proven that a linear combination of $\sigma_{u-\rm{out},\omega}$ and $\sigma_{u-\rm{in},\omega}$ gives the zero-mode of the BdG equation $\sigma_0$ with corrections $O(\omega)$. These results are easily extended to the modified vectors $\tilde{\sigma}$. Taking into account these points and using Cramer's rule for the system of $4$ equations arising from Eq. (\ref{eq:matching1delta}) in the same way as before, we find the expected behavior of the $S$-matrix in the limit $\omega\rightarrow0^+$.

\subsection{Waterfall configuration}\label{subsec:scatwaterfall}

This case is very similar to the single delta barrier since the GP wave function is qualitatively similar. Once more, for $x<0$, the GP wave function is a soliton wave so the BdG solutions are obtained from Eq. (\ref{eq:solitonspinors}) and for $x>0$ we have another supersonic plane wave so the BdG solutions are plane waves. The matching between subsonic and supersonic solutions is also made at $x=0$ but now there are no discontinuities in the derivatives of the BdG solutions and we do not need to use the $\tilde{\sigma}$ vectors. Following a similar reasoning to that of the delta barrier, it can be proven the correct limiting behavior of the scattering matrix near $\omega=0$ without using Poincare's theorem.

\subsection{Resonant configurations}\label{subsec:scatresonant}

For the resonant flat profile described in Sec. \ref{subsec:resonant}, all the BdG solutions are plane waves. We match the different BdG solutions at the discontinuities at $x_u=0$ and $x_d=d$ in Eq. (\ref{eq:scatteringlinearsystem}). Once more, we can prove the limiting behavior of the scattering matrix near $\omega=0$ without using Poincare's theorem since the fundamental matrix $F(x)$ is now analytic in the intermediate region $0<x<d$ and, as explained before, two modes are proportional to the zero mode vector $\sigma_0(x)$ for $\omega\rightarrow0^+$.

In respect to the double delta-barrier configuration, similar to the single barrier case, for $x>d$ the solutions of the BdG equations are those of Eq. (\ref{eq:PlaneWaveSpinors}) and for $x<0$, those of Eq. (\ref{eq:solitonspinors}). In the region $0<x<d$, there are no analytical solutions and the matrix $F(x)$ has to be computed numerically. The matching between the different solutions is made in an analog way as before, setting $x_u=0$ and $x_d=d$ in Eq. (\ref{eq:scatteringlinearsystem}) but now
\begin{eqnarray}\label{eq:2deltalinearsystem}
d&=&b_{u-\rm{in}}\left[\begin{array}{c}\tilde{\sigma}_{u-\rm{in},\omega}\left(x_u\right)\\ 0 \end{array}\right]\\
\nonumber &+&b_{d1-\rm{in}}\left[\begin{array}{c}0\\ \sigma_{d1-\rm{in},\omega}\left(x_d\right)\end{array}\right]+b_{d2-\rm{in}}\left[\begin{array}{c}0\\ \sigma_{d2-\rm{in},\omega}\left(x_d\right)\end{array}\right],\\
\nonumber A&=&\left[\begin{array}{ccccc}-\tilde{\sigma}_{u-\rm{out},\omega}\left(x_u\right)&-\tilde{\sigma}_{\rm{ev},\omega}\left(x_u\right)&F(x_u)&0&0\\ 0&0&\tilde{F}(x_d)&-\sigma_{d1-\rm{out},\omega}\left(x_d\right)&-\sigma_{d2-\rm{out},\omega}\left(x_d\right)\end{array}\right]
\end{eqnarray}
where $\tilde{F}(x)$ is the fundamental matrix with the $4$ independent solutions $\sigma_i(x)$ replaced by $\tilde{\sigma}_i$ as given by Eq. (\ref{eq:matchspinors}). We can check explicitly again the behavior near $\omega=0$ of the scattering matrix by using Cramer's rule, Eq. (\ref{eq:outevcombination}) and associated results, and Poincare's theorem for the region $0<x<d$.

These arguments and techniques can be straightforwardly extended to any black-hole created by an arbitrary combination of delta-barriers, waterfall potentials, etc. since in that case, the asymptotic GP solutions are either the soliton or the homogeneous plane wave in the subsonic region and the homogeneous plane wave in the supersonic region and then the asymptotic BdG solutions can be still computed explicitly.

\chapter{Initial configuration of the confined condensate} \label{app:confinedconfi}

We compute here the initial configuration of the confined condensate considered in Chapter \ref{chapter:MELAFO} by computing the corresponding GP wave function and the associated BdG solutions.

\section{Initial Gross-Pitaevskii wave function}\label{app:confinedGP}
We compute the wave function of the initial condensate, $\Psi_0(x)$, following the techniques developed in Appendix \ref{app:1DGP}.
We require a hard-wall boundary condition at $x=0$, $\Psi_0(0)=0$, which implies, via continuity equation [see discussion after Eq. (\ref{eq:1DPhaseamplitude})], that the phase of the condensate is a constant since $J=0$ and then $v(x)=0$. For simplicity, we take the wave function as real. The amplitude of the wave function $A(x)=|\Psi_0(x)|$ satisfies the equation
\begin{equation}
\left[-\frac{\hbar^{2}}{2m}\partial_{x}^{2}+gA^{2}\right]A=\mu_{0}A \, .
\end{equation}
in the region where there is no potential. The associated amplitude potential $W(A)$, Eq. (\ref{eq:GP-potential}), is now:
\begin{eqnarray} \label{eq:confinedGP-potential}
\frac{1}{2}A'^{2}+W(A) &=& E_A \\
\nonumber W(A)&=&\frac{m}{\hbar^{2}}\left(\mu A^{2}-\frac{g}{2}A^{4}\right)
\end{eqnarray}
We can rewrite the previous identity in a similar fashion to Eq. (\ref{eq:ellipticdensityenergy}):
\begin{equation}\label{eq:confinedellipticdensityenergy}
n'^2=4\tilde{g}n(n-n_2)(n-n_3)
\end{equation}
where now $0=n_1<n(x)<n_2<n_3$ and $n_{2,3}$ are the roots of:
\begin{equation}\label{eq:rootshardwall}
gn^2-2\mu n+2\frac{\hbar^2}{m}E_A=0
\end{equation}

As shown in Appendix \ref{app:1DGP}, the solution can be expressed in terms of elliptic functions. Imposing
the boundary condition $n(0)=A(0)=0$, one obtains a solution of the form
\begin{equation}\label{eq:generalGPsolutionin}
n(x)=n_2\text{sn}^{2}(\sqrt{\tilde{g}n_{3}}x,\nu),~\nu=\frac{n_{2}}{n_{3}},
\end{equation}
which is the same as Eq. (\ref{eq:ellipticdensity}) with $n_1=0$ and $x_0=0$.

In order to determine $n_{2,3}$, another boundary condition is needed, together with the total particle number normalization $\int\mathrm{d}x\, n(x)=N$. From (\ref{eq:confinedGP-potential}) and (\ref{eq:rootshardwall}), the chemical potential can be obtained as
\begin{equation}\label{eq:muzeros}
\mu_{0}=g\frac{n_{2}+n_{3}}{2}~.
\end{equation}

\subsubsection{Ideal confinement}\label{subsec:caseidealconfi}

The ideal confinement boundary condition is defined as $A(L)=0$, with no potential present, so the condensate is confined by hard walls between $0$ and $L$. Thus, the solution for the whole interval $[0,L]$ is given by Eq. (\ref{eq:generalGPsolutionin}). The condition $A(L)=0$ implies
\begin{equation}
\sqrt{\tilde{g}n_{3}}L=2nK(\nu), \, \, n\in\mathbb{N} \, ,
\label{eq:nquantization}
\end{equation}
where we remember that $K(\nu)$ is the complete elliptic integral of the first kind:
\begin{equation}\label{eq:ellipticfirstkind}
K(\nu)\equiv F(\frac{\pi}{2},\nu)=\int_0^{\frac{\pi}{2}}\frac{\mathrm{d}\varphi}{\sqrt{1-\nu\sin^2\varphi}}
\end{equation}
Hereafter we work with the ground state, corresponding to $n=1$ in Eq. (\ref{eq:nquantization}). The particle number normalization is
\begin{equation} \label{eq:N-integr}
N=\int_{0}^{L}\mathrm{d}x~n_{2}~\text{sn}^{2}\left(\sqrt{\tilde{g}n_{3}}x,\nu\right).
\end{equation}
By performing the integral in Eq. (\ref{eq:N-integr}) and using (\ref{eq:nquantization}), we have
\begin{equation}
N=\frac{n_{2}}{\sqrt{\tilde{g}n_{3}}}\frac{2}{\nu}\left[K(\nu)-E(\nu)\right],\label{eq:Nellipticalintegral-1}
\end{equation}
where $E(\nu)$ is the complete elliptic integral of the second kind \cite{Byrd1971}:
\begin{equation}\label{eq:ellipticsecondkind}
E(\nu)\equiv\int_0^{\frac{\pi}{2}}\mathrm{d}\varphi\sqrt{1-\nu\sin^2\varphi}
\end{equation}

Equations (\ref{eq:nquantization}) and (\ref{eq:Nellipticalintegral-1}) lead to
\begin{equation}
N\tilde{g}L=4K(\nu)\left[K(\nu)-E(\nu)\right],\label{eq:Nellipticalintegral-2}
\end{equation}
or
\begin{equation}
4K(\nu)\left[K(\nu)-E(\nu)\right]=\left(\frac{L}{\xi}\right)^{2}\, ,\label{eq:mequation}
\end{equation}
where we define the mean healing length $\xi\equiv \sqrt{\hbar^{2}L/mgN}$ as the healing length corresponding to the mean density $N/L$. We note that $\xi$ is not identical to $\xi_0$, defined in section \ref{sec:themodel}, which is the healing length associated to the chemical potential $\mu_0$.

After eventually obtaining $\nu$, $n_{2}$ and $n_{3}$,  Eq. (\ref{eq:generalGPsolutionin}) can be rewritten as
\begin{equation}
A(x)=\sqrt{n_{2}}~\text{sn}\left(2K(\nu)\frac{x}{L},\nu\right)\label{eq:idealconfinementsolution}
\, .
\end{equation}

For the chemical potential, we obtain, using (\ref{eq:muzeros})-(\ref{eq:nquantization})
\begin{equation}
\mu_{0}=\frac{2\hbar^2}{mL^2}(1+\nu)\left[K(\nu)\right]^2~.
\label{eq:muelliptic}
\end{equation}

\begin{figure}[tb!]
\includegraphics[width=1\columnwidth]{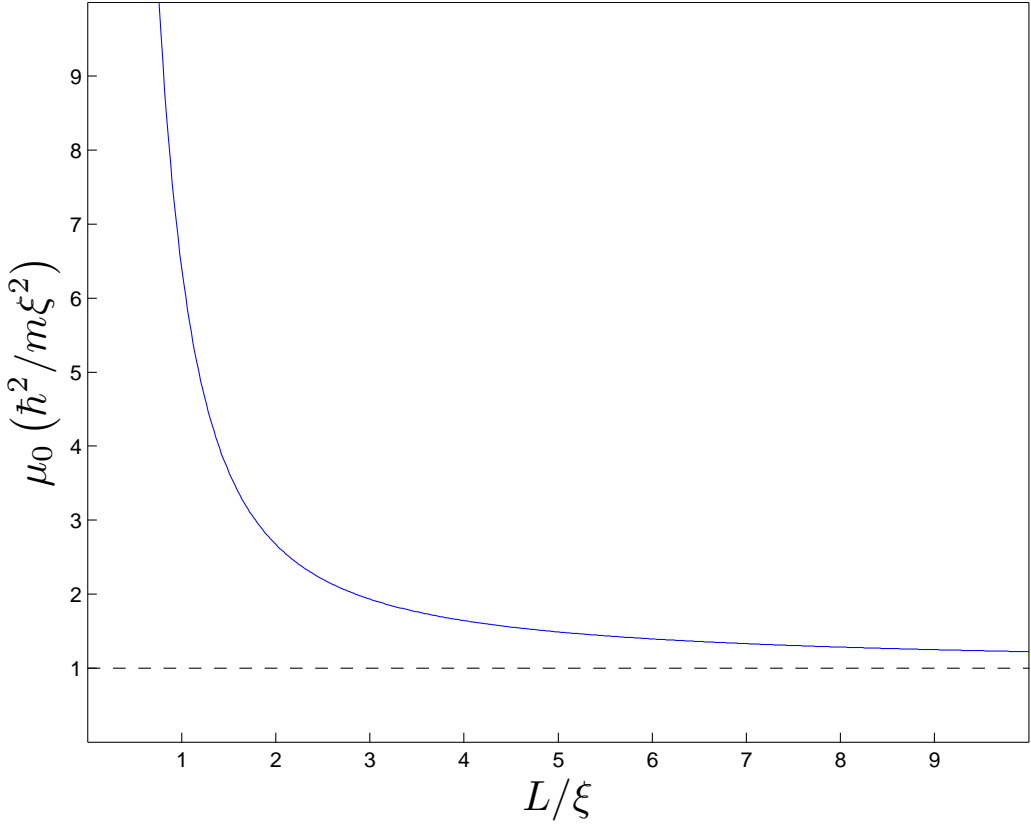} \caption{Computation of the chemical potential as a function of $L/\xi$ using Eqs. (\ref{eq:mequation}) and (\ref{eq:muelliptic}), for an ideal lattice confined between hard walls and in equilibrium. When $L/\xi\ll1$, we are in the Schr\"odinger limit in which $\mu_0\sim 1/L^2$ and when $L/\xi\gg1$, we are in the Thomas-Fermi regime and then $\mu_0\simeq\hbar^2/m\xi^2$.}
\label{fig:MuLxi}
\end{figure}

Taking into account that $\nu$ is a function of $L/\xi$, as given by Eq. (\ref{eq:mequation}), we plot Eq. (\ref{eq:muelliptic}) in Fig. \ref{fig:MuLxi}. In order to find $\nu$, Eq. (\ref{eq:mequation}) must be solved numerically. However, very good approximate solutions can be found. We can clearly distinguish two different regimes:  $L\ll\xi$ and  $L\gg\xi$. The physical interpretation of both conditions is straightforward because
the ratio between the kinetic energy and the interaction energy is precisely
\begin{equation}
\frac{E_{\rm int}}{E_{\rm kin}}
\sim\frac{gN/L}{\hbar^2/mL^2}\sim\frac{\hbar^2/m\xi^2}{\hbar^2/mL^2}=\left(\frac{L}{\xi}\right)^2 \, .
\end{equation}

Then, $L\ll\xi$ is the Schr\"odinger limit in which we have $\nu\simeq0$. In that limit we arrive at the well-known result $\sqrt{\tilde{g}n_{3}}L=\pi$ and $n(x)=n_2~\sin^2(\pi x/L)$. In all the cases considered in this work, $L\gg\xi$, so we work in the limit in which interactions represent the main contribution to the chemical potential (Thomas-Fermi regime). $K(\nu)$ diverges when $\nu\rightarrow1$
while $E(\nu)$ remains finite. From (\ref{eq:mequation}) this means that $\nu\simeq1$. In
fact, there are cases in which $1-\nu$ is so small that it falls below computers floating-point relative accuracy. In those cases, the only way to obtain the solution is through asymptotic expansion. One can prove the following relation in that limit \cite{Byrd1971}
\begin{equation}
K(\nu)\simeq\ln\frac{4}{\sqrt{1-\nu}},~E(\nu)\simeq 1.
\end{equation}
Thus, Eq. (\ref{eq:mequation}) is rewritten as
\begin{equation}
K^{2}-K-\frac{r^{2}}{4}=0,~r=\frac{L}{\xi},
\end{equation}
and from its solution we get $2K=1+\sqrt{1+r^{2}}$. Therefore, $1-\nu=16e^{-2K}\ll1$ which implies both
$n_{2}\simeq n_{3}$ and $\mu_{0}\simeq gn_{2}$. From Eq. (\ref{eq:nquantization}), we find
\begin{equation}
n_{3}=\frac{4K^{2}}{r^{2}}\frac{N}{L}\simeq\left(1+2\frac{\xi}{L}\right)\frac{N}{L}\, ,
\end{equation}
and then
\begin{eqnarray}\label{eq:muxi}
\frac{\mu_{0}}{\hbar^2/m \xi^2}
&=&  \frac{n_{0}}{N/L}=\left(\frac{\xi}{\xi_0}\right)^2  =\nonumber \\
&=& 2(1+\nu)\left[\frac{K(\nu)}{r}\right]^2 \nonumber \\
&\simeq& 1+2\frac{\xi}{L}+2\left(\frac{\xi}{L}\right)^2~.
\end{eqnarray}
We see that $n_0$ is equal to the mean density $N/L$ with corrections $O(\xi/L)$.

Collecting all these results, the wave function can be effectively approximated by
\begin{equation}\label{eq:effectivepsi0}
\Psi_{0}(x)\simeq \left\{ \begin{array}{cc}
\sqrt{n_0}\tanh\left(\frac{x}{\xi_0}\right), & 0\leq x\leq\frac{L}{2}\\ \\
\sqrt{n_0}\tanh\left(\frac{L-x}{\xi_0}\right), & \frac{L}{2}\leq x\leq L
\end{array}\right.
\end{equation}
since $\text{sn}(x,1)=\tanh(x)$ and we have taken $gn_2=gn_3=\mu_0$. As argued before, these approximations are extremely good since the corrections $O(1-\nu)$ are exponentially small. The function $\tanh$ quickly reaches the asymptotic value $1$, which means that in the central zone
of the confinement region (several healing lengths away from the boundaries), the solution is essentially flat.

\subsubsection{Ideal optical lattice potential}\label{subsec:caseidealol}

For computational purposes, the initial rightmost boundary condition is also taken $A(L)=0$ but now half a period of the lattice potential lies inside the confinement region, as explained in the main text. This artificial boundary condition does not create a problem because we take $V_{0}\gg gN/L$ so the function inside the lattice potential is exponentially small. In order to compute the stationary solution in the region where the potential is present, a numerical solution of the GP equation has to be performed. In the situations considered in the present work, $L\gg d$, so the wave function is very similar to that of the ideal confinement case. For numerical convenience, instead of fixing $N$ and then obtaining the chemical potential, we first set $n_0$ to a typical experimental value of the density. Then, using $\mu_0=gn_0$, we compute the number of particles $N$ by integrating the resultant GP wave function. The computed number of particles satisfies $N/L=n_0\left[1+O\left(\xi_0/L\right)+O\left(d/L\right)\right]$
since the wave function is flat in the central region, with density $n(x)=n_0$ and it is only modified near the boundaries of the confined region. This is because in the region where there is no potential the wave function is still given by Eq. (\ref{eq:generalGPsolutionin}) and, as we are in the Thomas-Fermi regime, it can be approximated in a similar way to Eq. (\ref{eq:effectivepsi0}).

\subsubsection{Realistic optical lattice potential}\label{subsec:caserealistic}

In order to simulate a more realistic scenario, we replace the hard-wall boundary condition at the left side by a Gaussian barrier centered at $x=0$ of the form $V_{L}(x)=U\exp(-2x^2/w_L^2)$, with $w_L=2~\mu\text{m}$. For the confinement at the right side, we replace the actual optical lattice by the  realistic Gaussian-enveloped optical lattice centered at $x=L$, which has the form $V(x)=V_0\cos^{2}\left[k_L(x-L)\right]\exp\left[-2(x-L)^2/\tilde{w}^2\right]$. We take the amplitudes of the confining potentials much larger than the chemical potential. We also take $L\gg\tilde{w}\gg w_L$, so that both confinement potentials are well separated in space and there is a large region in which the potential is negligible in order to have a well-defined homogeneous subsonic region. The width $w_L$ does not play a significant role in our simulations so we fix it $w_L=2~\mu\text{m}$. In this way, we can set as boundary conditions for the numerical computation $A(0)=0$ and $A(L_{\rm{bc}})=0$, with $L_{\rm{bc}}$ sufficiently deep in the region where $V_0\exp[-2(x-L)^2/\tilde{w}^2]\geq\mu_{0}$. Once we have set the potential and the boundary conditions for the numerical calculation, we repeat the same process of the previous subsection by fixing $n_0$ to a typical experimental value and using the resulting value of $\mu_0$ to compute the actual number of particles $N$. The same arguments of the discussion in the previous subsection guarantee that the wave function is essentially flat in the region where there is no potential, with $n(x)=n_0$.

\section{Initial Bogoliubov-de Gennes solutions}\label{app:confinedBdG}

We calculate now the Bogoliubov-de Gennes modes associated to the previously computed wave function. For illustrative purposes, we focus on the ideal confinement case, which gives analytic results. Following the results of Appendix \ref{app:BdGsolitonsolutions}, we consider the stationary solutions to the Lax pair of equations (\ref{eq:Laxstationary}):

\begin{eqnarray}\label{eq:Laxpairdimensions}
\frac{\partial}{\partial x}\left[\begin{array}{c} w_1(x) \\ w_2(x) \end{array}\right]&=&\left[\begin{array}{cc}-i\lambda & \Psi_0^*(x)\\
\Psi_0(x)& i\lambda \end{array}\right]\left[\begin{array}{c} w_1(x) \\ w_2(x) \end{array}\right]
\end{eqnarray}
where $\Psi_0(x)$ is the stationary wave function computed in the previous section. In order to match the dimensionless notation of the Lax equations, we have set units such that $\hbar=m=\mu_0=1$ and we have rescaled the wave function by extracting the factor $\sqrt{n_0}$.

By defining $w_{\pm}(x)=w_1(x)\pm w_2(x)$, we find:
\begin{equation}
w''_{\pm}(x)=(\Psi^2_0\pm\Psi'_0-\lambda^2)w_{\pm}(x)
\end{equation}
The previous equation is an exact result. Now, we consider the effective Thomas-Fermi approximation of Eq. (\ref{eq:effectivepsi0}), which in these units amounts to

\begin{equation}\label{eq:effectivepsi0dimensionless}
\Psi_{0}(x)= \left\{ \begin{array}{cc}
\tanh(x), & 0\leq x\leq\frac{L}{2}\\ \\
\tanh(L-x), & \frac{L}{2}\leq x\leq L
\end{array}\right.
\end{equation}
Computing the derivative, we obtain:
\begin{equation}
\Psi'_0(x)=\left\{\begin{array}{cc}
               1-\Psi_0^2, & 0\leq x\leq \frac{L}{2} \\
               \Psi_0^2-1, & \frac{L}{2} \leq x \leq L
             \end{array} \right.
\end{equation}
Interestingly, this implies that we can find plane wave type solutions since:

\begin{eqnarray}
\nonumber w_{+}''&=&(1-\lambda^2)w_{+},~0\leq x\leq \frac{L}{2}\\
w_{-}''&=&(1-\lambda^2)w_{-},~ \frac{L}{2} \leq x \leq L
\end{eqnarray}

After defining $k=\pm 2\sqrt{\lambda^2-1}$, $k$ being the total momentum of the final BdG spinor, and imposing the continuity of the functions, we arrive at:
\begin{eqnarray}\label{eq:laxspinors}
\left[\begin{array}{c} w_1(x) \\ w_2(x) \end{array}\right]&=&Ne^{i\frac{k}{2}x}\left[\begin{array}{c} \lambda-\frac{k}{2}-i\Psi_0(x) \\ \lambda+\frac{k}{2}+i\Psi_0(x) \end{array}\right], ~0\leq x\leq \frac{L}{2}\\
\nonumber \left[\begin{array}{c} w_1(x) \\ w_2(x) \end{array}\right]&=&Ne^{i\frac{k}{2}x}e^{i\phi_k}\left[\begin{array}{c} -\lambda+\frac{k}{2}-i\Psi_0(x) \\ \lambda+\frac{k}{2}-i\Psi_0(x)\end{array}\right],~ \frac{L}{2} \leq x \leq L
\end{eqnarray}
with $\phi_k=\arctan{2/k}$ and $N$ some normalization constant. Using the temporal Lax equation of (\ref{eq:Laxstationary}), we find $\omega=\lambda k$.

In the flat zone of the GP wave function, these plane wave solutions correspond to the usually considered BdG solutions for a condensate with zero momentum. For given $\omega$, we have $4$ values of $k$ that come in pairs $(k,-k)$. One pair corresponds to plane wave type solutions and the other pair to exponential type solutions. Now, we set the hard-wall boundary condition $u(0)=v(0)=0$ which reduces the $4$ independent solutions to $2$. These new solutions are linear combinations of the $(k,-k)$ pairs. Specifically, the real exponential type solutions give rise to:

\begin{eqnarray}
\nonumber &~&\left[
  \begin{array}{c}
    u(x) \\
    v(x) \\
  \end{array}
\right]=C\left[
            \begin{array}{c}
              \left(\left(\frac{\omega}{\kappa}-\frac{\kappa}{2}\right)^2+\Psi^2_0(x)\right)\sinh(\kappa x)+2\Psi_0(x)\left(\frac{\omega}{\kappa}-\frac{\kappa}{2}\right)\cosh(\kappa x) \\
              \left(\left(\frac{\omega}{\kappa}+\frac{\kappa}{2}\right)^2+\Psi^2_0(x)\right)\sinh(\kappa x)-2\Psi_0(x)\left(\frac{\omega}{\kappa}+\frac{\kappa}{2}\right)\cosh(\kappa x)\\
            \end{array}
          \right]
 , ~0\leq x\leq\frac{L}{2}\\
\nonumber &~& \left[
  \begin{array}{c}
    u(x) \\
    v(x) \\
  \end{array}
\right]=\\
\nonumber &~& =C\left[
            \begin{array}{c}
              \left(\left(\frac{\omega}{\kappa}-\frac{\kappa}{2}\right)^2+\Psi^2_0(x)\right)\sinh(\kappa x+u_\kappa)-2\Psi_0(x)\left(\frac{\omega}{\kappa}-\frac{\kappa}{2}\right)\cosh(\kappa x+u_\kappa)\\
              \left(\left(\frac{\omega}{\kappa}+\frac{\kappa}{2}\right)^2+\Psi^2_0(x)\right)\sinh(\kappa x+u_\kappa)+2\Psi_0(x)\left(\frac{\omega}{\kappa}+\frac{\kappa}{2}\right)\cosh(\kappa x+u_\kappa)\\
            \end{array}
          \right]
 , ~\frac{L}{2} \leq x\leq L\\
 &~&e^{u_\kappa}=\frac{\frac{\kappa}{2}-1}{\frac{\kappa}{2}+1}
\end{eqnarray}

while the stationary wave solution is:

\begin{eqnarray}\label{eq:stationarywaveBdG}
\nonumber &~&\left[
  \begin{array}{c}
    u(x) \\
    v(x) \\
  \end{array}
\right]=C\left[
            \begin{array}{c}
              \left(\left(\frac{\omega}{k}+\frac{k}{2}\right)^2-\Psi^2_0(x)\right)\sin(k x)+2\Psi_0(x)\left(\frac{\omega}{k}+\frac{k}{2}\right)\cos(k x) \\
              \left(\left(\frac{\omega}{k}-\frac{k}{2}\right)^2-\Psi^2_0(x)\right)\sin(k x)-2\Psi_0(x)\left(\frac{\omega}{k}-\frac{k}{2}\right)\cos(k x)\\
            \end{array}
          \right]
 , ~0\leq x\leq\frac{L}{2}\\
\nonumber &~& \left[
  \begin{array}{c}
    u(x) \\
    v(x) \\
  \end{array}
\right]=\\
\nonumber &~&= C\left[
            \begin{array}{c}
              \left(\left(\frac{\omega}{k}+\frac{k}{2}\right)^2-\Psi^2_0(x)\right)\sin(k x+ 2\phi_k)-2\Psi_0(x)\left(\frac{\omega}{k}+\frac{k}{2}\right)\cos(k x+2\phi_k) \\
              \left(\left(\frac{\omega}{k}-\frac{k}{2}\right)^2-\Psi^2_0(x)\right)\sin(k x+ 2\phi_k)+2\Psi_0(x)\left(\frac{\omega}{k}-\frac{k}{2}\right)\cos(k x+2\phi_k) \\
            \end{array}
          \right]
 , ~\frac{L}{2} \leq x\leq L\\
\end{eqnarray}
with $C$ some normalization constant. We compute now the BdG solutions in the ideal confinement case. Using only the stationary wave solution and taking into account the other boundary condition $u(L)=v(L)=0$, we find that the energies and the solutions are quantized through:

\begin{eqnarray}\label{eq:stationaryquantized}
\nonumber \left[
  \begin{array}{c}
    u_n(x) \\
    v_n(x) \\
  \end{array}
\right]&=&C_n\times\\
\nonumber &\times&\left[
            \begin{array}{c}
              \left(\left(\frac{\omega_n}{k_n}+\frac{k_n}{2}\right)^2-\Psi^2_0(x)\right)\sin(k_n x)+2\Psi_0(x)\left(\frac{\omega_n}{k_n}+\frac{k_n}{2}\right)\cos(k_n x) \\
              \left(\left(\frac{\omega_n}{k_n}-\frac{k_n}{2}\right)^2-\Psi^2_0(x)\right)\sin(k_n x)-2\Psi_0(x)\left(\frac{\omega_n}{k_n}-\frac{k_n}{2}\right)\cos(k_n x)\\
            \end{array}
          \right]
 , ~0\leq x\leq\frac{L}{2}\\
\nonumber \left[
  \begin{array}{c}
    u_n(x) \\
    v_n(x) \\
  \end{array}
\right]&=&\left[
            \begin{array}{c}
              (-1)^nu_n(L-x) \\
              (-1)^nv_n(L-x) \\
            \end{array}
          \right]
 , ~ \frac{L}{2}\leq x \leq L\\
\nonumber k_n&=&(n+1)\frac{\pi}{L}-\frac{2}{L}\arctan\frac{2}{k_n}, n=0,1,2\ldots  \\
\omega_n&=&\sqrt{k_n^2+\frac{k_n^4}{4}}~,
\end{eqnarray}
where $C_n$ is a normalization constant chosen such that $\int_0^L \mathrm{d}x |u_n(x)|^2-|v_n(x)|^2=1$. Then, $k_n$ is obtained by solving the previous transcendental equation. We have made explicit that the functions obey the usual parity properties of Sturm-Liouville problems.

We briefly discuss the case of non-ideal confinement. When the confinement is provided by the ideal optical lattice, as discussed in Sec. \ref{subsec:caseidealol}, the stationary wave solution (\ref{eq:stationarywaveBdG}) is still valid in the region where there is no potential. As the potential is present only in a small region of size $d\ll L$, the final quantized confined solutions can be expected to be given by Eq. (\ref{eq:stationaryquantized}), with small corrections of order $d/L\ll 1$. In the case of the realistic confinement, Sec. \ref{subsec:caserealistic}, the solutions in the region where there is no potential are still combinations of the plane waves $(k,-k)$. In particular, we can expect that the stationary wave solution (\ref{eq:stationarywaveBdG}) still gives a good approximation for the BdG modes in the free region once we have replaced the hard-wall boundary condition at $x=0$ by the actual finite size confining potential since its size satisfy $\tilde{w}_L\ll L$. We can also expect that the quantization given in Eq. (\ref{eq:stationaryquantized}) describes with some accuracy the actual quantization of the solutions to the BdG equations as we also have $\tilde{w}\ll L$.

\chapter{Flowing condensate in a nonlinear optical lattice} \label{app:nonlinearol}

\section{General discussion}

We review here the theory of a condensate in an optical lattice. Most of the results are extracted from Ref. \cite{Wu2003}. The time-independent GP equation in an ideal {\it infinite} optical lattice whose potential has the same form of the long-time potential of Eq. (\ref{eq:TDPotential}), $V(x)= V_{\infty} \cos^2(k_L x)$, reads
\begin{equation}\label{eq:dimensionOLGPequation}
-\frac{\hbar^2}{2m}\frac{d^{2}\Psi_{0}}{dx^{2}}+V_{\infty}\cos^2(k_L x )\Psi_{0}(x)+g|\Psi_{0}(x)|^{2}\Psi_{0} = \mu \Psi_{0}
\end{equation}
For simplifying the calculations, we define the dimensionless length $z\equiv 2k_L x$ and rescale the wave function as $\Psi_0\rightarrow \sqrt{n_r}\Psi_0$, with $n_r$ the average atomic density, so it become dimensionless. Using these quantities, we can rewrite the GP equation as:
\begin{equation}\label{eq:dimensionlessOLGPequation}
-\frac{1}{2}\frac{d^{2}\Psi_{0}}{dz^{2}}+v\cos(z)\Psi_{0}+c^{2}|\Psi_{0}(z)|^{2}\Psi_{0} = \alpha \Psi_{0}
\end{equation}
with
\begin{equation}
v=\frac{V_{\infty}}{2E_L},~c^{2}=
\frac{gn_{r}}{E_L},~ \alpha =\frac{\mu-V_{\infty}/2}{E_L},~ E_L=\frac{4\hbar^2k_L^2}{m}=8E_R
\end{equation}
Here, $\alpha$ is the dimensionless chemical potential, displaced $-V_{\infty}/2$, and $c$ is the dimensionless speed of sound of the lattice in the limit of vanishing amplitude $v\rightarrow 0$.

We look for Bloch type solutions of the form
\begin{equation}
\Psi_{0}(z)=e^{i\tilde{q}z}y_{\tilde{q}}(z),
\end{equation}
with $y_{\tilde{q}}(z+2\pi)=y_{\tilde{q}}(z)$ periodic since the non-linear term is periodic
for a Bloch-wave type solution. The normalization condition reads now
\begin{equation}
\frac{1}{2\pi}\int_{0}^{2\pi}\mathrm{d}z~|\Psi_{0}(z)|^{2}=\frac{1}{2\pi}\int_{0}^{2\pi}\mathrm{d}z~|y_{\tilde{q}}(z)|^{2}=1.\label{eq:olnormalization}
\end{equation}
The Brillouin zone is placed in the region
 $-1/2 < \tilde{q} < 1/2$. The equation for $y_{\tilde{q}}$ reads:
\begin{equation}\label{eq:dimensionlessOLGPBlochequation}
-\frac{1}{2}\left(\frac{\partial}{\partial z}+i\tilde{q}\right)^{2}y_{\tilde{q}}+v\cos(z)y_{\tilde{q}}+c^{2}|y_{\tilde{q}}|^{2}y_{\tilde{q}}=\alpha_{\tilde{q}}y_{\tilde{q}},
\end{equation}
where we have allowed for a $\tilde{q}$ - dependence of $\alpha$. The linear Schr\"odinger regime is obtained when $c=0$. For $c^{2}>v$, some extra non-linear Bloch waves appear. This generates a loop structure in the conduction band. In the systems analyzed in the present work, $c^{2}\sim10^{-3}-10^{-4}$ and $v\sim10^{-1}$, hence $v\gg c^{2}$. As a consequence: (a) loops do not appear; (b) the system is close to the linear Schr\"odinger regime.

To compute the Bloch energy eigenvalues, we follow the method developed
in Ref. \cite{Wu2003}. First, we perform a finite Fourier expansion of the periodic
function $y_{\tilde{q}}(z)$ of the form:
\begin{equation}\label{eq:Fouriercut}
y_{\tilde{q}}(z)=\sum_{n=-M}^{M}c_{n}e^{inz},
\end{equation}
where $M$ is a numerically enforced cut-off. After substitution of this solution in (\ref{eq:dimensionlessOLGPBlochequation})
and in (\ref{eq:olnormalization}), we get $2M+2$ equations for
$2M+2$ variables (the $2M+1$ values of the Fourier coefficients $c_{n}$ plus the eigenvalue $\alpha$). Instead of directly solving these non-linear
equations, it is more efficient to minimize the quadratic sum of the $2M+2$ equations,
\begin{equation}
S=\sum_{j=1}^{2M+2}f_{j}^{2}\label{eq:NonLinearOLQuadraticSum},
\end{equation}
where $f_{j}(c_{n},\alpha)=0$ (with $j=1,2\ldots2M+2$) are the equations to be solved. It is easy to see that all these equations are
real, so the coefficients $c_{n}$ can be chosen as real numbers and there is no need to take the modulus in Eq. (\ref{eq:NonLinearOLQuadraticSum}).

A given Bloch solution can be unstable, either dynamically or in the sense of Landau, as explained in Sec. \ref{subsec:timeindependent}.
We rewrite the functional (\ref{eq:Kfunctional}) in our system of units as:
\begin{equation}
K\left[\Psi\right]=\int\mathrm{d}z~ \left[ 
\frac{1}{2} \left| \frac{\partial \Psi}{\partial z}  \right|^2
+v\cos(z)|\Psi|^{2}+\frac{c^{2}}{2}|\Psi|^{4}-\alpha|\Psi|^{2} \right] \label{eq:olgrandcanonicalhamiltonian} \, .
\end{equation}
Considering linear perturbations around the stationary solution $\Psi_0$ as in Eq. (\ref{eq:Landauperturbation}), we find:
\begin{eqnarray}
\delta K & = & \frac{1}{2}\int\mathrm{d}z~[\delta\Psi^{*}~\delta\Psi]\Lambda\left[\begin{array}{c}
\delta\Psi\\
\delta\Psi^{*}
\end{array}\right]\nonumber \\
\Lambda & = & \left[\begin{array}{cc}
H' & L\\
L^{*} & H'
\end{array}\right]\nonumber \\
H' & = & -\frac{1}{2}\frac{\partial^{2}}{\partial z^{2}}+v\cos(z)+2c^{2}|\Psi_{0}|^{2}-\alpha_{\tilde{q}}\nonumber \\
L & = & c^{2}\Psi_{0}^{2} \, .
\end{eqnarray}
The appearance of negative eigenvalues in the Hermitian operator $\Lambda$ indicates that the wave function is energetically unstable. The corresponding eigenvalue equation reads
\begin{equation}\label{eq:landaueig}
\Lambda\left[\begin{array}{c}
u\\
v
\end{array}\right]=\tilde{\lambda}\left[\begin{array}{c}
u\\
v
\end{array}\right] \, .
\end{equation}
By absorbing the exponential plane-wave factor of
the Bloch-type GP solution, $u(z)=e^{i\tilde{q}z}u_{\tilde{q}}(z)$
and $v(z)=e^{-i\tilde{q}z}v_{\tilde{q}}(z)$, we arrive at a new matrix
operator $\Lambda_{\tilde{q}}$ which is periodic. Applying Bloch's theorem in the form $u_{\tilde{q}}(z)=e^{i\tilde{k}z}u_{\tilde{q},\tilde{k}}(z)$ and $v_{\tilde{q}}(z)=e^{i\tilde{k}z}v_{\tilde{q},\tilde{k}}(z)$
with $u_{\tilde{q},\tilde{k}}(z)$ and $v_{\tilde{q},\tilde{k}}(z)$ periodic in $[0,2\pi]$, the
final eigenvalue equation reads
\begin{eqnarray}\label{eq:landaublocheig}
\Lambda_{\tilde{q},\tilde{k}}\left[\begin{array}{c}
u_{\tilde{q},\tilde{k}}\\
v_{\tilde{q},\tilde{k}}
\end{array}\right] & = & \tilde{\lambda}_{\tilde{q},\tilde{k}}\left[\begin{array}{c}
u_{\tilde{q},\tilde{k}}\\
v_{\tilde{q},\tilde{k}}
\end{array}\right]\nonumber \\
\Lambda_{\tilde{q},\tilde{k}} & = & \left[\begin{array}{cc}
H^{''}_{\tilde{k}+\tilde{q}} & L_{\tilde{q}}\\
L_{\tilde{q}}^{*} & H^{''}_{\tilde{k}-\tilde{q}}
\end{array}\right]\nonumber \\
H^{''}_{\tilde{k}} & = & -\frac{1}{2}\left(\frac{\partial}{\partial z}+i\tilde{k}\right)^{2}+v\cos(z)+2c^{2}|y_{\tilde{q}}|^{2}-\alpha\nonumber \\
L_{\tilde{q}} & = & c^{2}y_{\tilde{q}}^{2} \, .
\end{eqnarray}

Dynamical instabilities correspond to modes that grow exponentially with time. They are computed by looking for non-real eigenvalues of the BdG equations, formally similar to Eq. (\ref{eq:landaueig}):
\begin{equation}
M\left[\begin{array}{c}
u\\
v
\end{array}\right]=\tilde{\epsilon}\left[\begin{array}{c}
u\\
v
\end{array}\right] \, ,
\end{equation}
with $M=\sigma_z \Lambda$ and $\tilde{\epsilon}$ is the dimensionless energy (in units of $E_L$). Bloch's theorem also applies here and after a computation similar to that which has led to Eq. (\ref{eq:landaublocheig}), the eigenvalue equation reads
\begin{equation}\label{eq:bdgblocheig}
M_{\tilde{q},\tilde{k}}\left[\begin{array}{c}
u_{\tilde{q},\tilde{k}}\\
v_{\tilde{q},\tilde{k}}
\end{array}\right] = \tilde{\epsilon}_{\tilde{q},\tilde{k}}\left[\begin{array}{c}
u_{\tilde{q},\tilde{k}}\\
v_{\tilde{q},\tilde{k}}
\end{array}\right],
\end{equation}
with $M_{\tilde{q},\tilde{k}}=\sigma_z \Lambda_{\tilde{q},\tilde{k}}$. As shown in the main text, the system can only be dynamically unstable whenever it is already Landau unstable. A plot of the parameter regions where the system is Landau and dynamically unstable can be found in Ref. \cite{Wu2003}.

In addition to the dynamical stability analysis, the real eigenvalues of this operator can be used to compute the speed of sound in the optical lattice. When $\tilde{q}=0$, it can be proven that for small (dimensionless) wave-vector $\tilde{k}$, the eigenvalues go like $\tilde{\epsilon}=\pm \tilde{s}|\tilde{k}|$, with $\tilde{s}$ the dimensionless speed of sound (here, in units of $2\hbar k_L/m$). On the other hand, the form of the Bloch-type solution of the GP equation as a function of both Bloch momentum and density can be directly used to compute the sound speed without the need to solve for the BdG equations \cite{Pethick2008}. Restoring dimensions with $q=2k_L\tilde{q}$, the sound speed $s$ in the lattice is
\begin{equation}\label{eq:OLsound}
s=\frac{\sqrt{\partial^2_n \mathcal{E} \partial^2_{q} \mathcal{E}}}{\hbar}
\end{equation}
where $\partial_n$ denotes derivative with respect the mean density, $n_r$, $\partial_q$ is the derivative with respect the Bloch momentum $q$, with both derivatives evaluated at $q=0$. In the previous expression, $\mathcal{E}$ is the average energy density, given by

\begin{eqnarray}\label{eq:energydensity}
\mathcal{E}&=&\frac{n_r}{d}\int_0^{d}\mathrm{d}x~y^*_{\tilde{q}}(x)\left[-\frac{\hbar^{2}}{2m}\left(\frac{\partial}{\partial x}+iq\right)^{2} + V_{\infty}\cos^2(k_L x) +\frac{gn_r}{2}|y_{\tilde{q}}(x)|^2 \right] y_{\tilde{q}}(x) \nonumber \\
&=&n_r\mu-\frac{gn^2_r}{4\pi}\int_0^{2\pi}\mathrm{d}z~|y_{\tilde{q}}(z)|^4.
\end{eqnarray}

\section{Perturbative treatment}

We consider now a perturbative treatment of the results presented in the previous section. In Eq. (\ref{eq:dimensionlessOLGPequation}) there are two dimensionless parameters, $v$ and $c^2$. The former is essentially the amplitude of the potential ($v\gg1$ is the tight-binding regime while $v\ll1$ corresponds the nearly-free-particle regime) and the latter is a measure of the strength of the interaction or nonlinearity. In the cases studied in this work, $v\ll 1$ but the following discussion applies to arbitrary values of $v$. The quantity
$\delta \equiv c^2 \ll 1$ is the small parameter of our perturbation theory. We expand in powers of $\delta$ both the Bloch wave $y_{\tilde{q}}(z)$ and the displaced and dimensionless chemical potential $\alpha_{\tilde{q}}$, which solve Eqs. (\ref{eq:olnormalization})-(\ref{eq:dimensionlessOLGPBlochequation}). We obtain
\begin{eqnarray}
y_{\tilde{q}}(z) & = & \sum_{n=0}^\infty \delta^n y_{\tilde{q}}^{(n)}(z) \, , \nonumber \\
\alpha_{\tilde{q}} & = & \sum_{n=0}^\infty \delta^n \alpha_{\tilde{q}}^{(n)}\, , \label{eq:alpha-q}
\end{eqnarray}
which transforms Eq. (\ref{eq:dimensionlessOLGPBlochequation}) into
\begin{eqnarray}
&& H^{(0)}_{\tilde{q}} y_{\tilde{q}}+\delta|y_{\tilde{q}}|^{2}y_{\tilde{q}} = \alpha_{\tilde{q}}y_{\tilde{q}} \nonumber \\
&& H^{(0)}_{\tilde{q}}\equiv  -\frac{1}{2}\left(\frac{\partial}{\partial z}+i\tilde{q}\right)^{2}+v\cos(z) \, .
\end{eqnarray}
In the following, we focus on the lowest Bloch band and keep terms up to $O(\delta^{2})$.
The lowest-order term solves the linear Schr\"odinger equation, $H^{(0)}_{\tilde{q}}y_{\tilde{q}}^{(0)}=\alpha_{\tilde{q}}^{(0)}y_{\tilde{q}}^{(0)}$. Therefore, $y_{\tilde{q}}^{(0)}=\phi_{\tilde{q},0}$ and $\alpha_{\tilde{q}}^{(0)}=\tilde{\varepsilon}_{\tilde{q},0}$, where $\phi_{\tilde{q},0}(z)$
is the corresponding eigenfunction for the lowest band, which involves Mathieu functions, and $\tilde{\varepsilon}_{\tilde{q},0}$ its dimensionless eigenvalue (note the use of the index $0$ for two different purposes: perturbative order in the superindex and band index in the subindex). $\phi_{\tilde{q},0}$ is normalized according to Eq. (\ref{eq:olnormalization}).

The first-order corrections satisfy
\begin{equation}
H^{(0)}_{\tilde{q}}y_{\tilde{q}}^{(1)}+|y_{\tilde{q}}^{(0)}|^2y_{\tilde{q}}^{(0)}=\alpha_{\tilde{q}}^{(0)}y_{\tilde{q}}^{(1)}+\alpha_{\tilde{q}}^{(1)}y_{\tilde{q}}^{(0)},
\end{equation}
which, using standard perturbation techniques, leads to
\begin{eqnarray}
\label{eq:perturbativeSolution}
\alpha_{\tilde{q}}^{(1)} & = & \frac{1}{2\pi}\int_0^{2\pi}\mathrm{d}z|\phi_{\tilde{q},0}|^4 \nonumber \\
y_{\tilde{q}}^{(1)} & = & \sum_{n=1}^{\infty} \beta_n \phi_{\tilde{q},n} \nonumber \\
\beta_n & = & \frac{\frac{1}{2\pi}\int_0^{2\pi}\mathrm{d}z~\phi^*_{\tilde{q},n}|\phi_{\tilde{q},0}|^2\phi_{\tilde{q},0}}{\tilde{\varepsilon}_{\tilde{q},0}-\tilde{\varepsilon}_{\tilde{q},n}},
\end{eqnarray}
where $\{\phi_{\tilde{q},n}\}_{n=1}^\infty$ are the Schr\"odinger eigenvectors of the rest of bands and $\tilde{\varepsilon}_{\tilde{q},n}$ its corresponding eigenvalues, for a given value of $\tilde{q}$.

The second-order equation reads

\begin{equation}
H^{(0)}_{\tilde{q}}y_{\tilde{q}}^{(2)}+2|y_{\tilde{q}}^{(0)}|^2y_{\tilde{q}}^{(1)}+y_{\tilde{q}}^{(0)2}y^{(1)*}_{\tilde{q}}=\alpha _{\tilde{q}}^{(0)}y_{\tilde{q}}^{(2)}+\alpha _{\tilde{q}}^{(1)}y_{\tilde{q}}^{(1)}+\alpha _{\tilde{q}}^{(2)}y_{\tilde{q}}^{(0)}.
\end{equation}

We note that $y_{\tilde{q}}^{(2)}$ is not needed to compute $\alpha_{\tilde{q}}^{(2)}$. Specifically,
\begin{equation}\label{eq:perturbativeSolutionW2}
\alpha _{\tilde{q}}^{(2)}=\frac{1}{2\pi}\int_0^{2\pi}\mathrm{d}z|\phi_{\tilde{q},0}|^{2}(2\phi^*_{\tilde{q},0}y_{\tilde{q}}^{(1)}+\phi_{\tilde{q},0}y_{\tilde{q}}^{(1)*})
=-3\sum_{n=1}^{\infty}|\beta_n|^2(\tilde{\varepsilon}_{\tilde{q},n}-\tilde{\varepsilon}_{\tilde{q},0}),
\end{equation}
which is always negative.

Instead of invoking Mathieu functions and integrals of them, the numerical computation of the formulae presented in this Appendix  [Eqs. (\ref{eq:perturbativeSolution}), (\ref{eq:perturbativeSolutionW2})] can be easily performed in a Fourier representation.
As $y_{\tilde{q}}, \phi_{\tilde{q},n}$ are periodic functions in $[0,2\pi]$, their Fourier expansion read $y_{\tilde{q}}(z)=\sum_{m=-\infty}^{\infty}c_me^{imz}$ and $\phi_{\tilde{q},n}(z) = \sum_{m=-\infty}^{\infty}a_{n,m}e^{imz} $. Both the $c_m$ and the $a_{n,m}$ coefficients can be chosen real [see discussion after Eq. (\ref{eq:NonLinearOLQuadraticSum})]. In this Fourier representation, $H^{(0)}_{\tilde{q}}$ is a tridiagonal matrix with elements $\mathbf{H}_{m,m\pm 1}=v/2$ and $\mathbf{H}_{m,m}=(m+\tilde{q})^2 /2$. Multiplication by $|\phi_{\tilde{q},0}|^{2}$ is represented by the matrix $\mathbf{r}_{m,p}=\sum_l a_{0,l}a_{0,l+m-p}$. The perturbative expansion of the Fourier components of solution to the GP equation reads $c_m=\sum_{n=0}^\infty \delta^n c_m^{(n)}$. In this Fourier basis, the Schr\"odinger equation for all the bands is written as an eigenvalue-eigenvector matrix equation
\begin{equation}
\mathbf{H} \mathbf{a}_n  =  \tilde{\varepsilon}_{\tilde{q},n} \mathbf{a}_n,
\end{equation}
where matrix multiplication is understood. We can also rewrite Eqs. (\ref{eq:perturbativeSolution}), (\ref{eq:perturbativeSolutionW2}) as:
\begin{eqnarray} \label{eq:fourier-pert}
\alpha_{\tilde{q}}^{(1)}&=& \mathbf{a}^{T}_0\mathbf{r}\mathbf{a}_0\nonumber \\
\mathbf{c}^{(1)} & = & \sum_{n=1}^{\infty} \beta_n \mathbf{a}_n \nonumber \\
\beta_n & = & \frac{\mathbf{a}^{T}_n\mathbf{r}\mathbf{a}_0}{\tilde{\varepsilon}_{\tilde{q},0}-\tilde{\varepsilon}_{\tilde{q},n}} \nonumber \\
\alpha_{\tilde{q}}^{(2)}&=&3\mathbf{a}^{T}_0\mathbf{r}\mathbf{c}^{(1)}=-3\sum_{n=1}^{\infty}|\beta_n|^2(\tilde{\varepsilon}_{\tilde{q},n}-\tilde{\varepsilon}_{\tilde{q},0}).
\end{eqnarray}
In practice, we impose a numerical cut-off in the Fourier series as that of Eq. (\ref{eq:Fouriercut}) to compute the previous magnitudes.
Now we can use the previous perturbative results to give approximate closed expressions for some magnitudes of the optical lattice. To first order in $\delta$, the energy density is given by Eq. (\ref{eq:energydensity})
\begin{equation} \label{eq:E-1st-order}
\mathcal{E}
\simeq n_r\left(\mu-\frac{gn_r}{2}\alpha_{\tilde{q}}^{(1)}\right).
\end{equation}
We can write
\begin{equation}\label{eq:oldmu}
\mu = E_L\alpha+\frac{V_{\infty}}{2}\simeq E_L\alpha_{\tilde{q}}^{(0)}+\frac{V_{\infty}}{2}
+E_L\alpha_{\tilde{q}}^{(1)}\delta= \mu^{(0)}+E_{L}\alpha_{\tilde{q}}^{(1)}\delta
\end{equation}
On the other hand, for the non-interacting chemical potential we have (assuming $\tilde{q}\ll1/2$):
\begin{equation}
\mu^{(0)}=\mu^{(0)}(q) \simeq E_{\rm{min}}+\frac{\hbar^{2}q^2}{2m^*} \, ,
\end{equation}
where $E_{\rm{min}}$ is the bottom of the Schr\"odinger conduction band as defined in the main text, $m^*$ is the effective mass, and we recall $q=2k_L\tilde{q}$. Thus,
\begin{equation}\label{eq:olmu}
\mu=E_{\rm{min}}+\frac{\hbar^{2}q^2}{2m^*}+gn_r\alpha_{\tilde{q}}^{(1)} \, .
\end{equation}
Using (\ref{eq:olmu}) we can rewrite (\ref{eq:E-1st-order}) as
\begin{equation}
\mathcal{E}\simeq n_rE_{\rm{min}}+n_r\frac{\hbar^{2}q^2}{2m^*}+\frac{gn^2_r}{2}\alpha_{\tilde{q}}^{(1)} \, .
\end{equation}
Computing the derivatives of Eq. (\ref{eq:OLsound}) to lowest order in $\delta$, we obtain:
\begin{eqnarray}
\left(\frac{\partial^2 \mathcal{E} }{\partial q^2}\right)_{q=0} &\simeq& n_r\frac{\hbar^2}{m^*} \, , \nonumber \\
\left(\frac{\partial^2 \mathcal{E}}{\partial n_r^2}\right)_{q=0}&\simeq& g\alpha_{0}^{(1)} \, .
\end{eqnarray}
and the speed of sound then reads
\begin{equation}\label{eq:PerturbativeSound}
s=\sqrt{\frac{gn_r}{m^*}\alpha_{0}^{(1)}}=\sqrt{\frac{gn_r}{m}}\sqrt{\frac{m}{m^*}\alpha_{0}^{(1)}} \, .
\end{equation}
Similar results appear in Ref. \cite{Kramer2003} and references therein. The first square root is the speed of sound in the absence of the optical lattice. The second factor on the right takes into account the presence of the optical lattice and is practically unity for $v\ll 1$. Specifically, we can write:
\begin{eqnarray}
m^*&=&m\left[1+8v^2+O(v^4)\right]\nonumber \\
\alpha_{\tilde{q}}^{(1)}&=&1+\frac{8v^2}{(1-4q^2)^2}+O(v^4),
\label{m-alpha-1}
\end{eqnarray}
and then $\sqrt{m\alpha_{0}^{(1)}/m^*}=1+O(v^4)$, so $s\simeq\sqrt{gn_r/m}$, which is the usual expression for the speed of sound. In particular, we can write the previous expression in dimensionless form as $\tilde{s}=c$, which motivates the election of this label for the non-linear dimensionless parameter. We note that Eq. (\ref{eq:PerturbativeSound}) can also be interpreted as the sound velocity arising in a system with an effective constant coupling $g_{\rm{eff}}=g\alpha_{0}^{(1)}$ and effective mass $m^{*}$ \cite{Carusotto2002}.

The current is conserved a stationary solution of the GP equation and is given, to lowest order in $\delta$ and $q$, by \cite{Pethick2008}:
\begin{equation}\label{eq:olcurrent}
j=\frac{1}{\hbar}\frac{\partial\mathcal{E}}{\partial q}=n_r\frac{\hbar q}{m^*} \, .
\end{equation}
In the nearly-free atom approximation, where the relative oscillations of the density around the mean value are small, we can write $j\simeq n_r\bar{v}$ were $\bar{v}$ is a locally averaged flow velocity (not to be confused with the dimensionless parameter $v$). Then, we have $\bar{v}\simeq\hbar q/m^*$ and we can rewrite Eq. (\ref{eq:olmu}) in a more appealing form:
\begin{equation}\label{eq:olhdmu}
\mu=E_{\rm{min}}+\frac{1}{2}m^*\bar{v}^2+m^*s^2 \, .
\end{equation}
The physical interpretation of this equation is straightforward: the chemical potential in the optical lattice is the sum of the energy of the bottom of the conduction band plus the contribution of the kinetic energy and the interaction energy, both with $m^*$ instead of $m$.

As explained at the end of general discussion of the previous section, the same result for the speed of sound can be obtained by solving the BdG equations (\ref{eq:bdgblocheig}) perturbatively to first order in $\delta$ for $q=0$. In that case, the GP solution $\Psi_0(z)=y_0(z)$ can be taken as real. We perform an expansion of the spinors in terms of solutions to the Schr\"odinger equation,
\begin{equation}
\left[\begin{array}{c}
u_{0,\tilde{k}}(z)\\
v_{0,\tilde{k}}(z)
\end{array}\right] =\sum_{n=0}^\infty \phi_{\tilde{k},n}(z)\chi_{\tilde{k},n},
\end{equation}
where $\chi_{\tilde{k},n}$ are spinors of constant coefficients. The matrix operator $M_{0,\tilde{k}}$
introduced in (\ref{eq:bdgblocheig}) can be written, to first order in $\delta$, as $M_{0,\tilde{k}}=M^{(0)}_{\tilde{k}}+ M^{(1)}\delta$  with:
\begin{eqnarray}
M^{(0)}_{\tilde{k}} & = & \left[\begin{array}{cc}
H^{(0)}_{\tilde{k}}-\tilde{\varepsilon}_{0,0} & 0\\
0 & -H^{(0)}_{\tilde{k}}+\tilde{\varepsilon}_{0,0}
\end{array}\right] \\
M^{(1)} & = & \left[\begin{array}{cc}
2\phi^2_{0,0}(z)-\alpha_{0}^{(1)} & \phi^2_{0,0}(z)\\
-\phi^2_{0,0}(z) & -2\phi^2_{0,0}(z)+\alpha_{0}^{(1)}
\end{array}\right] \, , \nonumber
\end{eqnarray}
$\phi_{0,0}(z)$ being the Schr\"odinger solution for the bottom of the lowest band. Note that $\phi_{\tilde{k},n}(z)$ are eigenfunctions of $H^{(0)}_{\tilde{k}}$. A matrix equation for the perturbative expansion of the $\chi_{\tilde{k},0}$ spinors to first order in $\delta$ can be obtained by projecting onto the lowest Bloch eigenfunction, $\phi^*_{\tilde{k},0}(z)$:

\begin{eqnarray}\label{eq:effectiveolbdg}
\tilde{\epsilon}(\tilde{q}=0,{\tilde{k}}) \chi_{\tilde{k},0} & = & \left[\begin{array}{cc}
\tilde{\varepsilon}_{\tilde{k},0}-\tilde{\varepsilon}_{0,0}+(2J(\tilde{k})-\alpha^{(1)}_{0})\delta & J(\tilde{k}) \delta \\
-J(\tilde{k}) \delta & -\tilde{\varepsilon}_{\tilde{k},0}+\tilde{\varepsilon}_{0,0}-(2J(\tilde{k})-\alpha^{(1)}_{0}) \delta
\end{array}\right] \chi_{\tilde{k},0} \nonumber \\
J(\tilde{k}) & = & \frac{1}{2\pi}\int_0^{2\pi}\mathrm{d}z|\phi_{\tilde{k},0}|^2|\phi_{0,0}|^2.
\end{eqnarray}

Restoring units for $\tilde{k}$ by using $k=2k_L\tilde{k}$, expanding to lowest order near $k=0$, and neglecting corrections $O(\delta)$ to $m^{*}$, the eigenvalues are approximated as
\begin{equation}
\epsilon(0,k) = \pm\left[
\left(\frac{\hbar^2k^2}{2m^*}\right)^2+gn_r\alpha_{0}^{(1)}\frac{\hbar^2k^2}{m^*}
\right]^{\frac{1}{2}} \, ,
\end{equation}
which for small $k$ gives $\epsilon(0,k)\simeq \hbar sk$ with $s$ given by (\ref{eq:PerturbativeSound}). As the optical lattice is subsonic, for given $\epsilon$, apart from the two solutions with $(k,-k)$, we have another two solutions with complex wave vector, similar to the subsonic region of a black-hole.

The previous results can be used to compute the corrections to the width of the lowest band, which by virtue of Eq. (\ref{eq:effectiveolbdg}) give
\begin{equation}
\Delta^{\rm BdG}_c \simeq \Delta_{c}+\left[2J(1/2)-\alpha_{0}^{(1)}\right] gn_r \, ,
\end{equation}
where $\Delta_{c}=E_L(\tilde{\varepsilon}_{\frac{1}{2},0}-\tilde{\varepsilon}_{0,0})$ is the Schr\"odinger bandwidth and
we have used $\Delta_{c}\gg gn_{r}$, which is true in all the situations considered in the present work. We see that in the limit of negligible interactions, the BdG bandwidth is equal to the linear bandwidth.

\chapter{Numerical methods}\label{app:numerical}

Several numerical methods have been used across the work to integrate both the GP and BdG equations. We now explain in detail those numerical schemes.

\section{Birth of a black hole in an optical lattice}\label{sec:numericalolmethods}

In Chapter \ref{chapter:MELAFO}, we integrate the time-dependent GP equation (\ref{eq:TDGPOL}) to obtain the time evolution of the system. After that, once in the quasi-stationary regime, we solve the resulting effective time-independent BdG equations in order to obtain the Hawking spectrum of the formed black-hole.

\subsection{Time-dependent GP equation} \label{subsec:GPOLAT}

\subsubsection{Crank-Nicolson method} \label{subsec:crnic}

The numerical computation of the time evolution of the GP wave function has been performed using the Crank-Nicolson method, as in Ref. \cite{PhysRevA.76.063605}. The spatial interval $[0,L_{g}]$ is divided into $N+2$ equally spaced points separated by a distance
$\Delta x=L_{g}/(N+1)$, and the time interval $[0,t]$ into steps of size $\Delta t$. Hence, we
write the grid points as
\begin{eqnarray}
x_{j} & = & j\Delta x~j=0,1\ldots N+1\nonumber \, \\
t_{k} & = & k\Delta t~k=0,1\ldots n \, .
\end{eqnarray}
For computational purposes, we work with a dimensionless GP equation so we set units such that $\hbar=m=\xi_0=1$, and rescale the wave function by extracting the factor $\sqrt{n_0}$. The GP equation can be then written as
\begin{eqnarray}\label{eq:numericalGP}
i\frac{\partial\Psi (x,t)}{\partial t}& = & H(x,t)\Psi(x,t) \\
H(x,t) & = & -\frac{1}{2}\frac{\partial^{2}}{\partial x^{2}}+V(x,t)+|\Psi(x,t)|^{2}-1 \nonumber
\end{eqnarray}
where the "$-1$" comes from subtracting the initial chemical potential $\mu_0$ and $V(x,t)$ is the total external time-dependent potential.
Here, $H(x,t)$ plays the role of an effective Hamiltonian. If we define the
spatial vector with the discrete values of the wave function at a given time $t_{k}$ as $\mathbf{\Psi}_{k}$, with
components $\Psi_{k}^{j}=\Psi(x_{j},t_{k})$, and using

\begin{eqnarray}
\Psi\left(x,t+\frac{\Delta t}{2}\right) & = & \frac{\Psi\left(x,t+\Delta t\right)+\Psi(x,t)}{2}+O\left(\Delta t^{2}\right)\nonumber \\
\frac{\partial\Psi}{\partial t}\left(x,t+\frac{\Delta t}{2}\right) & = & \frac{\Psi\left(x,t+\Delta t\right)-\Psi(x,t)}{\Delta t}+O\left(\Delta t^{2}\right) \\
\nonumber \frac{\partial^{2}\Psi}{\partial x^{2}}\left(x,t\right) & = & \frac{\Psi\left(x+\Delta x,t\right)+\Psi\left(x-\Delta x,t\right)-2\Psi(x,t)}{\Delta x^{2}}+O\left(\Delta x^{2}\right),
\end{eqnarray}
we can write, up to second order in $\Delta x$ and $\Delta t$, a
discrete version of Eq. (\ref{eq:numericalGP})
\begin{eqnarray}\label{eq:discreteGP}
i\frac{\mathbf{\Psi}_{k+1}-\mathbf{\Psi}_{k}}{\Delta t} & = & \mathbf{H}_{k+\frac{1}{2}}\frac{\mathbf{\Psi}_{k+1}+\mathbf{\Psi}_{k}}{2}\nonumber \\
\left(\mathbf{H}_{k+\frac{1}{2}}\mathbf{\Psi}\right)^{j} & = & -\frac{\Psi^{j+1}+\Psi^{j-1}-2\Psi^{j}}{2(\Delta x)^{2}}+V_{k+\frac{1}{2}}^{j}\Psi^{j}+|\Psi_{k+\frac{1}{2}}^{j}|^{2}\Psi^{j}-\Psi^{j}\nonumber \\
V_{k+\frac{1}{2}}^{j} & = & V\left(x_{j},t_{k}+\frac{\Delta t}{2}\right) \, .
\end{eqnarray}

The previous equation can be rewritten as a linear system of equations:
\begin{eqnarray}\label{eq:crnicimplicit}
\mathbf{M}_{2}\mathbf{\Psi}_{k+1} & = & \mathbf{M}_{1}\mathbf{\Psi}_{k}\nonumber \\
\mathbf{M}_{1,2} & = & 1\mp i\mathbf{H}_{k+\frac{1}{2}}\frac{\Delta t}{2} \, ,
\end{eqnarray}
where
\begin{equation}\label{eq:mmatrices}
\mathbf{M}_{1,2}=\left[\begin{array}{cccccc}
\ddots & \ddots & \ddots\\
 & \pm A & 1\mp B_{k}^{j} & \pm A\\
 &  & \pm A & 1\mp B_{k}^{j+1} & \pm A\\
 &  &  & \ddots & \ddots & \ddots
\end{array}\right] \, ,
\end{equation}
with
\begin{equation}
A=\frac{i\Delta t}{4\Delta x^{2}},~B_{k}^{j}=i\frac{\Delta t}{2}\left(\frac{1}{\Delta x^{2}}+V_{k+\frac{1}{2}}^{j}+|\Psi_{k+\frac{1}{2}}^{j}|^{2}-1\right) \, .
\end{equation}
The previous method is equivalent to approximate the time evolution operator between $t$ and $t+\Delta t$ by its Cayley's form \cite{Press2007}, given by the unitary operator:
\begin{equation}\label{eq:cayleyform}
U(t+\Delta t, t) \simeq \left[1+iH\left(x,t+\frac{\Delta t}{2}\right)\frac{\Delta t}{2}\right]^{-1}\left[1-iH\left(x,t+\frac{\Delta t}{2}\right)\frac{\Delta t}{2}\right]
\end{equation}
Because we ignore the value of $\Psi_{k+\frac{1}{2}}^{j}$ in the non-linear term,
we use a corrector-predictor method, which consists in performing an additional
step in every time iteration. In the first iteration, we use $\Psi_{k}^{j}$
instead of $\Psi_{k+\frac{1}{2}}^{j}$ in order to obtain
a value $\bar{\Psi}_{k+1}^{j}$. Next, we perform a new iteration
taking $\Psi_{k+\frac{1}{2}}^{j}=\left(\bar{\Psi}_{k+1}^{j}+\Psi_{k}^{j}\right)/2$
to obtain the final value $\Psi_{k+1}^{j}$.

The main advantage of this integration scheme is that the obtention of $\Psi_{k+1}^{j}$
only requires the resolution of a tridiagonal system of equations, which is computationally very efficient when using the Thomas algorithm \cite{Ames1977,Quarteroni2000} to solve it, since in that case the number of operations grows like $N$. Another important advantage of using the Crank-Nicolson method is that it is unconditionally unstable and then we can take higher values of $\Delta x, \Delta t$ than usual.

\subsubsection{Absorbing boundary conditions}\label{subsec:ABC}

We now discuss the implementation of the boundary conditions, which affect to the first and last points of the numerical grid. For instance, periodic boundary conditions are given by:
\begin{equation}\label{eq:PBC}
\Psi_{k}^{0}=\Psi_{k}^{N+1},\,\,~l=0,N+1,
\end{equation}
so the point $N+1$ is identified with the point $0$ and does not play any role on the dynamics. The periodic boundary conditions are simply implemented by writing the first and last rows of the $M$ matrices in Eq. (\ref{eq:mmatrices}) as:
\begin{equation}\label{eq:PBCmmatrices}
\mathbf{M}_{1,2}=\left[\begin{array}{cccc}
1\mp B_{k}^{0} & \pm A & \ldots & \pm A\\
~ & ~ & \ddots & \ddots\\
\pm A & \ldots & \pm A & 1\mp B_{k}^{N}
\end{array}\right] \, ,
\end{equation}
so the $M$ matrices are $(N+1)\times (N+1)$. More relevant boundary conditions for our case are the so-called hard-wall boundary conditions, expressed by the condition
\begin{equation}\label{eq:hardwall}
\Psi_{k}^{l}=0,\,\,~l=0,N+1,
\end{equation}
These boundary conditions are easily implemented by suppressing the first and the last
columns of the $M$ matrices in Eq. (\ref{eq:mmatrices}), arriving to $N \times N$ matrices of the form:
\begin{equation}\label{eq:hardwallcrnic}
\mathbf{M}_{1,2}=\left[\begin{array}{ccc}
1\mp B_{k}^{1} & \pm A\\
 & \ddots & \ddots\\
 & \pm A & 1\mp B_{k}^{N}
\end{array}\right]
\end{equation}
As explained in the main text, we implement a hard-wall boundary condition at $x=0$ in order to simulate the confinement of the potential. Even in the realistic case of Sec. \ref{sec:Gaussian-shaped}, where the left confinement is provided by a finite barrier, this boundary condition is set to truncate the otherwise infinite numerical grid. As explained in Sec. \ref{subsec:caserealistic}, this does not pose any problem since the amplitude of the confining potential is taken sufficiently large so the condensate has exponentially small amplitude within the barrier.

On the other hand, imposing any of the previous boundary conditions at the final point of the grid ($x=L_{g}$)
will induce unwanted reflections because our goal is to simulate a semi-infinite supersonic region. To minimize those spurious reflections, one can use complex absorbing potentials (CAP) at the grid boundaries \cite{Muga2004357}. Other possible method could be the implementation of a diffusive scheme near the boundaries of the grid; we use this method for integrating the GP equation corresponding to the BH laser scenario of Chapter \ref{chapter:BHL}, see Sec. \ref{sec:numericalbhl}. Instead of those methods, we make use of the so-called ABC (Absorbing Boundary Conditions) \cite{PhysRevE.74.037704,PhysRevE.78.026709}. The ABC are based on the linearization of the dispersion relation at the boundary in order to achieve the relation
corresponding to an outgoing plane wave. All the above mentioned methods are very useful because
they not only prevent the artificial reflection of the waves but also permit to reduce the size of the supersonic zone.

The point $x=L_{g}$ is placed in the supersonic zone, where there is no potential.
In addition, we expect that the non-linear term in Eq. (\ref{eq:numericalGP}) is sufficiently small to be neglected. This
means that the effective Hamiltonian $H$ in this region is the usual free (Schr\"odinger)
Hamiltonian and Eq. (\ref{eq:numericalGP}) can be written as
\begin{equation}
i\frac{\partial\Psi}{\partial t}=-\frac{1}{2}\frac{\partial^{2}\Psi}{\partial x^{2}}-\Psi\label{eq:spGP}
\end{equation}
which implies the dispersion relation
\begin{equation}
\omega=\frac{k^{2}}{2}-1
\end{equation}
On the supersonic side and in the quasi-stationary regime, the wave
function is well peaked in momentum space around a value $k_{0}\simeq\sqrt{2E_{\rm min}}$, where $E_{\rm min}$ is the energy of the bottom of the first
conduction band. By linearizing the dispersion relation around $k_{0}$ and expressing this relation
in terms of derivatives, one gets
\begin{equation}\label{eq:eqABC}
i\frac{\partial\Psi}{\partial t}=-ik_{0}\frac{\partial\Psi}{\partial x}-
\left(\frac{k_{0}^{2}}{2}+1\right)\Psi \, .
\end{equation}
We replace the hard-wall boundary condition at $j=N+1$, Eq. (\ref{eq:hardwall}), by the discrete version of Eq. (\ref{eq:eqABC}) at $j=N$. Then there are $N+1$ variables ($\Psi_{k}^{j}, j=1\ldots N+1$) and $N+1$ equations, corresponding to $N$ equations resulting from Eq. (\ref{eq:discreteGP}) for $j=1\ldots N$ and the ABC equation. We can regard the point $x_{N+1}$ as a ghost point because the GP equation is not properly defined there and the ABC (\ref{eq:eqABC}) is the corresponding equation for this point \cite{PhysRevE.78.026709}.
Following these considerations, we can easily implement the ABC condition by adding a new row to the matrices $M$ of Eq. (\ref{eq:hardwallcrnic}), which now are of size $\left(N+1\right)\times\left(N+1\right)$
and of the form
\begin{equation}
\mathbf{M}_{1,2}=\left[\begin{array}{cccc}
\ddots & \ddots\\
 & \pm A & 1\mp B_{k}^{N} & \pm A\\
 & \pm C & 1\pm D & \mp C
\end{array}\right]
\end{equation}
with
\begin{equation}
C=\frac{k_{0}\Delta t}{4\Delta x},~D=i\frac{\Delta t}{2}\left(\frac{k_{0}^{2}}{2}+1\right)
\end{equation}

Due to the finite size of the grid and to the nonzero width of the momentum distribution in the supersonic region, the absorption is not perfect. Nevertheless, we have found that, in practice, the small spurious reflections have no effect on the final results.

\subsection{Numerical integration of the quasi-stationary BdG equations}\label{subsec:BdGNum}

In order to obtain the scattering matrix in the quasi-stationary regime, we need to solve the system of equations (\ref{eq:scatteringlinearsystem}). The numerical computation of the fundamental matrix $F(x)$ presents some problems in the optical lattice as one of the eigenvalues of the fundamental matrix is exponentially large, corresponding to one of the local Bloch wave solutions with complex wave vector (see discussion at the end of Appendix \ref{app:nonlinearol}). For sufficiently large optical lattices, this makes that the fundamental matrix becomes singular within computer's relative accuracy, spoiling the calculation. In order to deal with this problem, we consider two methods: QR decomposition \cite{Slevin2014} and the Global Matrix method \cite{Lowe1995}. Both methods are based on the following property of the transfer matrix of Eq. (\ref{eq:transfermatrix}):
\begin{equation}\label{eq:transferdecom}
T(x_{n+1},x_1)=\prod_{i=1}^{n}T_i,~T_i\equiv T(x_{i+1},x_{i})
\end{equation}
with $x_1<x_2<\ldots<x_{n+1}$, i.e., the total transfer matrix can be decomposed as the product of intermediate transfer matrices. In our problem, we have to compute the transfer matrix $T(x_d,x_u)$ between the two asymptotic regions so $x_1=x_u$ and $x_{n+1}=x_d$. The two techniques here considered try to compute the previous product without finally arriving at a numerical singular matrix. The matrices $T_i$ are obtained by integrating the effective BdG equations (\ref{eq:effectiveTIBdG}) on top of the quasi-stationary GP wave function between the intermediate points $x_{i}$ and $x_{i+1}$. The distance between consecutive points $x_{i+1}-x_{i}$ cannot be taken very large since in that case the local transfer matrices $T_i$ also become numerically singular. For the integration of the BdG equations, we use a Runge-Kutta method of fourth-order where the numerical grid is chosen such it coincides with the numerical grid corresponding to the quasi-stationary GP wave function previously computed.

We have checked that both methods work correctly by comparing the results with the brute force result, in which the total transfer matrix is computed by direct multiplication of Eq. (\ref{eq:transferdecom}) after increasing the accuracy to 1000 digits. Of course, brute force represents a much slower method than the chosen methods, motivating their use. We have also checked the accuracy of the previous methods to obtain the scattering matrix by confirming the smallness of the elements of the residual matrix $\Delta \eta\equiv S^{\dagger}\eta S-\eta$, which measures the pseudo-unitarity of the $S$ matrix. Finally, we wish to note that there are no significant differences in the computational speed of both techniques so they represent good choices to do the job.

\subsubsection{QR decomposition}

As well known from linear algebra, a given matrix $M$ can be decomposed as $M=QR$ with $Q$ orthogonal and $R$ an upper-diagonal matrix. This is called the QR decomposition. The point is that we can perform the QR decomposition at every step of the product of Eq. (\ref{eq:transferdecom}) in order to compute the QR decomposition of the total transfer matrix. If $T(x_{k},x_{1})=Q_kR_k$, with $k<n+1$, we have $T(x_{k+1},x_{1})=T_{k+1}T(x_{k},x_{1})=T_{k+1}Q_kR_k$ and we can obtain the QR decomposition of $T(x_{k+1},x_{1})=Q_{k+1}R_{k+1}$ from the following QR decomposition $T_{k+1}Q_k=\tilde{Q}_k\tilde{R}_k$, which gives $Q_{k+1}=\tilde{Q}_k$ and $R_{k+1}=\tilde{R}_kR_{k}$. We can start this recursive process from $k=2$ until reaching $k=n$.

Once we get the $Q,R$ matrices of the global transfer matrix, $T(x_d,x_u)=QR$, we can compute the fundamental matrix in the supersonic region from that of the subsonic region as $F(x_d)=QRF(x_u)$. If we directly multiply $QR$ we would get a singular matrix; instead of that, we first diagonalize the $R$ matrix $R=MDM^{-1}$ with $D$ a diagonal matrix (this diagonalization does not present any associated numerical problem since $R$ is an upper-diagonal matrix). As $Q$ is orthogonal, $\det R=\det D=1$. The matrix $D$ has two eigenvalues that are exponentially large (small), which we denote as $\lambda_{+,-}$ (associated to the complex Bloch wave vectors) and other two eigenvalues of order $\sim 1$ (associated to the propagating Bloch wave vectors). The numerical problem arises from mixing the columns of $D$, since $\lambda_{+}$ crushes the eigenvalues of order $1$ as they are so relatively small that they fall under computers relative accuracy.

For fixing this numerical issue, we first rearrange $D$ so the column corresponding to $\lambda_+$ is the first one. We then choose $F(x_u)$ such $F(x_u)=M I_+$, where $I_+$ is the $4\times 4$ identity but with the first diagonal element replaced by $\lambda^{-1}_+$. As a consequence, we obtain $F(x_d)=QM\tilde{D}$ where the matrix $\tilde{D}$ is the same as $D$ but with the first diagonal element replaced by $1$. We note that, in this form, the columns of $F(x_{u,d})$ are not mixed and the Wronskian is conserved so the possible singular behavior disappears. However, for extremely long optical lattices, a problem of overflow can appear due to the exponentially large value of $\lambda_+$. In that limit case, there is no problem since we make use of the Global Matrix method.

\subsubsection{Global Matrix method}

The previous method just rearranges the product of Eq. (\ref{eq:transferdecom}) in a clever way in order to obtain a non-singular matrix. This method avoids the explicit computation of the total transfer matrix; instead of that, we simply rewrite Eq. (\ref{eq:scatteringlinearsystem}) as a $(4n+4)\times(4n+4)$ system of equations by also considering the matching of the solutions at the intermediate points $x=x_{i}$, $i=2,3,\ldots,n$

\begin{eqnarray}\label{eq:GlobalMatrix}
\nonumber d&=&A\cdot b\\
b&=&\left[b_{u-\rm{out}},b_{\rm{ev}},b_{1},b_{2},\ldots,b_{4n-1},b_{4n},b_{d1-\rm{out}},b_{d2-\rm{out}}\right]^{T}\\
\nonumber d&=&b_{u-\rm{in}}\left[\begin{array}{c}\sigma_{u-\rm{in},\omega}\left(x_u\right)\\ 0\\ \vdots \\ 0  \end{array}\right]+b_{d1-\rm{in}}\left[\begin{array}{c}0\\ \vdots \\ 0 \\   \sigma_{d1-\rm{in},\omega}\left(x_d\right)\end{array}\right]+b_{d2-\rm{in}}\left[\begin{array}{c}0\\ \vdots \\ 0 \\ \sigma_{d2-\rm{in},\omega}\left(x_d\right)\end{array}\right],\\
\nonumber A&=&\left[\begin{array}{cccccccc}-\sigma_{u-\rm{out},\omega}&-\sigma_{\rm{ev},\omega}&I&0&0&\ldots&0&0\\
0&0&T_1&-I&0&\ldots&0&0\\
0&0&0&T_2&-I&0&\ldots&0\\
~&~&~&~&\ddots&\ddots&\ddots&~\\
0&0&0&0&T_{n-1}&-I&0&0\\
0&0&0&0&0&T_n&-\sigma_{d1-\rm{out},\omega}&-\sigma_{d2-\rm{out},\omega}\end{array}\right]
\end{eqnarray}
where the spatial dependence of the vectors $\sigma$ inside the matrix $A$ is understood. In this way, we prevent the appearance of singular behavior in the total transfer matrix by solving at one time the {\it global} problem.

\section{Time evolution of the black-hole laser}\label{sec:numericalbhl}

We discuss here the numerical method used to integrate the time dependent GP equation (\ref{eq:GPNumerical}) which consists of a time-splitting finite-difference (TSFD) scheme \cite{Hua2012,Bao2013}. We start by rewriting the corresponding GP equation as:
\begin{eqnarray}\label{eq:GPsplitting}
\nonumber i\frac{\partial \Psi(x,t)}{\partial t}&=&H(x,t)\Psi(x,t)\\
&=&\left[H_0(x,t)+H_C(x,t)\right]\Psi(x,t)
\end{eqnarray}
where $H_0$ is the kinetic term and $H_C=g(x)(|\Psi(x,t)|^2-1)$ is the term associated with the contact interaction. We divide the spatial interval $[-L_g/2,L_g/2]$ in $N+1$ steps of size $\Delta x=L_g/(N+1)$ and impose periodic boundary conditions. The time interval $[0,t]$ is divided in $n$ steps of size $\Delta t=t/n$. Within this scheme, the (non-linear) time evolution operator from $t_k=k\Delta t$ to $t_{k+1}=(k+1)\Delta t$ is computed using a split-step approximation
\begin{equation}\label{eq:TSFD}
U_k\simeq e^{-iH_0\frac{\Delta t}{2}}e^{-iH^k_C\Delta t}e^{-iH_0\frac{\Delta t}{2}}
\end{equation}
where $H^k_C=g(x)(|\tilde{\Psi}(x,t_k)|^2-1)$, $\tilde{\Psi}(x,t_k)=e^{-iH_0\frac{\Delta t}{2}}\Psi(x,t_k)$. We refer to Sec. \ref{subsec:LDAjust} for a detailed theoretical discussion on the split-step method. The total evolution operator then reads
\begin{eqnarray}\label{eq:TimeOperator}
\Psi(x,t)&=&U(t)\Psi(x,0)\\
\nonumber U(t)&=&e^{-iH_0\frac{\Delta t}{2}}e^{-iH^{n-1}_C\Delta t}\left(\prod_{k=0}^{n-2}e^{-iH_0\Delta t}e^{-iH^k_C\Delta t}\right)e^{-iH_0\frac{\Delta t}{2}}
\end{eqnarray}

In the TSFD, instead of using the usual Fourier transform to compute the kinetic evolution operator $e^{-iH_0\Delta t}$, we use the Crank-Nicolson method discussed in Sec. \ref{subsec:crnic}. The corresponding Cayley's form [see Eq. (\ref{eq:cayleyform}] is:
\begin{equation}\label{eq:cayleyfreeform}
e^{-iH_0\Delta t} \simeq \left(1+iH_0\frac{\Delta t}{2}\right)^{-1}\left(1-iH_0\frac{\Delta t}{2}\right)
\end{equation}
When the previous operator acts on a given wave function $\Psi$, we obtain a new wave function $\Psi'=e^{-iH_0\Delta t}\Psi$. In order to compute $\Psi'$, we rewrite the previous equation in the same implicit form of Eq. (\ref{eq:crnicimplicit}):
\begin{eqnarray}
\mathbf{M}_{2}\mathbf{\Psi}' & = & \mathbf{M}_{1}\mathbf{\Psi}'\nonumber \\
\mathbf{M}_{1,2} & = & 1\mp i\mathbf{H}_{0}\frac{\Delta t}{2} \, ,
\end{eqnarray}
$\mathbf{\Psi},\mathbf{\Psi}'$ being the spatial vectors with the discrete values of the wave functions $\Psi,\Psi'$ and $\mathbf{H}_{0}$ the discrete matrix corresponding to the kinetic operator. Following the results of Sec. \ref{subsec:crnic}, we find:
\begin{equation}\label{eq:mmatricesfree}
\mathbf{M}_{1,2}=\left[\begin{array}{cccccc}
1\mp 2A & \pm A & ~ & \ldots & \ldots & \pm A\\
\pm A & 1\mp 2A & \pm A & ~ & ~ & ~\\
 &  & \ddots & \ddots & \ddots & \\
 &  &  & \pm A & 1\mp 2A & \pm A\\
 \pm A & \ldots & \ldots & ~ &  \pm A & 1\mp 2A
\end{array}\right] \, ,
\end{equation}
where we have taken into account that we are working with periodic boundary conditions, see Eq. (\ref{eq:PBCmmatrices}). As explained in Sec. \ref{subsec:crnic}, the Crank-Nicolson method is unconditionally unstable and very cheap from the computational point of view. Importantly, we note that the matrix $H_D$ is constant in time, so we can develop an optimized Thomas algorithm for the associated Crank-Nicolson scheme. In particular, the specific Thomas algorithm developed in this work is able to defeat the standard linear solver of MATLAB! Another advantage is that this scheme keeps working even when the kinetic term involves derivatives with a spatially dependent coefficient, see the discussion below.

Since we are focusing on long time dynamics, we add a diffusive term at the boundaries of the system in order to absorb the emitted waves and solitons so they do not return to the central supersonic region and disturb the analysis of the simulations. Specifically, we replace $H$ by $H_T=H+iH_A$ in Eq. (\ref{eq:GPsplitting}), where the absorbing term $H_A$ is given by
\begin{equation}\label{eq:diffusion}
H_A(x,t)=G(x)\left(D_0\partial_{x}^{2}e^{-ivx}-F_0[|\Psi(x,t)|^2-1]\right)~.
\end{equation}
The function $G(x)$ is a function localized on the boundary of the grid (typically a Gaussian distribution) and it is taken sufficiently smooth in order to avoid reflections. The first term of the r.h.s of Eq. (\ref{eq:diffusion}) is the diffusive term that absorbs the Fourier components outside the plane wave $e^{ivx}$ while the second term is a source term that equilibrates the loss of atoms due to the diffusive term. We note that the non-linear operator $H_A$ vanishes when acting on the homogeneous plane wave $e^{ivx}$.

The new diffusive GP equation can still be integrated using the mentioned TSFD method by just replacing $H_0$ and $H_C$ by $H_D$ and $H_X$ in Eqs. (\ref{eq:GPsplitting})-(\ref{eq:TimeOperator}), where $H_D$ is the term that contains the derivatives (the kinetic and diffusive terms) and $H_X$ is the term that takes into account the non-linearities (the interaction and the source term), i.e.,
\begin{eqnarray}\label{eq:DiffusiveOperators}
H_D&=&-\frac{1}{2}\frac{\partial^2}{\partial x^2}+iD_0G(x)\frac{\partial^2}{\partial x^2}e^{-ivx}\\
\nonumber H_X&=&[g(x)-iF_0G(x)][|\Psi(x,t)|^2-1]
\end{eqnarray}

The only modification that we have to implement in the TSFD method is the introduction of a predictor-corrector method, similar to that of Sec. \ref{subsec:crnic}, in order to improve the estimation of the non-linear term in $H_X$ due to the non-conserving norm of the source term. At every step, we perform a first iteration to compute $\tilde{\Psi}'(x,t_k)=e^{-iH^k_X\Delta t}\tilde{\Psi}(x,t_k)$ where $\tilde{\Psi}(x,t_k)$ is now $\tilde{\Psi}(x,t_k)=e^{-iH_D\frac{\Delta t}{2}}\Psi(x,t_k)$ and $H^k_X=[g(x)-iF_0G(x)](|\tilde{\Psi}(x,t_k)|^2-1)$. We replace $\tilde{\Psi}(x,t_k)$ by $(\tilde{\Psi}'(x,t_k)+\tilde{\Psi}(x,t_k))/2$ in $H^k_X$ and we perform a second iteration to obtain the final value of $\tilde{\Psi}'(x,t_k)$.

We have taken the values of $F_0,D_0$ in the range $1-10$ and we have not seen any significant difference in the result of the simulations. We have checked that the diffusive scheme does not introduce any spurious result by integrating the GP equation without diffusion in a larger spatial grid. Finally, we comment that the observation of the CES regime of Sec. \ref{subsec:CES} can require, in the most unstable cases, to place the boundaries sufficiently far away in order to avoid reflections that spoil the perfect periodicity of soliton emission.

\chapter{Parametrization of the scattering matrix and correlation functions}\label{app:parametrization}
\chaptermark{Parametrization}

\section{Parametrization of the scattering matrix}\label{sec:parametrizationsmatrix}

We discuss in this Appendix the parameters needed to compute the correlation functions defined in Sec. \ref{chapter:CSGPH}. A related discussion appeared in Ref. \cite{Busch2014}. The scattering matrix $S$ relates the ``out'' scattering states with the ``in'' scattering states through Eq. (\ref{eq:inoutmodesrelation}) and, as explained in the main text, it satisfies the pseudounitary relation (\ref{eq:pseudounitarity}) which means that $S\in U(2,1)$. From there, it follows that $S^{-1}=\eta S^{\dagger} \eta$ and, using standard linear algebra, it can be proven that:
\begin{equation}
S^*_{ij}=\frac{\eta_{ii} m_{ij} \eta_{jj}}{\textrm{det}~S}
\end{equation}
where $m_{ij}$ is the minor associated to the $S$ matrix element $S_{ij}$. In order to simplify the notation, in this section we relabel the indices $u,d1,d2$ as $1,2,3$ in order to match the matrix indices ordering. With the help of the previous results, it can be proven that any $U(2,1)$ matrix satisfies:
\begin{eqnarray}\label{eq:param}
BSA=S_{\rm PA}=&\left[\begin{array}{ccc} \det S &0&0\\
0&S^*_{33}&N\\
0&N&S_{33}\end{array}\right]&\\
\nonumber B\equiv \frac{1}{N}&\left[\begin{array}{ccc} -S_{23}&S_{13}&0\\
S^*_{13}&S^*_{23}&0\\
0&0&N\end{array}\right]&\\
\nonumber A\equiv \frac{1}{N}&\left[\begin{array}{ccc} -S_{32}&S^*_{31}&0\\
S_{31}&S^*_{32}&0\\
0&0&N\end{array}\right]&
\end{eqnarray}
where $N=\sqrt{|S_{13}|^2+|S_{23}|^2}=\sqrt{|S_{31}|^2+|S_{32}|^2}=\sqrt{|S_{33}|^2-1}$. We see that the matrices $A,B$ are unitary. Inverting the previous relation, we obtain:

\begin{equation}\label{eq:Sparam}
S=B^{\dagger}S_{\rm PA}A^{\dagger}
\end{equation}

Equations (\ref{eq:param}),(\ref{eq:Sparam}) show explicitly that we only need $9$ real parameters to parametrize the $S$ matrix: $|S_{31}|,|S_{13}|,|S_{33}|,G,\phi_{13},\phi_{23},\phi_{31},\phi_{32},\phi_{33}$, with $\phi_{ij}$ the phase of the element $S_{ij}$, $S_{ij}=|S_{ij}|e^{i\phi_{ij}}$, and $G$ the phase of the determinant of the scattering matrix, $\det S=e^{iG}$. That is, we use $3$ amplitudes and $6$ phases to characterize the complete scattering matrix. Interestingly, the matrix $S_{\rm PA}$ shows explicitly that Hawking radiation acts as a non-degenerate parametric amplifier \cite{Walls2008}.

\section{Parametrization of correlations for incoming incoherent states}\label{sec:parametrizationincoherent}
\sectionmark{Correlations for incoming incoherent states}
As for practical purposes we focus only on incoherent incoming Gaussian states, we do not need all $9$ parameters for computing the physical quantities considered in this work. In that case, we consider a more convenient way to write the $S$ matrix:
\begin{eqnarray}\label{eq:inoutphases}
S&=&U^{\dagger}_{\rm{out}}\tilde{S}U_{\rm{in}}\\
\nonumber U_{k}&=&{\rm diag}[e^{i\delta_{1-k}},e^{i\delta_{2-k}},e^{i\delta_{3-k}}],~k=\rm{in,out},
\end{eqnarray}
since we can fix the elements of $U_{\rm{in,out}}$ in such a way that they absorb the phases $\phi_{31},\phi_{32},\phi_{13},\phi_{23},\phi_{33}$ so $\tilde{S}_{31},\tilde{S}_{32},\tilde{S}_{13},\tilde{S}_{23},\tilde{S}_{33}$ become purely real. This means that the matrix $\tilde{S}$ is characterized by only $4$ parameters, $|S_{31}|,|S_{13}|,|S_{33}|, \tilde{G}$, with
\begin{equation}
\det \tilde{S}=e^{i\tilde{G}}=\det S~e^{-i(\phi_{13}+\phi_{23}+\phi_{13}+\phi_{23}-\phi_{33})}.
\end{equation}

We note that the CS violations of Eq. (\ref{eq:CSGamma}) are invariant under phase transformations of the ``out'' states. It can be also shown that the GPH function is invariant under these transformations, see Eq. (\ref{eq:GPHcomplete}). As we are considering incoherent incoming states, all the quantities are also invariant under phase transformations of the incoming modes. From Eq. (\ref{eq:inoutphases}), and taking into account these observations, we conclude that all the requested quantities depend only on the matrix $\tilde{S}$, which is characterized by $4$ parameters, three amplitudes, $|S_{31}|,|S_{13}|,|S_{33}|$, and one phase, $\tilde{G}$. Using Eqs. (\ref{eq:pseudounitarity}),(\ref{eq:Sparam}) we can write $\tilde{S}$ as:

\begin{equation}\label{eq:definitiveparametrization}
\tilde{S}=\left[\begin{array}{ccc} -\sin\alpha&\cos\alpha&0\\
\cos\alpha&\sin\alpha&0\\
0&0&1\end{array}\right]\left[\begin{array}{ccc} e^{i\tilde{G}} &0&0\\
0&\cosh \gamma&\sinh \gamma\\
0&\sinh \gamma&\cosh \gamma\end{array}\right]\left[\begin{array}{ccc} -\sin\beta&\cos\beta&0\\
\cos\beta&\sin\beta&0\\
0&0&1\end{array}\right]
\end{equation}

with $|S_{33}|=\cosh \gamma, |S_{13}|=\sinh \gamma \cos\alpha, |S_{31}|=\sinh \gamma \cos\beta$. Thus, we only need $4$ parameters in order to parametrize the scattering matrix for the calculations of this work: $\alpha,\beta,\gamma, \tilde{G}$.  We note that this parametrization differs slightly from that used in Ref. \cite{Busch2014}, where $4$ amplitudes were used to characterize the state of the system. Nevertheless, that election is equivalent to that used here.

Finally, taking into account that the state of the system $\hat{\rho}$ is characterized by only $3$ numbers, $n_i$ [with $i=1,2,3$, see Eq. (\ref{eq:incoherent})], we have that the whole problem is completely determined by $7$ parameters, $4$ arising from the $S$ matrix and $3$ arising from the specification of the incoherent incoming Gaussian state.

\chapter{Black-hole laser solutions}\label{app:technicalBH}

We briefly present here the technical details of the theoretical model used for studying BH lasers in Sec. \ref{sec:BHLaserintro}. We first start by showing how to compute the dynamical instability frequencies and after that, we compute explicitly the family of the sh-sh solutions. For a more detailed discussion, see Refs. \cite{Michel2013,Michel2015}. We follow the same notation of the main text.

\section{Dynamical instabilities}\label{sec:technicalinstabilities}

We compute the BdG solutions associated to the homogeneous plane wave solution of Eq. (\ref{eq:stationaryhomogeneousplanewave}). For that purpose, we divide the system in three regions: the upstream subsonic region ($x<-X/2$), labeled here as $u$, the supersonic central region ($|x|<X/2$), labeled as int, and the downstream subsonic region ($x>X/2$), labeled as $d$. In each region, the associated BdG equation is homogeneous so the solutions for a given $\omega$ are linear combination of plane waves with wave vector $k$ obtained from the dispersion relation (\ref{eq:dispersionrelationBHL}).

We first consider the case where $\omega$ is real and positive. In the asymptotic subsonic regions, we have two propagating modes and two real exponential type solutions, one of them is an evanescent wave and the other one an exploding wave, which we discard as it is unphysical. Specifically, in the upstream (downstream) region the solution with positive (negative) group velocity is an incoming scattering channel and that with negative (positive) group velocity represents an outgoing scattering channel; at the same time, the evanescent exponential solution corresponds to that with negative (positive) imaginary part of the complex wave vector. In the central supersonic region, for $\omega<\omega_{\rm max}$, the four plane wave solutions are propagating. For $\omega>\omega_{\rm max}$, only two modes are propagating. In a similar way to Appendix \ref{app:SMatrixBehav}, we obtain the global scattering states by matching the supersonic solutions with the subsonic solutions at the boundaries of the central region, $x=\pm X/2$. We can further simplify the calculations by removing the phase of the condensate in the plane wave spinors of Eq. (\ref{eq:PlaneWaveSpinors}):
\begin{equation}
s^r_{a,\omega}(x)\equiv\left[\begin{array}{c}u^r_{a,\omega}(x)\\
v^r_{a,\omega}(x)
\end{array}\right],~s_{a,\omega}(x)=\left[\begin{array}{c}e^{ivx}u^r_{a,\omega}(x)\\
e^{-ivx}v^r_{a,\omega}(x)
\end{array}\right]
\end{equation}
Setting $x_u=-\frac{X}{2}$ and $x_d=\frac{X}{2}$, we can write the matching equations for the previous spinors and their derivatives as:
\begin{eqnarray}\label{eq:BHBdGmatching}
\nonumber b_{u-\rm{in}}\sigma^r_{u-\rm{in},\omega}\left(x_u\right)&+&b_{u-\rm{out}}\sigma^r_{u-\rm{out},\omega}\left(x_u\right)
+b_{u-\rm{ev}}\sigma^r_{u-\rm{ev},\omega}\left(x_u\right)=\sum_{j=1}^{4}b_{\rm{int},j}\sigma^r_{\rm{int},\omega}\left(x_u\right)\\
\sum_{j=1}^{4}b_{j}\sigma^r_{i,\omega}\left(x_d\right)&=&b_{d-\rm{in}}\sigma^r_{d-\rm{in},\omega}\left(x_d\right)\\
\nonumber &+&b_{d-\rm{out}}\sigma^r_{d-\rm{out},\omega}\left(x_d\right)+b_{d-\rm{ev}}\sigma^r_{d-\rm{ev},\omega}\left(x_d\right)
\end{eqnarray}
where the four-component vectors $\sigma^r_{a,\omega}$ are the same as the $\sigma_{a,\omega}$ defined after Eq. (\ref{eq:udmatching}) but replacing $s$ by $s^r$. We have $10$ amplitudes and $8$ restrictions so there are only $2$ degrees of freedom which, as in BH configurations, we can choose as the amplitudes of the ``in" scattering channels. Writing Eq. (\ref{eq:BHBdGmatching}) as a linear system of $8$ equations gives
\begin{eqnarray}\label{eq:BHBdGlinearsystem}
\nonumber d&=&\tilde{A}\cdot b\\
b&=&\left[b_{u-\rm{out}},b_{u-\rm{ev}},b_{\rm{int},1},b_{\rm{int},2},b_{\rm{int},3},b_{\rm{int},4},b_{d-\rm{out}},b_{d-\rm{ev}}\right]^{T}\\
\nonumber d&=&b_{u-\rm{in}}\left[\begin{array}{c}\sigma^r_{u-\rm{in},\omega}\left(x_u\right)\\ 0 \end{array}\right]+b_{d-\rm{in}}\left[\begin{array}{c}0\\ \sigma^r_{d-\rm{in},\omega}\left(x_d\right)\end{array}\right],\\
\nonumber \tilde{A}&=&\left[\begin{array}{ccccc}-\sigma^r_{u-\rm{out},\omega}\left(x_u\right)&-\sigma^r_{\rm{ev},\omega}\left(x_u\right)&F^r(x_u)&0&0\\ 0&0&F^r(x_d)&-\sigma^r_{d-\rm{out},\omega}\left(x_d\right)&-\sigma^r_{d-\rm{ev},\omega}\left(x_d\right)\end{array}\right]
\end{eqnarray}
$F^r(x)$ being now the fundamental matrix whose columns are the four independent plane wave solutions $\sigma^r_{\rm{int},i}(x)$ in the scattering region. Considering again Cramer's rule (\ref{eq:Cramer}), it can be shown that all the previous amplitudes are a linear function of the amplitude of the ``in'' modes. From the previous linear system, we can compute the coefficients of all the plane waves and obtain the global scattering state.

We can extend now the previous calculation to complex frequencies $\gamma=\omega+i\Gamma$. We are looking for dynamical instabilities so we only consider the case $\Gamma>0$. First, we study the situation in the subsonic regions. Imagine that we add a small positive imaginary part $\Delta \Gamma$ to a purely real frequency $\omega$. Then, the real wave vector of the propagating modes is modified by an amount $\Delta k =i\Delta \Gamma/w_{a}\left(\omega\right)$ with $w_{a}$ the corresponding group velocity. For the upstream (downstream) regions, only solutions with negative (positive) imaginary part of $k$ are valid since they have to be asymptotically bounded. We conclude that only the outgoing modes satisfy this criterion. We can extend the previous argument to the whole upper half of the complex frequency plane since for arbitrary complex frequency the wave vector is also complex and thus, the wave vector is confined by continuity to one semi-plane of the complex wave vector plane. Consequently, in the analysis of dynamical instabilities, we only take into account the analytical prolongation of the outgoing modes and we discard that of the incoming modes as they give rise to unbound solutions. The same reasoning holds for the solutions with complex wave vectors previously considered and then, we only take the analytical prolongation of the evanescent waves. For the supersonic solutions, we do not care about their specific character since they are restricted to a finite spatial region. We finally remark that the expression given for the BdG plane wave spinors, Eq. (\ref{eq:PlaneWaveSpinors}), is also valid for complex wave vector and frequency and moreover, as they are not scattering states, the normalization factors are no longer needed, simplifying in this way the calculations.

After the previous considerations, we see that there are $2$ amplitudes (the prolongations of the outgoing and evanescent modes) in each subsonic region and $4$ amplitudes in the supersonic region, making $8$ total degrees of freedom. Combined with the $8$ restrictions of the matching equations, we get the homogeneous version of Eq. (\ref{eq:BHBdGlinearsystem}):
\begin{equation}\label{eq:BHBdGlinearunstable}
0=\tilde{A}(\gamma)\cdot b(\gamma)
\end{equation}
where $\tilde{A}(\gamma),b(\gamma)$ are now the analytic prolongation for complex frequency $\gamma$ of $\tilde{A},b$. The previous homogeneous system only presents non-trivial solutions whenever
\begin{equation}\label{eq:BHBdGdeterminantunstable}
\det~\tilde{A}(\gamma)=0
\end{equation}
The values of the complex frequencies of the dynamically unstable modes are obtained from the previous equation. We note that we can also find solutions with $\Gamma<0$ to the previous equation, which correspond to quasinormal modes (QNMs). They are analytical prolongations to the lower half of the complex frequency plane of scattering states involving only outgoing and evanescent waves. We note that such solutions are unphysical as they are unbounded. An interesting point is that, taking into account Cramer's rule (\ref{eq:Cramer}), the QNMs or instabilities appear as poles of the scattering coefficients $b_i$ of Eq. (\ref{eq:BHBdGlinearsystem}). The previous arguments can be translated to resonant BH configurations, explaining the appearance of resonant peaks in the Hawking spectrum \cite{Zapata2011}.

\section{Stationary solutions}\label{sec:technicalstationary}

We compute now the different non-linear stationary solutions to the GP equation for the BH laser configuration, using the techniques developed in Sec. \ref{app:1DGP}. Specifically, we only consider those asymptotically matching the plane wave $e^{ivx}$, as explained in the main text. In each homogeneous region, we can use the potential amplitude $W(A)$ of Eq. (\ref{eq:GP-potential}) to integrate the equation for the amplitude:
\begin{equation}\label{eq:MechanicalEnergyConservation}
\frac{A'^2}{2}+\frac{v^2}{2A^2}-g_i\frac{A^4}{2}+\mu_i A^2=E_i
\end{equation}
where $E_i$ is the conserved amplitude energy for each $i=1,2$ region and $\mu_i\equiv\mu-V_i$, with $V_i$ the constant value of the external potential in the region $i$. We recall that $i=1$ stands for the subsonic region while $i=2$ does it for the supersonic one. As the solutions asymptotically match the plane wave $e^{ivx}$ in the subsonic regions, the current is fixed to $J=v$. In our units, $g_i=c_i^2$ and $\mu_i=g_i+v^2/2$; in particular, $g_1=1$ and $\mu_1=1+v^2/2$. Similar to Eq. (\ref{eq:subsonicfictionenergy}), we find that $E_1=1/2+v^2$.

From the analysis of Appendix \ref{app:1DGP}, we conclude that two choices exist for the solution in the subsonic regions: the shadow soliton, with amplitude $A(x)>1$, and the soliton solution, with amplitude $A(x)<1$. The homogeneous plane wave solution is discarded as it can only give rise to the trivial global homogeneous plane wave solution. As argued in the main text, the solutions with lower energy are those with higher amplitude, which implies taking the shadow soliton solution in both subsonic regions. Because of this reason, we denote this kind of solutions as the sh-sh solutions. Other type of solutions are the sol-sh solutions (soliton solution for $x\rightarrow -\infty$ and shadow solution for $x\rightarrow \infty$), the sh-sol solutions (the inverse situation to the sh-sol solutions) and the sol-sol solutions (with two soliton solutions in the subsonic regions), but they all represent higher energy families of solutions.

In the internal region $|x|<X/2$, the GP wave function of the sh-sh solution can only be an oscillating elliptic solution of the form of Eq. (\ref{eq:ellipticdensity}) because a blowing solution is forbidden as it monotonically increases. Joining all the pieces, we find:
\begin{eqnarray}\label{eq:groundnonlinearstates}
\nonumber \rho(x)&=&v^2+(1-v^2)\coth^2\left[\sqrt{1-v^2}(\delta x-L/2-x)\right],~ x<-\frac{X}{2}  \\
\nonumber \rho(x)&=&\rho_1+(\rho_2-\rho_1)\text{sn}^2(c_2\sqrt{\rho_3-\rho_1}x+(n+1)K(\nu),\nu),~\nu=\frac{\rho_2-\rho_1}{\rho_3-\rho_1},~|x|<X/2\\
\nonumber \rho(x)&=&v^2+(1-v^2)\coth^2\left[\sqrt{1-v^2}(\delta x-L/2+x)\right],~ x>\frac{X}{2}\\
\theta(x)&=&\int_0^x\mathrm{d}x'~\frac{v}{\rho(x')}
\end{eqnarray}
with $\rho(x)=A^2(x)$ the density and $\theta(x)$ the phase of the wave function, $\Psi_0(x)=A(x)e^{i\theta(x)}$, which can be obtained explicitly after following the calculations of Appendix \ref{app:1DGP}. The number $n=0,1,2\ldots$ characterizes the solution and gives the number of periods inside the supersonic region. The parameters $0<\rho_1<\rho_2<\rho_3$, in analogy to Eq. (\ref{eq:roots}), are obtained from the zeros of the equation $W(A)=E_2$ in the supersonic region, which amounts to obtain the roots of the polynomial $g_2\rho^3-2\mu_2\rho^2+2E_2\rho-v^2=0$.


The value of $E_2$ is obtained by imposing the continuity of the wave function and of its derivative at $x=\pm X/2$. On one hand, after using Eq. (\ref{eq:MechanicalEnergyConservation}), we get:
\begin{equation}\label{eq:nonlinearmatching}
\rho_m\equiv=n(-X/2)=n(X/2)=1+\sqrt{1-2\frac{E_1-E_2}{1-c^2_2}}
\end{equation}
On the other hand, using Eq. (\ref{eq:groundnonlinearstates}), one finds that the final values of $\delta x$ and $E_2$ are determined by:
\begin{eqnarray}\label{eq:constantequation}
\delta x&=&\frac{\coth^{-1}\left(\sqrt{\frac{\rho_m-v^2}{1-v^2}}\right)}{\sqrt{1-v^2}}\\
\nonumber \text{sn}^{-1}\left(\sqrt{\frac{\rho_m-\rho_1}{\rho_2-\rho_1}},\nu \right)&=&(n+1)K(\nu)-c_2\sqrt{\rho_3-\rho_1}\frac{X}{2}
\end{eqnarray}
The last equation is an implicit equation for $E_2$ as $\rho_m$ and the values of the three roots $\rho_i$ depend on $E_2$. As $K(\nu)$ is an increasing function of $\nu$, the minimum length at which a new solution appears is found in the limit $\nu\rightarrow0$ where the elliptic function reduces to $\text{sn}(u,\nu)\backsimeq\sin(u)$. From this property, it can be easily shown that the minimum length at which the $n$ non-linear solution appears coincides with the onset of the dynamical instability at $X=X_n$ as given by Eq. (\ref{Eq:UnstableLength}). The same result arises from a linear analysis \cite{Michel2013}.

\chapter{Movies}\label{app:Movies}

In this Appendix, we summarize the system parameters used for the Movies that are discussed in this work. In all panels, the time-evolving spatial density profile is shown as a thin blue line.
\begin{itemize}
 \item \href{https://www.youtube.com/watch?v=JyLDGYGep-I}{Movie 1}: $v=0.75$, $c_2=0.3$ and $X=2$ satisfying $X_0(v,c_2)<X<X_{1/2}(v,c_2)$. The thick black line shows the $n=0$ nonlinear stationary solution.
 \item \href{https://www.youtube.com/watch?v=OBHGV2xQ9Fo}{Movie 2}: Same parameters as Movie 1. Different initial noise.
 \item \href{https://www.youtube.com/watch?v=F681IomByYw}{Movie 3}: $v=0.75,c_2=0.5,X=5$, with $X_{1/2}(v,c_2)<X<X_1(v,c_2)$. We observe the oscillating behavior during the growth the unstable mode.
 \item \href{https://www.youtube.com/watch?v=SsC5OlM1dcc}{Movie 4}: $v=0.75,c_2=0.6,X=10$, with $X_{1}(v,c_2)<X<X_{3/2}(v,c_2)$.
 \item \href{https://www.youtube.com/watch?v=QO1RRSHeEWQ}{Movie 5}: Same parameters as Movie 4 but different initial noise.
 \item \href{https://www.youtube.com/watch?v=2_TyxbXALxQ}{Movie 6}: $v=0.9,c_2=0.75,X=20$, with $X_{5/2}(v,c_2)<X<X_3(v,c_2)$. The thick black line represents the $n=1$ stationary solution, around which the system oscillates at the end of the simulation.
 \item \href{https://www.youtube.com/watch?v=olTkoSW5eR4}{Movie 7}: $v=0.8,c_2=0.4,X=2.2$, with $X_{0}(v,c_2)<X<X_{1/2}(v,c_2)$. We observe that the system emits a soliton to the upstream region and reaches the ground state $n=0$ solution.
 \item \href{https://www.youtube.com/watch?v=CX-A-Rxy7MU}{Movie 8}: Same parameters as Movie 7, but with $X=2.4$. We observe the same behavior as in Movie 7, but emitting a slower soliton.
 \item \href{https://www.youtube.com/watch?v=CNPdeuaL0jk}{Movie 9}: Same parameters as Movie 7, but with $X=2.5$. In this case, the upstream traveling soliton bounces back and the system reaches the CES regime.
 \item \href{https://www.youtube.com/watch?v=7_H6jmMCfaI}{Movie 10}: $v=0.65,c_2=0.3,X=8$, with $X_{1}(v,c_2)<X<X_{3/2}(v,c_2)$. The ground state $n=0$ solution is depicted with a thick black line.
 \item \href{https://www.youtube.com/watch?v=1InNpUDHNVc}{Movie 11}: $v=0.65,c_2=0.1,X=8$, with $X_{1}(v,c_2)<X<X_{3/2}(v,c_2)$. We depict the non-linear $n=1$ stationary solution with a thick black line.
 \item \href{https://www.youtube.com/watch?v=6W3smK7QZcY}{Movie 12}: $v=0.9,c_2=0.2,X=20$, with $X_{5}(v,c_2)<X<X_{11/2}(v,c_2)$. The non-linear $n=5$ is plotted with a thick black line.
\end{itemize}

\chapter{Solution of the Hartree-Fock equations}\label{app:magneticFF}
\chaptermark{Solution of the Hartree-Fock equations}

\section{Basic results of the 2DEG}

As a first step, we define the magnetic form factors
\begin{equation}\label{eq:defmagneticFF}
A_{nn'}\left(\mathbf{k}\right)\equiv\int\mathrm{d}x\, \phi_n\left(x-\frac{k_yl^2_B}{2}\right)\phi_{n'}\left(x+\frac{k_yl^2_B}{2}\right)e^{-ik_xx}
\end{equation}
where the functions $\phi_n(x)$ are the usual harmonic oscillator functions
\begin{equation}\label{eq:oscillatorwavefunctions}
\braket{x|n}=\phi_n(x)=\frac{1}{\sqrt{2^nn!\sqrt{\pi}l_B}}H_n(\frac{x}{l_B})e^{-\frac{x^2}{2l^2_B}},
\end{equation}
corresponding to replace $l$ by $l_B$ in Eq. (\ref{eq:numberstates}). In order to compute the explicit expression of the magnetic form factors, we rewrite Eq. (\ref{eq:defmagneticFF}) as:
\begin{equation}\label{eq:magneticFFelement}
A_{nn'}\left(\mathbf{k}\right)=e^{-i\frac{k_xk_yl^2_B}{2}}\int\mathrm{d}x\, \phi_n\left(x\right)\phi_{n'}\left(x+k_yl^2_B\right)e^{-ik_xx}=e^{-i\frac{k_xk_yl^2B}{2}}\braket{n|e^{-ik_xx}e^{i\frac{P_x}{\hbar}k_yl^2_B}|n'}
\end{equation}
and we remind that $P_x$, the momentum operator in the $x$-direction, is the generator of translations in the $x$ direction. Rewriting the previous expression in terms of the usual harmonic oscillator destruction operator $a$ in the $x$-direction [obtained by replacing $l$ by $l_B$ in Eq. (\ref{eq:1DHO})] yields
\begin{eqnarray}\label{eq:destructors}
\nonumber A_{nn'}\left(\mathbf{k}\right)&=&\braket{n|e^{\frac{(-ik_x-k_y)}{\sqrt{2}}l_Ba^{\dagger}}e^{\frac{(-ik_x+k_y)}{\sqrt{2}}l_Ba}|n'}e^{-\frac{(kl_B)^2}{4}}\\
a&=&\frac{\frac{x}{l_B}+i\frac{p_xl_B}{\hbar}}{\sqrt{2}}
\end{eqnarray}
Then, after some straightforward manipulations, we obtain
\begin{eqnarray}\label{eq:magneticFF}
\nonumber A_{nn'}\left(\mathbf{k}\right)&=&\sqrt{\frac{n'!}{n!}}\left(\frac{(-ik_x-k_y)l_B}{\sqrt{2}}\right)^{n-n'} L^{n-n'}_{n'}\left(\frac{(kl_B)^2}{2}\right)e^{-\frac{(kl_B)^2}{4}},~n\geq n'\\
A_{nn'}\left(\mathbf{k}\right)&=&A^*_{n'n}\left(-\mathbf{k}\right),~n<n' \\
\nonumber L^{m}_n(x)&=&e^x \frac{x^{-m}}{n!} \frac{d^n}{dx^n}x^{n+m} e^{-x}=\sum^n_{j=0}\frac{(-1)^j(n+m)!}{j!(n-j)!(m+j)!}x^j
\end{eqnarray}
with $L^{m}_n(x)$ a generalized Laguerre polynomial. It is worth studying the dependence on $\mathbf{k}$ of the magnetic form factors. If we switch to the following polar coordinates, $(k_x,k_y)=k(\sin \varphi_{\mathbf{k}},\cos \varphi_{\mathbf{k}})$, we find that
\begin{eqnarray}\label{eq:magneticFFspherical}
\nonumber A_{nn'}\left(\mathbf{k}\right)&=&\sqrt{\frac{n'!}{n!}}(-1)^{n-n'}e^{i(n-n')\varphi_{\mathbf{k}}}\left(\frac{kl_B}{\sqrt{2}}\right)^{n-n'} L^{n-n'}_{n'}\left[\frac{(kl_B)^2}{2}\right]e^{-\frac{(kl_B)^2}{4}},~n\geq n'\\
A_{nn'}\left(\mathbf{k}\right)&=&A^*_{n'n}\left(-\mathbf{k}\right)\propto e^{i(n-n')\varphi_{\mathbf{k}}},~n<n'
\end{eqnarray}
and thus, $A_{nn'}\left(\mathbf{k}\right)\propto e^{i(n-n')\varphi_{\mathbf{k}}}$. As in practice we restrict to the ZLL, we give the explicit expression of the magnetic form factors involved in the calculations:
\begin{eqnarray}\label{eq:magneticFFZLL}
\nonumber A_{00}\left(\mathbf{k}\right)&=&e^{-\frac{(kl_B)^2}{4}}\\
A_{10}\left(\mathbf{k}\right)&=&\frac{-ik_x-k_y}{\sqrt{2}}e^{-\frac{(kl_B)^2}{4}},~A_{01}\left(\mathbf{k}\right)=\frac{-ik_x+k_y}{\sqrt{2}}e^{-\frac{(kl_B)^2}{4}}\\
\nonumber A_{11}\left(\mathbf{k}\right)&=&\left[1-\frac{(kl_B)^2}{2}\right]e^{-\frac{(kl_B)^2}{4}}
\end{eqnarray}
Interestingly, in the context of the Wigner function, the magnetic form factors are the Moyal functions of the harmonic oscillator \cite{Schleich2001}.

All the matrix elements and Green's functions of the main text involve the magnetic form factors and integrals of them. For simplicity, we first review the usual case of integer QH states in the 2D electron gas (2DEG), where the field operator only has two components corresponding to the spin polarizations $\sigma=\pm$. The extension to the graphene scenario is discussed in the next subsection. The non-interacting eigenfunctions of the 2DEG are:
\begin{eqnarray}\label{eq:magneticwavefunctions}
\nonumber \phi^0_{n,k,\sigma}(\mathbf{x})&=&\phi_{n,k}(\mathbf{x})\chi_{\sigma}\\
\phi_{n,k}(\mathbf{x})&\equiv&\braket{\mathbf{x}|n,k}=\frac{e^{iky}}{\sqrt{L_y}}\phi_n(x+kl^2_B),~n=0,1,2\ldots
\end{eqnarray}
where $\phi_{n,p}(\mathbf{x})$ are the orbital wave-functions of the magnetic levels and $\chi_{\sigma}$ the spin wave-function with polarization $\sigma$. The non-interacting eigenvalues of these wave functions are
\begin{equation}
\epsilon^0_{n,\sigma}=\left(n+\frac{1}{2}\right)\hbar\omega_B-\sigma\epsilon_Z
\end{equation}
with $\epsilon_Z$ the corresponding Zeeman energy.

We now suppose that the electrons interact through an arbitrary scalar potential $V$ (as it is, for instance, the Coulomb interaction $V_0$). Its spatial matrix elements are given in terms of the magnetic form factors:
\begin{eqnarray}\label{eq:scalarmatrixelements}
V^{n_ln_kn_jn_m}_{p_lp_kp_jp_m}&\equiv& \braket{n_lp_l~n_jp_j|V|n_kp_k~n_mp_m}\\
\nonumber &=&\int\mathrm{d}^2\mathbf{x}~\mathrm{d}^2\mathbf{x'}~\phi^*_{n_l,p_l}(\mathbf{x})\phi_{n_k,p_k}(\mathbf{x})V(\mathbf{x}-\mathbf{x'})\phi^*_{n_j,p_j}(\mathbf{x'})\phi_{n_m,p_m}(\mathbf{x'})\\
\nonumber &=&\frac{1}{S}\sum_{\mathbf{q}}\delta_{p_k-p_l,-q_y}\delta_{p_m-p_j,q_y}e^{-iq_x(p_l-p_j-q_y)l^2_B}V(\mathbf{q})A_{n_ln_k}(-\mathbf{q})A_{n_jn_m}(\mathbf{q})
\end{eqnarray}
The HF equations associated to the previous interacting potential are:
\begin{eqnarray}\label{eq:2DHFeqs}
\nonumber \epsilon_{n,\sigma}\phi_{n,k,\sigma}(\mathbf{x})&=&\frac{\mathbf{\pi}^2}{2m}\phi_{n,k,\sigma}(\mathbf{x})+\sum_{m,p,\sigma'}\nu_{m,\sigma'}\left(\left[\int\mathrm{d}^2\mathbf{x'}~V(\mathbf{x}-\mathbf{x'})\phi_{m,p,\sigma'}^{\dagger}(\mathbf{x'})\phi_{m,p,\sigma'}(\mathbf{x'}) \right]\phi_{n,k,\sigma}(\mathbf{x})\right.\\
\nonumber &-&\left.\int\mathrm{d}^2\mathbf{x'}~V(\mathbf{x}-\mathbf{x'})\phi_{m,p,\sigma'}(\mathbf{x})\phi_{m,p,\sigma'}^{\dagger}(\mathbf{x'})\phi_{n,k,\sigma}(\mathbf{x'})\right)-\epsilon_Z\sigma_z\phi_{n,k,\sigma}(\mathbf{x})\\
\end{eqnarray}
with $\nu_{m,\sigma'}$ the occupation number of each level and the components of the vector $\mathbf{\pi}$ given by Eq. (\ref{eq:peierlssub}). We analyze now the spatial structure of the previous equation. For that purpose, we define two associated mean-field potentials denoted as $V^{H,F}$, that take into account the Hartree (direct) and Fock (exchange) contributions from the potential $V$, respectively. We consider the {\it non-self-consistent} (NSC) problem, where the mean-field potential is created by the bare (non-interacting) wave functions $\phi^0_{n,k,\sigma}$. In that case, the above HF equations are:
\begin{eqnarray}\label{eq:2DNSCHFeqs}
\nonumber \epsilon_{n,\sigma}\phi_{n,k,\sigma}(\mathbf{x})&=&\frac{\mathbf{\pi}^2}{2m}\phi_{n,k,\sigma}(\mathbf{x})+\sum_{m,\sigma'}\nu_{m,\sigma'}\left[\int\mathrm{d}^2\mathbf{x'}~V_m^{H}(\mathbf{x},\mathbf{x'}) -V_m^{F}(\mathbf{x},\mathbf{x'})\chi_{\sigma'}\chi^{\dagger}_{\sigma'}\right]\phi_{n,k,\sigma}(\mathbf{x'})\\
&-&\epsilon_Z\sigma_z\phi_{n,k,\sigma}(\mathbf{x})
\end{eqnarray}
where $V_m^{H,F}$ are the contributions from each magnetic level to the mean-field Hartree and Fock potentials:
\begin{eqnarray}\label{eq:HFpotential}
\nonumber V_m^{H}(\mathbf{x},\mathbf{x'})&=&\int\mathrm{d}^2\mathbf{x''}~V(\mathbf{x}-\mathbf{x''})K_m(\mathbf{x''},\mathbf{x''})\delta(\mathbf{x}-\mathbf{x'})\\
V_m^{F}(\mathbf{x},\mathbf{x'})&=&V(\mathbf{x}-\mathbf{x'})K_m(\mathbf{x},\mathbf{x'})
\end{eqnarray}
The function $K_n(\mathbf{x},\mathbf{x'})\equiv K_{nn}(\mathbf{x},\mathbf{x'})$ is the spatial part of the Green's function, with
\begin{eqnarray}\label{eq:spatialGreenfunction}
K_{nn'}(\mathbf{x},\mathbf{x'})&=&\sum_p \phi_{n,p}(\mathbf{x})\phi^*_{n',p}(\mathbf{x'})
\end{eqnarray}
We can compute explicitly $K_{nn'}(\mathbf{x},\mathbf{x'})$ by transforming the discrete sum over $p$ into an integral and by making the transformation $x_p=X+pl^2_B$, with the center of mass and relative coordinates defined as $\mathbf{R}\equiv(\mathbf{x}+\mathbf{x'})/2$ and $\Delta \mathbf{x}\equiv\mathbf{x}-\mathbf{x'}$. The result is
\begin{equation}\label{eq:LLNonDiagonalGreenfunction}
K_{nn'}(\mathbf{x},\mathbf{x'})=\frac{e^{-i\frac{X\Delta y}{l^2_B}}}{2\pi l^2_B}A_{nn'}\left(-\frac{\Delta y}{l^2_B},-\frac{\Delta x}{l^2_B}\right)
\end{equation}
In particular,
\begin{equation}\label{eq:LLGreenfunction}
K_n(\mathbf{x},\mathbf{x'})=\frac{e^{-i\frac{X\Delta y}{l^2_B}}}{2\pi l^2_B}L_n\left(\frac{\Delta r^2}{2l^2_B}\right)e^{-\frac{\Delta r^2}{4 l^2_B}},
\end{equation}
with $\Delta r=|\Delta \mathbf{x}|$. Since $K_{nn'}(\mathbf{x},\mathbf{x})=\frac{\delta_{nn'}}{2\pi l^2_B}$, $(2\pi l^2_B)^{-1}$ being the homogeneous density corresponding to a completely filled magnetic level, the Hartree contribution is uniform as
\begin{equation}\label{eq:HFpotential}
V_m^{H}(\mathbf{x},\mathbf{x'})=\frac{V(0)}{2\pi l^2_B}\delta(\mathbf{x}-\mathbf{x'})~,
\end{equation}
$V(0)$ being the Fourier transform of the potential evaluated at $\mathbf{k}=0$. In the usual case of the Coulomb interaction, the Hartree potential is canceled by the positive charge background. For a short-range interaction, proportional to $\delta(\mathbf{x}-\mathbf{x'})$, the spatial part of the Hartree potential is equal to the Fock potential.

We now compute the matrix elements of the previous mean-field potentials. The Hartree potential is trivial as explained above so we focus on the Fock potential. For general purposes, we consider the following matrix elements
\begin{eqnarray}\label{eq:generalFockelements}
\braket{n k|V_{mm'}^{F}|n'k'} &\equiv& \int\mathrm{d}^2\mathbf{x}\mathrm{d}^2\mathbf{x'}~ \phi^*_{n,k}(\mathbf{x})V(\mathbf{x}-\mathbf{x'})K_{mm'}(\mathbf{x},\mathbf{x'})\phi_{n',k'}(\mathbf{x'})
\end{eqnarray}
The case of the Fock potential in Eq. (\ref{eq:HFpotential}) is obtained for $m=m'$. To compute the previous integral, we switch to the center of mass and relative coordinates defined before Eq. (\ref{eq:LLGreenfunction}) and integrate along the center of mass coordinates $X,Y$, obtaining:
\begin{equation}\label{eq:Fockgeneralized}
\braket{n k|V_{mm'}^{F}|n' k'}=\frac{\delta_{k,k'}}{2\pi l^2_B}\int\mathrm{d}^2 \mathbf{x}~A_{n'n}\left(\frac{y}{l^2_B},\frac{x}{l^2_B}\right)A_{mm'}\left(-\frac{y}{l^2_B},-\frac{x}{l^2_B} \right)V(\mathbf{x})
\end{equation}
Now, we assume the typical situation where the potential $V(\mathbf{x})$ is rotationally invariant (as the Coulomb potential, for example). The polar dependence described in Eq. (\ref{eq:magneticFFspherical}) gives the result
\begin{eqnarray}\label{eq:FockNonDiagonalmatrixelements}
\braket{n k|V_{mm'}^{F}|n' k'}&=&\delta_{kk'}\delta_{n-n',m-m'}F_{nmm'}
\end{eqnarray}
with $F_{nmm'}$ the {\it on-shell} value of the previous integral. In particular, for the case of the Fock potential, $m=m'$, we get
\begin{eqnarray}\label{eq:Fockmatrixelements}
\braket{n k|V_{m}^{F}|n' k'}&=&\delta_{kk'}\delta_{nn'}F_{nm}\\
\nonumber F_{nm}&=&\frac{1}{l^2_B}\int_0^{\infty}~\mathrm{d}r~ r ~V(r)L_{n}\left(\frac{r^2}{2l^2_B}\right)L_{m}\left(\frac{r^2}{2l^2_B}\right)e^{-\frac{r^2}{4l^2_B}}
\end{eqnarray}
Note that $F_{mn}=F_{nm}$. Amazingly, the Hartree and Fock potentials of Eq. (\ref{eq:HFpotential}) are diagonal in the magnetic base of the eigenfunctions of Eq. (\ref{eq:magneticwavefunctions}), which implies that the self-consistent orbital wave functions are indeed the same as the non-interacting eigenfunctions. Moreover, from Eq. (\ref{eq:2DNSCHFeqs}), it is easy to show that this result also holds for the spin part. Thus, the self-consistent wave functions are equal to the non-interacting ones.

It is useful to reproduce the previous results in Fourier space by writing the matrix elements of the Hartree and Fock potentials in terms of the matrix elements of the potential $V$ in Eq. (\ref{eq:scalarmatrixelements}). For instance, for the Hartree potential we trivially find:
\begin{equation}
\braket{n k|V_{m}^{H}|n' k'}=\sum_p V^{nn'mm}_{kk'pp}=\frac{V(0)}{2\pi l^2_B}\delta_{kk'}\delta_{nn'}
\end{equation}
On the other hand, for the generalized Fock potential of Eq. (\ref{eq:Fockgeneralized}), we obtain
\begin{eqnarray}\label{eq:HFPotentialFourier}
\braket{n k|V_{mm'}^{F}|n' k'}&=&\sum_p V^{nmm'n'}_{kppk'}=\frac{\delta_{kk'}}{(2\pi)^2}\int\mathrm{d}^2\mathbf{q}~V(\mathbf{q}) A^*_{mn}(\mathbf{q})A_{m'n'}(\mathbf{q})
\end{eqnarray}
Again, if we consider that the potential is rotationally invariant, its Fourier transform is also rotationally invariant, hence
\begin{eqnarray}\label{eq:FockFourierNonDiagonalmatrixelement}
\braket{n k|V_{mm'}^{F}|n' k'}&=&\delta_{kk'}\delta_{n-n',m-m'}F_{nmm'}~,
\end{eqnarray}
and for $m=m'$
\begin{eqnarray}\label{eq:FockFouriermatrixelement}
\braket{n k|V_m^{F}|n' k'}&=&\delta_{kk'}\delta_{nn'}F_{nm}\\
\nonumber F_{nm}&=&\frac{1}{(2\pi)^2}\int\mathrm{d}^2\mathbf{q}~|A_{mn}(\mathbf{q})|^2V(q)
\end{eqnarray}
Equations (\ref{eq:FockNonDiagonalmatrixelements}) and (\ref{eq:FockFourierNonDiagonalmatrixelement}) are related to each other through the identity:
\begin{equation}\label{eq:FFB}
\int\frac{\mathrm{d}^2\mathbf{q}}{(2\pi)^2}~e^{i(q_xk_y-q_yk_x)l^2_B}A_{n_jn_k}(-\mathbf{q})A_{n_ln_m}(\mathbf{q})=\frac{1}{2\pi l^2_B}A_{n_ln_k}(-\mathbf{k})A_{n_jn_m}(\mathbf{k})
\end{equation}
which can be proven by inserting the definition of the magnetic form factors, Eq. (\ref{eq:defmagneticFF}).

The resulting self-consistent Green's function of the 2DEG reads:
\begin{eqnarray}
\label{eq:2DGreenfunction}G^{\mu\nu}(\mathbf{x},\mathbf{x'},\omega)&=&\sum^{\infty}_{n=0}\sum_{\sigma}K_n(\mathbf{x},\mathbf{x'})P^{\mu\nu}_{\sigma}G_{n,\sigma}(\omega)\\
G_{n,\sigma}(\omega)&=&\frac{1-\nu_{n,\sigma}}{\omega-\omega_{n,\sigma}+i\eta}+\frac{\nu_{n,\sigma}}{\omega-\omega_{n,\sigma}-i\eta}~,
\end{eqnarray}
with $P_{\sigma}$ the projector corresponding to the spinor $\chi_{\sigma}$, $\hbar\omega_{n,\sigma}=\epsilon_{n,\sigma}$ and $\mu,\nu$ label the components in the spin subspace.

\section{Diagonalization of the Hartree-Fock equations in graphene}\label{app:SCHF}
\sectionmark{Diagonalization}

After the previous training, obtaining the HF solutions in graphene is an (relatively) easy job. We start by considering the non-interacting problem, where we compute the eigenfunctions of the first-quantization version of the Hamiltonian (\ref{eq:spBHamiltonian}):
\begin{equation}\label{eq:sp1QBHamiltonian}
H_0=H(\pi)+\epsilon_VT_{zz}-\epsilon_Z\sigma_z
\end{equation}
Its matrix elements, $\braket{n k\alpha|H_0|n'k'\alpha'}$, with $\braket{\mathbf{x}|n k\alpha}=\Psi^{0}_{n,k,\alpha}(\mathbf{x})$ the wave functions of Eqs. (\ref{eq:Landaueigenfunctions})-(\ref{eq:ZLL}), are given by:
\begin{eqnarray}\label{eq:spmnmn}
\nonumber\braket{n k\alpha|H_0|n'k'\alpha'}&=& \epsilon_n\delta_{nn'}\delta_{kk'}\delta_{\alpha\alpha'}-\epsilon_V\delta_{n,-n'}\delta_{kk'}(T_z)_{\alpha\alpha'}-\epsilon_Z\delta_{nn'}\delta_{kk'}(\sigma_z)_{\alpha\alpha'},~|n|\neq0,1\\
\braket{n k\alpha|H_0|n'k'\alpha'}&=&-\epsilon_V\delta_{nn'}\delta_{kk'}(T_z)_{\alpha\alpha'}-\epsilon_Z\delta_{nn'}\delta_{kk'}(\sigma_z)_{\alpha\alpha'},~n=0,1
\end{eqnarray}
with $(T_z)_{\alpha\alpha'}=\chi^{\dagger}_{\alpha}T_z\chi_{\alpha'}$. We see that the previous Hamiltonian is diagonal in the ZLL while it mixes the LLs $\pm n$ due to the voltage layer. The reason is that, outside the ZLL, the wave functions are not localized on one specific sublattice and thus they are able to experiment the effect of the voltage.
However, as long as $\epsilon_V\ll \hbar\omega_B$, it is a good approximation to consider that $H_0$ is diagonal within every LL so their corresponding eigenfunctions are still given by $\Psi^{0}_{n,p,\alpha}(\mathbf{x})$. The associated eigenvalues $H\ket{nk\alpha}=\hbar\omega^{0}_{n,\alpha}\ket{nk\alpha}$ are then
\begin{eqnarray}\label{eq:spmnmneigenvalues}
\hbar\omega^{0}_{n,\alpha}&\simeq&\epsilon_n-\epsilon_Z\sigma,~|n|\neq 0,1\\
\nonumber  \hbar\omega^{0}_{n,\alpha}&=&-\epsilon_Z\sigma-\epsilon_V n_z,~|n|= 0,1
\end{eqnarray}
where $n_z,\sigma$ label the valley and spin polarizations, $n_z=+1$ corresponding to $K$ valley and $n_z=-1$ to $K'$ valley. The non-interacting spinors are given by all possible orthogonal combinations of valley-spin polarizations, $\chi_{\alpha}^0=\ket{\pm n_z}\otimes\ket{\pm s_z}$; see Eq. (\ref{eq:Fphase}) for the notation of the wave-functions in valley-spin space.

The next step is to compute the non-interacting Green's function, which is obtained from the field operator (\ref{eq:fieldoperatorLL}),
\begin{eqnarray}\label{eq:GreenBarefunction}
G_0^{\mu\nu}(\mathbf{x},\mathbf{x'},\omega)&=&\sideset{}{'}\sum_{n,p,\alpha}\Psi^{0\mu}_{n,p,\alpha}(\mathbf{x})\Psi^{0\nu*}_{n,p,\alpha}(\mathbf{x'})G^{(0)}_{n,\alpha}(\omega)\\
\nonumber G^{(0)}_{n,\alpha}(\omega)&=&\frac{1-\nu^{0}_{n,\alpha}}{\omega-\omega^{0}_{n,\alpha}+i\eta}+\frac{\nu^{0}_{n,\alpha}}{\omega-\omega^{0}_{n,\alpha}-i\eta}
\end{eqnarray}
where the indices $\mu\nu$ now label the components of the wave functions in the total subspace $KK'\otimes\bar{A}\bar{B} \otimes s$ and $\nu^{0}_{n,\alpha}$ is the occupation factor. For the $\nu=0$ QH state, $\nu^{0}_{n,\alpha}=1$ for $n\leq-2$ and $\nu^{0}_{n,\alpha}=0$ for $n\geq 2$. In the ZLL, the energy only depends on the polarization in valley-spin space so the non-interacting occupation factors are $\nu^{0}_{n,\alpha}=1$ for the two non-interacting lowest-energy polarizations and $\nu^{0}_{n,\alpha}=0$ for the remaining ones. Following the notation of the main text, we label the occupied non-interacting spinors as $\chi^0_{a,b}$ and the empty ones as $\chi^0_{c,d}$. In particular, for $\epsilon_V<\epsilon_Z$, the ZLL is filled in the same F configuration of Eq. (\ref{eq:Fphase}) and for $\epsilon_V>\epsilon_Z$ it is filled in the FLP configuration of Eq. (\ref{eq:FLPphase}). Thus, in the non-interacting problem, these are the only two possible phases for the $\nu=0$ QH state.

Following the tricks developed in the previous section, we rewrite the Green's function as:
\begin{eqnarray}\label{eq:GreenBarefunction}
G_0^{\mu\nu}(\mathbf{x},\mathbf{x'},\omega)&=&\sideset{}{'}\sum_{n,\alpha}K^{\mu\nu}_{n,\alpha}(\mathbf{x},\mathbf{x'})G^{(0)}_{n,\alpha}(\omega)
\end{eqnarray}
with $K^{\mu\nu}_{n,\alpha}\mathbf{x},\mathbf{x'})$ the elements of the matrix:
\begin{eqnarray}\label{eq:GreenBarefunctionSpatial}
K_{n,\alpha}(\mathbf{x},\mathbf{x'})&\equiv&K^{(2)}_n(\mathbf{x},\mathbf{x'})\otimes P_{\alpha}\\
\nonumber K^{(2)}_n(\mathbf{x},\mathbf{x'})&=&\frac{1}{2}\left[\begin{array}{cc}
K_{|n|-2,|n|-2}(\mathbf{x},\mathbf{x'})& \textrm{sgn}\ n\ K_{|n|-2,|n|}(\mathbf{x},\mathbf{x'})\\
\textrm{sgn}\ n\  K_{|n|,|n|-2}(\mathbf{x},\mathbf{x'}) & K_{|n|,|n|}(\mathbf{x},\mathbf{x'})
\end{array}\right],~|n|\neq 0,1\\
\nonumber K^{(2)}_n(\mathbf{x},\mathbf{x'})&=&\left[\begin{array}{cc}
0 & 0\\
0 & K_{n,n}(\mathbf{x},\mathbf{x'})
\end{array}\right],~n=0,1
\end{eqnarray}
The matrix $K^{(2)}_n(\mathbf{x},\mathbf{x'})$ belongs to the subspace $\bar{A}\bar{B}$ while the matrix $P_{\alpha}$ is the projector onto the non-interacting valley-spin wave function $\chi^0_{\alpha}$, $P_{\alpha}=\chi^0_{\alpha}\chi^{0\dagger}_{\alpha}$. For $\mathbf{x}=\mathbf{x'}$,  $K^{(2)}_n(\mathbf{x},\mathbf{x'})=(4\pi l^2_B)^{-1}\text{diag}[1,1]$ for $|n|\neq 0,1$ and $K^{(2)}_n(\mathbf{x},\mathbf{x'})=(2\pi l^2_B)^{-1}P_{\bar{B}}$ for $n=0,1$, $P_{\bar{B}}=\text{diag}[0,1]$ being the projector onto the subspace $\bar{B}$.

\subsection{Non-self consistent problem}

The matrix $K^{(2)}_n(\mathbf{x},\mathbf{x'})$ is key to understand the orbital structure of the HF equations. With that finality, we compute as in the 2DEG case the {\it non-self-consistent} mean-field Hamiltonian. We write
\begin{equation}\label{eq:meanfieldDSHamiltonian}
H_{NSCHF}=H_0+H_{NSCDS}+H_{NSCZLL}
\end{equation}
with $H_{NSCDS}$ the mean-field potential created by the Dirac sea, formed by all the states with $n\leq -2$,
\begin{eqnarray}\label{eq:DiracZLLHFHamiltonian}
H_{NSCDS}&=&-\sum^{\infty}_{m=2}\left[(V_0)^{(2)}_{-m}(\mathbf{x},\mathbf{x'})+\bar{u}\right]\\
\nonumber (V_0)^{(2)}_{m}(\mathbf{x},\mathbf{x'})&\equiv&V_0(\mathbf{x}-\mathbf{x'})K^{(2)}_m(\mathbf{x},\mathbf{x'}),~~\bar{u}=\sideset{}{'}\sum_{i,j}g_{ij}\hbar\omega_B
\end{eqnarray}
The NSC mean-field potential created by the ZLL reads:
\begin{equation}\label{eq:NSCZLLHFHamiltonian}
H_{NSCZLL}=-\sum_{m=0,1}(V_0)^{F}_{m}(\mathbf{x},\mathbf{x'})P^0+\sum_{i}u_{i}\left([\text{tr}(P^0T_{i})]T_{i}-T_{i}P^0T_{i}\right)P_{\bar{B}}
-v_{i}T_{i}P^0T_{i}P_{\bar{A}}
\end{equation}
with $P^0=\chi^0_{a}\chi^{0\dagger}_{a}+\chi^0_{b}\chi^{0\dagger}_{b}$ the non-interacting analog of the matrix $P$ [defined after Eq. (\ref{eq:HFenergy})] for the non-interacting spinors and $P_{\bar{A}}$ the projector onto the subspace $\bar{A}$, $P_{\bar{A}}=\text{diag}[1,0]$. The coupling energy $v_{i}$ is $v_{i}=4\hbar\omega_B(g_{ix}
+g_{iy})$. As usual, the positive charge background cancels the Hartree term of the Coulomb interaction. We recall that $(V_0)^{F}_{m}$ is the Fock potential associated to the Coulomb interaction for a 2DEG, as given by Eq. (\ref{eq:HFpotential}). Since all the valley-spin polarizations are occupied in the Dirac sea, the Hartree term of the short-range interactions vanishes and the corresponding Fock term is just an scalar. The matrix elements of the Fock contribution of the Coulomb potential are
\begin{eqnarray}\label{eq:GraphenegeneralFockelements}
\braket{n k|(V_0)^{(2)}_{m}|n'k'} &\equiv& \int\mathrm{d}^2\mathbf{x}\mathrm{d}^2\mathbf{x'}~ \Psi^{0\dagger}_{n,k}(\mathbf{x})(V_0)^{(2)}_{m}(\mathbf{x},\mathbf{x'})\Psi^0_{n',k'}(\mathbf{x'})\\
\nonumber &=&\left[C_{nm}\delta_{n,n'}+D_{nm}\delta_{n,-n'}\right]\delta_{kk'},~|n|\neq 0,1\\
\nonumber \braket{n k|(V_0)^{(2)}_{m}|n'k'}&=&\frac{1+\delta_{m,0}+\delta_{m,1}}{2}\delta_{nn'}\delta_{kk'},~n=0,1~,
\end{eqnarray}
with $C_{nm},D_{nm}$ some coefficients that can be expressed in terms of the values of the generalized matrix elements of Eq. (\ref{eq:FockNonDiagonalmatrixelements}); their particular expression is not interesting for the current purposes. Thus, the Fock term of the Coulomb potential mixes $n$ with $-n$ outside the ZLL and it is diagonal within the ZLL, in the same fashion of the non-interacting Hamiltonian. It is easy to show that this behavior also persists in the matrix elements of the short-range terms; therefore, we conclude that the NSC Hamiltonian $H_{NSCHF}$ only couples the LLs $\pm n$ and leaves invariant the ZLL. Moreover, even when the mean-field potentials are self-consistently renormalized, this structure is preserved. Thus, the orbital part of the self-consistent wave functions $\Psi_{n,k,\alpha}$ is a linear combination of the non-interacting orbital wave functions $\Psi^0_{\pm n,k}$; in particular, the orbital wave functions in the ZLL are equal to the non-interacting ones. In the following, we neglect the coupling between LLs as we are supposing that interactions are weak enough to neglect LL mixing. A very nice discussion about the coupling of the $\pm n$ LLs for strong Coulomb interactions is presented in Ref. \cite{Toke2013}.

\subsection{Self consistent problem}

Although, as explained above, we neglect LL mixing in the HF equations and then the orbital part of the {\it self-consistent} wave functions is the same as that of the non-interacting ones, we still have to determine the spin-valley structure of the solutions. For that purpose, we switch to the actual {\it self-consistent} HF problem, where the mean-field potential is created by the self-consistent HF solutions,
$H_{HF}=H_0+H_{DS}+H_{ZLL}$. As there is no LL mixing, we can regard the Dirac sea as frozen and hence the self-consistent HF mean-field potential created by the Dirac sea, $H_{DS}$, is the same as the NSC one of Eq. (\ref{eq:DiracZLLHFHamiltonian}), $H_{DS}=H_{NSCDS}$. In respect to the mean-field potential created by the ZLL, Eq. (\ref{eq:NSCZLLHFHamiltonian}), we only have to replace the NSC projector $P^0$ by the self-consistent one, $P$, as the filling of the ZLL is of the same form (see Sec. \ref{subsec:HF}). Since we are focusing on the ZLL, we project $H_{HF}$ onto the sublattice $\bar{B}$, obtaining:
\begin{eqnarray}\label{eq:HFprojected}
H^{(0)}_{HF}&=&P_{\bar{B}}H_{HF}P_{\bar{B}}=H^{(0)}_{0}+H^{(0)}_{DS}+H^{(0)}_{ZLL}\\
\nonumber H^{(0)}_{0}&=&-\epsilon_VT_z-\epsilon_Z\sigma_z\\
\nonumber H^{(0)}_{DS}&\equiv&V_{DS}(\mathbf{x},\mathbf{x'})=P_{\bar{B}}H_{DS}P_{\bar{B}}=-\sum^{\infty}_{m=2}~\left[\frac{1}{2}(V_0)^{F}_{m}(\mathbf{x},\mathbf{x'})+\bar{u}\right]\\
\nonumber H^{(0)}_{ZLL}&=&-\sum_{m=0,1}(V_0)^{F}_{m}(\mathbf{x},\mathbf{x'})P+\sum_{i}u_{i}\left([\text{tr}(PT_{i})]T_{i}-T_{i}PT_{i}\right)
\end{eqnarray}
The term $\bar{u}$ in $V_{DS}$ provides an infinite constant shift that we can reabsorb in the Hamiltonian. The resulting self-consistent HF equations for the ZLL, $H^{(0)}_{HF}\Psi_{n,k,\alpha}=\epsilon_{n,\alpha}\Psi_{n,k,\alpha}$, are equivalent to those of Eq. (\ref{eq:HFeqs}) and read
\begin{eqnarray}\label{eq:HFanalytical}
\nonumber\epsilon_{n,\alpha}\Psi_{n,k,\alpha}(\mathbf{x})&=&-\frac{1}{2}\sum^{\infty}_{m=2}\int\mathrm{d}^2\mathbf{x'}(V_0)^{F}_{m}(\mathbf{x},\mathbf{x'})\Psi_{n,k,\alpha}(\mathbf{x'})
-\sum_{m=0,1}\int\mathrm{d}^2\mathbf{x'}(V_0)^{F}_{m}(\mathbf{x},\mathbf{x'})P\Psi_{n,k,\alpha}(\mathbf{x'})\\
\nonumber &+&\sum_{i}u_{i}\left([\text{tr}(PT_{i})]T_{i}-T_{i}PT_{i}\right)\Psi_{n,k,\alpha}(\mathbf{x})
-\epsilon_VT_z\Psi_{n,k,\alpha}(\mathbf{x})-\epsilon_Z\sigma_z\Psi_{n,k,\alpha}(\mathbf{x})\\
\end{eqnarray}
Since the mean-field HF potentials are invariant under unitary transformations of the filled states, the self-consistent problem is still diagonal in the orbital part as the two magnetic levels $n=0,1$ are filled exactly in the same way. Then, the self-consistent wave functions have the form $\Psi_{n,k,\alpha}(\mathbf{x})=\Psi^0_{n,k}(\mathbf{x})\chi_{\alpha}$. In order to arrive at an equation for the spinor $\chi_{\alpha}$, we multiply by $\Psi^{0,\dagger}_{n,k}(\mathbf{x})$ and integrate in Eq. (\ref{eq:HFanalytical}), obtaining:
\begin{equation}\label{eq:HFenergyDirac}
\epsilon_{n,\alpha}\chi_{\alpha}=-\frac{1}{2}\sum_{m=2}^{\infty}F_{nm}\chi_{\alpha}-(F_{n0}+F_{n1})P\chi_{\alpha}
+\sum_{i}u_{i}\left([\text{tr}(PT_{i})]T_{i}-T_{i}PT_{i}\right)\chi_{\alpha}-\epsilon_VT_z\chi_{\alpha}-\epsilon_Z\sigma_z\chi_{\alpha}
\end{equation}
with $F_{nm}$ the eigenvalues of the Fock potential associated to the Coulomb interaction, Eq. (\ref{eq:Fockmatrixelements}). The contribution from the mean-field interaction with the Dirac sea, which is the origin of the analog of the Lamb shift \cite{Shizuya2012}, corresponds to the first term at the l.h.s. of the above equation. We rearrange the associated series as:
\begin{equation}\label{eq:Diraccontribution}
\sum_{m=2}^{\infty}F_{nm}=\sum_{m=0}^{\infty}F_{nm}-F_{n0}-F_{n1}
\end{equation}
The completeness relation
\begin{equation}\label{eq:completenessrelation}
\sum_{m=0}^{\infty}|A_{nm}(\mathbf{k})|^2=1~,
\end{equation}
can be proven from the definition of the magnetic form factors Eq. (\ref{eq:defmagneticFF}) and the completeness relation of the oscillator wave functions. Using the relation (\ref{eq:completenessrelation}) and Eq. (\ref{eq:FockFouriermatrixelement}), we find that
\begin{equation}
\sum_{m=0}^{\infty}F_{nm}=\frac{1}{(2\pi)^2}\int\mathrm{d}^2\mathbf{q}~V(\mathbf{q})=V(\mathbf{x}=0)~,
\end{equation}
which represents a constant energy shift that can be absorbed in the Hamiltonian, as that arising from the short-range interactions [see discussion below Eq. (\ref{eq:HFprojected})]. Then, by defining
\begin{equation}\label{eq:HFCoulombEnergies}
F_{n}\equiv F_{n0}+F_{n1}
\end{equation}
we finally get Eq. (\ref{eq:HFenergy}). The values for the ZLL of the eigenvalues of the Fock potential associated to the Coulomb interaction, $F_{nm}$, are computed from Eqs. (\ref{eq:magneticFFZLL}), (\ref{eq:FockFouriermatrixelement}):
\begin{eqnarray}\label{eq:Fockeigenvalues}
\nonumber F_{00}&=&\sqrt{\frac{\pi}{2}}\frac{e^2_c}{\kappa l_B}\\
F_{01}&=&F_{10}=\frac{1}{2}F_{00}\\
\nonumber F_{11}&=&\frac{3}{4}F_{00}
\end{eqnarray}

All the above calculations can be trivially translated to the case of monolayer graphene.

\chapter{Analytical expression of the dispersion matrix}\label{app:analyticalTDHFA}

In this section, we compute analytically the elements of the dispersion matrix, $X_{kl,jm}(\mathbf{k})$, given by Eq. (\ref{eq:dispersionmatrix}). In particular, we focus on the non-trivial part arising from interactions, $W_{lk,jm}(\mathbf{k})$. For that purpose, we first give the expression of the matrix elements $V_{lk,jm}$, defined in Eq. (\ref{eq:potentialmatrixelements}). As we are restricting to the ZLL, where the orbital part is fully contained in the $\bar{B}$ component and is proportional to the magnetic wave function, the matrix elements can be straightforwardly calculated through Eq. (\ref{eq:scalarmatrixelements}) as
\begin{equation}\label{eq:matrixelementsexplicit}
V_{lk,jm}=(V_0)^{n_ln_kn_jn_m}_{p_lp_kp_jp_m}\delta_{\alpha_l\alpha_k}\delta_{\alpha_j\alpha_m}
+(V_{sr})^{n_ln_kn_jn_m}_{p_lp_kp_jp_m}\sum_i g_i  (T_i)_{\alpha_l\alpha_k}(T_i)_{\alpha_j\alpha_m},~V_{sr}(\mathbf{x})\equiv\frac{4\pi\hbar^2\delta(\mathbf{x})}{m}
\end{equation}
We remind that $V_0$ is the Coulomb interaction of Eq. (\ref{eq:coulombHamiltonian}) and $(T_i)_{\alpha_l\alpha_k}$ is defined after Eq. (\ref{eq:ladderRPA}). The matrix $W_{lk,jm}(\mathbf{k})$ is obtained by performing the summation of Eq. (\ref{eq:calculationladderRPA}) and the result is given in Eq. (\ref{eq:ladderRPA}). In order to give its explicit expression, we separate the Coulomb from the short-ranged terms
\begin{equation}\label{eq:interactingcontributions}
W_{lk,jm}(\mathbf{k})=W^{C}_{lk,jm}(\mathbf{k})+W^{sr}_{lk,jm}(\mathbf{k})
\end{equation}
The Coulomb term, $W^{C}_{lk,jm}(\mathbf{k})$, has not direct (RPA) contribution as it vanishes when considering the proper elements of the dispersion matrix because, within the ZLL, the empty levels have different valley-spin polarization from that of the filled ones, see discussion at the beginning of Sec. \ref{subsec:preliminar}. Thus, we only have to consider the exchange (ladder) contribution, which yields:
\begin{eqnarray}\label{eq:Coulombmatrices}
\nonumber W^{C}_{lk,jm}(\mathbf{k})&=&-U^{C}_{n_jn_k,n_ln_m}(\mathbf{k})\delta_{\alpha_j\alpha_k}\delta_{\alpha_l\alpha_m}\\
U^{C}_{n_jn_k,n_ln_m}(\mathbf{k})&=&\int\frac{\mathrm{d}^2\mathbf{q}}{(2\pi)^2}~e^{i(q_xk_y-q_yk_x)l^2_B}A_{n_jn_k}(-\mathbf{q})A_{n_ln_m}(\mathbf{q})V_0(\mathbf{q})
\end{eqnarray}
These energies are given in terms of modified Bessel functions. For instance, we compute the element $U^{C}_{00,00}(\mathbf{k})$. After changing to polar coordinates, taking into account that only the real part of the complex exponential survives and performing the integral over the radial coordinate, we find:
\begin{equation}\label{eq:Coulomb0000}
U^{C}_{00,00}(\mathbf{k})=F_{00}\frac{1}{2\pi}\int_0^{2\pi}\mathrm{d}\varphi ~e^{-\frac{(kl_B)^2\cos^2\varphi}{4}}=F_{00}I_0\left[\frac{(kl_B)^2}{4}\right]e^{-\frac{(kl_B)^2}{4}}
\end{equation}
with $F_{00}$ given in Eq. (\ref{eq:Fockeigenvalues}) and the modified Bessel functions $I_n(x)$ defined as
\begin{equation}\label{eq:BesselModifiedintegral}
I_n(x)\equiv(-i)^nJ_n(ix)=\frac{1}{2\pi}\int_0^{2\pi}\mathrm{d}\varphi ~e^{x\cos\varphi}e^{-in\varphi}=\sum_{k=0}^\infty\frac{1}{(n+k)!k!}\left(\frac{x}{2}\right)^{n+2k}~,
\end{equation}
$J_n(x)$ being the usual Bessel functions, defined in Eq. (\ref{eq:BesselDef}). The other elements $U^{C}_{n_jn_k,n_ln_m}(\mathbf{k})$ can also be computed analytically. Instead of using the complicated expression found in the tables, we simply take into account that $A_{n_ln_m}(\mathbf{q})$ is a polynomial in $q_x,q_y$ multiplied by $e^{-\frac{(ql_B)^2}{4}}$. Then,
\begin{equation}
U^{C}_{n_jn_k,n_ln_m}(\mathbf{k})=\int\frac{\mathrm{d}^2\mathbf{q}}{(2\pi)^2}~P_{n_jn_k,n_ln_m}(q_x,q_y)e^{i(q_xk_y-q_yk_x)l^2_B}V_0(\mathbf{q})e^{-\frac{(ql_B)^2}{2}}
\end{equation}
with $P_{n_jn_k,n_ln_m}(q_x,q_y)$ some polynomial. For $U^{C}_{00,00}(\mathbf{k})$, $P_{00,00}(q_x,q_y)=1$. Therefore, from the usual properties of Fourier transforms, one has that
\begin{eqnarray}\label{eq:Coulombmatricesanalytical}
U^{C}_{n_jn_k,n_ln_m}(\mathbf{k})=P_{n_jn_k,n_ln_m}\left(-i\frac{\partial}{\partial k_y},i\frac{\partial}{\partial k_x}\right)U^{C}_{00,00}(\mathbf{k})
\end{eqnarray}
Since the Bessel functions are solutions of a second order differential equation, their higher order derivatives can always be put in terms of themselves and their first derivatives. For $I_0(x)$, $\frac{dI_0}{dx}=I_1(x)$, which means that the elements $U^{C}_{n_jn_k,n_ln_m}(\mathbf{k})$ are expressed through combinations of $I_0,I_1$ and elemental functions. For instance, from Eq. (\ref{eq:magneticFFZLL}), we obtain
\begin{equation}
U^{C}_{10,00}=\frac{\frac{\partial}{\partial k_y}+i\frac{\partial}{\partial k_x}}{\sqrt{2}}U^{C}_{00,00}(\mathbf{k})=F_{00}\frac{k_y+ik_x}{2\sqrt{2}}\left(I_1\left[\frac{(kl_B)^2}{4}\right]-I_0\left[\frac{(kl_B)^2}{4}\right]\right)e^{-\frac{(kl_B)^2}{4}}
\end{equation}
and so on. For large $x$, a saddle-point approximation of the integral in Eq. (\ref{eq:BesselModifiedintegral}) gives $I_0(x)\sim\frac{e^x}{\sqrt {x}}$. Then, for large wave vector $k$, $kl_B\gg 1$, $U^{C}_{00,00}(\mathbf{k})\sim (kl_B)^{-1}$. As the other elements are obtained from derivatives of $U^{C}_{00,00}(\mathbf{k})$, the total matrix $U^C$ decays at least as $(kl_B)^{-1}$ for large $k$.

We now turn back our attention to the short-range interactions. Using the property (\ref{eq:FFB}), we find that the integrals in momenta of the RPA and ladder contributions are equal and then:
\begin{eqnarray}\label{eq:SRmatrices}
W^{sr}_{lk,jm}(\mathbf{k})=\frac{1}{2}A_{n_ln_k}(-\mathbf{k})A_{n_jn_m}(\mathbf{k})M^{\alpha_k\alpha_l,\alpha_j\alpha_m}
\end{eqnarray}
with $M^{\alpha_k\alpha_l,\alpha_j\alpha_m}$ defined in Eq. (\ref{eq:pseudospinmatrices}).

In order to simplify the notation, we define a matrix $C(\mathbf{k})$ in the magnetic subspace with the same index ordering of the main text [see Eq. (\ref{eq:Ymagneticstructure})]:
\begin{equation}\label{eq:Coulombmagneticmatrix}
C(\mathbf{k})=\left[\begin{array}{cccc}
C_{00,00}&C_{00,01}&C_{00,10}&C_{00,11}\\
C_{01,00}&C_{01,01}&C_{01,10}&C_{01,11}\\
C_{10,00}&C_{10,01}&C_{10,10}&C_{10,11}\\
C_{11,00}&C_{11,01}&C_{11,10}&C_{11,11}\\
\end{array}\right]
\end{equation}
and its elements containing the exchange Coulomb energies
\begin{equation}\label{eq:Coulombmagneticmatrices}
C_{n_kn_l,n_jn_m}(\mathbf{k})\equiv U^{C}_{n_jn_k,n_ln_m}(\mathbf{k})
\end{equation}
We may define an analog matrix $R(\mathbf{k})$ for the short-range interactions
\begin{equation}\label{eq:SRmagneticmatrices}
R_{n_kn_l,n_jn_m}(\mathbf{k})\equiv \frac{1}{2}A_{n_ln_k}(-\mathbf{k})A_{n_jn_m}(\mathbf{k})
\end{equation}
With the help of these matrices, we can rewrite the expression of $W_{lk,jm}(\mathbf{k})$ in the compact form:
\begin{eqnarray}\label{eq:CompactformPot}
W_{lk,jm}(\mathbf{k})=-C(\mathbf{k})\delta_{\alpha_j\alpha_k}\delta_{\alpha_l\alpha_m}+R(\mathbf{k})M^{\alpha_k\alpha_l,\alpha_j\alpha_m}
\end{eqnarray}

Finally, from these results, we show that the dispersion relation is isotropic. The key point is the polar structure of the magnetic form factors, displayed in Eq. (\ref{eq:magneticFFspherical}). As $(k_x,k_y)=k(\sin \varphi_\mathbf{k},\cos \varphi_\mathbf{k})$, $\varphi_{-\mathbf{k}}=\varphi_\mathbf{k}+\pi$ and then
\begin{equation}\label{eq:Rsphericalstructure}
R_{n_kn_l,n_jn_m}(\mathbf{k})=e^{i(n_l-n_k)\varphi_\mathbf{k}}e^{i(n_j-n_m)\varphi_\mathbf{k}}R_{n_kn_l,n_jn_m}(k)
\end{equation}
with $R_{n_kn_l,n_jn_m}(k)$ a function that only depends on $k=|\mathbf{k}|$. For the Coulomb interaction, we switch to the following polar coordinates in the integral of Eq. (\ref{eq:Coulombmatrices}), $\mathbf{q}=q(\cos \varphi_\mathbf{q},\sin \varphi_\mathbf{q})$. This gives
\begin{equation}\label{eq:Coulombmatrices}
U^{C}_{n_jn_k,n_ln_m}(\mathbf{k})=\int_0^{\infty}\mathrm{d}q~\left(\frac{1}{2\pi}\int_0^{2 \pi}\mathrm{d}\varphi_\mathbf{q}~e^{iqkl^2_B\cos(\varphi_\mathbf{q}-\varphi_\mathbf{k})}e^{-i(n_k+n_m-n_j-n_l)\varphi_\mathbf{q}}\right)w_{n_jn_k,n_ln_m}(q)
\end{equation}
with $w_{n_jn_k,n_ln_m}(q)$ some function that only depends on $q$. The polar integral gives
\begin{eqnarray}\label{eq:Coulombmatrices}
&~&\frac{1}{2\pi}\int_0^{2 \pi}\mathrm{d}\varphi_\mathbf{q}~e^{iqkl^2_B\cos(\varphi_\mathbf{q}-\varphi_\mathbf{k})}e^{-i(n_k+n_m-n_j-n_l)\varphi_\mathbf{q}}=\\
\nonumber &~&=i^{(n_k+n_m-n_j-n_l)}e^{-i(n_k+n_m-n_j-n_l)\varphi_\mathbf{k}}J_{n_k+n_m-n_j-n_l}(qkl^2_B)
\end{eqnarray}
where we have used the integral representation of the Bessel functions given in Eq. (\ref{eq:Besselintegral}). This implies that the matrix $C(\mathbf{k})$ has the same dependence on the polar angle of Eq. (\ref{eq:Rsphericalstructure}) and so the total matrix $W_{lk,jm}$. In particular, for the diagonal elements $(n_kn_l)=(n_jn_m)$ the phase factor $e^{i(n_l-n_k)\varphi_\mathbf{k}}e^{i(n_j-n_m)\varphi_\mathbf{k}}$ is the unity. Therefore, we infer from Eq. (\ref{eq:dispersionmatrix}) that we can write $X_{kl,jm}(\mathbf{k},\omega)$ as
\begin{equation}
X_{kl,jm}(\mathbf{k},\omega)=e^{-i(n_k-n_l)\varphi_\mathbf{q}}X_{kl,jm}(k,\omega)e^{i(n_j-n_m)\varphi_\mathbf{q}}
\end{equation}
with $X_{kl,jm}(k,\omega)$ depending only on $k=|\mathbf{k}|$. Since the collective modes are given by the condition (\ref{eq:detcollmod}), a unitary transformation does not vary the dispersion relation. Thus, by making an appropriated phase transformation in the magnetic indices, we get rid of the dependence on the polar angle of the momentum and obtain the matrix $X_{kl,jm}(k,\omega)$, which explicitly depends solely on $k=|\mathbf{k}|$. Hence, the resulting dispersion relation is isotropic, $\omega(\mathbf{k})=\omega(|\mathbf{k}|)$.

\bibliography{PhDJR}
\bibliographystyle{ieeetr}

\end{document}